%% file: From_Qubit_to_Qubit.tex
\documentclass[11pt]{book}

\usepackage[utf8]{inputenc}
\usepackage[T1]{fontenc}
\usepackage[margin=1in]{geometry}
\usepackage{amsmath, amssymb, amsfonts, amsthm}
\usepackage{braket}
\usepackage{physics}
\usepackage{array}
\renewcommand{\ket}[1]{|#1\rangle}
\renewcommand{\bra}[1]{\langle#1|}
\renewcommand{\braket}[2]{\langle#1|#2\rangle}
\usepackage{graphicx}
\usepackage{enumitem}
\usepackage{bbm}
\setlist[enumerate,1]{label=(\alph*)}
\setlist[enumerate,2]{label=(\roman*)}
\usepackage{fancyhdr}
\usepackage{titlesec}
\usepackage[most]{tcolorbox}
\usepackage{xcolor}
\usepackage{float}
\usepackage{caption}
\usepackage{subcaption}
\usepackage{mathtools}
\usepackage{comment}
\usepackage{multirow}
\usepackage{booktabs}
\usepackage{appendix}
\usepackage{makeidx}
\usepackage{tikz}
\usepackage{pgfplots}
\pgfplotsset{compat=1.18}
\usetikzlibrary{arrows.meta,decorations.markings,decorations.pathreplacing}

\usepackage{xurl}
\usepackage{hyperref}
\hypersetup{
    colorlinks=true,
    linkcolor=blue!60!black,
    filecolor=magenta,
    urlcolor=blue!60!black,
    citecolor=green!50!black,
    pdfborder={0 0 0},
    bookmarksopen=true
}

\definecolor{accentcolor}{RGB}{47,79,79}
\definecolor{lightgray}{RGB}{245,245,245}
\definecolor{darkgray}{RGB}{100,100,100}
\definecolor{problemcolor}{RGB}{139,69,19}
\definecolor{examplecolor}{RGB}{34,139,34}
\definecolor{notecolor}{RGB}{178,34,34}

\theoremstyle{definition}
\newtheorem{definition}{Definition}[section]

\tcbuselibrary{breakable, theorems}

\newcounter{example}[chapter]
\renewcommand{\theexample}{\thechapter.\arabic{example}}

\newcounter{hwproblem}[chapter]
\renewcommand{\thehwproblem}{\thechapter.\arabic{hwproblem}}
\newcommand{\problem}[1]{%
  \refstepcounter{hwproblem}%
  \vspace{1em}%
  \noindent\textbf{Problem \thehwproblem} \textit{(#1)}\\%
}

\newtcolorbox{keyidea}[1]{
    enhanced, breakable,
    colback=blue!5, colframe=blue!60!black,
    boxrule=1pt, arc=3pt,
    left=10pt, right=10pt, top=8pt, bottom=8pt,
    title={\textbf{Key Idea: #1}},
    fonttitle=\color{white}, coltitle=white, colbacktitle=blue!60!black
}

\newtcolorbox[use counter=example]{example}[1][]{
    enhanced, breakable,
    colback=examplecolor!8, colframe=examplecolor,
    boxrule=1.5pt, arc=3pt,
    left=10pt, right=10pt, top=10pt, bottom=10pt,
    before skip=12pt, after skip=12pt,
    title={\textbf{Example \theexample\ifx\relax#1\relax\else: #1\fi}},
    fonttitle=\color{white}, coltitle=white, colbacktitle=examplecolor,
    attach boxed title to top left={yshift=-2mm, xshift=3mm},
    boxed title style={sharp corners}
}

\newtcolorbox{example*}[1]{
    enhanced, breakable,
    colback=examplecolor!5, colframe=examplecolor,
    boxrule=1pt, arc=3pt,
    left=10pt, right=10pt, top=8pt, bottom=8pt,
    title={\textbf{Example: #1}},
    fonttitle=\color{white}, coltitle=white, colbacktitle=examplecolor
}

\newtcolorbox{mathematicalaside}[1]{
    enhanced, breakable,
    colback=lightgray, colframe=darkgray,
    boxrule=1pt, arc=3pt,
    left=10pt, right=10pt, top=8pt, bottom=8pt,
    title={\textbf{Mathematical Aside: #1}},
    fonttitle=\color{white}, coltitle=white, colbacktitle=darkgray
}

\newtcolorbox{physicalinsight}{
    enhanced, breakable,
    colback=yellow!10, colframe=orange!70!black,
    boxrule=1pt, arc=3pt,
    left=10pt, right=10pt, top=8pt, bottom=8pt,
    title={\textbf{Physical Insight}},
    fonttitle=\color{white}, coltitle=white, colbacktitle=orange!70!black
}

\newtcolorbox{blochcubeactivity}[1]{
    enhanced, breakable,
    colback=green!5, colframe=green!60!black,
    boxrule=1pt, arc=3pt,
    left=10pt, right=10pt, top=8pt, bottom=8pt,
    title={\textbf{Bloch Cube Activity: #1}},
    fonttitle=\color{white}, coltitle=white, colbacktitle=green!60!black
}

\titleformat{\chapter}[display]
{\Huge\bfseries\color{accentcolor}}
{\filleft\chaptertitlename\ \thechapter}
{20pt}
{\titlerule\vspace{2ex}\filleft}
[\vspace{2ex}\titlerule]

\titleformat{\section}
{\Large\bfseries\color{accentcolor}}
{\thesection}{1em}{}
[\titlerule]

\titleformat{\subsection}
{\large\bfseries\color{accentcolor!80}}
{\thesubsection}{1em}{}

\newcommand{\identity}{\mathbb{I}}
\newcommand{\hilbert}{\mathcal{H}}

\newcommand{\repeq}[1]{\underset{#1}{\doteq}}

\makeindex

\begin{document}

\frontmatter
\include{frontmatter/titlepage}
\include{frontmatter/preface}
\tableofcontents

\mainmatter
\include{chapters/ch01_foundations}
\include{chapters/ch02_tensor_products}
\include{chapters/ch03_lattice_qft}
\include{chapters/ch04_time}
\include{chapters/ch05_space}
\include{chapters/ch06_3d_angular}
\include{chapters/ch07_angular_addition}
\include{chapters/ch08_hydrogen_atom}
\include{chapters/ch09_perturbation_theory}
\include{chapters/ch10_entanglement_bell}
\include{chapters/ch11_dirac_equation}
\include{chapters/ch12_renormalization_group}
\include{chapters/epilogue}


\backmatter

\end{document}

%% file: frontmatter/titlepage.tex
\begin{titlepage}
    \centering
    \vspace*{2cm}
    
    {\Huge\bfseries\color{accentcolor} From  Qubit to  Qubit}\\[0.7cm]
    {\Large\color{accentcolor!80} A Graduate Course in Quantum Mechanics}\\[1cm]
    \vspace{1.5cm}

    {\Large\textbf{Jeremy Levy}}\\[0.5cm]
    {\large Department of Physics and Astronomy}\\
    {\large University of Pittsburgh}\\[1cm]
    \includegraphics[width=0.75\linewidth]{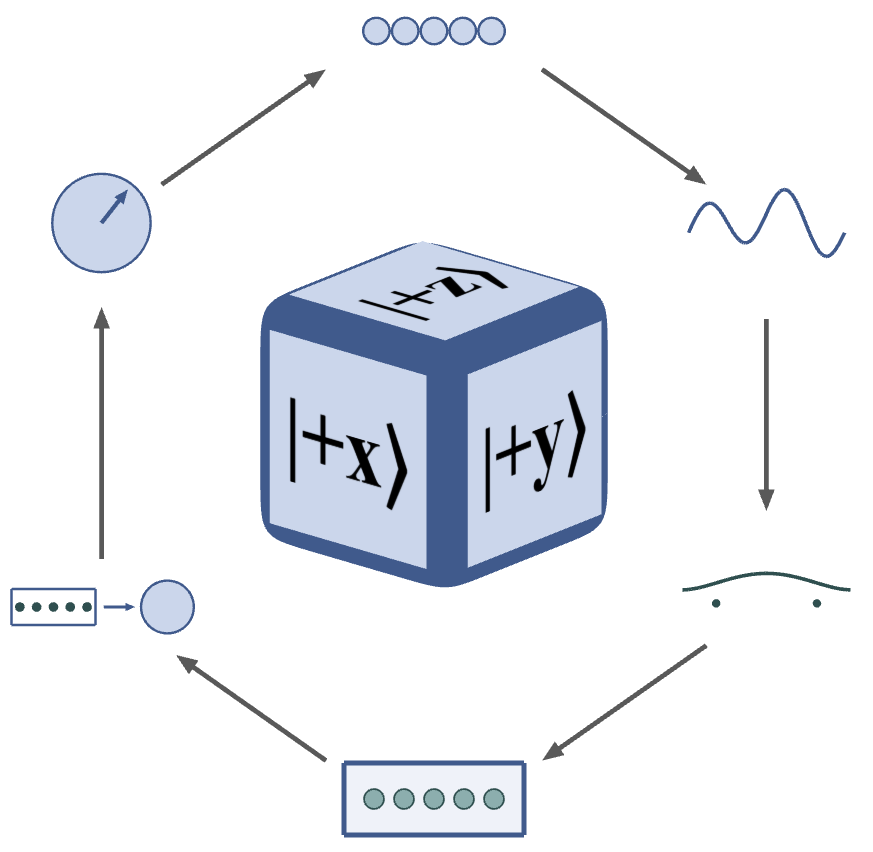}

\begin{center}
        \vspace{15cm}
        \includegraphics[width=0.9\linewidth]{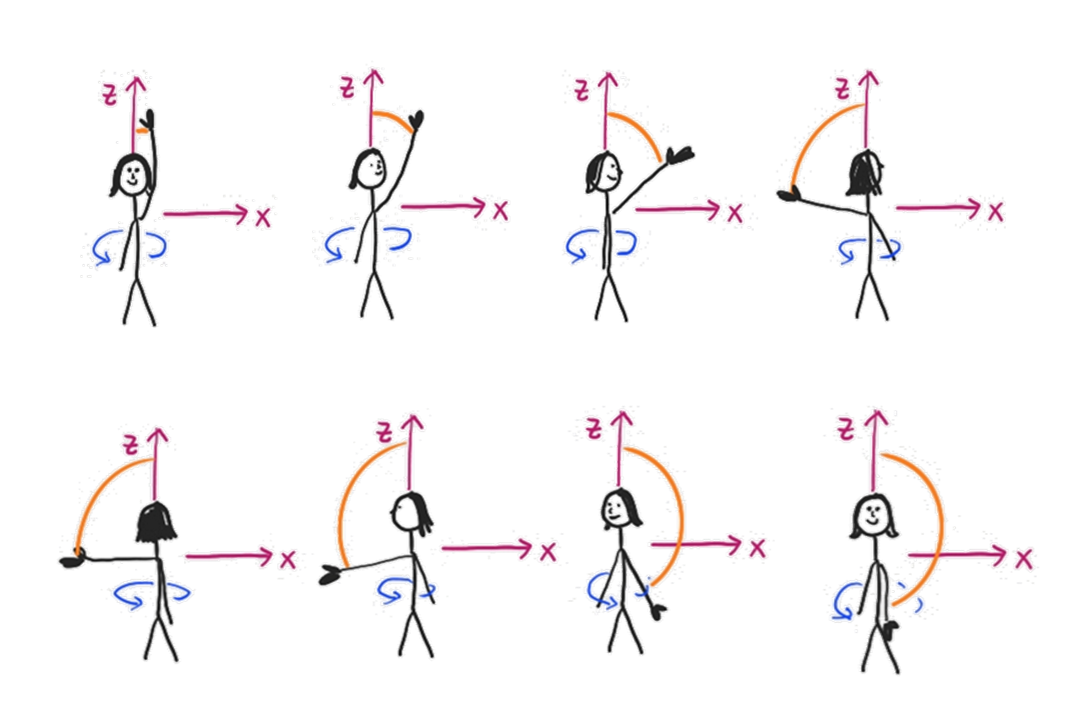}\\[1em]
    {\itshape Illustration by Natalia Valderrama.}
\end{center}
    


\end{titlepage}

%% file: frontmatter/preface.tex
\chapter*{Preface}
\addcontentsline{toc}{chapter}{Preface}

This book is the textbook for a two-semester graduate course in quantum mechanics taught at the University of Pittsburgh (PHYS 2565 and PHYS 2566). The notes have evolved over several years of teaching, and the present arrangement is the one I have settled on.

What distinguishes this treatment from other graduate texts is the order of presentation. The first two chapters develop quantum mechanics within the spin-1/2 system, with the Bloch cube, a six-sided physical object whose faces are the cardinal states of a single qubit, used as a concrete companion to the algebraic notation. Rotations of the cube enact unitary operations; pairs of cubes make entangled states visible; many of the basic calculations can be performed with a cube in hand. Quantum fields are introduced early on, in chapter 3, where the lattice version of a quantum field theory is the natural object once one has many qubits in a row.  Three-dimensional angular momentum, the hydrogen atom, and perturbation theory occupy the middle chapters and are developed in the standard way. Bell's theorem, the Dirac equation, and the renormalization group complete the year. The book closes by returning to the qubit, now as a logical qubit protected by quantum error correction; the threshold for fault-tolerant computation appears as an unstable fixed point of the renormalization group, which is what the title points at.

The discrete-to-continuous transition is a recurring theme: in Hilbert space, when finite- dimensional qubits give way to infinite-dimensional state spaces; in real space, when lattice models give way to wave mechanics; in time, when discrete unitary steps give way to the Schr\"odinger equation; and finally in scale, when block-spin renormalization group transformations give way to continuous rescaling and to continuous variation in the number of spatial dimensions.

Viewed this way, the book differs from a standard graduate text in its global topology rather than its local content. Each section taken individually is conventional and could appear in any other textbook; this is by design. The order, however, closes into a loop: the qubit at the start identifies with the qubit at the end, the way coordinates on a circle wrap back to themselves rather than running off to infinity on a line. The path between them, through lattice fields to wave mechanics to relativistic dynamics to scaling, is the rearrangement, and many of the connections that this rearrangement makes visible are harder to see when the same material is presented linearly.

The treatment is largely self-contained. A reader with a working command of linear algebra, complex analysis, multivariable calculus, and the standard undergraduate sequence of mechanics and electromagnetism should be able to follow it. Many sections are accessible to advanced undergraduates; the later material assumes more mathematical maturity. Researchers in quantum information, condensed matter, or atomic physics may find the integrated treatment useful as a single reference for material that is usually scattered.

Each chapter uses a small set of pedagogical elements. Key Idea boxes mark the central conceptual moves. Examples work calculations end to end. Bloch Cube Activities are short physical exercises meant to be done with a cube in hand. Mathematical Asides give technical derivations that can be skipped on a first reading. Physical Insights connect the formalism to experiment.

I am grateful to the students whose questions and frustrations have shaped this presentation, and to colleagues who have read earlier drafts. Special thanks are due to Chandralekha Singh for the collaborative development of the pedagogical approach, and to the American Journal of Physics for publishing our paper on the dot-vector formalism and the Bloch cube.

\vspace{1cm}
\begin{flushright}
Jeremy Levy\\
Department of Physics and Astronomy\\
University of Pittsburgh
\end{flushright}

%% file: chapters/ch01_foundations.tex
\chapter{Foundations via Spin-1/2 Systems}
\label{ch:foundations}

\section{Introduction: Quantum Foundations via Two-State Systems}

\begin{keyidea}{The Power of Two-State Systems}
The mathematical structure of quantum mechanics reveals itself most clearly through the study of two-state systems. These systems, despite their apparent simplicity, encode all the essential features of quantum theory: superposition, measurement, collapse, and unitary evolution. By developing a thorough understanding of two-state quantum mechanics, we establish a foundation that extends naturally to infinite-dimensional Hilbert spaces.
\end{keyidea}

This course takes a pedagogically distinct approach to graduate quantum mechanics. Rather than beginning with wave mechanics and the Schr\"odinger equation, we start with finite-dimensional systems where the mathematical structure is transparent and the physical interpretation is unambiguous. One important pedagogical tool will be the ``Bloch cube,'' a physical representation of quantum states that makes abstract mathematical operations tangible and intuitive.

\begin{figure}
    \centering
    \includegraphics[width=0.5\linewidth]{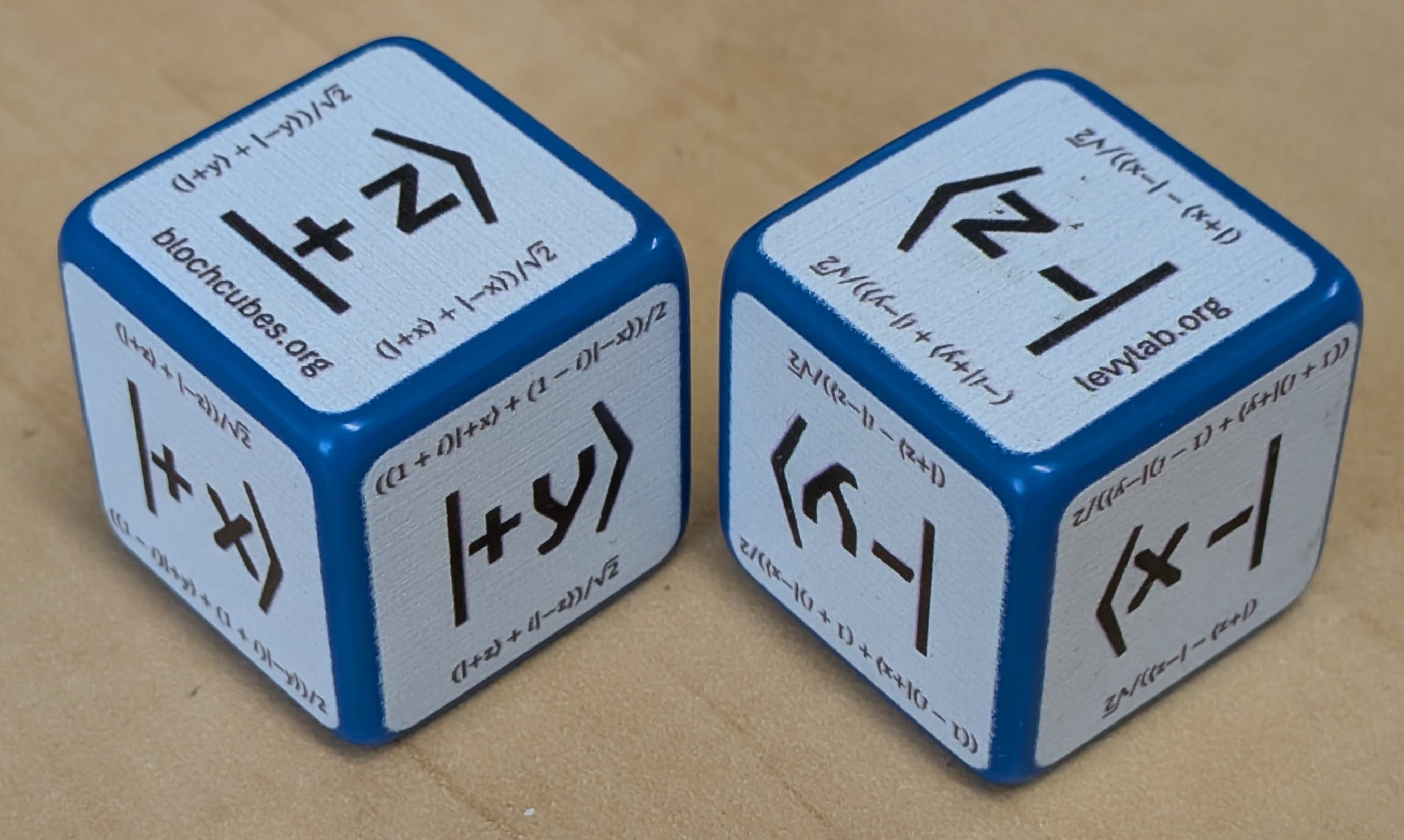}
    \caption{Two Bloch Cubes.}
    \label{fig:enter-label}
\end{figure}

The Bloch cube is a six-sided object, similar to a die, where each face represents a distinct quantum state of a two-level system. Through physical manipulation of these cubes, students can perform quantum operations, visualize superposition, and understand measurement in a concrete way before abstracting to the mathematical formalism. This approach bridges the gap between classical intuition and quantum formalism through what we call ``dot-vec notation,'' a mathematical framework that connects familiar vector operations to the Dirac notation of quantum mechanics.

The power of this approach lies in its ability to make the abstract concrete. When we say that quantum states can be "superposed," what does this really mean? When we claim that measurement "collapses" a quantum state, how can we visualize this process? The Bloch cube provides immediate, tactile answers to these questions. By the end of this chapter, you will understand quantum mechanics not just as a mathematical formalism, but as a physical theory with clear operational meaning.

\section{Mathematical Preliminaries: Vectors and Dot-Vectors}

\subsection{Review of Two-Dimensional Vector Spaces}

We begin with the familiar setting of two-dimensional Euclidean space. Consider orthonormal basis vectors $\hat{x}$ and $\hat{y}$ satisfying the standard inner product relations: $\hat{x} \cdot \hat{x} = 1$, $\hat{x} \cdot \hat{y} = 0$, $\hat{y} \cdot \hat{x} = 0$, and $\hat{y} \cdot \hat{y} = 1$. Any vector in this space can be expressed as a linear combination $\vec{V} = V_x \hat{x} + V_y \hat{y}$, where the components are extracted through the inner product: $V_x = \hat{x} \cdot \vec{V}$ and $V_y = \hat{y} \cdot \vec{V}$.

This elementary construction forms the basis for our approach to quantum mechanics. The key insight is that quantum states behave mathematically like vectors, while measurement processes behave like inner products. To make this connection explicit, we introduce a notational innovation that clarifies the dual roles played by vectors in quantum theory.

Let us explore this vector space more deeply. Consider a general vector $\vec{V} = 3\hat{x} + 4\hat{y}$. Its magnitude is $|\vec{V}| = \sqrt{3^2 + 4^2} = 5$, and we can normalize it to create a unit vector: $\hat{V} = \frac{3}{5}\hat{x} + \frac{4}{5}\hat{y}$. The angle this vector makes with the x-axis is $\theta = \arctan(4/3) \approx 53.1$ degrees. These familiar calculations from classical mechanics will have direct quantum mechanical analogs.

\subsection{The Dot-Vector Construction}

We now introduce a formal distinction between vectors and what we term ``dot-vectors.'' For each basis vector, we define a corresponding dot-vector by symbolically attaching the dot operation: $\hat{x} \rightarrow \hat{x}\cdot$ and $\hat{y} \rightarrow \hat{y}\cdot$. While this may appear to be merely a notational trick, it serves a profound pedagogical purpose by making explicit the dual nature of vectors in quantum mechanics.

A general dot-vector takes the form $\vec{W}\cdot = W_x \hat{x}\cdot + W_y \hat{y}\cdot$. The action of a dot-vector on a regular vector produces a scalar through the natural extension of the inner product: $\vec{W}\cdot\vec{V} = W_x V_x + W_y V_y$. This operation mirrors the action of a bra on a ket in quantum mechanics, providing students with a concrete mathematical bridge to the more abstract Dirac notation.

To illustrate the power of this notation, consider the dot-vector $\vec{W}\cdot = 2\hat{x}\cdot + 3\hat{y}\cdot$ acting on our previous vector $\vec{V} = 3\hat{x} + 4\hat{y}$:
\begin{align}
\vec{W}\cdot\vec{V} &= (2\hat{x}\cdot + 3\hat{y}\cdot)(3\hat{x} + 4\hat{y})\\
&= 2 \cdot 3 + 3 \cdot 4\\
&= 6 + 12 = 18
\end{align}

This calculation, while elementary, establishes the pattern for all quantum mechanical calculations involving state amplitudes and measurement probabilities.

\subsection{Operator Construction from Vectors and Dot-Vectors}

The true power of the dot-vector notation emerges when we consider products in the reverse order. The expression $\hat{x}\hat{x}\cdot$ represents an operator that acts on vectors. To understand its action, consider its effect on an arbitrary vector $\vec{V}$:

\begin{equation}
(\hat{x}\hat{x}\cdot)\vec{V} = \hat{x}(\hat{x}\cdot\vec{V}) = \hat{x}V_x = V_x\hat{x}
\end{equation}

This operator extracts the $x$-component of the vector and returns a vector pointing in the $x$-direction with that magnitude. Similarly, $\hat{y}\hat{y}\cdot$ projects onto the $y$-direction.

Let's examine these projection operators more carefully. Consider their action on our example vector $\vec{V} = 3\hat{x} + 4\hat{y}$:
\begin{align}
(\hat{x}\hat{x}\cdot)\vec{V} &= \hat{x}(\hat{x}\cdot(3\hat{x} + 4\hat{y})) = \hat{x}(3) = 3\hat{x}\\
(\hat{y}\hat{y}\cdot)\vec{V} &= \hat{y}(\hat{y}\cdot(3\hat{x} + 4\hat{y})) = \hat{y}(4) = 4\hat{y}
\end{align}

The sum of these projections recovers the original vector:
\begin{equation}
[(\hat{x}\hat{x}\cdot) + (\hat{y}\hat{y}\cdot)]\vec{V} = 3\hat{x} + 4\hat{y} = \vec{V}
\end{equation}

This demonstrates that the sum of projection operators yields the identity operator:
\begin{equation}
\hat{I} = \hat{x}\hat{x}\cdot + \hat{y}\hat{y}\cdot
\end{equation}

This decomposition of the identity will later correspond to the completeness relation in quantum mechanics. More generally, we can construct operators that transform between different directions. For instance, the operator $\hat{y}\hat{x}\cdot$ maps the $x$-direction to the $y$-direction: $(\hat{y}\hat{x}\cdot)\hat{x} = \hat{y}(\hat{x}\cdot\hat{x}) = \hat{y}$.

\section{The Quantum Correspondence}

\subsection{From Dot-Vec to Dirac Notation}

We now establish the formal correspondence between our dot-vector notation and the Dirac notation of quantum mechanics. This mapping transforms our concrete two-dimensional vector space into the abstract two-dimensional Hilbert space of a quantum two-state system:

\begin{center}
\begin{tabular}{|c|c|c|}
\hline
\textbf{Dot-Vec Notation} & \textbf{Dirac Notation} & \textbf{Physical Meaning} \\
\hline
$\hat{x}$ & $\ket{+z}$ & Spin up along z-axis \\
$\hat{y}$ & $\ket{-z}$ & Spin down along z-axis \\
$\hat{x}\cdot$ & $\bra{+z}$ & Measurement of spin up \\
$\hat{y}\cdot$ & $\bra{-z}$ & Measurement of spin down \\
$\hat{x}\cdot\hat{y}$ & $\braket{+z}{-z}$ & Transition amplitude \\
$\hat{x}\hat{y}\cdot$ & $\ketbra{+z}{-z}$ & Spin flip operator \\
\hline
\end{tabular}
\end{center}

This correspondence is not merely formal---it captures the deep mathematical structure shared by classical vectors and quantum states. The orthonormality of our basis vectors translates directly to the orthogonality of quantum basis states: $\braket{+z}{+z} = 1$, $\braket{-z}{-z} = 1$, and $\braket{+z}{-z} = \braket{-z}{+z} = 0$.

\subsection{Quantum Superposition}

The principle of superposition, fundamental to quantum mechanics, emerges naturally in our framework. Just as any two-dimensional vector can be expressed as a linear combination of basis vectors, any quantum state in our two-dimensional Hilbert space can be written as:

\begin{equation}
\ket{\psi} = \alpha\ket{+z} + \beta\ket{-z}
\end{equation}

Here, $\alpha$ and $\beta$ are complex numbers satisfying the normalization condition $|\alpha|^2 + |\beta|^2 = 1$. This condition ensures that the total probability of finding the system in some state equals unity. The complex nature of these coefficients, absent in our classical vector analogy, introduces the uniquely quantum phenomenon of phase.

To understand the role of complex coefficients, consider the state:
\begin{equation}
\ket{\psi} = \frac{1}{\sqrt{2}}\ket{+z} + \frac{e^{i\pi/4}}{\sqrt{2}}\ket{-z}
\end{equation}

The magnitudes $|\alpha| = |\beta| = 1/\sqrt{2}$ determine measurement probabilities, while the relative phase $e^{i\pi/4}$ affects interference between the components. This phase has no classical analog in our dot-vector notation---it represents genuinely quantum behavior.

The coefficients can be extracted using the inner product, just as in the classical case: $\alpha = \braket{+z}{\psi}$ and $\beta = \braket{-z}{\psi}$. These probability amplitudes encode all information about the quantum state and determine the probabilities of measurement outcomes through the Born rule.

\section{The Bloch Cube: A Tangible Quantum System}

\subsection{Construction and Concept}

We now introduce the central pedagogical tool of our approach: the Bloch cube. This physical object serves as a concrete representation of a quantum two-state system, making abstract quantum concepts tangible and manipulable. The Bloch cube is a regular cube (such as a die) where each of the six faces represents a specific quantum state.

The full Bloch sphere, which represents all possible states of a two-level quantum system, forms a sphere in three-dimensional space. The general state $\ket{\psi} = \cos(\theta/2)\ket{+z} + e^{i\phi}\sin(\theta/2)\ket{-z}$ corresponds to the point with spherical coordinates $(\theta, \phi)$ on this sphere. While this continuum of states is mathematically elegant, it presents pedagogical challenges for beginning students.

The Bloch cube simplifies this picture by restricting attention to six special states that form the vertices of a regular octahedron inscribed within the Bloch sphere. These states correspond to the intersections of the sphere with the three coordinate axes. By limiting ourselves to these six states initially, we can perform all fundamental quantum operations while maintaining a clear physical picture.

\subsection{The Six Fundamental States}

The six faces of the Bloch cube represent the following quantum states:

Along the z-axis, we have $\ket{+z}$ and $\ket{-z}$, which serve as our computational basis states. These represent definite values of the z-component of spin for a spin-1/2 particle.

Along the x-axis, we find $\ket{+x} = \frac{1}{\sqrt{2}}(\ket{+z} + \ket{-z})$ and $\ket{-x} = \frac{1}{\sqrt{2}}(\ket{+z} - \ket{-z})$. These states represent equal superpositions of the z-basis states, differing only in their relative phase.

Finally, along the y-axis, we have $\ket{+y} = \frac{1}{\sqrt{2}}(\ket{+z} + i\ket{-z})$ and $\ket{-y} = \frac{1}{\sqrt{2}}(\ket{+z} - i\ket{-z})$. The presence of the imaginary unit $i$ in these superpositions introduces a phase difference of $\pi/2$ between the components.

\subsection{Complete Basis Representations}

A crucial tool for quantum calculations is the ability to express any state in any basis. Here we provide the complete representation of all six Bloch cube states in all three bases:

\begin{tcolorbox}[enhanced, title={Bloch Cube States in All Bases}, colback=blue!5, colframe=blue!50!black, breakable]
\textbf{Z-basis representations:}
\begin{align}
\ket{+z} &= \ket{+z} \\
\ket{-z} &= \ket{-z} \\
\ket{+x} &= \frac{1}{\sqrt{2}}(\ket{+z} + \ket{-z}) \\
\ket{-x} &= \frac{1}{\sqrt{2}}(\ket{+z} - \ket{-z}) \\
\ket{+y} &= \frac{1}{\sqrt{2}}(\ket{+z} + i\ket{-z}) \\
\ket{-y} &= \frac{1}{\sqrt{2}}(\ket{+z} - i\ket{-z})
\end{align}

\textbf{X-basis representations:}
\begin{align}
\ket{+z} &= \frac{1}{\sqrt{2}}(\ket{+x} + \ket{-x}) \\
\ket{-z} &= \frac{1}{\sqrt{2}}(\ket{+x} - \ket{-x}) \\
\ket{+x} &= \ket{+x} \\
\ket{-x} &= \ket{-x} \\
\ket{+y} &= \frac{(1+i)\ket{+x} + (1-i)\ket{-x}}{2} \\
\ket{-y} &= \frac{(1-i)\ket{+x} + (1+i)\ket{-x}}{2}
\end{align}

\textbf{Y-basis representations:}
\begin{align}
\ket{+z} &= \frac{1}{\sqrt{2}}(\ket{+y} + \ket{-y}) \\
\ket{-z} &= \frac{1}{\sqrt{2}}(-i\ket{+y} + i\ket{-y}) \\
\ket{+x} &= \frac{(1-i)\ket{+y} + (1+i)\ket{-y}}{2} \\
\ket{-x} &= \frac{(1+i)\ket{+y} + (1-i)\ket{-y}}{2} \\
\ket{+y} &= \ket{+y} \\
\ket{-y} &= \ket{-y}
\end{align}
\end{tcolorbox}

These representations are essential for calculating transition probabilities between different bases and understanding how the same quantum state appears different when viewed from different measurement perspectives.

\subsection{Using Cross-Basis Representations}

Let's work through an example that demonstrates the utility of these cross-basis representations. Suppose we want to calculate $\braket{+x}{+y}$ directly using the Z-basis.

\begin{example}[Cross-Basis Inner Product]
Calculate $\braket{+x}{+y}$ using Z-basis representations:

From our table:
\begin{align}
\ket{+x} &= \frac{1}{\sqrt{2}}(\ket{+z} + \ket{-z})\\
\ket{+y} &= \frac{1}{\sqrt{2}}(\ket{+z} + i\ket{-z})
\end{align}

Therefore:
\begin{align}
\braket{+x}{+y} &= \frac{1}{\sqrt{2}}(\bra{+z} + \bra{-z}) \cdot \frac{1}{\sqrt{2}}(\ket{+z} + i\ket{-z})\\
&= \frac{1}{2}[\braket{+z}{+z} + i\braket{+z}{-z} + \braket{-z}{+z} + i\braket{-z}{-z}]\\
&= \frac{1}{2}[1 + 0 + 0 + i]\\
&= \frac{1}{2}(1 + i)
\end{align}

We can verify this is correct since $|\braket{+x}{+y}|^2 = \frac{1}{2}$, as expected for adjacent cube faces.
\end{example}

\subsection{Orthogonality and the Cube Geometry}

The geometric structure of the cube encodes the orthogonality relations between quantum states. States on opposite faces of the cube are orthogonal, meaning their inner product vanishes. This can be verified through direct calculation: $\braket{+z}{-z} = 0$, $\braket{+x}{-x} = 0$, and $\braket{+y}{-y} = 0$. This orthogonality has profound physical meaning---orthogonal states can be perfectly distinguished by appropriate measurements.

States on adjacent faces of the cube are not orthogonal. The inner products between adjacent states all have magnitude $1/\sqrt{2}$:
\begin{align}
|\braket{+z}{+x}| &= |\braket{+z}{+y}| = \frac{1}{\sqrt{2}}\\
|\braket{+x}{+y}| &= |\braket{+x}{-y}| = \frac{1}{\sqrt{2}}\\
|\braket{+y}{+z}| &= |\braket{+y}{-z}| = \frac{1}{\sqrt{2}}
\end{align}

This uniform overlap between adjacent states reflects the high symmetry of the cube arrangement on the Bloch sphere.

\begin{blochcubeactivity}{1.1: Constructing and Exploring the Bloch Cube}
To begin your hands-on exploration of quantum mechanics, first construct or obtain a Bloch cube. You can use a standard six-sided die or create a cube with labeled faces. Label the six faces according to the states described above: $\ket{+z}$, $\ket{-z}$, $\ket{+x}$, $\ket{-x}$, $\ket{+y}$, and $\ket{-y}$. Once you have your Bloch cube, verify the following properties through direct manipulation:
\begin{enumerate}
    \item Opposite faces represent orthogonal states
    \item Each state can be expressed as a superposition of any pair of orthogonal states
    \item The cube can be oriented to place any face ``up,'' representing the current quantum state
    \item Rotations of the cube correspond to unitary transformations of the quantum state
    \item Use the basis representation table to verify that $\ket{+x}$ is indeed an equal superposition of $\ket{+z}$ and $\ket{-z}$
\end{enumerate}
\end{blochcubeactivity}

\section{Measurement Theory and the Born Rule}

\subsection{Measurement as Fundamental Process}

Measurement in quantum mechanics differs fundamentally from classical measurement. While classical measurements merely reveal pre-existing properties, quantum measurements actively disturb the system and, in general, change its state. Our Bloch cube framework provides a concrete model for understanding this process.

A quantum measurement on our two-state system corresponds to asking a yes-no question about the state. For the Bloch cube, we have three natural measurement types, corresponding to the three spatial axes. The Z-measurement asks whether the system is in state $\ket{+z}$ or $\ket{-z}$. Similarly, X-measurements distinguish between $\ket{+x}$ and $\ket{-x}$, while Y-measurements distinguish between $\ket{+y}$ and $\ket{-y}$.

To understand how measurement procedures translate into mathematical operators, let's construct them step by step from our concrete measurement outcomes. We've already encountered the concept of eigenstates and eigenvalues in the context of rotations---here, the eigenvalues will correspond to the possible measurement outcomes and eigenstates correspond to the states that yield definite results.

Consider the Z-measurement procedure: we ask whether the system is in state $\ket{+z}$ or $\ket{-z}$. If we assign numerical scores to these outcomes---say +1 for finding $\ket{+z}$ and -1 for finding $\ket{-z}$---then we can construct a mathematical operator that encodes this measurement:

\begin{equation}
\hat{\sigma}_z = (+1)\ketbra{+z}{+z} + (-1)\ketbra{-z}{-z} = \ketbra{+z}{+z} - \ketbra{-z}{-z}
\end{equation}

This operator has the property that when it acts on the measurement eigenstates, it returns the corresponding measurement outcome: $\hat{\sigma}_z\ket{+z} = (+1)\ket{+z}$ and $\hat{\sigma}_z\ket{-z} = (-1)\ket{-z}$.

Similarly, we can construct operators for X and Y measurements:
\begin{align}
\hat{\sigma}_x &= \ketbra{+x}{+x} - \ketbra{-x}{-x}\\
\hat{\sigma}_y &= \ketbra{+y}{+y} - \ketbra{-y}{-y}
\end{align}

These measurement operators are known as the Pauli operators, which form the foundation of two-level quantum mechanics. Notice that the six Bloch cube states we introduced earlier are precisely the eigenstates of these three operators: $\ket{\pm z}$ are eigenstates of $\sigma_z$, $\ket{\pm x}$ are eigenstates of $\sigma_x$, and $\ket{\pm y}$ are eigenstates of $\sigma_y$.

Observe the pattern in our construction: each measurement operator was built by assigning real numerical values ($\pm 1$) to the possible outcomes and weighting the corresponding projection operators. This systematic construction ensures that the eigenvalues---which represent the measurement outcomes we can observe---are always real numbers. This is a general feature of all quantum measurements.

\subsection{The Born Rule and Probability}

The Born rule provides the fundamental connection between the mathematical formalism of quantum mechanics and experimental predictions. According to this rule, if a system is in state $\ket{\psi}$ and we measure an observable with eigenstates $\ket{\phi_i}$, the probability of obtaining the outcome corresponding to $\ket{\phi_i}$ is:

\begin{equation}
P(\ket{\psi}\rightarrow\ket{\phi_i}) = |\braket{\phi_i}{\psi}|^2
\end{equation}

This formula encodes one of the most profound aspects of quantum mechanics: even with complete knowledge of the quantum state, we can only predict probabilities of measurement outcomes, not the outcomes themselves.

Let's explore several examples to understand the Born rule deeply.

\begin{example}[Measurement Probabilities for $\ket{+x}$]
Consider the state $\ket{+x} = \frac{1}{\sqrt{2}}(\ket{+z} + \ket{-z})$. Let's calculate the probabilities for all three measurement types.

\textbf{Z-measurement:}
\begin{align}
P(\ket{+x}\rightarrow\ket{+z}) &= |\braket{+z}{+x}|^2 = \left|\frac{1}{\sqrt{2}}\right|^2 = \frac{1}{2}\\
P(\ket{+x}\rightarrow\ket{-z}) &= |\braket{-z}{+x}|^2 = \left|\frac{1}{\sqrt{2}}\right|^2 = \frac{1}{2}
\end{align}

\textbf{X-measurement:}
\begin{align}
P(\ket{+x}\rightarrow\ket{+x}) &= |\braket{+x}{+x}|^2 = |1|^2 = 1\\
P(\ket{+x}\rightarrow\ket{-x}) &= |\braket{-x}{+x}|^2 = |0|^2 = 0
\end{align}

\textbf{Y-measurement:}
Using $\ket{+x} = \frac{1}{\sqrt{2}}(\ket{+y} + \ket{-y})$ from our basis table:
\begin{align}
P(\ket{+x}\rightarrow\ket{+y}) &= |\braket{+y}{+x}|^2 = \left|\frac{1}{\sqrt{2}}\right|^2 = \frac{1}{2}\\
P(\ket{+x}\rightarrow\ket{-y}) &= |\braket{-y}{+x}|^2 = \left|\frac{1}{\sqrt{2}}\right|^2 = \frac{1}{2}
\end{align}
\end{example}

\subsection{State Collapse and Sequential Measurements}

Following a measurement, the quantum state ``collapses'' to the eigenstate corresponding to the observed outcome. This collapse is instantaneous and irreversible within the standard quantum formalism. If we measure $\ket{+x}$ along the z-axis and the system collapses to $\ket{+z}$, the post-measurement state is $\ket{+z}$, not $\ket{+x}$.

This collapse leads to one of the most striking features of quantum measurement: the order of measurements matters. Consider performing X and Z measurements in sequence. If we start with $\ket{+y}$, measure X (causing the system to collapse to, say, $\ket{+x}$), then measure Z, we get random Z results. However, if we first measure Z, then X, we get random X results. The non-commutativity of measurements reflects the non-commutativity of the corresponding operators.

\begin{example}[Sequential Measurements]
Start with $\ket{+y} = \frac{1}{\sqrt{2}}(\ket{+z} + i\ket{-z})$.

\textbf{Path 1: Measure Z, then X}
\begin{enumerate}
\item Initial state: $\ket{+y}$
\item Measure Z: The system collapses to $\ket{+z}$ with probability $1/2$ or $\ket{-z}$ with probability $1/2$
\item Suppose the system collapses to $\ket{+z}$
\item Now measure X on $\ket{+z} = \frac{1}{\sqrt{2}}(\ket{+x} + \ket{-x})$
\item The system collapses to $\ket{+x}$ or $\ket{-x}$ each with probability $1/2$
\end{enumerate}

\textbf{Path 2: Measure X, then Z}
\begin{enumerate}
\item Initial state: $\ket{+y} = \frac{(1+i)\ket{+x} +(1-i)\ket{-x}}{2}$
\item Measure X: The system collapses to $\ket{+x}$ with probability $1/2$ or $\ket{-x}$ with probability $1/2$
\item Suppose the system collapses to $\ket{+x}$
\item Now measure Z on $\ket{+x} = \frac{1}{\sqrt{2}}(\ket{+z} + \ket{-z})$
\item The system collapses to $\ket{+z}$ or $\ket{-z}$ each with probability $1/2$
\end{enumerate}

The final probability distributions are identical, but the intermediate states differ!
\end{example}

\begin{blochcubeactivity}{1.2: Simulating Quantum Measurement}
Using your Bloch cube, simulate the measurement process:
\begin{enumerate}
    \item Set the cube to show $\ket{+x}$ (facing up)
    \item To simulate a Z-measurement, calculate the probabilities for collapsing to $\ket{\pm z}$ states
    \item Use a random process (coin flip) to determine the outcome
    \item Rotate the cube to show the post-measurement state
    \item Repeat multiple times to verify the statistical predictions
    \item Try starting from $\ket{+y}$ and perform measurements in different orders
\end{enumerate}
\end{blochcubeactivity}

\subsection{Measurement Outcomes and Expectation Values}

So far, we have discussed measurements as producing states $\ket{\pm z}$, $\ket{\pm x}$, or $\ket{\pm y}$. To connect with observable physical quantities, we assign numerical values to these outcomes. We adopt a standard ``scoring'' system:

\begin{itemize}
\item Measuring $\ket{+w}$ yields outcome $+1$
\item Measuring $\ket{-w}$ yields outcome $-1$
\end{itemize}

where $W \in \{X, Y, Z\}$ labels the measurement type. This assignment is not arbitrary---it corresponds to measuring the component of spin angular momentum along the specified axis (in units of $\hbar/2$).

With this scoring system, each measurement operator can be written as:
\begin{align}
\hat{\sigma}_z &= \ketbra{+z}{+z} -\ketbra{-z}{-z}\\
\hat{\sigma}_x &= \ketbra{+x}{+x} -\ketbra{-x}{-x}\\
\hat{\sigma}_y &= \ketbra{+y}{+y} -\ketbra{-y}{-y}
\end{align}

These are the Pauli operators. Notice something important: unlike the rotation operators whose eigenvalues were complex phases ($1$ and $i$ for $\hat{Z}$), the Pauli operators have real eigenvalues ($\pm 1$) that represent the possible measurement outcomes. This is not a coincidence---measurement outcomes must be real numbers that we can observe in experiments.

This pattern---that measurement operators constructed from real-valued outcomes always have real eigenvalues---holds generally in quantum mechanics. Operators with this property are called \textit{Hermitian} operators. The requirement that measurement outcomes be real numbers forces measurement operators to be Hermitian, while the requirement that rotations preserve probability forces rotation operators to be \textit{unitary} (ensuring eigenvalues with magnitude 1).

Interestingly, there's a simple relationship between the rotation and Pauli operators: $\hat{\sigma}_W = \hat{W}^2$ for $W \in \{X, Y, Z\}$. That is, the Pauli operators correspond to 180\ensuremath{^\circ} rotations about their respective axes. This makes sense geometrically---a 180\ensuremath{^\circ} rotation takes each Bloch cube face to its opposite face, which is precisely what the Pauli operators encode through their $\pm 1$ eigenvalues.

For a system in state $\ket{\psi}$, the \textit{expectation value} of a measurement is the average outcome we would obtain if we performed the measurement on many identically prepared systems:

\begin{equation}
\langle \hat{\sigma}_W \rangle = P(\ket{\psi}\rightarrow\ket{+W}) \cdot (+1) + P(\ket{\psi}\rightarrow\ket{-W}) \cdot (-1) = P(\ket{\psi}\rightarrow\ket{+W}) - P(\ket{\psi}\rightarrow\ket{-W})
\end{equation}

where $P(\ket{\psi}\rightarrow\ket{\pm W})$ are the probabilities calculated using the Born rule.

\begin{example}[Expectation Values for $\ket{+x}$]
Calculate the expectation values of all three Pauli operators for the state $\ket{+x}$.

From Example 1.2, we found:
\begin{itemize}
\item Z-measurement: $P(\ket{+x}\rightarrow\ket{+z}) = P(\ket{+x}\rightarrow\ket{-z}) = 1/2$
\item X-measurement: $P(\ket{+x}\rightarrow\ket{+x}) = 1$, $P(\ket{+x}\rightarrow\ket{-x}) = 0$
\item Y-measurement: $P(\ket{+x}\rightarrow\ket{+y}) = P(\ket{+x}\rightarrow\ket{-y}) = 1/2$
\end{itemize}

Therefore:
\begin{align}
\langle \hat{\sigma}_z \rangle &= P(\ket{+x}\rightarrow\ket{+z}) - P(\ket{+x}\rightarrow\ket{-z}) = \frac{1}{2} - \frac{1}{2} = 0\\
\langle \hat{\sigma}_x \rangle &= P(\ket{+x}\rightarrow\ket{+x}) - P(\ket{+x}\rightarrow\ket{-x}) = 1 - 0 = 1\\
\langle \hat{\sigma}_y \rangle &= P(\ket{+x}\rightarrow\ket{+y}) - P(\ket{+x}\rightarrow\ket{-y}) = \frac{1}{2} - \frac{1}{2} = 0
\end{align}

The state $\ket{+x}$ has definite value $+1$ for X-measurements but gives random (zero average) results for Y and Z measurements.
\end{example}

Expectation values can also be calculated directly using the inner product:
\begin{equation}
\langle \hat{A} \rangle = \bra{\psi}\hat{A}\ket{\psi}
\end{equation}

This formula, fundamental to quantum mechanics, connects the abstract mathematical formalism to measurable physical quantities.

\begin{blochcubeactivity}{1.3: Measuring Expectation Values with Bloch Cubes}
Using your Bloch cube:
\begin{enumerate}
    \item Prepare the state $\ket{+y}$
    \item Perform 20 simulated Z-measurements, recording $+1$ for $\ket{+z}$ and $-1$ for $\ket{-z}$
    \item Calculate the average of your results
    \item Compare with the theoretical expectation value $\langle \hat{\sigma}_z \rangle = 0$
    \item Repeat for X and Y measurements
    \item Discuss how many measurements would be needed to distinguish $\langle \hat{\sigma}_z \rangle = 0$ from $\langle \hat{\sigma}_z \rangle = 0.1$
\end{enumerate}
\end{blochcubeactivity}

\section{Unitary Evolution and Quantum Dynamics}

\subsection{Discrete Rotations as Quantum Gates}

The time evolution of isolated quantum systems is governed by unitary transformations. For our Bloch cube, these transformations are implemented as physical rotations of the cube. This provides an immediate, tactile understanding of quantum dynamics that complements the mathematical formalism.

Consider rotations by 90 degrees about the three coordinate axes. These discrete rotations form a group and can be represented by unitary operators. The rotation by 90 degrees clockwise about the z-axis (when viewed from the positive z-direction) is represented by the operator:

\begin{equation}
\hat{Z} = \ketbra{+z}{+z} + i\ketbra{-z}{-z}
\end{equation}

Let's trace through how this operator acts on all six Bloch cube states:

\begin{example}[Action of $\hat{Z}$ on All Cube States]
\begin{align}
\hat{Z}\ket{+z} &= \ket{+z}\\
\hat{Z}\ket{-z} &= i\ket{-z}\\
\hat{Z}\ket{+x} &= \hat{Z}\frac{1}{\sqrt{2}}(\ket{+z} + \ket{-z}) = \frac{1}{\sqrt{2}}(\ket{+z} + i\ket{-z}) = \ket{+y}\\
\hat{Z}\ket{-x} &= \hat{Z}\frac{1}{\sqrt{2}}(\ket{+z} - \ket{-z}) = \frac{1}{\sqrt{2}}(\ket{+z} - i\ket{-z}) = \ket{-y}\\
\hat{Z}\ket{+y} &= \hat{Z}\frac{1}{\sqrt{2}}(\ket{+z} + i\ket{-z}) = \frac{1}{\sqrt{2}}(\ket{+z} - \ket{-z}) = \ket{-x}\\
\hat{Z}\ket{-y} &= \hat{Z}\frac{1}{\sqrt{2}}(\ket{+z} - i\ket{-z}) = \frac{1}{\sqrt{2}}(\ket{+z} + \ket{-z}) = \ket{+x}
\end{align}

This confirms that $\hat{Z}$ implements a 90\ensuremath{^\circ} rotation: $+x \rightarrow +y \rightarrow -x \rightarrow -y \rightarrow +x$.
\end{example}

Notice something remarkable in the calculations above: when $\hat{Z}$ acts on $\ket{+z}$ and $\ket{-z}$, these states are only multiplied by phase factors (1 and $i$ respectively). This illustrates a fundamental concept in quantum mechanics: \textit{eigenstates} and \textit{eigenvalues}.

\begin{keyidea}{Eigenstates and Eigenvalues}
A state $\ket{\psi}$ is called an eigenstate of an operator $\hat{A}$ if the operator's action on that state merely multiplies it by a complex number:
\begin{equation}
\hat{A}\ket{\psi} = \lambda\ket{\psi}
\end{equation}
The complex number $\lambda$ is called the eigenvalue. For our rotation operator $\hat{Z}$:
\begin{itemize}
\item $\ket{+z}$ is an eigenstate with eigenvalue $\lambda = 1$
\item $\ket{-z}$ is an eigenstate with eigenvalue $\lambda = i$
\end{itemize}
Physically, eigenstates represent states that are ``stable'' under the operation---they don't change direction, only acquire a phase. For rotations about the z-axis, it makes intuitive sense that states aligned with the z-axis ($\ket{\pm z}$) would be special in this way.
\end{keyidea}

The eigenvalues of rotation operators are always complex numbers with magnitude 1 (i.e., pure phases), which ensures that the operators preserve the normalization of quantum states. This is a consequence of unitarity. For measurement operators, as we've seen earlier, the eigenvalues will be real numbers corresponding to the possible measurement outcomes.

Similarly, we can define the X and Y rotation operators:
\begin{align}
\hat{X} &= \ketbra{+x}{+x} + i\ketbra{-x}{-x}\\
\hat{Y} &= \ketbra{+y}{+y} + i\ketbra{-y}{-y}
\end{align}

\subsection{The Non-Commutativity of Rotations}

A fundamental feature of quantum mechanics is that operations generally do not commute. This non-commutativity can be directly experienced through Bloch cube manipulations. Consider the unitary operators $\hat{X}$ and $\hat{Y}$, representing 90-degree rotations about the x and y axes respectively.

To understand this rigorously, we need to determine how these operators act on the state $\ket{+z}$. Using the operator definitions:
\begin{align}
\hat{X} &= \ketbra{+x}{+x} + i\ketbra{-x}{-x}\\
\hat{Y} &= \ketbra{+y}{+y} + i\ketbra{-y}{-y}
\end{align}

\begin{example}[Rigorous Derivation of Rotation Actions]
\textbf{Finding $\hat{X}\ket{+z}$:}

\begin{align}
\hat{X}\ket{+z} &= (\ketbra{+x}{+x} + i\ketbra{-x}{-x})\ket{+z}\\
&= \ket{+x}\braket{+x}{+z} + i\ket{-x}\braket{-x}{+z}
\end{align}

Using our basis representations: $\braket{+x}{+z} = \braket{-x}{+z} = \frac{1}{\sqrt{2}}$

\begin{align}
\hat{X}\ket{+z} &= \ket{+x}\frac{1}{\sqrt{2}} + i\ket{-x}\frac{1}{\sqrt{2}}\\
&= \frac{1}{\sqrt{2}}(\ket{+x} + i\ket{-x})
\end{align}

Converting to Z-basis using $\ket{+x} = \frac{1}{\sqrt{2}}(\ket{+z} + \ket{-z})$ and $\ket{-x} = \frac{1}{\sqrt{2}}(\ket{+z} - \ket{-z})$:

\begin{align}
\hat{X}\ket{+z} &= \frac{1}{\sqrt{2}}\left[\frac{1}{\sqrt{2}}(\ket{+z} + \ket{-z}) + i\frac{1}{\sqrt{2}}(\ket{+z} - \ket{-z})\right]\\
&= \frac{1}{2}[(1 + i)\ket{+z} + (1 - i)\ket{-z}]
\end{align}

Factoring out the phase: $(1 + i) = \sqrt{2}e^{i\pi/4}$, so:
\begin{equation}
\hat{X}\ket{+z} = e^{i\pi/4}\frac{1}{\sqrt{2}}(\ket{+z} - i\ket{-z}) = e^{i\pi/4}\ket{-y}
\end{equation}

\textbf{Finding $\hat{Y}\ket{+z}$:}

Similarly:
\begin{align}
\hat{Y}\ket{+z} &= (\ketbra{+y}{+y} + i\ketbra{-y}{-y})\ket{+z}\\
&= \ket{+y}\braket{+y}{+z} + i\ket{-y}\braket{-y}{+z}\\
&= \frac{1}{\sqrt{2}}(\ket{+y} + i\ket{-y})
\end{align}

Converting to Z-basis and following similar algebra:
\begin{equation}
\hat{Y}\ket{+z} = e^{i\pi/4}\ket{+x}
\end{equation}
\end{example}

Now let's trace through both possible orderings with the correct phase factors:

\textbf{Order 1: $\hat{X}$ then $\hat{Y}$}
$\ket{+z} \xrightarrow{\hat{X}} e^{i\pi/4}\ket{-y} \xrightarrow{\hat{Y}} ?$

To find $\hat{Y}\ket{-y}$:
\begin{align}
\hat{Y}\ket{-y} &= (\ketbra{+y}{+y} + i\ketbra{-y}{-y})\ket{-y}\\
&= \ket{+y}\underbrace{\braket{+y}{-y}}_{=0} + i\ket{-y}\underbrace{\braket{-y}{-y}}_{=1}\\
&= i\ket{-y}
\end{align}

Therefore: $\ket{+z} \xrightarrow{\hat{X}} e^{i\pi/4}\ket{-y} \xrightarrow{\hat{Y}} ie^{i\pi/4}\ket{-y} = e^{i3\pi/4}\ket{-y}$

\textbf{Order 2: $\hat{Y}$ then $\hat{X}$}
$\ket{+z} \xrightarrow{\hat{Y}} e^{i\pi/4}\ket{+x} \xrightarrow{\hat{X}} ?$

To find $\hat{X}\ket{+x}$:
\begin{align}
\hat{X}\ket{+x} &= (\ketbra{+x}{+x} + i\ketbra{-x}{-x})\ket{+x}\\
&= \ket{+x}\underbrace{\braket{+x}{+x}}_{=1} + i\ket{-x}\underbrace{\braket{-x}{+x}}_{=0}\\
&= \ket{+x}
\end{align}

Therefore: $\ket{+z} \xrightarrow{\hat{Y}} e^{i\pi/4}\ket{+x} \xrightarrow{\hat{X}} e^{i\pi/4}\ket{+x}$

\textbf{Comparison:}
\begin{itemize}
\item $\hat{Y}\hat{X}\ket{+z} = e^{i3\pi/4}\ket{-y}$
\item $\hat{X}\hat{Y}\ket{+z} = e^{i\pi/4}\ket{+x}$
\end{itemize}

The final states differ both in direction ($\ket{-y}$ vs $\ket{+x}$) and global phase ($e^{i3\pi/4}$ vs $e^{i\pi/4}$), demonstrating that $\hat{X}\hat{Y} \neq \hat{Y}\hat{X}$.

\begin{keyidea}{Phase Factors in Quantum Rotations}
The 90\ensuremath{^\circ} rotation operators naturally introduce phase factors that are often omitted in simplified treatments. These phases are essential for:
\begin{itemize}
\item Mathematical consistency with the operator definitions
\item Correct calculation of sequential transformations
\item Understanding the geometric structure of rotation groups
\item Proper treatment of commutation relations
\end{itemize}
While global phases don't affect measurement probabilities, they are crucial for understanding how operators compose and for advanced topics like geometric phases.
\end{keyidea}

This result can be verified through explicit matrix calculations using the matrix representations of these rotation operators. The non-trivial result provides a concrete demonstration that quantum rotations do not commute, with immediate physical consequences observable on the Bloch sphere.

\begin{blochcubeactivity}{1.4: Exploring Non-Commutativity}
Using your Bloch cube:
\begin{enumerate}
    \item Start with the cube showing $\ket{+z}$
    \item Perform a 90\ensuremath{^\circ} rotation about the X-axis, then about the Y-axis
    \item Record the final state
    \item Reset to $\ket{+z}$ and perform the rotations in reverse order
    \item Compare the final states
    \item Repeat for different initial states and rotation sequences
    \item Discuss the physical implications of non-commutativity
\end{enumerate}
\end{blochcubeactivity}

\section{The Density Operator Formalism}

\subsection{Pure States and Statistical Mixtures}

Thus far, we have represented quantum states as vectors in Hilbert space. This representation assumes maximal knowledge about the quantum system---what we call a pure state. However, in many practical situations, our knowledge is incomplete, necessitating a more general formalism.

The density operator provides this generalization. For a pure state $\ket{\psi}$, the density operator is simply:
\begin{equation}
\rho = \ketbra{\psi}{\psi}
\end{equation}

For example, the density operator for $\ket{+x} = \frac{1}{\sqrt{2}}(\ket{+z} + \ket{-z})$ is:
\begin{align}
\rho_{+x} &= \ketbra{+x}{+x}\\
&= \frac{1}{2}(\ket{+z} + \ket{-z})(\bra{+z} + \bra{-z})\\
&= \frac{1}{2}(\ketbra{+z}{+z} + \ketbra{+z}{-z} + \ketbra{-z}{+z} + \ketbra{-z}{-z})
\end{align}

This operator contains all information about the quantum state and can be used to calculate expectation values of observables: $\langle A \rangle = \text{Tr}(\rho A)$.

When our knowledge is incomplete, we have a statistical mixture of pure states. If the system is in state $\ket{\psi_i}$ with classical probability $p_i$, the density operator becomes:
\begin{equation}
\rho = \sum_i p_i \ketbra{\psi_i}{\psi_i}
\end{equation}

\subsection{Properties and Interpretation}

The density operator formalism reveals important distinctions between quantum and classical probability. Consider two preparations:
\begin{enumerate}
\item A pure state $\ket{\psi} = \sqrt{0.6}\ket{+z} + \sqrt{0.4}\ket{-z}$
\item A mixture with 60\% probability of $\ket{+z}$ and 40\% probability of $\ket{-z}$
\end{enumerate}

For Z-measurements, both preparations yield identical statistics. However, they differ dramatically for X-measurements. The pure state shows quantum interference through the off-diagonal terms $\ketbra{+z}{-z}$ and $\ketbra{-z}{+z}$ in its density operator, while the mixture does not have these terms.

This difference is captured by the purity $\text{Tr}(\rho^2)$. Pure states always have purity 1, while mixed states have purity less than 1.

\begin{example}[Calculating Expectation Values for a Mixed State Ensemble]
Consider the mixed state $\rho = 0.6\ketbra{+z}{+z} + 0.4\ketbra{-z}{-z}$. We'll calculate the expectation values $\langle\sigma_x\rangle$, $\langle\sigma_y\rangle$, and $\langle\sigma_z\rangle$ using a Bloch cube ensemble.

\textbf{Step 1: Create the ensemble}
\begin{itemize}
\item Take 5 Bloch cubes
\item Set 3 to show $\ket{+z}$ and 2 to show $\ket{-z}$
\item This ensemble represents our mixed state
\end{itemize}

\textbf{Step 2: Use pairing to simplify}
A key insight: opposite states contribute equally but with opposite signs to any measurement. Therefore, any pair of cubes showing $\ket{+z}$ and $\ket{-z}$ contributes zero to all expectation values.
\begin{itemize}
\item Pair up 2 $\ket{+z}$ cubes with the 2 $\ket{-z}$ cubes
\item These 4 cubes (2 pairs) contribute 0 to all expectation values
\item We're left with 1 unpaired $\ket{+z}$ cube
\end{itemize}

\textbf{Step 3: Calculate expectation values}
For the ensemble, the expectation value is the average over all cubes:

\textbf{Z-measurement:}
\begin{itemize}
\item 4 paired cubes: contribute 0
\item 1 unpaired $\ket{+z}$: contributes +1
\item Average over 5 cubes: $\langle\sigma_z\rangle = \frac{0 + 1}{5} = 0.2$
\end{itemize}

\textbf{X-measurement:}
\begin{itemize}
\item For $\ket{+z}$: equal probability of $\pm 1$, so expectation is 0
\item All 5 cubes have expectation 0 for X-measurement
\item $\langle\sigma_x\rangle = 0$
\end{itemize}

\textbf{Y-measurement:}
\begin{itemize}
\item Same reasoning as X-measurement
\item $\langle\sigma_y\rangle = 0$
\end{itemize}

\textbf{Verification:} We can check this matches the theoretical calculation:
$\langle\sigma_z\rangle = 0.6(+1) + 0.4(-1) = 0.6 - 0.4 = 0.2$ 

This example shows how the pairing trick simplifies calculations and provides physical intuition: mixed states with equal amounts of opposite states are "balanced" and give zero expectation values for all measurements.

\textbf{Contrast with pure state:} The pure state $\sqrt{0.6}\ket{+z} + \sqrt{0.4}\ket{-z}$ would also give $\langle\sigma_z\rangle = 0.2$, but would give non-zero values for $\langle\sigma_x\rangle$ and $\langle\sigma_y\rangle$ due to quantum interference. This mixture cannot show such interference because it's just a classical probability distribution over definite states.
\end{example}

\begin{blochcubeactivity}{1.5: The Pairing Principle for Mixed States}
Explore the pairing principle with different mixed states:
\begin{enumerate}
\item Create an ensemble with 4 cubes: 2 showing $\ket{+x}$ and 2 showing $\ket{-x}$
\item Verify that all expectation values are zero by pairing opposite states
\item Now add a 5th cube showing $\ket{+x}$. Calculate $\langle\sigma_x\rangle$, $\langle\sigma_y\rangle$, $\langle\sigma_z\rangle$
\item Create a more complex ensemble: 2 $\ket{+z}$, 2 $\ket{-z}$, 1 $\ket{+x}$, 1 $\ket{-x}$
\item Use pairing to simplify the calculation of expectation values
\item Discuss: Can you always use pairing? What happens with ensembles that have states from all three axes?
\end{enumerate}
\end{blochcubeactivity}

\section{Advanced Topics: Beyond the Cube}

\subsection{The Full Bloch Sphere}

While the Bloch cube restricts us to six special states, the full Bloch sphere encompasses all possible states of a two-level system. Any point on the unit sphere corresponds to a valid quantum state:
\begin{equation}
\ket{\psi(\theta,\phi)} = \cos(\theta/2)\ket{+z} + e^{i\phi}\sin(\theta/2)\ket{-z}
\end{equation}

The Bloch vector $\vec{r} = (r_x, r_y, r_z)$ provides an alternative parameterization:
\begin{equation}
\rho = \frac{1}{2}(\hat{I} + \vec{r} \cdot \vec{\sigma})
\end{equation}

For pure states, $|\vec{r}| = 1$ (on the sphere surface). For mixed states, $|\vec{r}| < 1$ (inside the sphere). The completely mixed state $\rho = \frac{1}{2}\hat{I}$ sits at the origin with $\vec{r} = 0$.

\subsection{Quantum Process Tomography}

Given an unknown quantum state, how can we determine it experimentally? This is the problem of quantum state tomography. For a two-level system, we need to measure in three different bases to fully characterize the state.

If we measure many copies of the state in the X, Y, and Z bases, obtaining expectation values:
\begin{align}
\langle\sigma_x\rangle &= \text{Tr}(\rho\sigma_x) = r_x\\
\langle\sigma_y\rangle &= \text{Tr}(\rho\sigma_y) = r_y\\
\langle\sigma_z\rangle &= \text{Tr}(\rho\sigma_z) = r_z
\end{align}

We can reconstruct the density operator from these Bloch vector components. This provides a complete experimental characterization of the quantum state.

\section{The Postulates of Quantum Mechanics}

Having developed the mathematical and physical tools, we can now state the fundamental postulates of quantum mechanics in a form particularly suited to our Bloch cube approach:

\textbf{Postulate 1 (State Space):} The state of a quantum system is represented by a normalized vector $\ket{\psi}$ in a complex Hilbert space. For our two-state system, this space is two-dimensional.

\textbf{Postulate 2 (Observables):} Physical quantities are represented by operators constructed from measurement procedures. The eigenvalues of these operators correspond to possible measurement outcomes. Since measurement outcomes must be real numbers, these operators are necessarily Hermitian.

\textbf{Postulate 3 (Born Rule):} Upon measuring an observable $\hat{A}$ for a system in state $\ket{\psi}$, the probability of obtaining eigenvalue $a_i$ is $P(\ket{\psi}\rightarrow\ket{a_i}) = |\braket{a_i}{\psi}|^2$, where $\ket{a_i}$ is the corresponding eigenstate.

\textbf{Postulate 4 (State Collapse):} After obtaining measurement result $a_i$, the system's state becomes $\ket{a_i}$.

\textbf{Postulate 5 (Unitary Evolution):} The time evolution of an isolated quantum system is governed by the unitary operator $\hat{U}(t) = e^{-i\hat{H}t/\hbar}$, where $\hat{H}$ is the Hamiltonian.

These postulates, while abstractly stated, become concrete through Bloch cube manipulations. Each postulate corresponds to physical operations: preparation, measurement, and rotation of the cube.

\begin{blochcubeactivity}{1.6: Verifying the Postulates}
Using your Bloch cube, verify each postulate:
\begin{enumerate}
    \item \textbf{State Space}: Show that any face can be expressed as a normalized superposition
    \item \textbf{Observables}: Identify the three measurement types with cube orientations
    \item \textbf{Born Rule}: Prepare $\ket{+x}$ and measure Z many times to verify 50-50 statistics
    \item \textbf{Collapse}: After measuring Z on $\ket{+x}$, verify the state is now $\ket{\pm z}$
    \item \textbf{Evolution}: Show that rotations preserve normalization and orthogonality
\end{enumerate}
\end{blochcubeactivity}

\section{Matrix Representations}

In the computational basis, quantum states are represented as column vectors and operators as matrices. This section formally introduces the matrix representations that complement our Bloch cube approach.

\subsection{Single-Qubit Matrix Representations}

For a single qubit, the computational (Z) basis states have the matrix representations:
\begin{equation}
\ket{+z}=\ket{0} \repeq{z} \begin{pmatrix} 1 \\ 0 \end{pmatrix}, \quad \ket{-z}=\ket{1} \repeq{z} \begin{pmatrix} 0 \\ 1 \end{pmatrix}
\end{equation}

The projection operators from earlier in this chapter can now be written explicitly as matrices. For example:
\begin{align}
\ketbra{0}{0} &\repeq{z} \begin{pmatrix} 1 & 0 \\ 0 & 0 \end{pmatrix}\\
\ketbra{1}{1} &\repeq{z} \begin{pmatrix} 0 & 0 \\ 0 & 1 \end{pmatrix}\\
\ketbra{0}{1} &\repeq{z} \begin{pmatrix} 0 & 1 \\ 0 & 0 \end{pmatrix}\\
\ketbra{1}{0} &\repeq{z} \begin{pmatrix} 0 & 0 \\ 1 & 0 \end{pmatrix}
\end{align}

The Pauli matrices, which represent measurements along the three coordinate axes, have the explicit forms:
\begin{align}
\sigma_x &\repeq{z} \begin{pmatrix} 0 & 1 \\ 1 & 0 \end{pmatrix}\\
\sigma_y &\repeq{z} \begin{pmatrix} 0 & -i \\ i & 0 \end{pmatrix}\\
\sigma_z &\repeq{z} \begin{pmatrix} 1 & 0 \\ 0 & -1 \end{pmatrix}
\end{align}

The rotation operators from earlier also have matrix representations. For example:
\begin{align}
\hat{R}_z(\theta) &= e^{-i\theta\sigma_z/2} \repeq{z} \begin{pmatrix} e^{-i\theta/2} & 0 \\ 0 & e^{i\theta/2} \end{pmatrix}\\
\hat{Z} &= \hat{R}_z(\pi/2) \repeq{z} \begin{pmatrix} 1 & 0 \\ 0 & i \end{pmatrix}
\end{align}

Similarly, the discrete rotation operators have the matrix forms:
\begin{align}
\hat{X} &\repeq{z} \frac{1}{2}\begin{pmatrix} 1+i & 1-i \\ 1-i & 1+i \end{pmatrix}\\
\hat{Y} &\repeq{z} \frac{1}{2}\begin{pmatrix} 1+i & -1-i \\ 1+i & 1+i \end{pmatrix}
\end{align}

\section{Comprehensive Basis Representations}

Before moving on to more advanced topics, we establish a comprehensive reference for the six Bloch cube states expressed in all three orthonormal bases. This reference will prove invaluable throughout our study of quantum mechanics, as it allows us to quickly read off the representation of any state in any basis.

\subsection{The Six Bloch Cube States in All Bases}

\begin{tcolorbox}[enhanced, title={Master Reference: All Bloch Cube States in All Bases}, colback=red!5, colframe=red!50!black, breakable]
\textbf{Z-basis representations:} $\{\ket{+z}, \ket{-z}\}$
\begin{align}
\ket{+x} &= \frac{1}{\sqrt{2}}(\ket{+z} + \ket{-z}) \repeq{z} \frac{1}{\sqrt{2}}\begin{pmatrix} 1 \\ 1 \end{pmatrix} \\
\ket{-x} &= \frac{1}{\sqrt{2}}(\ket{+z} - \ket{-z}) \repeq{z} \frac{1}{\sqrt{2}}\begin{pmatrix} 1 \\ -1 \end{pmatrix}\\
\ket{+y} &= \frac{1}{\sqrt{2}}(\ket{+z} + i\ket{-z}) \repeq{z} \frac{1}{\sqrt{2}}\begin{pmatrix} 1 \\ i \end{pmatrix}\\
\ket{-y} &= \frac{1}{\sqrt{2}}(\ket{+z} - i\ket{-z}) \repeq{z} \frac{1}{\sqrt{2}}\begin{pmatrix} 1 \\ -i \end{pmatrix}\\
\ket{+z} &= \ket{+z} \repeq{z} \begin{pmatrix} 1 \\ 0 \end{pmatrix}\\
\ket{-z} &= \ket{-z} \repeq{z} \begin{pmatrix} 0 \\ 1 \end{pmatrix}
\end{align}

\textbf{X-basis representations:} $\{\ket{+x}, \ket{-x}\}$
\begin{align}
\ket{+x} &= \ket{+x} \repeq{x} \begin{pmatrix} 1 \\ 0 \end{pmatrix}\\
\ket{-x} &= \ket{-x} \repeq{x} \begin{pmatrix} 0 \\ 1 \end{pmatrix}\\
\ket{+y} &= \frac{(1+i)\ket{+x} + (1-i)\ket{-x}}{2} \repeq{x} \frac{1}{2}\begin{pmatrix} 1+i \\ 1-i \end{pmatrix}\\
\ket{-y} &= \frac{(1-i)\ket{+x} + (1+i)\ket{-x}}{2} \repeq{x} \frac{1}{2}\begin{pmatrix} 1-i \\ 1+i \end{pmatrix}\\
\ket{+z} &= \frac{1}{\sqrt{2}}(\ket{+x} + \ket{-x}) \repeq{x} \frac{1}{\sqrt{2}}\begin{pmatrix} 1 \\ 1 \end{pmatrix}\\
\ket{-z} &= \frac{1}{\sqrt{2}}(\ket{+x} - \ket{-x}) \repeq{x} \frac{1}{\sqrt{2}}\begin{pmatrix} 1 \\ -1 \end{pmatrix}
\end{align}

\textbf{Y-basis representations:} $\{\ket{+y}, \ket{-y}\}$
\begin{align}
\ket{+x} &= \frac{(1-i)\ket{+y} + (1+i)\ket{-y}}{2} \repeq{y} \frac{1}{2}\begin{pmatrix} 1-i \\ 1+i \end{pmatrix}\\
\ket{-x} &= \frac{(1+i)\ket{+y} + (1-i)\ket{-y}}{2} \repeq{y} \frac{1}{2}\begin{pmatrix} 1+i \\ 1-i \end{pmatrix}\\
\ket{+y} &= \ket{+y} \repeq{y} \begin{pmatrix} 1 \\ 0 \end{pmatrix}\\
\ket{-y} &= \ket{-y} \repeq{y} \begin{pmatrix} 0 \\ 1 \end{pmatrix}\\
\ket{+z} &= \frac{1}{\sqrt{2}}(\ket{+y} + \ket{-y}) \repeq{y} \frac{1}{\sqrt{2}}\begin{pmatrix} 1 \\ 1 \end{pmatrix}\\
\ket{-z} &= \frac{1}{\sqrt{2}}(-i\ket{+y} + i\ket{-y}) \repeq{y} \frac{1}{\sqrt{2}}\begin{pmatrix} -i \\ i \end{pmatrix}
\end{align}
\end{tcolorbox}

\subsection{Using the Reference Table}

This comprehensive table enables several powerful calculations:

\begin{example}[Cross-Basis Inner Products]
Calculate $\braket{+x}{+y}$ using different basis representations.

\textbf{Method 1: Using Z-basis}
\begin{align}
\braket{+x}{+y} &= \left(\frac{1}{\sqrt{2}}(\bra{+z} + \bra{-z})\right)\left(\frac{1}{\sqrt{2}}(\ket{+z} + i\ket{-z})\right)\\
&= \frac{1}{2}(\braket{+z}{+z} + i\braket{+z}{-z} + \braket{-z}{+z} + i\braket{-z}{-z})\\
&= \frac{1}{2}(1 + 0 + 0 + i) = \frac{1+i}{2}
\end{align}

\textbf{Method 2: Using X-basis}
From the table: $\ket{+y} = \frac{(1+i)\ket{+x} + (1-i)\ket{-x}}{2}$
\begin{align}
\braket{+x}{+y} &= \bra{+x}\left(\frac{(1+i)\ket{+x} + (1-i)\ket{-x}}{2}\right)\\
&= \frac{1}{2}[(1+i)\braket{+x}{+x} + (1-i)\braket{+x}{-x}]\\
&= \frac{1}{2}[(1+i) \cdot 1 + (1-i) \cdot 0]\\
&= \frac{1+i}{2}
\end{align}

Both methods give the same result, confirming our calculation.

\textbf{Magnitude:} $|\braket{+x}{+y}| = \left|\frac{1+i}{2}\right| = \frac{\sqrt{2}}{2} = \frac{1}{\sqrt{2}}$

\textbf{Phase:} $\arg(\braket{+x}{+y}) = \arg(1+i) = \pi/4$
\end{example}

\subsection{Transformation Matrices Between Bases}

The basis transformations are most clearly understood through their basis-independent operator form, which then yields matrix representations in any chosen basis.

\subsubsection{Basis-Independent Form}

The transformation operators map basis states from one representation to another:

\begin{tcolorbox}[enhanced, title={Basis Transformation Operators}, colback=green!5, colframe=green!50!black]
\textbf{Z to X transformation:}
\begin{equation}
U_{xz} = \ketbra{+x}{+z} + \ketbra{-x}{-z}
\end{equation}
This operator maps: $\ket{+z} \mapsto \ket{+x}$ and $\ket{-z} \mapsto \ket{-x}$

\textbf{Z to Y transformation:}
\begin{equation}
U_{yz} = \ketbra{+y}{+z} + \ketbra{-y}{-z}
\end{equation}
This operator maps: $\ket{+z} \mapsto \ket{+y}$ and $\ket{-z} \mapsto \ket{-y}$

\textbf{X to Y transformation:}
\begin{equation}
U_{yx} = \ketbra{+y}{+x} + \ketbra{-y}{-x}
\end{equation}
This operator maps: $\ket{+x} \mapsto \ket{+y}$ and $\ket{-x} \mapsto \ket{-y}$
\end{tcolorbox}

These operators transform states according to their definition. To find the matrix representation of any operator $U$ in a given basis, we construct the matrix by computing all matrix elements in that basis.

\subsubsection{Deriving Matrix Representations}

To find the matrix representation of an operator $U$ in any basis, we compute the matrix elements $\langle i|U|j\rangle$ where $|i\rangle$ and $|j\rangle$ are the basis states. For example, in the Z-basis, the matrix elements are:
\begin{equation}
U \repeq{z} \begin{pmatrix} \langle +z|U|+z\rangle & \langle +z|U|-z\rangle \\ \langle -z|U|+z\rangle & \langle -z|U|-z\rangle \end{pmatrix}
\end{equation}

\begin{example}[Deriving $U_{xz}$ in the Z-basis]
The operator $U_{xz} = \ket{+x}\bra{+z} + \ket{-x}\bra{-z}$ has the following matrix elements in the Z-basis:

\begin{align}
\langle +z|U_{xz}|+z\rangle &= \langle +z|(\ket{+x}\bra{+z} + \ket{-x}\bra{-z})|+z\rangle\\
&= \langle +z|+x\rangle\underbrace{\langle +z|+z\rangle}_{=1} + \langle +z|-x\rangle\underbrace{\langle -z|+z\rangle}_{=0}\\
&= \langle +z|+x\rangle = \frac{1}{\sqrt{2}}
\end{align}

\begin{align}
\langle +z|U_{xz}|-z\rangle &= \langle +z|+x\rangle\underbrace{\langle +z|-z\rangle}_{=0} + \langle +z|-x\rangle\underbrace{\langle -z|-z\rangle}_{=1}\\
&= \langle +z|-x\rangle = \frac{1}{\sqrt{2}}
\end{align}

\begin{align}
\langle -z|U_{xz}|+z\rangle &= \langle -z|+x\rangle\underbrace{\langle +z|+z\rangle}_{=1} + \langle -z|-x\rangle\underbrace{\langle -z|+z\rangle}_{=0}\\
&= \langle -z|+x\rangle = \frac{1}{\sqrt{2}}
\end{align}

\begin{align}
\langle -z|U_{xz}|-z\rangle &= \langle -z|+x\rangle\underbrace{\langle +z|-z\rangle}_{=0} + \langle -z|-x\rangle\underbrace{\langle -z|-z\rangle}_{=1}\\
&= \langle -z|-x\rangle = \frac{-1}{\sqrt{2}}
\end{align}

Therefore:
\begin{equation}
U_{xz} \repeq{z} \frac{1}{\sqrt{2}}\begin{pmatrix} 1 & 1 \\ 1 & -1 \end{pmatrix}
\end{equation}

\textbf{Verification:} This operator transforms the Z-basis states to X-basis states:
\begin{align}
U_{xz}\ket{+z} &= \ket{+x} = \frac{1}{\sqrt{2}}(\ket{+z} + \ket{-z})\\
U_{xz}\ket{-z} &= \ket{-x} = \frac{1}{\sqrt{2}}(\ket{+z} - \ket{-z})
\end{align}

We can verify this using the matrix representation:
\begin{align}
U_{xz}\ket{+z} &\repeq{z} \frac{1}{\sqrt{2}}\begin{pmatrix} 1 & 1 \\ 1 & -1 \end{pmatrix}\begin{pmatrix} 1 \\ 0 \end{pmatrix} = \frac{1}{\sqrt{2}}\begin{pmatrix} 1 \\ 1 \end{pmatrix} = \ket{+x} \checkmark
\end{align}
\end{example}

\begin{keyidea}{Matrix Representations of Operators}
The matrix representation of any operator $U$ in a given basis is constructed from its matrix elements:
\begin{itemize}
\item In the Z-basis: $(U)_{ij} = \langle i|U|j\rangle$ where $i,j \in \{+z, -z\}$
\item The operator equation $U\ket{\psi} = \ket{\phi}$ becomes a matrix equation when expressed in any basis
\item Different bases give different matrix representations of the same operator
\item The physical content (eigenvalues, trace, determinant) remains invariant
\end{itemize}
\end{keyidea}

\begin{example}[Deriving $U_{yz}$ in the Z-basis]
The operator $U_{yz} = \ket{+y}\bra{+z} + \ket{-y}\bra{-z}$ has the following matrix elements in the Z-basis:

\begin{align}
\langle +z|U_{yz}|+z\rangle &= \langle +z|(\ket{+y}\bra{+z} + \ket{-y}\bra{-z})|+z\rangle\\
&= \langle +z|+y\rangle\underbrace{\langle +z|+z\rangle}_{=1} + \langle +z|-y\rangle\underbrace{\langle -z|+z\rangle}_{=0}\\
&= \langle +z|+y\rangle = \frac{1}{\sqrt{2}}
\end{align}

\begin{align}
\langle +z|U_{yz}|-z\rangle &= \langle +z|+y\rangle\underbrace{\langle +z|-z\rangle}_{=0} + \langle +z|-y\rangle\underbrace{\langle -z|-z\rangle}_{=1}\\
&= \langle +z|-y\rangle = \frac{1}{\sqrt{2}}
\end{align}

\begin{align}
\langle -z|U_{yz}|+z\rangle &= \langle -z|+y\rangle\underbrace{\langle +z|+z\rangle}_{=1} + \langle -z|-y\rangle\underbrace{\langle -z|+z\rangle}_{=0}\\
&= \langle -z|+y\rangle = \frac{i}{\sqrt{2}}
\end{align}

\begin{align}
\langle -z|U_{yz}|-z\rangle &= \langle -z|+y\rangle\underbrace{\langle +z|-z\rangle}_{=0} + \langle -z|-y\rangle\underbrace{\langle -z|-z\rangle}_{=1}\\
&= \langle -z|-y\rangle = \frac{-i}{\sqrt{2}}
\end{align}

Therefore:
\begin{equation}
U_{yz} \repeq{z} \frac{1}{\sqrt{2}}\begin{pmatrix} 1 & 1 \\ i & -i \end{pmatrix}
\end{equation}

\textbf{Verification:} This operator transforms:
\begin{align}
U_{yz}\ket{+z} &= \ket{+y} = \frac{1}{\sqrt{2}}(\ket{+z} + i\ket{-z})\\
U_{yz}\ket{-z} &= \ket{-y} = \frac{1}{\sqrt{2}}(\ket{+z} - i\ket{-z})
\end{align}
\end{example}

\begin{example}[Deriving $U_{yx}$ in the X-basis]
The operator $U_{yx} = \ket{+y}\bra{+x} + \ket{-y}\bra{-x}$ has the following matrix elements in the X-basis:

\begin{align}
\langle +x|U_{yx}|+x\rangle &= \langle +x|(\ket{+y}\bra{+x} + \ket{-y}\bra{-x})|+x\rangle\\
&= \langle +x|+y\rangle\underbrace{\langle +x|+x\rangle}_{=1} + \langle +x|-y\rangle\underbrace{\langle -x|+x\rangle}_{=0}\\
&= \langle +x|+y\rangle
\end{align}

From the correct expression $\ket{+y} = \frac{(1+i)\ket{+x} + (1-i)\ket{-x}}{2}$, we have $\braket{+x}{+y} = \frac{1+i}{2}$, so:
\begin{equation}
\langle +x|U_{yx}|+x\rangle = \frac{1+i}{2}
\end{equation}

Similarly:
\begin{align}
\langle +x|U_{yx}|-x\rangle &= \langle +x|+y\rangle\underbrace{\langle +x|-x\rangle}_{=0} + \langle +x|-y\rangle\underbrace{\langle -x|-x\rangle}_{=1} = \langle +x|-y\rangle = \frac{1-i}{2}\\
\langle -x|U_{yx}|+x\rangle &= \langle -x|+y\rangle\underbrace{\langle +x|+x\rangle}_{=1} + \langle -x|-y\rangle\underbrace{\langle -x|+x\rangle}_{=0} = \langle -x|+y\rangle = \frac{1-i}{2}\\
\langle -x|U_{yx}|-x\rangle &= \langle -x|+y\rangle\underbrace{\langle +x|-x\rangle}_{=0} + \langle -x|-y\rangle\underbrace{\langle -x|-x\rangle}_{=1} = \langle -x|-y\rangle = \frac{1+i}{2}
\end{align}

Therefore:
\begin{equation}
U_{yx} \repeq{x} \frac{1}{2}\begin{pmatrix} 1+i & 1-i \\ 1-i & 1+i \end{pmatrix}
\end{equation}
\end{example}

\subsubsection{Summary of Transformation Matrices}

The basis transformation operators have the following matrix representations:

\begin{align}
U_{xz} &\repeq{z} \frac{1}{\sqrt{2}}\begin{pmatrix} 1 & 1 \\ 1 & -1 \end{pmatrix} \quad \text{(transforms } \ket{+z} \to \ket{+x}, \ket{-z} \to \ket{-x}\text{)}\\
U_{yz} &\repeq{z} \frac{1}{\sqrt{2}}\begin{pmatrix} 1 & 1 \\ i & -i \end{pmatrix} \quad \text{(transforms } \ket{+z} \to \ket{+y}, \ket{-z} \to \ket{-y}\text{)}\\
U_{yx} &\repeq{x} \frac{1}{2}\begin{pmatrix} 1+i & 1-i \\ 1-i & 1+i \end{pmatrix} \quad \text{(transforms } \ket{+x} \to \ket{+y}, \ket{-x} \to \ket{-y}\text{)}
\end{align}

These matrices satisfy important properties:
\begin{itemize}
\item Unitarity: $U^\dagger U = U U^\dagger = I$
\item They implement the basis transformations $U_{xz}$, $U_{yz}$, and $U_{yx}$ defined above
\item Each matrix is constructed from the matrix elements $\langle i|U|j\rangle$ in the respective basis
\end{itemize}

\section{Chapter Summary}

This chapter has established the mathematical and conceptual foundations of quantum mechanics through the study of two-state systems. Beginning with the familiar mathematics of two-dimensional vectors, we developed the dot-vector notation as a bridge to Dirac's formalism. The introduction of the Bloch cube provided a tangible representation of quantum states, making abstract concepts physically manipulable.

We explored the fundamental features of quantum mechanics: superposition of states, the probabilistic nature of measurement described by the Born rule, state collapse following measurement, and unitary evolution. The non-commutativity of quantum operations emerged naturally from the non-commutativity of rotations. We provided complete basis representations for all six Bloch cube states, enabling calculations in any basis. Finally, we introduced the density operator formalism to handle situations of incomplete knowledge.

This foundation, built on the simplest non-trivial quantum system, contains all the essential features of quantum mechanics. The complete basis representation table will serve as a crucial reference throughout the course. In subsequent chapters, we will extend these concepts to multi-qubit systems, continuous variables, and ultimately to the wave mechanics of particles in space. The physical intuition developed through Bloch cube manipulations will guide us through increasingly abstract mathematical territories.

\input{book_problems/ch01_problems.tex}

\section*{References and Further Reading}
\addcontentsline{toc}{section}{References and Further Reading}

\begin{description}
\item[Levy, J., and Singh, C.] ``Teaching quantum formalism and postulates to first-year undergraduates.'' \emph{American Journal of Physics} \textbf{93}, 46--51 (2025). \href{https://doi.org/10.1119/5.0209945}{doi:10.1119/5.0209945}. The pedagogical paper introducing the dot-vec notation and Bloch cube approach used throughout this chapter.

\item[Nielsen, M.~A., and Chuang, I.~L.] \emph{Quantum Computation and Quantum Information}, 10th anniversary ed. Cambridge University Press, 2010. Chapters 1--2 develop the qubit and the postulates of quantum mechanics in the language used in this chapter; the standard reference for measurement formalism and the density operator.

\item[Sakurai, J.~J., and Napolitano, J.] \emph{Modern Quantum Mechanics}, 3rd ed. Cambridge University Press, 2017. Chapter 1 develops spin-1/2 systems and the Stern--Gerlach experiment as the foundation of quantum mechanics; the closest graduate-textbook parallel to the approach taken here.

\item[McIntyre, D.~H., Manogue, C.~A., and Tate, J.] \emph{Quantum Mechanics: A Paradigms Approach}. Pearson, 2012. Closest pedagogical sibling to this textbook: builds quantum mechanics from spin first, with Stern--Gerlach as the central experiment, before introducing wave mechanics.

\item[Feynman, R.~P., Leighton, R.~B., and Sands, M.] \emph{The Feynman Lectures on Physics}, Vol.~III: \emph{Quantum Mechanics}. Addison-Wesley, 1965. Available online at \href{https://www.feynmanlectures.caltech.edu/III_toc.html}{feynmanlectures.caltech.edu/III}. Qualitative two-state physics from a complementary point of view; particularly clear on superposition and amplitude addition.

\item[Townsend, J.~S.] \emph{A Modern Approach to Quantum Mechanics}, 2nd ed. University Science Books, 2012. Another spin-first textbook; useful for cross-reference, with extensive discussion of measurement, expectation values, and the density operator at the level of this chapter.
\end{description}

%% file: book_problems/ch01_problems.tex
\section{Problems}
\setcounter{hwproblem}{0}  

\problem{Dot-Vec Notation and Basic Operations}
Using dot-vec notation and the Bloch cube:
\begin{enumerate}[label=(\alph*)]
    \item Express the operator that exchanges $x$ and $y$ components of any vectors using operators composed of vectors and dot-vectors
    \item Apply this operator to $\hat{x}$ and verify the result
    \item Translate your result to Dirac notation for $\ket{+z}$ and $\ket{-z}$. That is, find an operator that swaps these two states
    \item Use your Bloch cube to verify this transformation physically
\end{enumerate}

\problem{Inner Products and Basis Representations}
For the quantum state $\ket{+y} = \frac{1}{\sqrt{2}}(\ket{+z} + i\ket{-z})$:
\begin{enumerate}[label=(\alph*)]
    \item Calculate the inner products $\braket{\pm x}{+y}$, $\braket{\pm z}{+y}$ using the Z-basis
    \item Verify orthogonality: $\braket{-y}{+y} = 0$ using the z-basis
    \item Determine measurement probabilities for X, Y, and Z measurements
    \item Express $\ket{+y}$ in the X-basis using the spectral resolution of identity
\end{enumerate}

\problem{Bloch Cube Rotations}
Perform the following sequence on your Bloch cube:
\begin{enumerate}[label=(\alph*)]
    \item Start with $\ket{+z}$ (top face up)
    \item Apply a $90^\circ$ rotation about the X-axis
    \item Apply a $90^\circ$ rotation about the Y-axis
    \item Identify the final state and express it in the Z-basis
    \item Calculate the expectation values for $\hat{X}$, $\hat{Y}$, and $\hat{Z}$ for your final state
\end{enumerate}

\problem{The Born Rule and Measurement Probabilities}
The Born rule states that for a system in state $\ket{\psi}$ and measurement eigenstate $\ket{\phi_i}$, the probability is $P(\ket{\psi}\rightarrow\ket{\phi_i}) = |\braket{\phi_i}{\psi}|^2$.
\begin{enumerate}[label=(\alph*)]
    \item Consider the Bloch cube state $\ket{+x} = \frac{1}{\sqrt{2}}(\ket{+z} + \ket{-z})$. Verify it is normalized, then use the Born rule to calculate the probabilities for obtaining $\ket{+z}$ and $\ket{-z}$ in a Z-measurement.
    \item For the state $\ket{+y} = \frac{1}{\sqrt{2}}(\ket{+z} + i\ket{-z})$, calculate all six probabilities: $P(\ket{+y}\rightarrow\ket{\pm z})$, $P(\ket{+y}\rightarrow\ket{\pm x})$, and $P(\ket{+y}\rightarrow\ket{\pm y})$.
    \item Using your Bloch cube: Create an ensemble of 8 cubes to represent the mixed state $\rho = 0.5\ketbra{+x}{+x} + 0.5\ketbra{-x}{-x}$. How many cubes should show each face? Calculate $\langle\sigma_z\rangle$ for this ensemble and compare to the pure state $\ket{+y}$.
    \item Conceptual: Explain in your own words why the Born rule gives probabilities (not deterministic outcomes) even when we have complete knowledge of the quantum state. How does this differ from classical probability?
\end{enumerate}

\problem{Properties of Bloch Cube States}
Analyze the six Bloch cube face states:
\begin{enumerate}[label=(\alph*)]
    \item Verify that each state is normalized: $\braket{\psi}{\psi} = 1$
    \item Show that opposite faces are orthogonal (verify for $\ket{\pm z}$ pair)
    \item Calculate the overlap $|\braket{+x}{+y}|^2$ and interpret physically
    \item Starting with $\ket{+y}$, compare final probability distributions for measuring Z then X versus measuring X then Z
\end{enumerate}

\problem{Born Rule Applications: Sequential Measurements and Expectation Values}
Explore deeper applications of the Born rule using only Bloch cube states:
\begin{enumerate}[label=(\alph*)]
    \item For each of the six Bloch cube states, calculate $\langle\sigma_z\rangle$ using the Born rule:
    \begin{enumerate}[label=(\roman*)]
        \item $\ket{+z}$ and $\ket{-z}$
        \item $\ket{+x}$ and $\ket{-x}$
        \item $\ket{+y}$ and $\ket{-y}$
        \item What pattern do you observe for opposite faces?
    \end{enumerate}
    \item Using your Bloch cube to simulate: Start with $\ket{+x}$. Perform 20 Z-measurements (using coin flips for the 50-50 probability). Record $+1$ for $\ket{+z}$ and $-1$ for $\ket{-z}$. Calculate the average and compare to the theoretical $\langle\sigma_z\rangle = 0$.
    \item Sequential measurements: Consider starting with $\ket{+y}$
    \begin{enumerate}[label=(\roman*)]
        \item Calculate the probability of measuring $\ket{+x}$ then $\ket{+z}$ in sequence
        \item Calculate the probability of measuring $\ket{+z}$ then $\ket{+x}$ in sequence
        \item Explain why these probabilities differ despite involving the same measurements
    \end{enumerate}
\end{enumerate}

\problem{Pauli Matrix Operations and Quantum States}
Working with the Pauli matrices and quantum states:
\begin{enumerate}[label=(\alph*)]
    \item Calculate $\sigma_x^2$, $\sigma_y^2$, and $\sigma_z^2$ explicitly
    \item Verify the commutation relation $[\sigma_x, \sigma_y] = 2i\sigma_z$
    \item Find how $\sigma_x$ acts on $\ket{+z}$ and $\ket{+y}$
    \item Compare the mixed state $\rho = 0.5\ket{+z}\bra{+z} + 0.5\ket{-z}\bra{-z}$ with the pure state $\ket{+x}$ by calculating $\langle\sigma_z\rangle$ for both
\end{enumerate}

\problem{Density Operator and Pure vs Mixed States}
The density operator $\hat{\rho}$ fully characterizes a quantum state. For a pure state $\ket{\psi}$, we have $\hat{\rho} = \ket{\psi}\bra{\psi}$ with $\text{Tr}(\hat{\rho}^2) = 1$. Mixed states have $\text{Tr}(\hat{\rho}^2) < 1$.
\begin{enumerate}[label=(\alph*)]
    \item Construct the density matrices for the six Bloch cube face states
    \item For each, verify that $\text{Tr}(\hat{\rho}) = 1$ and $\text{Tr}(\hat{\rho}^2) = 1$ (confirming purity)
    \item Consider the thermal state $\hat{\rho}_T = \frac{1}{2}\hat{I}$. Calculate $\text{Tr}(\hat{\rho}_T^2)$ and explain what this purity means physically
    \item Show that for any density matrix, the eigenvalues $p_i$ satisfy $0 \leq p_i \leq 1$ and $\sum_i p_i = 1$
    \item Construct a mixed state as a 50-50 mixture of $\ket{+z}$ and $\ket{+x}$: $\hat{\rho} = 0.5\ket{+z}\bra{+z} + 0.5\ket{+x}\bra{+x}$. Calculate its purity
\end{enumerate}

\problem{Matrix Exponential and Rotations on the Bloch Sphere}
Single-qubit rotations can be expressed as $U(\theta) = \exp(-i\theta\hat{n} \cdot \vec{\sigma}/2)$ where $\hat{n}$ is the rotation axis.
\begin{enumerate}[label=(\alph*)]
    \item Calculate $\exp(-i\theta\sigma_z/2)$ explicitly using the Taylor series
    \item Verify that your result matches the Z-rotation matrix $R_z(\theta)$
    \item Show that $\exp(-i\theta\sigma_x/2)\ket{+z}\exp(i\theta\sigma_x/2)$ rotates $\ket{+z}$ around the X-axis by angle $\theta$
    \item For $\theta = \pi/2$, compute the action of this operator on $\ket{+z}$ and identify the final state
\end{enumerate}

\problem{Commutation Relations and Observables}
Pauli matrices generate all of spin-1/2 quantum mechanics through their (anti)commutation relations:
\begin{enumerate}[label=(\alph*)]
    \item Verify all three commutation relations: $[\sigma_i, \sigma_j] = 2i\epsilon_{ijk}\sigma_k$
    \item Calculate the anticommutators $\{\sigma_i, \sigma_j\}$ for all pairs
    \item Show that $[\sigma_x, \sigma_z] = -[\sigma_z, \sigma_x]$ (anticommutativity of different axes)
    \item Use these relations to prove that $\sigma_x\sigma_y\sigma_z = i\hat{I}$
    \item Interpret: Why does the fact that $[\sigma_x, \sigma_z] \neq 0$ mean that X and Z measurements cannot be simultaneously sharp?
\end{enumerate}

\problem{Basis Transformations and Change of Representation}
The choice of basis is arbitrary in quantum mechanics. Different choices reveal different information:
\begin{enumerate}[label=(\alph*)]
    \item Write the six Bloch cube face states in all three basis choices: Z-basis, X-basis, and Y-basis
    \item For the state $\ket{+y}$, express it in the X-basis by finding its components in $\{\ket{+x}, \ket{-x}\}$
    \item Calculate the change-of-basis matrices: $U_{X \to Z}$, $U_{Y \to Z}$, and $U_{X \to Y}$
    \item Verify that these are unitary: $U^\dagger U = \hat{I}$
    \item Consider an operator $\hat{O} = \sigma_x$ in the Z-basis. What is its matrix representation in the X-basis?
\end{enumerate}

\problem{Expectation Values and Variance}
The expectation value $\langle\hat{A}\rangle = \text{Tr}(\hat{\rho}\hat{A})$ encodes measurement statistics:
\begin{enumerate}[label=(\alph*)]
    \item For the state $\ket{\psi} = \frac{1}{\sqrt{3}}(\ket{+z} + \ket{+x} + \ket{+y})$, calculate $\langle\sigma_x\rangle$, $\langle\sigma_y\rangle$, and $\langle\sigma_z\rangle$
    \item Define variance as $\Delta A = \sqrt{\langle A^2\rangle - \langle A\rangle^2}$. For the state $\ket{+x}$, calculate $\Delta\sigma_z$
    \item Show that for eigenstate measurements, the variance is zero
    \item For the equal superposition above, calculate the variances in all three directions
    \item Explain the uncertainty principle relationship between variances of non-commuting observables
\end{enumerate}

\problem{Quantum Interference and Superposition}
Quantum superposition exhibits interference effects unlike classical probabilities:
\begin{enumerate}[label=(\alph*)]
    \item Consider the state $\ket{\psi} = \frac{1}{\sqrt{2}}(\ket{+z} + e^{i\phi}\ket{-z})$ for arbitrary phase $\phi$. Show that the probability of measuring $\ket{+z}$ depends on $\phi$
    \item Calculate $P(\psi \to +z)$ for $\phi = 0$, $\phi = \pi/2$, and $\phi = \pi$ explicitly
    \item For $\phi = \pi$, simplify the state. What measurement basis makes this state an eigenstate?
    \item Show that the classical mixture $\rho = 0.5\ket{+z}\bra{+z} + 0.5\ket{-z}\bra{-z}$ gives the same probability for all phases $\phi$ (no interference)
    \item Use your Bloch cube to visualize why the superposition $\ket{+z} + e^{i\phi}\ket{-z}$ traces different paths on the sphere as $\phi$ varies, whereas the classical mixture is always at the center
\end{enumerate}

\problem{Projectors and Spectral Decomposition}
Projection operators $\hat{P}_i = \ket{\phi_i}\bra{\phi_i}$ extract components of quantum states:
\begin{enumerate}[label=(\alph*)]
    \item For the state $\ket{\psi} = \frac{1}{\sqrt{2}}(\ket{+z} + \ket{+x})$, construct the projection operators $\hat{P}_{+z}$ and $\hat{P}_{+x}$
    \item Apply these operators to $\ket{\psi}$ and interpret the results as measurement
    \item Show that $\hat{P}_{+z} + \hat{P}_{-z} = \hat{I}$ (completeness)
    \item For a basis $\{\ket{\phi_i}\}$, the spectral decomposition of an observable is $\hat{A} = \sum_i a_i \hat{P}_i$. Write $\sigma_z$ in this form using the Z-basis eigenstates
    \item Express the same operator in the X-basis and verify that the spectrum (eigenvalues) is unchanged
\end{enumerate}

\problem{Operator Polar Decomposition}
Any operator can be decomposed as $\hat{A} = \hat{U}\hat{P}$ where $\hat{U}$ is unitary and $\hat{P}$ is positive semidefinite:
\begin{enumerate}[label=(\alph*)]
    \item Show that $\sigma_x$ can be written in polar form (hint: $\sigma_x$ is both unitary and Hermitian)
    \item For the operator $\hat{A} = \hat{I} + i\sigma_y$ (not Hermitian), find its polar decomposition
    \item Construct the operator $\hat{A}^\dagger\hat{A}$ and verify that it is positive semidefinite
    \item For a general density matrix $\hat{\rho}$, explain why the polar decomposition naturally emerges from its spectral decomposition with $p_i \geq 0$
    \item Consider the operator $\hat{b} = \frac{1}{\sqrt{2}}(\sigma_x + i\sigma_y) = \ket{-z}\bra{+z}$. Compute $\hat{b}^\dagger\hat{b}$ and interpret as a projection
\end{enumerate}

\problem{Two-Level System Kinematics: State Space Geometry}
The space of all quantum states forms a geometric object---the Bloch ball:
\begin{enumerate}[label=(\alph*)]
    \item Show that any single-qubit density matrix can be written as $\hat{\rho} = \frac{1}{2}(\hat{I} + \vec{r} \cdot \vec{\sigma})$ with $|\vec{r}| \leq 1$
    \item For the pure states (surface of the Bloch ball), what is the magnitude of $\vec{r}$?
    \item For the completely mixed state $\hat{I}/2$ (center of the Bloch ball), what is $\vec{r}$?
    \item Show that the constraint $\text{Tr}(\hat{\rho}^2) \leq 1$ is equivalent to $|\vec{r}| \leq 1$
    \item For a state with $\vec{r} = (1/2, 0, 1/2)$ (normalized), construct the density matrix and verify its purity
\end{enumerate}

%% file: chapters/ch02_tensor_products.tex
\chapter{Tensor Products and Multi-Qubit Systems}
\label{ch:tensor_products}

\section{Introduction: The Mathematical Structure of Composite Systems}

\begin{keyidea}{Exponential Scaling in Quantum Mechanics}
The tensor product structure of quantum mechanics leads to an exponential growth in the dimension of Hilbert space as systems are combined. For $n$ two-level systems, the composite Hilbert space has dimension $2^n$. This exponential scaling underlies both the computational power of quantum systems and the difficulty of simulating them classically.
\end{keyidea}

Having established the quantum mechanics of single two-level systems in Chapter 1, we now face a fundamental question: how do we describe multiple quantum systems? The answer lies in the tensor product structure of quantum mechanics, a mathematical framework that governs how quantum systems combine while preserving their essential quantum properties.

This chapter develops the mathematical machinery needed to describe composite quantum systems. We emphasize that our focus remains purely on the kinematical structure---how to build larger Hilbert spaces from smaller ones. The rich physics of entanglement, which makes full use of this structure, is reserved for future study. Our goal here is to establish fluency with tensor products, multi-qubit states, and the operations that act upon them.

The journey from single qubits to multi-qubit systems represents more than a simple generalization. It reveals new phenomena that have no classical analog and opens the door to quantum information processing. By the end of this chapter, you will understand how to construct and manipulate quantum states of arbitrary numbers of qubits, setting the stage for quantum algorithms and quantum many-body physics.

\section{The Physical Setting: Multiple Distinguishable Systems}

\subsection{Combining Independent Quantum Systems}

Consider two Bloch cubes placed on a table, labeled ``1'' and ``2'' to distinguish them. Each cube represents an independent two-level quantum system, and each can be in any of the six face states we studied in Chapter 1. The fundamental question is: how do we mathematically describe the combined state of both cubes?

The answer employs the tensor product, denoted by $\otimes$. If cube 1 is in state $\ket{\psi}_1$ and cube 2 is in state $\ket{\phi}_2$, the combined system is described by the product state:
\begin{equation}
\ket{\Psi} = \ket{\psi}_1 \otimes \ket{\phi}_2
\end{equation}

This mathematical operation captures a profound physical principle: the state of a composite system contains complete information about all its constituents. The tensor product preserves the individual quantum properties of each subsystem while creating a unified description of the whole.

Several notational conventions exist for tensor products, all representing the same mathematical object. For the specific case where cube 1 shows $\ket{+z}$ and cube 2 shows $\ket{-x}$, we may write:
\begin{align}
\ket{+z}_1 \otimes \ket{-x}_2 &= \ket{+z}\ket{-x} \\
&= \ket{+z} \otimes \ket{-x}
\end{align}

The subscripts indicate which subsystem each state refers to, though these are often omitted when the context is clear. The convention of writing states adjacently without the explicit $\otimes$ symbol is particularly common in quantum information, where brevity is valued.

Let us examine a concrete example to solidify these concepts. Suppose cube 1 is in the state $\ket{+x}_1 = \frac{1}{\sqrt{2}}(\ket{+z}_1 + \ket{-z}_1)$ and cube 2 is in the state $\ket{+y}_2 = \frac{1}{\sqrt{2}}(\ket{+z}_2 + i\ket{-z}_2)$. The composite state is:

\begin{align}
\ket{\Psi} &= \ket{+x}_1 \otimes \ket{+y}_2\\
&= \frac{1}{\sqrt{2}}(\ket{+z}_1 + \ket{-z}_1) \otimes \frac{1}{\sqrt{2}}(\ket{+z}_2 + i\ket{-z}_2)\\
&= \frac{1}{2}(\ket{+z}_1 \otimes \ket{+z}_2 + i\ket{+z}_1 \otimes \ket{-z}_2 + \ket{-z}_1 \otimes \ket{+z}_2 + i\ket{-z}_1 \otimes \ket{-z}_2)\\
&= \frac{1}{2}(\ket{+z}\ket{+z} + i\ket{+z}\ket{-z} + \ket{-z}\ket{+z} + i\ket{-z}\ket{-z})
\end{align}

This expansion reveals that the composite system exists in a superposition of four different configurations, with specific phase relationships between them.

\subsection{The Computational Basis for Multiple Qubits}

For quantum information processing, we adopt the standard notation where $\ket{0} \equiv \ket{+z}$ and $\ket{1} \equiv \ket{-z}$. This choice, while arbitrary, has become universal in quantum computing. The historical reason for this convention relates to early implementations where the ground state (lower energy) was labeled $\ket{0}$ and the excited state (higher energy) was labeled $\ket{1}$.

For two qubits, the four computational basis states are:
\begin{align}
\ket{00} &= \ket{0}_1 \otimes \ket{0}_2 = \ket{+z}\ket{+z}\\
\ket{01} &= \ket{0}_1 \otimes \ket{1}_2 = \ket{+z}\ket{-z}\\
\ket{10} &= \ket{1}_1 \otimes \ket{0}_2 = \ket{-z}\ket{+z}\\
\ket{11} &= \ket{1}_1 \otimes \ket{1}_2 = \ket{-z}\ket{-z}
\end{align}

These four states form a complete orthonormal basis for the two-qubit Hilbert space. The orthonormality can be verified using the fundamental property that tensor products preserve inner products:
\begin{equation}
\braket{ij}{kl} = \braket{i}{k}_1 \braket{j}{l}_2 = \delta_{ik}\delta_{jl}
\end{equation}

For instance, $\braket{01}{10} = \braket{0}{1}_1\braket{1}{0}_2 = 0 \cdot 0 = 0$, confirming orthogonality.

The completeness of this basis means that any two-qubit state can be expressed as a linear combination:
\begin{equation}
\ket{\psi} = \alpha_{00}\ket{00} + \alpha_{01}\ket{01} + \alpha_{10}\ket{10} + \alpha_{11}\ket{11}
\end{equation}
where the complex coefficients satisfy the normalization condition $|\alpha_{00}|^2 + |\alpha_{01}|^2 + |\alpha_{10}|^2 + |\alpha_{11}|^2 = 1$.

\begin{blochcubeactivity}{2.1: Exploring Two-Qubit Basis States}
Using two Bloch cubes labeled 1 and 2:
\begin{enumerate}
    \item Arrange the cubes to represent each of the four computational basis states
    \item For each arrangement, verify that you can identify which qubit is in which state
    \item Count the total number of possible face-state combinations (36 = 6 \ensuremath{\times} 6)
    \item Create the state $\ket{+x}\ket{-y}$ and express it in the computational basis
    \item Calculate the probability of finding the system in each computational basis state
    \item Verify that your probabilities sum to 1
\end{enumerate}
\end{blochcubeactivity}

\subsection{Complete Basis Representations for Two-Qubit States}

Building on the single-qubit basis representations from Chapter 1, we now provide comprehensive tables for expressing common two-qubit states in different bases. This is essential for cross-basis calculations and understanding how measurements in different bases affect composite systems.

\begin{tcolorbox}[enhanced, breakable, title={Two-Qubit Computational Basis States in Product Bases}, colback=blue!5, colframe=blue!50!black]
\textbf{In the ZZ basis (computational basis):}
\begin{align}
\ket{00} &= \ket{00}\\
\ket{01} &= \ket{01}\\
\ket{10} &= \ket{10}\\
\ket{11} &= \ket{11}
\end{align}

\textbf{In the XX basis:}
\begin{align}
\ket{00} &= \frac{1}{2}(\ket{+x}\ket{+x} + \ket{+x}\ket{-x} + \ket{-x}\ket{+x} + \ket{-x}\ket{-x})\\
\ket{01} &= \frac{1}{2}(\ket{+x}\ket{+x} - \ket{+x}\ket{-x} + \ket{-x}\ket{+x} - \ket{-x}\ket{-x})\\
\ket{10} &= \frac{1}{2}(\ket{+x}\ket{+x} + \ket{+x}\ket{-x} - \ket{-x}\ket{+x} - \ket{-x}\ket{-x})\\
\ket{11} &= \frac{1}{2}(\ket{+x}\ket{+x} - \ket{+x}\ket{-x} - \ket{-x}\ket{+x} + \ket{-x}\ket{-x})
\end{align}

\textbf{In the YY basis:}
\begin{align}
\ket{00} &= \frac{1}{2}(\ket{+y}\ket{+y} - i\ket{+y}\ket{-y} - i\ket{-y}\ket{+y} - \ket{-y}\ket{-y})\\
\ket{01} &= \frac{1}{2}(\ket{+y}\ket{+y} + i\ket{+y}\ket{-y} - i\ket{-y}\ket{+y} + \ket{-y}\ket{-y})\\
\ket{10} &= \frac{1}{2}(\ket{+y}\ket{+y} - i\ket{+y}\ket{-y} + i\ket{-y}\ket{+y} + \ket{-y}\ket{-y})\\
\ket{11} &= \frac{1}{2}(\ket{+y}\ket{+y} + i\ket{+y}\ket{-y} + i\ket{-y}\ket{+y} - \ket{-y}\ket{-y})
\end{align}

\textbf{In mixed bases (e.g., XY basis):}
\begin{align}
\ket{00} &= \frac{1}{2}(\ket{+x}\ket{+y} - i\ket{+x}\ket{-y} + \ket{-x}\ket{+y} - i\ket{-x}\ket{-y})\\
\ket{01} &= \frac{1}{2}(\ket{+x}\ket{+y} + i\ket{+x}\ket{-y} + \ket{-x}\ket{+y} + i\ket{-x}\ket{-y})\\
\ket{10} &= \frac{1}{2}(\ket{+x}\ket{+y} - i\ket{+x}\ket{-y} - \ket{-x}\ket{+y} + i\ket{-x}\ket{-y})\\
\ket{11} &= \frac{1}{2}(\ket{+x}\ket{+y} + i\ket{+x}\ket{-y} - \ket{-x}\ket{+y} - i\ket{-x}\ket{-y})
\end{align}
\end{tcolorbox}

These representations enable us to calculate measurement probabilities in any basis. For example, if we have the state $\ket{00}$ and measure both qubits in the X basis, we can immediately see that all four outcomes $(\pm x, \pm x)$ occur with equal probability $1/4$.

\begin{example}[Cross-Basis Measurement Calculation]
Consider the state $\ket{\psi} = \frac{1}{\sqrt{2}}(\ket{00} + \ket{11})$. Calculate the probability of measuring $\ket{+x}\ket{+x}$.

Using the XX basis expansions:
\begin{align}
\ket{\psi} &= \frac{1}{\sqrt{2}}[\frac{1}{2}(\ket{+x}\ket{+x} + \ket{+x}\ket{-x} + \ket{-x}\ket{+x} + \ket{-x}\ket{-x})\\
&\quad + \frac{1}{2}(\ket{+x}\ket{+x} - \ket{+x}\ket{-x} - \ket{-x}\ket{+x} + \ket{-x}\ket{-x})]\\
&= \frac{1}{\sqrt{2}} \cdot \frac{1}{2}[2\ket{+x}\ket{+x} + 2\ket{-x}\ket{-x}]\\
&= \frac{1}{\sqrt{2}}(\ket{+x}\ket{+x} + \ket{-x}\ket{-x})
\end{align}

Therefore, $P(\ket{\psi}\rightarrow\ket{+x}\ket{+x}) = |(\bra{+x}\bra{+x})\ket{\psi}|^2 = \left|\frac{1}{\sqrt{2}}\right|^2 = \frac{1}{2}$.

We can verify this using the direct inner product:
\begin{align}
(\bra{+x}\bra{+x})\ket{\psi} &= \frac{1}{\sqrt{2}}[(\bra{+x}\bra{+x}) (\ket{+z}\ket{+z}) + (\bra{+x} \bra{+x}) (\ket{-z}\ket{-z})]\\
&= \frac{1}{\sqrt{2}}[\frac{1}{2} + \frac{1}{2}]\\
&= \frac{1}{\sqrt{2}}
\end{align}
\end{example}

\section{Mathematical Structure of Tensor Products}

\subsection{Vector Space Properties}

The tensor product of two vector spaces creates a new vector space whose dimension equals the product of the individual dimensions. For our two-qubit system, each qubit lives in a two-dimensional complex vector space $\mathcal{H}_1 \cong \mathcal{H}_2 \cong \mathbb{C}^2$, so the composite system inhabits a four-dimensional space $\mathcal{H}_1 \otimes \mathcal{H}_2 \cong \mathbb{C}^4$.

The tensor product operation satisfies several crucial properties that ensure mathematical consistency. First, it exhibits bilinearity:
\begin{align}
(a\ket{\psi} + b\ket{\phi}) \otimes \ket{\chi} &= a\ket{\psi} \otimes \ket{\chi} + b\ket{\phi} \otimes \ket{\chi}\\
\ket{\psi} \otimes (c\ket{\chi} + d\ket{\xi}) &= c\ket{\psi} \otimes \ket{\chi} + d\ket{\psi} \otimes \ket{\xi}
\end{align}

This bilinearity is not merely a mathematical convenience---it encodes the principle of quantum superposition for composite systems. If a subsystem can exist in a superposition of states, then the composite system inherits this superposition structure.

Second, the tensor product is associative:
\begin{equation}
(\ket{\psi} \otimes \ket{\phi}) \otimes \ket{\chi} = \ket{\psi} \otimes (\ket{\phi} \otimes \ket{\chi})
\end{equation}

This property ensures that the order of grouping doesn't matter when combining three or more systems, though we must maintain the sequential ordering of the systems themselves. For three qubits, we can write $\ket{\psi}_1 \otimes \ket{\phi}_2 \otimes \ket{\chi}_3$ without ambiguity about how to parse this expression.

Importantly, the tensor product is not commutative:
\begin{equation}
\ket{\psi} \otimes \ket{\phi} \neq \ket{\phi} \otimes \ket{\psi}
\end{equation}

The order matters because we're describing distinguishable systems. Swapping the order would mean exchanging which physical system is in which state. This non-commutativity reflects the physical reality that qubit 1 and qubit 2 are different objects, perhaps located at different positions in space or implemented using different physical degrees of freedom.

\subsection{Expanding Superposition States}

The bilinearity of tensor products becomes particularly important when dealing with superposition states. Consider the case where the first qubit is in the superposition state $\ket{+x}_1 = \frac{1}{\sqrt{2}}(\ket{0}_1 + \ket{1}_1)$ while the second qubit is in the definite state $\ket{0}_2$. The combined state is:

\begin{align}
\ket{\Psi} &= \ket{+x}_1 \otimes \ket{0}_2\\
&= \frac{1}{\sqrt{2}}(\ket{0}_1 + \ket{1}_1) \otimes \ket{0}_2\\
&= \frac{1}{\sqrt{2}}(\ket{0}_1 \otimes \ket{0}_2 + \ket{1}_1 \otimes \ket{0}_2)\\
&= \frac{1}{\sqrt{2}}(\ket{00} + \ket{10})
\end{align}

This expansion demonstrates how single-qubit superpositions lead to multi-qubit superpositions. The resulting state is a coherent superposition of two computational basis states, with the relative phase preserved from the original single-qubit superposition.

Let's examine a more complex example where both qubits are in superposition:

\begin{example}[Both Qubits in Superposition]
Consider $\ket{\psi}_1 = \frac{1}{\sqrt{2}}(\ket{0}_1 + i\ket{1}_1) = \ket{+y}_1$ and $\ket{\phi}_2 = \frac{1}{\sqrt{2}}(\ket{0}_2 - \ket{1}_2) = \ket{-x}_2$.

The composite state is:
\begin{align}
\ket{\Psi} &= \ket{+y}_1 \otimes \ket{-x}_2\\
&= \frac{1}{\sqrt{2}}(\ket{0}_1 + i\ket{1}_1) \otimes \frac{1}{\sqrt{2}}(\ket{0}_2 - \ket{1}_2)\\
&= \frac{1}{2}[(\ket{0}_1 + i\ket{1}_1) \otimes (\ket{0}_2 - \ket{1}_2)]\\
&= \frac{1}{2}[\ket{0}_1 \otimes \ket{0}_2 - \ket{0}_1 \otimes \ket{1}_2 + i\ket{1}_1 \otimes \ket{0}_2 - i\ket{1}_1 \otimes \ket{1}_2]\\
&= \frac{1}{2}(\ket{00} - \ket{01} + i\ket{10} - i\ket{11})
\end{align}

Note how the phases from both individual superpositions combine in the final expression. The phase structure becomes richer as we combine more complex superpositions.
\end{example}

\subsection{Inner Products in Composite Systems}

The inner product in the tensor product space inherits its structure from the inner products of the constituent spaces. For product states, we have:
\begin{equation}
\braket{\psi_1 \otimes \phi_2}{\chi_1 \otimes \xi_2} = \braket{\psi_1}{\chi_1}_1 \braket{\phi_2}{\xi_2}_2
\end{equation}

This factorization property is fundamental to many calculations in quantum information. It allows us to compute inner products efficiently by handling each subsystem separately.

For general states that are not simple tensor products, we must expand them in a common basis:

\begin{example}[Inner Product of General States]
Let $\ket{\Psi} = \frac{1}{\sqrt{2}}(\ket{00} + \ket{11})$ and $\ket{\Phi} = \frac{1}{\sqrt{2}}(\ket{01} + \ket{10})$.

Calculate $\braket{\Phi}{\Psi}$:
\begin{align}
\braket{\Phi}{\Psi} &= \frac{1}{\sqrt{2}}(\bra{01} + \bra{10}) \cdot \frac{1}{\sqrt{2}}(\ket{00} + \ket{11})\\
&= \frac{1}{2}[\braket{01}{00} + \braket{01}{11} + \braket{10}{00} + \braket{10}{11}]\\
&= \frac{1}{2}[0 + 0 + 0 + 0]\\
&= 0
\end{align}

The states are orthogonal! This is a special property---these particular superpositions form part of an orthonormal basis for the two-qubit space known as the Bell basis.
\end{example}

\section{Operations on Composite Systems}

\subsection{Local Operations}

A fundamental class of operations on composite systems consists of those that act on only one subsystem while leaving others unchanged. These local operations are represented mathematically by tensoring the desired operation with identity operators on the untouched subsystems.

For a two-qubit system, an operation $\hat{U}$ acting only on the first qubit is represented as $\hat{U}_1 = \hat{U} \otimes \hat{I}$, where $\hat{I}$ is the $2 \times 2$ identity matrix. Similarly, an operation acting only on the second qubit is $\hat{U}_2 = \hat{I} \otimes \hat{U}$.

The physical interpretation is straightforward: we perform a quantum operation on one qubit while doing nothing to the other. This might correspond to applying a magnetic field to one qubit while shielding the other, or addressing one qubit with a laser pulse in an optical system.

Consider the Pauli X operation (NOT gate) applied to the first qubit:
\begin{equation}
\hat{X}_1 = \hat{X} \otimes \hat{I} \repeq{z} \begin{pmatrix} 0 & 1 \\ 1 & 0 \end{pmatrix} \otimes \begin{pmatrix} 1 & 0 \\ 0 & 1 \end{pmatrix} \repeq{z} \begin{pmatrix}
0 & 0 & 1 & 0 \\
0 & 0 & 0 & 1 \\
1 & 0 & 0 & 0 \\
0 & 1 & 0 & 0
\end{pmatrix}
\end{equation}

This matrix acts on the four-dimensional two-qubit space, flipping the first qubit while leaving the second unchanged. We can verify this by applying it to basis states:
\begin{align}
\hat{X}_1\ket{00} &= \ket{10}\\
\hat{X}_1\ket{01} &= \ket{11}\\
\hat{X}_1\ket{10} &= \ket{00}\\
\hat{X}_1\ket{11} &= \ket{01}
\end{align}

Let's examine how local operations affect superposition states:

\begin{example}[Local Operation on Superposition]
Apply $\hat{Y}_2$ (Pauli Y on the second qubit) to the state $\ket{\psi} = \frac{1}{\sqrt{2}}(\ket{00} + \ket{11})$.

First, construct the operator:
\begin{equation}
\hat{Y}_2 = \hat{I} \otimes \hat{Y} \repeq{z} \begin{pmatrix} 1 & 0 \\ 0 & 1 \end{pmatrix} \otimes \begin{pmatrix} 0 & -i \\ i & 0 \end{pmatrix}
\repeq{z} \begin{pmatrix}
0 & -i & 0 & 0 \\
i & 0 & 0 & 0 \\
0 & 0 & 0 & -i \\
0 & 0 & i & 0
\end{pmatrix}
\end{equation}

Now apply it:
\begin{align}
\hat{Y}_2\ket{\psi} &= \hat{Y}_2 \cdot \frac{1}{\sqrt{2}}(\ket{00} + \ket{11})\\
&= \frac{1}{\sqrt{2}}(\hat{Y}_2\ket{00} + \hat{Y}_2\ket{11})\\
&= \frac{1}{\sqrt{2}}(i\ket{01} - i\ket{10})\\
&= \frac{i}{\sqrt{2}}(\ket{01} - \ket{10})
\end{align}

The operation has transformed our initial state into a different superposition with a global phase factor of $i$.
\end{example}

\subsection{Multi-Qubit Operations}

Beyond local operations, quantum mechanics allows operations that act on multiple qubits simultaneously. These operations cannot be decomposed into tensor products of single-qubit operations and represent genuinely multi-qubit phenomena. They are essential for creating correlations between qubits and form the backbone of quantum algorithms.

The controlled-NOT (CNOT) gate serves as the canonical example of a two-qubit operation. This gate flips the state of the second qubit (target) if and only if the first qubit (control) is in state $\ket{1}$. Its action on the computational basis is:
\begin{align}
\text{CNOT}\ket{00} &= \ket{00}\\
\text{CNOT}\ket{01} &= \ket{01}\\
\text{CNOT}\ket{10} &= \ket{11}\\
\text{CNOT}\ket{11} &= \ket{10}
\end{align}

In matrix form:
\begin{equation}
\text{CNOT} \repeq{z} \begin{pmatrix}
1 & 0 & 0 & 0 \\
0 & 1 & 0 & 0 \\
0 & 0 & 0 & 1 \\
0 & 0 & 1 & 0
\end{pmatrix}
\end{equation}

The CNOT gate can be understood as a coherent version of classical conditional logic. In classical computing, we might write ``IF (qubit 1 = 1) THEN flip qubit 2.'' The quantum version maintains coherence, so superpositions are preserved:

\begin{example}[CNOT on Superposition States]
Apply CNOT to $\ket{\psi} = \frac{1}{\sqrt{2}}(\ket{0} + \ket{1}) \otimes \ket{0} = \frac{1}{\sqrt{2}}(\ket{00} + \ket{10})$.

\begin{align}
\text{CNOT}\ket{\psi} &= \text{CNOT} \cdot \frac{1}{\sqrt{2}}(\ket{00} + \ket{10})\\
&= \frac{1}{\sqrt{2}}(\text{CNOT}\ket{00} + \text{CNOT}\ket{10})\\
&= \frac{1}{\sqrt{2}}(\ket{00} + \ket{11})
\end{align}

The result is an entangled state! The two qubits are now correlated---measuring the first qubit determines the state of the second. While we defer detailed discussion of entanglement to later chapters, this example shows how multi-qubit gates create quantum correlations.
\end{example}

The CNOT gate plays a fundamental role in quantum computation, serving as a basic building block for more complex operations. When combined with single-qubit gates, it forms a universal gate set capable of implementing any quantum computation.

Another important two-qubit gate is the controlled-Z (CZ) gate:
\begin{equation}
\text{CZ} \repeq{z} \begin{pmatrix}
1 & 0 & 0 & 0 \\
0 & 1 & 0 & 0 \\
0 & 0 & 1 & 0 \\
0 & 0 & 0 & -1
\end{pmatrix}
\end{equation}

This gate applies a phase of -1 if and only if both qubits are in state $\ket{1}$. Unlike CNOT, the CZ gate is symmetric with respect to its two inputs---there's no distinction between control and target.

\subsection{Measurement in Composite Systems}

Measurements on composite systems can be either local (measuring individual qubits) or global (measuring multi-qubit observables). Local measurements are the most common in practice and have particularly simple descriptions.

When we measure just the first qubit of a two-qubit system in the computational basis, we're asking: ``Is the first qubit in state $\ket{0}$ or $\ket{1}$?'' The measurement operators are:
\begin{align}
\hat{P}_0^{(1)} &= \ket{0}\bra{0} \otimes \hat{I} \repeq{z} \begin{pmatrix}
1 & 0 & 0 & 0 \\
0 & 1 & 0 & 0 \\
0 & 0 & 0 & 0 \\
0 & 0 & 0 & 0
\end{pmatrix}\\
\hat{P}_1^{(1)} &= \ket{1}\bra{1} \otimes \hat{I} \repeq{z} \begin{pmatrix}
0 & 0 & 0 & 0 \\
0 & 0 & 0 & 0 \\
0 & 0 & 1 & 0 \\
0 & 0 & 0 & 1
\end{pmatrix}
\end{align}

These projectors satisfy $\hat{P}_0^{(1)} + \hat{P}_1^{(1)} = \hat{I}$ and $\hat{P}_0^{(1)}\hat{P}_1^{(1)} = 0$, as required for a complete measurement.

\begin{example}[Partial Measurement]
Consider measuring the first qubit of $\ket{\psi} = \frac{1}{2}(\ket{00} + \ket{01} + \ket{10} + \ket{11})$.

The probability of finding the first qubit in state $\ket{0}$ is:
\begin{align}
P(\ket{\psi}\rightarrow\ket{0}_1) &= \bra{\psi}\hat{P}_0^{(1)}\ket{\psi}\\
&= \left|\frac{1}{2}(\ket{00} + \ket{01})\right|^2\\
&= \frac{1}{4}(2)^2 = \frac{1}{2}
\end{align}

If we obtain outcome 0, the post-measurement state is:
\begin{align}
\ket{\psi'} &= \frac{\hat{P}_0^{(1)}\ket{\psi}}{\sqrt{P(\ket{\psi}\rightarrow\ket{0}_1)}}\\
&= \frac{\frac{1}{2}(\ket{00} + \ket{01})}{\sqrt{1/2}}\\
&= \frac{1}{\sqrt{2}}(\ket{00} + \ket{01})\\
&= \ket{0} \otimes \frac{1}{\sqrt{2}}(\ket{0} + \ket{1})
\end{align}

The first qubit is now definitely in state $\ket{0}$, while the second qubit remains in superposition.
\end{example}

\section{Scaling to Many Qubits}

\subsection{The General Multi-Qubit Framework}

The tensor product structure extends naturally to systems with more than two qubits. For $n$ qubits, the computational basis consists of $2^n$ states, each specified by a binary string of length $n$. The basis states are written as $\ket{i_1 i_2 \ldots i_n}$ where each $i_j \in \{0,1\}$.

The dimension of the Hilbert space grows exponentially with the number of qubits:
\begin{itemize}
\item 1 qubit: 2 dimensions
\item 2 qubits: 4 dimensions
\item 3 qubits: 8 dimensions
\item 10 qubits: 1,024 dimensions
\item 20 qubits: 1,048,576 dimensions
\item 50 qubits: $\approx 10^{15}$ dimensions
\end{itemize}

This exponential scaling has profound implications. On one hand, it provides the computational power that makes quantum algorithms potentially more powerful than classical ones. On the other hand, it makes the classical simulation of quantum systems extremely challenging, as the amount of information needed to specify a general $n$-qubit state grows exponentially.

For three qubits, the eight computational basis states are:
\begin{align}
&\ket{000}, \ket{001}, \ket{010}, \ket{011},\\
&\ket{100}, \ket{101}, \ket{110}, \ket{111}
\end{align}

A general three-qubit state requires eight complex amplitudes (subject to normalization), meaning 15 real parameters are needed to fully specify the state:
\begin{equation}
\ket{\psi} = \sum_{i,j,k \in \{0,1\}} \alpha_{ijk}\ket{ijk}
\end{equation}
where $\sum_{i,j,k} |\alpha_{ijk}|^2 = 1$.

Let's explore a specific three-qubit state:

\begin{example}[Three-Qubit GHZ State]
Consider the three-qubit state $\ket{\text{GHZ}} = \frac{1}{\sqrt{2}}(\ket{000} + \ket{111})$.

This state exhibits three-party correlations. Let's calculate the probability of various measurement outcomes:

\textbf{Measuring all three qubits:}
\begin{align}
P(\ket{\text{GHZ}}\rightarrow\ket{000}) &= |\braket{000}{\text{GHZ}}|^2 = \left|\frac{1}{\sqrt{2}}\right|^2 = \frac{1}{2}\\
P(\ket{\text{GHZ}}\rightarrow\ket{111}) &= |\braket{111}{\text{GHZ}}|^2 = \left|\frac{1}{\sqrt{2}}\right|^2 = \frac{1}{2}\\
P(\ket{\text{GHZ}}\rightarrow\ket{\text{any other}}) &= 0
\end{align}

\textbf{Measuring just the first qubit:}
\begin{align}
P(\ket{\text{GHZ}}\rightarrow\ket{0}_1) &= |\braket{0**}{\text{GHZ}}|^2 = \left|\frac{1}{\sqrt{2}}\ket{000}\right|^2 = \frac{1}{2}\\
P(\ket{\text{GHZ}}\rightarrow\ket{1}_1) &= |\braket{1**}{\text{GHZ}}|^2 = \left|\frac{1}{\sqrt{2}}\ket{111}\right|^2 = \frac{1}{2}
\end{align}

If we measure the first qubit and get 0, the state collapses to $\ket{000}$. If we get 1, it collapses to $\ket{111}$. This demonstrates strong correlations among all three qubits.
\end{example}

\subsection{Operations on Selected Qubits}

In multi-qubit systems, we often need to apply operations to specific qubits while leaving others untouched. For an $n$-qubit system, applying a single-qubit operation $\hat{U}$ to the $k$-th qubit is represented as:

\begin{equation}
\hat{U}^{(k)} = \hat{I} \otimes \cdots \otimes \hat{I} \otimes \underbrace{\hat{U}}_{k\text{-th position}} \otimes \hat{I} \otimes \cdots \otimes \hat{I}
\end{equation}

This notation, while cumbersome, precisely specifies which qubit is affected. The operation acts as the identity on all qubits except the $k$-th, where it applies the transformation $\hat{U}$.

For example, in a three-qubit system, applying a Hadamard gate to the middle qubit:
\begin{equation}
\hat{H}^{(2)} = \hat{I} \otimes \hat{H} \otimes \hat{I}
\end{equation}

where $\hat{H} \repeq{z} \frac{1}{\sqrt{2}}\begin{pmatrix} 1 & 1 \\ 1 & -1 \end{pmatrix}$ is the Hadamard gate.

Multi-qubit gates can similarly be applied to specific subsets of qubits. For instance, a CNOT gate with qubit $j$ as control and qubit $k$ as target affects only those two qubits while acting as the identity on all others. The notation $\text{CNOT}_{j,k}$ specifies which qubits are involved.

\begin{example}[Circuit on Three Qubits]
Starting from $\ket{000}$, apply the following sequence:
1. $\hat{H}^{(1)}$ (Hadamard on first qubit)
2. $\text{CNOT}_{1,2}$ (CNOT from qubit 1 to qubit 2)
3. $\text{CNOT}_{2,3}$ (CNOT from qubit 2 to qubit 3)

Step 1:
\begin{align}
\hat{H}^{(1)}\ket{000} &= \frac{1}{\sqrt{2}}(\ket{0} + \ket{1}) \otimes \ket{00}\\
&= \frac{1}{\sqrt{2}}(\ket{000} + \ket{100})
\end{align}

Step 2:
\begin{align}
\text{CNOT}_{1,2}\frac{1}{\sqrt{2}}(\ket{000} + \ket{100}) &= \frac{1}{\sqrt{2}}(\ket{000} + \ket{110})
\end{align}

Step 3:
\begin{align}
\text{CNOT}_{2,3}\frac{1}{\sqrt{2}}(\ket{000} + \ket{110}) &= \frac{1}{\sqrt{2}}(\ket{000} + \ket{111})
\end{align}

We've created the GHZ state! This circuit demonstrates how local operations and two-qubit gates can create complex multi-qubit states.
\end{example}

\subsection{The Challenge of Classical Simulation}

The exponential growth of Hilbert space dimension presents both opportunities and challenges. For quantum computation, this exponential scaling potentially allows quantum computers to explore vast solution spaces efficiently. However, it also means that classical simulation of quantum systems quickly becomes intractable.

Consider the memory requirements for storing a general quantum state:
\begin{itemize}
\item 10 qubits: $2^{10} = 1,024$ complex numbers = 16 KB (assuming double precision)
\item 20 qubits: $2^{20} \approx 10^6$ complex numbers = 16 MB
\item 30 qubits: $2^{30} \approx 10^9$ complex numbers = 16 GB
\item 40 qubits: $2^{40} \approx 10^{12}$ complex numbers = 16 TB
\item 50 qubits: $2^{50} \approx 10^{15}$ complex numbers = 16 PB
\end{itemize}

Beyond memory, the computational cost of simulating quantum operations also scales exponentially. Applying a general $n$-qubit gate requires $O(2^n)$ operations. This exponential barrier motivates the development of actual quantum computers, which naturally handle this complexity.

However, not all quantum states require exponential resources to describe. States with special structure, such as product states or states with limited entanglement, can often be represented more efficiently. This observation underlies various classical simulation techniques for restricted classes of quantum computations.

\section{Basis Transformations in Composite Systems}

\subsection{Local Basis Changes}

Understanding how basis transformations work in composite systems is crucial for quantum information processing. If we know how to transform between bases for individual qubits, we can determine the transformation for the composite system using the tensor product structure.

Consider expressing the computational basis state $\ket{00}$ in terms of the X-basis states $\ket{\pm x}$. Since $\ket{0} = \frac{1}{\sqrt{2}}(\ket{+x} + \ket{-x})$, we have:

\begin{align}
\ket{00} &= \ket{0}_1 \otimes \ket{0}_2\\
&= \frac{1}{\sqrt{2}}(\ket{+x}_1 + \ket{-x}_1) \otimes \frac{1}{\sqrt{2}}(\ket{+x}_2 + \ket{-x}_2)\\
&= \frac{1}{2}(\ket{+x}\ket{+x} + \ket{+x}\ket{-x} + \ket{-x}\ket{+x} + \ket{-x}\ket{-x})
\end{align}

This expansion shows that measuring both qubits in the X-basis would yield each of the four possible outcomes with equal probability 1/4.

Let's work through a more complex basis transformation:

\begin{example}[Mixed Basis Measurements]
Express $\ket{\psi} = \frac{1}{\sqrt{2}}(\ket{00} - i\ket{11})$ in the basis where the first qubit is measured in the X-basis and the second in the Y-basis.

Using the single-qubit transformations from Chapter 1:
\begin{align}
\ket{0} &= \frac{1}{\sqrt{2}}(\ket{+x} + \ket{-x})\\
\ket{1} &= \frac{1}{\sqrt{2}}(\ket{+x} - \ket{-x})\\
\ket{0} &= \frac{1}{\sqrt{2}}(\ket{+y} + \ket{-y})\\
\ket{1} &= \frac{1}{\sqrt{2}}(-i\ket{+y} + i\ket{-y})
\end{align}

Therefore:
\begin{align}
\ket{00} &= \frac{1}{2}(\ket{+x} + \ket{-x})(\ket{+y} + \ket{-y})\\
&= \frac{1}{2}(\ket{+x}\ket{+y} + \ket{+x}\ket{-y} + \ket{-x}\ket{+y} + \ket{-x}\ket{-y})\\
\ket{11} &= \frac{1}{2}(\ket{+x} - \ket{-x})(-i\ket{+y} + i\ket{-y})\\
&= \frac{1}{2}(-i\ket{+x}\ket{+y} + i\ket{+x}\ket{-y} + i\ket{-x}\ket{+y} - i\ket{-x}\ket{-y})
\end{align}

Substituting into $\ket{\psi}$:
\begin{align}
\ket{\psi} &= \frac{1}{\sqrt{2}}\left[\frac{1}{2}(\ket{+x}\ket{+y} + \ket{+x}\ket{-y} + \ket{-x}\ket{+y} + \ket{-x}\ket{-y})\right.\\
&\quad \left.- i \cdot \frac{1}{2}(-i\ket{+x}\ket{+y} + i\ket{+x}\ket{-y} + i\ket{-x}\ket{+y} - i\ket{-x}\ket{-y})\right]\\
&= \frac{1}{2\sqrt{2}}[(1-1)\ket{+x}\ket{+y} + (1+1)\ket{+x}\ket{-y} + (1+1)\ket{-x}\ket{+y} + (1-1)\ket{-x}\ket{-y}]\\
&= \frac{1}{\sqrt{2}}(\ket{+x}\ket{-y} + \ket{-x}\ket{+y})
\end{align}

So measuring in the XY basis yields either $(+x,-y)$ or $(-x,+y)$, each with probability 1/2.
\end{example}

\subsection{The Role of Phase in Multi-Qubit Systems}

Phase relationships become increasingly important in multi-qubit systems. Consider the difference between the states:
\begin{align}
\ket{\psi_1} &= \frac{1}{2}(\ket{00} + \ket{01} + \ket{10} + \ket{11})\\
\ket{\psi_2} &= \frac{1}{2}(\ket{00} + \ket{01} + \ket{10} - \ket{11})
\end{align}

While both give identical probabilities for computational basis measurements (each outcome has probability 1/4), they behave differently under other measurements or further operations.

To see the difference, let's measure both qubits in the X-basis:

\begin{example}[Phase Effects in Measurement]
For $\ket{\psi_1} = \frac{1}{2}(\ket{00} + \ket{01} + \ket{10} + \ket{11})$:

First, note that this can be factored:
\begin{align}
\ket{\psi_1} &= \frac{1}{2}(\ket{0} + \ket{1}) \otimes (\ket{0} + \ket{1})\\
&= \ket{+x} \otimes \ket{+x}
\end{align}

So measuring in the X-basis gives $(+x,+x)$ with certainty.

For $\ket{\psi_2} = \frac{1}{2}(\ket{00} + \ket{01} + \ket{10} - \ket{11})$:

This cannot be factored as a product state. Using the X-basis expansions:
\begin{align}
\ket{\psi_2} &= \frac{1}{2}[\ket{00} + \ket{01} + \ket{10} - \ket{11}]
\end{align}

Working through the basis transformation (as in our table), we find:
\begin{align}
P(\ket{\psi_2}\rightarrow\ket{+x,+x}) &= \left|\frac{1}{2}(1 + 1 + 1 - 1)\right|^2 = \frac{1}{2}\\
P(\ket{\psi_2}\rightarrow\ket{+x,-x}) &= \left|\frac{1}{2}(1 - 1 + 1 + 1)\right|^2 = \frac{1}{2}\\
P(\ket{\psi_2}\rightarrow\ket{-x,+x}) &= P(\ket{\psi_2}\rightarrow\ket{-x,-x}) = 0
\end{align}

The phase difference dramatically changes the measurement statistics!
\end{example}

These phase relationships are not merely mathematical artifacts but have observable consequences. They determine interference patterns in quantum algorithms and affect the outcomes of measurements in rotated bases. Understanding and controlling these phases is crucial for quantum computation and quantum information processing.

\newpage
\section{The Reduced Density Matrix and Partial Trace}

\subsection{Describing Subsystems of Composite States}

\begin{keyidea}{From Pure to Mixed: The Cost of Ignoring Correlations}
When we have access to only part of an entangled system, we lose quantum coherence. The reduced density matrix captures this transition from pure quantum states to mixed statistical states, revealing how quantum information is stored non-locally in correlations.
\end{keyidea}

So far, we have described composite quantum systems using state vectors in the full tensor product space. But what if we only have access to one part of a composite system? How do we describe what we can observe about qubit A when qubit B is inaccessible? The answer involves the reduced density matrix, obtained through an operation called the partial trace.

\subsection{The Density Matrix Formalism}

Before introducing the reduced density matrix, we briefly review the density matrix representation. Any quantum state $\ket{\psi}$ can be represented by its density matrix:
\begin{equation}
\rho = \ket{\psi}\bra{\psi}
\end{equation}

For a pure state, the density matrix satisfies $\rho^2 = \rho$ and $\text{Tr}(\rho^2) = 1$. The density matrix formalism naturally extends to mixed states (statistical ensembles of pure states), where $\text{Tr}(\rho^2) < 1$.

\subsection{The Partial Trace Operation}

For a two-qubit system in state $\ket{\psi}_{AB}$, the reduced density matrix for qubit A is obtained by "tracing out" qubit B:
\begin{equation}
\rho_A = \text{Tr}_B(\ket{\psi}\bra{\psi}_{AB}) = \sum_j (\mathbb{I}_A \otimes \bra{j}_B) \ket{\psi}\bra{\psi}_{AB} (\mathbb{I}_A \otimes \ket{j}_B)
\end{equation}

where $\{\ket{j}_B\}$ is any orthonormal basis for qubit B's Hilbert space.

\begin{mathematicalaside}{Matrix Construction of Projection Operators}
To understand how the partial trace works in matrix form, let's explicitly construct the projection operators. For a two-qubit system with a $4 \times 4$ density matrix, we need:

\textbf{The operators $\mathbb{I}_A \otimes \ket{j}_B$ (4\ensuremath{\times}2 matrices):}

For $j = 0$:
\begin{align}
\mathbb{I}_A \otimes \ket{0}_B &= \begin{pmatrix} 1 & 0 \\ 0 & 1 \end{pmatrix} \otimes \begin{pmatrix} 1 \\ 0 \end{pmatrix}\\
&= \begin{pmatrix} 
1 \cdot \begin{pmatrix} 1 \\ 0 \end{pmatrix} & 0 \cdot \begin{pmatrix} 1 \\ 0 \end{pmatrix} \\
0 \cdot \begin{pmatrix} 1 \\ 0 \end{pmatrix} & 1 \cdot \begin{pmatrix} 1 \\ 0 \end{pmatrix}
\end{pmatrix}
= \begin{pmatrix} 1 & 0 \\ 0 & 0 \\ 0 & 1 \\ 0 & 0 \end{pmatrix}
\end{align}

For $j = 1$:
\begin{align}
\mathbb{I}_A \otimes \ket{1}_B &= \begin{pmatrix} 1 & 0 \\ 0 & 1 \end{pmatrix} \otimes \begin{pmatrix} 0 \\ 1 \end{pmatrix}\\
&= \begin{pmatrix} 0 & 0 \\ 1 & 0 \\ 0 & 0 \\ 0 & 1 \end{pmatrix}
\end{align}

\textbf{The operators $\mathbb{I}_A \otimes \bra{j}_B$ (2\ensuremath{\times}4 matrices):}

For $j = 0$:
\begin{align}
\mathbb{I}_A \otimes \bra{0}_B &= \begin{pmatrix} 1 & 0 \\ 0 & 1 \end{pmatrix} \otimes \begin{pmatrix} 1 & 0 \end{pmatrix}\\
&= \begin{pmatrix} 
1 \cdot \begin{pmatrix} 1 & 0 \end{pmatrix} & 0 \cdot \begin{pmatrix} 1 & 0 \end{pmatrix} \\
0 \cdot \begin{pmatrix} 1 & 0 \end{pmatrix} & 1 \cdot \begin{pmatrix} 1 & 0 \end{pmatrix}
\end{pmatrix}
= \begin{pmatrix} 1 & 0 & 0 & 0 \\ 0 & 0 & 1 & 0 \end{pmatrix}
\end{align}

For $j = 1$:
\begin{align}
\mathbb{I}_A \otimes \bra{1}_B &= \begin{pmatrix} 1 & 0 \\ 0 & 1 \end{pmatrix} \otimes \begin{pmatrix} 0 & 1 \end{pmatrix}\\
&= \begin{pmatrix} 0 & 1 & 0 & 0 \\ 0 & 0 & 0 & 1 \end{pmatrix}
\end{align}

\textbf{The complete operation:}
The partial trace performs the matrix multiplication:
\[
(\mathbb{I}_A \otimes \bra{j}_B) \cdot \rho_{AB} \cdot (\mathbb{I}_A \otimes \ket{j}_B) = (2\times 4) \cdot (4\times 4) \cdot (4\times 2) = (2\times 2)
\]

This sequence of multiplications selects and extracts a $2 \times 2$ submatrix from the original $4 \times 4$ density matrix, corresponding to the reduced state of qubit A when B is in state $\ket{j}$. Summing over all values of $j$ gives the final reduced density matrix.

\textbf{Verification:} Let's verify this works with our example density matrix. For $j=0$:
\begin{align}
&(\mathbb{I}_A \otimes \bra{0}_B) \cdot \rho_{AB} \cdot (\mathbb{I}_A \otimes \ket{0}_B)\\
&= \begin{pmatrix} 1 & 0 & 0 & 0 \\ 0 & 0 & 1 & 0 \end{pmatrix} 
\begin{pmatrix}
0.4 & 0 & 0 & 0.2 \\
0 & 0.1 & 0.05 & 0 \\
0 & 0.05 & 0.1 & 0 \\
0.2 & 0 & 0 & 0.4
\end{pmatrix}
\begin{pmatrix} 1 & 0 \\ 0 & 0 \\ 0 & 1 \\ 0 & 0 \end{pmatrix}\\
&= \begin{pmatrix} 1 & 0 & 0 & 0 \\ 0 & 0 & 1 & 0 \end{pmatrix}
\begin{pmatrix} 0.4 & 0 \\ 0 & 0 \\ 0 & 0.1 \\ 0.2 & 0 \end{pmatrix}\\
&= \begin{pmatrix} 0.4 & 0 \\ 0 & 0.1 \end{pmatrix}
\end{align}

This correctly extracts the elements at positions (0,0), (0,2), (2,0), and (2,2) from $\rho_{AB}$, arranging them into a $2 \times 2$ matrix!

\textbf{Key insight:} The projection operators $(\mathbb{I}_A \otimes \bra{j}_B)$ and $(\mathbb{I}_A \otimes \ket{j}_B)$ act as "selection matrices" that extract the relevant $2 \times 2$ block from the $4 \times 4$ density matrix. The dimensions work out perfectly: $(2\times4) \cdot (4\times4) \cdot (4\times2) = (2\times2)$, reducing the composite system's density matrix to the subsystem's density matrix.
\end{mathematicalaside}

\begin{physicalinsight}
\textbf{What "Tracing Out" Physically Means}

Tracing out is NOT measurement! It represents situations where:
\begin{itemize}
\item One qubit is spatially inaccessible (e.g., sent to a distant location)
\item One qubit is lost to the environment (decoherence)
\item Experimental apparatus can only measure one qubit
\item We choose to ignore one qubit and ask what we can observe about the other
\end{itemize}

The reduced density matrix answers: "What measurement statistics can I obtain from qubit A alone, without any access to qubit B?"
\end{physicalinsight}

\subsection{Computing Partial Traces: A Detailed Example}

Let's work through a concrete example that demonstrates the partial trace operation. Consider a two-qubit system with density matrix:

\begin{equation}
\rho_{AB} = \begin{pmatrix}
0.4 & 0 & 0 & 0.2 \\
0 & 0.1 & 0.05 & 0 \\
0 & 0.05 & 0.1 & 0 \\
0.2 & 0 & 0 & 0.4
\end{pmatrix}
\end{equation}

This matrix has several interesting features:
\begin{itemize}
    \item It is Hermitian: $\rho_{AB} = \rho_{AB}^\dagger$
    \item It has unit trace: $\text{Tr}(\rho_{AB}) = 0.4 + 0.1 + 0.1 + 0.4 = 1$
    \item Off-diagonal elements at (0,3) and (3,0) indicate quantum correlations between $\ket{00}$ and $\ket{11}$
    \item Off-diagonal elements at (1,2) and (2,1) indicate quantum correlations between $\ket{01}$ and $\ket{10}$
\end{itemize}

\begin{example}[Tracing Over Qubit B]
The partial trace over qubit $B$ is:
\begin{equation}
\rho_A = \text{Tr}_B(\rho_{AB}) = \sum_{j \in \{0,1\}} (\mathbb{I}_A \otimes \bra{j}_B) \rho_{AB} (\mathbb{I}_A \otimes \ket{j}_B)
\end{equation}

(See the Mathematical Aside above for the explicit matrix construction of these projection operators.)

\textbf{Step 1: Select elements for each value of $j$}

For $j = 0$ (selecting $\ket{0}_B$): We extract the submatrix corresponding to basis states where B is in state $\ket{0}$ (i.e., $\ket{00}$ and $\ket{10}$, which are indices 0 and 2):
\[
(\mathbb{I}_A \otimes \bra{0}_B) \rho_{AB} (\mathbb{I}_A \otimes \ket{0}_B) = \begin{pmatrix}
\rho_{00,00} & \rho_{00,10} \\
\rho_{10,00} & \rho_{10,10}
\end{pmatrix} = \begin{pmatrix}
0.4 & 0 \\
0 & 0.1
\end{pmatrix}
\]

For $j = 1$ (selecting $\ket{1}_B$): We extract the submatrix corresponding to basis states where B is in state $\ket{1}$ (i.e., $\ket{01}$ and $\ket{11}$, which are indices 1 and 3):
\[
(\mathbb{I}_A \otimes \bra{1}_B) \rho_{AB} (\mathbb{I}_A \otimes \ket{1}_B) = \begin{pmatrix}
\rho_{01,01} & \rho_{01,11} \\
\rho_{11,01} & \rho_{11,11}
\end{pmatrix} = \begin{pmatrix}
0.1 & 0 \\
0 & 0.4
\end{pmatrix}
\]

\textbf{Step 2: Sum the contributions}
\[
\rho_A = \begin{pmatrix}
0.4 & 0 \\
0 & 0.1
\end{pmatrix} + \begin{pmatrix}
0.1 & 0 \\
0 & 0.4
\end{pmatrix} = \begin{pmatrix}
0.5 & 0 \\
0 & 0.5
\end{pmatrix} = \frac{1}{2}\mathbb{I}
\]
\end{example}

\begin{example}[Tracing Over Qubit A]
Similarly, the partial trace over qubit $A$ is:
\begin{equation}
\rho_B = \text{Tr}_A(\rho_{AB}) = \sum_{i \in \{0,1\}} (\bra{i}_A \otimes \mathbb{I}_B) \rho_{AB} (\ket{i}_A \otimes \mathbb{I}_B)
\end{equation}

\textbf{Step 1: Identify the block structure}

When tracing over $A$, we can write $\rho_{AB}$ in block form:
\[
\rho_{AB} = \begin{pmatrix}
\begin{matrix} 0.4 & 0 \\ 0 & 0.1 \end{matrix} & \begin{matrix} 0 & 0.2 \\ 0.05 & 0 \end{matrix} \\[0.3cm]
\begin{matrix} 0 & 0.05 \\ 0.2 & 0 \end{matrix} & \begin{matrix} 0.1 & 0 \\ 0 & 0.4 \end{matrix}
\end{pmatrix}
\]

The upper-left block corresponds to $i = 0$ ($\ket{0}_A$ subspace), and the lower-right block corresponds to $i = 1$ ($\ket{1}_A$ subspace).

\textbf{Step 2: Extract and sum the diagonal blocks}

For $i = 0$: Upper-left $2 \times 2$ block = $\begin{pmatrix} 0.4 & 0 \\ 0 & 0.1 \end{pmatrix}$

For $i = 1$: Lower-right $2 \times 2$ block = $\begin{pmatrix} 0.1 & 0 \\ 0 & 0.4 \end{pmatrix}$

\textbf{Step 3: Sum the contributions}
\[
\rho_B = \begin{pmatrix}
0.4 & 0 \\
0 & 0.1
\end{pmatrix} + \begin{pmatrix}
0.1 & 0 \\
0 & 0.4
\end{pmatrix} = \begin{pmatrix}
0.5 & 0 \\
0 & 0.5
\end{pmatrix} = \frac{1}{2}\mathbb{I}
\]
\end{example}

\begin{mathematicalaside}{Index Notation Method}
We can also compute partial traces using index notation. For a density matrix with elements $\rho_{ij,kl}$ where $(i,j) \in \{0,1\}$ label qubit $A$ and $(k,l) \in \{0,1\}$ label qubit $B$:

\textbf{Tracing over B:}
\[
(\rho_A)_{ij} = \sum_{k \in \{0,1\}} \rho_{ik,jk}
\]

\textbf{Tracing over A:}
\[
(\rho_B)_{kl} = \sum_{i \in \{0,1\}} \rho_{ik,il}
\]

For our specific matrix, the mapping between basis states and matrix indices is:
\begin{align}
\ket{00} &\leftrightarrow \text{index } 0 \\
\ket{01} &\leftrightarrow \text{index } 1 \\
\ket{10} &\leftrightarrow \text{index } 2 \\
\ket{11} &\leftrightarrow \text{index } 3
\end{align}

Using this method for $\rho_B$:
\begin{itemize}
    \item $(\rho_B)_{00} = \rho_{00,00} + \rho_{10,10} = \rho[0,0] + \rho[2,2] = 0.4 + 0.1 = 0.5$
    \item $(\rho_B)_{01} = \rho_{00,01} + \rho_{10,11} = \rho[0,1] + \rho[2,3] = 0 + 0 = 0$
    \item $(\rho_B)_{10} = \rho_{01,00} + \rho_{11,10} = \rho[1,0] + \rho[3,2] = 0 + 0 = 0$
    \item $(\rho_B)_{11} = \rho_{01,01} + \rho_{11,11} = \rho[1,1] + \rho[3,3] = 0.1 + 0.4 = 0.5$
\end{itemize}
\end{mathematicalaside}

\subsection{The Block Matrix Method}

The block matrix method provides an intuitive way to compute partial traces by organizing the density matrix according to the slow and fast indices.

\begin{mathematicalaside}{Block Matrix Structure and Index Conventions}
\textbf{Index convention:} For the basis ordering $\ket{00}, \ket{01}, \ket{10}, \ket{11}$:
\begin{itemize}
\item A is the \textit{slow index} (leftmost bit) - changes every 2 elements
\item B is the \textit{fast index} (rightmost bit) - alternates with each element
\end{itemize}

\textbf{Block organization:} We can write any 4\ensuremath{\times}4 density matrix in 2\ensuremath{\times}2 blocks organized by the A index:
\[
\rho_{AB} = \begin{pmatrix}
\rho^{00} & \rho^{01} \\
\rho^{10} & \rho^{11}
\end{pmatrix}
\]

where each $\rho^{ij}$ is a 2\ensuremath{\times}2 matrix:
\begin{itemize}
\item $\rho^{00}$ contains elements for A=0\ensuremath{\to}A=0 transitions, B varies (rows/cols 0,1)
\item $\rho^{01}$ contains elements for A=0\ensuremath{\to}A=1 transitions, B varies (rows 0,1; cols 2,3)
\item $\rho^{10}$ contains elements for A=1\ensuremath{\to}A=0 transitions, B varies (rows 2,3; cols 0,1)
\item $\rho^{11}$ contains elements for A=1\ensuremath{\to}A=1 transitions, B varies (rows/cols 2,3)
\end{itemize}

\textbf{The simple rules:}
\begin{itemize}
\item \textbf{To trace out B (keep A):} $\rho_A = \begin{pmatrix} \text{Tr}(\rho^{00}) & \text{Tr}(\rho^{01}) \\ \text{Tr}(\rho^{10}) & \text{Tr}(\rho^{11}) \end{pmatrix}$ 
      
      Take the trace of each 2\ensuremath{\times}2 block (summing over B within each A sector)
      
\item \textbf{To trace out A (keep B):} $\rho_B = \rho^{00} + \rho^{11}$ 
      
      Sum the diagonal blocks (summing over A values while preserving B structure)
\end{itemize}

\textbf{Intuitive explanation:}
\begin{itemize}
\item To trace out the fast index (B): Take the trace within each block---this sums over B values while preserving the A structure
\item To trace out the slow index (A): Sum the diagonal blocks---this sums over A values while preserving the B structure
\item The pattern matches the index structure: trace out fast \ensuremath{\to} trace within blocks; trace out slow \ensuremath{\to} sum across blocks
\end{itemize}
\end{mathematicalaside}

Let's apply this method to our example density matrix:

\begin{example}[Block Matrix Method Application]
Using the same density matrix from Examples 2.12 and 2.13:
\[
\rho_{AB} = \begin{pmatrix}
0.4 & 0 & 0 & 0.2 \\
0 & 0.1 & 0.05 & 0 \\
0 & 0.05 & 0.1 & 0 \\
0.2 & 0 & 0 & 0.4
\end{pmatrix}
\]

\textbf{Step 1: Identify the 2\ensuremath{\times}2 blocks}
\[
\rho_{AB} = \begin{pmatrix}
\rho^{00} & \rho^{01} \\
\rho^{10} & \rho^{11}
\end{pmatrix} = \begin{pmatrix}
\begin{pmatrix} 0.4 & 0 \\ 0 & 0.1 \end{pmatrix} & \begin{pmatrix} 0 & 0.2 \\ 0.05 & 0 \end{pmatrix} \\[0.3cm]
\begin{pmatrix} 0 & 0.05 \\ 0.2 & 0 \end{pmatrix} & \begin{pmatrix} 0.1 & 0 \\ 0 & 0.4 \end{pmatrix}
\end{pmatrix}
\]

\textbf{Step 2: Trace out B (keep A)}

Take the trace of each block:
\begin{align}
\text{Tr}(\rho^{00}) &= 0.4 + 0.1 = 0.5 \\
\text{Tr}(\rho^{01}) &= 0 + 0 = 0 \\
\text{Tr}(\rho^{10}) &= 0 + 0 = 0 \\
\text{Tr}(\rho^{11}) &= 0.1 + 0.4 = 0.5
\end{align}

Therefore:
\[
\rho_A = \begin{pmatrix}
\text{Tr}(\rho^{00}) & \text{Tr}(\rho^{01}) \\
\text{Tr}(\rho^{10}) & \text{Tr}(\rho^{11})
\end{pmatrix} = \begin{pmatrix}
0.5 & 0 \\
0 & 0.5
\end{pmatrix} = \frac{1}{2}\mathbb{I}
\]

\textbf{Step 3: Trace out A (keep B)}

Sum the diagonal blocks:
\[
\rho_B = \rho^{00} + \rho^{11} = \begin{pmatrix} 0.4 & 0 \\ 0 & 0.1 \end{pmatrix} + \begin{pmatrix} 0.1 & 0 \\ 0 & 0.4 \end{pmatrix} = \begin{pmatrix} 0.5 & 0 \\ 0 & 0.5 \end{pmatrix} = \frac{1}{2}\mathbb{I}
\]

Both methods yield the same maximally mixed state, as expected!
\end{example}

\begin{physicalinsight}
\textbf{Why the Block Matrix Method Works}

The block matrix method reflects the hierarchical structure of the tensor product basis:
\begin{itemize}
\item The A index determines which 2\ensuremath{\times}2 block we're in
\item The B index determines position within each block
\item Tracing means summing over the traced index while keeping the other fixed
\item For the fast index (B): sum within blocks (trace operation)
\item For the slow index (A): sum across blocks (matrix addition)
\end{itemize}

This pattern generalizes: for $n$ qubits, organize the matrix hierarchically and trace/sum at the appropriate level!
\end{physicalinsight}

\begin{example}[Visual Method for Partial Trace]
Let's visualize the partial trace operation using the same density matrix, highlighting which elements contribute to each reduced density matrix.

\[
\rho_{AB} = \begin{pmatrix}
0.4 & 0 & 0 & 0.2 \\
0 & 0.1 & 0.05 & 0 \\
0 & 0.05 & 0.1 & 0 \\
0.2 & 0 & 0 & 0.4
\end{pmatrix}
\]

\textbf{Tracing over B (keeping A):}

We can visualize which elements contribute by coloring them:
\[
\rho_{AB} = \begin{pmatrix}
\colorbox{red}{0.4} & 0 & \colorbox{red}{0} & 0.2 \\
0 & \colorbox{blue!20}{0.1} & 0.05 & \colorbox{blue!20}{0} \\
\colorbox{red}{0} & 0.05 & \colorbox{red}{0.1} & 0 \\
0.2 & \colorbox{blue!20}{0} & 0 & \colorbox{blue!20}{0.4}
\end{pmatrix}
\]

\begin{itemize}
    \item \colorbox{red}{Red elements} ($j = 0$): Elements where B is in state $\ket{0}$ - from basis states $\ket{00}$ and $\ket{10}$ (matrix positions 0 and 2)
    \item \colorbox{blue!20}{Blue elements} ($j = 1$): Elements where B is in state $\ket{1}$ - from basis states $\ket{01}$ and $\ket{11}$ (matrix positions 1 and 3)
\end{itemize}

The partial trace collects these elements:
\begin{align}
\text{For } j=0: \quad &\begin{pmatrix} \colorbox{red}{0.4} & \colorbox{red}{0} \\ \colorbox{red}{0} & \colorbox{red}{0.1} \end{pmatrix} \\
\text{For } j=1: \quad &\begin{pmatrix} \colorbox{blue!20}{0.1} & \colorbox{blue!20}{0} \\ \colorbox{blue!20}{0} & \colorbox{blue!20}{0.4} \end{pmatrix}
\end{align}

Summing: $\rho_A = \begin{pmatrix} 0.5 & 0 \\ 0 & 0.5 \end{pmatrix}$

\textbf{Tracing over A (keeping B):}

Alternatively, we can view the matrix in $2 \times 2$ blocks:
\[
\rho_{AB} = \begin{pmatrix}
\colorbox{red}{$\begin{matrix} 0.4 & 0 \\ 0 & 0.1 \end{matrix}$} & \begin{matrix} 0 & 0.2 \\ 0.05 & 0 \end{matrix} \\[0.3cm]
\begin{matrix} 0 & 0.05 \\ 0.2 & 0 \end{matrix} & \colorbox{blue!20}{$\begin{matrix} 0.1 & 0 \\ 0 & 0.4 \end{matrix}$}
\end{pmatrix}
\]

\begin{itemize}
    \item \colorbox{red}{Red block}: Upper-left block for $i = 0$ ($\ket{0}_A$ subspace)
    \item \colorbox{blue!20}{Blue block}: Lower-right block for $i = 1$ ($\ket{1}_A$ subspace)
\end{itemize}

The partial trace sums the diagonal blocks:
\[
\rho_B = \colorbox{red}{$\begin{pmatrix} 0.4 & 0 \\ 0 & 0.1 \end{pmatrix}$} + \colorbox{blue!20}{$\begin{pmatrix} 0.1 & 0 \\ 0 & 0.4 \end{pmatrix}$} = \begin{pmatrix} 0.5 & 0 \\ 0 & 0.5 \end{pmatrix}
\]

\textbf{Key Insight:} The visual method shows that tracing over B selects matrix elements based on B's state value, while tracing over A sums diagonal blocks. Both methods yield the same maximally mixed state, confirming that the quantum correlations are hidden when we look at either subsystem alone.
\end{example}

\section{Physical Interpretation and Implementation}

\subsection{Spatial Arrangement of Qubits}

The tensor product structure naturally accommodates the physical reality that qubits exist at different spatial locations. When we write $\ket{\psi}_1 \otimes \ket{\phi}_2$, the subscripts can be interpreted as labeling physical locations in space. This spatial interpretation becomes particularly relevant when considering the implementation of quantum gates.

In most physical implementations of quantum computers, not all pairs of qubits can directly interact. The connectivity constraints depend on the physical platform:

\textbf{Superconducting qubits:} Typically arranged in 2D lattices with nearest-neighbor connectivity. Each qubit can directly interact only with 2-4 neighboring qubits.

\textbf{Trapped ions:} All-to-all connectivity is possible, but gates between distant ions may be slower or less accurate.

\textbf{Quantum dots:} Often arranged in linear or 2D arrays with nearest-neighbor interactions.

\textbf{Photonic systems:} Connectivity determined by the optical network topology.

These connectivity constraints influence quantum algorithm design and the compilation of quantum circuits. Operations between non-adjacent qubits must be decomposed into sequences of allowed gates, potentially introducing additional errors.

\begin{example}[SWAP Networks]
Consider a linear array of 4 qubits where only nearest-neighbor gates are allowed. To apply CNOT between qubits 1 and 4:

\begin{enumerate}
\item SWAP qubits 1 and 2
\item SWAP qubits 2 and 3 (qubit 1's state is now at position 3)
\item Apply CNOT between positions 3 and 4
\item SWAP qubits 2 and 3
\item SWAP qubits 1 and 2 (returning qubit 1's state to its original position)
\end{enumerate}

This requires 3 SWAP gates (each decomposable into 3 CNOTs) plus 1 CNOT, totaling 10 two-qubit gates instead of just 1.
\end{example}

\subsection{The Bridge to Many-Body Physics}

The connection to spatial arrangement also provides a bridge to condensed matter physics. An array of qubits at fixed positions resembles a lattice model, where quantum particles can occupy discrete sites. This correspondence, explored further in Chapter~\ref{ch:lattice_qft}, reveals deep connections between quantum information and many-body physics.

Consider the mapping:
\begin{itemize}
\item Qubit at site $i$ in state $\ket{0}$: No particle at site $i$
\item Qubit at site $i$ in state $\ket{1}$: One particle at site $i$
\end{itemize}

Under this mapping, the state $\ket{01001}$ represents particles at sites 2 and 5 of a 5-site lattice. The CNOT gate becomes related to particle hopping, and multi-qubit states describe many-body quantum systems.

This connection has proven fruitful in both directions. Quantum information concepts illuminate many-body physics, while condensed matter insights guide quantum algorithm development. The tensor product structure underlies both fields, emphasizing its fundamental importance in quantum mechanics.

\section{Chapter Summary}

This chapter has established the mathematical framework for describing composite quantum systems through the tensor product structure. We have seen how individual quantum systems combine to form larger systems, with the dimension of the composite Hilbert space growing exponentially with the number of constituents.

The key concepts developed include:
\begin{itemize}
\item The tensor product operation for states and operators, with complete basis representations
\item Local and multi-qubit operations, including the fundamental CNOT gate
\item Measurement theory for composite systems, including partial measurements
\item Basis transformations using the tensor product structure
\item The exponential scaling challenge and its implications
\item The reduced density matrix and partial trace operation for describing subsystems
\item The distinction between tracing out (ignoring) and measuring subsystems
\item Physical interpretations related to spatial arrangement and connectivity
\end{itemize}

We have provided comprehensive tables showing how two-qubit computational basis states appear in different measurement bases, enabling cross-basis calculations. Through numerous examples, we've demonstrated how phase relationships in multi-qubit systems lead to observable differences in measurement statistics.

The reduced density matrix formalism provides a crucial tool for understanding subsystems of composite quantum states. When we trace out part of an entangled system, the resulting mixed state reveals how quantum information is stored non-locally in correlations. This concept illuminates the fundamental difference between quantum and classical correlations and explains the mechanism of decoherence.

While we have touched on entanglement through examples like the Bell states, a full treatment of entanglement---those special quantum states that cannot be written as tensor products of individual qubit states---is reserved for advanced studies. The mathematical foundation developed here, including the reduced density matrix formalism, provides the tools needed to fully appreciate the subtleties of entanglement when encountered later.

The exponential growth of Hilbert space dimension emerges as both the key resource for quantum computation and the primary obstacle to classical simulation. This fundamental tension drives much of quantum information science and motivates the development of quantum technologies.

\subsection{Distinguishing Tracing from Measurement}

It's crucial to understand that tracing out is fundamentally different from measurement:

\begin{example}[Trace vs. Measure]
Starting with $\ket{\Phi^+} = \frac{1}{\sqrt{2}}(\ket{00} + \ket{11})$:

\textbf{Option 1: Trace out B (ignore it)}
\begin{itemize}
\item Result: $\rho_A = \frac{1}{2}\mathbb{I}$ (maximally mixed)
\item If we measure A in the Z-basis: 50\% chance of $\ket{0}$, 50\% chance of $\ket{1}$
\item Purity: $\text{Tr}(\rho_A^2) = \frac{1}{2}$
\end{itemize}

\textbf{Option 2: Measure B first, then look at A}
\begin{itemize}
\item If B measured as $\ket{0}$: A collapses to pure state $\ket{0}$
\item If B measured as $\ket{1}$: A collapses to pure state $\ket{1}$
\item Each outcome has purity: $\text{Tr}(\rho_A^2) = 1$
\end{itemize}

The key difference: Measurement collapses the entanglement, while tracing out preserves it (we just can't access it).
\end{example}

\subsection{Physical Applications}

The reduced density matrix appears naturally in several important contexts:

\textbf{1. Quantum Decoherence}: When a quantum system entangles with an inaccessible environment, tracing out the environment explains the apparent loss of quantum coherence.

\textbf{2. Quantum Communication}: Alice and Bob share entangled qubits. Alice's local measurement statistics are described by her reduced density matrix.

\textbf{3. Subsystem Thermodynamics}: The thermal state of a subsystem in a larger quantum system is given by the reduced density matrix after tracing out the rest.

\textbf{4. Entanglement Witnesses}: The purity $\text{Tr}(\rho_A^2)$ quantifies entanglement---it equals 1 for product states and less than 1 for entangled states.

\subsection{The Deeper Meaning}

The reduced density matrix reveals a profound aspect of quantum mechanics: information in quantum systems can be stored non-locally in correlations. When subsystems are entangled:

\begin{itemize}
\item Neither subsystem has a complete quantum state of its own
\item The "missing" information is stored in the quantum correlations
\item Ignoring one subsystem converts pure quantum superpositions into classical statistical mixtures
\item This is fundamentally different from classical correlations, where subsystems always have definite states
\end{itemize}

This concept bridges quantum information theory and statistical mechanics, explaining how quantum systems appear classical when we lack access to all degrees of freedom. It also quantifies the non-local nature of quantum information---a key resource in quantum computing and communication.

\input{book_problems/ch02_problems.tex}

\section*{References and Further Reading}
\addcontentsline{toc}{section}{References and Further Reading}

\begin{description}
\item[Nielsen, M.~A., and Chuang, I.~L.] \emph{Quantum Computation and Quantum Information}, 10th anniversary ed. Cambridge University Press, 2010. Sections 2.1--2.5 cover composite systems, density operators, and the partial trace at the level of this chapter; the standard reference for the formalism developed here.

\item[Preskill, J.] \emph{Lecture Notes on Quantum Computation}. California Institute of Technology. \href{http://theory.caltech.edu/~preskill/ph219/}{theory.caltech.edu/$\sim$preskill/ph219/}. Chapter 2 of the notes treats foundations of quantum theory, density operators, and reduced states; concise and freely available.

\item[Mermin, N.~D.] \emph{Quantum Computer Science: An Introduction}. Cambridge University Press, 2007. Student-friendly treatment of multi-qubit operations, with explicit circuit diagrams; the most accessible companion to the gate-level material in this chapter.

\item[Wilde, M.~M.] \emph{Quantum Information Theory}, 2nd ed. Cambridge University Press, 2017. \href{https://doi.org/10.1017/9781316809976}{doi:10.1017/9781316809976}. Chapter 4 develops density operators, reduced density matrices, and the partial trace with full mathematical rigor; the natural next step for the formalism.

\item[Watrous, J.] \emph{The Theory of Quantum Information}. Cambridge University Press, 2018. The most mathematically rigorous treatment of tensor products, channels, and operator structures; a long-term reference rather than a first read.
\end{description}

%% file: book_problems/ch02_problems.tex
\section{Problems}
\setcounter{hwproblem}{0}  

\problem{Two-Qubit Product States}
For two Bloch cubes representing a two-qubit system:
\begin{enumerate}[label=(\alph*)]
    \item Write the four computational basis states $\ket{00}$, $\ket{01}$, $\ket{10}$, $\ket{11}$ using tensor products
    \item If both qubits are in state $\ket{+x}$, express the composite state in the computational basis
    \item Calculate the probability of each measurement outcome for a Z-measurement on both qubits
    \item Calculate the probability of measuring the first qubit in $\ket{0}$ (regardless of the second qubit)
    \item Use your two Bloch cubes to verify these probability predictions
\end{enumerate}

\problem{Tensor Product Properties}
Prove the following basic properties:
\begin{enumerate}[label=(\alph*)]
    \item $(A \otimes B)(C \otimes D) = AC \otimes BD$ for $2 \times 2$ matrices
    \item $\text{Tr}(A \otimes B) = \text{Tr}(A) \cdot \text{Tr}(B)$
    \item If $\ket{\psi} = \alpha\ket{0} + \beta\ket{1}$ and $\ket{\phi} = \gamma\ket{0} + \delta\ket{1}$, write $\ket{\psi} \otimes \ket{\phi}$ in the computational basis
    \item Calculate $(\sigma_z \otimes I)\ket{+x}\ket{+x}$ and express the result in the computational basis
    \item Verify that $\braket{\psi \otimes \phi}{\psi \otimes \phi} = \braket{\psi}{\psi}\braket{\phi}{\phi}$
\end{enumerate}

\problem{CNOT Gate}
Analyze the controlled-NOT (CNOT) gate:
\begin{enumerate}[label=(\alph*)]
    \item Write out how CNOT acts on all four computational basis states
    \item Construct the $4 \times 4$ matrix representation of CNOT
    \item Apply CNOT to the state $\frac{1}{\sqrt{2}}(\ket{00} + \ket{10})$
    \item Apply CNOT to the state $\frac{1}{\sqrt{2}}(\ket{01} + \ket{11})$
    \item Show that CNOT$^2 = I$ (applying CNOT twice returns to the original state)
    \item Use two Bloch cubes to physically implement these CNOT operations
\end{enumerate}

\problem{Bell State Creation}
Create and analyze Bell states:
\begin{enumerate}[label=(\alph*)]
    \item Starting with $\ket{00}$, apply $H \otimes I$ where $H = \frac{1}{\sqrt{2}}(\sigma_x + \sigma_z)$
    \item Apply CNOT to the result to create $\ket{\Phi^+} = \frac{1}{\sqrt{2}}(\ket{00} + \ket{11})$
    \item Calculate the probability of each computational basis measurement outcome
    \item Show that measuring the first qubit always gives the same result as measuring the second qubit
    \item Express $\ket{\Phi^+}$ in the XX basis: $\{\ket{+x}\ket{+x}, \ket{+x}\ket{-x}, \ket{-x}\ket{+x}, \ket{-x}\ket{-x}\}$
    \item Calculate measurement probabilities in the XX basis
\end{enumerate}

\problem{Three-Qubit Systems}
Extend to three qubits:
\begin{enumerate}[label=(\alph*)]
    \item Write the eight computational basis states for three qubits
    \item Express the state $\ket{+x}\ket{+x}\ket{+x}$ in the computational basis
    \item Apply $\sigma_z \otimes I \otimes I$ (Z gate on first qubit only, corresponding to a $\pi$ rotation around the z-axis)
    \item Calculate the resulting amplitudes and verify normalization
    \item Find the probability of measuring the first qubit in $\ket{0}$
    \item Use three Bloch cubes to verify your calculation
\end{enumerate}

\problem{Matrix Representations Using Bit String Convention}
For $n$-qubit systems, we label computational basis states using the bit string to number convention:
$\ket{b_{n-1}b_{n-2}...b_1b_0} \equiv \ket{j}$ where $j = \sum_{i=0}^{n-1} b_i \cdot 2^i$.

For example, in a 2-qubit system: $\ket{00} \equiv \ket{0}$, $\ket{01} \equiv \ket{1}$, $\ket{10} \equiv \ket{2}$, $\ket{11} \equiv \ket{3}$.

For a 3-qubit system: $\ket{000} \equiv \ket{0}$, $\ket{001} \equiv \ket{1}$, ..., $\ket{111} \equiv \ket{7}$.

\begin{enumerate}[label=(\alph*)]
    \item Write the state $\ket{+x}\ket{+x}$ from Problem 2.1(b) as a 4-dimensional column vector using the ordered basis $\{\ket{0}, \ket{1}, \ket{2}, \ket{3}\}$

    \item Write the operator $\sigma_z \otimes I$ as a $4 \times 4$ matrix in this same basis. Verify your answer by applying it to the column vector from part (a) and checking that you get the result from Problem 2.2(d)

    \item Write the CNOT gate as a $4 \times 4$ matrix using the ordering $\{\ket{0}, \ket{1}, \ket{2}, \ket{3}\}$ and verify it matches your answer from Problem 2.3(b)

    \item Write the Bell state $\ket{\Phi^+} = \frac{1}{\sqrt{2}}(\ket{00} + \ket{11})$ as a 4-dimensional column vector

    \item For the three-qubit state $\ket{+x}\ket{+x}\ket{+x}$ from Problem 2.5(b), write it as an 8-dimensional column vector using the ordered basis $\{\ket{0}, \ket{1}, ..., \ket{7}\}$

    \item Write the operator $\sigma_z \otimes I \otimes I$ as an $8 \times 8$ matrix. Show explicitly the first two rows and the last two rows of this matrix (you may use ``...'' to indicate the middle rows)

    \item Verify that multiplying the $8 \times 8$ matrix from part (f) by the column vector from part (e) gives the correct result after applying the Z gate to the first qubit
\end{enumerate}

\problem{Partial Trace Fundamentals}
Consider a two-qubit density matrix:
\begin{equation}
\rho_{AB} = \begin{pmatrix}
0.3 & 0 & 0 & 0.1 \\
0 & 0.2 & 0.05 & 0 \\
0 & 0.05 & 0.2 & 0 \\
0.1 & 0 & 0 & 0.3
\end{pmatrix}
\end{equation}
\begin{enumerate}[label=(\alph*)]
    \item Use the block matrix method to compute $\rho_A = \text{Tr}_B(\rho_{AB})$
    \item Use the projection method to compute $\rho_B = \text{Tr}_A(\rho_{AB})$
    \item Calculate the purity $\text{Tr}(\rho_A^2)$ and $\text{Tr}(\rho_B^2)$
    \item What do the purity values tell you about entanglement in the original state?
    \item Using an ensemble of Bloch cubes, can you represent the reduced states $\rho_A$ and $\rho_B$? Explain how or why not.
\end{enumerate}

\problem{Pure vs Mixed States through Partial Trace}
Consider the following two-qubit states:
\begin{enumerate}[label=(\alph*)]
    \item Product state: $\ket{\psi_1} = \ket{+x}_A \otimes \ket{0}_B$
    \item Bell state: $\ket{\psi_2} = \frac{1}{\sqrt{2}}(\ket{00} + \ket{11})$
    \item Superposition: $\ket{\psi_3} = \frac{1}{\sqrt{2}}(\ket{01} + \ket{10})$
\end{enumerate}
For each state:
\begin{enumerate}[label=(\roman*)]
    \item Construct the full density matrix $\rho_{AB} = \ket{\psi}\bra{\psi}$
    \item Calculate both reduced density matrices $\rho_A$ and $\rho_B$
    \item Determine if each reduced state is pure or mixed
    \item Calculate measurement probabilities for the Z-basis on qubit A alone
    \item Compare your results: which states show entanglement and how can you tell?
\end{enumerate}

\problem{Measurement vs Tracing Out}
Start with the Bell state $\ket{\Phi^+} = \frac{1}{\sqrt{2}}(\ket{00} + \ket{11})$:
\begin{enumerate}[label=(\alph*)]
    \item Option 1 (Trace out qubit B): Calculate $\rho_A = \text{Tr}_B(\ket{\Phi^+}\bra{\Phi^+})$
    \item What are the probabilities of measuring qubit A in states $\ket{0}$ and $\ket{1}$?
    \item Option 2 (Measure qubit B first): If we measure qubit B in the Z-basis and obtain result $\ket{0}$, what is the post-measurement state of the full system?
    \item What is now the state of qubit A alone?
    \item If we measure qubit B and obtain result $\ket{1}$, what is the state of qubit A?
    \item Compare the measurement probabilities for qubit A in both options
    \item Option 3 (Measure qubit B in X-basis): Calculate what happens if we measure qubit B in the $\{\ket{+x}, \ket{-x}\}$ basis instead
    \item Explain the key physical difference between tracing out and measuring
\end{enumerate}

\problem{Schmidt Decomposition of Entangled States}
The Schmidt decomposition expresses any two-qubit state uniquely as $\ket{\psi} = \sum_i \sqrt{\lambda_i}\ket{\phi_i}_A \otimes \ket{\chi_i}_B$ where $\lambda_i$ are the Schmidt coefficients (eigenvalues of $\rho_A$).
\begin{enumerate}[label=(\alph*)]
    \item For the Bell state $\ket{\Phi^+} = \frac{1}{\sqrt{2}}(\ket{00} + \ket{11})$, find the Schmidt decomposition
    \item Calculate the Schmidt number (number of non-zero $\lambda_i$) and explain what it tells you about entanglement
    \item For a product state $\ket{\psi} = \ket{+x}_A \otimes \ket{-x}_B$, compute the Schmidt decomposition and Schmidt number
    \item Show that the Bell state and the product state have different Schmidt numbers, confirming their different entanglement properties
    \item For the superposition state $\ket{\psi} = \frac{1}{\sqrt{2}}(\ket{01} + \ket{10})$, find the Schmidt decomposition
\end{enumerate}

\problem{Controlled-Phase and Two-Qubit Gate Equivalences}
Two-qubit gates can often be expressed in multiple equivalent forms:
\begin{enumerate}[label=(\alph*)]
    \item The controlled-phase gate is $\hat{U}_{\text{CP}} = \begin{pmatrix} 1 & 0 & 0 & 0 \\ 0 & 1 & 0 & 0 \\ 0 & 0 & 1 & 0 \\ 0 & 0 & 0 & e^{i\phi} \end{pmatrix}$. Express this in terms of projectors on the $\ket{11}$ state
    \item Show that CNOT can be decomposed as CNOT $= (I \otimes H) \text{CZ} (I \otimes H)$ where CZ is the controlled-Z gate
    \item Calculate how CNOT and controlled-phase gates differ in their action on the Bell state $\ket{\Phi^+}$
    \item For $\phi = \pi$, show that the controlled-phase gate creates the $\ket{\Psi^-}$ Bell state when applied to $\ket{00}$ after a Hadamard
    \item Express the $\sqrt{\text{CNOT}}$ gate in the same matrix format (it should be a $4 \times 4$ unitary matrix)
\end{enumerate}

\problem{GHZ vs W State Distinguishability}
Multi-qubit entangled states can exhibit qualitatively different properties:
\begin{enumerate}[label=(\alph*)]
    \item Define the GHZ state (3-qubit): $\ket{\text{GHZ}} = \frac{1}{\sqrt{2}}(\ket{000} + \ket{111})$ and the W state: $\ket{W} = \frac{1}{\sqrt{3}}(\ket{100} + \ket{010} + \ket{001})$
    \item For the GHZ state, calculate all three single-qubit reduced density matrices
    \item For the W state, calculate all three single-qubit reduced density matrices
    \item Show that tracing out any single qubit from GHZ leaves the two remaining qubits in a separable state
    \item Show that tracing out any single qubit from W leaves the two remaining qubits in an entangled state
    \item Explain the physical meaning of these differences for quantum information and resilience to particle loss
\end{enumerate}

\problem{Entanglement Entropy and Information Theory}
The entanglement entropy quantifies bipartite entanglement: $S_A = -\text{Tr}(\rho_A \log_2 \rho_A)$:
\begin{enumerate}[label=(\alph*)]
    \item For a product state $\ket{\psi} = \ket{0}_A \ket{1}_B$, show that $S_A = 0$
    \item For the Bell state $\ket{\Phi^+}$, calculate $S_A$ and interpret the result
    \item For the separable mixed state $\rho = 0.5\ket{00}\bra{00} + 0.5\ket{11}\bra{11}$, calculate $S_A$
    \item Show that $S_A = S_B$ (entanglement entropy is symmetric)
    \item For a partially entangled state $\ket{\psi} = \sqrt{0.7}\ket{00} + \sqrt{0.3}\ket{11}$, calculate $S_A$
    \item Compare the entanglement entropy with the purity: which is a better measure of non-classicality?
\end{enumerate}

\problem{Partial Transpose and Entanglement Criterion}
The partial transpose (PPT) criterion detects some (but not all) entangled states. For a density matrix $\rho_{AB}$, the partial transpose with respect to system B is $\rho_{AB}^{T_B}$:
\begin{enumerate}[label=(\alph*)]
    \item Define the partial transpose operation: $\rho_{AB}^{T_B} = \sum_{i,j,k,l} \rho_{ijkl} \ket{i}\bra{j} \otimes \ket{l}\bra{k}$
    \item For the Bell state $\ket{\Phi^+} = \frac{1}{\sqrt{2}}(\ket{00} + \ket{11})$, construct $\rho = \ket{\Phi^+}\bra{\Phi^+}$ and compute $\rho^{T_B}$
    \item Show that $\rho^{T_B}$ has a negative eigenvalue, confirming entanglement via the PPT criterion
    \item For a product state $\ket{+x}_A \otimes \ket{-x}_B$, compute the partial transpose and verify it is positive semidefinite
    \item Explain why PPT is necessary but not sufficient for separability in $2 \times 2$ and $2 \times 3$ systems
\end{enumerate}

\problem{Two-Qubit State Tomography}
Quantum state tomography reconstructs an unknown state from measurements:
\begin{enumerate}[label=(\alph*)]
    \item For a general two-qubit state $\rho_{AB}$, how many independent measurement bases are needed to fully characterize it?
    \item For measurements in the bases $\{Z, X, Y\} \otimes \{Z, X, Y\}$, show that you obtain 9 measurement settings (for a $4 \times 4$ density matrix with $16 - 4 = 12$ independent real parameters)
    \item Given measurement outcomes for all nine settings on the Bell state $\ket{\Phi^+}$, how would you reconstruct $\rho = \ket{\Phi^+}\bra{\Phi^+}$?
    \item For the separable state $\rho = \ket{+x}\bra{+x} \otimes \ket{0}\bra{0}$, predict the measurement outcomes in all nine bases before explicit calculation
    \item Discuss how measurement noise affects the reconstructed state and when tomography becomes impractical
\end{enumerate}

%% file: chapters/ch03_lattice_qft.tex
\chapter{Quantum Fields on a Lattice}
\label{ch:lattice_qft}

\section{Introduction: From Spins to Fields}

\begin{keyidea}{Universal Quantum Evolution and the Spin-Particle Correspondence}
Two profound principles unite in this chapter. First, any unitary transformation on $N$ qubits can be decomposed into single-qubit rotations and two-qubit entangling gates---a generalization of our single Bloch sphere rotations to many-body systems. Second, these abstract qubits can represent physical particle occupation on lattice sites, transforming quantum information into quantum field theory. Together, these principles show that quantum fields can be understood through discrete gate operations.
\end{keyidea}

In the previous chapters, we developed the mathematical machinery for describing systems of qubits---abstract two-level quantum systems. We saw how a single qubit's evolution could be described by rotations on the Bloch sphere. We now extend this to $N$ qubits and reveal a fundamental physical interpretation: each qubit can represent a location in space that either contains a quantum particle or is empty. This reinterpretation is not merely a convenient analogy; it establishes an exact mathematical correspondence that bridges quantum information theory and quantum field theory.

\subsection{Universal Gate Sets}

Before diving into the physics, we establish a crucial mathematical fact:

\begin{tcolorbox}[enhanced, title={Universality for N Qubits}, colback=blue!5, colframe=blue!50!black]
Any unitary operator $U$ acting on $N$ qubits can be approximated to arbitrary precision using only:
\begin{itemize}
    \item Single-qubit rotations: $R_x(\theta)$, $R_y(\theta)$, $R_z(\phi)$
    \item Two-qubit entangling gates: CNOT or $\sqrt{\text{SWAP}}$
\end{itemize}
This is the many-body generalization of the fact that any single-qubit state can be reached by rotations around different axes of the Bloch sphere.
\end{tcolorbox}

This universality means that complex quantum evolution---whether it represents computation or physical dynamics---can always be broken down into simple building blocks.

\subsection{The Spin-Particle Mapping}

Consider a one-dimensional lattice of sites, each capable of being occupied by at most one particle. This "hard-core" constraint emerges naturally in many physical systems where strong repulsive interactions prevent multiple occupation. We establish the correspondence:
\begin{align}
\ket{\uparrow}_i \equiv \ket{0}_i &\leftrightarrow \text{site } i \text{ is empty}\\
\ket{\downarrow}_i \equiv \ket{1}_i &\leftrightarrow \text{site } i \text{ contains a particle}
\end{align}

This mapping immediately imbues our abstract quantum formalism with physical meaning. A superposition state $\frac{1}{\sqrt{2}}(\ket{0}_i + \ket{1}_i)$ no longer represents an abstract qubit in a superposition of computational basis states---it represents a lattice site in a quantum superposition of being empty and occupied.

\begin{blochcubeactivity}{3.1: Visualizing the Spin-Particle Correspondence}
Arrange three Bloch cubes in a line to represent a three-site lattice:
\begin{enumerate}
    \item Orient each cube so that $\ket{+z}$ (top) represents "empty" and $\ket{-z}$ (bottom) represents "occupied"
    \item Create the configuration with particles at sites 1 and 3: point cubes 1 and 3 downward ($\ket{-z}$), cube 2 upward ($\ket{+z}$)
    \item This represents the quantum state $\ket{101}$ in our occupation number notation
    \item Now rotate cube 2 to the $\ket{+x}$ position---this represents site 2 in superposition $\frac{1}{\sqrt{2}}(\ket{0}_2 + \ket{1}_2)$
    \item The full state is now $\ket{1} \otimes \frac{1}{\sqrt{2}}(\ket{0} + \ket{1}) \otimes \ket{1}$
    \item Discuss: What does measuring site 2's occupation mean physically? How does this differ from a classical probabilistic mixture?
\end{enumerate}
\end{blochcubeactivity}

\section{Quantum Fields from Ladder Operators}

\subsection{Two Types of Particles: Hardcore Bosons and Fermions}

We can realize this spin-particle correspondence with two different types of quantum statistics. The choice determines the physical system we're modeling:

\begin{tcolorbox}[enhanced, title={Two Realizations of Hardcore Particles}, colback=yellow!5, colframe=yellow!50!black]
\textbf{1. Hardcore Bosons:} Particles that cannot doubly occupy a site but have NO exchange phase
\begin{itemize}
\item Use operators $\hat{b}_i$ and $\hat{b}_i^\dagger$
\item Commute at different sites: $[\hat{b}_i, \hat{b}_j^\dagger] = 0$ for $i \neq j$
\item Physical examples: Cold atoms in optical lattices, photons in cavities, Rydberg atoms
\end{itemize}

\textbf{2. Spinless Fermions:} Particles with Pauli exclusion AND exchange antisymmetry
\begin{itemize}
\item Use operators $\hat{c}_i$ and $\hat{c}_i^\dagger$
\item Anticommute at different sites: $\{\hat{c}_i, \hat{c}_j^\dagger\} = 0$ for $i \neq j$
\item Physical examples: Electrons in quantum dots, cold fermionic atoms
\end{itemize}
\end{tcolorbox}

\subsection{Field Operators for Hardcore Bosons}

For hardcore bosons, we define the fundamental ladder operators. Let's first establish their matrix representations in the occupation number basis $\{\ket{0}, \ket{1}\}$:

\begin{tcolorbox}[enhanced, title={Matrix Representations of Basic Operators}, colback=gray!5, colframe=gray!50!black]
In the basis where $\ket{0} = \begin{pmatrix} 1 \\ 0 \end{pmatrix}$ (empty) and $\ket{1} = \begin{pmatrix} 0 \\ 1 \end{pmatrix}$ (occupied):

\textbf{Projection operators:}
\begin{align}
\ket{0}\bra{0} &= \begin{pmatrix} 1 & 0 \\ 0 & 0 \end{pmatrix} \quad \text{(projects onto empty state)}\\
\ket{1}\bra{1} &= \begin{pmatrix} 0 & 0 \\ 0 & 1 \end{pmatrix} \quad \text{(projects onto occupied state)}
\end{align}

\textbf{Transition operators:}
\begin{align}
\ket{1}\bra{0} &= \begin{pmatrix} 0 & 0 \\ 1 & 0 \end{pmatrix} \quad \text{(creates particle)}\\
\ket{0}\bra{1} &= \begin{pmatrix} 0 & 1 \\ 0 & 0 \end{pmatrix} \quad \text{(destroys particle)}
\end{align}
\end{tcolorbox}

Now we can define the hardcore boson operators with their explicit matrix forms:

\begin{tcolorbox}[enhanced, title={Hardcore Boson Operators}, colback=green!5, colframe=green!50!black]
\textbf{Creation operator at site $i$:}
\begin{equation}
\hat{b}_i^\dagger = \ket{1}_i\bra{0}_i = \ket{\downarrow}_i\bra{\uparrow}_i = \begin{pmatrix} 0 & 0 \\ 1 & 0 \end{pmatrix}_i
\end{equation}
Creates a particle at site $i$ if empty, gives zero if occupied.

\textbf{Annihilation operator at site $i$:}
\begin{equation}
\hat{b}_i = \ket{0}_i\bra{1}_i = \ket{\uparrow}_i\bra{\downarrow}_i = \begin{pmatrix} 0 & 1 \\ 0 & 0 \end{pmatrix}_i
\end{equation}
Removes a particle from site $i$ if occupied, gives zero if empty.

\textbf{Number operator at site $i$:}
\begin{equation}
\hat{n}_i = \hat{b}_i^\dagger \hat{b}_i = \ket{1}_i\bra{1}_i = \ket{\downarrow}_i\bra{\downarrow}_i = \begin{pmatrix} 0 & 0 \\ 0 & 1 \end{pmatrix}_i
\end{equation}
Counts particles at site $i$: eigenvalue 0 if empty, 1 if occupied.
\end{tcolorbox}

The algebra for hardcore bosons:
\begin{align}
\text{On-site:} \quad \{\hat{b}_i, \hat{b}_i^\dagger\} &= \hat{I}_i \quad \text{(anticommutator - hardcore constraint)}\\
(\hat{b}_i^\dagger)^2 &= 0, \quad (\hat{b}_i)^2 = 0 \quad \text{(no double occupation)}\\
\text{Different sites:} \quad [\hat{b}_i, \hat{b}_j^\dagger] &= 0 \quad \text{for } i \neq j \quad \text{(commutator - bosonic)}
\end{align}

\subsection{Field Operators for Spinless Fermions}

For spinless fermions, we define similar operators with the same matrix representations but different inter-site statistics:

\begin{tcolorbox}[enhanced, title={Spinless Fermion Operators}, colback=blue!5, colframe=blue!50!black]
\textbf{Creation operator at site $i$:}
\begin{equation}
\hat{c}_i^\dagger = \ket{1}_i\bra{0}_i = \ket{\downarrow}_i\bra{\uparrow}_i = \begin{pmatrix} 0 & 0 \\ 1 & 0 \end{pmatrix}_i
\end{equation}

\textbf{Annihilation operator at site $i$:}
\begin{equation}
\hat{c}_i = \ket{0}_i\bra{1}_i = \ket{\uparrow}_i\bra{\downarrow}_i = \begin{pmatrix} 0 & 1 \\ 0 & 0 \end{pmatrix}_i
\end{equation}

\textbf{Number operator at site $i$:}
\begin{equation}
\hat{n}_i = \hat{c}_i^\dagger \hat{c}_i = \ket{1}_i\bra{1}_i = \begin{pmatrix} 0 & 0 \\ 0 & 1 \end{pmatrix}_i
\end{equation}

Note: The matrix representations are IDENTICAL to bosons for single sites!
\end{tcolorbox}

The algebra for spinless fermions:
\begin{align}
\text{On-site:} \quad \{\hat{c}_i, \hat{c}_i^\dagger\} &= \hat{I}_i \quad \text{(same as bosons)}\\
(\hat{c}_i^\dagger)^2 &= 0, \quad (\hat{c}_i)^2 = 0 \quad \text{(same as bosons)}\\
\text{Different sites:} \quad \{\hat{c}_i, \hat{c}_j^\dagger\} &= 0 \quad \text{for } i \neq j \quad \text{(anticommutator - fermionic)}
\end{align}

\begin{example}[Ladder Operator Action on Lattice States]
Consider a three-site system. Let's examine how ladder operators act:

\textbf{For hardcore bosons:}
\begin{align}
\hat{b}_2^\dagger \ket{100} &= \ket{110} \quad \text{(creates particle at site 2)}\\
\hat{b}_1^\dagger \ket{100} &= 0 \quad \text{(site 1 already occupied)}\\
\hat{b}_3^\dagger \hat{b}_2^\dagger \hat{b}_1^\dagger \ket{000} &= \ket{111} \quad \text{(fills all sites)}
\end{align}

\textbf{For fermions (same local action):}
\begin{align}
\hat{c}_2^\dagger \ket{100} &= \ket{110}\\
\hat{c}_1^\dagger \ket{100} &= 0\\
\hat{c}_3^\dagger \hat{c}_2^\dagger \hat{c}_1^\dagger \ket{000} &= \ket{111}
\end{align}

The difference appears when we exchange operator order (see Section~\ref{sec:bosons_vs_fermions}).
\end{example}

\begin{blochcubeactivity}{3.2: Ladder Operations with Bloch Cubes}
Using four Bloch cubes in a line:
\begin{enumerate}
    \item Start with all cubes pointing up (empty lattice): $\ket{0000}$
    \item Apply $\hat{b}_2^\dagger$ (or $\hat{c}_2^\dagger$): flip cube 2 to point down. Result: $\ket{0100}$
    \item Apply $\hat{b}_4^\dagger$ (or $\hat{c}_4^\dagger$): flip cube 4 to point down. Result: $\ket{0101}$
    \item Apply $\hat{b}_1^\dagger \hat{b}_2$ (or $\hat{c}_1^\dagger \hat{c}_2$): moves particle from site 2 to site 1
    \item Show this by flipping cube 2 up and cube 1 down. Result: $\ket{1001}$
    \item Try applying $\hat{b}_1^\dagger$ again---nothing happens (cube 1 already points down)
    \item For single particles, bosons and fermions behave identically!
\end{enumerate}
\end{blochcubeactivity}

\section{Unitary Gates in Second Quantization}

\subsection{Building Gates from Field Operators}

We can construct unitary gates directly from creation and annihilation operators. The gate forms are the same for both particle types, using either $\hat{a} = \hat{b}$ (bosons) or $\hat{a} = \hat{c}$ (fermions). Let's first see how these gates look in matrix form:

\begin{tcolorbox}[enhanced, title={Matrix Representations of Two-Qubit Gates}, colback=gray!5, colframe=gray!50!black]
In the two-qubit computational basis $\{\ket{00}, \ket{01}, \ket{10}, \ket{11}\}$:

\textbf{Identity operator:}
\begin{equation}
\hat{I} \otimes \hat{I} = \begin{pmatrix}
1 & 0 & 0 & 0 \\
0 & 1 & 0 & 0 \\
0 & 0 & 1 & 0 \\
0 & 0 & 0 & 1
\end{pmatrix}
\end{equation}

\textbf{SWAP gate:}
\begin{equation}
\hat{\text{SWAP}} = \begin{pmatrix}
1 & 0 & 0 & 0 \\
0 & 0 & 1 & 0 \\
0 & 1 & 0 & 0 \\
0 & 0 & 0 & 1
\end{pmatrix}
\end{equation}
Exchanges the states of two qubits: $\ket{01} \leftrightarrow \ket{10}$

\textbf{Controlled-NOT (CNOT) gate:}
\begin{equation}
\hat{\text{CNOT}} = \begin{pmatrix}
1 & 0 & 0 & 0 \\
0 & 1 & 0 & 0 \\
0 & 0 & 0 & 1 \\
0 & 0 & 1 & 0
\end{pmatrix}
\end{equation}
Flips target qubit if control is 1: $\ket{10} \leftrightarrow \ket{11}$
\end{tcolorbox}

Now we can express these gates using field operators:

\begin{tcolorbox}[enhanced, title={Fundamental Unitary Gates}, colback=green!5, colframe=green!50!black]
\textbf{SWAP gate between sites $i$ and $j$:}
\begin{equation}
\hat{\text{SWAP}}_{ij} = (\hat{I}_i - \hat{n}_i)(\hat{I}_j - \hat{n}_j) + \hat{n}_i\hat{n}_j + \hat{a}_i^\dagger \hat{a}_j + \hat{a}_j^\dagger \hat{a}_i
\end{equation}

\textbf{Phase gate (particle-dependent):}
\begin{equation}
\hat{P}_i(\phi) = \hat{I}_i - \hat{n}_i + e^{i\phi}\hat{n}_i = \begin{pmatrix} 1 & 0 \\ 0 & e^{i\phi} \end{pmatrix}_i
\end{equation}

\textbf{Controlled-NOT (CNOT) gate:}
\begin{equation}
\hat{\text{CNOT}}_{ij} = \ket{00}\bra{00} + \ket{01}\bra{01} + \ket{11}\bra{10} + \ket{10}\bra{11}
\end{equation}

Note: $\hat{a}$ represents either $\hat{b}$ (bosons) or $\hat{c}$ (fermions)
\end{tcolorbox}

These gates act directly on our lattice states without reference to time evolution or Hamiltonians.

\subsection{The SWAP and Square-Root-of-SWAP Gates}

Following Loss and DiVincenzo (1998), we recognize that the SWAP gate and its square root are particularly important for quantum computation with quantum dots:

\begin{tcolorbox}[enhanced, title={The $\sqrt{\text{SWAP}}$ Gate}, colback=yellow!5, colframe=yellow!50!black]
The $\sqrt{\text{SWAP}}$ gate creates entanglement between neighboring sites:
\begin{equation}
\sqrt{\hat{\text{SWAP}}}_{ij} = \begin{pmatrix}
1 & 0 & 0 & 0 \\
0 & \frac{1+i}{2} & \frac{1-i}{2} & 0 \\
0 & \frac{1-i}{2} & \frac{1+i}{2} & 0 \\
0 & 0 & 0 & 1
\end{pmatrix}
\end{equation}

This gate, combined with single-qubit rotations, forms a universal gate set for quantum computation.
\end{tcolorbox}

\begin{example}[$\sqrt{\text{SWAP}}$ Gate Action]
For a three-site system with one particle (same for bosons and fermions):
\begin{align}
\sqrt{\hat{\text{SWAP}}}_{12}\ket{100} &= \frac{1+i}{2}\ket{100} + \frac{1-i}{2}\ket{010}\\
&= \frac{e^{i\pi/4}}{\sqrt{2}}(\ket{100} + e^{-i\pi/2}\ket{010})
\end{align}

The particle becomes delocalized across both sites with a specific phase relationship.

Applying it twice recovers the full SWAP:
\begin{align}
(\sqrt{\hat{\text{SWAP}}}_{12})^2\ket{100} = \hat{\text{SWAP}}_{12}\ket{100} = \ket{010}
\end{align}
\end{example}

\section{Building Quantum States from Creation Operators}

\subsection{The Vacuum and Fock States}

The vacuum state $\ket{\text{vac}} = \ket{00...0}$ represents the lattice with no particles anywhere. We build configurations by applying creation operators, but the statistics differ:

\textbf{For hardcore bosons:}
\begin{equation}
\ket{n_1n_2...n_N}_{\text{boson}} = \prod_{j: n_j=1} \hat{b}_j^\dagger \ket{\text{vac}}
\end{equation}
Order doesn't matter: $\hat{b}_i^\dagger \hat{b}_j^\dagger = \hat{b}_j^\dagger \hat{b}_i^\dagger$

\textbf{For spinless fermions:}
\begin{equation}
\ket{n_1n_2...n_N}_{\text{fermion}} = \prod_{j: n_j=1} \hat{c}_j^\dagger \ket{\text{vac}}
\end{equation}
Order matters: $\hat{c}_i^\dagger \hat{c}_j^\dagger = -\hat{c}_j^\dagger \hat{c}_i^\dagger$

\begin{example}[Building Many-Particle States]
\textbf{Single particle localized at site $j$ (same for both):}
\begin{align}
\text{Bosons:} \quad &\ket{j} = \hat{b}_j^\dagger \ket{\text{vac}}\\
\text{Fermions:} \quad &\ket{j} = \hat{c}_j^\dagger \ket{\text{vac}}
\end{align}

\textbf{Two particles at sites $i$ and $j$ (statistics matter!):}
\begin{align}
\text{Bosons:} \quad &\ket{i,j} = \hat{b}_i^\dagger \hat{b}_j^\dagger \ket{\text{vac}} = \hat{b}_j^\dagger \hat{b}_i^\dagger \ket{\text{vac}}\\
\text{Fermions:} \quad &\ket{i,j} = \hat{c}_i^\dagger \hat{c}_j^\dagger \ket{\text{vac}} = -\hat{c}_j^\dagger \hat{c}_i^\dagger \ket{\text{vac}}
\end{align}

\textbf{Delocalized single particle (same form):}
\begin{align}
\text{Bosons:} \quad &\ket{\psi} = \sum_{j=1}^N \psi_j \hat{b}_j^\dagger \ket{\text{vac}}\\
\text{Fermions:} \quad &\ket{\psi} = \sum_{j=1}^N \psi_j \hat{c}_j^\dagger \ket{\text{vac}}
\end{align}

\textbf{Example entangled two-particle states:}
\begin{align}
\text{Bosons:} \quad &\ket{\Phi^+} = \frac{1}{\sqrt{2}}(\hat{b}_1^\dagger \hat{b}_2^\dagger + \hat{b}_3^\dagger \hat{b}_4^\dagger)\ket{\text{vac}}\\
\text{Fermions:} \quad &\ket{\Phi^-} = \frac{1}{\sqrt{2}}(\hat{c}_1^\dagger \hat{c}_2^\dagger - \hat{c}_3^\dagger \hat{c}_4^\dagger)\ket{\text{vac}}
\end{align}

Note: These are just illustrative examples. Both particle types can form either $\pm$ combination---the sign is NOT enforced by the operator algebra! The operators themselves already encode the statistics through their (anti)commutation relations.
\end{example}

\subsection{The Total Particle Number and Conservation}

The total number operator counts all particles on the lattice (same form for both):
\begin{align}
\text{Bosons:} \quad \hat{N} &= \sum_{j=1}^N \hat{n}_j = \sum_{j=1}^N \hat{b}_j^\dagger \hat{b}_j\\
\text{Fermions:} \quad \hat{N} &= \sum_{j=1}^N \hat{n}_j = \sum_{j=1}^N \hat{c}_j^\dagger \hat{c}_j
\end{align}

This operator generates a decomposition of the full Hilbert space:
\begin{equation}
\mathcal{H} = \bigoplus_{m=0}^N \mathcal{H}_m
\end{equation}
where $\mathcal{H}_m$ contains all states with exactly $m$ particles. (Note: The direct sum $\oplus$ represents mutually exclusive subspaces---a state is in $\mathcal{H}_0$ OR $\mathcal{H}_1$ OR $\mathcal{H}_2$, etc., but never in multiple sectors simultaneously.)

\begin{example}[Structure of Particle Number Sectors]
For a 4-site lattice (same for bosons and fermions):

\textbf{Zero particles} ($\mathcal{H}_0$, dimension 1):
\begin{equation}
\{\ket{0000}\}
\end{equation}

\textbf{One particle} ($\mathcal{H}_1$, dimension 4):
\begin{equation}
\{\ket{1000}, \ket{0100}, \ket{0010}, \ket{0001}\}
\end{equation}

\textbf{Two particles} ($\mathcal{H}_2$, dimension 6):
\begin{equation}
\{\ket{1100}, \ket{1010}, \ket{1001}, \ket{0110}, \ket{0101}, \ket{0011}\}
\end{equation}

\textbf{Three particles} ($\mathcal{H}_3$, dimension 4):
\begin{equation}
\{\ket{1110}, \ket{1101}, \ket{1011}, \ket{0111}\}
\end{equation}

\textbf{Four particles} ($\mathcal{H}_4$, dimension 1):
\begin{equation}
\{\ket{1111}\}
\end{equation}

Total dimension: $1 + 4 + 6 + 4 + 1 = 16 = 2^4$ 

The dimensions follow Pascal's triangle: $\binom{4}{0}, \binom{4}{1}, \binom{4}{2}, \binom{4}{3}, \binom{4}{4}$
\end{example}

\section{Physical Evolution from Gate Sequences}

\subsection{Particle Hopping via Gates}

Instead of starting with a hopping Hamiltonian, we can directly implement particle motion using gate sequences:

\begin{tcolorbox}[enhanced, title={Hopping from Gates}, colback=green!5, colframe=green!50!black]
Nearest-neighbor particle hopping can be implemented by:
\begin{itemize}
\item Full hop: SWAP gate between adjacent sites (exchanges particle positions)
\item Partial hop: $\sqrt{\text{SWAP}}$ gate (creates quantum superposition)
\item Controlled hop: CNOT-based conditional movement
\item Multi-hop sequence: Chain of SWAP gates to move particles across the lattice
\end{itemize}

These gates preserve particle number and maintain the hardcore constraint automatically.
\end{tcolorbox}

\subsection{Universal Quantum Computation on the Lattice}

The combination of single-site operations and two-site gates forms a universal gate set:

\begin{tcolorbox}[enhanced, title={Universal Gate Set (Loss-DiVincenzo)}, colback=red!5, colframe=red!50!black]
Any unitary operation on $N$ qubits can be approximated to arbitrary precision using:
\begin{enumerate}
\item Single-qubit rotations: $\hat{R}_x(\theta)$, $\hat{R}_y(\theta)$, $\hat{R}_z(\phi)$
\item Two-qubit entangling gate: $\sqrt{\hat{\text{SWAP}}}$ or CNOT
\end{enumerate}

For quantum simulation of particles on a lattice, this means any physically realizable evolution can be implemented through sequences of these basic gates.
\end{tcolorbox}

\section{Hardcore Bosons versus Spinless Fermions: The Key Difference}
\label{sec:bosons_vs_fermions}

\subsection{The Statistical Difference}

While hardcore bosons and spinless fermions have identical on-site properties (both limited to occupation 0 or 1), they differ fundamentally in their multi-site behavior:

\begin{tcolorbox}[enhanced, title={Statistical Distinction}, colback=blue!5, colframe=blue!50!black]
\textbf{Hardcore Bosons:}
\begin{align}
[\hat{b}_i, \hat{b}_j^\dagger] &= 0 \quad \text{for } i \neq j \quad \text{(commute)}\\
[\hat{b}_i, \hat{b}_j] &= 0 \quad \text{for } i \neq j\\
\hat{b}_i^\dagger \hat{b}_j^\dagger &= \hat{b}_j^\dagger \hat{b}_i^\dagger \quad \text{(symmetric)}
\end{align}

\textbf{Spinless Fermions:}
\begin{align}
\{\hat{c}_i, \hat{c}_j^\dagger\} &= 0 \quad \text{for } i \neq j \quad \text{(anticommute)}\\
\{\hat{c}_i, \hat{c}_j\} &= 0 \quad \text{for } i \neq j\\
\hat{c}_i^\dagger \hat{c}_j^\dagger &= -\hat{c}_j^\dagger \hat{c}_i^\dagger \quad \text{(antisymmetric)}
\end{align}
\end{tcolorbox}

\subsection{Jordan-Wigner Transformation}

The Jordan-Wigner transformation maps between spin operators and fermionic operators:

\begin{tcolorbox}[enhanced, title={Jordan-Wigner Transformation}, colback=yellow!5, colframe=yellow!50!black]
\textbf{For hardcore bosons:} Direct mapping to spins!
\begin{align}
\hat{b}_j &= \hat{\sigma}_j^+ = \frac{1}{2}(\sigma_{x,j} + i\sigma_{y,j})\\
\hat{b}_j^\dagger &= \hat{\sigma}_j^- = \frac{1}{2}(\sigma_{x,j} - i\sigma_{y,j})\\
\hat{n}_j &= \frac{1}{2}(\hat{I}_j - \sigma_{z,j})
\end{align}

\textbf{For spinless fermions:} Need non-local string operators!
\begin{align}
\hat{c}_j &= \left(\prod_{k<j} e^{i\pi \hat{n}_k}\right) \hat{\sigma}_j^+ = \left(\prod_{k<j} \sigma_{z,k}\right) \hat{\sigma}_j^+\\
\hat{c}_j^\dagger &= \left(\prod_{k<j} \sigma_{z,k}\right) \hat{\sigma}_j^-\\
\hat{n}_j &= \frac{1}{2}(\hat{I}_j - \sigma_{z,j}) \quad \text{(same as bosons locally)}
\end{align}

The string operator $\prod_{k<j} \sigma_{z,k}$ counts particles to the left, introducing the fermionic sign.
\end{tcolorbox}

This transformation shows that fermions require non-local phase factors to maintain their anticommutation relations, while hardcore bosons map directly to local spin operators.

\subsection{Physical Realizations}

\textbf{Hardcore bosons naturally arise in:}
\begin{enumerate}
\item \textbf{Rydberg atoms}: Strong van der Waals interactions create a "blockade"
\item \textbf{Photons in cavity arrays}: Strong photon-photon interactions
\item \textbf{Josephson junction arrays}: Cooper pairs with strong charging energy
\item \textbf{Ultracold atoms in optical lattices}: Rb-87, Na-23 with strong repulsion
\end{enumerate}

\textbf{Spinless fermions naturally arise in:}
\begin{enumerate}
\item \textbf{Electrons in quantum dots}: The Loss-DiVincenzo proposal
\item \textbf{Cold fermionic atoms}: K-40, Li-6 in optical lattices
\item \textbf{Nuclear spins}: NMR quantum computing with fermionic statistics
\end{enumerate}

\section{From Many Particles to One: Emergence of the Wavefunction}

\subsection{The Single-Particle Sector and Traditional Quantum Mechanics}

When restricted to states with exactly one particle, our field theory reduces to traditional single-particle quantum mechanics. Since a single particle has no exchange partner, bosons and fermions behave identically:

\begin{equation}
\mathcal{H}_1 = \text{span}\{\hat{a}_j^\dagger\ket{\text{vac}} : j = 1, 2, ..., N\}
\end{equation}
where $\hat{a}$ can be either $\hat{b}$ (bosons) or $\hat{c}$ (fermions).

Any single-particle state can be written as:
\begin{equation}
\ket{\psi} = \sum_{j=1}^N \psi_j \hat{a}_j^\dagger \ket{\text{vac}}
\end{equation}

where the complex amplitudes $\psi_j$ form the discrete wavefunction.

\begin{tcolorbox}[enhanced, title={Connection to the Wavefunction}, colback=cyan!5, colframe=cyan!50!black]
The coefficients $\psi_j$ are precisely the values of the wavefunction at lattice sites:
\begin{equation}
\psi_j = \psi(x_j) \quad \text{where} \quad x_j = ja
\end{equation}

Physical observables emerge naturally:
\begin{itemize}
\item Probability density: $|\psi_j|^2 = |\psi(x_j)|^2$
\item Normalization: $\sum_{j=1}^N |\psi_j|^2 = 1$
\item Position expectation: $\langle x \rangle = a\sum_{j=1}^N j|\psi_j|^2$
\end{itemize}
\end{tcolorbox}

\subsection{Position and Momentum Representations}

The position basis states are simply:
\begin{equation}
\ket{x_j} = \hat{a}_j^\dagger\ket{\text{vac}} = \ket{0...010...0}
\end{equation}
with the particle at site $j$.

For a lattice with periodic boundary conditions, we can define momentum eigenstates:
\begin{equation}
\ket{k} = \frac{1}{\sqrt{N}} \sum_{j=1}^N e^{ikx_j} \hat{a}_j^\dagger \ket{\text{vac}}
\end{equation}

where the allowed momenta are quantized: $k = \frac{2\pi m}{Na}$ with $m = 0, 1, ..., N-1$.

\begin{example}[Fourier Transform on the Lattice]
The transformation between position and momentum representations is the discrete Fourier transform:
\begin{align}
\psi(x_j) &= \frac{1}{\sqrt{N}} \sum_{m=0}^{N-1} \tilde{\psi}(k_m) e^{ik_m x_j}\\
\tilde{\psi}(k_m) &= \frac{1}{\sqrt{N}} \sum_{j=1}^N \psi(x_j) e^{-ik_m x_j}
\end{align}

This can be implemented exactly using quantum gates---the Quantum Fourier Transform (QFT).
\end{example}

\subsection{Entanglement and Subspace Projections: When is a Superposition Entangled?}

We close this chapter with a subtle but fundamental observation about the nature of entanglement. Consider a two-site system with the state:
\begin{equation}
\ket{\psi} = \frac{1}{\sqrt{2}}(\ket{10} + \ket{01})
\end{equation}

Is this state entangled? The answer reveals a deep truth: \textit{entanglement depends on the Hilbert space context in which we view the state}.

\begin{tcolorbox}[enhanced, title={The Context-Dependence of Entanglement}, colback=purple!5, colframe=purple!50!black]
\textbf{Viewed in the single-particle sector} $\mathcal{H}_1 = \text{span}\{\ket{10}, \ket{01}\}$:
\begin{itemize}
    \item This is a 2-dimensional space describing one particle's position
    \item We can map: $\ket{10} \rightarrow \ket{0}_{\text{eff}}$ and $\ket{01} \rightarrow \ket{1}_{\text{eff}}$
    \item Then $\ket{\psi} \rightarrow \frac{1}{\sqrt{2}}(\ket{0}_{\text{eff}} + \ket{1}_{\text{eff}}) = \ket{+}_{\text{eff}}$
    \item This is NOT entangled---it's just a superposition of one particle at two locations
    \item Physically: a single particle wavefunction $\psi(x)$ spread over two sites
\end{itemize}

\textbf{Viewed in the full two-qubit space} $\mathcal{H} = \text{span}\{\ket{00}, \ket{01}, \ket{10}, \ket{11}\}$:
\begin{itemize}
    \item This is a 4-dimensional space with tensor product structure $\mathcal{H} = \mathcal{H}_{\text{site1}} \otimes \mathcal{H}_{\text{site2}}$
    \item The state $\ket{\psi} = \frac{1}{\sqrt{2}}(\ket{10} + \ket{01})$ is the Bell state $\ket{\Psi^+}$
    \item This IS entangled---it cannot be written as $\ket{\alpha}_1 \otimes \ket{\beta}_2$
    \item Physically: the two sites have perfectly correlated occupations
\end{itemize}
\end{tcolorbox}

The key insight is that entanglement is defined relative to a tensor product decomposition. When we restrict to the $n=1$ sector, we lose the tensor product structure---we're no longer asking about correlations between sites but about the spatial distribution of a single entity.

\begin{example}[The Resolution of the Paradox]
Consider the following apparent contradiction:
\begin{itemize}
    \item In the $n=1$ sector, no state can be entangled (there's only one particle!)
    \item The same mathematical object $\frac{1}{\sqrt{2}}(\ket{10} + \ket{01})$ is maximally entangled in the full space
\end{itemize}

The resolution: We're asking different questions!
\begin{itemize}
    \item \textbf{$n=1$ sector question}: "Is the particle's wavefunction entangled with itself?" \ensuremath{\to} No, that's meaningless
    \item \textbf{Full space question}: "Are the two sites entangled?" \ensuremath{\to} Yes, their occupations are correlated
\end{itemize}

This is analogous to the difference between first and second quantization:
\begin{itemize}
    \item \textbf{First quantization}: Single-particle wavefunction $\psi(x)$---no notion of entanglement
    \item \textbf{Second quantization}: Occupation number basis---entanglement between modes/sites
\end{itemize}
\end{example}

This phenomenon generalizes to $N$ sites:

\begin{tcolorbox}[enhanced, title={Generalization to N Sites}, colback=green!5, colframe=green!50!black]
For $N$ sites with one particle, any state $\ket{\psi} = \sum_{j=1}^N \psi_j \ket{j}$ in the $n=1$ sector:
\begin{itemize}
    \item \textbf{Within $\mathcal{H}_1$}: Never entangled (it's a single-particle wavefunction)
    \item \textbf{In full $2^N$-dimensional space}: Can be highly entangled between sites
\end{itemize}

Example for 3 sites:
\begin{equation}
\ket{W} = \frac{1}{\sqrt{3}}(\ket{100} + \ket{010} + \ket{001})
\end{equation}
\begin{itemize}
    \item In $\mathcal{H}_1$: Just a uniform superposition (not entangled)
    \item In full space: Genuine tripartite entanglement (W-state)
\end{itemize}
\end{tcolorbox}

\begin{mathematicalaside}{Why This Matters}
This distinction is crucial for understanding:
\begin{enumerate}
    \item \textbf{Quantum simulation}: When simulating fermions with qubits, are we creating "real" entanglement or just mimicking single-particle physics?
    \item \textbf{Quantum computation}: Subspace restrictions can make entangled states appear unentangled
    \item \textbf{Many-body physics}: The transition from few to many particles changes the relevant questions about correlations
    \item \textbf{Measurement theory}: What we measure depends on which subspace we probe
\end{enumerate}

The lesson: Quantum properties like entanglement are not absolute but emerge from our choice of description and the constraints we impose on the system.
\end{mathematicalaside}

\begin{physicalinsight}
In the two-particle sector, an even richer structure emerges. Two distinguishable particles can be entangled through their spatial wavefunctions, their spin degrees of freedom, or both. The interplay between spatial and spin entanglement, combined with fermionic antisymmetry, leads to phenomena like the spin-statistics connection and exchange interactions. This topic, while beyond our current scope, forms the foundation of quantum chemistry and condensed matter physics.

For further exploration of entanglement in identical particle systems, see the references on fermionic entanglement and mode entanglement at the end of this chapter.
\end{physicalinsight}

\section{Chapter Summary}

This chapter has revealed the profound connection between abstract quantum information and concrete physical systems. We began by establishing that any unitary transformation on $N$ qubits can be decomposed into single-qubit rotations and two-qubit entangling gates---a powerful generalization of single-qubit Bloch sphere rotations to many-body systems.

Through the spin-particle correspondence, we discovered that qubits can represent occupation numbers on a lattice, with spin up as empty sites and spin down as occupied sites. This transforms collections of two-level systems into quantum fields. We introduced two types of particles:
\begin{itemize}
    \item \textbf{Hardcore bosons} ($\hat{b}_i$, $\hat{b}_i^\dagger$): Commute at different sites, no exchange phase
    \item \textbf{Spinless fermions} ($\hat{c}_i$, $\hat{c}_i^\dagger$): Anticommute at different sites, acquire phase under exchange
\end{itemize}

We showed how quantum gates like SWAP and $\sqrt{\text{SWAP}}$ have direct physical interpretations as particle exchange and coherent tunneling operations. The Loss-DiVincenzo framework demonstrated that natural physical interactions (exchange coupling) can implement universal quantum computation, with the $\sqrt{\text{SWAP}}$ gate plus single-qubit rotations forming a complete gate set.

The distinction between hardcore bosons and spinless fermions emerged through their commutation relations---identical local physics but different global quantum statistics. The Jordan-Wigner transformation revealed that fermions require non-local string operators to maintain anticommutation, while bosons map directly to local spin operators. When restricted to single particles, both types behave identically, reducing to familiar quantum mechanics with creation operator coefficients forming the discrete wavefunction.

The framework developed here bridges quantum computation and many-body physics. We've seen that quantum fields can be understood through discrete gate operations, where any unitary evolution decomposes into elementary gates. The choice between bosons and fermions depends on the physical system being modeled, with bosons offering simpler mathematics and fermions capturing electronic physics.

In the following chapter, we will extend this framework to continuous time evolution, showing how Hamiltonians emerge as generators of unitary gates and connecting our discrete gate model to traditional quantum dynamics.

\input{book_problems/ch03_problems.tex}

\section*{References and Further Reading}
\addcontentsline{toc}{section}{References and Further Reading}

\begin{description}
\item[Loss, D., and DiVincenzo, D.~P.] ``Quantum computation with quantum dots.'' \emph{Physical Review A} \textbf{57}, 120--126 (1998). \href{https://doi.org/10.1103/PhysRevA.57.120}{doi:10.1103/PhysRevA.57.120}. The proposal that motivates the $\sqrt{\text{SWAP}}$-plus-single-qubit gate set used in this chapter; reads cleanly as physics, not just as a computing proposal.

\item[Jordan, P., and Wigner, E.] ``\"Uber das Paulische \"Aquivalenzverbot.'' \emph{Zeitschrift f\"ur Physik} \textbf{47}, 631--651 (1928). \href{https://doi.org/10.1007/BF01331938}{doi:10.1007/BF01331938}. The original transformation between spin and fermion operators; the source of the string operators that appear in our derivation.

\item[Lloyd, S.] ``Universal quantum simulators.'' \emph{Science} \textbf{273}, 1073--1078 (1996). \href{https://doi.org/10.1126/science.273.5278.1073}{doi:10.1126/science.273.5278.1073}. The proof that local Hamiltonians can be efficiently simulated on a quantum computer; the theoretical basis for everything in this chapter that goes by the name ``quantum simulation.''

\item[Hanson, R., Kouwenhoven, L.~P., Petta, J.~R., Tarucha, S., and Vandersypen, L.~M.~K.] ``Spins in few-electron quantum dots.'' \emph{Reviews of Modern Physics} \textbf{79}, 1217--1265 (2007). \href{https://doi.org/10.1103/RevModPhys.79.1217}{doi:10.1103/RevModPhys.79.1217}. The standard review of the experimental platform that the Loss--DiVincenzo proposal envisioned; concrete numbers for exchange couplings, gate times, and decoherence.

\item[Georgescu, I.~M., Ashhab, S., and Nori, F.] ``Quantum simulation.'' \emph{Reviews of Modern Physics} \textbf{86}, 153--185 (2014). \href{https://doi.org/10.1103/RevModPhys.86.153}{doi:10.1103/RevModPhys.86.153}. Modern review of digital and analog quantum simulation; surveys spin and fermionic systems together with current experimental platforms.

\item[Esslinger, T.] ``Fermi-Hubbard physics with atoms in an optical lattice.'' \emph{Annual Review of Condensed Matter Physics} \textbf{1}, 129--152 (2010). \href{https://doi.org/10.1146/annurev-conmatphys-070909-104059}{doi:10.1146/annurev-conmatphys-070909-104059}. The cleanest experimental realization of the lattice fermion physics derived here, in cold atomic systems where occupation numbers and hopping amplitudes are directly tunable.
\end{description}

%% file: book_problems/ch03_problems.tex
\section{Problems}
\setcounter{hwproblem}{0}  

\problem{Spin-Particle Correspondence and Ladder Operators}
Consider a three-site lattice with the spin-particle correspondence: $\ket{\uparrow}_i \equiv \ket{0}_i$ (empty) and $\ket{\downarrow}_i \equiv \ket{1}_i$ (occupied).
\begin{enumerate}[label=(\alph*)]
    \item Write the matrix representations for the creation operator $\hat{b}_2^\dagger$, annihilation operator $\hat{b}_2$, and number operator $\hat{n}_2$ at site 2
    \item Apply $\hat{b}_2^\dagger$ to the state $\ket{100}$ and $\ket{110}$. Explain the physical meaning of each result
    \item Calculate $(\hat{b}_2^\dagger)^2\ket{100}$ and explain why this result makes sense for hardcore particles
    \item Show that $\hat{n}_2 = \hat{b}_2^\dagger \hat{b}_2$ by explicit matrix multiplication
    \item Using Bloch cubes, demonstrate the action of $\hat{b}_2^\dagger$ on the state $\ket{100}$ by showing the cube orientations before and after
\end{enumerate}

\problem{Gate Operations on Lattice States}
Consider the SWAP gate between sites 1 and 2:
\begin{equation}
\hat{\text{SWAP}}_{12} = \begin{pmatrix}
1 & 0 & 0 & 0 \\
0 & 0 & 1 & 0 \\
0 & 1 & 0 & 0 \\
0 & 0 & 0 & 1
\end{pmatrix}
\end{equation}
\begin{enumerate}[label=(\alph*)]
    \item Apply this gate to the states $\ket{10}$, $\ket{01}$, and $\ket{11}$. Interpret the results physically in terms of particle positions
    \item Apply the gate to the superposition state $\frac{1}{\sqrt{2}}(\ket{10} + \ket{01})$. What type of state do you get?
    \item Now consider the $\sqrt{\text{SWAP}}$ gate. Apply it to $\ket{10}$ and show that applying it twice gives the full SWAP operation
    \item Express the action of $\hat{\text{SWAP}}_{12}$ using creation and annihilation operators (hint: think about what operators exchange particles between sites)
\end{enumerate}

\problem{Building Many-Body States}
Consider a four-site lattice with hardcore particles.
\begin{enumerate}[label=(\alph*)]
    \item Write the vacuum state $\ket{\text{vac}}$ and create the single-particle states at each site using creation operators
    \item Create a delocalized single-particle state: $\ket{\psi} = \frac{1}{2}(\hat{a}_1^\dagger + \hat{a}_2^\dagger + \hat{a}_3^\dagger + \hat{a}_4^\dagger)\ket{\text{vac}}$. What is this state in the occupation number basis?
    \item Calculate the total particle number operator $\hat{N} = \sum_{i=1}^4 \hat{n}_i$ and verify that $\ket{\psi}$ is an eigenstate with eigenvalue 1
    \item Create a two-particle state with particles at sites 1 and 3: $\ket{1,3} = \hat{a}_1^\dagger \hat{a}_3^\dagger \ket{\text{vac}}$
    \item How many total two-particle states are possible on this four-site lattice? List them all
    \item Using the binomial coefficient formula, verify that the total dimension of the Hilbert space is $2^4 = 16$
\end{enumerate}

\problem{Entanglement in Different Contexts}
Consider the state $\ket{\psi} = \frac{1}{\sqrt{2}}(\ket{10} + \ket{01})$ on a two-site lattice.
\begin{enumerate}[label=(\alph*)]
    \item Full Hilbert space view: Treat this as a two-qubit state in the 4-dimensional space $\{\ket{00}, \ket{01}, \ket{10}, \ket{11}\}$. Show that it cannot be written as a product state $\ket{a}_1 \otimes \ket{b}_2$
    \item Calculate the reduced density matrices $\rho_1$ and $\rho_2$ by tracing out the other site. What is the purity of each?
    \item Single-particle sector view: Now view this as a single-particle state in the 2-dimensional space $\{\ket{10}, \ket{01}\}$. Define $\ket{L} = \ket{10}$ (particle on left) and $\ket{R} = \ket{01}$ (particle on right)
    \item In this new basis, is $\ket{\psi} = \frac{1}{\sqrt{2}}(\ket{L} + \ket{R})$ entangled? Explain your reasoning
    \item Compare the physical interpretations: when is this state describing ``entanglement between sites'' vs ``superposition of a particle's position''?
    \item This demonstrates that entanglement depends on the tensor product structure we choose. Discuss the implications for quantum simulation and many-body physics
\end{enumerate}

\problem{Tight-Binding Dispersion Relation}
For a one-dimensional lattice with nearest-neighbor hopping amplitude $t$, the single-particle Hamiltonian is $\hat{H} = -t\sum_{\langle i,j\rangle} (\hat{a}_i^\dagger \hat{a}_j + \hat{a}_j^\dagger \hat{a}_i)$. In momentum space, this gives $E(k) = -2t\cos(ka)$ where $a$ is the lattice spacing.
\begin{enumerate}[label=(\alph*)]
    \item Plot the dispersion relation $E(k)$ for $k \in [-\pi/a, \pi/a]$ (first Brillouin zone)
    \item Find the minimum and maximum energy and explain their physical significance
    \item For a finite chain of $N=5$ sites with periodic boundary conditions, write the allowed $k$ values
    \item Calculate the density of states $\rho(E) = |dE/dk|^{-1}$ and explain where it diverges
    \item For a chain with half-filled occupation (2 particles on 4 sites), estimate the ground state energy using the tight-binding model
\end{enumerate}

\problem{Momentum Eigenstates on a Ring}
For a particle on a ring of $N$ sites with periodic boundary conditions, momentum eigenstates are $\ket{k} = \frac{1}{\sqrt{N}}\sum_{j=1}^N e^{ikx_j} \hat{a}_j^\dagger \ket{\text{vac}}$ with $k = 2\pi m/Na$ and $m = 0,1,...,N-1$.
\begin{enumerate}[label=(\alph*)]
    \item Show that $\ket{k}$ is an eigenstate of the total momentum operator $\hat{P} = \sum_j k_j \hat{n}_j$ with eigenvalue $\hbar k$ (for each particle)
    \item For $N=4$ sites and $a=1$, write out the explicit form of the $k=\pi/2$ eigenstate in real space
    \item Calculate $\braket{k}{k'}$ for two different momentum eigenstates and verify orthonormality
    \item Show that the momentum space wavefunction $\tilde{\psi}(k) = \frac{1}{\sqrt{N}}\sum_j e^{-ikx_j}\psi(x_j)$ is the discrete Fourier transform of the real-space wavefunction
    \item For a Gaussian wavepacket in momentum space centered at $k_0$, describe its behavior in real space on the lattice
\end{enumerate}

\problem{Particle-Hole Symmetry of Half-Filled Lattice}
For a half-filled lattice with $N$ sites and $N/2$ particles, particle-hole symmetry pairs empty sites with occupied sites. The transformation is $\hat{a}_i \leftrightarrow \hat{a}_i^\dagger$ (up to phases and careful ordering).
\begin{enumerate}[label=(\alph*)]
    \item For a two-site system with one particle, list all states and their particle-hole conjugates
    \item Show that the state $\frac{1}{\sqrt{2}}(\ket{10} + \ket{01})$ maps to itself under particle-hole transformation
    \item For the half-filled 4-site tight-binding model with $E(k) = -2t\cos(ka)$, show that $E(k) = -E(\pi - k)$ (particle-hole symmetry of the spectrum)
    \item Explain why this symmetry implies that the chemical potential at half-filling is always $\mu = 0$
    \item For a non-half-filled system, does particle-hole symmetry still hold? If not, what is broken?
\end{enumerate}

\problem{Jordan-Wigner Transformation for Fermions}
The Jordan-Wigner (JW) transformation maps spinless fermions to spins using non-local string operators: $\hat{c}_j = \left(\prod_{k<j} e^{i\pi \hat{n}_k}\right) \hat{\sigma}_j^+ = \left(\prod_{k<j} \sigma_{z,k}\right) \hat{\sigma}_j^+$.
\begin{enumerate}[label=(\alph*)]
    \item For a two-site system, explicitly write out $\hat{c}_1$ and $\hat{c}_2$ using Pauli matrices
    \item Verify that $\{\hat{c}_j, \hat{c}_k\} = 0$ for $j \neq k$ (anticommutation at different sites)
    \item Verify that $\{\hat{c}_j, \hat{c}_j^\dagger\} = \hat{I}$ (on-site anticommutation)
    \item Show that the occupation number operator $\hat{n}_j = \hat{c}_j^\dagger \hat{c}_j$ is local: it equals $\frac{1}{2}(\hat{I} - \sigma_{z,j})$
    \item Compute the hopping term $\hat{c}_i^\dagger \hat{c}_{i+1}$ in the spin language and explain why non-local string operators appear
\end{enumerate}

\problem{Hopping in 2D Square Lattice}
Extend tight-binding to a 2D square lattice with sites labeled $(m,n)$ and nearest-neighbor hopping $t$.
\begin{enumerate}[label=(\alph*)]
    \item Write the Hamiltonian: $\hat{H} = -t\sum_{\langle (m,n),(m',n')\rangle} (\hat{a}_{m,n}^\dagger \hat{a}_{m',n'} + \text{h.c.})$
    \item In momentum space with $k = (k_x, k_y)$, show that $E(\vec{k}) = -2t[\cos(k_x a) + \cos(k_y a)]$
    \item Plot the Fermi surface (constant energy contour) for a given filling
    \item For a half-filled 2D system, estimate the ground state energy
    \item Explain how the density of states differs at the van Hove singularities (where $dE/dk = 0$) compared to the 1D case
\end{enumerate}

\problem{Current Operator and Particle Flow}
The current operator represents particle flow between sites. Define $\hat{I}_{ij} = i\frac{t}{\hbar}(\hat{a}_i^\dagger \hat{a}_j - \hat{a}_j^\dagger \hat{a}_i)$ (current from $i$ to $j$).
\begin{enumerate}[label=(\alph*)]
    \item Show that current is Hermitian: $\hat{I}_{ij}^\dagger = -\hat{I}_{ij}$ (current reverses direction under conjugation)
    \item Verify current conservation: $\frac{d\hat{n}_i}{dt} + \sum_j \hat{I}_{ij} = 0$ (particle number at site $i$ changes due to net current flow)
    \item For a superposition $\ket{\psi} = \frac{1}{\sqrt{2}}(\ket{10} + \ket{01})$ on a two-site lattice, calculate $\langle \hat{I}_{12}\rangle$
    \item Explain the relationship between $\langle \hat{I}_{ij}\rangle$ and quantum coherence between sites
    \item In the tight-binding ground state, is there net particle current? Explain why or why not
\end{enumerate}

\problem{Commutator Algebra for Lattice Operators}
The commutation relations $[\hat{a}_i, \hat{a}_j^\dagger] = \delta_{ij}$ and $[\hat{a}_i, \hat{a}_j] = 0$ (for bosons/hardcore bosons) determine all operator algebra on the lattice.
\begin{enumerate}[label=(\alph*)]
    \item Show that $[\hat{n}_i, \hat{n}_j] = 0$ for all $i,j$ (particle numbers commute)
    \item Calculate $[\hat{a}_i^\dagger \hat{a}_j, \hat{a}_k^\dagger \hat{a}_l]$ for general sites (hint: use Jacobi identity and commutation relations)
    \item Show that hopping operators at different sites commute: $[\hat{a}_i^\dagger\hat{a}_{i+1}, \hat{a}_{j+1}^\dagger\hat{a}_j] = 0$ for non-overlapping pairs
    \item For the number operator $\hat{N} = \sum_i \hat{n}_i$, verify that $[\hat{N}, \hat{a}_i^\dagger] = \hat{a}_i^\dagger$ (creation increases particle count)
    \item Compute $[\hat{H}_{\text{hop}}, \hat{n}_i]$ for the hopping Hamiltonian and explain what symmetry this reveals
\end{enumerate}

%% file: chapters/ch04_time.tex
\chapter{Continuous Time Evolution and Dynamics}
\label{ch:time}

\section{Introduction: From Discrete Rotations to Continuous Time Evolution}

Throughout the previous chapters, we have worked exclusively with unitary operators---from the discrete 90-degree rotations of the Bloch cube to the universal gate sets for many-body quantum systems. These unitary transformations represent the fundamental mathematical structure of quantum mechanics. Now we explore how unitary evolution emerges from exponentiation of Hermitian operators, the infinitesimal generators of time translation, connecting our discrete manipulations to the flowing dynamics of physical systems.

This chapter bridges the gap between our discrete quantum toolkit and the continuous time evolution that governs real physical systems. We will see that the discrete rotations we have studied are special cases of a much richer continuous dynamics, where time evolution is generated by Hermitian operators called Hamiltonians. The Schr\"odinger equation emerges naturally as the differential form of unitary evolution, and the rotation matrices we constructed for the Bloch cube become specific instances of the general exponential formula $U(t) = e^{-iHt/\hbar}$.

The transition from discrete to continuous operations reveals new physical phenomena impossible in our discrete framework. Quantum interference patterns arise from precise phase relationships that evolve continuously. Resonance effects depend critically on timing and frequency matching. Conservation laws emerge from symmetries of the Hamiltonian. Most importantly, we will see how the non-commutativity we encountered in the Pauli matrix algebra underlies all quantum dynamics, from simple two-level systems to complex many-body physics.

\section{Unitary Operators and the Schr\"odinger Equation}

\subsection{The Fundamental Postulate of Quantum Dynamics}

Every quantum mechanical operation we have encountered---from Bloch cube rotations to quantum gates---has been described by a unitary operator. For any quantum system evolving from time $t = 0$ to time $t$, there exists a unitary operator $\hat{U}(t)$ such that:

\begin{equation}
\ket{\psi(t)} = \hat{U}(t)\ket{\psi(0)}
\end{equation}

The time evolution operator must satisfy several physical requirements that ensure the consistency of quantum mechanics:

First, it must be unitary: $\hat{U}^\dagger(t)\hat{U}(t) = \hat{I}$. This guarantees that quantum states remain normalized for all times. If we start with a normalized state $\langle\psi(0)|\psi(0)\rangle = 1$, then $\langle\psi(t)|\psi(t)\rangle = \langle\psi(0)|\hat{U}^\dagger(t)\hat{U}(t)|\psi(0)\rangle = 1$ for all times.

Second, it must satisfy the group property: $\hat{U}(t_1)\hat{U}(t_2) = \hat{U}(t_1 + t_2)$. Evolution for time $t_1$ followed by evolution for time $t_2$ must equal evolution for the total time $t_1 + t_2$. This property is essential for the consistency of quantum mechanics and ensures that the dynamics are Markovian---that is, future evolution depends only on the present state, not on the past history. A process is Markovian when the probability of transitioning to a future state depends solely on the current state, with no memory of how that state was reached.

Third, it must be continuous: $\hat{U}(0) = \hat{I}$ and $\hat{U}(t)$ varies smoothly with time. At $t = 0$, no evolution has occurred, so the evolution operator is the identity. The smoothness condition ensures that quantum states don't evolve discontinuously.

\subsection{Hamiltonian: The Generator of Time Evolution}

For infinitesimal time evolution, we can expand the unitary operator to first order in $dt$:
\begin{equation}
\hat{U}(dt) = \hat{I} + \hat{G} \, dt + O(dt^2)
\end{equation}

where $\hat{G}$ is the generator of time evolution. Since $\hat{U}(dt)$ must be unitary to first order, we have:
\begin{align}
\hat{U}^\dagger(dt)\hat{U}(dt) &= (\hat{I} + \hat{G}^\dagger dt)(\hat{I} + \hat{G} dt) + O(dt^2)\\
&= \hat{I} + (\hat{G}^\dagger + \hat{G})dt + O(dt^2) = \hat{I}
\end{align}

This requires $\hat{G}^\dagger + \hat{G} = 0$, so $\hat{G}$ must be anti-Hermitian. We conventionally write $\hat{G} = -i\hat{H}/\hbar$ where $\hat{H}$ is a Hermitian operator called the Hamiltonian. This gives:
\begin{equation}
\hat{U}(dt) = \hat{I} - i\hat{H}dt/\hbar
\end{equation}

\subsection{The Schr\"odinger Equation for Time-Independent Hamiltonians}

For finite time evolution with a \textbf{time-independent} Hamiltonian, the group property demands that:
\begin{equation}
\hat{U}(t + dt) = \hat{U}(dt)\hat{U}(t) = \left(\hat{I} - i\hat{H}dt/\hbar\right)\hat{U}(t)
\end{equation}

Rearranging and taking the limit $dt \to 0$:
\begin{align}
\frac{\hat{U}(t + dt) - \hat{U}(t)}{dt} &= -i\hat{H}\hat{U}(t)/\hbar\\
\frac{d\hat{U}(t)}{dt} &= -i\hat{H}\hat{U}(t)/\hbar
\end{align}

This differential equation has the solution:
\begin{equation}
\hat{U}(t) = e^{-i\hat{H}t/\hbar}
\end{equation}
\textbf{provided the Hamiltonian is time-independent}. This crucial assumption allows the exponential form and underlies much of elementary quantum mechanics.

Applying this to an arbitrary state gives us the time-dependent Schr\"odinger equation:
\begin{align}
\ket{\psi(t)} &= \hat{U}(t)\ket{\psi(0)} = e^{-i\hat{H}t/\hbar}\ket{\psi(0)}\\
\frac{d}{dt}\ket{\psi(t)} &= \frac{d}{dt}e^{-i\hat{H}t/\hbar}\ket{\psi(0)} = -i\hat{H}e^{-i\hat{H}t/\hbar}\ket{\psi(0)}/\hbar\\
&= -i\hat{H}\ket{\psi(t)}/\hbar
\end{align}

Rearranging yields the familiar form:
\begin{equation}
i\hbar\frac{d}{dt}\ket{\psi(t)} = \hat{H}\ket{\psi(t)}
\end{equation}

The Schr\"odinger equation is thus the differential formulation of unitary quantum evolution. Every discrete unitary operation we have studied can be understood as arising from continuous evolution under an appropriate Hamiltonian for a specific duration.

\section{Series Solution and Temporal Composition}

\subsection{Building Finite Evolution from Infinitesimal Steps}

The exponential form of the time evolution operator $\hat{U}(t) = e^{-i\hat{H}t/\hbar}$ has a profound physical interpretation that connects directly to our understanding of infinitesimal evolution. The fundamental definition of the exponential function:
\begin{equation}
e^x = \lim_{n \to \infty} \left(1 + \frac{x}{n}\right)^n
\end{equation}

Applied to our time evolution operator, this becomes:
\begin{equation}
\hat{U}(t) = e^{-i\hat{H}t/\hbar} = \lim_{n \to \infty} \left(\hat{I} - \frac{i\hat{H}t}{n\hbar}\right)^n
\end{equation}

This formula has immediate physical meaning: finite time evolution emerges from composing many infinitesimal steps. If we divide the total time $t$ into $n$ small intervals of duration $\Delta t = t/n$, then:
\begin{equation}
\hat{U}(t) = \lim_{n \to \infty} \left[\hat{U}(\Delta t)\right]^n = \lim_{n \to \infty} \left[\hat{I} - \frac{i\hat{H}\Delta t}{\hbar}\right]^n
\end{equation}

This shows that evolving for time $t$ is equivalent to applying the infinitesimal evolution operator $n$ times, where each application evolves the system for time $t/n$. As $n \to \infty$, the discrete steps become continuous evolution.

The exponential can also be understood through its series expansion:
\begin{equation}
\hat{U}(t) = \sum_{n=0}^\infty \frac{(-it/\hbar)^n}{n!}\hat{H}^n = \hat{I} - \frac{it}{\hbar}\hat{H} + \frac{1}{2!}\left(\frac{-it}{\hbar}\right)^2\hat{H}^2 + \cdots
\end{equation}

This series provides computational insight. The first few terms show how finite time evolution emerges from infinitesimal changes: the zeroth order term $\hat{I}$ represents no evolution, the first order term $-\frac{it}{\hbar}\hat{H}$ gives the instantaneous rate of change, and the second order term provides acceleration effects.

For two-level systems where $\hat{H} = \vec{h} \cdot \vec{\sigma}$, the series simplifies dramatically because $(\vec{h} \cdot \vec{\sigma})^{2n} = |\vec{h}|^{2n}\hat{I}$ and $(\vec{h} \cdot \vec{\sigma})^{2n+1} = |\vec{h}|^{2n}(\vec{h} \cdot \vec{\sigma})$. This allows us to separate even and odd powers:
\begin{align}
\hat{U}(t) &= \sum_{n=0}^\infty \frac{(-1)^n(|\vec{h}|t/\hbar)^{2n}}{(2n)!}\hat{I} + \sum_{n=0}^\infty \frac{(-1)^n(-i)(|\vec{h}|t/\hbar)^{2n+1}}{(2n+1)!}\frac{\vec{h} \cdot \vec{\sigma}}{|\vec{h}|}\\
&= \cos(|\vec{h}|t/\hbar)\hat{I} - i\sin(|\vec{h}|t/\hbar)\frac{\vec{h} \cdot \vec{\sigma}}{|\vec{h}|}
\end{align}

This recovers our rotation formula and demonstrates how the matrix exponential reduces to familiar trigonometric functions.

\subsection{Temporal Composition and the Trotter Formula}

A fundamental property of time evolution is that it composes: evolution for time $t_1$ followed by evolution for time $t_2$ equals evolution for total time $t_1 + t_2$. Mathematically:
\begin{equation}
\hat{U}(t_1 + t_2) = \hat{U}(t_2)\hat{U}(t_1) = e^{-i\hat{H}(t_1 + t_2)/\hbar}
\end{equation}

This composition property allows us to build long-time evolution from many short-time steps, as we saw in the limit definition of the exponential:
\begin{equation}
\hat{U}(T) = \lim_{N \to \infty} \left[\hat{U}(T/N)\right]^N = \lim_{N \to \infty} \left[e^{-i\hat{H}T/(N\hbar)}\right]^N
\end{equation}

For small time steps $\Delta t = T/N$, we can approximate:
\begin{equation}
\hat{U}(\Delta t) \approx \hat{I} - i\hat{H}\Delta t/\hbar
\end{equation}

This leads to the discrete time evolution:
\begin{equation}
\hat{U}(T) \approx \left(\hat{I} - i\hat{H}T/(N\hbar)\right)^N
\end{equation}

In the limit $N \to \infty$, this reproduces the continuous evolution operator, as we established earlier through the exponential limit definition. This is not just a mathematical curiosity---it shows that continuous quantum evolution can be approximated arbitrarily well by discrete time steps, a principle fundamental to quantum simulation.

When the Hamiltonian is a sum of non-commuting terms, $\hat{H} = \hat{H}_1 + \hat{H}_2$, the evolution operator cannot be written as a simple product $e^{-i\hat{H}_1 t/\hbar}e^{-i\hat{H}_2 t/\hbar}$ because the exponentials don't factor when the operators don't commute. However, for small time steps, the Trotter formula provides an approximation:
\begin{equation}
e^{-i(\hat{H}_1 + \hat{H}_2)\Delta t/\hbar} \approx e^{-i\hat{H}_1\Delta t/\hbar}e^{-i\hat{H}_2\Delta t/\hbar} + O((\Delta t)^2)
\end{equation}

This allows complex many-body evolution to be decomposed into simpler pieces that can be implemented individually. For quantum simulation, this means that complicated Hamiltonians can be built from elementary quantum gates, each implementing one term in the Hamiltonian decomposition. The Trotter approach bridges the gap between continuous quantum dynamics and discrete quantum computation, showing how any quantum evolution can be efficiently approximated using a quantum computer with sufficient gates and qubits.

\section{Bloch Sphere Rotations from Hamiltonians}

\subsection{From Discrete Faces to Continuous Sphere}

The six faces of the Bloch cube correspond to the eigenstates of the three Pauli matrices:
\begin{align}
\sigma_x\ket{\pm x} &= \pm\ket{\pm x}\\
\sigma_y\ket{\pm y} &= \pm\ket{\pm y}\\
\sigma_z\ket{\pm z} &= \pm\ket{\pm z}
\end{align}

These discrete states are special points on the Bloch sphere, but the sphere contains infinitely many other quantum states. Any pure state of a two-level system can be written as:
\begin{equation}
\ket{\psi(\theta,\phi)} = \cos(\theta/2)\ket{+z} + e^{i\phi}\sin(\theta/2)\ket{-z}
\end{equation}

where $\theta \in [0,\pi]$ and $\phi \in [0,2\pi)$ are the polar and azimuthal angles on the unit sphere.

\subsection{Pauli Matrices as Generators of Rotations}

The continuous rotations that connect all points on the Bloch sphere are generated by the Pauli matrices through exponentiation. For any unit vector $\hat{n} = (n_x, n_y, n_z)$ and rotation angle $\theta$, the rotation operator is:
\begin{equation}
\hat{R}_{\hat{n}}(\theta) = e^{-i\theta(\hat{n} \cdot \vec{\sigma})/2}
\end{equation}

where $\vec{\sigma} = (\sigma_x, \sigma_y, \sigma_z)$ is the vector of Pauli matrices.

The key mathematical property that makes this exponential tractable is that $(\hat{n} \cdot \vec{\sigma})^2 = \hat{I}$ for any unit vector $\hat{n}$. This allows us to evaluate the exponential using the identity:
\begin{equation}
e^{-i\theta(\hat{n} \cdot \vec{\sigma})/2} = \cos(\theta/2)\hat{I} - i\sin(\theta/2)(\hat{n} \cdot \vec{\sigma})
\end{equation}

This is the matrix analog of Euler's formula $e^{i\theta} = \cos\theta + i\sin\theta$.

For rotation about the z-axis by angle $\theta$:
\begin{align}
\hat{R}_z(\theta) &= e^{-i\theta\sigma_z/2} = \cos(\theta/2)\hat{I} - i\sin(\theta/2)\sigma_z\\
&= \begin{pmatrix} \cos(\theta/2) - i\sin(\theta/2) & 0 \\ 0 & \cos(\theta/2) + i\sin(\theta/2) \end{pmatrix}\\
&= \begin{pmatrix} e^{-i\theta/2} & 0 \\ 0 & e^{i\theta/2} \end{pmatrix}
\end{align}

For rotation about the x-axis by angle $\theta$:
\begin{align}
\hat{R}_x(\theta) &= \cos(\theta/2)\hat{I} - i\sin(\theta/2)\sigma_x\\
&= \begin{pmatrix} \cos(\theta/2) & -i\sin(\theta/2) \\ -i\sin(\theta/2) & \cos(\theta/2) \end{pmatrix}
\end{align}

The discrete rotations of the Bloch cube correspond to specific angles. For instance, a 90-degree rotation about the z-axis gives $\hat{Z} = \hat{R}_z(\pi/2)$, while a 90-degree rotation about the x-axis yields $\hat{X} = \hat{R}_x(\pi/2)$.

\textbf{Bloch Cube Visualization and Energy Conservation:} An important property of rotations about the z-axis is that they preserve the expectation value of energy when the Hamiltonian is proportional to $\sigma_z$. Consider a state represented on the Bloch cube undergoing rotation about the z-axis. The energy expectation value is given by $\langle\hat{H}\rangle = \langle\psi|\sigma_z|\psi\rangle$, which corresponds to the z-component of the Bloch vector. When we rotate the Bloch cube about the z-axis, the z-component remains unchanged---you can visualize this by imagining spinning the cube about its vertical axis. The x and y components change, but the height (z-component) of any point on the cube stays the same. This geometric invariance directly translates to conservation of energy during the rotation.

\subsection{Physical Interpretation: Spin in a Magnetic Field}

The most direct physical realization of these rotations is a spin-1/2 particle in a magnetic field. The interaction Hamiltonian is:
\begin{equation}
\hat{H} = -\vec{\mu} \cdot \vec{B} = -\gamma\vec{S} \cdot \vec{B} = -\frac{\gamma\hbar}{2}\vec{\sigma} \cdot \vec{B}
\end{equation}

where $\gamma$ is the gyromagnetic ratio and $\vec{S} = \frac{\hbar}{2}\vec{\sigma}$ is the spin angular momentum.

For a magnetic field $\vec{B} = B\hat{n}$ in direction $\hat{n}$, the time evolution operator becomes:
\begin{equation}
\hat{U}(t) = e^{-i\hat{H}t/\hbar} = e^{i\gamma Bt(\hat{n} \cdot \vec{\sigma})/2} = \hat{R}_{\hat{n}}(\gamma Bt)
\end{equation}

The spin precesses about the magnetic field direction with Larmor frequency $\omega_L = \gamma B$. The angle of rotation grows linearly with time, implementing precisely the continuous rotations we derived algebraically.

\begin{example}{Larmor Precession}
Consider a spin initially in the $\ket{+x}$ state in a magnetic field $\vec{B} = B\hat{z}$. The time evolution is:
\begin{align}
\ket{\psi(t)} &= \hat{R}_z(\omega_L t)\ket{+x}\\
&= e^{-i\omega_L t\sigma_z/2}\frac{1}{\sqrt{2}}(\ket{+z} + \ket{-z})\\
&= \frac{1}{\sqrt{2}}(e^{-i\omega_L t/2}\ket{+z} + e^{i\omega_L t/2}\ket{-z})\\
&= \frac{e^{-i\omega_L t/2}}{\sqrt{2}}(\ket{+z} + e^{i\omega_L t}\ket{-z})
\end{align}

Factoring out the global phase and converting back to Cartesian coordinates:
\begin{align}
\langle\sigma_x\rangle(t) &= \cos(\omega_L t)\\
\langle\sigma_y\rangle(t) &= \sin(\omega_L t)\\
\langle\sigma_z\rangle(t) &= 0
\end{align}

The spin precesses in the xy-plane with period $T = 2\pi/\omega_L$, tracing a circle of constant latitude on the Bloch sphere.
\end{example}

\section{Diagonalizing the Hamiltonian}

\subsection{Energy Eigenstates and Eigenvalues}

The exponential form of the time evolution operator $\hat{U}(t) = e^{-i\hat{H}t/\hbar}$ is most easily evaluated when the Hamiltonian is diagonal. For any Hermitian operator, we can find a basis of eigenstates where the matrix representation is diagonal.

For a general two-level Hamiltonian of the form $\hat{H} = \vec{h} \cdot \vec{\sigma}$, the eigenvalue equation is:
\begin{equation}
\hat{H}\ket{E_\pm} = E_\pm\ket{E_\pm}
\end{equation}

The eigenvalues are $E_\pm = \pm|\vec{h}|$, and the corresponding eigenstates are:
\begin{align}
\ket{E_+} &= \cos(\alpha/2)\ket{+z} + e^{i\beta}\sin(\alpha/2)\ket{-z}\\
\ket{E_-} &= -e^{-i\beta}\sin(\alpha/2)\ket{+z} + \cos(\alpha/2)\ket{-z}
\end{align}

where the angles $\alpha$ and $\beta$ are determined by the direction of $\vec{h}$:
\begin{align}
\cos\alpha &= \frac{h_z}{|\vec{h}|}\\
e^{i\beta} &= \frac{h_x + ih_y}{\sqrt{h_x^2 + h_y^2}}
\end{align}

\subsection{Similarity Transformations and Basis Changes}

The diagonalization process can be understood as a similarity transformation. If we define the unitary matrix $\hat{U}$ whose columns are the eigenstates, then:
\begin{equation}
\hat{U}^\dagger\hat{H}\hat{U} = \begin{pmatrix} E_+ & 0 \\ 0 & E_- \end{pmatrix}
\end{equation}

This similarity transformation rotates our coordinate system so that the Hamiltonian appears diagonal. In this "natural" basis, the time evolution becomes:
\begin{equation}
e^{-i\hat{H}t/\hbar} = \hat{U}\begin{pmatrix} e^{-iE_+t/\hbar} & 0 \\ 0 & e^{-iE_-t/\hbar} \end{pmatrix}\hat{U}^\dagger
\end{equation}

The eigenstates evolve with simple phase factors, while arbitrary superpositions develop more complex dynamics through the interference of these phases.

\begin{example}{Constructing the Diagonalization Matrix for a Bloch Cube State}
Consider a Hamiltonian representing a magnetic field along the x-direction:
\begin{equation}
\hat{H} = \frac{\hbar\omega}{2}\sigma_x = \frac{\hbar\omega}{2}\begin{pmatrix} 0 & 1 \\ 1 & 0 \end{pmatrix}
\end{equation}

The eigenvalues are $E_\pm = \pm\frac{\hbar\omega}{2}$. To find the eigenstates, we solve:
\begin{equation}
\sigma_x\ket{\psi} = \pm\ket{\psi}
\end{equation}

The eigenstates are the $\ket{\pm x}$ states from the Bloch cube:
\begin{align}
\ket{+x} &= \frac{1}{\sqrt{2}}(\ket{+z} + \ket{-z}) = \frac{1}{\sqrt{2}}\begin{pmatrix} 1 \\ 1 \end{pmatrix}\\
\ket{-x} &= \frac{1}{\sqrt{2}}(\ket{+z} - \ket{-z}) = \frac{1}{\sqrt{2}}\begin{pmatrix} 1 \\ -1 \end{pmatrix}
\end{align}

Now we construct the unitary matrix $\hat{U}$ with these eigenstates as columns:
\begin{equation}
\hat{U} = \frac{1}{\sqrt{2}}\begin{pmatrix} 1 & 1 \\ 1 & -1 \end{pmatrix}
\end{equation}

where the first column is $\ket{+x}$ and the second column is $\ket{-x}$.

The adjoint is:
\begin{equation}
\hat{U}^\dagger = \frac{1}{\sqrt{2}}\begin{pmatrix} 1 & 1 \\ 1 & -1 \end{pmatrix}
\end{equation}

Note that $\hat{U}^\dagger = \hat{U}$ in this case because the matrix is real and symmetric.

Let's verify the diagonalization explicitly:
\begin{align}
\hat{U}^\dagger\hat{H}\hat{U} &= \frac{1}{\sqrt{2}}\begin{pmatrix} 1 & 1 \\ 1 & -1 \end{pmatrix} \frac{\hbar\omega}{2}\begin{pmatrix} 0 & 1 \\ 1 & 0 \end{pmatrix} \frac{1}{\sqrt{2}}\begin{pmatrix} 1 & 1 \\ 1 & -1 \end{pmatrix}\\
&= \frac{\hbar\omega}{4}\begin{pmatrix} 1 & 1 \\ 1 & -1 \end{pmatrix} \begin{pmatrix} 1 & -1 \\ 1 & 1 \end{pmatrix}\\
&= \frac{\hbar\omega}{4}\begin{pmatrix} 2 & 0 \\ 0 & -2 \end{pmatrix}\\
&= \begin{pmatrix} \frac{\hbar\omega}{2} & 0 \\ 0 & -\frac{\hbar\omega}{2} \end{pmatrix}
\end{align}

This is indeed diagonal with eigenvalues $E_+ = \frac{\hbar\omega}{2}$ and $E_- = -\frac{\hbar\omega}{2}$.

\textbf{Time Evolution in Both Bases:}
In the original $\{\ket{+z}, \ket{-z}\}$ basis:
\begin{equation}
e^{-i\hat{H}t/\hbar} = e^{-i\omega t\sigma_x/2} = \cos(\omega t/2)\hat{I} - i\sin(\omega t/2)\sigma_x = \begin{pmatrix} \cos(\omega t/2) & -i\sin(\omega t/2) \\ -i\sin(\omega t/2) & \cos(\omega t/2) \end{pmatrix}
\end{equation}

Using the diagonalization:
\begin{align}
e^{-i\hat{H}t/\hbar} &= \hat{U}\begin{pmatrix} e^{-i\omega t/2} & 0 \\ 0 & e^{i\omega t/2} \end{pmatrix}\hat{U}^\dagger\\
&= \frac{1}{2}\begin{pmatrix} 1 & 1 \\ 1 & -1 \end{pmatrix}\begin{pmatrix} e^{-i\omega t/2} & 0 \\ 0 & e^{i\omega t/2} \end{pmatrix}\begin{pmatrix} 1 & 1 \\ 1 & -1 \end{pmatrix}\\
&= \frac{1}{2}\begin{pmatrix} e^{-i\omega t/2} + e^{i\omega t/2} & e^{-i\omega t/2} - e^{i\omega t/2} \\ e^{-i\omega t/2} - e^{i\omega t/2} & e^{-i\omega t/2} + e^{i\omega t/2} \end{pmatrix}\\
&= \begin{pmatrix} \cos(\omega t/2) & -i\sin(\omega t/2) \\ -i\sin(\omega t/2) & \cos(\omega t/2) \end{pmatrix}
\end{align}

Both methods give the same result, confirming our diagonalization. The key insight: in the $\{\ket{+x}, \ket{-x}\}$ basis (the eigenbasis), evolution is trivial---just phase factors. The complexity arises only when we express this evolution in a different basis like $\{\ket{+z}, \ket{-z}\}$.
\end{example}

\begin{example}{Diagonalizing a Tilted Field Hamiltonian}
Consider $\hat{H} = \frac{\hbar\omega}{2}(\sigma_x + \sigma_z)/\sqrt{2}$, representing a magnetic field at 45 degrees to the z-axis.

The field vector is $\vec{h} = \frac{\hbar\omega}{2}(1, 0, 1)/\sqrt{2}$, so $|\vec{h}| = \frac{\hbar\omega}{2}$ and the angles are:
\begin{align}
\cos\alpha &= \frac{1/\sqrt{2}}{1} = \frac{1}{\sqrt{2}} \Rightarrow \alpha = \pi/4\\
\beta &= 0 \text{ (since } h_y = 0 \text{ and } h_x > 0\text{)}
\end{align}

The eigenvalues are $E_+ = \frac{\hbar\omega}{2}$ and $E_- = -\frac{\hbar\omega}{2}$.

The eigenstates are:
\begin{align}
\ket{E_+} &= \cos(\pi/8)\ket{+z} + \sin(\pi/8)\ket{-z}\\
\ket{E_-} &= -\sin(\pi/8)\ket{+z} + \cos(\pi/8)\ket{-z}
\end{align}

To construct the diagonalization matrix $\hat{V}$ whose columns are the eigenstates:
\begin{equation}
\hat{V} = \begin{pmatrix} 
\cos(\pi/8) & -\sin(\pi/8) \\
\sin(\pi/8) & \cos(\pi/8)
\end{pmatrix}
\end{equation}

The time evolution operator in the energy eigenbasis is:
\begin{equation}
\hat{U}_{\text{diag}}(t) = \begin{pmatrix} 
e^{-i\omega t} & 0 \\
0 & e^{i\omega t}
\end{pmatrix}
\end{equation}

Following the pattern established in equation (4.51), the time evolution operator in the original basis is:
\begin{equation}
e^{-i\hat{H}t/\hbar} = \hat{V}\hat{U}_{\text{diag}}(t)\hat{V}^\dagger = \hat{V}\begin{pmatrix} 
e^{-i\omega t} & 0 \\
0 & e^{i\omega t}
\end{pmatrix}\hat{V}^\dagger
\end{equation}

Therefore, any initial state evolves as:
\begin{equation}
\ket{\psi(t)} = \hat{V}\begin{pmatrix} 
e^{-i\omega t} & 0 \\
0 & e^{i\omega t}
\end{pmatrix}\hat{V}^\dagger\ket{\psi(0)}
\end{equation}

This triple matrix product structure shows how the evolution decomposes into: (1) transformation to the eigenbasis via $\hat{V}^\dagger$, (2) simple phase evolution in the eigenbasis, and (3) transformation back to the original basis via $\hat{V}$. The rotation axis for this Hamiltonian is along $(1, 0, 1)/\sqrt{2}$, and states rotate about this axis with frequency $2\omega$.
\end{example}

\subsection{General Rotation Formula}

The diagonalization procedure leads to a general formula for rotations about an arbitrary axis. For any unit vector $\hat{n}$ and angle $\theta$:
\begin{equation}
\hat{R}_{\hat{n}}(\theta) = \cos(\theta/2)\hat{I} - i\sin(\theta/2)(\hat{n} \cdot \vec{\sigma})
\end{equation}

This formula encapsulates all possible single-qubit operations and shows how any rotation can be decomposed into rotations about the coordinate axes. It also reveals that rotations about parallel axes commute:
\begin{equation}
\hat{R}_{\hat{n}}(\theta_1)\hat{R}_{\hat{n}}(\theta_2) = \hat{R}_{\hat{n}}(\theta_1 + \theta_2)
\end{equation}

while rotations about different axes generally do not commute, reflecting the non-Abelian structure of the rotation group.

\section{Conservation Laws and Symmetries}

\subsection{The Connection Between Symmetry and Conservation}

One of the most profound insights in physics is Noether's theorem, which states that every continuous symmetry corresponds to a conserved quantity. In quantum mechanics, this manifests as the requirement that observables commute with the Hamiltonian to be conserved.

\textbf{Simple Proof for Quantum Systems:} Consider a unitary transformation $\hat{U}(\epsilon) = e^{-i\epsilon\hat{A}}$ generated by an observable $\hat{A}$, where $\epsilon$ is an infinitesimal parameter. If the Hamiltonian is invariant under this transformation for all $\epsilon$, then:
\begin{equation}
\hat{U}^\dagger(\epsilon)\hat{H}\hat{U}(\epsilon) = \hat{H}
\end{equation}

Expanding to first order in $\epsilon$:
\begin{align}
(\hat{I} + i\epsilon\hat{A})\hat{H}(\hat{I} - i\epsilon\hat{A}) &= \hat{H}\\
\hat{H} + i\epsilon(\hat{A}\hat{H} - \hat{H}\hat{A}) + O(\epsilon^2) &= \hat{H}\\
i\epsilon[\hat{A}, \hat{H}] &= 0
\end{align}

Since this must hold for all $\epsilon$, we have $[\hat{H}, \hat{A}] = 0$. An observable $\hat{A}$ is conserved if its expectation value remains constant under time evolution:
\begin{equation}
\frac{d}{dt}\langle\hat{A}\rangle = 0
\end{equation}

Using the Heisenberg equation of motion:
\begin{equation}
\frac{d}{dt}\langle\hat{A}\rangle = \frac{1}{i\hbar}\langle[\hat{A}, \hat{H}]\rangle
\end{equation}

This vanishes if and only if $[\hat{H}, \hat{A}] = 0$. Thus, symmetries of the Hamiltonian (transformations leaving $\hat{H}$ invariant) directly correspond to conserved quantities (observables that commute with $\hat{H}$).

The physical interpretation is that $\hat{A}$ generates a symmetry transformation that leaves the Hamiltonian invariant.

\subsection{Conservation Laws in Two-Level Systems}

For a general two-level Hamiltonian $\hat{H} = \vec{h} \cdot \vec{\sigma}$, the conserved quantity is the component of spin along the direction of $\vec{h}$:
\begin{equation}
\hat{A}_{\text{conserved}} = (\hat{h} \cdot \vec{\sigma}) = \frac{\vec{h} \cdot \vec{\sigma}}{|\vec{h}|}
\end{equation}

This can be verified by direct calculation:
\begin{align}
[\hat{H}, \hat{h} \cdot \vec{\sigma}] &= \vec{h} \cdot \vec{\sigma} \cdot [\vec{h} \cdot \vec{\sigma}, \hat{h} \cdot \vec{\sigma}]\\
&= |\vec{h}|^2[\hat{h} \cdot \vec{\sigma}, \hat{h} \cdot \vec{\sigma}] = 0
\end{align}

since any operator commutes with itself.

\begin{example}{Energy and Angular Momentum Conservation}
For a spin in a magnetic field $\vec{B} = B\hat{z}$, the Hamiltonian is $\hat{H} = \frac{\hbar\omega}{2}\sigma_z$.

The conserved quantities are energy ($\hat{H}$ itself, since $[\hat{H}, \hat{H}] = 0$) and the z-component of angular momentum ($\hat{S}_z = \hbar\sigma_z/2$, since $[\hat{H}, \sigma_z] = 0$). The non-conserved quantities are the x-component, where $[\hat{H}, \sigma_x] = \frac{\hbar\omega}{2}[\sigma_z, \sigma_x] = i\hbar\omega\sigma_y \neq 0$, and the y-component, where $[\hat{H}, \sigma_y] = \frac{\hbar\omega}{2}[\sigma_z, \sigma_y] = -i\hbar\omega\sigma_x \neq 0$.

The x and y components oscillate with frequency $2\omega$, while the z-component and total energy remain constant. This reflects the rotational symmetry about the z-axis.
\end{example}

\subsection{Noether's Theorem and Rotational Symmetry}

The rotational symmetry of space leads to conservation of angular momentum. For a two-level system, if the Hamiltonian is invariant under rotations about a particular axis $\hat{n}$, then the component of angular momentum along that axis is conserved.

Mathematically, if $\hat{H}$ commutes with the rotation operator $\hat{R}_{\hat{n}}(\theta)$ for all angles $\theta$, then it also commutes with the generator of that rotation:
\begin{equation}
[\hat{H}, \hat{R}_{\hat{n}}(\theta)] = 0 \Rightarrow [\hat{H}, \hat{n} \cdot \vec{\sigma}] = 0
\end{equation}

This principle extends to larger systems and other symmetries. Time translation symmetry leads to energy conservation, spatial translation symmetry leads to momentum conservation, and gauge symmetries lead to charge conservation. The two-level system serves as a laboratory for understanding these fundamental connections between symmetry and conservation in their simplest manifestation.

\section{Time-Dependence of Expectation Values and the Heisenberg Picture}

\subsection{General Formula for Time Evolution of Observables}

The time dependence of expectation values is governed by a general formula that encompasses both the intrinsic time dependence of operators and their evolution due to the system dynamics. For any observable $\hat{A}(t)$ that may have explicit time dependence:

\begin{equation}
\frac{d}{dt}\langle\hat{A}(t)\rangle = \frac{1}{i\hbar}\langle[\hat{A}(t), \hat{H}]\rangle + \left\langle\frac{\partial\hat{A}(t)}{\partial t}\right\rangle
\end{equation}

This master equation shows two sources of time dependence:
\begin{itemize}
\item The commutator term $\frac{1}{i\hbar}\langle[\hat{A}, \hat{H}]\rangle$ represents evolution due to the system dynamics
\item The partial derivative term $\left\langle\frac{\partial\hat{A}}{\partial t}\right\rangle$ accounts for explicit time dependence in the operator itself
\end{itemize}

When an observable commutes with the Hamiltonian and has no explicit time dependence, its expectation value is conserved. This provides the quantum mechanical foundation for conservation laws.

\subsection{The Heisenberg Picture}

In the Heisenberg picture, we transfer the time dependence from the states to the operators. While the Schr\"odinger picture keeps operators fixed and evolves states:
\begin{equation}
\ket{\psi_S(t)} = e^{-i\hat{H}t/\hbar}\ket{\psi_S(0)}
\end{equation}

The Heisenberg picture keeps states fixed and evolves operators:
\begin{equation}
\hat{A}_H(t) = e^{i\hat{H}t/\hbar}\hat{A}_S e^{-i\hat{H}t/\hbar}
\end{equation}

where subscripts S and H denote Schr\"odinger and Heisenberg pictures, respectively.

The expectation values remain the same in both pictures:
\begin{equation}
\langle\hat{A}\rangle(t) = \langle\psi_S(t)|\hat{A}_S|\psi_S(t)\rangle = \langle\psi_H|\hat{A}_H(t)|\psi_H\rangle
\end{equation}

The Heisenberg equation of motion for operators is:
\begin{equation}
\frac{d\hat{A}_H(t)}{dt} = \frac{1}{i\hbar}[\hat{A}_H(t), \hat{H}] + \frac{\partial\hat{A}_H(t)}{\partial t}
\end{equation}

This provides an alternative formulation of quantum dynamics where operators evolve while states remain fixed---often more convenient for calculating time-dependent expectation values.

\subsection{Computing Expectation Values During Evolution}

For a state evolving under a time-independent Hamiltonian, any observable $\hat{A}$ has time-dependent expectation value:
\begin{equation}
\langle\hat{A}\rangle(t) = \langle\psi(t)|\hat{A}|\psi(t)\rangle = \langle\psi(0)|e^{i\hat{H}t/\hbar}\hat{A}e^{-i\hat{H}t/\hbar}|\psi(0)\rangle
\end{equation}

This formula shows that the time dependence can be viewed in two equivalent ways: either the state evolves ($\ket{\psi(t)} = e^{-i\hat{H}t/\hbar}\ket{\psi(0)}$) while the operator $\hat{A}$ remains fixed, or equivalently, the operator evolves (via the sandwich $e^{i\hat{H}t/\hbar}\hat{A}e^{-i\hat{H}t/\hbar}$) while the state remains fixed. This second viewpoint is formalized as the Heisenberg picture, which we define in the next subsection.

\begin{example}{Spin Evolution in a Magnetic Field}
Consider a spin initially in the $\ket{+x}$ state evolving under $\hat{H} = \frac{\hbar\omega}{2}\sigma_z$. Let's compute the expectation values of all three spin components.

Starting state: $\ket{\psi(0)} = \ket{+x} = \frac{1}{\sqrt{2}}(\ket{+z} + \ket{-z})$

Time-evolved state: $\ket{\psi(t)} = e^{-i\omega t\sigma_z/2}\ket{+x} = \frac{1}{\sqrt{2}}(e^{-i\omega t/2}\ket{+z} + e^{i\omega t/2}\ket{-z})$

For the x-component:
\begin{align}
\langle\sigma_x\rangle(t) &= \langle\psi(t)|\sigma_x|\psi(t)\rangle\\
&= \frac{1}{2}(e^{i\omega t/2}\bra{+z} + e^{-i\omega t/2}\bra{-z})\sigma_x(e^{-i\omega t/2}\ket{+z} + e^{i\omega t/2}\ket{-z})\\
&= \frac{1}{2}(e^{i\omega t/2}\bra{+z} + e^{-i\omega t/2}\bra{-z})(e^{i\omega t/2}\ket{-z} + e^{-i\omega t/2}\ket{+z})\\
&= \frac{1}{2}(e^{i\omega t} + e^{-i\omega t}) = \cos(\omega t)
\end{align}

For the y-component:
\begin{align}
\langle\sigma_y\rangle(t) &= \frac{1}{2}(e^{i\omega t/2}\bra{+z} + e^{-i\omega t/2}\bra{-z})(-i\ket{-z} + i\ket{+z})(e^{-i\omega t/2}\ket{+z} + e^{i\omega t/2}\ket{-z})\\
&= \frac{i}{2}(e^{i\omega t} - e^{-i\omega t}) = \sin(\omega t)
\end{align}

For the z-component:
\begin{align}
\langle\sigma_z\rangle(t) &= \frac{1}{2}(e^{i\omega t/2}\bra{+z} + e^{-i\omega t/2}\bra{-z})(e^{-i\omega t/2}\ket{+z} - e^{i\omega t/2}\ket{-z})\\
&= \frac{1}{2}(1 - 1) = 0
\end{align}

The results show that $\langle\sigma_x\rangle$ and $\langle\sigma_y\rangle$ oscillate sinusoidally with frequency $\omega$, while $\langle\sigma_z\rangle$ remains constant at zero.
\end{example}

\subsection{Geometric Interpretation on the Bloch Sphere}

The time evolution of expectation values has a beautiful geometric interpretation. The Bloch vector $\vec{r}(t) = (\langle\sigma_x\rangle(t), \langle\sigma_y\rangle(t), \langle\sigma_z\rangle(t))$ traces a path on the Bloch sphere as the system evolves.

For a Hamiltonian $\hat{H} = \vec{h} \cdot \vec{\sigma}$, the Bloch vector rotates about the axis $\hat{h} = \vec{h}/|\vec{h}|$ with angular frequency $\omega = 2|\vec{h}|/\hbar$. States that are eigenstates of the Hamiltonian correspond to fixed points on the rotation axis, while all other states trace circles of constant latitude.

The key insight is that quantum evolution preserves angles on the Bloch sphere. If two states initially make an angle $\theta$, they will continue to subtend the same angle throughout their evolution under any unitary transformation. This geometric constraint underlies many conservation laws and provides intuition for understanding complex quantum dynamics.

\begin{example}{Understanding Quantum Interference Through Geometry}
Consider two states $\ket{\psi_1}$ and $\ket{\psi_2}$ with Bloch vectors making an angle $\theta$ with respect to each other. Their overlap is:
\begin{equation}
|\langle\psi_1|\psi_2\rangle|^2 = \cos^2(\theta/2)
\end{equation}

This relationship between the geometric angle and quantum overlap explains why orthogonal states (separated by 180\ensuremath{^\circ} on the Bloch sphere) have zero overlap, antipodal states differ by a relative phase but have the same measurement probabilities, and states separated by 90\ensuremath{^\circ} have overlap $1/2$, leading to maximum interference visibility.

During time evolution, the angle between states is preserved, so their interference pattern depends only on their initial relative orientation and any additional phases accumulated during evolution.
\end{example}

\section{Time-Dependent Hamiltonians and Nuclear Magnetic Resonance}

\subsection{The Problem with Time-Dependent Hamiltonians}

Up to this point, we have considered only time-independent Hamiltonians, where energy eigenstates evolve with simple exponential phases and the time evolution operator takes the form $\hat{U}(t) = e^{-i\hat{H}t/\hbar}$. However, many physical systems involve time-dependent external fields that drive transitions between energy levels.

When the Hamiltonian depends explicitly on time, $\hat{H}(t)$, naively writing
\begin{equation}
\hat{U}(t) = \exp\left(-\frac{i}{\hbar}\int_{0}^{t} \hat{H}(t') dt'\right) \quad \text{(INCORRECT in general)}
\end{equation}
is \textbf{wrong} unless $[\hat{H}(t_1), \hat{H}(t_2)] = 0$ for all times $t_1, t_2$.

\textbf{Why?} The exponential of an integral is not generally equal to integrating exponentials when the integrand doesn't commute with itself at different times. To see this, consider that for time-independent $\hat{H}$:
\begin{equation}
e^{-i(\hat{H}_1 + \hat{H}_2)t/\hbar} \neq e^{-i\hat{H}_1 t/\hbar}e^{-i\hat{H}_2 t/\hbar}
\end{equation}
unless $[\hat{H}_1, \hat{H}_2] = 0$. The same principle applies when comparing $\hat{H}(t_1)$ and $\hat{H}(t_2)$ at different times.

\subsection{Derivation of the Time-Ordered Solution}

The correct approach starts from the time-dependent Schr\"odinger equation:
\begin{equation}
i\hbar\frac{d}{dt}\ket{\psi(t)} = \hat{H}(t)\ket{\psi(t)}
\end{equation}

Integrating both sides from $t_0$ to $t$:
\begin{equation}
\ket{\psi(t)} = \ket{\psi(t_0)} - \frac{i}{\hbar}\int_{t_0}^{t} \hat{H}(t_1)\ket{\psi(t_1)} dt_1
\end{equation}

This is an integral equation since $\ket{\psi(t)}$ appears on both sides. We solve it iteratively by substituting the expression for $\ket{\psi(t_1)}$ back into itself:
\begin{align}
\ket{\psi(t)} &= \ket{\psi(t_0)} - \frac{i}{\hbar}\int_{t_0}^{t} \hat{H}(t_1)\ket{\psi(t_0)} dt_1\\
&\quad + \left(-\frac{i}{\hbar}\right)^2\int_{t_0}^{t} dt_1 \int_{t_0}^{t_1} dt_2\, \hat{H}(t_1)\hat{H}(t_2)\ket{\psi(t_0)} + \cdots
\end{align}

This is the \textbf{Dyson series}. The complete time evolution operator is:
\begin{equation}
\hat{U}(t,t_0) = \mathbbm{1} + \sum_{n=1}^{\infty}\left(-\frac{i}{\hbar}\right)^n \int_{t_0}^{t}dt_1\int_{t_0}^{t_1}dt_2\cdots\int_{t_0}^{t_{n-1}}dt_n\, \hat{H}(t_1)\hat{H}(t_2)\cdots\hat{H}(t_n)
\end{equation}

Note the crucial ordering: $t \geq t_1 \geq t_2 \geq \cdots \geq t_n \geq t_0$. The operators are naturally \textbf{time-ordered} with later times to the left.

\subsection{The Time-Ordering Operator}

The time-ordering operator $\hat{\mathcal{T}}$ (or simply $\mathcal{T}$) arranges operators chronologically with \textbf{later times to the left}:
\begin{equation}
\hat{\mathcal{T}}\{\hat{A}(t_1)\hat{B}(t_2)\} = \begin{cases} 
\hat{A}(t_1)\hat{B}(t_2) & \text{if } t_1 > t_2 \\
\hat{B}(t_2)\hat{A}(t_1) & \text{if } t_2 > t_1
\end{cases}
\end{equation}

More generally, for multiple operators:
\begin{equation}
\hat{\mathcal{T}}\{\hat{A}(t_1)\hat{B}(t_2)\hat{C}(t_3)\cdots\} = \text{product with operators ordered by decreasing time}
\end{equation}

Using the time-ordering operator, the Dyson series becomes:
\begin{equation}
\hat{U}(t,t_0) = \hat{\mathcal{T}}\exp\left(-\frac{i}{\hbar}\int_{t_0}^{t} \hat{H}(t') dt'\right)
\end{equation}

\begin{equation}
= \hat{\mathcal{T}}\left\{\sum_{n=0}^{\infty}\frac{1}{n!}\left(-\frac{i}{\hbar}\right)^n \left[\int_{t_0}^{t} \hat{H}(t') dt'\right]^n\right\}
\end{equation}

The time-ordering operator $\hat{\mathcal{T}}$ converts the unrestricted integral over all time orderings into the properly ordered Dyson series.

\subsection{Physical Interpretation: Causality}

Time ordering reflects \textbf{causality}: an operator at time $t_1$ can only affect the system before operators at later times $t_2 > t_1$ act. The time-ordered product ensures:

\begin{enumerate}
\item \textbf{Sequential application}: The Hamiltonian at earlier times acts first
\item \textbf{Non-commutativity}: When $[\hat{H}(t_1), \hat{H}(t_2)] \neq 0$, the order matters
\item \textbf{Perturbation theory}: Each term in the Dyson series represents $n$th-order processes
\end{enumerate}

\begin{example*}{Time Ordering in a Driven Two-Level System}
Consider a spin-1/2 system in a time-varying magnetic field $\vec{B}(t)$:
\begin{equation}
\hat{H}(t) = -\gamma \vec{B}(t) \cdot \hat{\vec{S}} = -\gamma[B_x(t)\hat{S}_x + B_y(t)\hat{S}_y + B_z(t)\hat{S}_z]
\end{equation}

If $\vec{B}(t)$ changes direction, then $[\hat{H}(t_1), \hat{H}(t_2)] \neq 0$ since different spin components don't commute. For example:
\begin{equation}
[\hat{S}_x, \hat{S}_y] = i\hbar\hat{S}_z \neq 0
\end{equation}

The evolution cannot be written as a simple exponential---time ordering is essential. To first order in perturbation theory:
\begin{equation}
\ket{\psi(t)} \approx \left[\mathbbm{1} - \frac{i}{\hbar}\int_0^t \hat{H}(t_1)dt_1\right]\ket{\psi(0)}
\end{equation}

To second order, we must respect time ordering:
\begin{align}
\ket{\psi(t)} &\approx \left[\mathbbm{1} - \frac{i}{\hbar}\int_0^t \hat{H}(t_1)dt_1\right.\\  
&\left.\quad + \left(-\frac{i}{\hbar}\right)^2\int_0^t dt_1 \int_0^{t_1} dt_2\, \hat{H}(t_1)\hat{H}(t_2)\right]\ket{\psi(0)}
\end{align}

The nested integral structure with $t_1 > t_2$ ensures that the Hamiltonian at time $t_2$ acts before the Hamiltonian at time $t_1$, preserving causality.
\end{example*}

\subsection{Practical Computation and Connection to Perturbation Theory}

For weak time-dependence or perturbative calculations, we truncate the Dyson series:

\begin{itemize}
\item \textbf{Zeroth order}: $\hat{U} \approx \mathbbm{1}$ (no evolution)
\item \textbf{First order}: $\hat{U} \approx \mathbbm{1} - \frac{i}{\hbar}\int_{t_0}^{t} \hat{H}(t') dt'$
\item \textbf{Second order}: Include the double integral term
\end{itemize}

This perturbative approach forms the foundation of time-dependent perturbation theory and Fermi's Golden Rule.

\textbf{Reduction to Time-Independent Case:} When $\hat{H}(t) = \hat{H}$ is time-independent, all operators at different times commute: $[\hat{H}(t_1), \hat{H}(t_2)] = [\hat{H}, \hat{H}] = 0$. Time ordering becomes trivial, and:
\begin{align}
\hat{U}(t,t_0) &= \mathcal{T}\exp\left(-\frac{i}{\hbar}\int_{t_0}^{t} \hat{H}dt'\right)\\
&= \exp\left(-\frac{i}{\hbar}\hat{H}(t-t_0)\right)
\end{align}
recovering our familiar exponential form.

\subsection{Connection to the Interaction Picture}

In the interaction picture with $\hat{H} = \hat{H}_0 + \hat{V}(t)$, time ordering appears naturally in the evolution operator:
\begin{equation}
\hat{U}_I(t,t_0) = \hat{\mathcal{T}}\exp\left(-\frac{i}{\hbar}\int_{t_0}^{t} \hat{V}_I(t') dt'\right)
\end{equation}

where $\hat{V}_I(t) = e^{i\hat{H}_0t/\hbar}\hat{V}(t)e^{-i\hat{H}_0t/\hbar}$ is the interaction in the interaction picture. This formulation is essential for quantum field theory and many-body physics.

\textbf{Key Takeaway}: Time ordering is not merely a mathematical technicality but a fundamental requirement when dealing with time-dependent quantum systems where the Hamiltonian at different times doesn't commute. It ensures causality and provides the correct perturbative expansion for quantum evolution.

\subsection{The Driven Two-Level System}

A fundamental example is a two-level system with a static energy splitting driven by an oscillating field:
\begin{equation}
\hat{H}(t) = \frac{\hbar\omega_0}{2}\sigma_z + \frac{\hbar\Omega}{2}[\sigma_x\cos(\omega t) + \sigma_y\sin(\omega t)]
\end{equation}

This describes numerous physical systems including an atom in a laser field, where $\omega_0$ is the atomic transition frequency and $\Omega$ is proportional to the electric field amplitude; a nuclear spin in an oscillating magnetic field (NMR), where $\omega_0$ is the Larmor frequency and $\Omega$ is the Rabi frequency; and a superconducting qubit driven by microwave pulses for quantum computation.

\subsection{The Rotating Frame, Interaction Picture, and Resonance}

The key insight for solving driven systems is to transform to the interaction picture (rotating frame). This is analogous to moving to a reference frame rotating at the drive frequency. 

\textbf{The Interaction Picture:} Recall from Section 4.7 that we have two equivalent formulations of quantum dynamics:
\begin{itemize}
\item \textbf{Schr\"odinger picture}: States evolve, operators are fixed: $\ket{\psi_S(t)} = e^{-i\hat{H}t/\hbar}\ket{\psi_S(0)}$
\item \textbf{Heisenberg picture}: Operators evolve, states are fixed: $\hat{A}_H(t) = e^{i\hat{H}t/\hbar}\hat{A}_S e^{-i\hat{H}t/\hbar}$
\end{itemize}

For time-dependent Hamiltonians, we introduce a third picture that combines features of both. When the Hamiltonian can be separated into a time-independent part $\hat{H}_0$ and a time-dependent perturbation $\hat{V}(t)$:
\begin{equation}
\hat{H}(t) = \hat{H}_0 + \hat{V}(t)
\end{equation}

we define the \textbf{interaction picture} (also called the \textbf{Dirac picture}) as follows:
\begin{align}
\textbf{Interaction picture states:} \quad \ket{\psi_I(t)} &= e^{i\hat{H}_0 t/\hbar}\ket{\psi_S(t)}\\
\textbf{Interaction picture operators:} \quad \hat{A}_I(t) &= e^{i\hat{H}_0 t/\hbar}\hat{A}_S e^{-i\hat{H}_0 t/\hbar}
\end{align}

In this picture, both states and operators evolve with time, but the time evolution of operators is governed only by $\hat{H}_0$, while the time evolution of states is governed by the transformed perturbation $\hat{V}_I(t)$. The interaction picture Schr\"odinger equation becomes:
\begin{equation}
i\hbar\frac{d}{dt}\ket{\psi_I(t)} = \hat{V}_I(t)\ket{\psi_I(t)}
\end{equation}
where $\hat{V}_I(t) = e^{i\hat{H}_0 t/\hbar}\hat{V}(t)e^{-i\hat{H}_0 t/\hbar}$ is the interaction picture perturbation.

\textbf{Connection to the Dyson Series:} The interaction picture provides the natural setting for the Dyson series developed in Section 4.8.3. The full Schr\"odinger picture evolution operator can be written as:
\begin{equation}
\hat{U}_S(t,t_0) = e^{-i\hat{H}_0(t-t_0)/\hbar}\hat{U}_I(t,t_0)
\end{equation}
where the interaction picture evolution operator is given by the Dyson series applied to $\hat{V}_I(t)$:
\begin{equation}
\hat{U}_I(t,t_0) = \mathcal{T}\exp\left(-\frac{i}{\hbar}\int_{t_0}^{t} \hat{V}_I(t') dt'\right)
\end{equation}

This separation is powerful because:
\begin{itemize}
\item The evolution due to $\hat{H}_0$ is trivial (simple exponential)
\item Only the perturbation $\hat{V}_I(t)$ requires time-ordering via the Dyson series
\item When $\hat{V}_I$ is small, the Dyson series converges rapidly, enabling perturbative calculations
\end{itemize}

For the driven two-level system below, when we work in the rotating frame and apply the rotating wave approximation, the interaction picture Hamiltonian becomes approximately time-independent, allowing us to avoid the Dyson series altogether and use a simple exponential.

\textbf{Application to the Driven Two-Level System:} For our system, we separate:
\begin{align}
\hat{H}_0 &= \frac{\hbar\omega_0}{2}\sigma_z \quad \text{(static field)}\\
\hat{V}(t) &= \frac{\hbar\Omega}{2}[\sigma_x\cos(\omega t) + \sigma_y\sin(\omega t)] \quad \text{(rotating field)}
\end{align}

The transformation to the interaction picture is:
\begin{equation}
\ket{\psi_I(t)} = e^{i\hat{H}_0 t/\hbar}\ket{\psi_S(t)} = e^{i\omega_0 t\sigma_z/2}\ket{\psi_S(t)}
\end{equation}

Applying the Schr\"odinger equation in the interaction picture:
\begin{align}
\hat{H}_I(t) &= e^{i\hat{H}_0 t/\hbar}\hat{V}(t)e^{-i\hat{H}_0 t/\hbar}\\
&= e^{i\omega_0 t\sigma_z/2}\frac{\hbar\Omega}{2}[\sigma_x\cos(\omega t) + \sigma_y\sin(\omega t)]e^{-i\omega_0 t\sigma_z/2}
\end{align}

Using the rotation properties $e^{i\theta\sigma_z/2}\sigma_x e^{-i\theta\sigma_z/2} = \sigma_x\cos\theta + \sigma_y\sin\theta$ and $e^{i\theta\sigma_z/2}\sigma_y e^{-i\theta\sigma_z/2} = -\sigma_x\sin\theta + \sigma_y\cos\theta$:

\begin{align}
\hat{H}_I(t) &= \frac{\hbar\Omega}{2}[(\sigma_x\cos\omega_0 t + \sigma_y\sin\omega_0 t)\cos\omega t\\
&\quad + (-\sigma_x\sin\omega_0 t + \sigma_y\cos\omega_0 t)\sin\omega t]\\
&= \frac{\hbar\Omega}{2}[\sigma_x(\cos\omega_0 t\cos\omega t - \sin\omega_0 t\sin\omega t)\\
&\quad + \sigma_y(\sin\omega_0 t\cos\omega t + \cos\omega_0 t\sin\omega t)]\\
&= \frac{\hbar\Omega}{2}[\sigma_x\cos((\omega_0 - \omega)t) + \sigma_y\sin((\omega_0 - \omega)t)]
\end{align}

This interaction picture Hamiltonian still has explicit time dependence, oscillating at the difference frequency $(\omega_0 - \omega)$. When the driving frequency $\omega$ is close to the natural frequency $\omega_0$, this oscillation is slow compared to the original oscillation at $\omega_0$.

\textbf{Transforming to the Doubly Rotating Frame:} To further simplify, we make a second transformation to a frame rotating at the drive frequency $\omega$. Define:
\begin{equation}
\ket{\psi_{rot}(t)} = e^{i\omega t\sigma_z/2}\ket{\psi_I(t)}
\end{equation}

The Hamiltonian in this doubly rotating frame becomes:
\begin{equation}
\hat{H}_{rot}(t) = e^{i\omega t\sigma_z/2}\hat{H}_I(t)e^{-i\omega t\sigma_z/2} - \frac{\hbar\omega}{2}\sigma_z
\end{equation}

\textbf{Derivation of the $-\frac{\hbar\omega}{2}\sigma_z$ term:} This term arises from the time derivative of the rotating frame transformation. When we define $\ket{\psi_{rot}(t)} = e^{i\omega t\sigma_z/2}\ket{\psi_I(t)}$, we need to find the equation of motion for $\ket{\psi_{rot}(t)}$. Taking the time derivative:
\begin{align}
\frac{d}{dt}\ket{\psi_{rot}(t)} &= \frac{d}{dt}\left[e^{i\omega t\sigma_z/2}\ket{\psi_I(t)}\right]\notag\\
&= \frac{d}{dt}\left[e^{i\omega t\sigma_z/2}\right]\ket{\psi_I(t)} + e^{i\omega t\sigma_z/2}\frac{d}{dt}\ket{\psi_I(t)}\notag\\
&= i\frac{\omega}{2}\sigma_z e^{i\omega t\sigma_z/2}\ket{\psi_I(t)} + e^{i\omega t\sigma_z/2}\frac{d}{dt}\ket{\psi_I(t)}
\end{align}

The interaction picture Schr\"odinger equation is:
\begin{equation}
i\hbar\frac{d}{dt}\ket{\psi_I(t)} = \hat{H}_I(t)\ket{\psi_I(t)}
\end{equation}

Substituting this into our expression:
\begin{align}
\frac{d}{dt}\ket{\psi_{rot}(t)} &= i\frac{\omega}{2}\sigma_z e^{i\omega t\sigma_z/2}\ket{\psi_I(t)} + e^{i\omega t\sigma_z/2}\left(-\frac{i}{\hbar}\hat{H}_I(t)\ket{\psi_I(t)}\right)\notag\\
&= i\frac{\omega}{2}\sigma_z \ket{\psi_{rot}(t)} - \frac{i}{\hbar}e^{i\omega t\sigma_z/2}\hat{H}_I(t)e^{-i\omega t\sigma_z/2}\ket{\psi_{rot}(t)}
\end{align}

Multiplying both sides by $i\hbar$:
\begin{equation}
i\hbar\frac{d}{dt}\ket{\psi_{rot}(t)} = -\frac{\hbar\omega}{2}\sigma_z\ket{\psi_{rot}(t)} + e^{i\omega t\sigma_z/2}\hat{H}_I(t)e^{-i\omega t\sigma_z/2}\ket{\psi_{rot}(t)}
\end{equation}

This shows that the effective Hamiltonian in the rotating frame is:
\begin{equation}
\hat{H}_{rot}(t) = e^{i\omega t\sigma_z/2}\hat{H}_I(t)e^{-i\omega t\sigma_z/2} - \frac{\hbar\omega}{2}\sigma_z
\end{equation}

The $-\frac{\hbar\omega}{2}\sigma_z$ term represents the "fictitious" energy shift due to the rotating frame, analogous to fictitious forces in classical rotating reference frames. This is why the effective detuning becomes $\Delta\omega = \omega_0 - \omega$ in the rotating frame.

To evaluate the transformation, we express the driving field in terms of ladder operators. Starting from:
\begin{align}
\hat{V}(t) &= \frac{\hbar\Omega}{2}[\sigma_x\cos(\omega t) + \sigma_y\sin(\omega t)]\\
&= \frac{\hbar\Omega}{4}[(\sigma_x - i\sigma_y)e^{i\omega t} + (\sigma_x + i\sigma_y)e^{-i\omega t}]\\
&= \frac{\hbar\Omega}{4}[\sigma_+ e^{-i\omega t} + \sigma_- e^{i\omega t}]
\end{align}

In the interaction picture with $\hat{H}_0 = \frac{\hbar\omega_0}{2}\sigma_z$:
\begin{align}
\hat{V}_I(t) &= e^{i\omega_0 t\sigma_z/2}\hat{V}(t)e^{-i\omega_0 t\sigma_z/2}\\
&= \frac{\hbar\Omega}{4}[\sigma_+ e^{i(\omega_0-\omega)t} + \sigma_- e^{-i(\omega_0-\omega)t}]
\end{align}

In the doubly rotating frame, the complete Hamiltonian is:
\begin{equation}
\hat{H}_{rot}(t) = e^{i\omega t\sigma_z/2}\left(\frac{\hbar\omega_0}{2}\sigma_z + \hat{V}_I(t)\right)e^{-i\omega t\sigma_z/2} - \frac{\hbar\omega}{2}\sigma_z
\end{equation}

Since $e^{i\omega t\sigma_z/2}\sigma_z e^{-i\omega t\sigma_z/2} = \sigma_z$:
\begin{equation}
\hat{H}_{rot}(t) = \frac{\hbar(\omega_0-\omega)}{2}\sigma_z + e^{i\omega t\sigma_z/2}\hat{V}_I(t)e^{-i\omega t\sigma_z/2}
\end{equation}

For the second term, we use the transformation properties:
\begin{align}
e^{i\omega t\sigma_z/2}\sigma_+ e^{-i\omega t\sigma_z/2} &= \sigma_+ e^{i\omega t}\\
e^{i\omega t\sigma_z/2}\sigma_- e^{-i\omega t\sigma_z/2} &= \sigma_- e^{-i\omega t}
\end{align}

Therefore:
\begin{equation}
\hat{H}_{rot}(t) = \frac{\hbar(\omega_0-\omega)}{2}\sigma_z + \frac{\hbar\Omega}{4}[\sigma_+ e^{i(\omega_0-\omega)t}e^{i\omega t} + \sigma_- e^{-i(\omega_0-\omega)t}e^{-i\omega t}]
\end{equation}

Defining the detuning $\Delta\omega = \omega_0 - \omega$ and expanding the exponentials:
\begin{equation}
\hat{H}_{rot}(t) = \frac{\hbar\Delta\omega}{2}\sigma_z + \frac{\hbar\Omega}{4}[\sigma_+ e^{i\Delta\omega t}e^{i\omega t} + \sigma_- e^{-i\Delta\omega t}e^{-i\omega t}]
\end{equation}

Near resonance ($\Delta\omega \ll \omega$), the exponentials $e^{\pm i\Delta\omega t}$ vary slowly compared to $e^{\pm i\omega t}$. We can separate the expression into slowly and rapidly oscillating parts by writing $e^{i\Delta\omega t} \approx 1$ for the slowly varying terms:
\begin{equation}
\hat{H}_{rot}(t) = \frac{\hbar\Delta\omega}{2}\sigma_z + \frac{\hbar\Omega}{4}[\sigma_+ + \sigma_-] + \frac{\hbar\Omega}{4}[\sigma_+ e^{2i\omega t} + \sigma_- e^{-2i\omega t}]
\end{equation}

\textbf{The Rotating Wave Approximation:} The complete Hamiltonian in the doubly rotating frame is:
\begin{equation}
\hat{H}_{rot}(t) = \frac{\hbar\Delta\omega}{2}\sigma_z + \frac{\hbar\Omega}{4}[\sigma_+ + \sigma_-] + \frac{\hbar\Omega}{4}[\sigma_+ e^{2i\omega t} + \sigma_- e^{-2i\omega t}]
\end{equation}

The last term oscillates rapidly at frequency $2\omega$. The rotating wave approximation (RWA) consists of neglecting these rapidly oscillating "counter-rotating" terms. The physical justification is that these terms oscillate much faster than any other timescale in the problem, so they average to zero and don't contribute to the long-time dynamics.

Dropping the rapidly oscillating terms and using $\sigma_+ + \sigma_- = 2\sigma_x$:
\begin{equation}
\hat{H}_{eff} = \frac{\hbar\Delta\omega}{2}\sigma_z + \frac{\hbar\Omega}{2}\sigma_x
\end{equation}

This is the effective Hamiltonian in the rotating wave approximation. It is now time-independent, allowing us to use simple exponential time evolution instead of the Dyson series.

\subsection{Rabi Oscillations and Bloch Vector Dynamics}

In the rotating frame with the rotating wave approximation, the effective Hamiltonian generates rotations about the axis $(\Omega, 0, \Delta\omega)$ with frequency $\Omega_{eff} = \sqrt{\Omega^2 + (\Delta\omega)^2}$. 

For exact resonance ($\Delta\omega = 0$), the effective Hamiltonian becomes $\hat{H}_{eff} = \hbar\Omega\sigma_x/2$, generating rotations about the x-axis. Starting from the ground state $\ket{+z}$:
\begin{align}
\ket{\psi(t)} &= e^{-i\Omega t\sigma_x/2}\ket{+z}\\
&= \cos(\Omega t/2)\ket{+z} - i\sin(\Omega t/2)\ket{-z}
\end{align}

The population oscillates between the two levels:
\begin{equation}
P_{excited}(t) = |\langle-z|\psi(t)\rangle|^2 = \sin^2(\Omega t/2)
\end{equation}

These are called Rabi oscillations, and they provide a precise method for controlling quantum states through the duration and amplitude of applied pulses.

\section{Adiabatic Evolution and Geometric Phases}

\subsection{Slow Parameter Variation and the Adiabatic Theorem}

When the parameters in a quantum system change slowly compared to the energy gaps, the system can remain in an instantaneous eigenstate while accumulating both dynamical and geometric phases. This adiabatic evolution provides a powerful tool for robust quantum state manipulation and reveals deep connections between quantum mechanics and geometry.

Consider a time-dependent Hamiltonian $\hat{H}(t)$ that changes slowly over time. At each instant $t$, we can define instantaneous eigenstates $\ket{n(t)}$ and eigenvalues $E_n(t)$:
\begin{equation}
\hat{H}(t)\ket{n(t)} = E_n(t)\ket{n(t)}
\end{equation}

The adiabatic theorem states that if the system starts in eigenstate $\ket{n(0)}$ and the Hamiltonian changes sufficiently slowly, the system will remain in the instantaneous eigenstate $\ket{n(t)}$ but acquire a phase:
\begin{equation}
\ket{\psi_n(t)} = e^{i\phi_n(t)}\ket{n(t)}
\end{equation}

where $\ket{\psi_n(t)}$ is the actual state of the system (which follows the $n$-th eigenstate adiabatically) and $\ket{n(t)}$ is the instantaneous eigenstate of $\hat{H}(t)$ at time $t$.

The total phase has two contributions:
\begin{align}
\phi_n(t) &= \phi_n^{(dyn)}(t) + \phi_n^{(geom)}(t)\\
\phi_n^{(dyn)}(t) &= -\frac{1}{\hbar}\int_0^t E_n(t')dt' \quad \text{(dynamical)}\\
\phi_n^{(geom)}(t) &= i\int_0^t \langle n(t')|\frac{d}{dt'}|n(t')\rangle dt' \quad \text{(geometric)}
\end{align}

\textbf{Origin of the Geometric Phase:} The geometric phase arises from the requirement that the state remain normalized during evolution. When we write $\ket{\psi_n(t)} = e^{i\phi_n(t)}\ket{n(t)}$, taking the time derivative and using the Schr\"odinger equation gives:
\begin{equation}
i\hbar\frac{d}{dt}\ket{\psi_n(t)} = \hat{H}(t)\ket{\psi_n(t)}
\end{equation}

Expanding the left side using the product rule:
\begin{equation}
i\hbar\left[i\frac{d\phi_n}{dt}e^{i\phi_n}\ket{n(t)} + e^{i\phi_n}\frac{d\ket{n(t)}}{dt}\right] = \hat{H}(t)e^{i\phi_n}\ket{n(t)}
\end{equation}

Since $\hat{H}(t)\ket{n(t)} = E_n(t)\ket{n(t)}$, this becomes:
\begin{equation}
i\hbar\frac{d\phi_n}{dt}e^{i\phi_n}\ket{n(t)} + i\hbar e^{i\phi_n}\frac{d\ket{n(t)}}{dt} = E_n(t)e^{i\phi_n}\ket{n(t)}
\end{equation}

Dividing through by $e^{i\phi_n}$ (which is just a phase factor):
\begin{equation}
i\hbar\frac{d\phi_n}{dt}\ket{n(t)} + i\hbar\frac{d\ket{n(t)}}{dt} = E_n(t)\ket{n(t)}
\end{equation}

Taking the inner product with $\bra{n(t)}$:
\begin{equation}
i\hbar\frac{d\phi_n}{dt} + i\hbar\langle n(t)|\frac{d}{dt}|n(t)\rangle = E_n(t)
\end{equation}

Rearranging:
\begin{equation}
\frac{d\phi_n}{dt} = -\frac{E_n(t)}{\hbar} - i\langle n(t)|\frac{d}{dt}|n(t)\rangle
\end{equation}

\textbf{Why the second term is purely real:} The inner product $\langle n(t)|\frac{d}{dt}|n(t)\rangle$ is purely imaginary, not the phase itself. Since $\langle n(t)|n(t)\rangle = 1$ for all $t$, taking the time derivative:
\begin{equation}
\frac{d}{dt}\langle n(t)|n(t)\rangle = \left\langle\frac{d}{dt}n(t)\middle|n(t)\right\rangle + \left\langle n(t)\middle|\frac{d}{dt}n(t)\right\rangle = 0
\end{equation}

This means:
\begin{equation}
\left\langle n(t)\middle|\frac{d}{dt}n(t)\right\rangle = -\left\langle\frac{d}{dt}n(t)\middle|n(t)\right\rangle = -\left\langle n(t)\middle|\frac{d}{dt}n(t)\right\rangle^*
\end{equation}

So $\langle n(t)|\frac{d}{dt}|n(t)\rangle$ is purely imaginary. When multiplied by $-i$ in the equation for $d\phi_n/dt$, the second term becomes purely real. Therefore, both terms on the right side contribute real values to the phase.

Integrating yields the two contributions: the first term gives the dynamical phase (energy-dependent), while the second term gives the geometric phase. This geometric phase depends only on how the instantaneous eigenstate $\ket{n(t)}$ evolves along its path in Hilbert space, not on the rate of evolution or the energy. It is this path-dependence that gives the phase its geometric character.

The dynamical phase depends on the energy, while the geometric phase depends only on the path traced by the state vector in Hilbert space.

\subsection{Berry Phase Equals Half the Solid Angle for Two-Level Systems}

For a two-level system undergoing adiabatic evolution around a closed path, the geometric Berry phase is exactly half the solid angle enclosed by the path on the Bloch sphere. We now derive this remarkable result.

\textbf{Setup:} Consider a two-level Hamiltonian where a magnetic field of constant magnitude traces a path on the unit sphere:
\begin{equation}
\hat{H}(t) = \frac{\hbar\omega_0}{2} \vec{B}(t) \cdot \vec{\sigma}
\end{equation}
where $\vec{B}(t)$ is a unit vector parameterized by spherical coordinates $(\theta(t), \phi(t))$:
\begin{equation}
\vec{B}(t) = (\sin\theta(t)\cos\phi(t), \sin\theta(t)\sin\phi(t), \cos\theta(t))
\end{equation}

The Hamiltonian depends on time through the magnetic field direction. The energy splitting between the two levels is $\hbar\omega_0$, with eigenvalues $\pm\hbar\omega_0/2$. As time evolves and the field traces a path on the sphere, the parameters $(\theta(t), \phi(t))$ change, causing the instantaneous eigenstate to evolve.

\textbf{Instantaneous Eigenstates:} At each instant, the ground state eigenstate (lower energy) is:
\begin{equation}
\ket{\psi_-(\theta, \phi)} = \cos(\theta/2)\ket{+z} - e^{i\phi}\sin(\theta/2)\ket{-z}
\end{equation}

This eigenstate points opposite to $\vec{B}(t)$ on the Bloch sphere (spin anti-aligned with the field).

\textbf{The Berry Connection:} From equation (4.174), the geometric phase is:
\begin{equation}
\phi^{(geom)} = i\int_0^t \langle \psi_- | \frac{d}{dt'} | \psi_- \rangle dt'
\end{equation}

Using the chain rule:
\begin{equation}
\frac{d}{dt}\ket{\psi_-} = \frac{\partial \ket{\psi_-}}{\partial \theta}\frac{d\theta}{dt} + \frac{\partial \ket{\psi_-}}{\partial \phi}\frac{d\phi}{dt}
\end{equation}

Substituting and changing variables from time to path parameters:
\begin{equation}
\phi^{(geom)} = i\int_{\text{path}} \left[\langle \psi_- | \frac{\partial}{\partial \theta} | \psi_- \rangle d\theta + \langle \psi_- | \frac{\partial}{\partial \phi} | \psi_- \rangle d\phi\right]
\end{equation}

We define the \textbf{Berry connection}:
\begin{equation}
\vec{A}(\theta, \phi) = i\langle \psi_- | \nabla | \psi_- \rangle = \left(A_\theta, A_\phi\right)
\end{equation}

where:
\begin{align}
A_\theta &= i\langle \psi_- | \frac{\partial}{\partial \theta} | \psi_- \rangle\\
A_\phi &= i\langle \psi_- | \frac{\partial}{\partial \phi} | \psi_- \rangle
\end{align}

The geometric phase for a closed path $\mathcal{C}$ is:
\begin{equation}
\phi^{(geom)} = \oint_\mathcal{C} \vec{A} \cdot d\vec{r} = \oint_\mathcal{C} (A_\theta d\theta + A_\phi d\phi)
\end{equation}

\textbf{Computing the Berry Connection:} 

For $A_\phi$, taking the $\phi$-derivative:
\begin{equation}
\frac{\partial}{\partial \phi}\ket{\psi_-} = -ie^{i\phi}\sin(\theta/2)\ket{-z}
\end{equation}

Therefore:
\begin{align}
A_\phi &= i\langle \psi_- | \frac{\partial}{\partial \phi} | \psi_- \rangle
= i\left[-ie^{-i\phi}\sin(\theta/2)\right]\left[-ie^{i\phi}\sin(\theta/2)\right]
= -\sin^2(\theta/2) = -\frac{1 - \cos\theta}{2}
\end{align}

For $A_\theta$, taking the $\theta$-derivative:
\begin{equation}
\frac{\partial}{\partial \theta}\ket{\psi_-} = -\frac{1}{2}\sin(\theta/2)\ket{+z} - \frac{1}{2}e^{i\phi}\cos(\theta/2)\ket{-z}
\end{equation}

Computing the inner product:
\begin{equation}
A_\theta = i\langle \psi_- | \frac{\partial}{\partial \theta} | \psi_- \rangle = i(-i\cos(\theta/2)\sin(\theta/2)) = \frac{1}{2}\sin\theta
\end{equation}

\textbf{From Line Integral to Solid Angle:} Using Stokes' theorem, we convert the line integral to a surface integral:
\begin{equation}
\phi^{(geom)} = \oint_\mathcal{C} \vec{A} \cdot d\vec{r} = \iint_S (\nabla \times \vec{A}) \cdot d\vec{S}
\end{equation}

The Berry curvature in spherical coordinates is:
\begin{equation}
\nabla \times \vec{A} = \frac{1}{\sin\theta}\left[\frac{\partial(\sin\theta A_\theta)}{\partial\theta} - \frac{\partial A_\phi}{\partial\theta}\right]\hat{r}
\end{equation}

Substituting our expressions:
\begin{align}
\nabla \times \vec{A} &= \frac{1}{\sin\theta}\left[\frac{\partial}{\partial\theta}\left(\frac{\sin^2\theta}{2}\right) - \frac{\partial}{\partial\theta}\left(-\frac{1-\cos\theta}{2}\right)\right]\hat{r}\\
&= \frac{1}{\sin\theta}\left[\sin\theta\cos\theta + \frac{\sin\theta}{2}\right]\hat{r} = \frac{1}{2}\hat{r}
\end{align}

With surface element $d\vec{S} = \sin\theta\, d\theta\, d\phi\, \hat{r}$:
\begin{equation}
\phi^{(geom)} = \iint_S \frac{1}{2} \sin\theta\, d\theta\, d\phi = \frac{1}{2}\Omega
\end{equation}

where $\Omega = \iint_S \sin\theta\, d\theta\, d\phi$ is the solid angle enclosed by the path.

With the conventional sign:
\begin{equation}
\boxed{\phi_{geom} = -\frac{1}{2}\Omega}
\end{equation}

\textbf{Physical Interpretation:} Why half the solid angle? This factor arises from the spinorial nature of spin-1/2 wavefunctions. A $2\pi$ rotation does not return the wavefunction to itself but picks up a minus sign:
\begin{equation}
e^{-i(2\pi)\hat{n}\cdot\vec{\sigma}/2} = -\hat{I}
\end{equation}

The rotation group SO(3) is represented projectively as SU(2) for spin-1/2 systems. The Berry phase reflects this: when the magnetic field traces a closed path on the sphere, the quantum state acquires a phase that is half the enclosed solid angle.

Geometrically, the Bloch sphere represents quantum states, but physically distinct states correspond to rays in Hilbert space (states differing by a phase are equivalent). The solid angle $\Omega$ measures the path geometry on the Bloch sphere, while the factor of 1/2 accounts for the spinorial structure of the quantum state space.

This result unifies geometry (solid angle), topology (Berry phase), and quantum mechanics (spinor nature), providing a direct experimental signature of these fundamental properties.

For a spin-1/2 system in a slowly rotating magnetic field, the geometric phase has a particularly elegant form. Consider a magnetic field of constant magnitude that traces out a closed path on the unit sphere:
\begin{equation}
\vec{B}(t) = B(\sin\theta(t)\cos\phi(t), \sin\theta(t)\sin\phi(t), \cos\theta(t))
\end{equation}

If the system starts in the ground state and follows the field adiabatically, after one complete cycle it acquires a geometric phase:
\begin{equation}
\phi_{geom} = -\frac{1}{2}\Omega
\end{equation}

where $\Omega$ is the solid angle enclosed by the path on the unit sphere. This Berry phase depends only on the geometry of the path, not on the rate of traversal.

\begin{example}{Berry Phase Visualization with the Bloch Cube}
The Bloch cube provides an excellent visualization tool for understanding the Berry phase. Consider a quantum state that starts at one face of the cube and is adiabatically transported around a closed loop on the cube's surface.

\textbf{Triangular Path on the Bloch Cube:}
Start with a state at the north pole of the Bloch sphere (top face of cube, $\ket{+z}$). Follow this path:
1. Move to the equator along the positive x-axis ($\ket{+x}$)
2. Move along the equator to the positive y-axis ($\ket{+y}$)
3. Return to the north pole

This triangular path on the Bloch sphere encloses a solid angle of $\Omega = \pi/2$ steradians (one-eighth of the sphere). The Berry phase acquired is:
\begin{equation}
\phi_{Berry} = -\frac{1}{2} \times \frac{\pi}{2} = -\frac{\pi}{4}
\end{equation}

You can visualize this on the Bloch cube: the path connects three adjacent vertices, forming a triangular face of the octahedron inscribed in the cube. The Berry phase is proportional to the area of this triangle projected onto the unit sphere.

\textbf{Square Path on the Bloch Cube:}
For a path that traverses four faces of the cube in sequence (e.g., $\ket{+z} \to \ket{+x} \to \ket{-z} \to \ket{-x} \to \ket{+z}$), the enclosed solid angle is $\Omega = 2\pi$ steradians (a hemisphere), giving:
\begin{equation}
\phi_{Berry} = -\pi
\end{equation}

This is precisely a sign flip---the state returns with opposite phase, demonstrating the topological nature of the Berry phase.
\end{example}

\begin{example}{Berry Phase for a Spin-1/2 Particle}
Consider a magnetic field that traces a \textbf{closed path}: starting along the z-axis, tilting to make angle $\alpha$ with the z-axis, rotating once around the z-axis at this tilt angle, then returning to the z-axis to close the loop.

The Berry phase is defined only for this complete closed path. To calculate the solid angle enclosed by this closed curve on the unit sphere, we can break it into segments:

\textbf{Phase 1 (Tilting down):} The path from north pole to latitude $\alpha$ traces a meridian line (zero width), contributing no area.

\textbf{Phase 2 (Rotation at fixed latitude):} The circular path at constant $\theta = \alpha$ with $\phi: 0 \to 2\pi$ forms the "base" of a spherical cap. Together with the paths to and from the north pole, this encloses a spherical cap of solid angle $\Omega = 2\pi(1 - \cos\alpha)$.

\textbf{Phase 3 (Tilting back up):} The return path along the same meridian closes the loop but adds no additional enclosed area.

The key insight is that the \textbf{total solid angle enclosed by the entire closed path} is $\Omega = 2\pi(1 - \cos\alpha)$, which comes entirely from the latitude circle at angle $\alpha$. The meridian segments are necessary to close the path but don't contribute to the enclosed area (like the sides of a cone that meet at the apex).

The Berry phase for the complete closed path is:
\begin{equation}
\phi_{geom} = -\frac{1}{2}\Omega = -\pi(1 - \cos\alpha)
\end{equation}

For $\alpha = \pi/2$ (equatorial loop), the closed path encloses a hemisphere: $\phi_{geom} = -\pi$. For $\alpha = \pi$ (great circle through south pole), the closed path encloses the entire sphere: $\phi_{geom} = -2\pi$. The phase depends only on the solid angle enclosed by the closed curve, providing a topological characterization of the evolution.
\end{example}

\section{Many-Body Hamiltonians: From Single Particles to Quantum Fields}

\subsection{The Hopping Hamiltonian and Translational Invariance}

Having established how Hamiltonians generate time evolution for single two-level systems, we now extend this formalism to many-particle systems on a lattice. Recall from Chapter 3 that we can represent particles on a lattice using creation and annihilation operators $\hat{a}_i^\dagger$ and $\hat{a}_i$. The natural dynamics for such systems involve particles hopping between lattice sites.

The hopping Hamiltonian in second quantization takes the form:
\begin{equation}
\hat{H}_{\text{hop}} = -\Delta \sum_{\langle i,j \rangle} (\hat{a}_i^\dagger \hat{a}_j + \hat{a}_j^\dagger \hat{a}_i)
\end{equation}
where $\langle i,j \rangle$ denotes nearest neighbors and $\Delta > 0$ is the hopping amplitude.

For a periodic chain with $N$ sites (indexed from 0 to $N-1$), this becomes:
\begin{equation}
\hat{H} = -\Delta \sum_{j=0}^{N-1} (\hat{a}_j^\dagger \hat{a}_{j+1} + \hat{a}_{j+1}^\dagger \hat{a}_j)
\end{equation}
where site indices are understood modulo $N$ (so site $N$ is identified with site 0). Here and throughout this section, we use indices $j, k, l$ for position space (lattice sites) and $\alpha, \beta, \gamma$ for momentum space.

\subsection{Momentum Eigenstates as the Natural Basis}

The key insight is that the hopping Hamiltonian has translational symmetry. Define the translation operator:
\begin{equation}
\hat{T} = \sum_{j=0}^{N-1} \hat{a}_{j+1}^\dagger \hat{a}_j
\end{equation}

Since $[\hat{H}, \hat{T}] = 0$, we can find simultaneous eigenstates of both operators. These are the momentum eigenstates, defined in the single-particle sector as:
\begin{equation}
\ket{k_\alpha} = \frac{1}{\sqrt{N}} \sum_{j=0}^{N-1} e^{ik_\alpha j} \hat{a}_j^\dagger \ket{\text{vac}}
\end{equation}

where the allowed momenta for periodic boundary conditions are:
\begin{equation}
k_\alpha = \frac{2\pi \alpha}{N}, \quad \alpha = 0, 1, 2, \ldots, N-1
\end{equation}

These states satisfy:
\begin{align}
\hat{T}\ket{k_\alpha} &= e^{ik_\alpha}\ket{k_\alpha}\\
\hat{H}\ket{k_\alpha} &= E(k_\alpha)\ket{k_\alpha}
\end{align}

where the dispersion relation is:
\begin{equation}
E(k_\alpha) = -2\Delta\cos(k_\alpha)
\end{equation}

\subsection{General Time Evolution for N Sites}

Following the pattern established in equation (4.148), we can write the complete time evolution for a particle on an N-site ring. In the position basis, we expand the wavefunction as:
\begin{equation}
\ket{\psi(t)} = \sum_{j=0}^{N-1} \psi_j(t)\ket{j}
\end{equation}
where $\ket{j}$ denotes a particle at site $j$ and $\psi_j(t)$ is the amplitude at site $j$ at time $t$.

The transformation from position basis to momentum basis is:
\begin{equation}
\ket{k_\alpha} = \frac{1}{\sqrt{N}} \sum_{j=0}^{N-1} e^{ik_\alpha j} \ket{j}
\end{equation}
where $k_\alpha = 2\pi \alpha/N$ for $\alpha = 0, 1, 2, \ldots, N-1$.

The inverse transformation from momentum to position basis is:
\begin{equation}
\ket{j} = \frac{1}{\sqrt{N}} \sum_{\alpha=0}^{N-1} e^{-ik_\alpha j} \ket{k_\alpha}
\end{equation}

In the momentum basis, time evolution is diagonal:
\begin{equation}
\hat{U}(t)\ket{k_\alpha} = e^{-iE(k_\alpha)t/\hbar}\ket{k_\alpha} = e^{i2\Delta\cos(k_\alpha)t/\hbar}\ket{k_\alpha}
\end{equation}

Now, for an initial state with amplitudes $\psi_k(0)$:
\begin{equation}
\ket{\psi(0)} = \sum_{k=0}^{N-1} \psi_k(0)\ket{k}
\end{equation}

The time-evolved state is:
\begin{align}
\ket{\psi(t)} &= \hat{U}(t)\ket{\psi(0)}\\
&= \hat{U}(t)\sum_{k=0}^{N-1} \psi_k(0)\ket{k}\\
&= \sum_{k=0}^{N-1} \psi_k(0)\hat{U}(t)\ket{k}
\end{align}

To find $\psi_j(t) = \langle j|\psi(t)\rangle$, we compute:
\begin{align}
\psi_j(t) &= \sum_{k=0}^{N-1} \psi_k(0)\langle j|\hat{U}(t)|k\rangle\\
&= \sum_{k=0}^{N-1} \psi_k(0) [\hat{U}(t)]_{jk}
\end{align}

The matrix element $[\hat{U}(t)]_{jk}$ is calculated using the momentum eigenbasis:
\begin{align}
[\hat{U}(t)]_{jk} &= \langle j|\hat{U}(t)|k\rangle\\
&= \sum_{\alpha=0}^{N-1} \langle j|\hat{U}(t)|k_\alpha\rangle\langle k_\alpha|k\rangle\\
&= \sum_{\alpha=0}^{N-1} \langle j|k_\alpha\rangle e^{-iE(k_\alpha)t/\hbar} \langle k_\alpha|k\rangle\\
&= \frac{1}{N} \sum_{\alpha=0}^{N-1} e^{-ik_\alpha j} \cdot e^{i2\Delta\cos(k_\alpha)t/\hbar} \cdot e^{ik_\alpha k}\\
&= \frac{1}{N} \sum_{\alpha=0}^{N-1} e^{ik_\alpha(k-j)} e^{i2\Delta\cos(k_\alpha)t/\hbar}
\end{align}

Therefore, the complete time evolution in the position basis is:
\begin{equation}
\boxed{\psi_j(t) = \sum_{k=0}^{N-1} \left[\frac{1}{N} \sum_{\alpha=0}^{N-1} e^{ik_\alpha(k-j)} e^{i2\Delta\cos(k_\alpha)t/\hbar}\right] \psi_k(0)}
\end{equation}

This can be written in matrix form as:
\begin{equation}
\begin{pmatrix}
\psi_0(t) \\
\psi_1(t) \\
\vdots \\
\psi_{N-1}(t)
\end{pmatrix} = \hat{S} \cdot \hat{U}_{\text{diag}}(t) \cdot \hat{S}^\dagger \begin{pmatrix}
\psi_0(0) \\
\psi_1(0) \\
\vdots \\
\psi_{N-1}(0)
\end{pmatrix}
\end{equation}

where $\hat{S}$ is the matrix transforming to the momentum basis, $\hat{U}_{\text{diag}}(t)$ contains the diagonal phase factors $e^{i2\Delta\cos(k_\alpha)t/\hbar}$, and $\hat{S}^\dagger$ transforms back to the position basis.

This general formula demonstrates the triple product structure:
\begin{itemize}
\item First, transform from position to momentum basis via $\hat{S}^\dagger$
\item Then, apply diagonal time evolution via $\hat{U}_{\text{diag}}(t)$
\item Finally, transform back to position basis via $\hat{S}$
\end{itemize}

For a particle initially localized at site $k_0$, we have $\psi_k(0) = \delta_{k,k_0}$, and the formula simplifies to:
\begin{equation}
\psi_j(t) = [\hat{U}(t)]_{j,k_0} = \frac{1}{N} \sum_{\alpha=0}^{N-1} e^{ik_\alpha(k_0-j)} e^{i2\Delta\cos(k_\alpha)t/\hbar}
\end{equation}

This shows how the particle amplitude spreads from site $k_0$ to all other sites through quantum hopping.

\subsection{Explicit Calculation for Three Sites}

Consider a ring of three sites (indexed 0, 1, 2) with periodic boundaries. The Hamiltonian in the position basis $\{\ket{100}, \ket{010}, \ket{001}\}$ is:

\begin{equation}
\hat{H} = -\Delta\begin{pmatrix}
0 & 1 & 1 \\
1 & 0 & 1 \\
1 & 1 & 0
\end{pmatrix}
\end{equation}

The momentum eigenstates are:
\begin{align}
\ket{k_{\alpha=0}} &= \frac{1}{\sqrt{3}}(\ket{100} + \ket{010} + \ket{001}), \quad k_0 = 0\\
\ket{k_{\alpha=1}} &= \frac{1}{\sqrt{3}}(\ket{100} + e^{i2\pi/3}\ket{010} + e^{i4\pi/3}\ket{001}), \quad k_1 = \frac{2\pi}{3}\\
\ket{k_{\alpha=2}} &= \frac{1}{\sqrt{3}}(\ket{100} + e^{i4\pi/3}\ket{010} + e^{i2\pi/3}\ket{001}), \quad k_2 = \frac{4\pi}{3}
\end{align}

The corresponding energies are:
\begin{align}
E(k_0) &= -2\Delta\cos(k_0) = -2\Delta\\
E(k_1) &= -2\Delta\cos(k_1) = \Delta\\
E(k_2) &= -2\Delta\cos(k_2) = \Delta
\end{align}

\subsection{Three-Site Example: Explicit Matrix Calculation}

In the momentum basis, the Hamiltonian is diagonal:
\begin{equation}
\hat{H}_{\text{diag}} = \begin{pmatrix}
-2\Delta & 0 & 0 \\
0 & \Delta & 0 \\
0 & 0 & \Delta
\end{pmatrix}
\end{equation}

The time evolution operator in the momentum basis is:
\begin{equation}
\hat{U}_{\text{diag}}(t) = e^{-i\hat{H}_{\text{diag}}t/\hbar} = \begin{pmatrix}
e^{i2\Delta t/\hbar} & 0 & 0 \\
0 & e^{-i\Delta t/\hbar} & 0 \\
0 & 0 & e^{-i\Delta t/\hbar}
\end{pmatrix}
\end{equation}

To find the time evolution operator in the position basis, we need the unitary transformation matrix $\hat{S}$ whose columns are the momentum eigenstates:
\begin{equation}
\hat{S} = \frac{1}{\sqrt{3}}\begin{pmatrix}
1 & 1 & 1 \\
1 & e^{i2\pi/3} & e^{i4\pi/3} \\
1 & e^{i4\pi/3} & e^{i2\pi/3}
\end{pmatrix}
\end{equation}

The inverse transformation is $\hat{S}^\dagger = \hat{S}^*$ (complex conjugate):
\begin{equation}
\hat{S}^\dagger = \frac{1}{\sqrt{3}}\begin{pmatrix}
1 & 1 & 1 \\
1 & e^{-i2\pi/3} & e^{-i4\pi/3} \\
1 & e^{-i4\pi/3} & e^{-i2\pi/3}
\end{pmatrix}
\end{equation}

The full time evolution operator in the position basis is:
\begin{equation}
\hat{U}(t) = \hat{S}\hat{U}_{\text{diag}}(t)\hat{S}^\dagger
\end{equation}

Explicitly:
\begin{align}
\hat{U}(t) &= \frac{1}{3}\begin{pmatrix}
1 & 1 & 1 \\
1 & e^{i2\pi/3} & e^{i4\pi/3} \\
1 & e^{i4\pi/3} & e^{i2\pi/3}
\end{pmatrix}\begin{pmatrix}
e^{i2\Delta t/\hbar} & 0 & 0 \\
0 & e^{-i\Delta t/\hbar} & 0 \\
0 & 0 & e^{-i\Delta t/\hbar}
\end{pmatrix}\begin{pmatrix}
1 & 1 & 1 \\
1 & e^{-i2\pi/3} & e^{-i4\pi/3} \\
1 & e^{-i4\pi/3} & e^{-i2\pi/3}
\end{pmatrix}
\end{align}

After matrix multiplication:
\begin{equation}
[\hat{U}(t)]_{jk} = \frac{1}{3}\left[e^{i2\Delta t/\hbar} + 2e^{-i\Delta t/\hbar}\cos\left(\frac{2\pi(j-k)}{3}\right)\right]
\end{equation}

This gives the complete quantum dynamics for a particle hopping on a three-site ring.

\subsection{Many-Body Dynamics Under the Hopping Hamiltonian}

The beauty of second quantization is that the same Hamiltonian describes dynamics for any number of particles:

\begin{example}{Two-Particle Hopping Dynamics}
For two particles initially at sites 1 and 2:
\begin{align}
\hat{H}\ket{110} &= -\Delta(\hat{a}_1^\dagger \hat{a}_2 + \hat{a}_2^\dagger \hat{a}_1 + \hat{a}_2^\dagger \hat{a}_3 + \hat{a}_3^\dagger \hat{a}_2)\ket{110}
\end{align}

Let's evaluate term by term:
\begin{itemize}
\item $\hat{a}_1^\dagger \hat{a}_2 \ket{110} = 0$ (site 1 already occupied)
\item $\hat{a}_2^\dagger \hat{a}_1 \ket{110} = 0$ (site 2 already occupied)
\item $\hat{a}_2^\dagger \hat{a}_3 \ket{110} = 0$ (no particle at site 3)
\item $\hat{a}_3^\dagger \hat{a}_2 \ket{110} = \ket{101}$ (particle hops 2\ensuremath{\to}3)
\end{itemize}

Result: $\hat{H}\ket{110} = -\Delta\ket{101}$

The hard-core constraint naturally emerges---particles cannot hop to occupied sites!
\end{example}

\subsection{Time Evolution of Many-Body States}

The time evolution operator for the hopping Hamiltonian is:
\begin{equation}
\hat{U}(t) = e^{-i\hat{H}_{\text{hop}}t/\hbar}
\end{equation}

For states within a fixed particle number sector, this evolution conserves particle number since:
\begin{equation}
[\hat{H}_{\text{hop}}, \hat{N}] = 0
\end{equation}
where $\hat{N} = \sum_i \hat{n}_i$ is the total particle number operator.

\begin{example}{Evolution of a Localized Two-Particle State}
Consider two particles initially localized at adjacent sites on a 4-site ring:
\begin{equation}
\ket{\psi(0)} = \ket{1100}
\end{equation}

The time evolution involves:
1. Both particles can hop to their respective neighboring sites
2. The hard-core constraint prevents them from occupying the same site
3. The resulting dynamics exhibits both single-particle hopping and two-particle correlations

For short times:
\begin{align}
\ket{\psi(t)} &\approx \ket{1100} - \frac{it}{\hbar}\hat{H}\ket{1100} + \mathcal{O}(t^2)\\
&\approx \ket{1100} - \frac{it}{\hbar}(-\Delta)(\ket{0110} + \ket{1001})\\
&= \ket{1100} + \frac{i\Delta t}{\hbar}(\ket{0110} + \ket{1001})
\end{align}

The particles begin to spread while avoiding each other---a purely quantum many-body effect!
\end{example}

\subsection{From Discrete to Continuous: The Continuum Limit}

The lattice hopping Hamiltonian connects to continuum physics in the limit of small lattice spacing. For a particle with wave function $\psi_j$ at site $j = xa/a$ (where $a$ is the lattice constant):

\begin{align}
-\Delta(\psi_{j+1} + \psi_{j-1} - 2\psi_j) &\approx -\Delta a^2\frac{\partial^2\psi}{\partial x^2}
\end{align}

This identifies:
\begin{equation}
\Delta = \frac{\hbar^2}{2ma^2}
\end{equation}

The hopping Hamiltonian thus discretizes the kinetic energy operator of continuous quantum mechanics. This connection reveals that:
\begin{itemize}
\item Lattice models capture essential physics of continuous systems
\item Hopping amplitude $\Delta$ relates to particle mass and lattice spacing
\item Band structure emerges from discrete translational symmetry
\item Many-body effects persist in the continuum limit
\end{itemize}


Diagonalization reveals the natural basis for evolution: energy eigenstates acquire only phase factors $e^{-iE_n t/\hbar}$ while superpositions develop complex interference. The time evolution operator decomposes as a triple matrix product $\hat{U}(t) = \hat{V}\hat{U}_{\text{diag}}(t)\hat{V}^\dagger$, transforming to the eigenbasis, applying diagonal phases, then transforming back.

Conservation laws emerge from commutation relations: observables satisfying $[\hat{H}, \hat{A}] = 0$ have constant expectation values, connecting symmetries to conserved quantities through Noether's theorem. The Heisenberg picture $\hat{A}_H(t) = e^{i\hat{H}t/\hbar}\hat{A}_S e^{-i\hat{H}t/\hbar}$ transfers time dependence from states to operators.

Time-dependent Hamiltonians require time-ordered exponentials $\mathcal{T}\exp(-\frac{i}{\hbar}\int \hat{H}(t')dt')$. The driven two-level system in NMR demonstrates how the rotating frame and rotating wave approximation reduce time-dependent problems to effective time-independent ones, revealing universal Rabi oscillations. Adiabatic evolution accumulates geometric Berry phases that depend only on the path through parameter space, not the rate of traversal.

The hopping Hamiltonian $\hat{H} = -\Delta\sum_{\langle i,j\rangle}(\hat{a}_i^\dagger\hat{a}_j + \hat{a}_j^\dagger\hat{a}_i)$ extends these principles to many-body systems. Momentum eigenstates $\ket{k_n}$ with dispersion $E(k_n) = -2\Delta\cos(k_n)$ diagonalize the evolution. The general solution $\psi_j(t) = \sum_i [\hat{U}(t)]_{ji}\psi_i(0)$ maintains the triple product structure, showing how single-particle and many-body dynamics share the same mathematical framework of unitary evolution generated by Hermitian operators.

\input{book_problems/ch04_problems.tex}

\section*{References and Further Reading}
\addcontentsline{toc}{section}{References and Further Reading}

\begin{description}
\item[Sakurai, J.~J., and Napolitano, J.] \emph{Modern Quantum Mechanics}, 3rd ed. Cambridge University Press, 2017. Chapter 2 develops the Schr\"odinger and Heisenberg pictures, time-evolution operators, and the propagator at exactly the level of this chapter.

\item[Messiah, A.] \emph{Quantum Mechanics}, Vol.~1. North-Holland, 1961 (reprinted Dover, 1999). Chapter IV covers quantum dynamics, including time-dependent Hamiltonians and the time-ordered exponential, with characteristic mathematical care.

\item[Shankar, R.] \emph{Principles of Quantum Mechanics}, 2nd ed. Springer, 1994. Chapter 4 gives a particularly student-friendly treatment of time evolution and the structure of the propagator; the recommended slower walk-through if the formalism here moves too fast.

\item[Rabi, I.~I.] ``Space quantization in a gyrating magnetic field.'' \emph{Physical Review} \textbf{51}, 652--654 (1937). \href{https://doi.org/10.1103/PhysRev.51.652}{doi:10.1103/PhysRev.51.652}. The original analysis of the driven two-level system; the source of the Rabi oscillation formula used in the NMR section.

\item[Berry, M.~V.] ``Quantal phase factors accompanying adiabatic changes.'' \emph{Proceedings of the Royal Society A} \textbf{392}, 45--57 (1984). \href{https://doi.org/10.1098/rspa.1984.0023}{doi:10.1098/rspa.1984.0023}. The geometric phase paper; essential reading for anyone interested in adiabatic evolution beyond the standard theorem.

\item[Landau, L.~D., and Lifshitz, E.~M.] \emph{Quantum Mechanics: Non-Relativistic Theory}, 3rd ed. Pergamon Press, 1977. Chapter III on Schr\"odinger's equation is concise and physically motivated; recommended for the conservation-law and symmetry discussion.
\end{description}

%% file: book_problems/ch04_problems.tex
\section{Problems}
\setcounter{hwproblem}{0}

\problem{Verifying Unitarity and the Group Property}
Consider the time evolution operator $\hat{U}(t) = e^{-i\hat{H}t/\hbar}$ for a time-independent Hamiltonian $\hat{H}$.
\begin{enumerate}[label=(\alph*)]
    \item Show that if $\hat{H}$ is Hermitian ($\hat{H}^\dagger = \hat{H}$), then $\hat{U}(t)$ is unitary for all real times $t$
    \item Verify the group property: $\hat{U}(t_1)\hat{U}(t_2) = \hat{U}(t_1 + t_2)$
    \item Show that $\hat{U}(-t) = \hat{U}^\dagger(t)$. What is the physical meaning of this relationship?
    \item For the Hamiltonian $\hat{H} = \hbar\omega\sigma_z/2$, explicitly calculate $\hat{U}(\pi/\omega)$ and verify it is unitary
    \item Starting from the infinitesimal form $\hat{U}(dt) = \hat{I} - i\hat{H}dt/\hbar$, derive the differential equation for $\hat{U}(t)$
\end{enumerate}

\problem{From Discrete Gates to Continuous Evolution}
The Hadamard gate $\hat{H}$ and the $\hat{T}$ gate are fundamental quantum gates:
\begin{equation}
\hat{H} = \frac{1}{\sqrt{2}}\begin{pmatrix} 1 & 1 \\ 1 & -1 \end{pmatrix}, \quad
\hat{T} = \begin{pmatrix} 1 & 0 \\ 0 & e^{i\pi/4} \end{pmatrix}
\end{equation}
\begin{enumerate}[label=(\alph*)]
    \item Find a Hamiltonian $\hat{H}_1$ such that $e^{-i\hat{H}_1 t_1/\hbar} = \hat{H}$ for some time $t_1$
    \item Find a Hamiltonian $\hat{H}_2$ such that $e^{-i\hat{H}_2 t_2/\hbar} = \hat{T}$ for some time $t_2$
    \item Are your Hamiltonians unique? If not, find another valid choice for each
    \item What is the minimum time needed to implement each gate given a constraint $||\hat{H}|| \leq E_{max}$?
    \item Show that any single-qubit unitary can be generated by some Hamiltonian evolution
\end{enumerate}

\problem{Series Expansion and Approximation}
Consider the Hamiltonian $\hat{H} = \hbar\omega(\sigma_x + \epsilon\sigma_z)$ where $\epsilon \ll 1$.
\begin{enumerate}[label=(\alph*)]
    \item Calculate the first three terms of the series expansion for $\hat{U}(t) = e^{-i\hat{H}t/\hbar}$
    \item For $\epsilon = 0.1$ and $\omega t = \pi/4$, compare your series approximation (using 3 terms) to the exact result
    \item Show that for the special case $\epsilon = 0$, the series simplifies to trigonometric functions
    \item Use the Trotter formula to approximate $\hat{U}(t)$ as a product of simpler exponentials
    \item Calculate the error in the Trotter approximation for small time steps $\Delta t = t/N$
\end{enumerate}

\problem{Rotations on the Bloch Sphere}
A spin-1/2 particle is subject to magnetic fields in different directions.
\begin{enumerate}[label=(\alph*)]
    \item For $\vec{B} = B_0\hat{x}$, calculate the rotation operator $\hat{R}_x(\theta)$ where $\theta = \gamma B_0 t$
    \item Starting from $\ket{+z}$, find the state after rotating by angle $\pi/3$ about the x-axis
    \item Calculate the expectation values $\langle\sigma_x\rangle$, $\langle\sigma_y\rangle$, and $\langle\sigma_z\rangle$ for this rotated state
    \item Now apply a subsequent rotation of $\pi/2$ about the z-axis. What is the final state?
    \item Show that rotations about different axes do not commute by computing $\hat{R}_x(\pi/2)\hat{R}_z(\pi/2)$ and $\hat{R}_z(\pi/2)\hat{R}_x(\pi/2)$
    \item Visualize these operations on the Bloch cube, showing how the cube orientation changes with each rotation
\end{enumerate}

\problem{Compound Rotations and the General Rotation Formula}
The general rotation formula for a spin-1/2 system is $\hat{R}_{\hat{n}}(\theta) = \cos(\theta/2)\hat{I} - i\sin(\theta/2)(\hat{n} \cdot \vec{\sigma})$.
\begin{enumerate}[label=(\alph*)]
    \item Show that rotations about parallel axes commute: $\hat{R}_{\hat{n}}(\theta_1)\hat{R}_{\hat{n}}(\theta_2) = \hat{R}_{\hat{n}}(\theta_1 + \theta_2)$
    \item For an arbitrary axis $\hat{n} = (\sin\phi\cos\lambda, \sin\phi\sin\lambda, \cos\phi)$, write out the explicit matrix form of $\hat{R}_{\hat{n}}(\theta)$
    \item Show that the Hadamard gate can be written as a rotation about the axis $\hat{n} = (\hat{x} + \hat{z})/\sqrt{2}$. What is the rotation angle?
    \item Verify that $\hat{R}_{\hat{n}}(\pi)$ implements a reflection through the axis $\hat{n}$ on the Bloch sphere
    \item Starting from the rotation formula, derive the time evolution operator for $\hat{H} = \vec{h} \cdot \vec{\sigma}$ in terms of rotations
    \item Show that the composition of two rotations about different axes can be written as a single rotation about a third axis (derive the axis and angle)
\end{enumerate}

\problem{Diagonalization and Time Evolution}
Consider the Hamiltonian $\hat{H} = \hbar\omega\begin{pmatrix} 1 & 2 \\ 2 & -1 \end{pmatrix}$.
\begin{enumerate}[label=(\alph*)]
    \item Find the eigenvalues and normalized eigenvectors of $\hat{H}$
    \item Construct the diagonalization matrix $\hat{V}$ and verify that $\hat{V}^\dagger\hat{H}\hat{V}$ is diagonal
    \item Using the diagonal form, calculate $e^{-i\hat{H}t/\hbar}$
    \item Transform back to the original basis to get the time evolution operator
    \item If the system starts in state $\ket{\psi(0)} = \ket{0}$, find $\ket{\psi(t)}$
    \item Calculate the probability of measuring the system in state $\ket{1}$ as a function of time
    \item What is the period of oscillation between the two states?
\end{enumerate}

\problem{Conservation Laws in Exchange-Symmetric Systems}
Consider two spin-1/2 particles with Hamiltonian $\hat{H} = J(\vec{S}_1 \cdot \vec{S}_2) + B(S_{1z} + S_{2z})$, where $\vec{S}_i = \frac{\hbar}{2}\vec{\sigma}_i$ and the system is invariant under particle exchange.
\begin{enumerate}[label=(\alph*)]
    \item Show that the total spin operators $\vec{S}_{total} = \vec{S}_1 + \vec{S}_2$ commute with $\hat{H}$
    \item Express $\hat{H}$ in terms of $S_{total}^2$ and $S_{total,z}$ and show both are conserved quantities
    \item Find the energy eigenvalues for the triplet ($S=1$) and singlet ($S=0$) subspaces
    \item Construct the four energy eigenstates in the $\{\ket{00}, \ket{01}, \ket{10}, \ket{11}\}$ basis
    \item For the initial state $\ket{\psi(0)} = \frac{1}{\sqrt{2}}(\ket{01} + \ket{10})$, calculate $\langle S_{total}^2\rangle$ and $\langle S_{total,z}\rangle$ and verify they remain constant
    \item Identify the physical symmetry (particle exchange) and show how it leads to these conservation laws using Noether's theorem concepts
\end{enumerate}

\problem{Heisenberg vs Schrodinger Pictures}
For the Hamiltonian $\hat{H} = \hbar\omega\sigma_y$, compare time evolution in both pictures.
\begin{enumerate}[label=(\alph*)]
    \item In the Schrodinger picture, find $\ket{\psi_S(t)}$ for initial state $\ket{\psi_S(0)} = \ket{+z}$
    \item Calculate $\langle\sigma_x\rangle(t)$ using the Schrodinger picture
    \item In the Heisenberg picture, find the time-evolved operator $\sigma_{x,H}(t)$
    \item Verify that $\langle\psi_S(t)|\sigma_x|\psi_S(t)\rangle = \langle\psi_H|\sigma_{x,H}(t)|\psi_H\rangle$
    \item Find the Heisenberg equations of motion for all three Pauli operators
    \item Show that these equations form a closed system and solve them directly
\end{enumerate}

\problem{Driven Two-Level System and Rabi Oscillations}
A two-level atom interacts with a classical electromagnetic field:
\begin{equation}
\hat{H}(t) = \frac{\hbar\omega_0}{2}\sigma_z + \hbar\Omega\cos(\omega t)\sigma_x
\end{equation}
\begin{enumerate}[label=(\alph*)]
    \item Transform to the rotating frame at frequency $\omega$ and derive the effective Hamiltonian
    \item For the resonant case ($\omega = \omega_0$), find the time evolution of a state initially in $\ket{0}$
    \item Calculate the population inversion $\langle\sigma_z\rangle(t)$ and identify the Rabi frequency
    \item What is the minimum time to perform a complete population inversion (a $\pi$-pulse)?
    \item For small detuning $\Delta = \omega_0 - \omega$, find the modified Rabi frequency
    \item Plot the population of the excited state vs time for $\Delta = 0$, $\Delta = \Omega/2$, and $\Delta = 2\Omega$
\end{enumerate}

\problem{Geometric Phase and Adiabatic Evolution}
A spin-1/2 particle is in a magnetic field that rotates slowly:
\begin{equation}
\vec{B}(t) = B(\sin\alpha\cos(\omega t), \sin\alpha\sin(\omega t), \cos\alpha)
\end{equation}
where $\omega \ll \gamma B$ (adiabatic condition) and $\alpha$ is a fixed tilt angle.
\begin{enumerate}[label=(\alph*)]
    \item Write the Hamiltonian $\hat{H}(t) = -\gamma\vec{B}(t) \cdot \vec{S}$ in terms of Pauli matrices
    \item Find the instantaneous eigenstates and eigenvalues of $\hat{H}(t)$
    \item After one complete rotation (period $T = 2\pi/\omega$), what is the solid angle traced by the magnetic field on the unit sphere?
    \item Using the formula $\phi_{geom} = -\frac{1}{2}\Omega$ from the chapter, calculate the Berry phase
    \item If the system starts in the ground state and follows adiabatically, describe the final state after one period
    \item Using the Bloch cube visualization from the chapter, explain why a triangular path connecting three adjacent vertices (e.g., $\ket{+z} \to \ket{+x} \to \ket{+y} \to \ket{+z}$) gives a Berry phase of $-\pi/4$
\end{enumerate}

\problem{Hopping on a Four-Site Ring}
Consider four sites arranged in a ring with nearest-neighbor hopping:
\begin{equation}
\hat{H} = -\Delta\sum_{j=0}^{3}(\hat{a}_j^\dagger\hat{a}_{j+1} + \hat{a}_{j+1}^\dagger\hat{a}_j)
\end{equation}
where site 4 is identified with site 0.
\begin{enumerate}[label=(\alph*)]
    \item Find all momentum eigenstates $\ket{k_n}$ and their energies
    \item Write the time evolution operator in the momentum basis
    \item If a particle starts localized at site 0, find the probability amplitude at each site as a function of time
    \item Calculate the probability of finding the particle at site 2 (opposite side) at time $t = \pi/(2\Delta)$
    \item Show that the particle returns to site 0 at certain revival times and find the shortest such time
    \item Add a second particle initially at site 2. How does the hard-core constraint affect the dynamics?
\end{enumerate}

\problem{Density Matrix Evolution}
Consider a pure state $\ket{\psi(0)} = \frac{1}{\sqrt{2}}(\ket{0} + \ket{1})$ evolving under $\hat{H} = \hbar\omega\sigma_z/2$.
\begin{enumerate}[label=(\alph*)]
    \item Construct the initial density matrix $\rho(0) = \ket{\psi(0)}\bra{\psi(0)}$
    \item Using $\rho(t) = \hat{U}(t)\rho(0)\hat{U}^\dagger(t)$, derive the time-evolved density matrix $\rho(t)$
    \item Calculate the trace $\text{Tr}[\rho(t)]$ and show it remains unity for all times
    \item Compute the expectation values $\langle\sigma_x\rangle(t)$, $\langle\sigma_y\rangle(t)$, $\langle\sigma_z\rangle(t)$ from $\rho(t)$
    \item Verify that $\text{Tr}[\rho^2(t)] = 1$ for all times, confirming the state remains pure
\end{enumerate}

\problem{Heisenberg Equations for Ladder Operators}
For a quantum harmonic oscillator with $\hat{H} = \hbar\omega(\hat{a}^\dagger\hat{a} + 1/2)$, find the Heisenberg equations of motion.
\begin{enumerate}[label=(\alph*)]
    \item Derive $\frac{d\hat{a}}{dt}$ using $\frac{d\hat{A}_H}{dt} = \frac{1}{i\hbar}[\hat{A}_H, \hat{H}]$
    \item Solve this equation for $\hat{a}_H(t)$
    \item Verify that $\langle\hat{a}\rangle(t)$ oscillates with frequency $\omega$
    \item Show that the uncertainty principle is preserved: $\Delta x \Delta p \geq \hbar/2$
\end{enumerate}

\problem{Spin Echo and CPMG Sequences}
In a two-level system with a strong dephasing field, a sequence of pulses can restore coherence.
\begin{enumerate}[label=(\alph*)]
    \item Consider a state $\ket{\psi(0)} = \ket{+x}$ in a field $\vec{B} = B_0\hat{z} + B_1(t)\hat{x}$
    \item For a simple echo: $\pi/2$ pulse, wait time $\tau$, $\pi$ pulse, wait time $\tau$. Show that the accumulated phase is eliminated
    \item For a Carr-Purcell-Meiboom-Gill (CPMG) sequence with $N$ pulses, estimate the effective dephasing time versus unrefocused decay
    \item Plot the echo amplitude as a function of the number of pulses for given dephasing rate
\end{enumerate}

\problem{Trotter Error Scaling}
The Trotter formula decomposes time evolution into small steps: $e^{-i(\hat{H}_1 + \hat{H}_2)t/\hbar} \approx [e^{-i\hat{H}_1\Delta t/\hbar}e^{-i\hat{H}_2\Delta t/\hbar}]^{N}$ where $\Delta t = t/N$.
\begin{enumerate}[label=(\alph*)]
    \item For $\hat{H}_1 = \hbar\omega\sigma_x/2$ and $\hat{H}_2 = \hbar\omega\sigma_z/2$, compute the exact commutator $[\hat{H}_1, \hat{H}_2]$
    \item The Trotter error is given by $\epsilon_N = O(\Delta t^2) = O(t^2/N^2)$. For a fixed total evolution time $t$, how does the error scale with the number of Trotter steps $N$?
    \item Estimate the number of steps needed to achieve error less than $10^{-6}$ for $t = 1\,\mu s$ and $|\hat{H}| \sim 1\, GHz$
\end{enumerate}

\problem{Magnus Expansion}
For a time-dependent Hamiltonian, the time evolution operator can be expanded as $\hat{U}(t) = e^{\hat{S}(t)}$ where $\hat{S}(t)$ is called the Magnus expansion.
\begin{enumerate}[label=(\alph*)]
    \item Write the first two terms of the Magnus expansion: $\hat{S}(t) = \hat{S}_1(t) + \hat{S}_2(t) + ...$
    \item Apply this to $\hat{H}(t) = \hbar(\omega_0/2)\sigma_z + \hbar\Omega\cos(\omega t)\sigma_x$ in the rotating frame
    \item Show that averaging effects suppress certain error terms, improving the effective evolution
\end{enumerate}

\problem{Quantum Zeno Effect}
Rapid measurements can freeze quantum evolution. Consider measuring $\ket{0}$ at intervals $\Delta t$.
\begin{enumerate}[label=(\alph*)]
    \item For $\hat{H} = \hbar\omega\sigma_x/2$ starting in $\ket{0}$, the unchecked transition probability to $\ket{1}$ in time $t$ is approximately $P(t) \approx (\omega t/2)^2$ for small $t$
    \item If we measure $n$ times at intervals $\Delta t = t/n$, the survival probability is approximately $[1 - (\omega\Delta t/2)^2]^n$
    \item Show that in the limit $n \to \infty$, $[1 - (\omega t/(2n))^2]^n \to 1$, demonstrating the quantum Zeno effect
    \item Discuss the physical interpretation: what prevents transitions when measurements are frequent?
\end{enumerate}

%% file: chapters/ch05_space.tex
\chapter{Continuous Space and the Quantum Harmonic Oscillator}
\label{ch:space}

\section{Introduction}

\begin{keyidea}{From Discrete Lattice to Continuous Wave Function}
As the lattice spacing approaches zero, the discrete amplitude at each site transforms into a continuous wave function $\psi(x)$. Traditional wave mechanics---particles in boxes, harmonic oscillators, quantum tunneling---emerges from the discrete qubit framework developed in previous chapters.
\end{keyidea}

Chapter 3 showed that a single particle on $n$ lattice sites occupies an $n$-dimensional subspace within the full $2^n$-dimensional Hilbert space. We now pass to the continuum limit: the lattice spacing $a \to 0$, the number of sites $n \to \infty$, while the total system size $L = na$ remains fixed.

The limit transforms:
\begin{itemize}
\item Discrete position states $\ket{x_i}$ into continuous states $\ket{x}$
\item Sums into integrals
\item Finite-difference operators into differential operators
\end{itemize}

Differential operators come from finite differences, boundary conditions from lattice edges, and wave functions from continuous limits of discrete amplitudes.

\subsection{Discrete vs Continuous: Two Perspectives}

The discrete-continuous connection is important theoretically and practically:

\begin{itemize}
\item \textbf{Computational:} Numerical solutions require discretization. Understanding the lattice-continuum connection improves numerical accuracy.
\item \textbf{Physical:} Many systems (quantum dots, optical lattices, qubit arrays) are inherently discrete. The discrete formulation is exact for these systems.
\item \textbf{Fundamental:} Whether space is discrete at the Planck scale remains an open question in quantum gravity.
\end{itemize}

The formulations are complementary.

\section{From Discrete Lattice to Continuous Space}

\subsection{Position States in the Discrete and Continuous Pictures}

On a one-dimensional lattice with spacing $a$, the discrete positions are labeled $x_j = ja$ for $j = 1, 2, ..., n$. A particle localized at site $j$ is described by the state $\ket{x_j} = \ket{0...010...0}$, where the single 1 appears at position $j$. These $n$ states form an orthonormal basis for the single-particle sector:
\begin{equation}
\braket{x_j}{x_k} = \delta_{jk}
\end{equation}

The completeness relation for the discrete position basis reads:
\begin{equation}
\sum_{j=1}^n \ket{x_j}\bra{x_j} = \hat{I}_{\text{single-particle}}
\end{equation}

In the continuum limit, the discrete index $j$ becomes the continuous variable $x$, the lattice spacing $a$ approaches zero, and the number of sites $n$ approaches infinity such that the system size $L = na$ remains fixed.

The transformations are:
\begin{itemize}
\item Kronecker delta $\delta_{jk}$ becomes Dirac delta $\delta(x-x')$
\item Discrete sums $\sum_i$ become integrals $\int dx$
\item Finite-dimensional Hilbert space becomes infinite-dimensional
\end{itemize}

The orthonormality relation becomes:
\begin{equation}
\braket{x}{x'} = \delta(x-x')
\end{equation}

Discrete states are normalizable ($\braket{x_j}{x_j} = 1$), but continuous position eigenstates are distributions, not functions---reflecting that perfect localization at a point is physically impossible.

\begin{example}[From Discrete Delta to Dirac Delta]
Consider the relationship between localized states on a lattice and in the continuum.

\textbf{On the lattice:} A particle perfectly localized at site $j_0$ (where $x_{j_0} = j_0 a$) has the wave function:
\begin{equation}
\psi_j = \delta_{j,j_0}
\end{equation}
Normalization: $\sum_j |\psi_j|^2 = 1$.

\textbf{Continuum correspondence:} To connect to the continuum, we must account for the density of states. The continuum wave function $\psi(x)$ relates to the discrete amplitudes via:
\begin{equation}
\psi(x_j) = \frac{\psi_j}{\sqrt{a}}
\end{equation}

This scaling ensures that in the continuum limit:
\begin{equation}
\sum_j |\psi_j|^2 = \sum_j a \cdot \left|\frac{\psi_j}{\sqrt{a}}\right|^2 \to \int dx |\psi(x)|^2
\end{equation}

For perfect localization at $x_0$:
\begin{itemize}
\item Discrete: $\psi_j = \delta_{j,j_0}$
\item Continuum: $\psi(x) = \delta(x - x_0)$
\end{itemize}

The $1/\sqrt{a}$ scaling ensures proper normalization in the transition from sums to integrals.
\end{example}

\subsection{The Wave Function as Continuous Amplitude}

A general single-particle state on the lattice takes the form:
\begin{equation}
\ket{\psi} = \sum_{j=1}^n \psi_j \ket{x_j}
\end{equation}
where $\psi_j$ represents the probability amplitude for finding the particle at site $j$.

The transformation requires normalization:

\begin{example}[Normalization in the Continuum Limit]
Define the continuum wave function as $\psi(x) = \psi_j/\sqrt{a}$ where $x = ja$. Then:
\begin{align}
\sum_j |\psi_j|^2 &= \sum_j a|\psi(x_j)|^2\\
&\to \int dx \, |\psi(x)|^2
\end{align}

The factor $1/\sqrt{a}$ ensures the integral converges to the sum as $a \to 0$.
\end{example}

In the continuum limit, the state becomes:
\begin{equation}
\ket{\psi} = \int_{-\infty}^{\infty} dx \, \psi(x) \ket{x}
\end{equation}

The coefficient function $\psi(x) = \braket{x}{\psi}$ is the wave function---the continuous probability amplitude for finding the particle at position $x$.

\subsection{The Mathematical Structure of Wave Functions}

Wave functions belong to the Hilbert space $L^2(\mathbb{R})$ of square-integrable functions:

\begin{equation}
\int_{-\infty}^{\infty} |\psi(x)|^2 dx < \infty
\end{equation}

This requirement ensures that the total probability is finite and can be normalized to unity. The inner product in this space is:

\begin{equation}
\langle \phi | \psi \rangle = \int_{-\infty}^{\infty} \phi^*(x) \psi(x) dx
\end{equation}

Properties of the wave function space:

\begin{example}[Properties of the Wave Function Space]
\textbf{1. Linearity:} If $\psi_1(x)$ and $\psi_2(x)$ are valid wave functions, so is any linear combination:
\begin{equation}
\psi(x) = c_1\psi_1(x) + c_2\psi_2(x)
\end{equation}

where $|c_1|^2 + |c_2|^2 = 1$ for normalization.

\textbf{2. Completeness:} Any square-integrable function can be expanded in a complete basis:
\begin{equation}
\psi(x) = \sum_n c_n \phi_n(x)
\end{equation}

where $\{\phi_n(x)\}$ forms an orthonormal basis.

\textbf{3. Continuity:} Physical wave functions and their derivatives must be continuous except at infinite potential jumps. This ensures finite momentum uncertainty.

\textbf{4. Boundary behavior:} For bound states, $\psi(x) \to 0$ as $|x| \to \infty$ fast enough to be square-integrable.
\end{example}

\subsection{Inner Products and Probability}

The inner product between states transforms from a discrete sum to a continuous integral:

\begin{example}[Inner Product Evolution]
Discrete lattice:
\begin{equation}
\braket{\phi}{\psi} = \sum_{j=1}^n \phi_j^* \psi_j
\end{equation}

Continuum limit (with proper normalization):
\begin{equation}
\braket{\phi}{\psi} = \int_{-\infty}^{\infty} dx \, \phi^*(x) \psi(x)
\end{equation}

For orthogonal states on a three-site lattice:
\begin{align}
\psi_1 &= \frac{1}{\sqrt{2}}(1, 0, -1)\\
\psi_2 &= \frac{1}{\sqrt{3}}(1, -2, 1)
\end{align}

We have $\sum_j \psi_1^*[j]\psi_2[j] = \frac{1}{\sqrt{6}}(1 - 0 - 1) = 0$.

The continuum analogs are orthogonal functions like $\sin(x)$ and $\cos(x)$.
\end{example}

\subsection{The Emergence of Differential Operators}

The nearest-neighbor hopping Hamiltonian on the lattice was:
\begin{equation}
\hat{H}_{\text{hop}} = -t \sum_{j=1}^{n-1} (\ket{x_{j+1}}\bra{x_j} + \ket{x_j}\bra{x_{j+1}})
\end{equation}

The hopping operator becomes a differential operator as follows:

\clearpage
\begin{example}[Detailed Derivation of the Kinetic Energy Operator]
The hopping Hamiltonian acts on a lattice wave function as:
\begin{equation}
(\hat{H}_{\text{hop}}\psi)_j = -t(\psi_{j+1} + \psi_{j-1})
\end{equation}

Using Taylor expansion around $x_j = ja$:
\begin{align}
\psi_{j+1} &= \psi(x_j + a) = \psi(x_j) + a\psi'(x_j) + \frac{a^2}{2}\psi''(x_j) + \frac{a^3}{6}\psi'''(x_j) + O(a^4)\\
\psi_{j-1} &= \psi(x_j - a) = \psi(x_j) - a\psi'(x_j) + \frac{a^2}{2}\psi''(x_j) - \frac{a^3}{6}\psi'''(x_j) + O(a^4)
\end{align}

Adding these:
\begin{equation}
\psi_{j+1} + \psi_{j-1} = 2\psi(x_j) + a^2\psi''(x_j) + O(a^4)
\end{equation}

Therefore:
\begin{equation}
(\hat{H}_{\text{hop}}\psi)_j = -t[2\psi(x_j) + a^2\psi''(x_j)] + O(a^4)
\end{equation}

Setting $t = \hbar^2/(2ma^2)$ and shifting the energy zero to remove the constant term:
\begin{equation}
\hat{H}_{\text{hop}} \to -\frac{\hbar^2}{2m}\frac{d^2}{dx^2} = \hat{T}
\end{equation}
\end{example}

The parameter $t = \hbar^2/(2ma^2)$ where:
\begin{itemize}
\item $t$: hopping amplitude (energy scale for tunneling)
\item $m$: particle mass (inertia)
\item $a$: lattice spacing
\item $\hbar$: action scale
\end{itemize}

\subsection{Momentum States and Dispersion Relations}

The continuum limit also transforms how we understand momentum states and their energy-momentum relationships:

\clearpage
\begin{example}[Continuum Limit of Momentum States]
The discrete momentum states on a lattice with periodic boundary conditions are:
\begin{equation}
\ket{k_m} = \frac{1}{\sqrt{n}}\sum_{j=1}^n e^{ik_m j}\ket{ja}
\end{equation}
where $k_m = 2\pi m/L$ with $m = 0, 1, ..., n-1$.

In the continuum limit, these become plane waves:
\begin{equation}
\psi_k(x) = \frac{1}{\sqrt{L}}e^{ikx}
\end{equation}

The dispersion relation transforms from On the lattice:
\begin{equation}
E_m = -2t\cos(k_m a)
\end{equation}

For small $k_m a$ (long wavelengths compared to lattice spacing):
\begin{align}
E_m &= -2t\cos(k_m a) \\
&\approx -2t\left(1 - \frac{(k_m a)^2}{2}\right) \quad \text{for } k_m a \ll 1\\
&= -2t + \frac{\hbar^2 k_m^2}{2m}
\end{align}

Dropping the constant energy shift, we recover the free particle dispersion:
\begin{equation}
E = \frac{\hbar^2 k^2}{2m} = \frac{p^2}{2m}
\end{equation}

The quadratic free-particle dispersion emerges from the lattice cosine dispersion at long wavelengths.
\end{example}

\subsection{The Position Representation of Quantum Mechanics}

In the position representation, quantum mechanical operators become differential operators acting on wave functions.

\begin{tcolorbox}[enhanced, title={Operator Dictionary: Discrete to Continuous}, colback=blue!5, colframe=blue!50!black]
\begin{center}
\begin{tabular}{|l|l|l|}
\hline
\textbf{Physical Quantity} & \textbf{Discrete (Lattice)} & \textbf{Continuous (Position)} \\
\hline
Position & $\hat{x} = \sum_j x_j \ket{x_j}\bra{x_j}$ & $\hat{x} = x$ (multiplication) \\
\hline
Translation by $a$ & $\hat{T}_a = \sum_j \ket{x_{j+1}}\bra{x_j}$ & $\hat{T}_a = e^{-i\hat{p}a/\hbar}$ \\
\hline
Momentum & $\hat{p} \approx -i\hbar \frac{\hat{T}_a - \hat{T}_{-a}}{2a}$ & $\hat{p} = -i\hbar\frac{d}{dx}$ \\
\hline
Kinetic Energy & $\hat{T} = -t(\hat{T}_a + \hat{T}_{-a} - 2\hat{I})$ & $\hat{T} = -\frac{\hbar^2}{2m}\frac{d^2}{dx^2}$ \\
\hline
Number at site & $\hat{n}_j = \ket{x_j}\bra{x_j}$ & $\delta(x - x_j)$ \\
\hline
\end{tabular}
\end{center}
\end{tcolorbox}

The momentum operator emerges from considering infinitesimal translations:

\begin{example}[Deriving the Momentum Operator]
For a small translation by distance $\epsilon$:
\begin{equation}
\psi(x - \epsilon) = \psi(x) - \epsilon\frac{d\psi}{dx} + O(\epsilon^2)
\end{equation}

The translation operator is:
\begin{equation}
\hat{T}_\epsilon = 1 - \epsilon\frac{d}{dx} = 1 - \frac{i\epsilon}{\hbar}(-i\hbar\frac{d}{dx})
\end{equation}

Identifying the generator of translations as momentum:
\begin{equation}
\hat{T}_\epsilon = 1 - \frac{i\epsilon}{\hbar}\hat{p} \quad \Rightarrow \quad \hat{p} = -i\hbar\frac{d}{dx}
\end{equation}

For finite translations: $\hat{T}_a = e^{-i\hat{p}a/\hbar}$.
\end{example}

\subsection{Commutation Relations}

The fundamental commutation relation emerges naturally:

\clearpage
\begin{example}[Position-Momentum Commutator]
\begin{align}
[\hat{x}, \hat{p}]\psi(x) &= x(-i\hbar\frac{d\psi}{dx}) - (-i\hbar\frac{d}{dx})(x\psi)\\
&= -i\hbar x\frac{d\psi}{dx} - (-i\hbar)\left(\psi + x\frac{d\psi}{dx}\right)\\
&= -i\hbar x\frac{d\psi}{dx} + i\hbar\psi + i\hbar x\frac{d\psi}{dx}\\
&= i\hbar\psi
\end{align}

Therefore: $[\hat{x}, \hat{p}] = i\hbar$.

This can be verified on the lattice for large $n$:
\begin{equation}
[\hat{x}_j, \hat{p}_k]_{\text{lattice}} \approx i\hbar\delta_{jk} + O(a)
\end{equation}

\textbf{Verification on the lattice:} On the lattice, position acts as multiplication by $x_j = ja$, while momentum implements a finite difference:
\begin{align}
(\hat{x}\psi)_j &= x_j \psi_j\\
(\hat{p}\psi)_j &= -i\hbar \frac{\psi_{j+1} - \psi_{j-1}}{2a}
\end{align}

Computing the commutator:
\begin{align}
[\hat{x}, \hat{p}]\psi_j &= x_j \cdot \left(-i\hbar \frac{\psi_{j+1} - \psi_{j-1}}{2a}\right) - \hat{p}(x_j\psi_j)
\end{align}

For the second term, noting that $x_{j\pm1} = x_j \pm a$:
\begin{align}
\hat{p}(x\psi)_j &= -i\hbar \frac{x_{j+1}\psi_{j+1} - x_{j-1}\psi_{j-1}}{2a}\\
&= -i\hbar \left[\frac{x_j(\psi_{j+1} - \psi_{j-1})}{2a} + \frac{\psi_{j+1} + \psi_{j-1}}{2}\right]
\end{align}

Therefore:
\begin{align}
[\hat{x}, \hat{p}]\psi_j &= i\hbar \frac{\psi_{j+1} + \psi_{j-1}}{2}
\end{align}

Using Taylor expansion $\psi_{j\pm1} = \psi_j + O(a)$, we get $[\hat{x}, \hat{p}] = i\hbar \cdot \mathbb{I} + O(a^2)$ on the lattice, confirming that the canonical commutation relation emerges exactly in the continuum limit.
\end{example}

\subsection{The Baker-Campbell-Hausdorff Formula}

The commutation relation $[\hat{x}, \hat{p}] = i\hbar$ has profound consequences that extend far beyond the uncertainty principle. Before proceeding to multiple representations and specific examples, we introduce a powerful mathematical tool that will be essential throughout our study of continuous quantum systems.

\subsubsection{Physical Intuition: Commutators as Derivatives}

The Baker-Campbell-Hausdorff (BCH) formula describes how operators transform under conjugation by exponentials. This is deeply connected to time evolution in quantum mechanics.

Consider the Heisenberg picture, where operators evolve in time while states remain constant. The key equation is:
\begin{equation}
\frac{d\hat{B}}{dt} = \frac{i}{\hbar}[\hat{H}, \hat{B}]
\end{equation}

The commutator $[\hat{H}, \cdot]$ acts like a derivative operator in the space of operators!

\subsubsection{Formal Solution: From Schr\"odinger to Heisenberg Picture}

We know from the Schr\"odinger picture that states evolve as:
\begin{equation}
\ket{\psi(t)} = e^{-i\hat{H}t/\hbar}\ket{\psi(0)}
\end{equation}

Expectation values must be independent of which picture we use. In the Schr\"odinger picture:
\begin{equation}
\langle \hat{B} \rangle = \bra{\psi(t)}\hat{B}\ket{\psi(t)} = \bra{\psi(0)}e^{i\hat{H}t/\hbar}\hat{B}e^{-i\hat{H}t/\hbar}\ket{\psi(0)}
\end{equation}

This suggests defining a time-evolved operator in the Heisenberg picture:
\begin{equation}
\boxed{\hat{B}(t) = e^{i\hat{H}t/\hbar}\hat{B}(0)e^{-i\hat{H}t/\hbar}}
\end{equation}

This is the formal solution to the operator equation of motion.

\textbf{Verification:} Let us verify that this exponential form satisfies $\frac{d\hat{B}}{dt} = \frac{i}{\hbar}[\hat{H}, \hat{B}]$.

Taking the time derivative using the product rule:
\begin{align}
\frac{d\hat{B}}{dt} &= \frac{d}{dt}\left(e^{i\hat{H}t/\hbar}\hat{B}(0)e^{-i\hat{H}t/\hbar}\right)\nonumber\\
&= \frac{i\hat{H}}{\hbar}e^{i\hat{H}t/\hbar}\hat{B}(0)e^{-i\hat{H}t/\hbar} + e^{i\hat{H}t/\hbar}\hat{B}(0)\left(-\frac{i\hat{H}}{\hbar}\right)e^{-i\hat{H}t/\hbar}\nonumber\\
&= \frac{i}{\hbar}\hat{H}\hat{B}(t) - \frac{i}{\hbar}\hat{B}(t)\hat{H}\nonumber\\
&= \frac{i}{\hbar}[\hat{H}, \hat{B}(t)]
\end{align}

The exponential form indeed satisfies the operator equation of motion. This formal solution is the foundation of the BCH formula.

\subsubsection{Taylor Series Expansion with Nested Commutators}

We can expand the formal solution as a Taylor series. Just as for ordinary functions:
\begin{equation}
f(t) = f(0) + t f'(0) + \frac{t^2}{2!}f''(0) + \frac{t^3}{3!}f'''(0) + \cdots
\end{equation}

we have for operators:
\begin{equation}
\hat{B}(t) = \hat{B}(0) + t\frac{d\hat{B}}{dt}\bigg|_{t=0} + \frac{t^2}{2!}\frac{d^2\hat{B}}{dt^2}\bigg|_{t=0} + \frac{t^3}{3!}\frac{d^3\hat{B}}{dt^3}\bigg|_{t=0} + \cdots
\end{equation}

Now substitute the expressions for the time derivatives:
\begin{align}
\frac{d\hat{B}}{dt} &= \frac{i}{\hbar}[\hat{H}, \hat{B}]\nonumber\\
\frac{d^2\hat{B}}{dt^2} &= \frac{i}{\hbar}[\hat{H}, \frac{d\hat{B}}{dt}] = \left(\frac{i}{\hbar}\right)^2[\hat{H}, [\hat{H}, \hat{B}]]\nonumber\\
\frac{d^3\hat{B}}{dt^3} &= \left(\frac{i}{\hbar}\right)^3[\hat{H}, [\hat{H}, [\hat{H}, \hat{B}]]]
\end{align}

This gives:
\begin{equation}
\hat{B}(t) = \hat{B} + \frac{it}{\hbar}[\hat{H}, \hat{B}] + \frac{(it/\hbar)^2}{2!}[\hat{H}, [\hat{H}, \hat{B}]] + \frac{(it/\hbar)^3}{3!}[\hat{H}, [\hat{H}, [\hat{H}, \hat{B}]]] + \cdots
\end{equation}

\subsubsection{Connection to the General BCH Formula}

Setting $\hat{A} = i\hat{H}t/\hbar$, the series becomes:
\begin{equation}
\hat{B}(t) = \hat{B} + [\hat{A}, \hat{B}] + \frac{1}{2!}[\hat{A}, [\hat{A}, \hat{B}]] + \frac{1}{3!}[\hat{A}, [\hat{A}, [\hat{A}, \hat{B}]]] + \cdots
\end{equation}

But we also showed that $\hat{B}(t) = e^{\hat{A}}\hat{B}e^{-\hat{A}}$. Therefore, we have discovered:
\begin{equation}
e^{\hat{A}}\hat{B}e^{-\hat{A}} = \hat{B} + [\hat{A}, \hat{B}] + \frac{1}{2!}[\hat{A}, [\hat{A}, \hat{B}]] + \frac{1}{3!}[\hat{A}, [\hat{A}, [\hat{A}, \hat{B}]]] + \cdots
\end{equation}

This is the BCH conjugation formula! The operator $e^{\hat{A}}$ "flows" $\hat{B}$ along the direction generated by repeated commutators with $\hat{A}$. Each term involves one more nested commutator, exactly as each term in a Taylor series involves one more derivative.

\subsubsection{The General Conjugation Formula}

More generally, for any operators $\hat{A}$ and $\hat{B}$:
\begin{equation}
\boxed{e^{\hat{A}}\hat{B}e^{-\hat{A}} = \hat{B} + [\hat{A},\hat{B}] + \frac{1}{2!}[\hat{A},[\hat{A},\hat{B}]] + \frac{1}{3!}[\hat{A},[\hat{A},[\hat{A},\hat{B}]]] + \cdots}
\end{equation}

Think of this as: the operator $e^{\hat{A}}$ "flows" $\hat{B}$ along the direction generated by repeated commutators with $\hat{A}$. Each term involves one more nested commutator, just as each term in a Taylor series involves one more derivative. The commutator $[\hat{A}, \cdot]$ is the "derivative" in operator space.

\begin{example}[BCH for Position Under Momentum Translation]
Consider translating position by momentum: $\hat{A} = -i\epsilon\hat{p}/\hbar$, $\hat{B} = \hat{x}$.

The first commutator:
\begin{equation}
[\hat{A}, \hat{B}] = [-i\epsilon\hat{p}/\hbar, \hat{x}] = -i\epsilon[\hat{p}, \hat{x}]/\hbar = -i\epsilon(-i\hbar)/\hbar = -\epsilon
\end{equation}

This is a c-number! Therefore all higher commutators vanish:
\begin{equation}
[\hat{A}, [\hat{A}, \hat{B}]] = [\hat{A}, -\epsilon] = 0
\end{equation}

The BCH series terminates:
\begin{equation}
e^{-i\epsilon\hat{p}/\hbar}\hat{x}e^{i\epsilon\hat{p}/\hbar} = \hat{x} - \epsilon
\end{equation}

Momentum translation shifts position by $\epsilon$, exactly as expected classically!
\end{example}

\subsubsection{The Key Simplification: C-Number Commutators}

When $[\hat{A},\hat{B}]$ is a c-number (a multiple of the identity), all higher commutators automatically vanish:
\begin{equation}
[\hat{A},[\hat{A},\hat{B}]] = [\hat{A}, c\hat{I}] = c[\hat{A}, \hat{I}] = 0
\end{equation}

The infinite series becomes just two terms:
\begin{equation}
\boxed{e^{\hat{A}}\hat{B}e^{-\hat{A}} = \hat{B} + [\hat{A},\hat{B}]} \quad \text{(when $[\hat{A},\hat{B}]$ is a c-number)}
\end{equation}

This dramatic simplification is why position and momentum operators have such simple transformation properties under displacement: their commutator $[\hat{x}, \hat{p}] = i\hbar$ is a c-number. We will exploit this repeatedly when constructing coherent states.

\subsubsection{The Exponential Product Formula}

A related BCH result concerns products of exponentials. For two operators $\hat{A}$ and $\hat{B}$ that don't commute:
\begin{equation}
e^{\hat{A} + \hat{B}} \neq e^{\hat{A}} e^{\hat{B}}
\end{equation}

The general BCH formula for exponential products is an infinite series involving nested commutators. However, when $[\hat{A}, \hat{B}]$ commutes with both $\hat{A}$ and $\hat{B}$ (automatically satisfied if $[\hat{A}, \hat{B}]$ is a c-number), the formula simplifies to:
\begin{equation}
\boxed{e^{\hat{A} + \hat{B}} = e^{\hat{A}} e^{\hat{B}} e^{-[\hat{A}, \hat{B}]/2}}
\end{equation}

This form is particularly useful when constructing coherent states of the harmonic oscillator.

\begin{example}[BCH for Harmonic Oscillator Displacement Operator]
For the harmonic oscillator displacement operator $\hat{D}(\alpha) = e^{\alpha\hat{a}^\dagger - \alpha^*\hat{a}}$, we have $\hat{A} = \alpha\hat{a}^\dagger$ and $\hat{B} = -\alpha^*\hat{a}$.

The commutator is:
\begin{equation}
[\hat{A}, \hat{B}] = [\alpha\hat{a}^\dagger, -\alpha^*\hat{a}] = -\alpha\alpha^*[\hat{a}^\dagger, \hat{a}] = |\alpha|^2
\end{equation}

Since this is a c-number, we can apply the exponential product formula:
\begin{equation}
e^{\alpha\hat{a}^\dagger - \alpha^*\hat{a}} = e^{\alpha\hat{a}^\dagger} e^{-\alpha^*\hat{a}} e^{-|\alpha|^2/2} = e^{-|\alpha|^2/2} e^{\alpha\hat{a}^\dagger} e^{-\alpha^*\hat{a}}
\end{equation}

This factorization is essential for constructing and analyzing coherent states.
\end{example}

\subsubsection{Summary: Why BCH Matters}

The Baker-Campbell-Hausdorff formula is not just a mathematical curiosity---it reveals the fundamental algebraic structure of quantum mechanics:

\begin{itemize}
\item \textbf{Commutators are derivatives:} The commutator $[\hat{A}, \cdot]$ acts like a directional derivative in operator space
\item \textbf{Exponentials generate flows:} $e^{\hat{A}}$ generates a "flow" along the direction defined by $\hat{A}$
\item \textbf{C-numbers simplify everything:} When commutators yield c-numbers, infinite series truncate to simple expressions
\item \textbf{Classical-quantum connection:} BCH shows how quantum operators reduce to classical transformations in appropriate limits
\end{itemize}

We will use BCH extensively in constructing coherent states, analyzing displacement operators, and understanding squeezed states. The formula provides both computational power and conceptual insight into the structure of quantum theory.

\subsection{Connections Between Discrete and Continuous}

\subsubsection{Finite Differences and Differential Operators}

The continuum limit can be reversed to recover discrete approximations:

\begin{example}[Finite Difference Approximations]
Second derivative:
\begin{equation}
\frac{d^2\psi}{dx^2} \approx \frac{\psi(x+a) - 2\psi(x) + \psi(x-a)}{a^2} + O(a^2)
\end{equation}

First derivative (symmetric):
\begin{equation}
\frac{d\psi}{dx} \approx \frac{\psi(x+a) - \psi(x-a)}{2a} + O(a^2)
\end{equation}

The discrete Schr\"odinger equation on a lattice:
\begin{equation}
-\frac{\hbar^2}{2ma^2}(\psi_{j+1} - 2\psi_j + \psi_{j-1}) + V_j\psi_j = E\psi_j
\end{equation}

This can be written as a matrix eigenvalue problem:
\begin{equation}
\begin{pmatrix}
V_1 + 2t & -t & 0 & \cdots \\
-t & V_2 + 2t & -t & \cdots \\
0 & -t & V_3 + 2t & \cdots \\
\vdots & \vdots & \vdots & \ddots
\end{pmatrix}
\begin{pmatrix}
\psi_1 \\
\psi_2 \\
\psi_3 \\
\vdots
\end{pmatrix}
= E
\begin{pmatrix}
\psi_1 \\
\psi_2 \\
\psi_3 \\
\vdots
\end{pmatrix}
\end{equation}

where $t = \hbar^2/(2ma^2)$.
\end{example}

\subsubsection{The Deep Connection}

The discrete-continuous relationship:

\begin{physicalinsight}
The discrete and continuous formulations represent complementary perspectives, neither more fundamental than the other. Each approach has its natural domain of application. Some quantum gravity theories propose discrete spacetime at the Planck scale, while others maintain continuity down to arbitrarily small distances. Condensed matter systems naturally exhibit both discrete descriptions at the atomic level and continuous effective field theories at larger scales. Mathematical physics moves fluidly between discrete algebraic methods and continuous analytic techniques, choosing the approach best suited to each problem.

The practical implications are profound. Lattice models can capture essential physics with finite computational resources, making them ideal for numerical simulations. Continuous formulations provide analytical insights and exact solutions that reveal general principles. Some phenomena like band structure and Wannier states in crystals are naturally discrete, while others like scattering theory and field propagation are naturally continuous. Modern quantum simulation exploits this duality, using discrete quantum systems to model both discrete and continuous physics.

The conceptual unity between these pictures is striking. Wave functions and discrete amplitudes transform into each other via the continuum limit or discretization. Differential operators and difference operators are mutually derivable through systematic expansion procedures. Boundary conditions in the continuum correspond to edge effects on the lattice. Analytical solutions complement numerical methods, each illuminating aspects the other cannot easily capture. This deep connection between discrete and continuous descriptions lies at the heart of modern quantum theory.
\end{physicalinsight}

\section{Examples of Continuous Hilbert Spaces}

\subsection{Multiple Representations}

\subsubsection{The Power of Multiple Perspectives}

Quantum systems can be described in multiple representations. Position and momentum space, related by the Fourier transform, are the most fundamental. Problems intractable in one representation often become simple in the other.

\subsubsection{Position and Momentum Space}

Different representations offer complementary insights:

\begin{tcolorbox}[enhanced, title={Position vs Momentum Representation}, colback=blue!5, colframe=blue!50!black]
\begin{center}
\begin{tabular}{|l|l|l|}
\hline
\textbf{Property} & \textbf{Position Space} & \textbf{Momentum Space} \\
\hline
State & $\psi(x)$ & $\tilde{\psi}(p)$ \\
\hline
Position operator & $\hat{x} = x$ & $\hat{x} = i\hbar\frac{d}{dp}$ \\
\hline
Momentum operator & $\hat{p} = -i\hbar\frac{d}{dx}$ & $\hat{p} = p$ \\
\hline
Kinetic energy & $\hat{T} = -\frac{\hbar^2}{2m}\frac{d^2}{dx^2}$ & $\hat{T} = \frac{p^2}{2m}$ \\
\hline
Free particle & Differential equation & Algebraic equation \\
\hline
Harmonic oscillator & Differential equation & Differential equation \\
\hline
\end{tabular}
\end{center}
\end{tcolorbox}

The relationship between representations is given by the Fourier transform:

\begin{tcolorbox}[enhanced, title={Master Reference: Wave Function Transformations}, colback=red!5, colframe=red!50!black, breakable]
\textbf{Fundamental Transformations:}

\textbf{Position to Momentum:}
\begin{equation}
\tilde{\psi}(p) = \frac{1}{\sqrt{2\pi\hbar}} \int_{-\infty}^{\infty} dx \, e^{-ipx/\hbar} \psi(x)
\end{equation}

\textbf{Momentum to Position:}
\begin{equation}
\psi(x) = \frac{1}{\sqrt{2\pi\hbar}} \int_{-\infty}^{\infty} dp \, e^{ipx/\hbar} \tilde{\psi}(p)
\end{equation}

\textbf{Key States in Both Representations:}
\begin{center}
\begin{tabular}{|l|l|l|}
\hline
\textbf{State} & \textbf{Position Space} & \textbf{Momentum Space} \\
\hline
Position eigenstate & $\delta(x-x_0)$ & $\frac{1}{\sqrt{2\pi\hbar}}e^{-ipx_0/\hbar}$ \\
\hline
Momentum eigenstate & $\frac{1}{\sqrt{2\pi\hbar}}e^{ip_0x/\hbar}$ & $\delta(p-p_0)$ \\
\hline
Gaussian & $(\frac{1}{2\pi\sigma_x^2})^{1/4}e^{-x^2/(4\sigma_x^2)}$ & $(\frac{2\sigma_x^2}{\pi\hbar^2})^{1/4}e^{-\sigma_x^2p^2/\hbar^2}$ \\
\hline
Square well ground & $\sqrt{\frac{2}{L}}\sin(\frac{\pi x}{L})$ & $\frac{\sqrt{2L}\pi}{p^2L^2-\pi^2\hbar^2}[(-1)^ne^{ipL/\hbar}-1]$ \\
\hline
Harmonic ground & $(\frac{1}{\pi x_0^2})^{1/4}e^{-x^2/(2x_0^2)}$ & $(\frac{1}{\pi p_0^2})^{1/4}e^{-p^2/(2p_0^2)}$ \\
\hline
\end{tabular}
\end{center}

\textbf{Operator Actions:}
\begin{center}
\begin{tabular}{|l|l|l|}
\hline
\textbf{Operator} & \textbf{Position Space} & \textbf{Momentum Space} \\
\hline
$\hat{x}$ & $x$ (multiply) & $i\hbar\frac{d}{dp}$ \\
\hline
$\hat{p}$ & $-i\hbar\frac{d}{dx}$ & $p$ (multiply) \\
\hline
$\hat{x}^2$ & $x^2$ & $-\hbar^2\frac{d^2}{dp^2} + i\hbar\frac{d}{dp}$ \\
\hline
$\hat{p}^2$ & $-\hbar^2\frac{d^2}{dx^2}$ & $p^2$ \\
\hline
$\hat{T} = \frac{\hat{p}^2}{2m}$ & $-\frac{\hbar^2}{2m}\frac{d^2}{dx^2}$ & $\frac{p^2}{2m}$ \\
\hline
$V(\hat{x})$ & $V(x)$ & $V(i\hbar\frac{d}{dp})$ \\
\hline
\end{tabular}
\end{center}
\end{tcolorbox}

\subsubsection{The Uncertainty Principle Revisited}

The Fourier transform relationship provides deep insight into the uncertainty principle:

\begin{example}[Gaussian Wave Packet Analysis]
Position space Gaussian:
\begin{equation}
\psi(x) = \left(\frac{1}{2\pi\sigma_x^2}\right)^{1/4} e^{-x^2/(4\sigma_x^2)}
\end{equation}

Fourier transform gives momentum space:
\begin{equation}
\tilde{\psi}(p) = \left(\frac{2\sigma_x^2}{\pi\hbar^2}\right)^{1/4} e^{-\sigma_x^2p^2/\hbar^2}
\end{equation}

Widths:
\begin{align}
\Delta x &= \sqrt{\langle x^2 \rangle} = \sigma_x\\
\Delta p &= \sqrt{\langle p^2 \rangle} = \frac{\hbar}{2\sigma_x}
\end{align}

Uncertainty product:
\begin{equation}
\Delta x \Delta p = \sigma_x \cdot \frac{\hbar}{2\sigma_x} = \frac{\hbar}{2}
\end{equation}

The Gaussian saturates the uncertainty bound. For other wave functions:
\begin{itemize}
\item Narrower in $x$ $\rightarrow$ Broader in $p$
\item Cannot be arbitrarily narrow in both
\item $\Delta x \Delta p \geq \hbar/2$ always
\end{itemize}
\end{example}

\subsubsection{Basis Transformations}

Different representations are connected by basis transformations:

\begin{example}[Energy Representation]
For a system with energy eigenstates $\{\psi_n(x)\}$ and eigenvalues $\{E_n\}$:

Position representation:
\begin{equation}
\psi(x,t) = \sum_n c_n e^{-iE_nt/\hbar} \psi_n(x)
\end{equation}

Energy representation:
\begin{equation}
c_n(t) = c_n(0) e^{-iE_nt/\hbar}
\end{equation}

The transformation:
\begin{align}
c_n &= \int dx \, \psi_n^*(x) \psi(x,0) \quad \text{(analysis)}\\
\psi(x,0) &= \sum_n c_n \psi_n(x) \quad \text{(synthesis)}
\end{align}

In energy representation, the Hamiltonian becomes diagonal with matrix elements $H_{nm} = E_n\delta_{nm}$. This diagonality makes time evolution trivial, reducing to simple phase factors for each energy eigenstate. However, the position operator becomes complicated in this basis, with matrix elements $x_{nm} = \int \psi_n^* x \psi_m dx$ that mix different energy levels.
\end{example}

\subsection{The Time-Independent Schr\"odinger Equation}

\subsubsection{Energy Eigenstates in Position Space}

For a particle in potential $V(x)$:
\begin{equation}
\hat{H} = \frac{\hat{p}^2}{2m} + V(\hat{x}) = -\frac{\hbar^2}{2m}\frac{d^2}{dx^2} + V(x)
\end{equation}

Energy eigenstates satisfy $\hat{H}\ket{\psi} = E\ket{\psi}$, which in position representation becomes:
\begin{equation}
-\frac{\hbar^2}{2m}\frac{d^2\psi(x)}{dx^2} + V(x)\psi(x) = E\psi(x)
\end{equation}

This can be rewritten as:

\begin{example}[Local de Broglie Wavelength]
Rearranging the Schr\"odinger equation:
\begin{equation}
\frac{d^2\psi}{dx^2} = -\frac{2m}{\hbar^2}[E - V(x)]\psi = -k^2(x)\psi
\end{equation}

where $k(x) = \sqrt{2m[E - V(x)]}/\hbar$ is the local wave number.

In classically allowed regions where $E > V(x)$, the wave number $k(x)$ is real and solutions oscillate with local wavelength $\lambda(x) = 2\pi/k(x)$. The wavelength decreases as the kinetic energy increases, reflecting the particle's higher local momentum. In classically forbidden regions where $E < V(x)$, the wave number becomes imaginary: $k(x) = i\kappa(x)$ where $\kappa(x) = \sqrt{2m[V(x) - E]}/\hbar$. Here solutions decay exponentially as $\psi \sim e^{-\kappa x}$ with a characteristic penetration depth of order $1/\kappa$.
\end{example}

\subsubsection{General Properties of Solutions}

General constraints on wave functions:

\begin{example}[Node Theorem]
For bound states in one dimension with a single-well potential, there is a direct relationship between the excitation level and the number of nodes. The ground state ($n=0$) has no nodes, the first excited state ($n=1$) has one node, and in general the $n$-th excited state has exactly $n$ nodes.

The reasoning is simple: each node represents a change of sign, and between nodes the wave function maintains definite sign. The kinetic energy $\int |\psi'|^2 dx$ increases with the number of sign changes because more oscillations require larger derivatives. This increased kinetic energy leads to higher total energy. The node theorem thus provides a way to identify excited states without solving the full equation---simply count the zeros of the wave function.
\end{example}

\subsection{The Free Particle}

The free particle ($V(x) = 0$) illustrates continuous spectrum states, the energy-momentum relationship, boundary conditions, and wave packets.

The free particle Schr\"odinger equation:
\begin{equation}
-\frac{\hbar^2}{2m}\frac{d^2\psi}{dx^2} = E\psi
\end{equation}

\begin{example}[Free Particle Solutions in Position Space]
Setting $k = \sqrt{2mE}/\hbar$, the equation becomes:
\begin{equation}
\frac{d^2\psi}{dx^2} + k^2\psi = 0
\end{equation}

This has the general solution:
\begin{equation}
\psi(x) = Ae^{ikx} + Be^{-ikx}
\end{equation}

Alternatively, using real functions:
\begin{equation}
\psi(x) = C\cos(kx) + D\sin(kx)
\end{equation}

The complex exponential form has clear physical interpretation. The term $e^{ikx}$ represents a right-moving wave with momentum $p = \hbar k$, while $e^{-ikx}$ represents a left-moving wave with momentum $p = -\hbar k$. The energy-momentum relation is $E = p^2/(2m) = \hbar^2k^2/(2m)$, and the de Broglie wavelength is $\lambda = 2\pi/k = h/p$.

\textbf{The normalization problem:} These plane wave states extend over all space with constant amplitude, making the integral $\int |\psi|^2 dx$ diverge. This reflects the unphysical nature of states with perfectly defined momentum---they are completely delocalized in position.
\end{example}

\begin{example}[Free Particle in Momentum Space]
In momentum space, the free particle problem becomes trivial. The Hamiltonian is:
\begin{equation}
\hat{H} = \frac{\hat{p}^2}{2m} = \frac{p^2}{2m}
\end{equation}

This is already diagonal! The momentum eigenstates are:
\begin{equation}
\tilde{\psi}_p(p') = \delta(p' - p)
\end{equation}

with eigenvalues:
\begin{equation}
E_p = \frac{p^2}{2m}
\end{equation}

Choice of representation affects problem difficulty.

\textbf{Connection to position space:} The position space wave function is obtained by inverse Fourier transform:
\begin{align}
\psi_p(x) &= \frac{1}{\sqrt{2\pi\hbar}} \int_{-\infty}^{\infty} dp' \, e^{ip'x/\hbar} \delta(p' - p)\\
&= \frac{1}{\sqrt{2\pi\hbar}}e^{ipx/\hbar}
\end{align}

This recovers our plane wave solution.
\end{example}

Free particle states can be normalized in different ways:

\begin{example}[Box Normalization vs Delta Normalization]
\textbf{Box normalization} (periodic boundary conditions on $[0,L]$):
\begin{equation}
\psi_k(x) = \frac{1}{\sqrt{L}}e^{ikx}, \quad k = \frac{2\pi n}{L}, \quad n \in \mathbb{Z}
\end{equation}

\textbf{Delta function normalization} (continuum):
\begin{equation}
\psi_k(x) = \frac{1}{\sqrt{2\pi}}e^{ikx}, \quad \int_{-\infty}^{\infty} \psi_k^*(x)\psi_{k'}(x)dx = \delta(k-k')
\end{equation}

The box normalization approaches delta normalization as $L \to \infty$.
\end{example}

\subsubsection{Wave Packet Dynamics}

Physical free particle states are wave packets:

\begin{example}[Wave Packets and Physical States]
Physical free particle states are wave packets---superpositions of different momentum components:
\begin{equation}
\psi(x,t) = \frac{1}{\sqrt{2\pi\hbar}} \int_{-\infty}^{\infty} dp \, A(p) e^{i(px - E_p t)/\hbar}
\end{equation}

where $A(p)$ is the momentum amplitude distribution and $E_p = p^2/(2m)$.

For a Gaussian momentum distribution centered at $p_0$ with width $\sigma_p$:
\begin{equation}
A(p) = \left(\frac{2\pi\sigma_p^2}{\pi}\right)^{1/4} \exp\left[-\frac{(p-p_0)^2}{2\sigma_p^2}\right]
\end{equation}

The resulting position space wave packet at $t = 0$ is:
\begin{equation}
\psi(x,0) = \left(\frac{2\sigma_p^2}{\pi\hbar^2}\right)^{1/4} \exp\left[-\frac{\sigma_p^2 x^2}{\hbar^2} + \frac{ip_0x}{\hbar}\right]
\end{equation}

This is a Gaussian with width $\sigma_x = \hbar/(2\sigma_p)$, demonstrating the uncertainty relation $\sigma_x \sigma_p = \hbar/2$.
\end{example}

\begin{example}[Gaussian Wave Packet Evolution]
Initial Gaussian centered at $x_0$ with average momentum $p_0$:
\begin{equation}
\psi(x,0) = \left(\frac{1}{2\pi\sigma_0^2}\right)^{1/4} \exp\left[-\frac{(x-x_0)^2}{4\sigma_0^2} + \frac{ip_0x}{\hbar}\right]
\end{equation}

For a free particle, the time-evolved state is:
\begin{equation}
\psi(x,t) = \left(\frac{1}{2\pi\sigma^2(t)}\right)^{1/4} \exp\left[-\frac{(x-x_0-v_0t)^2}{4\sigma^2(t)} + i\phi(x,t)\right]
\end{equation}

where:
\begin{align}
v_0 &= \frac{p_0}{m} \quad \text{(group velocity)}\\
\sigma^2(t) &= \sigma_0^2\left(1 + \frac{\hbar^2t^2}{4m^2\sigma_0^4}\right) \quad \text{(spreading)}\\
\phi(x,t) &= \frac{p_0}{\hbar}\left(x - \frac{p_0t}{2m}\right) + \text{other terms}
\end{align}

Key features:
\begin{itemize}
\item Center moves classically: $\langle x \rangle = x_0 + v_0t$
\item Width increases: quantum spreading
\item Spreading rate: $\frac{d\sigma}{dt} \sim \frac{\hbar}{2m\sigma_0}$ for $t \gg \frac{2m\sigma_0^2}{\hbar}$
\item Narrower initial packet spreads faster
\end{itemize}
\end{example}

\subsubsection{Ehrenfest's Theorem}

Expectation values follow classical laws:

\begin{example}[Expectation Values Follow Classical Laws]
For any quantum state:
\begin{align}
\frac{d\langle x \rangle}{dt} &= \frac{\langle p \rangle}{m}\\
\frac{d\langle p \rangle}{dt} &= -\left\langle \frac{dV}{dx} \right\rangle
\end{align}

For a harmonic oscillator where $V = \frac{1}{2}m\omega^2x^2$:
\begin{equation}
\frac{d^2\langle x \rangle}{dt^2} = -\omega^2\langle x \rangle
\end{equation}

The expectation value oscillates exactly like a classical oscillator!

However, for general potentials:
\begin{equation}
\left\langle \frac{dV}{dx} \right\rangle \neq \frac{dV}{dx}\bigg|_{\langle x \rangle}
\end{equation}

The difference depends on the wave packet width and potential nonlinearity.
\end{example}

\subsection{The Infinite Square Well}

The infinite square well demonstrates bound states and energy quantization.

The potential is defined as:
\begin{equation}
V(x) = \begin{cases}
0 & \text{for } 0 < x < L \\
\infty & \text{otherwise}
\end{cases}
\end{equation}

\clearpage
\begin{example}[Solving the Infinite Square Well in Position Space]
Inside the well ($0 < x < L$), the Schr\"odinger equation is:
\begin{equation}
-\frac{\hbar^2}{2m}\frac{d^2\psi}{dx^2} = E\psi
\end{equation}

This has the general solution:
\begin{equation}
\psi(x) = A\sin(kx) + B\cos(kx)
\end{equation}
where $k = \sqrt{2mE}/\hbar$.

Boundary conditions require the wave function to vanish at the walls:
\begin{align}
\psi(0) = 0 &\Rightarrow B = 0\\
\psi(L) = 0 &\Rightarrow A\sin(kL) = 0
\end{align}

The condition $\sin(kL) = 0$ requires $kL = n\pi$:
\begin{align}
k_n &= \frac{n\pi}{L}\\
E_n &= \frac{\hbar^2k_n^2}{2m} = \frac{n^2\pi^2\hbar^2}{2mL^2}\\
\psi_n(x) &= \sqrt{\frac{2}{L}}\sin\left(\frac{n\pi x}{L}\right)
\end{align}

The normalization factor $\sqrt{2/L}$ ensures $\int_0^L |\psi_n|^2 dx = 1$.

\textbf{Key physical features:}

The energy levels scale as $n^2$, reflecting the kinetic energy of a wave with $n$ half-wavelengths fitting in the box. The ground state energy $E_1 = \pi^2\hbar^2/(2mL^2)$ is non-zero, demonstrating zero-point energy---a purely quantum effect with no classical analog. The level spacing $E_{n+1} - E_n = (2n+1)\pi^2\hbar^2/(2mL^2)$ increases linearly with $n$, unlike the harmonic oscillator where spacing is constant. Each eigenfunction has exactly $n-1$ nodes inside the well, consistent with the node theorem.
\end{example}
\clearpage

\begin{example}[Orthogonality and Completeness]
The infinite well eigenfunctions form a complete orthonormal set on $[0, L]$:

\textbf{Orthogonality:}
\begin{align}
\int_0^L \psi_n^*(x)\psi_m(x)dx &= \frac{2}{L}\int_0^L \sin\left(\frac{n\pi x}{L}\right)\sin\left(\frac{m\pi x}{L}\right)dx\\
&= \frac{2}{L} \cdot \frac{L}{2}\delta_{nm} = \delta_{nm}
\end{align}

\textbf{Completeness:}
Any function $f(x)$ defined on $[0, L]$ with $f(0) = f(L) = 0$ can be expanded as:
\begin{equation}
f(x) = \sum_{n=1}^{\infty} c_n \psi_n(x)
\end{equation}

where the coefficients are:
\begin{equation}
c_n = \int_0^L \psi_n^*(x)f(x)dx = \sqrt{\frac{2}{L}}\int_0^L \sin\left(\frac{n\pi x}{L}\right)f(x)dx
\end{equation}

This is precisely the Fourier sine series, showing the deep connection between quantum mechanics and Fourier analysis.
\end{example}

Connection to the discrete lattice:

\begin{example}[Lattice Approximation of the Infinite Well]
For an $N$-site lattice with hard-wall boundaries:
\begin{equation}
\psi_n^{\text{lattice}}(j) = \sqrt{\frac{2}{N+1}}\sin\left(\frac{n\pi j}{N+1}\right)
\end{equation}

for $j = 1, 2, ..., N$ and $n = 1, 2, ..., N$.

As $N \to \infty$ with $L = Na$ fixed, the discrete variables transform continuously: the discrete ratio $j/N$ becomes the continuous variable $x/L$, the lattice wave function $\psi_n^{\text{lattice}}(j)$ converges to the continuum wave function $\psi_n(x)$, and the lowest $N$ modes of the lattice converge to the corresponding continuum results. This explicit construction demonstrates how the continuum emerges from the discrete.
\end{example}
\clearpage
\section{The Quantum Harmonic Oscillator}

The harmonic oscillator appears in molecular vibrations, electromagnetic field modes, phonons, and quantum field theory. It serves as a paradigm for understanding energy quantization, coherent states, and the classical-quantum transition.

\begin{equation}
V(x) = \frac{1}{2}m\omega^2x^2
\end{equation}

\subsection{The Operator Method: Creation and Annihilation Operators}

The harmonic oscillator Hamiltonian can be written as:
\begin{equation}
\hat{H} = \frac{\hat{p}^2}{2m} + \frac{1}{2}m\omega^2\hat{x}^2
\end{equation}

Define characteristic scales and ladder operators:
\begin{align}
x_0 &= \sqrt{\frac{\hbar}{m\omega}} \quad \text{(oscillator length scale)} \\
p_0 &= \sqrt{m\omega\hbar} \quad \text{(oscillator momentum scale)}
\end{align}

The position and momentum operators can be written as:
\begin{align}
\hat{x} &= x_0 \frac{\hat{a} + \hat{a}^\dagger}{\sqrt{2}} = \sqrt{\frac{\hbar}{2m\omega}}(\hat{a} + \hat{a}^\dagger) \\
\hat{p} &= p_0 \frac{\hat{a} - \hat{a}^\dagger}{i\sqrt{2}} = -i\sqrt{\frac{m\omega\hbar}{2}}(\hat{a} - \hat{a}^\dagger)
\end{align}

Inverting these relations gives the ladder operators:
\begin{align}
\hat{a} &= \frac{1}{\sqrt{2}}\left(\frac{\hat{x}}{x_0} + i\frac{\hat{p}}{p_0}\right) = \sqrt{\frac{m\omega}{2\hbar}}\hat{x} + \frac{i}{\sqrt{2m\omega\hbar}}\hat{p} \\
\hat{a}^\dagger &= \frac{1}{\sqrt{2}}\left(\frac{\hat{x}}{x_0} - i\frac{\hat{p}}{p_0}\right) = \sqrt{\frac{m\omega}{2\hbar}}\hat{x} - \frac{i}{\sqrt{2m\omega\hbar}}\hat{p}
\end{align}

\clearpage
\begin{example}[Properties of Ladder Operators]
The ladder operators satisfy the fundamental commutation relation:
\begin{equation}
[\hat{a}, \hat{a}^\dagger] = 1
\end{equation}

Using $[\hat{x}, \hat{p}] = i\hbar$:
\begin{align}
[\hat{a}, \hat{a}^\dagger] &= \frac{1}{2}\left[\frac{\hat{x}}{x_0} + i\frac{\hat{p}}{p_0}, \frac{\hat{x}}{x_0} - i\frac{\hat{p}}{p_0}\right] \\
&= \frac{1}{2}\left(-\frac{2i}{x_0 p_0}[\hat{x}, \hat{p}]\right) \\
&= \frac{1}{2}\left(-\frac{2i}{x_0 p_0} \cdot i\hbar\right) \\
&= \frac{\hbar}{x_0 p_0} = \frac{\hbar}{\sqrt{\hbar/(m\omega)} \cdot \sqrt{m\omega\hbar}} = 1
\end{align}

Expressing the Hamiltonian in ladder operators:
\begin{align}
\hat{x}^2 &= x_0^2 \frac{(\hat{a} + \hat{a}^\dagger)^2}{2} = \frac{\hbar}{2m\omega}(\hat{a}^2 + \hat{a}^\dagger 2 + \hat{a}\hat{a}^\dagger + \hat{a}^\dagger\hat{a}) \\
\hat{p}^2 &= p_0^2 \frac{(\hat{a} - \hat{a}^\dagger)^2}{-2} = \frac{m\omega\hbar}{2}(\hat{a}^\dagger\hat{a} + \hat{a}\hat{a}^\dagger - \hat{a}^2 - \hat{a}^\dagger 2)
\end{align}

Substituting into the Hamiltonian:
\begin{align}
\hat{H} &= \frac{1}{2m} \cdot \frac{m\omega\hbar}{2}(\hat{a}^\dagger\hat{a} + \hat{a}\hat{a}^\dagger - \hat{a}^2 - \hat{a}^\dagger 2) \\
&\quad + \frac{m\omega^2}{2} \cdot \frac{\hbar}{2m\omega}(\hat{a}^2 + \hat{a}^\dagger 2 + \hat{a}\hat{a}^\dagger + \hat{a}^\dagger\hat{a}) \\
&= \frac{\hbar\omega}{4}(\hat{a}^\dagger\hat{a} + \hat{a}\hat{a}^\dagger - \hat{a}^2 - \hat{a}^\dagger 2 + \hat{a}^2 + \hat{a}^\dagger 2 + \hat{a}\hat{a}^\dagger + \hat{a}^\dagger\hat{a}) \\
&= \frac{\hbar\omega}{4}(2\hat{a}^\dagger\hat{a} + 2\hat{a}\hat{a}^\dagger) \\
&= \frac{\hbar\omega}{2}(\hat{a}^\dagger\hat{a} + \hat{a}\hat{a}^\dagger)
\end{align}

Using $\hat{a}\hat{a}^\dagger = \hat{a}^\dagger\hat{a} + 1$, we get:
\begin{equation}
\hat{H} = \hbar\omega\left(\hat{a}^\dagger\hat{a} + \frac{1}{2}\right)
\end{equation}

Defining the number operator $\hat{n} = \hat{a}^\dagger\hat{a}$:
\begin{equation}
\hat{H} = \hbar\omega\left(\hat{n} + \frac{1}{2}\right)
\end{equation}
\end{example}

The eigenstates of the number operator $\hat{n}$ are denoted $\ket{n}$ with eigenvalues $n = 0, 1, 2, ...$:
\begin{align}
\hat{n}\ket{n} &= n\ket{n} \\
\hat{H}\ket{n} &= \hbar\omega\left(n + \frac{1}{2}\right)\ket{n}
\end{align}

\begin{example}[Action of Ladder Operators]
The operators act as:
\begin{align}
\hat{a}^\dagger\ket{n} &= \sqrt{n+1}\ket{n+1} \quad \text{(raises by one quantum)} \\
\hat{a}\ket{n} &= \sqrt{n}\ket{n-1} \quad \text{(lowers by one quantum)} \\
\hat{a}\ket{0} &= 0 \quad \text{(defines ground state)}
\end{align}

Any excited state can be built from the ground state:
\begin{equation}
\ket{n} = \frac{(\hat{a}^\dagger)^n}{\sqrt{n!}}\ket{0}
\end{equation}

The position representation of these states involves Hermite polynomials:
\begin{equation}
\psi_n(x) = \left(\frac{1}{\pi x_0^2}\right)^{1/4} \frac{1}{\sqrt{2^n n!}} H_n\left(\frac{x}{x_0}\right) e^{-x^2/(2x_0^2)}
\end{equation}
where $H_n$ are Hermite polynomials and $x_0 = \sqrt{\hbar/(m\omega)}$ is the oscillator length scale.
\end{example}

\begin{example}[Harmonic Oscillator Ground State in Position Representation]
The ground state is defined by $\hat{a}\ket{0} = 0$. In position representation:
\begin{equation}
\hat{a}\psi_0(x) = \frac{1}{\sqrt{2}}\left(\frac{x}{x_0} + \frac{x_0}{\hbar}\hat{p}\right)\psi_0(x) = 0
\end{equation}

Substituting $\hat{p} = -i\hbar d/dx$:
\begin{equation}
\left(\frac{x}{x_0} - ix_0\frac{d}{dx}\right)\psi_0(x) = 0
\end{equation}

This gives:
\begin{equation}
\frac{d\psi_0}{dx} = -\frac{x}{x_0^2}\psi_0 = -\frac{m\omega}{\hbar}x\psi_0
\end{equation}

Solving: $\psi_0(x) = A e^{-x^2/(2x_0^2)}$

Normalizing: $\psi_0(x) = \left(\frac{1}{\pi x_0^2}\right)^{1/4} e^{-x^2/(2x_0^2)}$

The ground state is indeed a minimum uncertainty Gaussian with width $x_0/\sqrt{2}$.
\end{example}

\subsection{Coherent States: Quantum States Most Like Classical Motion}

Coherent states are eigenstates of the annihilation operator and represent the quantum states that most closely resemble classical harmonic oscillator motion.

\begin{definition}
A coherent state $\ket{\alpha}$ is defined by:
\begin{equation}
\hat{a}\ket{\alpha} = \alpha\ket{\alpha}
\end{equation}
where $\alpha$ is a complex number.
\end{definition}

\subsubsection{Construction and Properties of Coherent States}

Coherent states can be generated from the vacuum state by the displacement operator:
\begin{equation}
\ket{\alpha} = \hat{D}(\alpha)\ket{0} = e^{\alpha\hat{a}^\dagger - \alpha^*\hat{a}}\ket{0}
\end{equation}

The complex parameter $\alpha$ encodes the displacement in phase space. We can parameterize it as:
\begin{equation}
\alpha = x_\alpha + ip_\alpha
\end{equation}
where $x_\alpha$ and $p_\alpha$ are real dimensionless parameters that determine the position and momentum displacements.

The displacement operator can be expressed in terms of position and momentum operators. Substituting the ladder operators:
\begin{align}
\hat{a} &= \sqrt{\frac{m\omega}{2\hbar}}\hat{x} + \frac{i}{\sqrt{2m\omega\hbar}}\hat{p}\\
\hat{a}^\dagger &= \sqrt{\frac{m\omega}{2\hbar}}\hat{x} - \frac{i}{\sqrt{2m\omega\hbar}}\hat{p}
\end{align}

into the displacement operator and working through the algebra, one finds that the displacement operator performs a phase space translation. The action of this operator shifts both position and momentum expectation values:
\begin{align}
\langle \hat{x} \rangle &= x_0\sqrt{2}\text{Re}[\alpha]\\
\langle \hat{p} \rangle &= p_0\sqrt{2}\text{Im}[\alpha]
\end{align}

where $x_0 = \sqrt{\hbar/(m\omega)}$ and $p_0 = \sqrt{m\omega\hbar}$ are the characteristic oscillator scales. The complex parameter $\alpha = x_\alpha + ip_\alpha$ thus encodes the classical phase space coordinates in dimensionless form.

\clearpage
\begin{example}[How Displacement Acts on Wave Functions]
The operator $e^{-ix_\alpha\hat{p}/\hbar}$ is the translation operator that shifts any wave function by $x_\alpha$:
\begin{equation}
e^{-ix_\alpha\hat{p}/\hbar}\psi(x) = \psi(x - x_\alpha)
\end{equation}

To verify this, expand the exponential:
\begin{align}
e^{-ix_\alpha \hat{p}/\hbar}\psi(x) &= \sum_{n=0}^\infty \frac{(-ix_\alpha/\hbar)^n}{n!}\hat{p}^n\psi(x)\\
&= \sum_{n=0}^\infty \frac{(-ix_\alpha)^n}{n!}\left(-i\hbar\frac{d}{dx}\right)^n\psi(x)\\
&= \sum_{n=0}^\infty \frac{(x_\alpha)^n}{n!}\left(-\frac{d}{dx}\right)^n\psi(x)\\
&= \psi(x - x_\alpha)
\end{align}

This is the Taylor series for $\psi(x - x_\alpha)$! The momentum operator generates spatial translations.

Similarly, $e^{ip_\alpha\hat{x}/\hbar}$ acts as a momentum boost. The momentum boost adds a phase to the wave function:
\begin{equation}
e^{ip_\alpha\hat{x}/\hbar}\psi(x) = e^{ip_\alpha x/\hbar}\psi(x)
\end{equation}

This multiplies the wave function by a plane wave, effectively "kicking" it with momentum $p_\alpha$.
\end{example}

\subsubsection{Decomposition of the Displacement Operator}

To better understand the physical meaning of the displacement operator, we can decompose it using the Baker-Campbell-Hausdorff formula. With $\alpha = x_\alpha + ip_\alpha$, where $x_\alpha$ and $p_\alpha$ are real dimensionless parameters, the displacement operator can be written as:
\begin{equation}
\hat{D}(\alpha) = \exp\left[\frac{i}{\hbar}(p_\alpha\hat{x} - x_\alpha\hat{p})\right] = e^{ip_\alpha\hat{x}/\hbar} e^{-ix_\alpha\hat{p}/\hbar} e^{-ip_\alpha x_\alpha/(2\hbar)}
\end{equation}

The last term is merely a c-number phase factor. Up to this phase, the displacement operator factorizes into two simple operations:
\begin{equation}
\hat{D}(\alpha) \approx e^{ip_\alpha\hat{x}/\hbar} \cdot e^{-ix_\alpha\hat{p}/\hbar}
\end{equation}

This factorization reveals that the displacement operator performs two sequential operations:
\begin{enumerate}
\item First, it translates the state by $x_\alpha$ in position: $e^{-ix_\alpha\hat{p}/\hbar}\psi(x) = \psi(x-x_\alpha)$
\item Then, it boosts the momentum by $p_\alpha$: $e^{ip_\alpha\hat{x}/\hbar}\psi(x) = e^{ip_\alpha x/\hbar}\psi(x)$
\end{enumerate}

The displacement operator thus performs a complete phase space translation. We can verify this explicitly by computing how it transforms the position and momentum operators:
\begin{align}
\hat{D}^\dagger(\alpha)\hat{x}\hat{D}(\alpha) &= \hat{x} + x_\alpha\\
\hat{D}^\dagger(\alpha)\hat{p}\hat{D}(\alpha) &= \hat{p} + p_\alpha
\end{align}

To verify these transformations, we use the commutation relations:
\begin{align}
[\hat{x}, p_\alpha\hat{x}/\hbar] &= 0\\
[\hat{x}, -x_\alpha\hat{p}/\hbar] &= -x_\alpha[\hat{x},\hat{p}]/\hbar = -ix_\alpha\\
[\hat{p}, p_\alpha\hat{x}/\hbar] &= p_\alpha[\hat{p},\hat{x}]/\hbar = -ip_\alpha\\
[\hat{p}, -x_\alpha\hat{p}/\hbar] &= 0
\end{align}

Using the operator identity $e^{\hat{A}}\hat{B}e^{-\hat{A}} = \hat{B} + [\hat{A},\hat{B}] + \frac{1}{2!}[\hat{A},[\hat{A},\hat{B}]] + ...$, and noting that higher commutators vanish in this case:
\begin{align}
\hat{D}^\dagger\hat{x}\hat{D} &= e^{-i(p_\alpha\hat{x} - x_\alpha\hat{p})/\hbar}\hat{x}e^{i(p_\alpha\hat{x} - x_\alpha\hat{p})/\hbar}\\
&= \hat{x} + \frac{i}{\hbar}[p_\alpha\hat{x} - x_\alpha\hat{p}, \hat{x}]\\
&= \hat{x} + \frac{i}{\hbar}(-x_\alpha[\hat{p},\hat{x}])\\
&= \hat{x} + x_\alpha
\end{align}

Similarly, for the momentum operator:
\begin{align}
\hat{D}^\dagger\hat{p}\hat{D} &= \hat{p} + \frac{i}{\hbar}[p_\alpha\hat{x} - x_\alpha\hat{p}, \hat{p}]\\
&= \hat{p} + \frac{i}{\hbar}(p_\alpha[\hat{x},\hat{p}])\\
&= \hat{p} + p_\alpha
\end{align}

This decomposition reveals the fundamental nature of coherent states: they are quantum states that have been displaced from the vacuum state in phase space. The complex parameter $\alpha = x_\alpha + ip_\alpha$ encodes this displacement, with the real part determining the position shift and the imaginary part determining the momentum shift. This dual action---combining position displacement with momentum boost---makes coherent states the quantum analog of classical phase space points, bridging the quantum and classical descriptions of harmonic oscillator motion.

Using the Baker-Campbell-Hausdorff formula:
\begin{equation}
\ket{\alpha} = e^{-|\alpha|^2/2} e^{\alpha\hat{a}^\dagger}\ket{0}
\end{equation}

Expanding in the number state basis:
\begin{equation}
\ket{\alpha} = e^{-|\alpha|^2/2} \sum_{n=0}^{\infty} \frac{\alpha^n}{\sqrt{n!}}\ket{n}
\end{equation}

The probability of finding $n$ quanta is Poisson distributed:
\begin{equation}
P(n) = |\langle n|\alpha\rangle|^2 = e^{-|\alpha|^2} \frac{|\alpha|^{2n}}{n!}
\end{equation}
with mean $\bar{n} = |\alpha|^2$ and variance $\Delta n^2 = |\alpha|^2$.

\textbf{Overlap of coherent states:}

The amplitude for overlap between two coherent states is
\begin{equation}
\braket{\alpha|\beta} = e^{-|\alpha|^2/2}\, e^{-|\beta|^2/2}\, e^{\alpha^*\beta},
\end{equation}
which is generally complex. The gauge-invariant magnitude takes a particularly transparent form:
\begin{equation}
\boxed{\,|\braket{\alpha|\beta}|^2 = e^{-|\alpha-\beta|^2}\,}
\end{equation}

\textbf{Derivation.} Using $|\alpha-\beta|^2 = |\alpha|^2 + |\beta|^2 - \alpha\beta^* - \alpha^*\beta$ and $\alpha\beta^* + \alpha^*\beta = 2\,\mathrm{Re}(\alpha^*\beta)$,
\begin{align}
|\braket{\alpha|\beta}|^2 &= e^{-|\alpha|^2}\,e^{-|\beta|^2}\,e^{\alpha^*\beta + \alpha\beta^*}\nonumber\\
&= e^{-|\alpha|^2 - |\beta|^2 + 2\,\mathrm{Re}(\alpha^*\beta)} = e^{-|\alpha-\beta|^2}.
\end{align}
The full amplitude differs from $e^{-|\alpha-\beta|^2/2}$ by a phase $e^{i\,\mathrm{Im}(\alpha^*\beta)}$, but the measurable overlap probability is given by the magnitude squared above.

The form $|\braket{\alpha|\beta}|^2 = e^{-|\alpha-\beta|^2}$ shows that:
\begin{itemize}
\item Coherent states are never orthogonal unless $|\alpha-\beta| \to \infty$
\item The overlap probability decreases as a Gaussian in phase-space distance
\item Unlike number states, coherent states form an over-complete basis
\end{itemize}

\begin{example}[Physical Properties of Coherent States]
Writing $\alpha = |\alpha|e^{i\phi}$, the coherent state has:

\textbf{Position expectation:}
\begin{align}
\langle \hat{x} \rangle &= \langle \alpha | \hat{x} | \alpha \rangle = x_0 \frac{\langle \alpha | (\hat{a} + \hat{a}^\dagger) | \alpha \rangle}{\sqrt{2}} \\
&= x_0 \frac{\alpha + \alpha^*}{\sqrt{2}} = x_0\sqrt{2}|\alpha|\cos(\phi)
\end{align}

\textbf{Momentum expectation:}
\begin{align}
\langle \hat{p} \rangle &= \langle \alpha | \hat{p} | \alpha \rangle = p_0 \frac{\langle \alpha | (\hat{a} - \hat{a}^\dagger) | \alpha \rangle}{i\sqrt{2}} \\
&= p_0 \frac{\alpha - \alpha^*}{i\sqrt{2}} = p_0\sqrt{2}|\alpha|\sin(\phi)
\end{align}

\textbf{Time evolution:}
\begin{equation}
\ket{\alpha(t)} = e^{-i\hat{H}t/\hbar}\ket{\alpha} = e^{-i\omega t/2}\ket{\alpha e^{-i\omega t}}
\end{equation}

The state remains coherent with $\alpha(t) = \alpha e^{-i\omega t}$. The expectation values oscillate:
\begin{align}
\langle \hat{x} \rangle_t &= x_0\sqrt{2}|\alpha|\cos(\omega t - \phi) \\
\langle \hat{p} \rangle_t &= -p_0\sqrt{2}|\alpha|\sin(\omega t - \phi)
\end{align}

This is exactly classical motion! The wave packet maintains its shape (minimum uncertainty) while oscillating.
\end{example}

\begin{example}[Wave Function of a Coherent State]
The position space wave function is:
\begin{equation}
\psi_\alpha(x) = \left(\frac{1}{\pi x_0^2}\right)^{1/4} \exp\left[-\frac{(x - \langle x \rangle)^2}{2x_0^2} + \frac{i\langle p \rangle x}{\hbar} - \frac{i(\langle p \rangle \langle x \rangle + \phi)}{2\hbar}\right]
\end{equation}

where $\langle x \rangle = x_0\sqrt{2}|\alpha|\cos\phi$ and $\langle p \rangle = p_0\sqrt{2}|\alpha|\sin\phi$.

This is a Gaussian wave packet with position width $\Delta x = x_0/\sqrt{2}$ (identical to the ground state), momentum width $\Delta p = p_0/\sqrt{2}$, and uncertainty product $\Delta x \Delta p = x_0 p_0/2 = \hbar/2$ achieving the minimum uncertainty bound.
\end{example}

\subsubsection{Time Dependence of Coherent States}

Coherent states remain coherent under time evolution, executing classical harmonic motion.

\begin{example}[Equation of Motion for Coherent States]
Time evolution from the Schr\"odinger equation:
\begin{equation}
i\hbar\frac{d}{dt}\ket{\alpha(t)} = \hat{H}\ket{\alpha(t)}
\end{equation}

For the harmonic oscillator Hamiltonian $\hat{H} = \hbar\omega(\hat{a}^\dagger\hat{a} + 1/2)$, we can write the formal solution:
\begin{equation}
\ket{\alpha(t)} = e^{-i\hat{H}t/\hbar}\ket{\alpha(0)}
\end{equation}

Expanding the initial coherent state in the number basis:
\begin{equation}
\ket{\alpha(0)} = e^{-|\alpha(0)|^2/2}\sum_{n=0}^{\infty}\frac{[\alpha(0)]^n}{\sqrt{n!}}\ket{n}
\end{equation}

The time evolution operator acts on each number state as:
\begin{equation}
e^{-i\hat{H}t/\hbar}\ket{n} = e^{-i\omega(n+1/2)t}\ket{n}
\end{equation}

Therefore:
\begin{align}
\ket{\alpha(t)} &= e^{-|\alpha(0)|^2/2}\sum_{n=0}^{\infty}\frac{[\alpha(0)]^n}{\sqrt{n!}}e^{-i\omega(n+1/2)t}\ket{n}\\[8pt]
&= e^{-i\omega t/2}e^{-|\alpha(0)|^2/2}\sum_{n=0}^{\infty}\frac{[\alpha(0)e^{-i\omega t}]^n}{\sqrt{n!}}\ket{n}\\[8pt]
&= e^{-i\omega t/2}\ket{\alpha(0)e^{-i\omega t}}
\end{align}

\textbf{Key result:} The coherent state remains a coherent state with a time-dependent complex amplitude:
\begin{equation}
\boxed{\alpha(t) = \alpha(0)e^{-i\omega t}}
\end{equation}

The global phase $e^{-i\omega t/2}$ is physically unobservable and is often dropped. The complex parameter $\alpha$ rotates in the complex plane with angular frequency $\omega$---the same frequency as the classical oscillator!
\end{example}

\begin{example}[Position and Momentum Expectation Values]
To find $\langle x(t)\rangle$ and $\langle p(t)\rangle$, we use the relations:
\begin{align}
\hat{x} &= x_0\frac{\hat{a} + \hat{a}^\dagger}{\sqrt{2}} = \sqrt{\frac{\hbar}{2m\omega}}(\hat{a} + \hat{a}^\dagger)\\[6pt]
\hat{p} &= p_0\frac{\hat{a} - \hat{a}^\dagger}{i\sqrt{2}} = -i\sqrt{\frac{m\omega\hbar}{2}}(\hat{a} - \hat{a}^\dagger)
\end{align}

Since $\hat{a}\ket{\alpha(t)} = \alpha(t)\ket{\alpha(t)}$ and $\bra{\alpha(t)}\hat{a}^\dagger = \bra{\alpha(t)}\alpha^*(t)$, we have:
\begin{align}
\langle x(t)\rangle &= \langle\alpha(t)|\hat{x}|\alpha(t)\rangle\\[4pt]
&= x_0\frac{\alpha(t) + \alpha^*(t)}{\sqrt{2}}\\[4pt]
&= x_0\frac{\alpha(0)e^{-i\omega t} + \alpha^*(0)e^{i\omega t}}{\sqrt{2}}
\end{align}

Writing $\alpha(0) = |\alpha_0|e^{i\phi_0}$ where $|\alpha_0|$ and $\phi_0$ are real:
\begin{align}
\langle x(t)\rangle &= x_0\frac{|\alpha_0|[e^{i(\phi_0-\omega t)} + e^{-i(\phi_0-\omega t)}]}{\sqrt{2}}\\[4pt]
&= x_0\sqrt{2}|\alpha_0|\cos(\omega t - \phi_0)
\end{align}

Similarly for momentum:
\begin{align}
\langle p(t)\rangle &= \langle\alpha(t)|\hat{p}|\alpha(t)\rangle\\[4pt]
&= p_0\frac{\alpha(t) - \alpha^*(t)}{i\sqrt{2}}\\[4pt]
&= p_0\frac{|\alpha_0|[e^{i(\phi_0-\omega t)} - e^{-i(\phi_0-\omega t)}]}{i\sqrt{2}}\\[4pt]
&= -p_0\sqrt{2}|\alpha_0|\sin(\omega t - \phi_0)
\end{align}

\textbf{Final results:}
\begin{align}
\boxed{\langle x(t)\rangle = x_0\sqrt{2}|\alpha_0|\cos(\omega t - \phi_0)}\\[8pt]
\boxed{\langle p(t)\rangle = -p_0\sqrt{2}|\alpha_0|\sin(\omega t - \phi_0)}
\end{align}

where $x_0 = \sqrt{\hbar/(m\omega)}$ and $p_0 = \sqrt{m\omega\hbar}$.
\end{example}

\begin{example}[Classical Correspondence]
The expectation values satisfy the classical equations of motion. Taking time derivatives:
\begin{align}
\frac{d\langle x\rangle}{dt} &= -x_0\sqrt{2}|\alpha_0|\omega\sin(\omega t - \phi_0) = \frac{\langle p\rangle}{m}\\[6pt]
\frac{d\langle p\rangle}{dt} &= -p_0\sqrt{2}|\alpha_0|\omega\cos(\omega t - \phi_0) = -m\omega^2\langle x\rangle
\end{align}

These are exactly the equations of motion for a classical harmonic oscillator! We can verify:
\begin{equation}
\frac{d^2\langle x\rangle}{dt^2} = -\omega^2\langle x\rangle
\end{equation}

Furthermore, the total energy expectation value is:
\begin{align}
\langle E(t)\rangle &= \frac{\langle p^2\rangle}{2m} + \frac{1}{2}m\omega^2\langle x^2\rangle\\[4pt]
&= \hbar\omega\left(|\alpha_0|^2 + \frac{1}{2}\right)
\end{align}

This is constant in time, as expected for energy conservation. The amplitude $|\alpha_0|$ determines the classical oscillation amplitude:
\begin{align}
A_x &= x_0\sqrt{2}|\alpha_0| = \sqrt{\frac{2\hbar}{m\omega}}|\alpha_0|\\[4pt]
A_p &= p_0\sqrt{2}|\alpha_0| = \sqrt{2m\omega\hbar}|\alpha_0|
\end{align}

The phase $\phi_0$ sets the initial phase of oscillation.
\end{example}

\begin{physicalinsight}
Coherent states provide the most direct connection between quantum and classical mechanics for the harmonic oscillator. Their expectation values follow classical trajectories exactly---not approximately, but exactly. The quantum state maintains its Gaussian shape throughout the motion, with constant uncertainties $\Delta x = x_0/\sqrt{2}$ and $\Delta p = p_0/\sqrt{2}$ that satisfy the minimum uncertainty relation.

This behavior explains why laser light, which consists of coherent photon states, exhibits classical-like electromagnetic field oscillations despite being fundamentally quantum mechanical. The coherent state bridges the quantum-classical divide by encoding classical phase space coordinates $(x, p)$ into the complex parameter $\alpha$ through:
\begin{align}
\mathrm{Re}[\alpha\sqrt{2}] &= \frac{\langle x\rangle}{x_0}\\[4pt]
\mathrm{Im}[\alpha\sqrt{2}] &= \frac{\langle p\rangle}{p_0}
\end{align}

The rotation of $\alpha(t)$ in the complex plane directly corresponds to circular motion in classical phase space. This geometric picture---where quantum evolution becomes rotation in a complex plane---extends far beyond the harmonic oscillator and underlies much of modern quantum optics and quantum information theory.
\end{physicalinsight}

\subsection{Squeezed States: Beyond Classical Limits}

While coherent states maintain the symmetric uncertainty of the ground state, squeezed states redistribute the uncertainty between position and momentum while preserving the minimum uncertainty product.

\begin{definition}
A squeezed state is generated by the squeeze operator:
\begin{equation}
\hat{S}(z) = \exp\left[\frac{z^*}{2}\hat{a}^2 - \frac{z}{2}(\hat{a}^\dagger)^2\right]
\end{equation}
where $z = re^{i\theta}$ is the squeeze parameter.
\end{definition}

\subsubsection{Operator Transformations Under Squeezing}

To understand how squeezing affects position and momentum, we need to compute how the squeeze operator transforms $\hat{x}$ and $\hat{p}$. For real squeeze parameter $r$, we use the Baker-Campbell-Hausdorff formula with $\hat{A} = \frac{r}{2}(\hat{a}^2 - (\hat{a}^\dagger)^2)$.

\begin{example}[Deriving the Position Transformation]
Since $\hat{S}^\dagger(r) = e^{-\hat{A}}$, we compute $e^{-\hat{A}}\hat{a}e^{\hat{A}}$ using BCH:
\begin{equation}
e^{-\hat{A}}\hat{a}e^{\hat{A}} = \hat{a} + [-\hat{A}, \hat{a}] + \frac{1}{2!}[-\hat{A}, [-\hat{A}, \hat{a}]] + \cdots
\end{equation}

\textbf{Key commutators:}
\begin{align}
[\hat{a}^2, \hat{a}] &= 0\nonumber\\
[(\hat{a}^\dagger)^2, \hat{a}] &= -2\hat{a}^\dagger\nonumber\\
[\hat{a}^2, \hat{a}^\dagger] &= 2\hat{a}
\end{align}

Computing the nested commutators with $-\hat{A}$:
\begin{align}
[-\hat{A}, \hat{a}] &= -\frac{r}{2}(0 - (-2\hat{a}^\dagger)) = -r\hat{a}^\dagger\nonumber\\
[-\hat{A}, [-\hat{A}, \hat{a}]] &= [-\hat{A}, -r\hat{a}^\dagger] = -r[-\hat{A}, \hat{a}^\dagger]\nonumber
\end{align}

where $[-\hat{A}, \hat{a}^\dagger] = -\frac{r}{2}(2\hat{a} - 0) = -r\hat{a}$, giving:
\begin{equation}
[-\hat{A}, [-\hat{A}, \hat{a}]] = r^2\hat{a}
\end{equation}

Third commutator:
\begin{equation}
[-\hat{A}, [-\hat{A}, [-\hat{A}, \hat{a}]]] = r^2[-\hat{A}, \hat{a}] = -r^3\hat{a}^\dagger
\end{equation}

The pattern alternates with signs! The BCH series becomes:
\begin{align}
\hat{S}^\dagger\hat{a}\hat{S} &= \hat{a} - r\hat{a}^\dagger + \frac{r^2}{2!}\hat{a} - \frac{r^3}{3!}\hat{a}^\dagger + \cdots\nonumber\\
&= \hat{a}\left(1 + \frac{r^2}{2!} + \frac{r^4}{4!} + \cdots\right) - \hat{a}^\dagger\left(r + \frac{r^3}{3!} + \cdots\right)
\end{align}

Recognizing the Taylor series for hyperbolic functions:
\begin{equation}
\boxed{\hat{S}^\dagger\hat{a}\hat{S} = \hat{a}\cosh r - \hat{a}^\dagger\sinh r}
\end{equation}

Similarly:
\begin{equation}
\boxed{\hat{S}^\dagger\hat{a}^\dagger\hat{S} = \hat{a}^\dagger\cosh r - \hat{a}\sinh r}
\end{equation}

Therefore, for position $\hat{x} = x_0(\hat{a} + \hat{a}^\dagger)/\sqrt{2}$:
\begin{align}
\hat{S}^\dagger\hat{x}\hat{S} &= \frac{x_0}{\sqrt{2}}\hat{S}^\dagger(\hat{a} + \hat{a}^\dagger)\hat{S}\nonumber\\
&= \frac{x_0}{\sqrt{2}}[(\hat{a} + \hat{a}^\dagger)\cosh r - (\hat{a}^\dagger + \hat{a})\sinh r]\nonumber\\
&= \frac{x_0}{\sqrt{2}}(\hat{a} + \hat{a}^\dagger)(\cosh r - \sinh r)
\end{align}

Using $\cosh r - \sinh r = e^{-r}$:
\begin{equation}
\boxed{\hat{S}^\dagger(r)\hat{x}\hat{S}(r) = e^{-r}\hat{x}}
\end{equation}

Similarly for momentum $\hat{p} = p_0(\hat{a} - \hat{a}^\dagger)/(i\sqrt{2})$:
\begin{align}
\hat{S}^\dagger\hat{p}\hat{S} &= \frac{p_0}{i\sqrt{2}}[(\hat{a} - \hat{a}^\dagger)\cosh r - (\hat{a}^\dagger - \hat{a})\sinh r]\nonumber\\
&= \frac{p_0}{i\sqrt{2}}(\hat{a} - \hat{a}^\dagger)(\cosh r + \sinh r)
\end{align}

Using $\cosh r + \sinh r = e^r$:
\begin{equation}
\boxed{\hat{S}^\dagger(r)\hat{p}\hat{S}(r) = e^r\hat{p}}
\end{equation}

For $r > 0$, position is squeezed by factor $e^{-r}$ while momentum is anti-squeezed by $e^r$.
\end{example}

\begin{example}[Properties of Squeezed States]
The squeezed vacuum state $\ket{z} = \hat{S}(z)\ket{0}$ has remarkable properties:

\textbf{Position variance:}
\begin{equation}
(\Delta x)^2 = \frac{x_0^2}{2}e^{-2r}
\end{equation}

\textbf{Momentum variance:}
\begin{equation}
(\Delta p)^2 = \frac{p_0^2}{2}e^{2r}
\end{equation}

\textbf{Uncertainty product:}
\begin{equation}
\Delta x \Delta p = \frac{x_0 p_0}{2} = \frac{\hbar}{2}
\end{equation}

For $r > 0$: position uncertainty is "squeezed" below the ground state value while momentum uncertainty increases. The converse occurs for $r < 0$.

\textbf{Number state expansion:}
\begin{equation}
\ket{z} = \frac{1}{\sqrt{\cosh r}} \sum_{n=0}^{\infty} \frac{\sqrt{(2n)!}}{2^n n!}(-e^{i\theta}\tanh r)^n\ket{2n}
\end{equation}

Only even number states appear due to the parity symmetry of the squeeze operator.
\end{example}

\begin{example}[Displaced Squeezed States]
The most general Gaussian state is a displaced squeezed state:
\begin{equation}
\ket{\alpha, z} = \hat{D}(\alpha)\hat{S}(z)\ket{0}
\end{equation}

This state has:
\begin{itemize}
\item Displaced mean position and momentum (from $\alpha$)
\item Squeezed uncertainties (from $z$)
\item Minimum uncertainty product $\Delta x \Delta p = \hbar/2$
\end{itemize}

The wave function is:
\begin{equation}
\psi_{\alpha,z}(x) = \left(\frac{1}{\pi\sigma_x^2}\right)^{1/4} \exp\left[-\frac{(x-\langle x \rangle)^2}{2\sigma_x^2} + \frac{i\langle p \rangle x}{\hbar} + i\phi(x,t)\right]
\end{equation}

where $\sigma_x^2 = x_0^2 e^{-2r}$ and $(\langle x \rangle, \langle p \rangle)$ are determined by $\alpha$.
\end{example}

\begin{physicalinsight}
The different quantum states of the harmonic oscillator form a hierarchy of increasing sophistication. Number states $\ket{n}$ are the energy eigenstates with quantized energy levels $E_n = \hbar\omega(n + 1/2)$. These are highly non-classical states with definite energy but completely uncertain phase. Coherent states $\ket{\alpha}$ are the most classical-like quantum states, maintaining minimum uncertainty while their expectation values oscillate exactly as a classical particle would. Squeezed states $\ket{z}$ exhibit non-classical uncertainty distributions where position and momentum uncertainties can be redistributed while preserving the minimum uncertainty product. Finally, displaced squeezed states $\ket{\alpha,z}$ represent the most general Gaussian states, combining both displacement and squeezing.

Coherent states appear throughout physics. Laser light consists of coherent photon states, and driven classical oscillators are well-described by coherent states in the quantum limit. They provide the foundation for semiclassical approximations and are central to quantum optics and cavity quantum electrodynamics. Squeezed states have become essential tools in precision measurement. The LIGO gravitational wave detector uses squeezed light to reduce quantum noise below the standard quantum limit. Squeezed states enable sub-shot-noise measurements in quantum metrology and provide the basis for continuous variable quantum computing. They also serve as testbeds for quantum mechanics at increasingly macroscopic scales.
\end{physicalinsight}

\subsection{Physical Implementation of Squeezing}

How can we physically create squeezed states in the laboratory? The key insight is that we need an interaction Hamiltonian that creates or destroys photons in pairs, not individually. This naturally leads to terms proportional to $(\hat{a}^\dagger)^2$ and $\hat{a}^2$.

\subsubsection{The Parametric Interaction}

In quantum optics experiments, squeezed states are typically created using nonlinear crystals where a strong classical "pump" field at frequency $2\omega$ drives the creation of photon pairs at frequency $\omega$. The interaction between the pump field and the quantum field mode can be described by the Hamiltonian:
\begin{equation}
\hat{H}_{\text{int}} = i\hbar g \left[(\hat{a}^\dagger)^2 e^{-2i\omega t} - \hat{a}^2 e^{2i\omega t}\right]
\end{equation}
where $g$ is a coupling constant proportional to the pump field amplitude and the nonlinear susceptibility of the crystal.

To understand this Hamiltonian, consider that the pump field can be approximated as a classical field $E_{\text{pump}} \propto e^{-2i\omega t}$. The nonlinear crystal mediates an interaction where this pump can create two photons at frequency $\omega$ (energy conservation requires $2\omega = \omega + \omega$). The term $(\hat{a}^\dagger)^2$ creates two photons simultaneously, while $\hat{a}^2$ annihilates two photons.

\subsubsection{Connection to the Squeeze Operator}

Moving to an interaction picture where we remove the fast time dependence, the effective Hamiltonian becomes:
\begin{equation}
\hat{H}_{\text{eff}} = i\hbar g \left[(\hat{a}^\dagger)^2 - \hat{a}^2\right]
\end{equation}

The time evolution under this Hamiltonian for a time $\tau$ gives:
\begin{equation}
\hat{U}(\tau) = \exp\left(-\frac{i\hat{H}_{\text{eff}}\tau}{\hbar}\right) = \exp\left[g\tau \left(\hat{a}^2 - (\hat{a}^\dagger)^2\right)\right]
\end{equation}

Comparing with our squeeze operator $\hat{S}(z) = \exp\left[\frac{z^*}{2}\hat{a}^2 - \frac{z}{2}(\hat{a}^\dagger)^2\right]$, we see that:
\begin{equation}
\hat{U}(\tau) = \hat{S}(z) \quad \text{with} \quad z = -2g\tau
\end{equation}

The interaction time $\tau$ and coupling strength $g$ determine the squeeze parameter. For real $g$ and $\tau$, we get $z = r$ (real), corresponding to squeezing along the position or momentum quadrature.

\subsubsection{Why Only Even Number States?}

A crucial feature of squeezed states is that when starting from vacuum, only even number states appear. We can understand this from the structure of the Hamiltonian. Consider the action of the evolution operator on the vacuum state:
\begin{equation}
\hat{S}(r)\ket{0} = \exp\left[\frac{r}{2}\left(\hat{a}^2 - (\hat{a}^\dagger)^2\right)\right]\ket{0}
\end{equation}

Expanding the exponential:
\begin{align}
\hat{S}(r)\ket{0} &= \sum_{n=0}^{\infty} \frac{1}{n!}\left[\frac{r}{2}\left(\hat{a}^2 - (\hat{a}^\dagger)^2\right)\right]^n \ket{0}\\
&= \ket{0} + \frac{r}{2}\left(\hat{a}^2 - (\hat{a}^\dagger)^2\right)\ket{0} + \frac{r^2}{8}\left(\hat{a}^2 - (\hat{a}^\dagger)^2\right)^2\ket{0} + \cdots
\end{align}

The key observation is that $\hat{a}^2\ket{0} = 0$ (you cannot annihilate two photons from vacuum), so only the $(\hat{a}^\dagger)^2$ terms survive when acting on $\ket{0}$. Each application of $(\hat{a}^\dagger)^2$ creates two photons:
\begin{align}
(\hat{a}^\dagger)^2\ket{0} &= \sqrt{2!}\ket{2}\\
[(\hat{a}^\dagger)^2]^2\ket{0} &= \sqrt{4!}\ket{4}\\
[(\hat{a}^\dagger)^2]^3\ket{0} &= \sqrt{6!}\ket{6}
\end{align}

Therefore, the squeezed vacuum state contains only even photon numbers:
\begin{equation}
\ket{r} = \hat{S}(r)\ket{0} = \frac{1}{\sqrt{\cosh r}} \sum_{n=0}^{\infty} \frac{\sqrt{(2n)!}}{2^n n!}(\tanh r)^n\ket{2n}
\end{equation}

This even-number-only structure is a direct consequence of the pairwise creation of photons. The physics is clear: the parametric process creates photons in pairs to conserve energy (two photons at $\omega$ from one at $2\omega$), leading to only even photon numbers in the output state.

Experimentally, this process is realized in nonlinear crystals pumped by strong laser fields. The squeezed light emerges at half the pump frequency, with the amount of squeezing controlled by the pump power and interaction length. Typical experiments achieve 10-15 dB of squeezing, meaning the noise in one quadrature is reduced by a factor of 3-5 below the vacuum level. These squeezed states are now routinely used in gravitational wave detectors like LIGO to surpass the standard quantum limit for position measurements.

\section{Chapter Summary}

This chapter developed the continuum limit of discrete lattice models. Taking lattice spacing $a \to 0$ while keeping system size fixed transforms discrete amplitudes into the continuous wave function $\psi(x)$.

The harmonic oscillator emerged as a paradigmatic system bridging classical and quantum mechanics through coherent states---eigenstates of the annihilation operator that maintain minimum uncertainty while following classical trajectories. Squeezed states demonstrate quantum control of uncertainty beyond classical limits, with applications in precision measurement.

This connection provides the foundation for numerical methods and applies directly to quantum dots, optical lattices, and qubit arrays. The framework extends to three dimensions, many-particle systems, and quantum field theory.

\input{book_problems/ch05_problems.tex}

\section*{References and Further Reading}
\addcontentsline{toc}{section}{References and Further Reading}

\begin{description}
\item[Griffiths, D.~J., and Schroeter, D.~F.] \emph{Introduction to Quantum Mechanics}, 3rd ed. Cambridge University Press, 2018. Chapters 2--3 give the standard undergraduate-level entry into the wave equation, the infinite well, and the harmonic oscillator; the gentlest companion to this chapter.

\item[Cohen-Tannoudji, C., Diu, B., and Lalo\"e, F.] \emph{Quantum Mechanics}, Vol.~1. Wiley-VCH, 1977. Chapters I--V cover the continuum formalism, position and momentum representations, and the harmonic oscillator with worked examples; the most thorough single textbook treatment.

\item[Shankar, R.] \emph{Principles of Quantum Mechanics}, 2nd ed. Springer, 1994. Chapters 1 and 7 develop the discrete-to-continuum limit, Fourier analysis, and the harmonic oscillator with unusual clarity; particularly recommended for the structure of the position-momentum Fourier relationship.

\item[Walls, D.~F., and Milburn, G.~J.] \emph{Quantum Optics}, 2nd ed. Springer, 2008. Chapters 2--3 develop coherent and squeezed states with the formalism extended at the end of this chapter; the natural next reference for the optical applications.

\item[Aasi, J.\ et~al.\ (LIGO Scientific Collaboration)] ``Enhanced sensitivity of the LIGO gravitational wave detector by using squeezed states of light.'' \emph{Nature Photonics} \textbf{7}, 613--619 (2013). \href{https://doi.org/10.1038/nphoton.2013.177}{doi:10.1038/nphoton.2013.177}. The squeezed states of this chapter, used to beat the standard quantum limit at LIGO; the most concrete demonstration that the formalism here is experimentally consequential.
\end{description}

%% file: book_problems/ch05_problems.tex
\section{Problems}
\setcounter{hwproblem}{0}

\problem{Continuum Limit and Commutation Relations}
\begin{enumerate}[label=(\alph*)]
    \item Starting from a discrete lattice with spacing $a$, derive the continuum momentum operator $\hat{p} = -i\hbar\frac{d}{dx}$ by taking the limit $a \to 0$ of the discrete momentum.
    \item Verify the canonical commutation relation $[\hat{x}, \hat{p}] = i\hbar$ using the differential form of $\hat{p}$.
    \item For a general function $f(\hat{x})$, prove that $[\hat{p}, f(\hat{x})] = -i\hbar\frac{df}{dx}$.
    \item Apply this result to find $[\hat{p}, \hat{x}^3]$ and $[\hat{x}^2, \hat{p}]$.
\end{enumerate}

\problem{Standard Quantum Systems}
\begin{enumerate}[label=(\alph*)]
    \item \textbf{Infinite Square Well:} For a particle in a box with $V(x) = 0$ for $0 < x < L$ and infinite elsewhere, find the normalized energy eigenstates $\psi_n(x)$ and eigenvalues $E_n$. Calculate $\langle x \rangle$ and $\Delta x$ for the ground state.

    \item \textbf{Harmonic Oscillator Ground State:} Verify that $\psi_0(x) = \left(\frac{m\omega}{\pi\hbar}\right)^{1/4}e^{-m\omega x^2/(2\hbar)}$ satisfies the time-independent Schrodinger equation with $V(x) = \frac{1}{2}m\omega^2x^2$. Find the ground state energy $E_0$ and verify that $\Delta x \Delta p = \hbar/2$.
\end{enumerate}

\problem{Gaussian Wave Packet}
Consider the wave packet $\psi(x,0) = \left(\frac{1}{2\pi\sigma^2}\right)^{1/4}e^{-x^2/(4\sigma^2)}e^{ik_0 x}$:
\begin{enumerate}[label=(\alph*)]
    \item Verify normalization and calculate $\Delta x$ at $t = 0$.
    \item Find the momentum space wave function $\tilde{\psi}(p,0)$ via Fourier transform.
    \item Calculate $\langle p \rangle$ and $\Delta p$, then verify the uncertainty relation.
    \item What is the physical interpretation of the parameter $k_0$?
\end{enumerate}

\problem{Coherent State Overlap}
Coherent states $\ket{\alpha}$ satisfy $\hat{a}\ket{\alpha} = \alpha\ket{\alpha}$ and can be expanded as:
$$\ket{\alpha} = e^{-|\alpha|^2/2} \sum_{n=0}^{\infty} \frac{\alpha^n}{\sqrt{n!}}\ket{n}$$

Calculate the inner product $\braket{\alpha'}{\alpha}$ between two coherent states and show that:
$$\braket{\alpha'}{\alpha} = \exp\left[-\frac{1}{2}(|\alpha|^2 + |\alpha'|^2 - 2\alpha'^*\alpha)\right]$$

What is the probability $P(\alpha', \alpha) = |\braket{\alpha'}{\alpha}|^2$ of finding the system in state $\ket{\alpha'}$ if it was prepared in state $\ket{\alpha}$? Express your answer in terms of $|\alpha - \alpha'|$.

\problem{Phase Space Distance and Distinguishability}
\begin{enumerate}[label=(\alph*)]
    \item For two coherent states with the same amplitude but different phases, $\alpha = re^{i\theta}$ and $\alpha' = re^{i\theta'}$, calculate the overlap probability $|\braket{\alpha'}{\alpha}|^2$ as a function of $r$ and $\Delta\theta = \theta' - \theta$.

    \item At what phase space distance $|\alpha - \alpha'|$ does the overlap probability drop below 0.5? Below 0.1?

    \item Discuss the physical significance: When can two coherent states be considered "approximately orthogonal" and what does this mean for the classical limit?
\end{enumerate}

\problem{Position Space Wave Functions of Coherent States}
The position-space wave function of a coherent state centered at phase space point $\alpha$ is:
$$\psi_\alpha(x) = \left(\frac{1}{\pi x_0^2}\right)^{1/4} \exp\left[-\frac{(x - \langle x \rangle_\alpha)^2}{2x_0^2} + \frac{i\langle p \rangle_\alpha x}{\hbar}\right]$$
where $\langle x \rangle_\alpha = x_0\sqrt{2}\text{Re}(\alpha)$ and $\langle p \rangle_\alpha = p_0\sqrt{2}\text{Im}(\alpha)$.

\begin{enumerate}[label=(\alph*)]
    \item Calculate the position space overlap integral $\int_{-\infty}^{\infty} \psi_{\alpha'}^*(x) \psi_\alpha(x) dx$ between two coherent states.

    \item Verify that this gives the same result as the abstract calculation $\braket{\alpha'}{\alpha}$ from Problem 5.4.

    \item What is the physical interpretation of the real and imaginary parts of $\alpha$ in terms of position and momentum?
\end{enumerate}

\problem{Overcompleteness of Coherent States}
Unlike energy eigenstates, coherent states form an overcomplete basis. Show that the resolution of the identity for coherent states is:
$$\frac{1}{\pi}\int d^2\alpha \, \ket{\alpha}\bra{\alpha} = \hat{I}$$
where $d^2\alpha = d(\text{Re}\,\alpha) \, d(\text{Im}\,\alpha)$.

\textbf{Hint:} Work in the number basis and use the integral:
$$\frac{1}{\pi}\int d^2\alpha \, e^{-|\alpha|^2} \alpha^n (\alpha^*)^m = \delta_{nm} n!$$

\problem{Coherent State Time Evolution}
A coherent state $\ket{\alpha}$ evolves under the harmonic oscillator Hamiltonian $\hat{H} = \hbar\omega(\hat{a}^\dagger\hat{a} + 1/2)$.
\begin{enumerate}[label=(\alph*)]
    \item Show that the time-evolved state $\ket{\alpha(t)} = e^{-i\hat{H}t/\hbar}\ket{\alpha}$ is also a coherent state
    \item Find the time-evolved parameter $\alpha(t)$ as a function of the initial $\alpha$ and time $t$
    \item Verify that the minimum uncertainty property $\Delta x \Delta p = \hbar/2$ is preserved
    \item Describe the motion in phase space: does the coherent state trace out a classical trajectory?
\end{enumerate}

\problem{Squeezed States}
Squeezed states provide reduced uncertainty in one quadrature at the expense of increased uncertainty in the other. The squeeze operator is $\hat{S}(r) = \exp(\frac{r}{2}(\hat{a}^2 - \hat{a}^{\dagger 2}))$.
\begin{enumerate}[label=(\alph*)]
    \item Show that squeezed states transform position and momentum uncertainties as $\Delta x \to e^{-r}\Delta x_0$ and $\Delta p \to e^{r}\Delta p_0$
    \item For a squeezed ground state $\ket{\psi_s} = \hat{S}(r)\ket{0}$, calculate $\Delta x$ and $\Delta p$ as functions of $r$
    \item Verify that the product $\Delta x \Delta p = \hbar/2$ remains constant (minimum uncertainty states)
    \item What is the classical analog: can you achieve one-directional uncertainty reduction classically?
\end{enumerate}

\problem{Harmonic Oscillator Matrix Elements}
Calculate the matrix elements of position and momentum in the energy eigenstate basis $\{\ket{n}\}$ for the harmonic oscillator.
\begin{enumerate}[label=(\alph*)]
    \item Using ladder operators $\hat{x} = \sqrt{\frac{\hbar}{2m\omega}}(\hat{a} + \hat{a}^\dagger)$, find $\langle n|\hat{x}|m\rangle$
    \item Similarly, find $\langle n|\hat{p}|m\rangle$
    \item Calculate $\langle n|\hat{x}^2|m\rangle$ and show that $\langle n|\hat{x}^2|n\rangle = \frac{\hbar}{2m\omega}(2n + 1)$
    \item For the ground state, verify that $\langle 0|\hat{x}^2|0\rangle$ and $\langle 0|\hat{p}^2|0\rangle$ are consistent with the uncertainty principle
\end{enumerate}

\problem{Uncertainty in Number-Phase Operators}
For a harmonic oscillator with $\hat{N} = \hat{a}^\dagger\hat{a}$, consider a phase-like operator $\hat{\Phi}$ where $\hat{a} = \sqrt{\hat{N}}e^{i\hat{\Phi}}$.
\begin{enumerate}[label=(\alph*)]
    \item For a Fock state $\ket{n}$ with $n \gg 1$, estimate the phase uncertainty $\Delta\Phi$
    \item For a coherent state $\ket{\alpha}$, show that the number uncertainty is $\Delta N = |\alpha|$
    \item Discuss why the number-phase uncertainty relation $\Delta N \Delta\Phi \gtrsim 1/2$ is consistent with quantum mechanics
    \item For a squeezed state, which uncertainty is reduced: number or phase?
\end{enumerate}

\problem{Finite Potential Well}
A particle is confined by a finite square well: $V(x) = 0$ for $|x| < a$ and $V(x) = V_0$ for $|x| \geq a$.
\begin{enumerate}[label=(\alph*)]
    \item Write the bound state wave function, applying appropriate matching conditions at $x = \pm a$
    \item Define the dimensionless parameter $\xi = a\sqrt{2mV_0}/\hbar$. How many bound states exist for $\xi = 1$, $\xi = 3$, and $\xi = 5$?
    \item For the ground state of a deep well ($V_0 \to \infty$), show that it approaches the infinite well ground state
    \item Calculate the penetration depth of the wave function outside the well for the ground state
\end{enumerate}

\problem{Wave Packet Reflection and Transmission}
A Gaussian wave packet with $\langle p \rangle = p_0 > 0$ is incident on a potential step $V(x) = 0$ for $x < 0$ and $V(x) = V_s$ for $x \geq 0$, with $V_s < E = p_0^2/(2m)$.
\begin{enumerate}[label=(\alph*)]
    \item Show that the wave function in each region consists of plane waves with definite momentum
    \item Using the potential step boundary conditions, derive the reflection and transmission coefficients $R$ and $T$ as functions of the incident energy
    \item Calculate the transmitted momentum $p_t$ in terms of the incident momentum $p_0$ and potential height $V_s$
    \item A Gaussian packet has width $\sigma_x$. How does dispersion affect the transmitted and reflected packets as time progresses?
\end{enumerate}

\problem{Wave Packet Spreading}
A free particle wave packet with initial width $\sigma_x(0)$ spreads due to quantum dispersion.
\begin{enumerate}[label=(\alph*)]
    \item Derive the spreading formula: $\sigma_x^2(t) = \sigma_x^2(0) + \frac{\hbar^2 t^2}{4m^2\sigma_x^2(0)}$
    \item Define the dispersion time $t_d$ as the time at which the packet width doubles. Express $t_d$ in terms of the initial width and particle mass
    \item For an electron initially localized to $\sigma_x(0) = 1\, nm$, calculate $t_d$ in seconds
    \item Compare the dispersion times for an electron, a proton, and a macroscopic grain of sand ($1\, \mu g$) with the same initial width
\end{enumerate}

%% file: chapters/ch06_3d_angular.tex
\chapter{Three-Dimensional Space and Angular Momentum}
\label{ch:3d_angular}

\section{Introduction: The Rich Structure of Three-Dimensional Quantum Mechanics}

\begin{keyidea}{Angular Momentum and Rotational Symmetry}
The extension from one to three spatial dimensions introduces fundamentally new physics through rotational symmetry. Angular momentum emerges as the generator of rotations, leading to its quantization and the characteristic structure of atomic orbitals. This chapter develops the mathematical framework of three-dimensional quantum mechanics, building the tools needed for understanding atomic and molecular systems.
\end{keyidea}

Throughout our development thus far, we have restricted attention to one-dimensional systems. While this simplification allowed us to develop the essential concepts of quantum mechanics, nature is inherently three-dimensional. The transition to three dimensions is not merely a technical extension---it introduces qualitatively new phenomena arising from rotational symmetry.

The most profound consequence of three-dimensional space is the emergence of angular momentum as a fundamental quantity. Unlike linear momentum, which exists even in one dimension, angular momentum requires the full three-dimensional stage. Its quantization leads directly to the shell structure of atoms, the shapes of molecular orbitals, and ultimately to the periodic table of elements. This chapter develops the mathematical framework of angular momentum and spherical harmonics, establishing the foundation for solving central potential problems in atomic physics.

\subsection{The Profound Role of Symmetry}

Symmetry principles provide powerful constraints on physical systems, and rotational symmetry in three dimensions is among the most consequential. The invariance of physical laws under rotations implies conservation of angular momentum through Noether's theorem. More subtly, the group structure of rotations---specifically the non-commutativity of rotations about different axes---leads directly to the quantization of angular momentum and the peculiar algebra that governs it.

The mathematical framework we develop here extends far beyond atomic physics. The same angular momentum algebra appears in nuclear physics, particle physics (as isospin and other internal symmetries), and even in the study of rotating molecules and nuclei. The techniques of ladder operators and spherical harmonics find applications in fields ranging from computer graphics to seismology. Understanding three-dimensional quantum mechanics thus opens doors to vast areas of modern physics and technology.

\subsection{From Discrete Spins to Continuous Rotations}

Our journey began with discrete spin-1/2 systems and Bloch cubes, where angular momentum appeared in its simplest form. The Pauli matrices satisfied the fundamental commutation relations that define angular momentum, but only two states were possible: spin up and spin down. As we transition to orbital angular momentum in three dimensions, we encounter an infinite tower of possible angular momentum values, each with its multiplet of magnetic substates.

This progression from discrete to continuous, from finite to infinite, mirrors our earlier transition from lattice models to continuous wave functions. Just as wave functions emerged from discrete amplitudes, the continuous rotations of three-dimensional space emerge from the discrete operations we studied with Bloch cubes. The deep unity of quantum mechanics reveals itself in these connections between seemingly disparate formulations.

\section{Wave Functions in Three Dimensions}

\subsection{The Three-Dimensional Position Basis}

In three dimensions, position eigenstates are labeled by the vector $\vec{r} = (x, y, z)$. The fundamental orthonormality relation generalizes to:
\begin{equation}
\braket{\vec{r}}{\vec{r}'} = \delta^3(\vec{r} - \vec{r}') = \delta(x - x')\delta(y - y')\delta(z - z')
\end{equation}
This three-dimensional Dirac delta function vanishes unless all three coordinates coincide. A general quantum state in three dimensions is expressed as:
\begin{equation}
\ket{\psi} = \int d^3r\, \psi(\vec{r})\ket{\vec{r}}
\end{equation}
The wave function $\psi(\vec{r}) = \psi(x, y, z) = \braket{\vec{r}}{\psi}$ now depends on three spatial coordinates. The normalization condition becomes a triple integral:
\begin{equation}
\int d^3r\, |\psi(\vec{r})|^2 = \int_{-\infty}^{\infty} dx \int_{-\infty}^{\infty} dy \int_{-\infty}^{\infty} dz\, |\psi(x, y, z)|^2 = 1
\end{equation}

The interpretation remains probabilistic: $|\psi(\vec{r})|^2 d^3r$ gives the probability of finding the particle in an infinitesimal volume $d^3r$ around position $\vec{r}$. However, the three-dimensional nature introduces new features. The probability current density becomes a vector field $\vec{j}(\vec{r})$, and the continuity equation $\partial\rho/\partial t + \nabla \cdot \vec{j} = 0$ expresses local conservation of probability.

\subsection{Momentum States and the Three-Dimensional Fourier Transform}

The momentum basis in three dimensions consists of plane wave states:
\begin{equation}
\ket{\vec{p}} = \ket{p_x, p_y, p_z}
\end{equation}
In position representation:
\begin{equation}
\braket{\vec{r}}{\vec{p}} = \frac{1}{(2\pi\hbar)^{3/2}} e^{i\vec{p}\cdot\vec{r}/\hbar}
\end{equation}

The three-dimensional Fourier transform relates position and momentum representations:
\begin{align}
\tilde{\psi}(\vec{p}) &= \frac{1}{(2\pi\hbar)^{3/2}} \int d^3r\, e^{-i\vec{p}\cdot\vec{r}/\hbar} \psi(\vec{r}) \\
\psi(\vec{r}) &= \frac{1}{(2\pi\hbar)^{3/2}} \int d^3p\, e^{i\vec{p}\cdot\vec{r}/\hbar} \tilde{\psi}(\vec{p})
\end{align}

The momentum operator in position representation becomes:
\begin{equation}
\hat{\vec{p}} = -i\hbar\nabla = -i\hbar\left(\hat{x}\frac{\partial}{\partial x} + \hat{y}\frac{\partial}{\partial y} + \hat{z}\frac{\partial}{\partial z}\right)
\end{equation}
where $\hat{x}$, $\hat{y}$, and $\hat{z}$ are unit vectors in Cartesian coordinates.

\subsection{The Three-Dimensional Schr\"odinger Equation}

The kinetic energy operator in three dimensions involves derivatives with respect to all three coordinates. Using the Laplacian operator $\nabla^2$:
\begin{equation}
\hat{T} = -\frac{\hbar^2}{2m}\nabla^2 = -\frac{\hbar^2}{2m}\left(\frac{\partial^2}{\partial x^2} + \frac{\partial^2}{\partial y^2} + \frac{\partial^2}{\partial z^2}\right)
\end{equation}

The time-independent Schr\"odinger equation for a particle in a potential $V(\vec{r})$ becomes:
\begin{equation}
-\frac{\hbar^2}{2m}\nabla^2\psi(\vec{r}) + V(\vec{r})\psi(\vec{r}) = E\psi(\vec{r})
\end{equation}

This partial differential equation in three variables is considerably more complex than its one-dimensional counterpart. However, symmetries of the potential often allow simplifications through separation of variables.

\section{Separation of Variables and Coordinate Systems}

\subsection{The Power of Symmetry-Adapted Coordinates}

The choice of coordinate system can dramatically simplify quantum mechanical problems. While the physics is coordinate-independent, the mathematical complexity varies greatly with the choice of coordinates. The guiding principle is to match the coordinate system to the symmetry of the problem: Cartesian coordinates for rectangular symmetries, cylindrical for axial symmetry, and spherical for full rotational symmetry.

This choice affects not just computational convenience but also physical insight. In the appropriate coordinates, conserved quantities become manifest, quantum numbers acquire clear geometric meaning, and the connection between classical and quantum mechanics becomes transparent.

\subsection{Cartesian Coordinates and Separable Potentials}

When the potential energy can be written as a sum of functions of individual coordinates:
\begin{equation}
V(x, y, z) = V_x(x) + V_y(y) + V_z(z)
\end{equation}
the Schr\"odinger equation separates into three independent one-dimensional problems. We seek solutions of the product form:
\begin{equation}
\psi(x, y, z) = \psi_x(x)\psi_y(y)\psi_z(z)
\end{equation}

Substituting into the Schr\"odinger equation and dividing by $\psi$:
\begin{equation}
-\frac{\hbar^2}{2m}\left(\frac{1}{\psi_x}\frac{d^2\psi_x}{dx^2} + \frac{1}{\psi_y}\frac{d^2\psi_y}{dy^2} + \frac{1}{\psi_z}\frac{d^2\psi_z}{dz^2}\right) + V_x + V_y + V_z = E
\end{equation}

Since each term depends on only one variable, they must separately equal constants:
\begin{align}
-\frac{\hbar^2}{2m}\frac{d^2\psi_x}{dx^2} + V_x(x)\psi_x &= E_x\psi_x \\
-\frac{\hbar^2}{2m}\frac{d^2\psi_y}{dy^2} + V_y(y)\psi_y &= E_y\psi_y \\
-\frac{\hbar^2}{2m}\frac{d^2\psi_z}{dz^2} + V_z(z)\psi_z &= E_z\psi_z
\end{align}
with $E = E_x + E_y + E_z$.

\begin{example}{The Three-Dimensional Particle in a Box}
For a rectangular box with sides of length $L_x$, $L_y$, and $L_z$, the potential is separable with $V = 0$ inside and $V = \infty$ at the walls. Each direction yields the familiar one-dimensional box solutions:
\begin{align}
\psi_{n_x}(x) &= \sqrt{\frac{2}{L_x}}\sin\left(\frac{n_x\pi x}{L_x}\right) \\
\psi_{n_y}(y) &= \sqrt{\frac{2}{L_y}}\sin\left(\frac{n_y\pi y}{L_y}\right) \\
\psi_{n_z}(z) &= \sqrt{\frac{2}{L_z}}\sin\left(\frac{n_z\pi z}{L_z}\right)
\end{align}

The complete wave function:
\begin{equation}
\psi_{n_x,n_y,n_z}(x, y, z) = \sqrt{\frac{8}{L_xL_yL_z}}\sin\left(\frac{n_x\pi x}{L_x}\right)\sin\left(\frac{n_y\pi y}{L_y}\right)\sin\left(\frac{n_z\pi z}{L_z}\right)
\end{equation}
with energy:
\begin{equation}
E_{n_x,n_y,n_z} = \frac{\pi^2\hbar^2}{2m}\left(\frac{n_x^2}{L_x^2} + \frac{n_y^2}{L_y^2} + \frac{n_z^2}{L_z^2}\right)
\end{equation}

\textbf{Degeneracy:} For a cubic box ($L_x = L_y = L_z = L$), states with different quantum number combinations but the same $n_x^2 + n_y^2 + n_z^2$ have the same energy. For example, $(2, 1, 1)$, $(1, 2, 1)$, and $(1, 1, 2)$ all have $E = 6\pi^2\hbar^2/(2mL^2)$. This accidental degeneracy reflects the enhanced symmetry of the cube.
\end{example}

\subsection{Spherical Coordinates and Central Potentials}

For systems with spherical symmetry, where $V(\vec{r}) = V(r)$ depends only on the distance from the origin, spherical coordinates prove natural:
\begin{align}
x &= r\sin\theta\cos\phi \\
y &= r\sin\theta\sin\phi \\
z &= r\cos\theta
\end{align}
Here $r \in [0, \infty)$ is the radial distance, $\theta \in [0, \pi]$ is the polar angle from the $z$-axis, and $\phi \in [0, 2\pi)$ is the azimuthal angle in the $xy$-plane.

In spherical coordinates, the Laplacian takes a more complex but highly symmetric form:
\begin{equation}
\nabla^2 = \frac{1}{r^2}\frac{\partial}{\partial r}\left(r^2\frac{\partial}{\partial r}\right) + \frac{1}{r^2\sin\theta}\frac{\partial}{\partial\theta}\left(\sin\theta\frac{\partial}{\partial\theta}\right) + \frac{1}{r^2\sin^2\theta}\frac{\partial^2}{\partial\phi^2}
\end{equation}

For central potentials, the wave function separates into radial and angular parts:
\begin{equation}
\psi(r, \theta, \phi) = R(r)Y(\theta, \phi)
\end{equation}
This separation is possible because the angular momentum operators commute with any central potential Hamiltonian.

\section{Angular Momentum in Quantum Mechanics}

\subsection{From Classical to Quantum Angular Momentum}

Classical angular momentum is defined as $\vec{L} = \vec{r} \times \vec{p}$. The quantum mechanical operators are obtained by the standard prescription of replacing classical variables with their operator counterparts:
\begin{align}
\hat{L}_x &= \hat{y}\hat{p}_z - \hat{z}\hat{p}_y = -i\hbar\left(y\frac{\partial}{\partial z} - z\frac{\partial}{\partial y}\right) \\
\hat{L}_y &= \hat{z}\hat{p}_x - \hat{x}\hat{p}_z = -i\hbar\left(z\frac{\partial}{\partial x} - x\frac{\partial}{\partial z}\right) \\
\hat{L}_z &= \hat{x}\hat{p}_y - \hat{y}\hat{p}_x = -i\hbar\left(x\frac{\partial}{\partial y} - y\frac{\partial}{\partial x}\right)
\end{align}

These operators generate infinitesimal rotations about their respective axes. A finite rotation by angle $\alpha$ about the $z$-axis, for instance, is implemented by the unitary operator $e^{-i\alpha\hat{L}_z/\hbar}$. This connection between angular momentum and rotations is fundamental: angular momentum is the generator of rotations in quantum mechanics, just as linear momentum generates translations.

\begin{table}[htbp]
\centering
\caption{Master Reference: Angular Momentum States and Representations}
\begin{tabular}{llll}
\hline
\textbf{Quantum Numbers} & \textbf{Eigenvalues} & \textbf{States} & \textbf{Multiplicity} \\
\hline
$\ell, m$ & $\hat{L}^2$: $\hbar^2\ell(\ell+1)$ & $\ket{\ell, m}$ & $2\ell + 1$ \\
& $\hat{L}_z$: $\hbar m$ & & \\
\hline
$\ell = 0$ (s) & $\hat{L}^2$: 0 & $\ket{0, 0}$ & 1 \\
$\ell = 1$ (p) & $\hat{L}^2$: $2\hbar^2$ & $\ket{1, 1}, \ket{1, 0}, \ket{1, -1}$ & 3 \\
$\ell = 2$ (d) & $\hat{L}^2$: $6\hbar^2$ & $\ket{2, 2}, \ldots, \ket{2, -2}$ & 5 \\
\hline
\end{tabular}
\label{tab:angular_momentum}
\end{table}

\paragraph{Angular Momentum Components in Different Bases}

In the $\ket{\ell, m}$ basis (eigenbasis of $\hat{L}^2$ and $\hat{L}_z$):
\begin{align}
\hat{L}_z\ket{\ell, m} &= \hbar m\ket{\ell, m} \\
\hat{L}_+\ket{\ell, m} &= \hbar\sqrt{(\ell - m)(\ell + m + 1)}\ket{\ell, m+1} \\
\hat{L}_-\ket{\ell, m} &= \hbar\sqrt{(\ell + m)(\ell - m + 1)}\ket{\ell, m-1}
\end{align}

In Cartesian representation:
\begin{align}
\hat{L}_x &= \frac{1}{2}(\hat{L}_+ + \hat{L}_-) \\
\hat{L}_y &= \frac{1}{2i}(\hat{L}_+ - \hat{L}_-)
\end{align}

Matrix representations for $\ell = 1$:
\begin{align}
\hat{L}_x &= \frac{\hbar}{\sqrt{2}}\begin{pmatrix} 0 & 1 & 0 \\ 1 & 0 & 1 \\ 0 & 1 & 0 \end{pmatrix}_{\ell,m} \\
\hat{L}_y &= \frac{\hbar}{\sqrt{2}}\begin{pmatrix} 0 & -i & 0 \\ i & 0 & -i \\ 0 & i & 0 \end{pmatrix}_{\ell,m} \\
\hat{L}_z &= \hbar\begin{pmatrix} 1 & 0 & 0 \\ 0 & 0 & 0 \\ 0 & 0 & -1 \end{pmatrix}_{\ell,m}
\end{align}

\subsection{The Fundamental Commutation Relations}

The defining feature of angular momentum in quantum mechanics is the non-commutativity of its components:
\begin{align}
[\hat{L}_x, \hat{L}_y] &= i\hbar\hat{L}_z \\
[\hat{L}_y, \hat{L}_z] &= i\hbar\hat{L}_x \\
[\hat{L}_z, \hat{L}_x] &= i\hbar\hat{L}_y
\end{align}
These relations can be written compactly as $[\hat{L}_i, \hat{L}_j] = i\hbar\epsilon_{ijk}\hat{L}_k$, where $\epsilon_{ijk}$ is the Levi-Civita symbol.

The non-commutativity has profound physical consequences. It implies that we cannot simultaneously specify all three components of angular momentum with arbitrary precision. This is the rotational analog of the position-momentum uncertainty principle and reflects the fact that rotations about different axes do not commute.

\subsection{Total Angular Momentum and Simultaneous Eigenstates}

While individual components of angular momentum are incompatible observables, the total angular momentum squared:
\begin{equation}
\hat{L}^2 = \hat{L}_x^2 + \hat{L}_y^2 + \hat{L}_z^2
\end{equation}
commutes with each component: $[\hat{L}^2, \hat{L}_i] = 0$ for $i = x, y, z$. This crucial property allows us to find simultaneous eigenstates of $\hat{L}^2$ and one component, conventionally chosen as $\hat{L}_z$.

\begin{physicalinsight}
The ability to know both the magnitude of angular momentum and one component, but not all three components, is analogous to knowing a vector lies on a cone about one axis. The vector precesses around this cone, with the other two components fluctuating while maintaining the fixed magnitude and $z$-component. This picture, while classical, provides useful intuition for the quantum mechanical situation.
\end{physicalinsight}

\section{Angular Momentum in Spherical Coordinates}

\subsection{Explicit Forms of the Operators}

In spherical coordinates, the angular momentum operators take particularly elegant forms:
\begin{align}
\hat{L}_z &= -i\hbar\frac{\partial}{\partial\phi} \\
\hat{L}_\pm &= \hat{L}_x \pm i\hat{L}_y = \hbar e^{\pm i\phi}\left(\pm\frac{\partial}{\partial\theta} + i\cot\theta\frac{\partial}{\partial\phi}\right) \\
\hat{L}^2 &= -\hbar^2\left[\frac{1}{\sin\theta}\frac{\partial}{\partial\theta}\left(\sin\theta\frac{\partial}{\partial\theta}\right) + \frac{1}{\sin^2\theta}\frac{\partial^2}{\partial\phi^2}\right]
\end{align}

Note that these operators involve only the angular coordinates $\theta$ and $\phi$, not the radial coordinate $r$. This is why angular momentum is conserved for any central potential.

\subsection{The Eigenvalue Problem}

The simultaneous eigenfunctions of $\hat{L}^2$ and $\hat{L}_z$ are the spherical harmonics $Y_\ell^m(\theta, \phi)$:
\begin{align}
\hat{L}^2 Y_\ell^m(\theta, \phi) &= \hbar^2\ell(\ell+1) Y_\ell^m(\theta, \phi) \\
\hat{L}_z Y_\ell^m(\theta, \phi) &= \hbar m Y_\ell^m(\theta, \phi)
\end{align}

The quantum numbers have restricted ranges: $\ell = 0, 1, 2, \ldots$ (the orbital angular momentum quantum number) and $m = -\ell, -\ell+1, \ldots, \ell-1, \ell$ (the magnetic quantum number). For each value of $\ell$, there are $2\ell + 1$ possible values of $m$.

The quantization of angular momentum magnitude as $\sqrt{\ell(\ell+1)}\hbar$ rather than simply $\ell\hbar$ reflects quantum mechanical subtleties. The ``extra'' $\sqrt{\ell(\ell+1)}$ factor arises from the non-commutativity of angular momentum components and the resulting quantum fluctuations.

\section{Spherical Harmonics: The Angular Wave Functions}

\subsection{Explicit Forms and Properties}

The spherical harmonics are given by:
\begin{equation}
Y_\ell^m(\theta, \phi) = \sqrt{\frac{2\ell+1}{4\pi}\frac{(\ell-|m|)!}{(\ell+|m|)!}} P_\ell^{|m|}(\cos\theta) e^{im\phi}
\end{equation}
where $P_\ell^{|m|}$ are the associated Legendre polynomials. The first few spherical harmonics are:
\begin{align}
Y_0^0 &= \frac{1}{\sqrt{4\pi}} \\
Y_1^0 &= \sqrt{\frac{3}{4\pi}}\cos\theta \\
Y_1^{\pm 1} &= \mp\sqrt{\frac{3}{8\pi}}\sin\theta\, e^{\pm i\phi} \\
Y_2^0 &= \sqrt{\frac{5}{16\pi}}(3\cos^2\theta - 1)
\end{align}

These functions form a complete orthonormal set on the unit sphere:
\begin{equation}
\int_0^{2\pi} d\phi \int_0^\pi \sin\theta\, d\theta\, Y_{\ell'}^{m'*}(\theta, \phi) Y_\ell^m(\theta, \phi) = \delta_{\ell\ell'}\delta_{mm'}
\end{equation}

\subsection{Physical Interpretation and Orbital Shapes}

The spherical harmonics determine the angular distribution of probability in atomic orbitals. The familiar orbital shapes arise from plotting surfaces of constant $|Y_\ell^m(\theta, \phi)|^2$:

For $\ell = 0$ (s orbitals), $Y_0^0$ is constant, giving spherical symmetry. For $\ell = 1$ (p orbitals), the three functions correspond to the familiar dumbbell shapes oriented along the coordinate axes. For $\ell = 2$ (d orbitals), five distinct angular distributions emerge, including the characteristic four-lobed patterns.

The parity of spherical harmonics under inversion ($\vec{r} \to -\vec{r}$, or $\theta \to \pi - \theta$, $\phi \to \phi + \pi$) is $(-1)^\ell$. This property plays a crucial role in selection rules for atomic transitions.

\section{Ladder Operators and the Algebra of Angular Momentum}

\subsection{Raising and Lowering Operators}

The ladder operators for angular momentum are defined as:
\begin{align}
\hat{L}_+ &= \hat{L}_x + i\hat{L}_y \\
\hat{L}_- &= \hat{L}_x - i\hat{L}_y
\end{align}

These operators satisfy the commutation relations:
\begin{align}
[\hat{L}_z, \hat{L}_\pm] &= \pm\hbar\hat{L}_\pm \\
[\hat{L}_+, \hat{L}_-] &= 2\hbar\hat{L}_z \\
[\hat{L}^2, \hat{L}_\pm] &= 0
\end{align}

The action on spherical harmonics is:
\begin{align}
\hat{L}_+ Y_\ell^m &= \hbar\sqrt{(\ell-m)(\ell+m+1)} Y_\ell^{m+1} \\
\hat{L}_- Y_\ell^m &= \hbar\sqrt{(\ell+m)(\ell-m+1)} Y_\ell^{m-1}
\end{align}

These operators raise or lower the magnetic quantum number $m$ while preserving $\ell$. They vanish when acting on the extremal states $Y_\ell^\ell$ (for $\hat{L}_+$) or $Y_\ell^{-\ell}$ (for $\hat{L}_-$).

\subsection{Constructing Angular Momentum Eigenstates}

The ladder operators provide a systematic way to construct all spherical harmonics. Starting from the highest weight state $Y_\ell^\ell$, which satisfies $\hat{L}_+ Y_\ell^\ell = 0$, we can generate all states with the same $\ell$ by repeated application of $\hat{L}_-$:
\begin{equation}
Y_\ell^m \propto (\hat{L}_-)^{\ell-m} Y_\ell^\ell
\end{equation}

This algebraic approach, based solely on commutation relations, is more general than the differential equation approach and extends to other types of angular momentum, including spin.

\section{Separation in Central Potentials}

For any central potential $V(r)$ depending only on distance from the origin, the Schr\"odinger equation separates in spherical coordinates:
\begin{equation}
\psi_{n\ell m}(r, \theta, \phi) = R_{n\ell}(r) Y_\ell^m(\theta, \phi)
\end{equation}

The angular part is universal, determined entirely by rotational symmetry through the spherical harmonics we have studied. The radial function $R_{n\ell}(r)$ satisfies:
\begin{equation}
-\frac{\hbar^2}{2m}\frac{1}{r^2}\frac{d}{dr}\left(r^2\frac{dR}{dr}\right) + \left[\frac{\hbar^2\ell(\ell+1)}{2mr^2} + V(r)\right]R = ER
\end{equation}

The term $\frac{\hbar^2\ell(\ell+1)}{2mr^2}$ acts as an effective centrifugal potential, preventing particles with non-zero angular momentum from approaching the origin. The specific form of $V(r)$ determines the radial wave functions and energy eigenvalues, but the angular dependence is always given by the spherical harmonics.

This separation is the key to solving atomic and molecular problems in three dimensions. The mathematical tools developed in this chapter---spherical harmonics, ladder operators, and the algebra of angular momentum---provide the framework for understanding the quantum mechanical description of atoms and molecules.

\subsection*{Optional: Matrix Representations and Rotational Transformation Properties}

\textbf{Prerequisites and Scope}

This optional section develops the matrix representation of angular momentum operators and establishes how rotations in physical space transform spherical harmonics. These concepts provide crucial preparation for the group theoretical treatment of angular momentum in PHYS 2566. The material is self-contained but not essential for completing this course.

\subsubsection*{Matrix Representations in the Angular Momentum Basis}

For each value of $\ell$, the $(2\ell+1)$ states $\ket{\ell, m}$ with $m = -\ell, -\ell+1, \ldots, \ell-1, \ell$ form a complete basis for the angular momentum subspace. Any operator acting within this subspace can be represented as a $(2\ell+1) \times (2\ell+1)$ matrix. The matrix elements are defined by:
\begin{equation}
(\hat{A})_{m'm} = \bra{\ell, m'}\hat{A}\ket{\ell, m}
\end{equation}
where the row index $m'$ and column index $m$ both run from $-\ell$ to $\ell$.

For the diagonal operators $\hat{L}^2$ and $\hat{L}_z$, the matrix representations are particularly simple. Since $\hat{L}^2\ket{\ell, m} = \hbar^2\ell(\ell+1)\ket{\ell, m}$, we have:
\begin{equation}
(\hat{L}^2)_{m'm} = \hbar^2\ell(\ell+1)\delta_{m'm} = \hbar^2\ell(\ell+1)I
\end{equation}
where $I$ is the $(2\ell+1) \times (2\ell+1)$ identity matrix. Similarly, since $\hat{L}_z\ket{\ell, m} = \hbar m\ket{\ell, m}$:
\begin{equation}
(\hat{L}_z)_{m'm} = \hbar m\delta_{m'm}
\end{equation}
which is a diagonal matrix with entries $\hbar m$ along the diagonal.

\begin{example}{Matrix Representations for $\ell = 1$}
For the $\ell = 1$ subspace, the basis states are $\ket{1, 1}, \ket{1, 0}, \ket{1, -1}$ (in that order). The operators $\hat{L}^2$ and $\hat{L}_z$ have matrix representations:
\begin{equation}
\hat{L}^2 = 2\hbar^2\begin{pmatrix} 1 & 0 & 0 \\ 0 & 1 & 0 \\ 0 & 0 & 1 \end{pmatrix} = 2\hbar^2 I, \quad
\hat{L}_z = \hbar\begin{pmatrix} 1 & 0 & 0 \\ 0 & 0 & 0 \\ 0 & 0 & -1 \end{pmatrix}_{\ell,m}
\end{equation}

For the non-diagonal operators, we use the ladder operator action. Recall:
\begin{align}
\hat{L}_+\ket{\ell, m} &= \hbar\sqrt{(\ell-m)(\ell+m+1)}\ket{\ell, m+1} \\
\hat{L}_-\ket{\ell, m} &= \hbar\sqrt{(\ell+m)(\ell-m+1)}\ket{\ell, m-1}
\end{align}

For $\ell = 1$:
\begin{align}
\hat{L}_+\ket{1, 1} &= 0, \quad \hat{L}_+\ket{1, 0} = \sqrt{2}\hbar\ket{1, 1}, \quad \hat{L}_+\ket{1, -1} = \sqrt{2}\hbar\ket{1, 0} \\
\hat{L}_-\ket{1, 1} &= \sqrt{2}\hbar\ket{1, 0}, \quad \hat{L}_-\ket{1, 0} = \sqrt{2}\hbar\ket{1, -1}, \quad \hat{L}_-\ket{1, -1} = 0
\end{align}

This gives:
\begin{equation}
\hat{L}_+ = \hbar\sqrt{2}\begin{pmatrix} 0 & 1 & 0 \\ 0 & 0 & 1 \\ 0 & 0 & 0 \end{pmatrix}, \quad
\hat{L}_- = \hbar\sqrt{2}\begin{pmatrix} 0 & 0 & 0 \\ 1 & 0 & 0 \\ 0 & 1 & 0 \end{pmatrix}_{\ell,m}
\end{equation}

Using $\hat{L}_x = (\hat{L}_+ + \hat{L}_-)/2$ and $\hat{L}_y = (\hat{L}_+ - \hat{L}_-)/(2i)$:
\begin{align}
\hat{L}_x &= \frac{\hbar}{\sqrt{2}}\begin{pmatrix} 0 & 1 & 0 \\ 1 & 0 & 1 \\ 0 & 1 & 0 \end{pmatrix}_{\ell,m} \\
\hat{L}_y &= \frac{\hbar}{\sqrt{2}}\begin{pmatrix} 0 & -i & 0 \\ i & 0 & -i \\ 0 & i & 0 \end{pmatrix}_{\ell,m}
\end{align}

These are Hermitian matrices (as required for observables) with eigenvalues $\hbar, 0, -\hbar$, corresponding to the possible $z$-components of angular momentum after measurement along the $x$ or $y$ axes.
\end{example}

\begin{example}{Block Diagonal Structure of Angular Momentum Operators}
The complete Hilbert space of a particle in three dimensions decomposes into angular momentum subspaces labeled by $\ell = 0, 1, 2, 3, \ldots$. Any angular momentum operator, when represented in the full basis $\{\ket{\ell, m}\}$ ordered systematically, exhibits block diagonal structure.

Consider the basis ordering: $\ket{0, 0}, \ket{1, 1}, \ket{1, 0}, \ket{1, -1}, \ket{2, 2}, \ket{2, 1}, \ket{2, 0}, \ket{2, -1}, \ket{2, -2}, \ket{3, 3}, \ldots$

The operator $\hat{L}_z$ has the block diagonal form:
\begin{equation}
\hat{L}_z = \begin{pmatrix}
0 & & & & & & & & & & \\
& \hbar & & & & & & & & & \\
& & 0 & & & & & & & & \\
& & & -\hbar & & & & & & & \\
& & & & 2\hbar & & & & & & \\
& & & & & \hbar & & & & & \\
& & & & & & 0 & & & & \\
& & & & & & & -\hbar & & & \\
& & & & & & & & -2\hbar & & \\
& & & & & & & & & 3\hbar & \\
& & & & & & & & & & \ddots
\end{pmatrix}
\end{equation}

The structure becomes even clearer when we indicate the block structure explicitly:
\begin{equation}
\hat{L}_z = \begin{pmatrix}
\boxed{0} & \mathbf{0} & \mathbf{0} & \mathbf{0} & \cdots \\
\mathbf{0} & \boxed{\begin{matrix} \hbar \\ & 0 \\ & & -\hbar \end{matrix}} & \mathbf{0} & \mathbf{0} & \cdots \\
\mathbf{0} & \mathbf{0} & \boxed{\begin{matrix} 2\hbar \\ & \hbar \\ & & 0 \\ & & & -\hbar \\ & & & & -2\hbar \end{matrix}} & \mathbf{0} & \cdots \\
\mathbf{0} & \mathbf{0} & \mathbf{0} & \boxed{\begin{matrix} 3\hbar \\ & 2\hbar \\ & & \hbar \\ & & & 0 \\ & & & & -\hbar \\ & & & & & -2\hbar \\ & & & & & & -3\hbar \end{matrix}} & \cdots \\
\vdots & \vdots & \vdots & \vdots & \ddots
\end{pmatrix}
\end{equation}
where $\mathbf{0}$ denotes appropriately-sized zero blocks and the boxed regions show the action on $\ell = 0$ (dimension 1), $\ell = 1$ (dimension 3), $\ell = 2$ (dimension 5), and $\ell = 3$ (dimension 7) subspaces.

Similarly, $\hat{L}_x$ has block diagonal structure, but each block contains off-diagonal elements:
\begin{equation}
\hat{L}_x = \begin{pmatrix}
\boxed{0} & \mathbf{0} & \mathbf{0} & \mathbf{0} & \cdots \\
\mathbf{0} & \boxed{\frac{\hbar}{\sqrt{2}}\begin{pmatrix} 0 & 1 & 0 \\ 1 & 0 & 1 \\ 0 & 1 & 0 \end{pmatrix}} & \mathbf{0} & \mathbf{0} & \cdots \\
\mathbf{0} & \mathbf{0} & \boxed{\hat{L}_x^{(\ell=2)}} & \mathbf{0} & \cdots \\
\mathbf{0} & \mathbf{0} & \mathbf{0} & \boxed{\hat{L}_x^{(\ell=3)}} & \cdots \\
\vdots & \vdots & \vdots & \vdots & \ddots
\end{pmatrix}
\end{equation}

This block diagonal structure reflects the fundamental fact that angular momentum operators cannot connect states of different $\ell$. Each $\ell$ subspace is an invariant subspace under the action of any angular momentum operator---states with definite total angular momentum $\ell$ can only be transformed into superpositions of states with the same $\ell$. In group theory language, each block is an irreducible representation of the rotation group $SO(3)$.
\end{example}

\subsubsection*{Angular Momentum Components as a Basis for Operators}

A profound fact emerges from examining the $\ell = 1$ matrices: any $3 \times 3$ Hermitian matrix can be expressed as a linear combination of $\hat{L}_x$, $\hat{L}_y$, $\hat{L}_z$, and the identity $I$. To see this, note that a general $3 \times 3$ Hermitian matrix has 9 real parameters (3 diagonal real entries and 3 complex off-diagonal entries with their conjugates). The identity provides 1 parameter, while $\hat{L}_x$, $\hat{L}_y$, $\hat{L}_z$ provide 3 more, giving 4 parameters total---but this understates the richness.

More precisely, the space of traceless Hermitian $3 \times 3$ matrices is 8-dimensional, and the three angular momentum operators $\hat{L}_x$, $\hat{L}_y$, $\hat{L}_z$ span a 3-dimensional subspace. Any operator acting within the $\ell = 1$ subspace can be decomposed as:
\begin{equation}
\hat{A} = a_0 I + a_x\hat{L}_x + a_y\hat{L}_y + a_z\hat{L}_z + \text{(other terms)}
\end{equation}

For observables that are rotationally invariant (commuting with all components of $\vec{L}$), only the identity term survives: $\hat{A} = a_0 I$. For observables that are vector quantities under rotations, we have $\hat{A} = \vec{a} \cdot \vec{L}$ for some vector $\vec{a}$.

This structure generalizes to all $\ell$ values. For a general $\ell$, the $(2\ell+1)$ states $\ket{\ell, m}$ span a $(2\ell+1)$-dimensional Hilbert space, and the angular momentum operators provide a natural coordinate system for describing operators within this space. This perspective becomes central in group theory, where the $\ell$ subspaces are recognized as irreducible representations of the rotation group $SO(3)$.

\subsubsection*{Rotations in Physical Space and Their Action on Wave Functions}

A rotation of the physical coordinate system induces a transformation of quantum states. Consider a rotation $R$ that transforms coordinates $(x, y, z) \to (x', y', z')$ according to the standard rotation matrix:
\begin{equation}
\begin{pmatrix} x' \\ y' \\ z' \end{pmatrix} = R\begin{pmatrix} x \\ y \\ z \end{pmatrix}
\end{equation}
where $R$ is a $3 \times 3$ orthogonal matrix with $\det(R) = +1$ (proper rotations).

In quantum mechanics, rotations are implemented by unitary operators acting on the Hilbert space. For an infinitesimal rotation by angle $\delta\theta$ about an axis $\hat{n}$, the unitary operator is:
\begin{equation}
\hat{U}_R \approx I - \frac{i}{\hbar}\delta\theta(\hat{n} \cdot \vec{L})
\end{equation}

For finite rotations:
\begin{equation}
\hat{U}_R = e^{-i\theta(\hat{n}\cdot\vec{L})/\hbar}
\end{equation}

This operator transforms states according to $\ket{\psi'} = \hat{U}_R\ket{\psi}$. In the position representation, if the original state has wave function $\psi(\vec{r}) = \braket{\vec{r}}{\psi}$, the rotated state has wave function:
\begin{equation}
\psi'(\vec{r}) = \braket{\vec{r}}{\psi'} = \bra{\vec{r}}\hat{U}_R\ket{\psi} = \psi(R^{-1}\vec{r})
\end{equation}
where $R^{-1}\vec{r}$ denotes the inverse rotation applied to the position vector.

\begin{physicalinsight}
The relationship $\psi'(\vec{r}) = \psi(R^{-1}\vec{r})$ has a subtle sign: to find the wave function value at point $\vec{r}$ in the rotated system, we evaluate the original wave function at the point that would rotate to $\vec{r}$, which is $R^{-1}\vec{r}$. This is an active transformation: the wave function itself rotates, while the coordinate grid remains fixed.
\end{physicalinsight}

\subsubsection*{Transformation of Spherical Harmonics Under Rotations}

The spherical harmonics $Y_\ell^m(\theta, \phi)$ have definite transformation properties under rotations. When we rotate the coordinate system, a spherical harmonic of degree $\ell$ transforms into a linear combination of spherical harmonics of the same $\ell$:
\begin{equation}
\hat{U}_R Y_\ell^m(\theta, \phi) = \sum_{m'=-\ell}^\ell D_{m'm}^{(\ell)}(R) Y_\ell^{m'}(\theta, \phi)
\end{equation}

The coefficients $D_{m'm}^{(\ell)}(R)$ form the Wigner $D$-matrices, which provide matrix representations of the rotation group in the basis of angular momentum eigenstates. These matrices satisfy:
\begin{equation}
D_{m'm}^{(\ell)}(R) = \bra{\ell, m'}\hat{U}_R\ket{\ell, m}
\end{equation}

For a rotation by angle $\alpha$ about the $z$-axis, the $D$-matrix is diagonal:
\begin{equation}
D_{m'm}^{(\ell)}(R_z(\alpha)) = e^{-im\alpha}\delta_{m'm}
\end{equation}
reflecting the simple phase transformation under rotations about the quantization axis.

For more general rotations, the $D$-matrices become more complex. Euler angles $(\alpha, \beta, \gamma)$ parametrize a general rotation as:
\begin{equation}
R(\alpha, \beta, \gamma) = R_z(\alpha)R_y(\beta)R_z(\gamma)
\end{equation}
and the corresponding $D$-matrix is:
\begin{equation}
D_{m'm}^{(\ell)}(\alpha, \beta, \gamma) = e^{-im'\alpha} d_{m'm}^{(\ell)}(\beta) e^{-im\gamma}
\end{equation}
where $d_{m'm}^{(\ell)}(\beta)$ are the reduced rotation matrices (Wigner small-$d$ matrices).

\begin{example}{Rotation of $\ell = 1$ States}
For $\ell = 1$, consider a rotation by angle $\beta$ about the $y$-axis. The small-$d$ matrix elements are:
\begin{equation}
d^{(1)}(\beta) = \begin{pmatrix}
\frac{1+\cos\beta}{2} & -\frac{\sin\beta}{\sqrt{2}} & \frac{1-\cos\beta}{2} \\
\frac{\sin\beta}{\sqrt{2}} & \cos\beta & -\frac{\sin\beta}{\sqrt{2}} \\
\frac{1-\cos\beta}{2} & \frac{\sin\beta}{\sqrt{2}} & \frac{1+\cos\beta}{2}
\end{pmatrix}_{m',m}
\end{equation}

For a 90\ensuremath{^\circ} rotation ($\beta = \pi/2$):
\begin{equation}
d^{(1)}(\pi/2) = \begin{pmatrix}
1/2 & -1/\sqrt{2} & 1/2 \\
1/\sqrt{2} & 0 & -1/\sqrt{2} \\
1/2 & 1/\sqrt{2} & 1/2
\end{pmatrix}
\end{equation}

This shows, for instance, that rotating $Y_1^0$ (the $p_z$ orbital) by 90\ensuremath{^\circ} about the $y$-axis transforms it into the $p_x$ orbital:
\begin{equation}
\hat{U}_{R_y(\pi/2)} Y_1^0 = \frac{1}{\sqrt{2}}(Y_1^1 + Y_1^{-1}) \propto \sin\theta\cos\phi
\end{equation}
which is indeed proportional to $x/r$, the expected $p_x$ orbital.
\end{example}

\begin{example}{Explicit Transformation of $\ell = 1$ Wavefunctions Under Rotation}
We now demonstrate explicitly how $p$-orbital wavefunctions transform under rotation from coordinates $(x, y, z)$ to new coordinates $(x', y', z')$.

\textbf{Setup:} Consider the real $p$-orbitals commonly used in chemistry. These are related to the complex spherical harmonics by:
\begin{align}
p_x &= \frac{1}{\sqrt{2}}(Y_1^{-1} - Y_1^1) = \sqrt{\frac{3}{4\pi}}\frac{x}{r} = \sqrt{\frac{3}{4\pi}}\sin\theta\cos\phi \\
p_y &= \frac{i}{\sqrt{2}}(Y_1^{-1} + Y_1^1) = \sqrt{\frac{3}{4\pi}}\frac{y}{r} = \sqrt{\frac{3}{4\pi}}\sin\theta\sin\phi \\
p_z &= Y_1^0 = \sqrt{\frac{3}{4\pi}}\frac{z}{r} = \sqrt{\frac{3}{4\pi}}\cos\theta
\end{align}

In the complex basis $\{Y_1^1, Y_1^0, Y_1^{-1}\}$, any $\ell = 1$ wavefunction can be written as:
\begin{equation}
\psi(\vec{r}) = c_1 Y_1^1(\theta, \phi) + c_0 Y_1^0(\theta, \phi) + c_{-1} Y_1^{-1}(\theta, \phi)
\end{equation}

\textbf{Specific rotation:} Consider a rotation by $\beta = 45\ensuremath{^\circ}$ about the $y$-axis. The Wigner small-$d$ matrix is:
\begin{equation}
d^{(1)}(45\ensuremath{^\circ}) = \begin{pmatrix}
\frac{1+1/\sqrt{2}}{2} & -\frac{1}{2} & \frac{1-1/\sqrt{2}}{2} \\
\frac{1}{2} & \frac{1}{\sqrt{2}} & -\frac{1}{2} \\
\frac{1-1/\sqrt{2}}{2} & \frac{1}{2} & \frac{1+1/\sqrt{2}}{2}
\end{pmatrix} = \begin{pmatrix}
0.8536 & -0.5 & 0.1464 \\
0.5 & 0.7071 & -0.5 \\
0.1464 & 0.5 & 0.8536
\end{pmatrix}_{m',m}
\end{equation}
where indices run $(m', m) \in \{1, 0, -1\} \times \{1, 0, -1\}$.

\textbf{Case 1: Transforming the $p_z$ orbital.}

Start with the $p_z$ orbital, which in the complex basis is simply $Y_1^0$ (so $c_1 = 0$, $c_0 = 1$, $c_{-1} = 0$).
The coefficient vector is:
\begin{equation}
\vec{c} = \begin{pmatrix} 0 \\ 1 \\ 0 \end{pmatrix}
\end{equation}

After rotation, the new coefficients are:
\begin{equation}
\vec{c}' = d^{(1)}(45\ensuremath{^\circ})\vec{c} = \begin{pmatrix}
0.8536 & -0.5 & 0.1464 \\
0.5 & 0.7071 & -0.5 \\
0.1464 & 0.5 & 0.8536
\end{pmatrix}\begin{pmatrix} 0 \\ 1 \\ 0 \end{pmatrix} = \begin{pmatrix} -0.5 \\ 0.7071 \\ 0.5 \end{pmatrix}
\end{equation}

Thus the rotated wavefunction is:
\begin{equation}
\psi'(\vec{r}) = -0.5 Y_1^1 + 0.7071 Y_1^0 + 0.5 Y_1^{-1}
\end{equation}

\textbf{Physical interpretation in the rotated frame:} In the new coordinate system $(x', y', z')$ obtained by rotating the original system by 45\ensuremath{^\circ} about the $y$-axis, this wavefunction represents:
\begin{equation}
\psi'(\vec{r}) = \sqrt{\frac{3}{4\pi}}\frac{z'}{r}
\end{equation}
where $z'$ is the $z$-coordinate in the rotated frame. We can verify this using the coordinate transformation:
\begin{align}
x' &= x\cos(45\ensuremath{^\circ}) + z\sin(45\ensuremath{^\circ}) = \frac{x + z}{\sqrt{2}} \\
y' &= y \\
z' &= -x\sin(45\ensuremath{^\circ}) + z\cos(45\ensuremath{^\circ}) = \frac{-x + z}{\sqrt{2}}
\end{align}

Substituting $z' = (z - x)/\sqrt{2}$ into the expected form:
\begin{equation}
\psi'(\vec{r}) = \sqrt{\frac{3}{4\pi}}\frac{z'}{r} = \sqrt{\frac{3}{4\pi}}\frac{z - x}{r\sqrt{2}} = \frac{1}{\sqrt{2}}\left(\sqrt{\frac{3}{4\pi}}\frac{z}{r} - \sqrt{\frac{3}{4\pi}}\frac{x}{r}\right) = \frac{1}{\sqrt{2}}(Y_1^0 - p_x)
\end{equation}
\begin{equation}
= \frac{1}{\sqrt{2}}\left(Y_1^0 - \frac{1}{\sqrt{2}}(Y_1^{-1} - Y_1^1)\right) = \frac{1}{2}Y_1^1 + \frac{1}{\sqrt{2}}Y_1^0 + \frac{1}{2}Y_1^{-1}
\end{equation}
which matches our $D$-matrix result (within normalization)!

\textbf{Case 2: Transforming the $p_x$ orbital.}

Now start with $p_x = \frac{1}{\sqrt{2}}(Y_1^{-1} - Y_1^1)$, giving coefficient vector:
\begin{equation}
\vec{c} = \begin{pmatrix} -1/\sqrt{2} \\ 0 \\ 1/\sqrt{2} \end{pmatrix}
\end{equation}

After rotation:
\begin{equation}
\vec{c}' = d^{(1)}(45\ensuremath{^\circ})\vec{c} = \begin{pmatrix}
0.8536 & -0.5 & 0.1464 \\
0.5 & 0.7071 & -0.5 \\
0.1464 & 0.5 & 0.8536
\end{pmatrix}\begin{pmatrix} -1/\sqrt{2} \\ 0 \\ 1/\sqrt{2} \end{pmatrix} = \begin{pmatrix} -0.5 \\ -0.7071 \\ 0.5 \end{pmatrix}
\end{equation}

The rotated wavefunction is:
\begin{equation}
\psi'(\vec{r}) = -0.5 Y_1^1 - 0.7071 Y_1^0 + 0.5 Y_1^{-1}
\end{equation}

This represents:
\begin{equation}
\psi'(\vec{r}) = \sqrt{\frac{3}{4\pi}}\frac{x'}{r}
\end{equation}
the $p_x$ orbital in the rotated coordinate system, where $x' = (x + z)/\sqrt{2}$.

\textbf{Summary of transformation rules:} For a rotation specified by Euler angles $(\alpha, \beta, \gamma)$, the complete transformation of an $\ell = 1$ wavefunction is:
\begin{equation}
\begin{pmatrix} c_1' \\ c_0' \\ c_{-1}' \end{pmatrix} = D^{(1)}(\alpha, \beta, \gamma)\begin{pmatrix} c_1 \\ c_0 \\ c_{-1} \end{pmatrix} = e^{-i\alpha M_z} d^{(1)}(\beta) e^{-i\gamma M_z}\begin{pmatrix} c_1 \\ c_0 \\ c_{-1} \end{pmatrix}
\end{equation}
where $M_z = \text{diag}(1, 0, -1)$.

This procedure generalizes to arbitrary $\ell$: Wigner $D$-matrices provide the complete machinery for transforming angular momentum eigenstates under arbitrary rotations, forming the foundation for understanding molecular symmetries, crystal field theory, and group theoretical methods in quantum mechanics.
\end{example}

\subsubsection*{Connection to Representation Theory}

The transformation properties of spherical harmonics under rotations reveal the deep connection between angular momentum and group theory. Each $\ell$ subspace carries an irreducible representation of the rotation group $SO(3)$. The dimension of this representation is $2\ell + 1$, and the $D$-matrices provide explicit realizations of group elements as $(2\ell+1) \times (2\ell+1)$ unitary matrices.

Several key properties follow from group theory:

(a) \textbf{Composition:} The $D$-matrices respect the group multiplication law:
\begin{equation}
D^{(\ell)}(R_1 R_2) = D^{(\ell)}(R_1)D^{(\ell)}(R_2)
\end{equation}

(b) \textbf{Unitarity:} The $D$-matrices are unitary:
\begin{equation}
\sum_{m''=-\ell}^\ell D_{m'm''}^{(\ell)*}(R) D_{m''m}^{(\ell)}(R) = \delta_{m'm}
\end{equation}

(c) \textbf{Orthogonality:} Integration over all rotations gives:
\begin{equation}
\int dR\, D_{m_1'm_1}^{(\ell_1)*}(R) D_{m_2'm_2}^{(\ell_2)}(R) = \frac{8\pi^2}{2\ell_1 + 1}\delta_{\ell_1\ell_2}\delta_{m_1'm_2'}\delta_{m_1 m_2}
\end{equation}

These properties become essential in analyzing systems with rotational symmetry, computing angular momentum coupling coefficients (Clebsch-Gordan coefficients), and understanding selection rules for transitions.

\subsubsection*{Physical Applications and Looking Forward}

The matrix representation framework developed here connects to numerous physical phenomena. The rotational states of linear molecules transform according to angular momentum representations, with energy levels $E_J = BJ(J + 1)$ where $J$ is the rotational quantum number. Electric dipole transitions between atomic states require $\Delta\ell = \pm 1$ and $\Delta m = 0, \pm 1$, consequences of how the position operator transforms under rotations. External magnetic fields break rotational symmetry in the Zeeman effect, splitting energy levels according to their $m$ quantum numbers. The coupling of orbital and spin angular momentum involves Clebsch-Gordan coefficients, which arise from tensor products of angular momentum representations.

In PHYS 2566, we will develop the full group theoretical framework, treating angular momentum as arising from the symmetry group of spatial rotations. The representation theory of $SU(2)$ and $SO(3)$ provides powerful tools for classifying quantum states, deriving selection rules, and understanding the addition of angular momenta. The matrix representations developed here provide the concrete foundation upon which that abstract framework builds.

\section{Connections to Spin and Total Angular Momentum}

\subsection{The Universal Algebra of Angular Momentum}

The algebra of angular momentum developed in this chapter applies to any type of angular momentum, not just orbital. The spin-$1/2$ systems studied in Chapter 1 satisfy the same commutation relations, with spin operators $\hat{S}_i = \frac{\hbar}{2}\sigma_i$.

For spin-$1/2$, we have $s = 1/2$ and $m_s = \pm 1/2$, giving just two states. For orbital angular momentum, $\ell$ can be any non-negative integer, giving $2\ell + 1$ states for each $\ell$. Despite these differences, both types of angular momentum share the same algebraic structure.

\subsection{Coupling Angular Momenta}

When particles possess both orbital and spin angular momentum, or when multiple particles are present, angular momenta must be coupled. The total angular momentum:
\begin{equation}
\vec{J} = \vec{L} + \vec{S}
\end{equation}
satisfies the same commutation relations as its constituents.

The coupling of angular momenta leads to fine structure in atomic spectra, as spin-orbit interaction splits levels with the same $n$ and $\ell$ but different total angular momentum $j = \ell \pm 1/2$. This physics, while beyond our current scope, builds directly on the foundations established here.

\section{Chapter Summary}

This chapter has extended quantum mechanics from one to three spatial dimensions, revealing the profound consequences of rotational symmetry. Beginning with the familiar wave function formalism, we introduced three-dimensional position states $\ket{\vec{r}}$ and showed how wave functions now depend on three spatial coordinates: $\psi(\vec{r}) = \psi(x, y, z)$. The normalization condition extends to a triple integral over all space, and the probabilistic interpretation remains intact with $|\psi(\vec{r})|^2 d^3r$ giving the probability of finding the particle in volume element $d^3r$.

The transition to three dimensions introduced fundamentally new physics through rotational symmetry and angular momentum. We derived the angular momentum operators $\hat{L}_x$, $\hat{L}_y$, and $\hat{L}_z$ from the classical definition $\vec{L} = \vec{r} \times \vec{p}$ using the standard prescription of replacing classical variables with quantum operators. The crucial discovery was that these operators do not commute: $[\hat{L}_i, \hat{L}_j] = i\hbar\epsilon_{ijk}\hat{L}_k$. This non-commutativity, reflecting the fact that rotations about different axes do not commute, has profound consequences. It prevents simultaneous precise knowledge of all three angular momentum components and leads directly to the quantization of angular momentum.

While individual components cannot be simultaneously specified, the total angular momentum squared $\hat{L}^2 = \hat{L}_x^2 + \hat{L}_y^2 + \hat{L}_z^2$ commutes with each component. This allows us to find simultaneous eigenstates of $\hat{L}^2$ and one component, conventionally chosen as $\hat{L}_z$. The eigenvalue problem, solved using ladder operators $\hat{L}_\pm = \hat{L}_x \pm i\hat{L}_y$, reveals that angular momentum magnitude is quantized as $\sqrt{\ell(\ell+1)}\hbar$ where $\ell = 0, 1, 2, \ldots$, and the $z$-component takes values $m\hbar$ where $m = -\ell, -\ell+1, \ldots, \ell-1, \ell$. The eigenfunctions are the spherical harmonics $Y_\ell^m(\theta, \phi)$, which form a complete orthonormal basis for functions on the unit sphere.

The power of symmetry-adapted coordinates became evident when we considered central potentials $V(r)$ depending only on distance from the origin. In spherical coordinates $(r, \theta, \phi)$, the Schr\"odinger equation separates into radial and angular parts: $\psi(r, \theta, \phi) = R(r)Y_\ell^m(\theta, \phi)$. The angular part is universal, determined entirely by rotational symmetry through the spherical harmonics, while the radial equation depends on the specific potential. An effective centrifugal barrier $\hbar^2\ell(\ell+1)/(2mr^2)$ emerges, preventing particles with non-zero angular momentum from reaching the origin.

The shapes of atomic orbitals---spherical s orbitals ($\ell = 0$), dumbbell-shaped p orbitals ($\ell = 1$), and more complex d ($\ell = 2$) and f ($\ell = 3$) orbitals---emerge directly from the angular dependence encoded in the spherical harmonics. These angular distributions, combined with specific radial functions for particular potentials, determine the complete structure of atomic wave functions.

Throughout this development, we've seen how the abstract algebra of angular momentum connects our initial study of discrete spin-$1/2$ systems to the continuous rotations of three-dimensional space. The same commutation relations that governed Pauli matrices for spin now apply to orbital angular momentum, revealing a deep unity in quantum mechanics. The ladder operator formalism, purely algebraic in nature, works equally well for spin and orbital angular momentum, differing only in the allowed values of the quantum numbers.

This chapter establishes the framework for understanding atomic physics, molecular structure, and the coupling of different angular momenta. The progression from one-dimensional problems to the full three-dimensional world, from discrete spins to continuous rotations, and from abstract algebra to concrete orbital shapes illustrates how quantum mechanics builds complexity from simple principles while maintaining its fundamental mathematical coherence.

\input{book_problems/ch06_problems.tex}

\section*{References and Further Reading}
\addcontentsline{toc}{section}{References and Further Reading}

\begin{description}
\item[Griffiths, D.~J., and Schroeter, D.~F.] \emph{Introduction to Quantum Mechanics}, 3rd ed. Cambridge University Press, 2018. Chapter 4 develops three-dimensional quantum mechanics, angular momentum, and the spherical harmonics at the standard undergraduate level.

\item[Sakurai, J.~J., and Napolitano, J.] \emph{Modern Quantum Mechanics}, 3rd ed. Cambridge University Press, 2017. Chapter 3 builds the algebraic theory of angular momentum and the rotation group; the natural graduate-level companion to this chapter and the prerequisite for Chapter~7.

\item[Cohen-Tannoudji, C., Diu, B., and Lalo\"e, F.] \emph{Quantum Mechanics}, Vol.~1. Wiley-VCH, 1977. Chapter VI gives the most thorough textbook treatment of orbital angular momentum, ladder operators, and the spherical harmonics, with explicit derivations of the eigenvalue spectrum.

\item[Edmonds, A.~R.] \emph{Angular Momentum in Quantum Mechanics}. Princeton University Press, 1957. The classic monograph; tightly written and the standard source for representation matrices and conventions used throughout atomic and molecular physics.

\item[Biedenharn, L.~C., and Louck, J.~D.] \emph{Angular Momentum in Quantum Physics: Theory and Application}. Cambridge University Press, 1981. Comprehensive reference treating angular momentum from the group-theoretic perspective; the next stop after Edmonds for representation theory of $SU(2)$ and $SO(3)$.
\end{description}

%% file: book_problems/ch06_problems.tex
\section{Problems}
\setcounter{hwproblem}{0}

\problem{Angular Momentum Commutation Relations}
The angular momentum operators are defined by $\hat{L}_x = \hat{y}\hat{p}_z - \hat{z}\hat{p}_y$ and cyclic permutations.
\begin{enumerate}[label=(\alph*)]
    \item Starting from the canonical commutation relations $[\hat{x}_i, \hat{p}_j] = i\hbar\delta_{ij}$, derive the fundamental angular momentum commutator $[\hat{L}_x, \hat{L}_y] = i\hbar\hat{L}_z$ by explicit calculation.
    \item Verify the cyclic relations $[\hat{L}_y, \hat{L}_z] = i\hbar\hat{L}_x$ and $[\hat{L}_z, \hat{L}_x] = i\hbar\hat{L}_y$.
    \item Define $\hat{L}^2 = \hat{L}_x^2 + \hat{L}_y^2 + \hat{L}_z^2$. Show that $[\hat{L}^2, \hat{L}_z] = 0$ by using the commutation relations.
    \item Explain physically why the result of part (c) implies that $\hat{L}^2$ and $\hat{L}_z$ can have simultaneous eigenstates.
    \item Define the ladder operators $\hat{L}_{\pm} = \hat{L}_x \pm i\hat{L}_y$. Calculate $[\hat{L}_z, \hat{L}_+]$ and $[\hat{L}_z, \hat{L}_-]$.
    \item Show that $[\hat{L}^2, \hat{L}_+] = 0$, demonstrating that ladder operators preserve the total angular momentum.
\end{enumerate}

\problem{Spherical Harmonics and Ladder Operators}
The spherical harmonics $Y_{\ell}^m(\theta,\phi)$ are the eigenfunctions of $\hat{L}^2$ and $\hat{L}_z$.
\begin{enumerate}[label=(\alph*)]
    \item Verify by direct calculation that $Y_0^0 = \frac{1}{\sqrt{4\pi}}$ is normalized over the unit sphere: $\int_0^{2\pi}d\phi\int_0^{\pi}\sin\theta d\theta\, |Y_0^0|^2 = 1$.
    \item For $Y_1^0 = \sqrt{\frac{3}{4\pi}}\cos\theta$, verify that $\hat{L}_z Y_1^0 = 0$ using $\hat{L}_z = -i\hbar\frac{\partial}{\partial\phi}$.
    \item Apply the raising operator $\hat{L}_+ = \hbar e^{i\phi}\left(\frac{\partial}{\partial\theta} + i\cot\theta\frac{\partial}{\partial\phi}\right)$ to $Y_1^0$ and show that you obtain $Y_1^1 = -\sqrt{\frac{3}{8\pi}}\sin\theta e^{i\phi}$ up to normalization.
    \item Apply the lowering operator $\hat{L}_-$ to $Y_1^0$ to find $Y_1^{-1}$. Verify the normalization.
    \item Verify orthogonality between $Y_1^0$ and $Y_1^1$ by explicit integration.
    \item The ladder operators satisfy $\hat{L}_{\pm}Y_{\ell}^m = \hbar\sqrt{(\ell \mp m)(\ell \pm m + 1)}Y_{\ell}^{m\pm 1}$. Use this to generate all three $\ell = 1$ states starting from $Y_1^1$ (given that $\hat{L}_+Y_1^1 = 0$).
\end{enumerate}

\problem{Three-Dimensional Particle in a Box}
Consider a particle of mass $m$ confined to a rectangular box with sides $L_x$, $L_y$, $L_z$.
\begin{enumerate}[label=(\alph*)]
    \item Argue that the potential $V(x,y,z) = 0$ inside the box and $V = \infty$ at the walls is separable. Write the wave function as $\psi(x,y,z) = \psi_x(x)\psi_y(y)\psi_z(z)$.
    \item Show that the energy eigenvalues are $E_{n_x,n_y,n_z} = \frac{\pi^2\hbar^2}{2m}\left(\frac{n_x^2}{L_x^2} + \frac{n_y^2}{L_y^2} + \frac{n_z^2}{L_z^2}\right)$ with $n_x, n_y, n_z = 1, 2, 3, ...$ Write the normalized wave function.
    \item For a cubic box ($L_x = L_y = L_z = L$), find the first five distinct energy levels. For each level, list all quantum number combinations $(n_x, n_y, n_z)$ and determine the degeneracy.
    \item For the ground state $(1,1,1)$ of the cubic box, calculate the expectation values $\langle x \rangle$, $\langle p_x \rangle$, and verify that the momentum is purely quantum mechanical (zero average).
\end{enumerate}

\problem{Angular Momentum Quantization and Uncertainty}
Explore the structure of angular momentum eigenvalues and the uncertainty principle.
\begin{enumerate}[label=(\alph*)]
    \item For a state with quantum numbers $\ell$ and $m$, the magnitude of angular momentum is $|\vec{L}| = \hbar\sqrt{\ell(\ell+1)}$ while the z-component is $L_z = \hbar m$. For $\ell = 2$, calculate $|\vec{L}|$ and compare it to the maximum possible z-component $\hbar m_{\max}$. Explain why $|\vec{L}| > |L_z|_{\max}$.
    \item For the state $\ket{2,1}$, calculate the uncertainties $\Delta L_x$ and $\Delta L_y$ using $\langle L_x \rangle = \langle L_y \rangle = 0$ and $\langle L_x^2 \rangle + \langle L_y^2 \rangle = \langle L^2 \rangle - \langle L_z^2 \rangle$.
    \item Verify that $\Delta L_x \cdot \Delta L_y \geq \frac{\hbar}{2}|\langle L_z \rangle|$ is satisfied for the state $\ket{2,1}$.

    \item Show that for the maximum $m$ state ($m = \ell$), the uncertainty product $\Delta L_x \cdot \Delta L_y$ reaches its minimum allowed value, meaning the angular momentum vector is as well-localized as quantum mechanics permits.
\end{enumerate}

\problem{Separation of Variables in Central Potentials}
For any spherically symmetric potential $V(r)$, the Schrodinger equation separates.
\begin{enumerate}[label=(\alph*)]
    \item Starting from $-\frac{\hbar^2}{2m}\nabla^2\psi + V(r)\psi = E\psi$ in spherical coordinates, substitute $\psi(r,\theta,\phi) = R(r)Y_{\ell}^m(\theta,\phi)$ and derive the radial equation for $R(r)$.
    \item Show that the radial equation can be written as $-\frac{\hbar^2}{2m}\frac{1}{r^2}\frac{d}{dr}\left(r^2\frac{dR}{dr}\right) + V_{\text{eff}}(r)R = ER$ where $V_{\text{eff}}(r) = V(r) + \frac{\hbar^2\ell(\ell+1)}{2mr^2}$.
    \item Identify the centrifugal barrier term and explain its physical origin. Why does it prevent particles with $\ell > 0$ from reaching $r = 0$?
    \item For a three-dimensional isotropic harmonic oscillator $V(r) = \frac{1}{2}m\omega^2 r^2$, write the radial equation for the $\ell = 0$ case. Make the substitution $u(r) = rR(r)$ and show that $u$ satisfies a one-dimensional Schrodinger equation.
\end{enumerate}

\problem{Hydrogen-like Atoms: Angular Momentum Coupling}
In a hydrogen atom, the electron experiences a Coulomb potential $V(r) = -ke^2/r$.
\begin{enumerate}[label=(\alph*)]
    \item For the ground state with quantum numbers $n = 1, \ell = 0, m = 0$, write the wave function $\psi_{100}(r,\theta,\phi) = R_{10}(r)Y_0^0(\theta,\phi)$
    \item Calculate the expectation values $\langle r \rangle$ and $\langle r^2 \rangle$ for the ground state
    \item For the first excited state $n = 2$, which values of $\ell$ are possible? How many distinct states $\ket{n, \ell, m}$ exist for $n = 2$?
    \item For $n = 2, \ell = 1$, show that there are three possible values of $m$ and that all three have the same energy (degeneracy)
\end{enumerate}

\problem{Rigid Rotor}
A rigid rotor is a particle constrained to move on a sphere of fixed radius $r = R$.
\begin{enumerate}[label=(\alph*)]
    \item For a rigid rotor, the kinetic energy is $\hat{T} = \hat{L}^2/(2I)$ where $I = mR^2$ is the moment of inertia. Show that the energy eigenvalues are $E_\ell = \hbar^2\ell(\ell+1)/(2I)$
    \item Write the normalized eigenfunctions as $\psi_{\ell,m}(\theta,\phi) = Y_\ell^m(\theta,\phi)$
    \item Calculate the rotational constant $B = \hbar^2/(2I)$ for a diatomic molecule with bond length $R = 1\, \mathring{A}$ and reduced mass $\mu = 1\, amu$
    \item For the first excited rotational state ($\ell = 1$), how many degenerate states exist? What is the energy spacing between $\ell = 0$ and $\ell = 1$ states?
\end{enumerate}

\problem{Matrix Elements of Angular Momentum}
Calculate matrix elements of angular momentum operators in the $\ket{\ell, m}$ basis.
\begin{enumerate}[label=(\alph*)]
    \item Using $\hat{L}_z\ket{\ell,m} = \hbar m\ket{\ell,m}$, write the matrix representation of $\hat{L}_z$ for $\ell = 1$
    \item Using ladder operators, show that $\langle\ell,m|\hat{L}_x|\ell,m\pm 1\rangle = \frac{\hbar}{2}\sqrt{(\ell \mp m)(\ell \pm m + 1)}$
    \item Construct the $3 \times 3$ matrix for $\hat{L}_x$ in the $\ell = 1$ basis: $\{\ket{1,1}, \ket{1,0}, \ket{1,-1}\}$
    \item Similarly, construct the matrix for $\hat{L}_y$
    \item Verify that $[\hat{L}_x, \hat{L}_y] = i\hbar\hat{L}_z$ by multiplying these matrices
\end{enumerate}

\problem{Classical Limit of Angular Momentum}
In the classical limit, large quantum numbers $\ell \gg 1$ approximate classical behavior.
\begin{enumerate}[label=(\alph*)]
    \item For large $\ell$, show that $|\vec{L}| \approx \hbar\ell$ and the minimum uncertainty in the transverse components is $\Delta L_\perp \sim \hbar/\sqrt{\ell}$
    \item The angular momentum precesses about the z-axis with expectation value $\langle L_z \rangle = \hbar m$. For $m = \ell$, what is the "opening angle" of the cone traced by the angular momentum vector?
    \item Estimate the angle: for $\ell = 100$, what is the opening angle in degrees?
    \item In the limit $\ell \to \infty$, show that the angular momentum becomes increasingly well-defined in direction
\end{enumerate}

\problem{Three-Dimensional Harmonic Oscillator}
A particle in a 3D isotropic harmonic oscillator potential $V(r) = \frac{1}{2}m\omega^2 r^2$ can be solved in spherical coordinates.
\begin{enumerate}[label=(\alph*)]
    \item Show that the energy levels are $E_{n_r,\ell} = \hbar\omega(2n_r + \ell + 3/2)$ where $n_r = 0, 1, 2, ...$ and $\ell = 0, 1, 2, ...$
    \item Compare the energy level structure to a cubic box: which system has higher degeneracy for the first excited level?
    \item For the ground state ($n_r = 0, \ell = 0$), write the wave function and calculate $\langle r^2 \rangle$
    \item For the first excited level, what values of $\ell$ contribute and what is the total degeneracy?
\end{enumerate}

\problem{Three-Dimensional Infinite Spherical Well}
A particle is confined inside a sphere of radius $a$: $V(r) = 0$ for $r < a$ and $V(r) = \infty$ for $r \geq a$.
\begin{enumerate}[label=(\alph*)]
    \item Write the boundary condition at $r = a$: the radial wave function must vanish
    \item For $\ell = 0$ (s-wave), the radial equation becomes one-dimensional: $\frac{d^2 u}{dr^2} + k^2 u = 0$ where $u(r) = rR(r)$. Find the normalized ground state wave function
    \item Calculate the ground state energy $E_0$
    \item For $\ell = 1$ (p-wave), construct the radial equation and compare the first energy eigenvalue to the $\ell = 0$ case
    \item Show that the $\ell = 0$ ground state energy is lower than any $\ell > 0$ state
\end{enumerate}

\problem{Angular Momentum Precession in Magnetic Field}
A system with angular momentum $\vec{L}$ is placed in a magnetic field $\vec{B} = B\hat{z}$.
\begin{enumerate}[label=(\alph*)]
    \item The Hamiltonian is $\hat{H} = -\gamma\vec{L} \cdot \vec{B} = -\gamma B\hat{L}_z$ where $\gamma$ is the gyromagnetic ratio
    \item Write the energy eigenstates and eigenvalues for this system
    \item For initial state $\ket{\psi(0)} = \ket{\ell, m_0}$, show that $\ket{\psi(t)} = e^{i\gamma Bt m_0}\ket{\ell, m_0}$ (up to global phase)
    \item Prepare an initial state $\ket{\psi(0)} = \frac{1}{\sqrt{2}}(\ket{\ell,\ell} + \ket{\ell,-\ell})$ and describe the time evolution of $\langle L_x \rangle$ and $\langle L_y \rangle$
    \item What is the precession frequency in terms of $\gamma$, $B$, and $\ell$?
\end{enumerate}

\problem{Spin-Orbit Coupling}
In heavy atoms, spin-orbit coupling becomes important. The interaction Hamiltonian is $\hat{H}_{so} = \lambda\vec{L} \cdot \vec{S}/\hbar^2$.
\begin{enumerate}[label=(\alph*)]
    \item Show that $\vec{L} \cdot \vec{S} = L_z S_z + \frac{1}{2}(L_+ S_- + L_- S_+)$ where $L_\pm$ and $S_\pm$ are ladder operators
    \item For a state with fixed $m_\ell$ and $m_s$, calculate the first-order energy correction due to spin-orbit coupling
    \item Discuss why spin-orbit coupling leads to fine structure splitting of atomic levels
\end{enumerate}

%% file: chapters/ch07_angular_addition.tex
\chapter{Addition of Angular Momentum}
\label{ch:angular_addition}

\section{Introduction: Addition is Multiplication}

In Chapter~\ref{ch:3d_angular}, we developed the quantum theory of angular momentum for a single particle, discovering that angular momentum operators satisfy the fundamental commutation relations $[\hat{J}_i, \hat{J}_j] = i\hbar\epsilon_{ijk}\hat{J}_k$. We found that states can be labeled by quantum numbers $j$ and $m$, with eigenvalues $\hbar^2 j(j+1)$ for $\hat{J}^2$ and $\hbar m$ for $\hat{J}_z$. The Hilbert space for a single angular momentum $j$ has dimension $2j+1$, corresponding to the $2j+1$ possible values of $m$.

Physical systems often involve multiple sources of angular momentum: orbital and spin angular momenta of a single particle, angular momenta of multiple particles in atoms or molecules, or nuclear spins coupled to electronic angular momentum. How do we describe such composite systems?

\textbf{A note on group theory:} Throughout this chapter, we will discuss several results from the representation theory of rotation groups. These results - including the decomposition of tensor products, the relationship between SU(2) and SO(3), and the structure of Wigner symbols - can be understood and applied without rigorous proof. We are using group theory as a framework to organize and understand quantum angular momentum, but we will not provide full mathematical justifications for the group-theoretic statements. Readers interested in the underlying proofs should consult texts on group representation theory.

\begin{keyidea}{Addition of Angular Momentum is Multiplication of Hilbert Spaces}
When we "add" two angular momenta $\vec{J}_1$ and $\vec{J}_2$, the underlying mathematical structure is the \textbf{tensor product} of Hilbert spaces:
\begin{equation}
\mathcal{H} = \mathcal{H}_1 \otimes \mathcal{H}_2
\end{equation}

If angular momentum $j_1$ has Hilbert space dimension $2j_1+1$ and angular momentum $j_2$ has dimension $2j_2+1$, the combined system has dimension:
\begin{equation}
\text{dim}(\mathcal{H}) = (2j_1+1)(2j_2+1)
\end{equation}

This is multiplication, not addition. The "addition" of angular momenta refers to the vector sum $\vec{J} = \vec{J}_1 + \vec{J}_2$, but the state space structure is multiplicative.
\end{keyidea}

\subsection{Why Two Descriptions of the Same Space?}

The tensor product space $\mathcal{H}_1 \otimes \mathcal{H}_2$ can be spanned by different bases. The choice of basis reflects which physical quantities we consider most natural to specify:

\textbf{The uncoupled basis} is expressed as a tensor product of angular momentum ket states: $\ket{j_1, m_1} \otimes \ket{j_2, m_2}$. For notational convenience, we compress this into a single ket $\ket{j_1, m_1; j_2, m_2}$, but it is essential to remember these are product states from two different Hilbert spaces. This is the natural product basis constructed from eigenstates of $\hat{J}_1^2$, $\hat{J}_{1z}$, $\hat{J}_2^2$, and $\hat{J}_{2z}$. There are $(2j_1+1)(2j_2+1)$ such states.

\textbf{The coupled basis} $\ket{j_1, j_2; j, m}$ instead uses the total angular momentum $\vec{J} = \vec{J}_1 + \vec{J}_2$. These are eigenstates of $\hat{J}_1^2$, $\hat{J}_2^2$, $\hat{J}^2$, and $\hat{J}_z$. Remarkably, the same $(2j_1+1)(2j_2+1)$-dimensional space can be organized as a direct sum of subspaces with definite total angular momentum:
\begin{equation}
\label{eq:direct_sum_decomposition}
\mathcal{H}_1 \otimes \mathcal{H}_2 = \bigoplus_{j=|j_1-j_2|}^{j_1+j_2} \mathcal{H}_j = \mathcal{H}_{|j_1-j_2|} \oplus \mathcal{H}_{|j_1-j_2|+1} \oplus \cdots \oplus \mathcal{H}_{j_1+j_2-1} \oplus \mathcal{H}_{j_1+j_2}
\end{equation}
where each $\mathcal{H}_j$ has dimension $2j+1$.

This decomposition reflects a deep mathematical structure from group theory that we'll explore in the next section. The key point is that the same Hilbert space admits two complementary descriptions.

\subsection{Physical Motivation: Why Total Angular Momentum?}

Why do we care about reorganizing our basis to use total angular momentum? The answer lies in the symmetries and interactions that govern quantum systems.

First, consider conservation laws. If the Hamiltonian of a system is rotationally invariant, meaning the physics doesn't depend on the orientation of our coordinate system, then total angular momentum $\vec{J}$ is a conserved quantity. This has profound implications for time evolution. States with definite total angular momentum quantum number $j$ evolve without mixing with states of different $j$. The energy eigenstates can be labeled by $j$, and transitions between energy levels must satisfy selection rules involving $j$. This makes the coupled basis the natural language for describing dynamics in rotationally symmetric systems.

Second, many fundamental interactions depend explicitly on how the constituent angular momenta are oriented relative to each other. The canonical example is spin-orbit coupling, where the interaction energy depends on $\vec{L} \cdot \vec{S}$, the dot product of orbital and spin angular momenta. This interaction can be rewritten using the total angular momentum:
\begin{equation}
\vec{L} \cdot \vec{S} = \frac{1}{2}(\hat{J}^2 - \hat{L}^2 - \hat{S}^2)
\end{equation}
In this form, the interaction is diagonal in the coupled basis: each state $\ket{\ell, s; j, m}$ has a definite value of $\vec{L} \cdot \vec{S}$ determined entirely by the quantum numbers $\ell$, $s$, and $j$. In the uncoupled basis $\ket{\ell, m_\ell; s, m_s}$, by contrast, the same interaction mixes different states, requiring diagonalization to find energy eigenstates. The coupled basis thus diagonalizes the physically relevant Hamiltonian.

Third, spectroscopy provides the empirical foundation for atomic physics, and experimentally observed atomic energy levels are labeled using total angular momentum quantum numbers. Spectroscopic notation like $^2P_{3/2}$ specifies $n$ (principal quantum number), $\ell$ (orbital angular momentum), total electronic spin $S$, and total angular momentum $j = \ell \pm s$. This notation reflects the physical reality that spin-orbit coupling splits energy levels according to $j$, not according to the individual projections $m_\ell$ and $m_s$. Reading and interpreting experimental spectra requires fluency in the coupled basis.

\begin{physicalinsight}
The choice between uncoupled and coupled bases reflects the physics of the problem. Use the uncoupled basis when $\vec{J}_1$ and $\vec{J}_2$ are nearly independent (weak coupling). Use the coupled basis when interactions couple them strongly or when total angular momentum is a good quantum number. The Clebsch-Gordan coefficients allow us to translate between these descriptions as needed.
\end{physicalinsight}

\section{The Rotation Group: SU(2) and SO(3)}

Before diving into the detailed mechanics of angular momentum addition, we need to understand the mathematical structure underlying rotations in quantum mechanics. This involves concepts from group theory - the mathematical theory of symmetry.

\subsection{A Minimal Introduction to Group Theory}

A \textbf{group} is a set of elements together with an operation (often called "multiplication") that combines any two elements to produce a third element in the set. The operation must satisfy four properties: closure (the product of any two elements is in the group), associativity, existence of an identity element, and existence of inverses.

The rotations in three-dimensional space form a group: the product of two rotations is another rotation, there is an identity rotation (do nothing), each rotation has an inverse (rotate by the opposite angle), and composition of rotations is associative.

A \textbf{representation} of a group assigns to each group element a matrix (or more generally, a linear operator) in such a way that the group multiplication corresponds to matrix multiplication. If group elements $g_1$ and $g_2$ are represented by matrices $D(g_1)$ and $D(g_2)$, then the product $g_1 g_2$ must be represented by $D(g_1)D(g_2)$.

Different representations can have different dimensions. For rotations, the dimension tells us how many components an object has. Scalars have one component (dimension 1), vectors have three components (dimension 3), and more complex objects can have higher dimensions.

A representation is \textbf{irreducible} if it cannot be decomposed into smaller representations. When we write a large matrix in block-diagonal form, each independent block is an irreducible representation. Finding irreducible representations is central to understanding how objects transform under symmetries.

These concepts provide the framework for understanding angular momentum in quantum mechanics. Angular momentum quantum numbers label irreducible representations of the rotation group.

\subsection{SO(3): The Rotation Group in Three Dimensions}

The group SO(3) consists of all rotations in three-dimensional space. Each rotation can be specified by an axis $\hat{n}$ and an angle $\theta$, giving a rotation operator:
\begin{equation}
R(\hat{n}, \theta) = e^{-i\theta \hat{n} \cdot \vec{J}/\hbar}
\end{equation}

These rotations form a group under composition: the product of two rotations is another rotation, there is an identity rotation ($\theta = 0$), and each rotation has an inverse (rotation by $-\theta$).

A complete rotation by $2\pi$ about any axis returns to the identity:
\begin{equation}
R(\hat{n}, 2\pi) = e^{-2\pi i \hat{n} \cdot \vec{J}/\hbar} = I
\end{equation}

This holds for all classical vectors and orbital angular momentum states, which transform according to representations of SO(3). The allowed values of angular momentum are integers: $\ell = 0, 1, 2, \ldots$

\subsection{SU(2): The Covering Group}

Quantum mechanics requires a more subtle structure. The group SU(2) consists of $2 \times 2$ unitary matrices with determinant one. Every SU(2) matrix can be written as:
\begin{equation}
U = e^{-i\theta \hat{n} \cdot \vec{\sigma}/2}
\end{equation}
where $\vec{\sigma} = (\sigma_x, \sigma_y, \sigma_z)$ are the Pauli matrices.

The crucial difference from SO(3) is that a rotation by $2\pi$ gives:
\begin{equation}
U(\hat{n}, 2\pi) = e^{-2\pi i \hat{n} \cdot \vec{\sigma}/2} = -I
\end{equation}

A $2\pi$ rotation produces a minus sign, not the identity. Only a rotation by $4\pi$ returns to the identity:
\begin{equation}
U(\hat{n}, 4\pi) = I
\end{equation}

This is the defining property of a spinor: an object that picks up a minus sign under a $2\pi$ rotation.

\subsection{The Double Cover}

SU(2) is the double cover of SO(3). There is a two-to-one homomorphism from SU(2) to SO(3):
\begin{equation}
\phi: \text{SU(2)} \to \text{SO(3)}
\end{equation}

Both $U$ and $-U$ in SU(2) map to the same rotation $R$ in SO(3). This means that for every physical rotation in three-dimensional space, there are two quantum mechanical phase factors that could represent it, differing by a sign.

This mathematical structure has profound physical consequences. SU(2) admits representations labeled by half-integers $j = 0, 1/2, 1, 3/2, 2, \ldots$, while SO(3) allows only integers $\ell = 0, 1, 2, \ldots$ The half-integer representations of SU(2) cannot be represented as rotations of vectors in three-dimensional space. They are intrinsically quantum mechanical objects.

\subsection{Physical Implications for Spin}

Electrons, protons, neutrons, and many other fundamental particles have spin-1/2. Their quantum states transform according to the spin-1/2 representation of SU(2), not SO(3). This has several profound consequences. A spin-1/2 particle rotated by $2\pi$ acquires a phase factor of $-1$, not $+1$. Two identical spin-1/2 particles (fermions) must have antisymmetric total wavefunctions. The Pauli exclusion principle holds for half-integer spin particles. Spinors require special mathematical treatment distinct from vectors.

The existence of half-integer angular momentum is not an accident or a peculiarity of quantum mechanics. It is a direct consequence of the group structure of rotations in quantum theory, which requires SU(2) rather than SO(3).

\subsection{Connection to Angular Momentum Addition}

The decomposition of tensor products of angular momentum representations,
\begin{equation}
j_1 \otimes j_2 = \bigoplus_{j=|j_1-j_2|}^{j_1+j_2} j
\end{equation}
is a statement about how representations of SU(2) combine. The Clebsch-Gordan coefficients are precisely the matrix elements that accomplish the change of basis between the product representation and the decomposed direct sum.

When we add two half-integer angular momenta, we can obtain either integer or half-integer results. For example, $1/2 \otimes 1/2 = 0 \oplus 1$ gives both the singlet (integer) and triplet (integer). But $1/2 \otimes 1 = 1/2 \oplus 3/2$ gives half-integer results. This mixing of integer and half-integer representations under tensor products is a feature of SU(2) that has no analog in the purely integer representations of SO(3).

\section{Cartesian and Spherical Tensors}

Having established the group-theoretic foundation, we now examine how composite objects transform under rotations. This connects directly to the spherical harmonics we studied in Chapter~\ref{ch:3d_angular} and provides concrete examples of how tensor products decompose into irreducible representations.

\subsection{Cartesian Tensors}

A Cartesian tensor of rank $n$ has $3^n$ components $T_{i_1 i_2 \ldots i_n}$ where each index runs over $x, y, z$. Under a rotation $R$, the components transform as:
\begin{equation}
T'_{i_1 i_2 \ldots i_n} = R_{i_1 j_1} R_{i_2 j_2} \cdots R_{i_n j_n} T_{j_1 j_2 \ldots j_n}
\end{equation}

Examples include rank 0 (scalars with 1 component such as mass, charge, and temperature), rank 1 (vectors with 3 components such as position, momentum, and electric field), and rank 2 ($3 \times 3$ matrices with 9 components such as the stress tensor, moment of inertia, and polarizability).

\subsection{Rotations of Vector Components: A Concrete Example}

From Chapter~\ref{ch:3d_angular}, we know that a rotation by angle $\theta$ about the $z$-axis transforms a vector according to:
\begin{equation}
\label{eq:rotation_matrix_z}
R_z(\theta) \underset{c}{\overset{\cdot}{=}} \begin{pmatrix} \cos\theta & -\sin\theta & 0 \\ \sin\theta & \cos\theta & 0 \\ 0 & 0 & 1 \end{pmatrix}
\end{equation}

Here we use the notation introduced in Chapter~\ref{ch:foundations}: the symbol $\overset{\cdot}{=}$ with a dot over the equals sign denotes a matrix representation, while the label underneath indicates which basis we're using. The label $c$ denotes the Cartesian basis $(x, y, z)$. This is a $3 \times 3$ matrix acting on the three components of a vector. The vector components mix under rotation, but they mix only among themselves - a rotated vector is still a vector.

\subsection{Connection to Spherical Harmonics}

In Chapter~\ref{ch:3d_angular}, we encountered the spherical harmonics $Y_{\ell m}(\theta, \phi)$ as eigenfunctions of $\hat{L}^2$ and $\hat{L}_z$ for orbital angular momentum. The simplest spherical harmonics are the $\ell = 1$ functions, which are directly related to the Cartesian coordinates.

For $\ell = 1$, the three spherical harmonics $Y_{1,m}$ with $m = -1, 0, +1$ can be expressed in terms of the Cartesian unit vector components $x/r$, $y/r$, and $z/r$:
\begin{align}
Y_{1,0}(\theta, \phi) &\propto \frac{z}{r} = \cos\theta\\
Y_{1,\pm 1}(\theta, \phi) &\propto \frac{x \pm iy}{r} = \sin\theta e^{\pm i\phi}
\end{align}

The $x$, $y$, and $z$ coordinates (divided by $r$ to make them dimensionless) transform as a rank-1 Cartesian tensor - that is, as a vector. But these same three components can be reorganized into the spherical basis $(Y_{1,+1}, Y_{1,0}, Y_{1,-1})$, which forms an irreducible representation of the rotation group with $\ell = 1$.

This connection is fundamental: the spherical harmonics $Y_{\ell m}$ provide the spherical tensor basis for objects with angular momentum $\ell$. The Cartesian components $(x, y, z)$ and spherical components $(Y_{1,+1}, Y_{1,0}, Y_{1,-1})$ describe the same vector space, just in different bases.

\subsection{Tensor Products Under Rotations}

Now consider the tensor product of two vectors $\vec{A}$ and $\vec{B}$, which could represent position vectors $\vec{r}_1$ and $\vec{r}_2$ of two particles. The 9 components $T_{ij} = A_i B_j$ form a rank-2 Cartesian tensor. How does this object transform under rotations?

Under a rotation $R$, each vector transforms:
\begin{align}
A'_i &= R_{ij} A_j\\
B'_k &= R_{k\ell} B_\ell
\end{align}

The tensor product components therefore transform as:
\begin{equation}
\label{eq:tensor_transformation}
T'_{ij} = A'_i B'_j = R_{im} A_m R_{jn} B_n = R_{im} R_{jn} T_{mn}
\end{equation}

We can rewrite this as a matrix equation by treating the 9 components as a vector. Label the components as $(T_{xx}, T_{xy}, T_{xz}, T_{yx}, T_{yy}, T_{yz}, T_{zx}, T_{zy}, T_{zz})$. The transformation becomes:
\begin{equation}
\vec{T}' \underset{c}{\overset{\cdot}{=}} (R \otimes R) \vec{T}
\end{equation}
where $R \otimes R$ is the tensor product of the $3 \times 3$ rotation matrix with itself, giving a $9 \times 9$ matrix. The label $c$ again indicates we're working in the Cartesian basis.

\subsection{\texorpdfstring{The $9\times 9$ Rotation Matrix}{The 9 x 9 Rotation Matrix}}

For a rotation about the $z$-axis, the $9 \times 9$ matrix $R_z(\theta) \otimes R_z(\theta)$ explicitly shows how the 9 tensor components mix. This matrix cannot be diagonalized in general - rotations don't have 9 independent eigenvectors with eigenvalue 1.

However, the matrix can be block diagonalized. The key observation is that certain combinations of the 9 components transform only among themselves, forming invariant subspaces. The $9 \times 9$ matrix breaks into blocks:
\begin{equation}
R_z(\theta) \otimes R_z(\theta) \underset{s}{\overset{\cdot}{=}} \begin{pmatrix}
[5 \times 5] & 0 & 0 \\
0 & [3 \times 3] & 0 \\
0 & 0 & [1 \times 1]
\end{pmatrix}
\end{equation}
where the label $s$ now denotes the spherical tensor basis. This change of basis transforms the matrix representation from the Cartesian form (label $c$) to the spherical form (label $s$), which reveals the block diagonal structure.

These blocks correspond to:
\begin{align}
5 \times 5 \text{ block:} &\quad \text{Symmetric traceless combinations (}\ell=2\text{)}\\
3 \times 3 \text{ block:} &\quad \text{Antisymmetric combinations (}\ell=1\text{)}\\
1 \times 1 \text{ block:} &\quad \text{Trace (}\ell=0\text{)}
\end{align}

This block structure is not an accident. It reflects the decomposition $1 \otimes 1 = 2 \oplus 1 \oplus 0$ from angular momentum addition.

\subsection{Irreducible Representations and Block Diagonalization}

A representation of the rotation group is called \textbf{irreducible} if it cannot be further decomposed into smaller blocks. The $3 \times 3$ vector representation is irreducible - you cannot find a change of basis that makes rotation matrices block diagonal for vectors.

The $9 \times 9$ tensor product representation is \textbf{reducible}. By choosing the right basis (symmetric traceless, antisymmetric, and trace combinations), we decompose it into three irreducible blocks. Each block transforms independently under rotations. The 5-component symmetric traceless tensor transforms by a $5 \times 5$ matrix (the $\ell=2$ representation), the 3-component antisymmetric tensor transforms by a $3 \times 3$ matrix (the $\ell=1$ representation), and the 1-component trace transforms by a $1 \times 1$ matrix, meaning it's invariant (the $\ell=0$ representation).

This is the essence of irreducible decomposition: finding the natural building blocks that don't mix under rotations.

\subsection{Example: Product of Position Vectors}

Let's be concrete. Consider two position vectors $\vec{r}_1 = (x_1, y_1, z_1)$ and $\vec{r}_2 = (x_2, y_2, z_2)$. Their tensor product gives 9 terms:
\begin{equation}
x_1 x_2, \quad x_1 y_2, \quad x_1 z_2, \quad y_1 x_2, \quad y_1 y_2, \quad y_1 z_2, \quad z_1 x_2, \quad z_1 y_2, \quad z_1 z_2
\end{equation}

How do these 9 objects decompose under rotations?

\textbf{The $\ell=0$ piece: Rotationally invariant (1 component)}

The scalar product is rotationally invariant:
\begin{equation}
\vec{r}_1 \cdot \vec{r}_2 = x_1 x_2 + y_1 y_2 + z_1 z_2
\end{equation}

Under any rotation, this combination remains unchanged. This is the unique $\ell=0$ component.

\textbf{The $\ell=1$ piece: Transforms like a vector (3 components)}

The antisymmetric combinations give a pseudovector (the cross product):
\begin{align}
(\vec{r}_1 \times \vec{r}_2)_x &= y_1 z_2 - z_1 y_2\\
(\vec{r}_1 \times \vec{r}_2)_y &= z_1 x_2 - x_1 z_2\\
(\vec{r}_1 \times \vec{r}_2)_z &= x_1 y_2 - y_1 x_2
\end{align}

These three components transform like a vector under rotations, forming the $\ell=1$ representation.

\textbf{The $\ell=2$ piece: Quadrupole tensor (5 components)}

The remaining 5 independent components form a symmetric traceless tensor. Examples include:
\begin{align}
&x_1 x_2 - \frac{1}{3}(\vec{r}_1 \cdot \vec{r}_2)\\
&y_1 y_2 - \frac{1}{3}(\vec{r}_1 \cdot \vec{r}_2)\\
&\frac{1}{2}(x_1 y_2 + y_1 x_2)\\
&\frac{1}{2}(x_1 z_2 + z_1 x_2)\\
&\frac{1}{2}(y_1 z_2 + z_1 y_2)
\end{align}

These transform according to the $\ell=2$ representation - the same symmetry as $d$-orbitals in atomic physics.

\subsubsection{Reading the Decomposition from Spherical Harmonics}

The decomposition of tensor products can be understood directly from the Cartesian expressions of spherical harmonics. The spherical harmonics $Y_{\ell m}(\theta, \phi)$ for low $\ell$ have simple Cartesian forms that reveal how products of coordinates naturally organize into irreducible representations.

For a single position vector $\vec{r} = (x, y, z)$, the $\ell = 1$ spherical harmonics in Cartesian form are:
\begin{align}
Y_{1,0} &\propto \frac{z}{r}\\
Y_{1,+1} &\propto -\frac{x+iy}{r}\\
Y_{1,-1} &\propto \frac{x-iy}{r}
\end{align}

For the product of two position vectors $\vec{r}_1$ and $\vec{r}_2$, the nine terms $x_1 x_2$, $x_1 y_2$, etc., reorganize into combinations that transform like spherical harmonics with $\ell = 0, 1, 2$. The table below shows this correspondence explicitly:

\begin{center}
\renewcommand{\arraystretch}{1.5}
\begin{tabular}{c|l|l}
\hline
$\ell$ & Spherical Harmonic Structure & Tensor Product Combination \\
\hline
\multirow{1}{*}{$\ell=0$} 
& $Y_{0,0} \propto 1$ (constant) 
& $x_1 x_2 + y_1 y_2 + z_1 z_2$ \\
& & (rotationally invariant) \\
\hline
\multirow{3}{*}{$\ell=1$} 
& $Y_{1,0} \propto z$ 
& $x_1 y_2 - y_1 x_2$ \\
& $Y_{1,+1} \propto -(x+iy)$ 
& $y_1 z_2 - z_1 y_2 - i(z_1 x_2 - x_1 z_2)$ \\
& $Y_{1,-1} \propto (x-iy)$ 
& $y_1 z_2 - z_1 y_2 + i(z_1 x_2 - x_1 z_2)$ \\
\hline
\multirow{5}{*}{$\ell=2$} 
& $Y_{2,0} \propto 3z^2 - r^2$ 
& $z_1 z_2 - \frac{1}{3}(x_1 x_2 + y_1 y_2 + z_1 z_2)$ \\
& $Y_{2,+1} \propto -(x+iy)z$ 
& $(x_1 z_2 + z_1 x_2) + i(y_1 z_2 + z_1 y_2)$ \\
& $Y_{2,-1} \propto (x-iy)z$ 
& $(x_1 z_2 + z_1 x_2) - i(y_1 z_2 + z_1 y_2)$ \\
& $Y_{2,+2} \propto (x+iy)^2$ 
& $(x_1 x_2 - y_1 y_2) + i(x_1 y_2 + y_1 x_2)$ \\
& $Y_{2,-2} \propto (x-iy)^2$ 
& $(x_1 x_2 - y_1 y_2) - i(x_1 y_2 + y_1 x_2)$ \\
\hline
\end{tabular}
\end{center}

The pattern is clear: products of coordinates organize into the same mathematical structure as spherical harmonics. For $\ell=0$, we get the unique rotationally invariant combination (the dot product). For $\ell=1$, the three components correspond to the cross product written in complex spherical basis. For $\ell=2$, the five symmetric traceless combinations have the same transformation properties as the $d$-orbitals.

This correspondence is not accidental. When we form the tensor product $\vec{r}_1 \otimes \vec{r}_2$, we are combining two $\ell=1$ objects. The decomposition $1 \otimes 1 = 0 \oplus 1 \oplus 2$ tells us that the resulting object can be reorganized into pieces that transform like $\ell=0$, $\ell=1$, and $\ell=2$ spherical harmonics. The Clebsch-Gordan coefficients provide the precise linear combinations needed to construct these irreducible components from the original Cartesian products.

This decomposition $3 \otimes 3 = 5 \oplus 3 \oplus 1$ corresponds precisely to the angular momentum addition formula $1 \otimes 1 = 2 \oplus 1 \oplus 0$. The Cartesian tensor product decomposes into spherical tensors of definite angular momentum.

\section{The Tensor Product Structure}

With this group-theoretic and geometric foundation in place, we now turn to the detailed structure of composite angular momentum systems in quantum mechanics.

\subsection{Hilbert Space Dimensions}

Consider two angular momentum systems with quantum numbers $j_1$ and $j_2$. Each lives in its own Hilbert space:
\begin{align}
\mathcal{H}_1: \quad &\text{dim} = 2j_1 + 1 \quad \text{(states with } m_1 = -j_1, \ldots, +j_1\text{)}\\
\mathcal{H}_2: \quad &\text{dim} = 2j_2 + 1 \quad \text{(states with } m_2 = -j_2, \ldots, +j_2\text{)}
\end{align}

The composite system lives in the tensor product space:
\begin{equation}
\mathcal{H} = \mathcal{H}_1 \otimes \mathcal{H}_2
\end{equation}

with dimension:
\begin{equation}
\label{eq:tensor_product_dimension}
\text{dim}(\mathcal{H}) = (2j_1+1) \times (2j_2+1)
\end{equation}

\textbf{Example:} Two spin-1/2 particles have $j_1 = j_2 = 1/2$, so each has a 2-dimensional Hilbert space. The combined system has dimension $2 \times 2 = 4$, with basis states $\ket{\uparrow\uparrow}$, $\ket{\uparrow\downarrow}$, $\ket{\downarrow\uparrow}$, $\ket{\downarrow\downarrow}$.

\subsection{Decomposition into Total Angular Momentum Subspaces}

The $(2j_1+1)(2j_2+1)$-dimensional tensor product space can be decomposed into subspaces of definite total angular momentum, as stated in Eq.~\eqref{eq:direct_sum_decomposition}. We can verify that the dimensions match by summing over all allowed $j$ values:
\begin{equation}
\label{eq:dimension_sum}
\sum_{j=|j_1-j_2|}^{j_1+j_2} (2j+1) = (2j_1+1)(2j_2+1)
\end{equation}

Comparing Eq.~\eqref{eq:dimension_sum} with Eq.~\eqref{eq:tensor_product_dimension}, we see both expressions give the same total dimension, confirming that the direct sum decomposition in Eq.~\eqref{eq:direct_sum_decomposition} accounts for all states in the tensor product space.

\section{The Uncoupled and Coupled Bases}

\subsection{The Uncoupled Basis}

Consider a system with two angular momenta $\vec{J}_1$ and $\vec{J}_2$. These could represent physically distinct scenarios. They might describe the orbital angular momentum $\vec{L}$ and spin angular momentum $\vec{S}$ of a single electron, where the orbital motion and intrinsic spin represent independent degrees of freedom that can be coupled by spin-orbit interaction. Alternatively, they could represent the angular momenta of two spatially separated particles, for instance, the spins of two electrons in different atoms, or the orbital angular momenta of two nucleons in a nucleus. More generally, $\vec{J}_1$ and $\vec{J}_2$ can be any two angular momentum operators that commute with each other, meaning they act on different parts of the Hilbert space or represent different physical degrees of freedom.

The key property is that operators acting on different subsystems commute:
\begin{equation}
[\hat{J}_{1i}, \hat{J}_{2j}] = 0 \quad \text{for all } i, j
\end{equation}

Within each subsystem, the standard angular momentum commutation relations hold:
\begin{align}
[\hat{J}_{1i}, \hat{J}_{1j}] &= i\hbar\epsilon_{ijk}\hat{J}_{1k}\\
[\hat{J}_{2i}, \hat{J}_{2j}] &= i\hbar\epsilon_{ijk}\hat{J}_{2k}
\end{align}

Since the operators $\hat{J}_1^2$, $\hat{J}_{1z}$, $\hat{J}_2^2$, and $\hat{J}_{2z}$ all mutually commute, they form a complete set of commuting observables. We can construct simultaneous eigenstates as tensor products:
\begin{equation}
\ket{j_1, m_1; j_2, m_2} = \ket{j_1, m_1} \otimes \ket{j_2, m_2}
\end{equation}

Often the tensor product symbol $\otimes$ is suppressed, but it's important to remember that these are product states from two different Hilbert spaces.

These states satisfy:
\begin{align}
\hat{J}_1^2 \ket{j_1, m_1; j_2, m_2} &= \hbar^2 j_1(j_1+1) \ket{j_1, m_1; j_2, m_2}\\
\hat{J}_{1z} \ket{j_1, m_1; j_2, m_2} &= \hbar m_1 \ket{j_1, m_1; j_2, m_2}\\
\hat{J}_2^2 \ket{j_1, m_1; j_2, m_2} &= \hbar^2 j_2(j_2+1) \ket{j_1, m_1; j_2, m_2}\\
\hat{J}_{2z} \ket{j_1, m_1; j_2, m_2} &= \hbar m_2 \ket{j_1, m_1; j_2, m_2}
\end{align}

The uncoupled basis provides $(2j_1+1)(2j_2+1)$ orthonormal product states that span the full tensor product space $\mathcal{H}_1 \otimes \mathcal{H}_2$. This basis is particularly natural when the two angular momenta are nearly independent, such as in weak spin-orbit coupling or widely separated particles.

\subsection{The Total Angular Momentum}

The total angular momentum is defined as the vector sum:
\begin{equation}
\vec{J} = \vec{J}_1 + \vec{J}_2
\end{equation}

or in components:
\begin{equation}
\hat{J}_i = \hat{J}_{1i} + \hat{J}_{2i} \quad \text{for } i = x, y, z
\end{equation}

\begin{physicalinsight}
The total angular momentum $\vec{J}$ is itself an angular momentum operator, satisfying the standard commutation relations $[\hat{J}_i, \hat{J}_j] = i\hbar\epsilon_{ijk}\hat{J}_k$. This can be verified using the individual commutation relations of $\vec{J}_1$ and $\vec{J}_2$ and the fact that operators from different subsystems commute.
\end{physicalinsight}

\subsection{The Coupled Basis}

An alternative description uses the total angular momentum. We can show that:
\begin{align}
[\hat{J}^2, \hat{J}_1^2] &= 0\\
[\hat{J}^2, \hat{J}_2^2] &= 0\\
[\hat{J}^2, \hat{J}_z] &= 0
\end{align}

Therefore, $\hat{J}_1^2$, $\hat{J}_2^2$, $\hat{J}^2$, and $\hat{J}_z$ form another complete set of commuting observables. The corresponding eigenstates are:
\begin{equation}
\ket{j_1, j_2; j, m}
\end{equation}

These states live in the \textbf{same} tensor product space $\mathcal{H}_1 \otimes \mathcal{H}_2$ as the uncoupled basis, but organized differently. They satisfy:
\begin{align}
\hat{J}_1^2 \ket{j_1, j_2; j, m} &= \hbar^2 j_1(j_1+1) \ket{j_1, j_2; j, m}\\
\hat{J}_2^2 \ket{j_1, j_2; j, m} &= \hbar^2 j_2(j_2+1) \ket{j_1, j_2; j, m}\\
\hat{J}^2 \ket{j_1, j_2; j, m} &= \hbar^2 j(j+1) \ket{j_1, j_2; j, m}\\
\hat{J}_z \ket{j_1, j_2; j, m} &= \hbar m \ket{j_1, j_2; j, m}
\end{align}

The coupled basis groups the $(2j_1+1)(2j_2+1)$ states according to their total angular momentum $j$. Each value of $j$ gives a $(2j+1)$-dimensional subspace of states with $m = -j, \ldots, +j$. States within each $j$ subspace transform among themselves under rotations but never mix with other $j$ values.

\subsection{Constraints on Total Angular Momentum}

Given fixed $j_1$ and $j_2$, what values can the total angular momentum $j$ take? Since $\hat{J}_z = \hat{J}_{1z} + \hat{J}_{2z}$:
\begin{equation}
m = m_1 + m_2
\end{equation}

The maximum value of $m$ occurs when both components are aligned: $m_{\text{max}} = j_1 + j_2$. Since $m$ ranges from $-j$ to $+j$, we must have:
\begin{equation}
j_{\text{max}} = j_1 + j_2
\end{equation}

The minimum value arises when the angular momenta are anti-aligned. The allowed values of $j$ follow the triangle inequality:
\begin{equation}
|j_1 - j_2| \leq j \leq j_1 + j_2
\end{equation}
with $j$ taking integer steps.

The complete list of allowed values is:
\begin{equation}
j \in \{|j_1 - j_2|, |j_1 - j_2| + 1, |j_1 - j_2| + 2, \ldots, j_1 + j_2 - 1, j_1 + j_2\}
\end{equation}

This gives a total of:
\begin{equation}
N_j = 2\min(j_1, j_2) + 1
\end{equation}
different values of $j$.

\section{Clebsch-Gordan Coefficients}

\subsection{Definition and Notation}

The uncoupled and coupled bases are related by a unitary transformation. We can expand coupled states in terms of uncoupled states:
\begin{equation}
\ket{j_1, j_2; j, m} = \sum_{m_1, m_2} \braket{j_1, m_1; j_2, m_2}{j_1, j_2; j, m} \ket{j_1, m_1; j_2, m_2}
\end{equation}

The expansion coefficients are called Clebsch-Gordan (CG) coefficients:
\begin{equation}
\braket{j_1, m_1; j_2, m_2}{j_1, j_2; j, m} \equiv \langle j_1, m_1; j_2, m_2 | j, m \rangle
\end{equation}

An alternative notation uses the symbol:
\begin{equation}
C^{j,m}_{j_1,m_1;j_2,m_2} = \langle j_1, m_1; j_2, m_2 | j, m \rangle
\end{equation}

Yet another common notation is:
\begin{equation}
\langle j_1, m_1; j_2, m_2 | j, m \rangle = \begin{pmatrix} j_1 & j_2 & j \\ m_1 & m_2 & m \end{pmatrix}
\end{equation}

\subsection{Properties of Clebsch-Gordan Coefficients}

The CG coefficients have several important properties:

\textbf{1. Conservation of $m$:}
\begin{equation}
\langle j_1, m_1; j_2, m_2 | j, m \rangle = 0 \quad \text{unless } m = m_1 + m_2
\end{equation}

\textbf{2. Triangle inequality:}
\begin{equation}
\langle j_1, m_1; j_2, m_2 | j, m \rangle = 0 \quad \text{unless } |j_1 - j_2| \leq j \leq j_1 + j_2
\end{equation}

\textbf{3. Orthonormality:}
\begin{align}
\sum_{m_1, m_2} \langle j_1, m_1; j_2, m_2 | j, m \rangle \langle j_1, m_1; j_2, m_2 | j', m' \rangle &= \delta_{jj'}\delta_{mm'}\\
\sum_{j, m} \langle j_1, m_1; j_2, m_2 | j, m \rangle \langle j_1, m'_1; j_2, m'_2 | j, m \rangle &= \delta_{m_1 m'_1}\delta_{m_2 m'_2}
\end{align}

\textbf{4. Symmetry relations:}
\begin{align}
\langle j_1, m_1; j_2, m_2 | j, m \rangle &= (-1)^{j_1+j_2-j} \langle j_2, m_2; j_1, m_1 | j, m \rangle\\
\langle j_1, m_1; j_2, m_2 | j, m \rangle &= (-1)^{j_1-m_1}\sqrt{\frac{2j+1}{2j_2+1}} \langle j_1, -m_1; j, m | j_2, -m_2 \rangle
\end{align}

\textbf{5. Reality (phase convention):}
With the Condon-Shortley phase convention, CG coefficients are real:
\begin{equation}
\langle j_1, m_1; j_2, m_2 | j, m \rangle \in \mathbb{R}
\end{equation}

\subsection{Computing Clebsch-Gordan Coefficients}

\subsubsection{The Ladder Operator Method}

The most systematic approach uses angular momentum ladder operators. Start from the state with maximum $m$:
\begin{equation}
\ket{j_1, j_1; j_2, j_2; j, j} = \ket{j_1, j_1; j_2, j_2}
\end{equation}

where $j = j_1 + j_2$ and $m = j$. This gives:
\begin{equation}
\langle j_1, j_1; j_2, j_2 | j_1 + j_2, j_1 + j_2 \rangle = 1
\end{equation}

Apply the lowering operator $\hat{J}_- = \hat{J}_{1-} + \hat{J}_{2-}$ repeatedly to generate states with smaller $m$. The action of ladder operators:
\begin{align}
\hat{J}_\pm \ket{j, m} &= \hbar\sqrt{j(j+1) - m(m\pm 1)} \ket{j, m \pm 1}\\
&= \hbar\sqrt{(j \mp m)(j \pm m + 1)} \ket{j, m \pm 1}
\end{align}

This generates all CG coefficients for a given $j$ systematically. States with different $j$ values are found using orthogonality.

\subsection{Reading Clebsch-Gordan Tables}

In practice, CG coefficients are typically looked up in standard tables rather than computed from scratch. Understanding how to read these tables is an essential skill.

\subsubsection{Standard Table Format}

CG tables are organized by the values of $j_1$ and $j_2$ being combined. For each pair $(j_1, j_2)$, the table shows the coefficients for all allowed combinations of $m_1$, $m_2$, and resulting $j$, $m$ values.

A typical table entry looks like:

\begin{center}
\renewcommand{\arraystretch}{1.3}
\begin{tabular}{c|ccc}
\multicolumn{4}{c}{\textbf{$j_1 = 1/2$, $j_2 = 1/2$}}\\[2pt]
\hline
$m_1 \quad m_2$ & $j=1$ & & $j=0$ \\
\hline
$+\frac{1}{2} \quad +\frac{1}{2}$ & $1$ & & \\[3pt]
$+\frac{1}{2} \quad -\frac{1}{2}$ & $\frac{1}{\sqrt{2}}$ & & $\frac{1}{\sqrt{2}}$ \\[3pt]
$-\frac{1}{2} \quad +\frac{1}{2}$ & $\frac{1}{\sqrt{2}}$ & & $-\frac{1}{\sqrt{2}}$ \\[3pt]
$-\frac{1}{2} \quad -\frac{1}{2}$ & $1$ & & \\
\hline
\end{tabular}
\end{center}

To read this table: the entry in row $(m_1, m_2)$ and column $j$ gives the CG coefficient $\langle j_1, m_1; j_2, m_2 | j, m \rangle$ where $m = m_1 + m_2$. Empty entries correspond to forbidden combinations (where $m \neq m_1 + m_2$ or $|m| > j$).

\subsubsection{Example: Using the Table}

From the table above, we can read off the coupled states:
\begin{align}
\ket{j=1, m=1} &= \ket{+\tfrac{1}{2}, +\tfrac{1}{2}}\\
\ket{j=1, m=0} &= \frac{1}{\sqrt{2}}\ket{+\tfrac{1}{2}, -\tfrac{1}{2}} + \frac{1}{\sqrt{2}}\ket{-\tfrac{1}{2}, +\tfrac{1}{2}}\\
\ket{j=0, m=0} &= \frac{1}{\sqrt{2}}\ket{+\tfrac{1}{2}, -\tfrac{1}{2}} - \frac{1}{\sqrt{2}}\ket{-\tfrac{1}{2}, +\tfrac{1}{2}}
\end{align}

\subsubsection{Table for $j_1 = 1$, $j_2 = 1/2$}

\begin{center}
\renewcommand{\arraystretch}{1.3}
\begin{tabular}{cc|cc}
\multicolumn{4}{c}{\textbf{$j_1 = 1$, $j_2 = 1/2$}}\\[2pt]
\hline
$m_1$ & $m_2$ & $j=3/2$ & $j=1/2$ \\
\hline
$+1$ & $+\frac{1}{2}$ & $1$ & \\[3pt]
$+1$ & $-\frac{1}{2}$ & $\sqrt{\frac{1}{3}}$ & $\sqrt{\frac{2}{3}}$ \\[3pt]
$0$ & $+\frac{1}{2}$ & $\sqrt{\frac{2}{3}}$ & $-\sqrt{\frac{1}{3}}$ \\[3pt]
$0$ & $-\frac{1}{2}$ & $\sqrt{\frac{2}{3}}$ & $\sqrt{\frac{1}{3}}$ \\[3pt]
$-1$ & $+\frac{1}{2}$ & $\sqrt{\frac{1}{3}}$ & $-\sqrt{\frac{2}{3}}$ \\[3pt]
$-1$ & $-\frac{1}{2}$ & $1$ & \\
\hline
\end{tabular}
\end{center}

This table is essential for spin-orbit coupling calculations. For example, to find $\ket{j=3/2, m=1/2}$:
\begin{equation}
\ket{\tfrac{3}{2}, \tfrac{1}{2}} = \sqrt{\tfrac{1}{3}}\ket{1, -\tfrac{1}{2}} + \sqrt{\tfrac{2}{3}}\ket{0, +\tfrac{1}{2}}
\end{equation}
where we use shorthand $\ket{m_1, m_2}$ for the uncoupled states.

\subsubsection{Square Root Convention}

Many tables display entries as fractions under a square root sign. For example, $\sqrt{2/3}$ might appear as simply $2/3$ with the understanding that the square root is implied. Some tables show the sign separately, with the magnitude squared in the table. Always check the conventions used in a particular reference.

\subsection{Explicit Examples}

\subsubsection{Adding $j_1 = 1/2$ and $j_2 = 1/2$}

For two spin-1/2 systems, the uncoupled basis has 4 states:
\begin{equation}
\ket{\uparrow\uparrow}, \quad \ket{\uparrow\downarrow}, \quad \ket{\downarrow\uparrow}, \quad \ket{\downarrow\downarrow}
\end{equation}

where we use shorthand $\ket{\uparrow} = \ket{1/2, 1/2}$ and $\ket{\downarrow} = \ket{1/2, -1/2}$.

\textbf{Triplet states ($j = 1$):}
\begin{align}
\ket{1, 1} &= \ket{\uparrow\uparrow}\\
\ket{1, 0} &= \frac{1}{\sqrt{2}}\left(\ket{\uparrow\downarrow} + \ket{\downarrow\uparrow}\right)\\
\ket{1, -1} &= \ket{\downarrow\downarrow}
\end{align}

\textbf{Singlet state ($j = 0$):}
\begin{equation}
\ket{0, 0} = \frac{1}{\sqrt{2}}\left(\ket{\uparrow\downarrow} - \ket{\downarrow\uparrow}\right)
\end{equation}

The non-zero CG coefficients are:
\begin{align}
\langle 1/2, 1/2; 1/2, 1/2 | 1, 1 \rangle &= 1\\
\langle 1/2, 1/2; 1/2, -1/2 | 1, 0 \rangle &= \frac{1}{\sqrt{2}}\\
\langle 1/2, -1/2; 1/2, 1/2 | 1, 0 \rangle &= \frac{1}{\sqrt{2}}\\
\langle 1/2, -1/2; 1/2, -1/2 | 1, -1 \rangle &= 1\\
\langle 1/2, 1/2; 1/2, -1/2 | 0, 0 \rangle &= \frac{1}{\sqrt{2}}\\
\langle 1/2, -1/2; 1/2, 1/2 | 0, 0 \rangle &= -\frac{1}{\sqrt{2}}
\end{align}

\subsubsection{Adding $j_1 = 1$ and $j_2 = 1/2$}

This case is relevant for adding orbital angular momentum $\ell = 1$ to spin $s = 1/2$, giving $j \in \{1/2, 3/2\}$.

For $j = 3/2$, $m = 3/2$:
\begin{equation}
\ket{3/2, 3/2} = \ket{1, 1; 1/2, 1/2}
\end{equation}

For $j = 3/2$, $m = 1/2$:
\begin{equation}
\ket{3/2, 1/2} = \sqrt{\frac{2}{3}}\ket{1, 1; 1/2, -1/2} + \sqrt{\frac{1}{3}}\ket{1, 0; 1/2, 1/2}
\end{equation}

For $j = 1/2$, $m = 1/2$ (orthogonal to above):
\begin{equation}
\ket{1/2, 1/2} = \sqrt{\frac{1}{3}}\ket{1, 1; 1/2, -1/2} - \sqrt{\frac{2}{3}}\ket{1, 0; 1/2, 1/2}
\end{equation}

The complete table of CG coefficients for this case appears above.

\subsubsection{Adding Two Integer Angular Momenta: $j_1 = 1$ and $j_2 = 1$}

For two $\ell = 1$ systems, we get $j \in \{0, 1, 2\}$, giving 9 states total.

For $j = 2$ (quintet):
\begin{align}
\ket{2, 2} &= \ket{1, 1; 1, 1}\\
\ket{2, 1} &= \frac{1}{\sqrt{2}}\left(\ket{1, 1; 1, 0} + \ket{1, 0; 1, 1}\right)\\
\ket{2, 0} &= \frac{1}{\sqrt{6}}\left(\ket{1, 1; 1, -1} + 2\ket{1, 0; 1, 0} + \ket{1, -1; 1, 1}\right)
\end{align}

For $j = 1$ (triplet):
\begin{equation}
\ket{1, 0} = \frac{1}{\sqrt{2}}\left(\ket{1, 1; 1, -1} - \ket{1, -1; 1, 1}\right)
\end{equation}

For $j = 0$ (singlet):
\begin{equation}
\ket{0, 0} = \frac{1}{\sqrt{3}}\left(\ket{1, 1; 1, -1} - \ket{1, 0; 1, 0} + \ket{1, -1; 1, 1}\right)
\end{equation}

\section{Physical Applications}

\subsection{Spin-Orbit Coupling}

For a single electron with orbital angular momentum $\vec{L}$ and spin $\vec{S}$, the total angular momentum is:
\begin{equation}
\vec{J} = \vec{L} + \vec{S}
\end{equation}

With $s = 1/2$, the possible values of $j$ are:
\begin{equation}
j = \ell \pm \frac{1}{2}
\end{equation}

This seemingly simple addition has profound consequences for atomic spectra. The spin-orbit interaction energy depends on $\vec{L} \cdot \vec{S}$, which as we showed earlier can be written as $(\hat{J}^2 - \hat{L}^2 - \hat{S}^2)/2$. This means states with different $j$ have different energies even when they share the same $n$ and $\ell$. The result is fine structure splitting: what would be a single energy level in the absence of spin splits into two levels separated by the spin-orbit coupling energy.

Consider the hydrogen $n=2$, $\ell=1$ level as a concrete example. Without spin-orbit coupling, this is the $2p$ level with six degenerate states (three from $m_\ell = -1, 0, 1$ times two from $m_s = \pm 1/2$). With spin-orbit coupling, these six states reorganize into two multiplets. The $j = 3/2$ level contains four states with magnetic quantum numbers $m = -3/2, -1/2, 1/2, 3/2$. The $j = 1/2$ level contains two states with $m = -1/2, 1/2$. These two levels have slightly different energies, producing the famous sodium D-line doublet at 589 nm, which corresponds to transitions from the $3s$ level to the $3p_{1/2}$ and $3p_{3/2}$ levels.

We will study spin-orbit coupling in detail when we analyze the hydrogen atom in the next chapter.

\subsection{Nuclear Hyperfine Structure}

Nuclear spin $\vec{I}$ couples to electronic angular momentum $\vec{J}$:
\begin{equation}
\vec{F} = \vec{J} + \vec{I}
\end{equation}

This produces hyperfine splitting in atomic spectra, used in atomic clocks and precision measurements.

\subsection{Multi-Electron Atoms}

In atoms with multiple electrons, the situation becomes more complex because we must couple several angular momenta. The appropriate coupling scheme depends on the relative strength of different interactions. In light atoms, LS coupling (Russell-Saunders coupling) applies when electron-electron repulsion dominates over spin-orbit effects. In heavy atoms, jj coupling becomes more appropriate when spin-orbit coupling is strong.

We will return to multi-electron atoms after developing the theory of the hydrogen atom, which will provide the foundation for understanding atomic structure more generally.

\section{Wigner 3j and 6j Symbols}

When working with angular momentum in more advanced contexts, alternative notations for CG coefficients prove useful. These symbols have superior symmetry properties and appear naturally in certain physical calculations. We also introduce the general formalism of spherical tensors and their transformation properties under rotations.

\subsection{The 3j Symbols}

The Wigner 3j symbols are related to CG coefficients by:
\begin{equation}
\begin{pmatrix} j_1 & j_2 & j_3 \\ m_1 & m_2 & m_3 \end{pmatrix} = \frac{(-1)^{j_1-j_2-m_3}}{\sqrt{2j_3+1}} \langle j_1, m_1; j_2, m_2 | j_3, -m_3 \rangle
\end{equation}

Note the sign of $m_3$ on the right-hand side: it appears as $-m_3$, not $+m_3$. This ensures the 3j symbol has enhanced symmetry.

\subsubsection{Symmetry Properties}

The 3j symbols have remarkable symmetry properties that make them preferable for theoretical work:

\textbf{1. Even permutations of columns:}
Any even permutation of the three columns leaves the 3j symbol unchanged:
\begin{align}
\begin{pmatrix} j_1 & j_2 & j_3 \\ m_1 & m_2 & m_3 \end{pmatrix} &= \begin{pmatrix} j_2 & j_3 & j_1 \\ m_2 & m_3 & m_1 \end{pmatrix} = \begin{pmatrix} j_3 & j_1 & j_2 \\ m_3 & m_1 & m_2 \end{pmatrix}
\end{align}

\textbf{2. Odd permutations introduce a phase:}
Odd permutations multiply the 3j symbol by $(-1)^{j_1+j_2+j_3}$:
\begin{equation}
\begin{pmatrix} j_1 & j_2 & j_3 \\ m_1 & m_2 & m_3 \end{pmatrix} = (-1)^{j_1+j_2+j_3} \begin{pmatrix} j_2 & j_1 & j_3 \\ m_2 & m_1 & m_3 \end{pmatrix}
\end{equation}

\textbf{3. Sign reversal of all $m$ values:}
\begin{equation}
\begin{pmatrix} j_1 & j_2 & j_3 \\ m_1 & m_2 & m_3 \end{pmatrix} = (-1)^{j_1+j_2+j_3} \begin{pmatrix} j_1 & j_2 & j_3 \\ -m_1 & -m_2 & -m_3 \end{pmatrix}
\end{equation}

\textbf{4. Selection rules:}
The 3j symbol vanishes unless three conditions are satisfied: $m_1 + m_2 + m_3 = 0$ (which differs from CG coefficients where $m = m_1 + m_2$), the triangle inequality $|j_1 - j_2| \leq j_3 \leq j_1 + j_2$ holds, and $|m_i| \leq j_i$ for all $i$.

\subsubsection{Physical Interpretation}

The 3j symbols describe the coupling of three angular momenta to give zero total angular momentum:
\begin{equation}
\vec{J}_1 + \vec{J}_2 + \vec{J}_3 = 0
\end{equation}

This symmetry in treating all three angular momenta equally (rather than singling out two to be added) makes 3j symbols natural for describing processes with three-fold symmetry, such as matrix elements of operators between angular momentum states, radiation transition probabilities in atoms, nuclear beta decay angular correlations, and scattering processes with three participating particles.

\subsubsection{Relation to Physical Matrix Elements}

Matrix elements of spherical tensor operators $T^{(k)}_q$ between angular momentum states have the form:
\begin{equation}
\langle j', m' | T^{(k)}_q | j, m \rangle \propto \begin{pmatrix} j & k & j' \\ m & q & -m' \end{pmatrix}
\end{equation}

The 3j symbol automatically encodes the selection rules for such transitions: $\Delta m = m' - m = q$ and $|j - j'| \leq k \leq j + j'$. This makes 3j symbols essential for calculating electromagnetic transition rates and selection rules in atomic physics.

\subsection{The 6j Symbols}

When coupling three angular momenta, different coupling orders can be used. Consider $\vec{J}_1$, $\vec{J}_2$, and $\vec{J}_3$. We could form $(\vec{J}_1 + \vec{J}_2) + \vec{J}_3$ with intermediate $\vec{J}_{12} = \vec{J}_1 + \vec{J}_2$, or alternatively $\vec{J}_1 + (\vec{J}_2 + \vec{J}_3)$ with intermediate $\vec{J}_{23} = \vec{J}_2 + \vec{J}_3$.

The 6j symbols relate these different coupling schemes:
\begin{equation}
\ket{(j_1 j_2) j_{12}; j_3; j m} = \sum_{j_{23}} \sqrt{(2j_{12}+1)(2j_{23}+1)} \begin{Bmatrix} j_1 & j_2 & j_{12} \\ j_3 & j & j_{23} \end{Bmatrix} \ket{j_1; (j_2 j_3) j_{23}; j m}
\end{equation}

These arise in recoupling problems and are essential for complex atomic structure calculations involving more than two angular momenta. Examples include LS coupling versus jj coupling in multi-electron atoms and multi-photon processes in quantum optics.

\subsection{Spherical Tensors: The General Picture}

Spherical tensors provide a powerful framework for understanding how composite objects transform under rotations, generalizing the concrete examples we saw in the Cartesian tensor section.

A spherical tensor of rank $k$ has $2k+1$ components $T^{(k)}_q$ where $q = -k, -k+1, \ldots, k-1, k$. These components transform among themselves according to:
\begin{equation}
\label{eq:spherical_tensor_transformation}
T'^{(k)}_q \underset{s}{\overset{\cdot}{=}} \sum_{q'=-k}^{k} D^{(k)}_{qq'}(R) T^{(k)}_{q'}
\end{equation}
where $D^{(k)}_{qq'}(R)$ is a $(2k+1) \times (2k+1)$ matrix called the Wigner D-matrix for the $\ell=k$ representation. The label $s$ indicates this transformation law is written in the spherical basis. The D-matrices are the matrix representations of rotations in the basis of spherical harmonics or, more generally, in the basis that diagonalizes $\hat{J}^2$ and $\hat{J}_z$.

The key advantage is that spherical tensors transform according to irreducible representations. Each rank $k$ forms a separate, unmixed representation. Components of different $k$ never mix under rotations, just as we saw in the block diagonal structure of tensor products.

\subsection{Wigner D-Matrices: Explicit Examples}

The Wigner D-matrices provide the explicit rotation matrices for each angular momentum representation. For a general rotation specified by Euler angles $(\alpha, \beta, \gamma)$, the D-matrix elements are:
\begin{equation}
D^{(j)}_{m'm}(\alpha, \beta, \gamma) = e^{-im'\alpha} d^{(j)}_{m'm}(\beta) e^{-im\gamma}
\end{equation}
where $d^{(j)}_{m'm}(\beta)$ are the Wigner small-d matrices that depend only on the rotation angle $\beta$ about the $y$-axis.

For rotations about the $z$-axis by angle $\theta$, the D-matrices simplify dramatically:
\begin{equation}
D^{(j)}_{m'm}(R_z(\theta)) = \delta_{m'm} e^{-im\theta}
\end{equation}
The matrix is diagonal, with each diagonal element a phase factor.

\subsubsection{Example: $j=1/2$ (Spin-1/2)}

For spin-1/2, the $2 \times 2$ D-matrix for rotation about the $z$-axis is:
\begin{equation}
D^{(1/2)}(R_z(\theta)) \underset{s}{\overset{\cdot}{=}} \begin{pmatrix}
e^{-i\theta/2} & 0 \\
0 & e^{+i\theta/2}
\end{pmatrix}
\end{equation}

For rotation about the $y$-axis by angle $\beta$, the small-d matrix is:
\begin{equation}
d^{(1/2)}(\beta) = \begin{pmatrix}
\cos(\beta/2) & -\sin(\beta/2) \\
\sin(\beta/2) & \cos(\beta/2)
\end{pmatrix}
\end{equation}

This is precisely the matrix representation of rotations acting on spinors.

\subsubsection{Example: $j=1$ (Vectors)}

For $j=1$, the $3 \times 3$ D-matrix for rotation about the $z$-axis is:
\begin{equation}
D^{(1)}(R_z(\theta)) \underset{s}{\overset{\cdot}{=}} \begin{pmatrix}
e^{-i\theta} & 0 & 0 \\
0 & 1 & 0 \\
0 & 0 & e^{+i\theta}
\end{pmatrix}
\end{equation}

Note that this is written in the spherical basis with components labeled by $m = +1, 0, -1$ (in that order). This differs from the Cartesian rotation matrix we saw in Eq.~\eqref{eq:rotation_matrix_z}.

For rotation about the $y$-axis, the small-d matrix elements are:
\begin{equation}
d^{(1)}(\beta) = \begin{pmatrix}
\frac{1+\cos\beta}{2} & -\frac{\sin\beta}{\sqrt{2}} & \frac{1-\cos\beta}{2} \\
\frac{\sin\beta}{\sqrt{2}} & \cos\beta & -\frac{\sin\beta}{\sqrt{2}} \\
\frac{1-\cos\beta}{2} & \frac{\sin\beta}{\sqrt{2}} & \frac{1+\cos\beta}{2}
\end{pmatrix}
\end{equation}

\subsubsection{Example: $j=2$ (Quadrupole)}

For $j=2$, the $5 \times 5$ D-matrix for rotation about the $z$-axis is:
\begin{equation}
D^{(2)}(R_z(\theta)) \underset{s}{\overset{\cdot}{=}} \begin{pmatrix}
e^{-2i\theta} & 0 & 0 & 0 & 0 \\
0 & e^{-i\theta} & 0 & 0 & 0 \\
0 & 0 & 1 & 0 & 0 \\
0 & 0 & 0 & e^{+i\theta} & 0 \\
0 & 0 & 0 & 0 & e^{+2i\theta}
\end{pmatrix}
\end{equation}

The components are ordered by $m = +2, +1, 0, -1, -2$.

\subsection{Using Wigner D-Matrix Tables}

Wigner D-matrices are tabulated in the same references that provide Clebsch-Gordan coefficients. Standard references include Edmonds (\textit{Angular Momentum in Quantum Mechanics}, Princeton, 1957), Varshalovich, Moskalev, and Khersonskii (\textit{Quantum Theory of Angular Momentum}, World Scientific, 1988), Rose (\textit{Elementary Theory of Angular Momentum}, Wiley, 1957), and the online NIST Digital Library of Mathematical Functions (https://dlmf.nist.gov/).

\subsubsection{Reading D-Matrix Tables}

D-matrix tables typically provide the small-d matrix elements $d^{(j)}_{m'm}(\beta)$ for rotation about the $y$-axis. A typical table entry looks like:

\begin{center}
\renewcommand{\arraystretch}{1.3}
\begin{tabular}{c|ccccc}
\multicolumn{6}{c}{\textbf{$d^{(2)}(\beta)$ matrix elements}}\\[2pt]
\hline
$m' \backslash m$ & $+2$ & $+1$ & $0$ & $-1$ & $-2$ \\
\hline
$+2$ & $\cos^4(\beta/2)$ & $-2\cos^3(\beta/2)\sin(\beta/2)$ & $\sqrt{6}\cos^2(\beta/2)\sin^2(\beta/2)$ & $\cdots$ & $\cdots$ \\[3pt]
$+1$ & $2\cos^3(\beta/2)\sin(\beta/2)$ & $\cos(\beta)(2\cos^2(\beta/2)-1)$ & $\cdots$ & $\cdots$ & $\cdots$ \\[3pt]
$0$ & $\sqrt{6}\cos^2(\beta/2)\sin^2(\beta/2)$ & $\cdots$ & $\frac{3\cos(2\beta)+1}{4}$ & $\cdots$ & $\cdots$ \\[3pt]
\hline
\end{tabular}
\end{center}

To read this table: the entry in row $m'$ and column $m$ gives $d^{(j)}_{m'm}(\beta)$. Due to symmetry properties, many tables only show half the entries.

\subsubsection{Symmetry Properties}

The d-matrices satisfy several useful symmetries:
\begin{align}
d^{(j)}_{m'm}(\beta) &= (-1)^{m'-m} d^{(j)}_{mm'}(\beta) \quad \text{(transpose symmetry)}\\
d^{(j)}_{m'm}(\beta) &= (-1)^{m'-m} d^{(j)}_{-m,-m'}(\beta) \quad \text{(reflection symmetry)}\\
d^{(j)}_{m'm}(0) &= \delta_{m'm} \quad \text{(identity at $\beta=0$)}\\
d^{(j)}_{m'm}(\pi) &= (-1)^{j+m} \delta_{m',-m} \quad \text{(at $\beta=\pi$)}
\end{align}

These properties allow you to compute missing table entries from the ones provided.

\subsubsection{Practical Usage}

When calculating how a spherical tensor component transforms under a rotation:
\begin{enumerate}
\item Identify the rank $j$ of your tensor
\item Determine the Euler angles $(\alpha, \beta, \gamma)$ for your rotation
\item Look up the $d^{(j)}_{m'm}(\beta)$ elements from tables
\item Apply the full D-matrix formula: $D^{(j)}_{m'm}(\alpha, \beta, \gamma) = e^{-im'\alpha} d^{(j)}_{m'm}(\beta) e^{-im\gamma}$
\item Sum over $m$ according to Eq.~\eqref{eq:spherical_tensor_transformation}
\end{enumerate}

For simple rotations about coordinate axes, use the simplified forms we derived above.

\subsection{Tensor Products and Angular Momentum}

The tensor product of two spherical tensors decomposes into spherical tensors of definite rank:
\begin{equation}
T^{(k_1)} \otimes T^{(k_2)} = \bigoplus_{k=|k_1-k_2|}^{k_1+k_2} T^{(k)}
\end{equation}

This is precisely the angular momentum addition formula. The rank $k$ of a spherical tensor is its angular momentum quantum number. The $(2k_1+1)(2k_2+1)$-dimensional reducible representation decomposes into irreducible blocks of dimensions $2k+1$ for $k = |k_1-k_2|, \ldots, k_1+k_2$.

\subsection{Example: Tensor Product of Two Vectors (\texorpdfstring{$\ell=1$}{l=1})}

Consider two vectors $\vec{A}$ and $\vec{B}$. In Cartesian coordinates, their tensor product gives 9 components:
\begin{equation}
T_{ij} = A_i B_j \quad \text{for } i, j \in \{x, y, z\}
\end{equation}

This is a rank-2 Cartesian tensor with 9 components. However, not all these components transform independently under rotations. We can decompose this tensor into three irreducible parts.

\subsubsection{Decomposition: $3 \otimes 3 = 5 \oplus 3 \oplus 1$}

In terms of angular momentum: $1 \otimes 1 = 2 \oplus 1 \oplus 0$. The 9 Cartesian components decompose into three pieces: $\ell = 2$ gives the symmetric traceless tensor (5 components), $\ell = 1$ gives the antisymmetric tensor (3 components), and $\ell = 0$ gives the trace (1 component).

\textbf{The $\ell=0$ piece: Dot product (1 component)}

The trace of the tensor gives a scalar:
\begin{equation}
T^{(0)} = \frac{1}{3}(A_x B_x + A_y B_y + A_z B_z) = \frac{1}{3}\vec{A} \cdot \vec{B}
\end{equation}

This is the unique rotationally invariant combination of two vectors. Under rotations, $\vec{A} \cdot \vec{B}$ remains unchanged, as expected for an $\ell=0$ object.

\textbf{The $\ell=1$ piece: Cross product (3 components)}

The antisymmetric part of the tensor corresponds to the cross product:
\begin{equation}
T^{(1)}_i = \frac{1}{2}(A_j B_k - A_k B_j) = \frac{1}{2}(\vec{A} \times \vec{B})_i
\end{equation}

where $(i,j,k)$ is a cyclic permutation of $(x,y,z)$. This gives three independent components:
\begin{align}
T^{(1)}_x &= A_y B_z - A_z B_y\\
T^{(1)}_y &= A_z B_x - A_x B_z\\
T^{(1)}_z &= A_x B_y - A_y B_x
\end{align}

These three components form a vector (an $\ell=1$ spherical tensor) that transforms like $\vec{A} \times \vec{B}$ under rotations.

\textbf{The $\ell=2$ piece: Symmetric traceless tensor (5 components)}

The symmetric traceless part is:
\begin{equation}
T^{(2)}_{ij} = \frac{1}{2}(A_i B_j + A_j B_i) - \frac{1}{3}\delta_{ij} \vec{A} \cdot \vec{B}
\end{equation}

This tensor is symmetric, $T^{(2)}_{ij} = T^{(2)}_{ji}$, which reduces 9 components to 6, and traceless, $T^{(2)}_{xx} + T^{(2)}_{yy} + T^{(2)}_{zz} = 0$, which further reduces 6 components to 5. The 5 independent components correspond to the $2\ell+1 = 5$ components of a spherical tensor with $\ell=2$. Examples include the quadrupole moment tensor and the $d$-orbitals in atomic physics.

\subsubsection{Physical Examples}

This decomposition appears throughout physics. For a charge distribution $\rho(\vec{r})$, the components $x_i x_j$ form a rank-2 tensor whose symmetric traceless part gives the electric quadrupole moment tensor $Q_{ij}$ with $\ell=2$, creating quadrupole electric fields. In atomic physics, combining two $p$-orbitals ($\ell=1$) gives $s$ ($\ell=0$), $p$ ($\ell=1$), and $d$ ($\ell=2$) orbitals, where the $s$-orbital is spherically symmetric (the dot product), $p$-orbitals are dipolar (related to the cross product structure), and $d$-orbitals have quadrupole symmetry. The interaction of light (photon angular momentum $\ell=1$) with matter involves the tensor product of the electromagnetic field with atomic transition operators, and the selection rules for electric dipole ($\ell=1$), quadrupole ($\ell=2$), and higher multipole transitions follow from this tensor decomposition. The polarizability tensor $\alpha_{ij}$ relates the induced dipole moment to an applied electric field, where its symmetric traceless part describes anisotropic polarizability (different response along different axes), while its trace gives the average isotropic polarizability.

The block diagonal structure we found for $\ell=1 \otimes \ell=1$ generalizes to all tensor products. The Wigner D-matrices provide the explicit rotation matrices for each irreducible block, and Clebsch-Gordan coefficients provide the change of basis that achieves the block diagonalization.

\subsection{Connection to Clebsch-Gordan Coefficients}

The decomposition of tensor products is accomplished by Clebsch-Gordan coefficients. For two vectors in spherical tensor form:
\begin{equation}
T^{(k)}_q = \sum_{q_1, q_2} \langle 1, q_1; 1, q_2 | k, q \rangle A^{(1)}_{q_1} B^{(1)}_{q_2}
\end{equation}

For $k=0$ (scalar), $k=1$ (vector), and $k=2$ (rank-2 tensor), the CG coefficients automatically project out the dot product, cross product, and symmetric traceless combinations, respectively.

\section{Chapter Summary}

The addition of angular momenta is a central problem in quantum mechanics, arising whenever a system possesses multiple sources of angular momentum. We developed two complementary descriptions:

The \textbf{uncoupled basis} $\ket{j_1, m_1; j_2, m_2}$ specifies each angular momentum separately and is natural when the subsystems are weakly interacting. The \textbf{coupled basis} $\ket{j_1, j_2; j, m}$ specifies the total angular momentum and is preferred when interactions couple the subsystems or when total angular momentum is conserved.

The tensor product of two angular momentum spaces decomposes into a direct sum of subspaces with definite total $j$:
\begin{equation}
\mathcal{H}_1 \otimes \mathcal{H}_2 = \mathcal{H}_{|j_1-j_2|} \oplus \mathcal{H}_{|j_1-j_2|+1} \oplus \cdots \oplus \mathcal{H}_{j_1+j_2}
\end{equation}

The transformation between bases is accomplished by \textbf{Clebsch-Gordan coefficients}, which encode the overlap between uncoupled and coupled states. These coefficients satisfy conservation of $m$, requiring $m = m_1 + m_2$, reflecting the additivity of angular momentum projections. They obey the triangle inequality $|j_1 - j_2| \leq j \leq j_1 + j_2$, which determines the allowed values of total angular momentum for given constituent angular momenta. The coefficients also satisfy orthonormality relations that ensure the transformation between bases is unitary, along with various symmetry relations that reflect the underlying rotational symmetry of the problem.

We computed explicit examples for common cases: two spin-1/2 particles (singlet and triplet), adding $\ell = 1$ to $s = 1/2$ (fine structure), and two $\ell = 1$ systems. Standard CG tables provide a practical resource for looking up these coefficients.

Physical applications include spin-orbit coupling, which produces fine structure splitting in atomic spectra, and nuclear hyperfine structure arising from the coupling of nuclear spin to electronic angular momentum. The 3j and 6j symbols provide alternative notations with enhanced symmetry properties, useful when combining more than two angular momenta.

The next chapter applies these angular momentum addition techniques to the hydrogen atom, where we'll see how spin-orbit coupling produces fine structure splitting of energy levels and how the coupled basis emerges as the natural description for atomic states.

\input{book_problems/ch07_problems.tex}

\section*{References and Further Reading}
\addcontentsline{toc}{section}{References and Further Reading}

\begin{description}
\item[Sakurai, J.~J., and Napolitano, J.] \emph{Modern Quantum Mechanics}, 3rd ed. Cambridge University Press, 2017. Sections 3.7--3.10 give the canonical graduate treatment of angular momentum addition, Clebsch--Gordan coefficients, and tensor operators; the natural next stop after this chapter.

\item[Edmonds, A.~R.] \emph{Angular Momentum in Quantum Mechanics}. Princeton University Press, 1957. The standard monograph; tightly written and the source most often cited for the conventions and symmetry relations used here.

\item[Varshalovich, D.~A., Moskalev, A.~N., and Khersonskii, V.~K.] \emph{Quantum Theory of Angular Momentum}. World Scientific, 1988. Comprehensive reference with explicit tables of 3-j, 6-j, and 9-j symbols; the first place to look when a CG coefficient is needed in practice.

\item[Wigner, E.~P.] \emph{Group Theory and Its Application to the Quantum Mechanics of Atomic Spectra}. Academic Press, 1959. Translation of the 1931 original. The historical foundation: angular momentum addition derived from the representation theory of $SU(2)$.

\item[Zare, R.~N.] \emph{Angular Momentum: Understanding Spatial Aspects in Chemistry and Physics}. Wiley, 1988. Particularly clear on spherical tensor operators and the Wigner--Eckart theorem; recommended for the rank-2 tensor decomposition discussed at the end of this chapter.
\end{description}

%% file: book_problems/ch07_problems.tex
\section{Problems}
\setcounter{hwproblem}{0}

\problem{Hilbert Space Dimensions and Tensor Products}
Consider two angular momentum systems with quantum numbers $j_1$ and $j_2$.
\begin{enumerate}[label=(\alph*)]
    \item A spin-1/2 particle (electron) and a spin-1 particle are combined into a composite system. What is the dimension of the total Hilbert space $\mathcal{H} = \mathcal{H}_1 \otimes \mathcal{H}_2$? Write out all basis states in the uncoupled basis $\ket{j_1, m_1; j_2, m_2}$ explicitly.
    \item Using the decomposition formula $\mathcal{H}_1 \otimes \mathcal{H}_2 = \bigoplus_{j=|j_1-j_2|}^{j_1+j_2} \mathcal{H}_j$, determine which total angular momentum values $j$ are allowed for this system. For each value of $j$, state the dimension of the corresponding subspace $\mathcal{H}_j$ and verify that the total dimension matches your answer from part (a).
    \item Two spin-1 particles are combined. Determine the allowed values of total angular momentum $j$ and their multiplicities. Identify which $j$ value corresponds to a symmetric state under particle exchange, which to an antisymmetric state, and which are mixed symmetry.
\end{enumerate}

\problem{Uncoupled versus Coupled Bases}
For a system of two spin-1/2 particles, we can use either the uncoupled basis $\ket{m_1, m_2}$ or the coupled basis $\ket{j, m}$.
\begin{enumerate}[label=(\alph*)]
    \item Write out the four uncoupled basis states explicitly. For each state, determine the eigenvalues of $\hat{J}_{1z}$, $\hat{J}_{2z}$, and $\hat{J}_z = \hat{J}_{1z} + \hat{J}_{2z}$.
    \item The state with maximum $m$ value and the state with minimum $m$ value can be immediately identified as coupled basis eigenstates. Identify these two states and explain why no calculation is needed.
    \item Consider the operator $\vec{J}_1 \cdot \vec{J}_2 = \frac{1}{2}(\hat{J}^2 - \hat{J}_1^2 - \hat{J}_2^2)$. Using this expression, determine the eigenvalue of $\vec{J}_1 \cdot \vec{J}_2$ for each coupled basis state $\ket{j, m}$. What are the numerical values (in units of $\hbar^2$) for the singlet and triplet states?
\end{enumerate}

\problem{SU(2), SO(3), and Spinors}
The distinction between SU(2) and SO(3) leads to fundamental differences between integer and half-integer angular momentum.
\begin{enumerate}[label=(\alph*)]
    \item A rotation by angle $\theta$ about the $z$-axis acts on a spin-1/2 state as $\hat{R}_z(\theta) = e^{-i\theta\hat{J}_z/\hbar} = e^{-i\theta\sigma_z/2}$. Calculate the matrix representation of $\hat{R}_z(2\pi)$ and $\hat{R}_z(4\pi)$ in the basis $\{\ket{\uparrow}, \ket{\downarrow}\}$. What is the physical interpretation?
    \item For orbital angular momentum with $\ell = 1$, show that $\hat{R}_z(2\pi) = \identity$ by considering how $Y_{\ell m}(\theta,\phi) \propto e^{im\phi}$ transforms under $\phi \to \phi + 2\pi$. Compare this result with part (a).
    \item The tensor product $\frac{1}{2} \otimes \frac{1}{2} = 0 \oplus 1$ gives both a singlet ($j=0$, integer) and a triplet ($j=1$, integer). Verify the dimension formula $(2j_1+1)(2j_2+1) = \sum_j(2j+1)$ for $\frac{1}{2} \otimes 1 = \frac{1}{2} \oplus \frac{3}{2}$. What pattern governs whether tensor products of half-integer angular momenta give integer or half-integer results?
\end{enumerate}

\problem{Cartesian and Spherical Tensor Decomposition}
Consider two position vectors $\vec{r}_1 = (x_1, y_1, z_1)$ and $\vec{r}_2 = (x_2, y_2, z_2)$.
\begin{enumerate}[label=(\alph*)]
    \item The tensor product $\vec{r}_1 \otimes \vec{r}_2$ has 9 components that decompose into irreducible parts with $\ell = 0, 1, 2$. Write explicit expressions for: (i) the $\ell=0$ scalar, (ii) the three components of the $\ell=1$ antisymmetric tensor, (iii) one example component from the $\ell=2$ symmetric traceless tensor. Verify that $1 + 3 + 5 = 9$.
    \item Show that the $\ell=0$ scalar combination $\vec{r}_1 \cdot \vec{r}_2$ is invariant under rotation about the $z$-axis by angle $\theta$.
    \item The spherical harmonics for $\ell=1$ are proportional to: $Y_{1,0} \propto \cos\theta$, $Y_{1,\pm 1} \propto \sin\theta e^{\pm i\phi}$. The Cartesian coordinates can be expressed as linear combinations of $(Y_{1,+1}, Y_{1,0}, Y_{1,-1})$. Invert these relations to express $x/r$, $y/r$, and $z/r$ in terms of the $Y_{1,m}$ and identify the coefficients.
\end{enumerate}

\problem{Triangle Inequality and Selection Rules}
The allowed values of total angular momentum must satisfy $|j_1 - j_2| \leq j \leq j_1 + j_2$.
\begin{enumerate}[label=(\alph*)]
    \item Consider the addition of orbital angular momentum $\ell = 2$ (a $d$-orbital) and spin $s = 1/2$. List all allowed values of total angular momentum $j$. For each allowed $j$, determine how many states (values of $m$) belong to that multiplet. Verify the total count matches $(2\ell+1)(2s+1)$.
    \item In spectroscopic notation $^{2S+1}L_J$, for a single electron in a $d$-orbital ($\ell=2$) with spin $s=1/2$, the term symbol is $^2D_{J}$. What values can $J$ take? Write out the complete term symbols.
    \item The spin-orbit Hamiltonian is $H_{\text{SO}} = \xi(\vec{r}) \vec{L} \cdot \vec{S}$. Using the identity $\vec{L} \cdot \vec{S} = \frac{1}{2}(\hat{J}^2 - \hat{L}^2 - \hat{S}^2)$, show that the coupled basis states $\ket{\ell, s; j, m}$ are eigenstates of $H_{\text{SO}}$ (assuming $\xi$ depends only on $r$). For the $^2D$ states, determine the energy splitting pattern if $\xi$ is constant. Which state has lower energy for $\xi > 0$?
\end{enumerate}

\problem{Clebsch-Gordan Structure and Orthogonality}
Although we have not yet calculated Clebsch-Gordan coefficients explicitly, we can understand their structure.
\begin{enumerate}[label=(\alph*)]
    \item The coupled and uncoupled bases are related by $\ket{j_1, j_2; j, m} = \sum_{m_1, m_2} C_{j_1 m_1, j_2 m_2}^{j m} \ket{j_1, m_1; j_2, m_2}$. The constraint $m = m_1 + m_2$ follows from conservation of the $z$-component of angular momentum. Explain why the Clebsch-Gordan coefficient $C_{j_1 m_1, j_2 m_2}^{j m} = 0$ whenever $m \neq m_1 + m_2$.
    \item For two spin-1/2 particles, the triplet state with $m=0$ is $\ket{j=1, m=0} = \frac{1}{\sqrt{2}}(\ket{\uparrow\downarrow} + \ket{\downarrow\uparrow})$. The singlet state must be orthogonal to this. Write down the singlet state $\ket{j=0, m=0}$ and verify orthogonality explicitly.
    \item The coupled basis forms a complete orthonormal set. For two spin-1/2 particles, verify the completeness relation $\sum_{j,m} \ket{j, m}\bra{j, m} = \identity$ by expanding both sides in the uncoupled basis and recovering the $4 \times 4$ identity matrix.
\end{enumerate}

\problem{Construction of Clebsch-Gordan Coefficients for Spin-1/2 + Spin-1/2}
Construct the Clebsch-Gordan coefficients for $1/2 \otimes 1/2$ using ladder operators.
\begin{enumerate}[label=(\alph*)]
    \item The highest state $\ket{1, +1}$ (triplet, $j=1$, $m=1$) is the unique state with maximum $m$. Write $\ket{1, +1} = \ket{\uparrow\uparrow}$.
    \item Apply the lowering operator $\hat{J}_- = \hat{J}_{1-} + \hat{J}_{2-}$ to $\ket{1, +1}$ to obtain $\ket{1, 0}$ (triplet with $m=0$). Normalize correctly.
    \item Apply $\hat{J}_-$ again to obtain $\ket{1, -1}$ (triplet with $m=-1$).
    \item The fourth state with $m=0$ must be the singlet $\ket{0, 0}$. Using orthogonality with the triplet state from part (b), write down $\ket{0, 0}$ explicitly.
    \item Verify that your four coupled states form an orthonormal basis by computing inner products.
\end{enumerate}

\problem{Addition of Three Spin-1/2 Particles}
Three spin-1/2 particles are combined into a composite system.
\begin{enumerate}[label=(\alph*)]
    \item First, couple particles 1 and 2: $\mathcal{H}_{12} = \mathcal{H}_1 \otimes \mathcal{H}_2 = 0 \oplus 1$. Then couple the result to particle 3. Show that the total decomposition is $(0 \oplus 1) \otimes \frac{1}{2} = \frac{1}{2} \oplus \frac{3}{2}$.
    \item Alternatively, first couple particles 1 and 3: $\mathcal{H}_{13} = 0 \oplus 1$. Then couple to particle 2, giving the same decomposition. Why must both orderings give the same result?
    \item How many total states are there? Verify that $2 \times 2 \times 2 = 8 = 2 + 6$ (two doublets and one sextet).
    \item Identify the state with maximum total spin projection $m_{\text{max}} = 3/2$. What is the total angular momentum quantum number for this state?
    \item Construct the fully antisymmetric three-spin state (the Fermi sea state for spin-3/2 fermions). This state is completely antisymmetric under all particle exchanges.
\end{enumerate}

\problem{Wigner-Eckart Theorem and Reduced Matrix Elements}
The Wigner-Eckart theorem relates matrix elements to reduced matrix elements and Clebsch-Gordan coefficients.
\begin{enumerate}[label=(\alph*)]
    \item For a spherical tensor operator $T_q^{(k)}$ acting between $\ell \to \ell'$ states, the matrix element is $\bra{\ell', m'} T_q^{(k)} \ket{\ell, m} = C_{\ell m, k q}^{\ell' m'} \langle \ell' \| T^{(k)} \| \ell \rangle$. For the dipole operator $z \propto T_0^{(1)} \propto Y_{1,0}$, verify the selection rules $\Delta\ell = \pm 1$, $\Delta m = 0$ by noting which Clebsch-Gordan coefficients are nonzero.
    \item In hydrogen, the selection rule for allowed transitions is $\Delta\ell = \pm 1$. Using the Wigner-Eckart theorem, explain why transitions $2s \to 1s$ and $3d \to 2d$ are forbidden, while $2p \to 1s$ is allowed.
    \item For the magnetic moment operator $\mu_z = -g \mu_B S_z / \hbar$ acting on spin-1/2, compute the reduced matrix element $\langle 1/2 \| \mu_z \| 1/2 \rangle$ and verify it is independent of $m$.
\end{enumerate}

\problem{Spin-Orbit Coupling and the Lande g-Factor}
The spin-orbit interaction for a single electron is $H_{\text{SO}} = \lambda \vec{L} \cdot \vec{S}$.
\begin{enumerate}[label=(\alph*)]
    \item Using $\vec{L} \cdot \vec{S} = \frac{1}{2}(\hat{J}^2 - \hat{L}^2 - \hat{S}^2)$, show that the energy shifts for a state $\ket{\ell, s; j, m}$ are $E_{\text{SO}} = \frac{\lambda\hbar^2}{2}[j(j+1) - \ell(\ell+1) - s(s+1)]$.
    \item For the $^2P$ term ($\ell = 1$, $s = 1/2$), determine the two levels $j = 1/2$ and $j = 3/2$ and their energy splitting.
    \item The magnetic moment in a magnetic field is $\mu_z = -g_J \mu_B m_j$ where $g_J$ is the Lande g-factor. Show that:
    $$g_J = 1 + \frac{j(j+1) + s(s+1) - \ell(\ell+1)}{2j(j+1)}$$
    \item For the $^2P_{3/2}$ state, calculate $g_J$ numerically. What is the energy shift in a magnetic field $B = 1$ Tesla (in eV)?
\end{enumerate}

\problem{Isospin Symmetry and Nucleon Doublets}
In nuclear physics, the nucleon (proton and neutron) forms an isospin doublet with $T = 1/2$.
\begin{enumerate}[label=(\alph*)]
    \item Write the nucleon isospin states as $\ket{1/2, +1/2}$ (proton) and $\ket{1/2, -1/2}$ (neutron). Under isospin rotations, these transform as a spin-1/2 system. What is the physical analog of the spin raising and lowering operators?
    \item For two nucleons, construct the coupled isospin states: singlet $\ket{0, 0}$ and triplet $\ket{1, T_z}$ with $T_z = -1, 0, +1$. Identify which states correspond to $(pp)$, $(pn)$, $(np)$, and $(nn)$.
    \item The strong nuclear force treats protons and neutrons identically and is invariant under isospin rotations. Under the Pauli exclusion principle for identical fermions, which two-nucleon states are allowed: even isospin, odd isospin, or both? Explain.
    \item The electric dipole operator breaks isospin symmetry. It transforms as $T = 1$ (like a vector). Which isospin transitions does it allow between states of different nuclei?
\end{enumerate}

\problem{Tensor Operators and Magnetic Quadrupole Moments}
The electric quadrupole moment is a rank-2 spherical tensor.
\begin{enumerate}[label=(\alph*)]
    \item A rank-2 tensor has $2k+1=5$ components. The electric quadrupole moment tensor is $Q_q^{(2)} \propto (3z_i z_j - \delta_{ij}r^2)$ in Cartesian form. What are the five spherical components?
    \item For a state with $j = 1$ (like $^3P$), compute the expectation value $\langle j=1 \| Q^{(2)} \| j=1 \rangle$ using the Wigner-Eckart theorem.
    \item The second-order Stark effect in hydrogen involves the polarizability tensor, which is rank-2. Can this tensor mix states with $\ell = 0$ to $\ell = 0$? Why or why not?
\end{enumerate}

%% file: chapters/ch08_hydrogen_atom.tex
\chapter{The Hydrogen Atom}
\label{ch:hydrogen_atom}

\section{Introduction}

The hydrogen atom represents one of the greatest triumphs of quantum mechanics. As the simplest atom with only one electron bound to a proton, it serves as the perfect testing ground for quantum mechanical principles while being exactly solvable. In this chapter, we solve the Schr\"{o}dinger equation for the hydrogen atom using the techniques developed thus far, particularly the angular momentum formalism from Chapters~\ref{ch:3d_angular} and~\ref{ch:angular_addition}.

\begin{keyidea}{Central Force Problems in Quantum Mechanics}
The hydrogen atom is the prototypical central force problem in quantum mechanics. The spherical symmetry allows us to separate the radial and angular parts of the wave function, leading to the famous quantum numbers $n$, $\ell$, and $m$.
\end{keyidea}

Before diving into the solution, we pause to discuss unit systems. The choice of units dramatically affects the appearance of our equations, and graduate students must be fluent in multiple conventions.

\section{Units in Electromagnetic and Atomic Physics}

\subsection{Why Units Matter}

The choice of units in physics is more than bookkeeping---it reflects what we consider fundamental and can dramatically simplify or complicate our equations. Consider the hydrogen atom Hamiltonian. In SI units:
\begin{equation}
    \hat{H} = -\frac{\hbar^2}{2m_e}\nabla^2 - \frac{e^2}{4\pi\epsilon_0 r}
\end{equation}
In atomic units:
\begin{equation}
    \hat{H} = -\frac{1}{2}\nabla^2 - \frac{1}{r}
\end{equation}
The physics is identical, but the second form lets you focus on the structure without the constant clutter. Different unit systems are optimized for different purposes.

\subsection{SI Units}

SI units (Syst\`{e}me International), also called MKS units, are the modern international standard. The base units are the meter, kilogram, second, and ampere. The permittivity of free space $\epsilon_0$ and permeability $\mu_0$ appear explicitly in electromagnetic equations.

Coulomb's law reads:
\begin{equation}
    F = \frac{1}{4\pi\epsilon_0}\frac{q_1 q_2}{r^2}
\end{equation}

The factors of $4\pi\epsilon_0$ appear throughout, which can obscure the physics but makes dimensional analysis straightforward. SI is standard in engineering, experimental physics, and undergraduate courses. Griffiths' textbooks use SI.

\subsection{Gaussian Units}

Gaussian units (CGS) were the standard in theoretical physics for most of the 20th century. The defining choice is that Coulomb's law takes its simplest form:
\begin{equation}
    F = \frac{q_1 q_2}{r^2}
\end{equation}
with no $4\pi\epsilon_0$ factor. The unit of charge, the \textit{statcoulomb} or \textit{esu}, is defined to make this work.

The trade-off is that factors of $c$ appear explicitly in Maxwell's equations and in relationships between electric and magnetic quantities. Gaussian units remain standard in plasma physics, astrophysics, and much of theoretical physics. Jackson's \textit{Classical Electrodynamics}, Landau \& Lifshitz, and most graduate quantum mechanics texts (Sakurai, Shankar) use Gaussian conventions.

\subsection{Heaviside-Lorentz Units}

Heaviside-Lorentz units modify Gaussian units by absorbing factors of $4\pi$ into the definitions of fields and charges. This ``rationalizes'' Maxwell's equations: the differential forms contain no $4\pi$ factors, while they appear in Coulomb's law:
\begin{equation}
    F = \frac{1}{4\pi}\frac{q_1 q_2}{r^2}
\end{equation}

This system is standard in quantum field theory and particle physics, where the clean form of Maxwell's equations is advantageous.

\subsection{Atomic Units}

Atomic units are constructed by setting the fundamental constants of atomic physics equal to unity:
\begin{equation}
    \hbar = m_e = e = 1
\end{equation}
(building on Gaussian conventions for electromagnetic quantities). This is not merely a notational trick---it defines a complete system of units with natural scales for atomic phenomena.

\textbf{The Bohr radius.} The \textbf{Bohr radius} $a_0$ is defined as:
\begin{equation}
    a_0 = \frac{\hbar^2}{m_e e^2} = 1 \quad \text{(atomic units)}
\end{equation}
In SI, $a_0 = 5.29 \times 10^{-11}$ m $= 0.529$ \AA. This is the natural length scale for atoms: the most probable electron-proton distance in the hydrogen ground state is exactly one Bohr radius.

The Bohr radius emerges from balancing kinetic and potential energy. The uncertainty principle requires $\Delta p \sim \hbar/r$ for confinement to radius $r$, giving kinetic energy $\sim \hbar^2/(m_e r^2)$. The potential energy is $\sim -e^2/r$. Minimizing the total energy yields $r \sim \hbar^2/(m_e e^2) = a_0$.

\textbf{The Hartree.} The \textbf{Hartree} $E_h$ (named after Douglas Hartree of Hartree-Fock fame) is the atomic unit of energy, defined as the potential energy of two elementary charges separated by one Bohr radius:
\begin{equation}
    E_h = \frac{e^2}{a_0} = 1 \quad \text{(atomic units)}
\end{equation}
In SI, $1 \, E_h = 27.2$ eV $= 4.36 \times 10^{-18}$ J. Equivalently, $E_h = m_e v_0^2$ where $v_0 = e^2/\hbar$ is the characteristic atomic velocity.

\textbf{The Rydberg.} The \textbf{Rydberg} (symbol Ry) is defined as half a Hartree:
\begin{equation}
    1 \text{ Ry} = \frac{1}{2} E_h = 13.6 \text{ eV}
\end{equation}
This is precisely the ionization energy of hydrogen---the energy required to remove the electron from the ground state. The Rydberg is often more convenient for spectroscopy since hydrogen energy levels are $E_n = -\text{Ry}/n^2$.

In atomic units, energies are measured in Hartrees, so the hydrogen ground state energy is $E_1 = -1/2$ (meaning $-\frac{1}{2} E_h = -1$ Ry $= -13.6$ eV).

\textbf{Other atomic units.} Setting $\hbar = m_e = e = 1$ determines units for all other quantities; see Table~\ref{tab:atomic_constants} in Section~\ref{sec:comparison}.

\textbf{The speed of light in atomic units.} In atomic units, $c$ is \textit{not} equal to 1. Instead:
\begin{equation}
    c = \frac{1}{\alpha} \approx 137
\end{equation}
where $\alpha = e^2/(\hbar c) \approx 1/137$ is the fine structure constant. This large value of $c$ reflects the fact that typical atomic velocities ($v \sim 1$ in atomic units) are much smaller than the speed of light, justifying non-relativistic quantum mechanics for atoms. Relativistic corrections are suppressed by factors of $(v/c)^2 = \alpha^2 \sim 10^{-4}$.

\subsection{Using Atomic Units in Practice}

A common source of confusion: if $a_0 = 1$ in atomic units, what do we write when a distance equals one Bohr radius? Do we write ``$r = 1$'' or ``$r = a_0$''?

The answer depends on context:

\textbf{During calculations}, use pure numbers. When working in atomic units, all quantities are dimensionless numbers. The ground state wave function is $\psi = \pi^{-1/2} e^{-r}$, the most probable radius is $r^* = 1$, and the ground state energy is $E_1 = -1/2$. This is the whole point of atomic units: no constants cluttering the algebra.

\textbf{When reporting results}, restore the units for clarity. Write ``the most probable radius is $r^* = 1 \, a_0 = 0.529$ \AA'' or ``the ground state energy is $E_1 = -1/2 \, E_h = -13.6$ eV.'' This makes the physical scale explicit and allows comparison with experiment.

\textbf{In equations that define the system}, you can go either way. The hydrogen Hamiltonian can be written as
\begin{equation}
    \hat{H} = -\frac{1}{2}\nabla^2 - \frac{1}{r}
\end{equation}
with the understanding that $r$ is measured in Bohr radii and $H$ in Hartrees, or equivalently
\begin{equation}
    \hat{H} = -\frac{1}{2}\nabla^2 - \frac{1}{r/a_0} \cdot E_h
\end{equation}
but the first form is cleaner and standard.

The key insight is that atomic units work exactly like setting $c = 1$ in special relativity or $\hbar = 1$ in quantum mechanics: during calculations you drop the constants, and when comparing to experiment you restore them by dimensional analysis.

\begin{physicalinsight}
Think of atomic units as measuring everything in ``natural'' atomic scales. Saying ``$r = 2$'' means the distance is twice the Bohr radius. Saying ``$E = -0.5$'' means the energy is half a Hartree (one Rydberg). The numbers directly tell you the physics in atomic terms.
\end{physicalinsight}

\subsection{Comparison of Key Formulas}
\label{sec:comparison}

Table~\ref{tab:formulas_units} shows how key formulas appear in different unit systems.

\begin{table}[ht]
\centering
\caption{Formulas in different unit systems.}
\label{tab:formulas_units}
\renewcommand{\arraystretch}{2.0}
\begin{tabular}{lccc}
\hline
\textbf{Expression} & \textbf{SI} & \textbf{Gaussian} & \textbf{Atomic} \\
\hline
Coulomb potential & $\displaystyle -\frac{e^2}{4\pi\epsilon_0 r}$ & $\displaystyle -\frac{e^2}{r}$ & $\displaystyle -\frac{1}{r}$ \\[12pt]
H Hamiltonian & $\displaystyle -\frac{\hbar^2}{2m_e}\nabla^2 - \frac{e^2}{4\pi\epsilon_0 r}$ & $\displaystyle -\frac{\hbar^2}{2m_e}\nabla^2 - \frac{e^2}{r}$ & $\displaystyle -\frac{1}{2}\nabla^2 - \frac{1}{r}$ \\[12pt]
Fine structure constant & $\displaystyle \frac{e^2}{4\pi\epsilon_0\hbar c}$ & $\displaystyle \frac{e^2}{\hbar c}$ & $\displaystyle \frac{1}{c}$ \\[8pt]
\hline
\end{tabular}
\end{table}

Table~\ref{tab:atomic_constants} lists the fundamental atomic constants.

\begin{table}[ht]
\centering
\caption{Fundamental atomic constants.}
\label{tab:atomic_constants}
\renewcommand{\arraystretch}{1.4}
\begin{tabular}{llccc}
\hline
\textbf{Constant} & \textbf{Symbol} & \textbf{Definition} & \textbf{Atomic value} & \textbf{SI value} \\
\hline
Bohr radius & $a_0$ & $\hbar^2/(m_e e^2)$ & 1 & 0.529 \AA \\
Hartree & $E_h$ & $e^2/a_0$ & 1 & 27.2 eV \\
Rydberg & Ry & $E_h/2$ & $1/2$ & 13.6 eV \\
Atomic time & $t_0$ & $\hbar/E_h$ & 1 & 24.2 as \\
Atomic velocity & $v_0$ & $a_0 E_h/\hbar$ & 1 & $2.19 \times 10^6$ m/s \\
Speed of light & $c$ & --- & $1/\alpha \approx 137$ & $3 \times 10^8$ m/s \\
\hline
\end{tabular}
\end{table}

\subsection{Maxwell's Equations in Different Units}

For completeness, Tables~\ref{tab:maxwell_si_gauss} and \ref{tab:maxwell_hl_atomic} show Maxwell's equations in all four unit systems, illustrating why different communities prefer different conventions.

\begin{table}[ht]
\centering
\caption{Maxwell's equations in SI and Gaussian units.}
\label{tab:maxwell_si_gauss}
\renewcommand{\arraystretch}{1.8}
\begin{tabular}{lcc}
\hline
\textbf{Equation} & \textbf{SI} & \textbf{Gaussian} \\
\hline
Gauss (E) & $\displaystyle \nabla \cdot \vec{E} = \frac{\rho}{\epsilon_0}$ & $\displaystyle \nabla \cdot \vec{E} = 4\pi\rho$ \\[10pt]
Gauss (B) & $\displaystyle \nabla \cdot \vec{B} = 0$ & $\displaystyle \nabla \cdot \vec{B} = 0$ \\[10pt]
Faraday & $\displaystyle \nabla \times \vec{E} = -\frac{\partial \vec{B}}{\partial t}$ & $\displaystyle \nabla \times \vec{E} = -\frac{1}{c}\frac{\partial \vec{B}}{\partial t}$ \\[10pt]
Amp\`{e}re-Maxwell & $\displaystyle \nabla \times \vec{B} = \mu_0\vec{J} + \mu_0\epsilon_0\frac{\partial \vec{E}}{\partial t}$ & $\displaystyle \nabla \times \vec{B} = \frac{4\pi}{c}\vec{J} + \frac{1}{c}\frac{\partial \vec{E}}{\partial t}$ \\[8pt]
\hline
\end{tabular}
\end{table}

\begin{table}[ht]
\centering
\caption{Maxwell's equations in Heaviside-Lorentz and atomic units.}
\label{tab:maxwell_hl_atomic}
\renewcommand{\arraystretch}{1.8}
\begin{tabular}{lcc}
\hline
\textbf{Equation} & \textbf{Heaviside-Lorentz} & \textbf{Atomic} \\
\hline
Gauss (E) & $\displaystyle \nabla \cdot \vec{E} = \rho$ & $\displaystyle \nabla \cdot \vec{E} = 4\pi\rho$ \\[10pt]
Gauss (B) & $\displaystyle \nabla \cdot \vec{B} = 0$ & $\displaystyle \nabla \cdot \vec{B} = 0$ \\[10pt]
Faraday & $\displaystyle \nabla \times \vec{E} = -\frac{1}{c}\frac{\partial \vec{B}}{\partial t}$ & $\displaystyle \nabla \times \vec{E} = -\alpha\frac{\partial \vec{B}}{\partial t}$ \\[10pt]
Amp\`{e}re-Maxwell & $\displaystyle \nabla \times \vec{B} = \frac{1}{c}\vec{J} + \frac{1}{c}\frac{\partial \vec{E}}{\partial t}$ & $\displaystyle \nabla \times \vec{B} = 4\pi\alpha\vec{J} + \alpha\frac{\partial \vec{E}}{\partial t}$ \\[8pt]
\hline
\end{tabular}
\end{table}

Heaviside-Lorentz has the cleanest form: no $4\pi$ factors in the differential equations. Atomic units are awkward for electrodynamics because $\alpha \approx 1/137$ appears where $1/c$ would be in Gaussian units---atomic units are optimized for non-relativistic quantum mechanics, where $c$ doesn't naturally appear.

\subsection{Converting Between Systems}

The key conversion from SI to Gaussian is:
\begin{equation}
    \frac{e^2}{4\pi\epsilon_0} \quad \text{(SI)} \longleftrightarrow \quad e^2 \quad \text{(Gaussian)}
\end{equation}
To convert any formula, replace every $e^2/(4\pi\epsilon_0)$ with $e^2$.

To convert from Gaussian to atomic units, set $\hbar = m_e = e = 1$. Lengths become pure numbers (measured in units of $a_0$); energies become pure numbers (measured in units of $E_h$).

The fine structure constant $\alpha$ is dimensionless and therefore identical in all unit systems:
\begin{equation}
    \alpha = \frac{e^2}{4\pi\epsilon_0\hbar c} = \frac{e^2}{\hbar c} = \frac{1}{c} \approx \frac{1}{137}
\end{equation}

\subsection{Which System to Use?}

\textbf{SI units}: Engineering, experimental physics, undergraduate teaching, numerical comparisons with experiment.

\textbf{Gaussian units}: Classic texts (Jackson, Landau), plasma physics, astrophysics, condensed matter theory, avoiding $4\pi\epsilon_0$ clutter.

\textbf{Heaviside-Lorentz units}: Quantum field theory, particle physics, when you want clean Maxwell equations.

\textbf{Atomic units}: Atomic/molecular/optical physics, quantum chemistry, whenever you want the simplest Schr\"{o}dinger equation.

\begin{keyidea}{The Fine Structure Constant}
The fine structure constant $\alpha \approx 1/137$ is the one truly dimensionless number characterizing electromagnetism. It measures the strength of the electromagnetic interaction: $\alpha = e^2/(\hbar c)$ in Gaussian units. Its smallness ($\alpha \ll 1$) is why perturbation theory works for atomic physics and why atoms exist at all. In atomic units, $\alpha = 1/c$, making the non-relativistic nature of atomic physics manifest.
\end{keyidea}

For this chapter and most atomic physics applications, we use \textbf{atomic units}: $\hbar = m_e = e = 1$. Lengths are measured in Bohr radii ($a_0 = 1$), energies in Hartrees ($E_h = 1$), and the Coulomb potential is simply $V(r) = -1/r$.

\section{The Schr\"{o}dinger Equation in Spherical Coordinates}

For a particle of mass $\mu$ (in units of $m_e$) in a central potential $V(r)$, the time-independent Schr\"{o}dinger equation is:
\begin{equation}
    -\frac{1}{2\mu}\nabla^2\psi + V(r)\psi = E\psi
\end{equation}

For hydrogen with an infinitely heavy nucleus, $\mu = 1$. In spherical coordinates $(r, \theta, \phi)$, the Laplacian takes the form:
\begin{equation}
    \nabla^2 = \frac{1}{r^2}\frac{\partial}{\partial r}\left(r^2\frac{\partial}{\partial r}\right) + \frac{1}{r^2\sin\theta}\frac{\partial}{\partial\theta}\left(\sin\theta\frac{\partial}{\partial\theta}\right) + \frac{1}{r^2\sin^2\theta}\frac{\partial^2}{\partial\phi^2}
\end{equation}

\subsection{Separation of Variables}

We seek solutions of the form:
\begin{equation}
    \psi(r,\theta,\phi) = R(r)Y_{\ell}^m(\theta,\phi)
\end{equation}
where $Y_{\ell}^m(\theta,\phi)$ are the spherical harmonics we studied in Chapter~\ref{ch:3d_angular}.

\section{The Radial Equation}

After separation, the radial equation becomes:
\begin{equation}
    \frac{1}{r^2}\frac{d}{dr}\left(r^2\frac{dR}{dr}\right) + \left[2\mu(E - V(r)) - \frac{\ell(\ell+1)}{r^2}\right]R = 0
\end{equation}

For the Coulomb potential $V(r) = -1/r$, this becomes the hydrogen radial equation.

\subsection{The Effective Potential}

\begin{physicalinsight}
The term $\frac{\ell(\ell+1)}{2\mu r^2}$ acts as a centrifugal barrier, preventing the electron from approaching the nucleus too closely for $\ell > 0$. This is a purely quantum mechanical effect arising from the angular momentum of the electron.
\end{physicalinsight}

\section{Energy Levels and Wave Functions}

\subsection{The Hydrogen Hamiltonian}

In atomic units, the hydrogen Hamiltonian takes its simplest possible form:
\begin{equation}
    \hat{H} = -\frac{1}{2}\nabla^2 - \frac{1}{r}
\end{equation}

This elegant expression contains no fundamental constants at all---just the kinetic energy operator and the Coulomb attraction.

\subsection{Energy Eigenvalues}

The energy eigenvalues are:
\begin{equation}
    E_n = -\frac{1}{2n^2}
\end{equation}
where $n = 1, 2, 3, \ldots$ is the principal quantum number. These energies are measured in Hartrees.

In terms of Rydbergs (1 Ry $= \frac{1}{2} E_h = 13.6$ eV):
\begin{equation}
    E_n = -\frac{\text{Ry}}{n^2}
\end{equation}

The ground state energy is exactly $E_1 = -\frac{1}{2} E_h = -1$ Ry $= -13.6$ eV.

\subsection{Radial Wave Functions}

The normalized radial wave functions are:
\begin{equation}
    R_{n\ell}(r) = \sqrt{\left(\frac{2}{n}\right)^3 \frac{(n-\ell-1)!}{2n(n+\ell)!}} e^{-r/n} \left(\frac{2r}{n}\right)^\ell L_{n-\ell-1}^{2\ell+1}\left(\frac{2r}{n}\right)
\end{equation}
where $r$ is measured in Bohr radii and $L_k^p(x)$ are the associated Laguerre polynomials.

\textbf{Explicit forms:}

Ground state ($1s$):
\begin{equation}
    R_{10}(r) = 2 e^{-r}
\end{equation}

$2s$ state:
\begin{equation}
    R_{20}(r) = \frac{1}{2\sqrt{2}} \left(1 - \frac{r}{2}\right) e^{-r/2}
\end{equation}

$2p$ state:
\begin{equation}
    R_{21}(r) = \frac{1}{2\sqrt{6}} r e^{-r/2}
\end{equation}

$3s$ state:
\begin{equation}
    R_{30}(r) = \frac{2}{81\sqrt{3}} \left(27 - 18r + 2r^2\right) e^{-r/3}
\end{equation}

\section{Quantum Numbers and Degeneracy}

\begin{keyidea}{Quantum Numbers of Hydrogen}
The hydrogen atom wave functions are characterized by three quantum numbers: the principal quantum number $n = 1, 2, 3, \ldots$ (which determines energy), the orbital angular momentum $\ell = 0, 1, 2, \ldots, n-1$, and the magnetic quantum number $m = -\ell, -\ell+1, \ldots, \ell-1, \ell$. The degeneracy of energy level $n$ is $n^2$.
\end{keyidea}

The total degeneracy is:
\begin{equation}
    g_n = \sum_{\ell=0}^{n-1}(2\ell + 1) = n^2
\end{equation}

Including electron spin: $g_n = 2n^2$.

\section{Probability Densities and Orbitals}

\subsection{Radial Probability Distribution}

The probability of finding the electron between $r$ and $r + dr$ is:
\begin{equation}
    P(r)dr = |R_{n\ell}(r)|^2 r^2 dr
\end{equation}

For the ground state:
\begin{equation}
    P_{1s}(r) = 4r^2 e^{-2r}
\end{equation}

The most probable radius (where $dP/dr = 0$) is $r^* = 1$ (one Bohr radius).

The expectation value is $\langle r \rangle = 3/2$ (in units of $a_0$).

\subsection{Angular Distributions and Orbital Shapes}

The angular part determines the shape of orbitals. The $s$ orbitals ($\ell = 0$) are spherically symmetric, $p$ orbitals ($\ell = 1$) are dumbbell shaped, and $d$ orbitals ($\ell = 2$) have more complex four-lobed shapes.

\section{The Effective Potential}

The effective potential in atomic units is:
\begin{equation}
    V_{\text{eff}}(r) = -\frac{1}{r} + \frac{\ell(\ell+1)}{2r^2}
\end{equation}

For $\ell > 0$, this has a minimum at:
\begin{equation}
    r_{\min} = \ell(\ell+1)
\end{equation}
with value:
\begin{equation}
    V_{\text{eff}}(r_{\min}) = -\frac{1}{2\ell(\ell+1)}
\end{equation}

\section{Hydrogen-like Atoms}

For ions with nuclear charge $Z$ (He$^+$, Li$^{2+}$, etc.), the potential becomes $V(r) = -Z/r$. The energy levels and wave functions scale simply:
\begin{align}
    E_n^{(Z)} &= -\frac{Z^2}{2n^2}\\
    R_{n\ell}^{(Z)}(r) &= Z^{3/2} R_{n\ell}(Zr)
\end{align}

The effective Bohr radius shrinks: $a_0^{(Z)} = 1/Z$.

\textbf{Examples:}
\begin{center}
\begin{tabular}{lccc}
\hline
Ion & $Z$ & $E_1$ (in $E_h$) & $E_1$ (in eV) \\
\hline
H & 1 & $-1/2$ & $-13.6$ \\
He$^+$ & 2 & $-2$ & $-54.4$ \\
Li$^{2+}$ & 3 & $-9/2$ & $-122.4$ \\
\hline
\end{tabular}
\end{center}

\section{Beyond the Bohr Model: Outlook}

The energy levels we have derived represent the leading-order solution to the hydrogen atom. Several small corrections modify these energies.

In atomic units, the speed of light is $c = 1/\alpha \approx 137$. This large value reflects the fact that typical atomic velocities ($v \sim 1$ in atomic units) are much smaller than $c$, justifying our non-relativistic treatment. Relativistic corrections scale as $(v/c)^2 \sim \alpha^2 \sim 10^{-4}$.

\textbf{Fine structure} ($\sim \alpha^2 E_n$): relativistic kinetic energy, spin-orbit coupling, Darwin term.

\textbf{Hyperfine structure} ($\sim (m_e/m_p)\alpha^2 E_n$): electron-nuclear spin interaction.

\textbf{Lamb shift} ($\sim \alpha^3 \ln\alpha \cdot E_n$): quantum electrodynamic vacuum fluctuations.

These corrections will be treated systematically using perturbation theory in Chapter~\ref{ch:perturbation_theory}.

\section{Chapter Summary}

In atomic units, the hydrogen atom achieves maximal simplicity:

\begin{center}
\renewcommand{\arraystretch}{1.5}
\begin{tabular}{lc}
\hline
\textbf{Quantity} & \textbf{Atomic Units} \\
\hline
Hamiltonian & $\displaystyle \hat{H} = -\frac{1}{2}\nabla^2 - \frac{1}{r}$ \\[6pt]
Energy levels & $\displaystyle E_n = -\frac{1}{2n^2}$ \\[6pt]
Ground state energy & $E_1 = -\frac{1}{2} E_h = -1$ Ry \\
Ground state wave function & $\psi_{100} = \pi^{-1/2} e^{-r}$ \\
Most probable radius & $r^* = 1 \, a_0$ \\
Mean radius (ground state) & $\langle r \rangle = \frac{3}{2} \, a_0$ \\
Degeneracy & $g_n = n^2$ \\
Hydrogen-like ($Z$) & $E_n = -Z^2/(2n^2)$ \\
\hline
\end{tabular}
\end{center}

The absence of fundamental constants focuses attention on the physics: the competition between kinetic energy (which favors spreading out) and potential energy (which favors localization), resolved by the uncertainty principle into discrete bound states.

In the next chapter, we develop perturbation theory to handle atoms and systems that cannot be solved exactly like hydrogen.

\input{book_problems/ch08_problems.tex}

\section*{References and Further Reading}
\addcontentsline{toc}{section}{References and Further Reading}

\begin{description}
\item[Bethe, H.~A., and Salpeter, E.~E.] \emph{Quantum Mechanics of One- and Two-Electron Atoms}. Springer, 1957 (reprinted Dover, 2008). The definitive monograph; covers everything in this chapter and far beyond, including QED corrections to the hydrogen spectrum.

\item[Sakurai, J.~J., and Napolitano, J.] \emph{Modern Quantum Mechanics}, 3rd ed. Cambridge University Press, 2017. Section 3.8 develops the hydrogen atom at graduate level; complements the atomic-units treatment given here.

\item[Griffiths, D.~J., and Schroeter, D.~F.] \emph{Introduction to Quantum Mechanics}, 3rd ed. Cambridge University Press, 2018. Chapter 4 is the standard undergraduate-level derivation; useful as a slower walk through the radial equation if needed.

\item[Bohr, N.] ``On the constitution of atoms and molecules.'' \emph{Philosophical Magazine} \textbf{26}, 1--25 (1913). \href{https://doi.org/10.1080/14786441308634955}{doi:10.1080/14786441308634955}. The original quantization argument that produced $E_n = -1/(2n^2)$ E$_h$.

\item[Pauli, W.] ``\"Uber das Wasserstoffspektrum vom Standpunkt der neuen Quantenmechanik.'' \emph{Zeitschrift f\"ur Physik} \textbf{36}, 336--363 (1926). \href{https://doi.org/10.1007/BF01450175}{doi:10.1007/BF01450175}. Pauli's algebraic solution using the Runge--Lenz vector, derived before Schr\"odinger's wave-mechanical treatment; the cleanest way to see the $n^2$ degeneracy as an $SO(4)$ symmetry.

\item[Cohen-Tannoudji, C., Diu, B., and Lalo\"e, F.] \emph{Quantum Mechanics}, Vol.~1. Wiley-VCH, 1977. Chapter VII treats the hydrogen atom in conventional units; useful for cross-checking the atomic-units formulas of this chapter against the SI versions.
\end{description}

%% file: book_problems/ch08_problems.tex
\section{Problems}
\setcounter{hwproblem}{0}

\noindent\textbf{Note:} Throughout these problems, we use atomic units ($\hbar = m_e = e = 1$) unless otherwise specified.

\problem{The Schr\"{o}dinger Equation in Spherical Coordinates}
The time-independent Schr\"{o}dinger equation for a particle of mass $\mu$ in a central potential $V(r)$ is (in atomic units):
\begin{equation*}
    -\frac{1}{2\mu}\nabla^2\psi + V(r)\psi = E\psi
\end{equation*}
\begin{enumerate}[label=(\alph*)]
    \item Starting from the Laplacian in spherical coordinates, show that the angular part can be written as $-\hat{L}^2/r^2$.
    \item By substituting $\psi(r,\theta,\phi) = R(r)Y_\ell^m(\theta,\phi)$ into the Schr\"{o}dinger equation, derive the radial equation:
    \begin{equation*}
        \frac{1}{r^2}\frac{d}{dr}\left(r^2\frac{dR}{dr}\right) + \left[2\mu(E - V(r)) - \frac{\ell(\ell+1)}{r^2}\right]R = 0
    \end{equation*}
    \item Define $u(r) = rR(r)$ and show that the radial equation becomes:
    \begin{equation*}
        -\frac{1}{2\mu}\frac{d^2 u}{dr^2} + \left[V(r) + \frac{\ell(\ell+1)}{2\mu r^2}\right]u = Eu
    \end{equation*}
\end{enumerate}

\problem{Atomic Units and the Hydrogen Atom}
In atomic units, the hydrogen atom Hamiltonian is $\hat{H} = -\frac{1}{2}\nabla^2 - \frac{1}{r}$, and the energy eigenvalues are $E_n = -1/(2n^2)$.
\begin{enumerate}[label=(\alph*)]
    \item The Bohr radius is $a_0 = \hbar^2/(m_e e^2)$ in Gaussian units. Verify this has dimensions of length and calculate its numerical value.
    \item The Hartree is $E_h = e^2/a_0$. Express this in SI units and calculate its numerical value in eV.
    \item Express the hydrogen ground state energy $E_1 = -1/2$ in: (i) Hartrees, (ii) Rydbergs, (iii) electron volts.
    \item For hydrogen-like ions with nuclear charge $Z$, show that $E_n^{(Z)} = -Z^2/(2n^2)$. Calculate the ionization energy of He$^+$ in eV.
\end{enumerate}

\problem{Quantum Numbers and Degeneracy}
The hydrogen atom wave functions are characterized by three quantum numbers: $n$, $\ell$, and $m$.
\begin{enumerate}[label=(\alph*)]
    \item For a given principal quantum number $n$, list all allowed values of $\ell$ and explain why these are the only allowed values.
    \item For a given $\ell$, list all allowed values of $m$ and explain the physical origin of this constraint.
    \item Calculate the total degeneracy $g_n = n^2$ by counting all allowed combinations of $\ell$ and $m$ for a given $n$.
    \item Including electron spin ($s = 1/2$), what is the total degeneracy of energy level $E_n$? For $n = 3$, enumerate all states using the notation $\ket{n, \ell, m_\ell, m_s}$.
\end{enumerate}

\problem{Radial Wave Functions and Probability Distributions}
In atomic units, the normalized radial wave functions for hydrogen are:
\begin{equation*}
    R_{n\ell}(r) = \sqrt{\left(\frac{2}{n}\right)^3 \frac{(n-\ell-1)!}{2n(n+\ell)!}} e^{-r/n} \left(\frac{2r}{n}\right)^\ell L_{n-\ell-1}^{2\ell+1}\left(\frac{2r}{n}\right)
\end{equation*}
\begin{enumerate}[label=(\alph*)]
    \item Write out the explicit form of $R_{10}(r)$ for the ground state ($n=1$, $\ell=0$).
    \item Write out the explicit form of $R_{21}(r)$ for the $2p$ state.
    \item The radial probability density is $P(r) = |R_{n\ell}(r)|^2 r^2$. For the ground state, find the most probable radius $r^*$ by solving $dP/dr = 0$.
    \item Calculate the expectation value $\langle r \rangle$ for the ground state and compare it to $r^*$. Why are these different?
\end{enumerate}

\problem{The Effective Potential and Classical Turning Points}
In atomic units, the effective potential for the hydrogen radial equation is:
\begin{equation*}
    V_{\text{eff}}(r) = -\frac{1}{r} + \frac{\ell(\ell+1)}{2r^2}
\end{equation*}
\begin{enumerate}[label=(\alph*)]
    \item Sketch $V_{\text{eff}}(r)$ for $\ell = 0, 1, 2$ on the same axes. Describe how the centrifugal barrier affects the behavior near $r = 0$.
    \item For $\ell > 0$, find the radius $r_{\min}$ where $V_{\text{eff}}$ has its minimum.
    \item Calculate $V_{\text{eff}}(r_{\min})$ and show that it equals $-1/[2\ell(\ell+1)]$.
    \item For the $2p$ state ($n=2$, $\ell=1$), find the inner and outer classical turning points where $E_2 = V_{\text{eff}}(r)$.
\end{enumerate}

\problem{Angular Momentum and Orbital Shapes}
The angular part of the hydrogen wave function is given by the spherical harmonics $Y_\ell^m(\theta,\phi)$.
\begin{enumerate}[label=(\alph*)]
    \item The $s$ orbitals ($\ell = 0$) have $Y_0^0 = 1/\sqrt{4\pi}$. Explain why $s$ orbitals are spherically symmetric.
    \item The $p$ orbitals ($\ell = 1$) have complex spherical harmonics $Y_1^0 \propto \cos\theta$ and $Y_1^{\pm 1} \propto \sin\theta e^{\pm i\phi}$. The real combinations are $p_z \propto \cos\theta$, $p_x \propto \sin\theta\cos\phi$, $p_y \propto \sin\theta\sin\phi$. Show that $p_x$ and $p_y$ are linear combinations of $Y_1^{+1}$ and $Y_1^{-1}$.
    \item Calculate $\langle \hat{L}_z \rangle$ for the state $p_x$. Is $p_x$ an eigenstate of $\hat{L}_z$?
    \item For a $p_z$ orbital, the probability density is $|\psi|^2 \propto |R_{n1}(r)|^2 \cos^2\theta$. In which directions is the probability maximum? Where is it zero? Describe the orbital shape.
\end{enumerate}

\problem{Hydrogen Expectation Values and Virial Theorem}
\begin{enumerate}[label=(\alph*)]
    \item For the hydrogen ground state $\ket{1s}$, calculate $\langle r \rangle$, $\langle r^2 \rangle$, and $\langle 1/r \rangle$ in atomic units. Use the radial wave function $R_{10}(r) = 2e^{-r}$.
    \item The virial theorem for a power-law potential $V \propto r^k$ states: $2\langle T \rangle = k\langle V \rangle$. For the Coulomb potential ($k = -1$), show that $\langle T \rangle = -E$ and $\langle V \rangle = 2E$.
    \item Verify the virial theorem numerically for the $n=2$ state using $E_2 = -1/8$.
    \item What is the ratio $\langle 1/r \rangle / \langle r \rangle^{-1}$ for the ground state? Interpret this result.
\end{enumerate}

\problem{Relativistic Corrections and the Darwin Term}
The relativistic correction to the kinetic energy is $T_{\text{rel}} = -p^4/(8m^3c^2)$ in leading order.
\begin{enumerate}[label=(\alph*)]
    \item In atomic units with $c \approx 137$, the relativistic correction to energy is $E_{\text{rel}} = -\langle p^4 \rangle / (8c^2)$. For the ground state, use $\langle p^2 \rangle = 1$ to estimate $\langle p^4 \rangle$ by assuming $\langle p^4 \rangle \sim \langle p^2 \rangle^2$.
    \item The exact result is $E_{\text{rel}} = -\alpha^2 E_n / (2n^3)[n/\ell + 1/2)^{-1} - 3/4]$ where $\alpha = 1/c$. For $n=1$, $\ell=0$, calculate the fractional shift in energy.
    \item The Darwin term (QED effect) contributes $\Delta E_{\text{Darwin}} = \alpha^2 \pi \delta^3(\vec{r}) / 2$ for $s$ states. Using $\psi_{1s}(0) = 1/\sqrt{\pi}$, calculate this contribution.
    \item The total fine structure splitting between $2S_{1/2}$ and $2P_{1/2}$ (historically the Lamb shift) is about 1058 MHz. Convert to eV and compare to the splitting predicted by Dirac theory alone.
\end{enumerate}

\problem{Helium Ground State and Pauli Exclusion}
Two electrons in helium interact via Coulomb repulsion. In the simplest approximation, each electron is in a hydrogen-like orbital with effective nuclear charge $Z_{\text{eff}}$.
\begin{enumerate}[label=(\alph*)]
    \item The ground state has both electrons in the $1s$ orbital with opposite spins: $\psi(1,2) = \ket{1s}_1\ket{1s}_2 \times \text{singlet spin}$. The total energy is $E_{\text{tot}} = 2E_1(Z_{\text{eff}}) + \langle V_{ee} \rangle$ where $V_{ee} = 1/r_{12}$ is the electron-electron repulsion.
    \item Using the variational method with trial wave function $\psi_{1s}(r) = (Z^3/\pi)^{1/2}e^{-Zr}$ for both electrons, show that the total energy is $E(Z) = -2Z^2 + (5/8)Z$ (in atomic units).
    \item Minimize $E(Z)$ to find the optimal $Z_{\text{eff}}$. What is the resulting ground state energy and how does it compare to the experimental value of $-5.81$ eV?
    \item Explain why $Z_{\text{eff}} < 2$. What is the physical origin of the screening effect?
\end{enumerate}

\problem{Spectroscopic Transitions and Selection Rules}
Selection rules for electric dipole transitions are $\Delta\ell = \pm 1$, $\Delta m = 0, \pm 1$ (from $z$ and $x \pm iy$ components of the dipole operator).
\begin{enumerate}[label=(\alph*)]
    \item Using the selection rules, list all allowed transitions from the $n = 3$ level of hydrogen to the $n = 2$ level. Identify the emitted photon wavelengths using $E = -1/(2n^2)$.
    \item Why is the $2s \to 1s$ transition forbidden? Why is the $2p \to 1s$ transition allowed?
    \item The transition $1s \to 2p$ requires absorption of a photon with energy $\Delta E = E_2 - E_1 = 3/8$ (in atomic units). Calculate the wavelength in nm and identify which spectral line this is (Lyman-alpha).
    \item For hydrogen-like atoms with nuclear charge $Z$, the transition frequency scales as $\omega \propto Z^2$. Estimate the wavelength of the Lyman-alpha line for He$^+$.
\end{enumerate}

\problem{Fine Structure and Lamb Shift}
The fine structure arises from spin-orbit coupling and relativistic corrections. The Lamb shift is a QED effect that splits the $2S_{1/2}$ and $2P_{1/2}$ levels.
\begin{enumerate}[label=(\alph*)]
    \item The spin-orbit energy is $E_{\text{SO}} = \alpha^2 / (2n^3) \times f(j, \ell, s)$ where $f$ depends on quantum numbers. For the $2P$ level ($\ell=1$, $s=1/2$), the two levels are $2P_{1/2}$ and $2P_{3/2}$. Calculate the spin-orbit splitting.
    \item The relativistic mass correction is $\Delta E_{\text{rel}} = -\alpha^2 / (2n^3) \times [3/4 - n/(j+1/2)]$. Calculate this for $2S_{1/2}$ and $2P_{1/2}$.
    \item The Lamb shift (QED) adds about $+1.6$ to the $2S_{1/2}$ energy (in units of $10^{-3}$ eV relative to $n=2$) but not to $2P_{1/2}$. Estimate the total $2S_{1/2} - 2P_{1/2}$ splitting and compare to the measured $1058$ MHz.
\end{enumerate}

\problem{Finite Nuclear Size Effects}
Real nuclei are not point charges. The nuclear charge distribution affects the $1s$ orbital significantly.
\begin{enumerate}[label=(\alph*)]
    \item For a uniformly charged sphere of radius $R$ (nuclear radius), the potential is $V(r) = -Z/r$ for $r > R$ and $V(r) = -Z[3/(2R) - r^2/(2R^3)]$ for $r < R$.
    \item For hydrogen, $R \sim 10^{-5}$ in atomic units. The shift in $1s$ energy due to finite size is $\Delta E_{1s} \sim (Z\alpha)^2 (R/a_0)^2 |E_1|$. Estimate this numerically.
    \item The shift in the muonic hydrogen $1s$ level (muon mass 207$m_e$) is about 200 times larger. Explain why.
    \item How could atomic spectroscopy of muonic atoms be used to measure the nuclear radius?
\end{enumerate}

%% file: chapters/ch09_perturbation_theory.tex
\chapter{Perturbation Theory}
\label{ch:perturbation_theory}

\section{Introduction}

While the hydrogen atom represents a pinnacle of exact solvability in quantum mechanics, most physical systems cannot be solved analytically. Perturbation theory provides a systematic approach to finding approximate solutions when the Hamiltonian can be written as a sum of a solvable part and a small perturbation.

\begin{keyidea}{The Perturbative Approach}
Perturbation theory treats a complicated problem as a simple, solvable problem plus a small correction. We expand solutions in powers of a small parameter, obtaining increasingly accurate approximations.
\end{keyidea}

\section{Time-Independent Perturbation Theory}

\subsection{Non-Degenerate Perturbation Theory}

Consider a Hamiltonian of the form:
\begin{equation}
    H = H_0 + \lambda V
\end{equation}
where $H_0$ is the unperturbed Hamiltonian with known eigenstates $\ket{n^{(0)}}$ and eigenvalues $E_n^{(0)}$, and $\lambda V$ is a small perturbation.

We expand the perturbed eigenstates and energies as power series in $\lambda$:
\begin{align}
    \ket{n} &= \ket{n^{(0)}} + \lambda\ket{n^{(1)}} + \lambda^2\ket{n^{(2)}} + \cdots \label{eq:state_expansion}\\
    E_n &= E_n^{(0)} + \lambda E_n^{(1)} + \lambda^2 E_n^{(2)} + \cdots \label{eq:energy_expansion}
\end{align}

\subsection{Derivation of First-Order Corrections}

Substituting the expansions~\eqref{eq:state_expansion} and~\eqref{eq:energy_expansion} into the eigenvalue equation $(H_0 + \lambda V)\ket{n} = E_n\ket{n}$:
\begin{equation}
    (H_0 + \lambda V)(\ket{n^{(0)}} + \lambda\ket{n^{(1)}} + \cdots) = (E_n^{(0)} + \lambda E_n^{(1)} + \cdots)(\ket{n^{(0)}} + \lambda\ket{n^{(1)}} + \cdots)
\end{equation}

Expanding and collecting terms by powers of $\lambda$:
\begin{align}
    \lambda^0: \quad & H_0\ket{n^{(0)}} = E_n^{(0)}\ket{n^{(0)}} \label{eq:zeroth_order}\\
    \lambda^1: \quad & H_0\ket{n^{(1)}} + V\ket{n^{(0)}} = E_n^{(0)}\ket{n^{(1)}} + E_n^{(1)}\ket{n^{(0)}} \label{eq:first_order}
\end{align}

Equation~\eqref{eq:zeroth_order} is just the unperturbed eigenvalue equation, which is satisfied by assumption.

\textbf{First-order energy correction:} To find $E_n^{(1)}$, take the inner product of equation~\eqref{eq:first_order} with $\bra{n^{(0)}}$:
\begin{equation}
    \bra{n^{(0)}}H_0\ket{n^{(1)}} + \bra{n^{(0)}}V\ket{n^{(0)}} = E_n^{(0)}\braket{n^{(0)}}{n^{(1)}} + E_n^{(1)}\braket{n^{(0)}}{n^{(0)}}
\end{equation}

Since $H_0$ is Hermitian, $\bra{n^{(0)}}H_0 = E_n^{(0)}\bra{n^{(0)}}$, so the first term on the left equals $E_n^{(0)}\braket{n^{(0)}}{n^{(1)}}$. This cancels the first term on the right, leaving:
\begin{equation}
    \boxed{E_n^{(1)} = \bra{n^{(0)}}V\ket{n^{(0)}}}
\end{equation}

\textbf{First-order state correction:} To find $\ket{n^{(1)}}$, expand it in the complete basis of unperturbed states:
\begin{equation}
    \ket{n^{(1)}} = \sum_k c_k^{(1)}\ket{k^{(0)}}
\end{equation}

Take the inner product of equation~\eqref{eq:first_order} with $\bra{m^{(0)}}$ for $m \neq n$:
\begin{equation}
    \bra{m^{(0)}}H_0\ket{n^{(1)}} + \bra{m^{(0)}}V\ket{n^{(0)}} = E_n^{(0)}\braket{m^{(0)}}{n^{(1)}} + E_n^{(1)}\underbrace{\braket{m^{(0)}}{n^{(0)}}}_{=0}
\end{equation}

Using $\bra{m^{(0)}}H_0 = E_m^{(0)}\bra{m^{(0)}}$ and $\braket{m^{(0)}}{n^{(1)}} = c_m^{(1)}$:
\begin{equation}
    E_m^{(0)} c_m^{(1)} + \bra{m^{(0)}}V\ket{n^{(0)}} = E_n^{(0)} c_m^{(1)}
\end{equation}

Solving for $c_m^{(1)}$:
\begin{equation}
    c_m^{(1)} = \frac{\bra{m^{(0)}}V\ket{n^{(0)}}}{E_n^{(0)} - E_m^{(0)}} \quad (m \neq n)
\end{equation}

Therefore:
\begin{equation}
    \boxed{\ket{n^{(1)}} = \sum_{m \neq n} \frac{\bra{m^{(0)}}V\ket{n^{(0)}}}{E_n^{(0)} - E_m^{(0)}}\ket{m^{(0)}}}
\end{equation}

\begin{physicalinsight}
The first-order energy shift is just the expectation value of the perturbation in the unperturbed state. States mix in proportion to their coupling strength divided by their energy separation.
\end{physicalinsight}

\begin{example}{Spin-1/2 in a Tilted Magnetic Field}
Consider a spin-1/2 particle in a strong magnetic field along $z$ with a weak transverse component along $x$. The Hamiltonian is:
\begin{equation}
    H = -\gamma B_z S_z - \gamma B_x S_x = \underbrace{-\gamma B_z S_z}_{H_0} + \underbrace{(-\gamma B_x S_x)}_{V}
\end{equation}
where $B_x \ll B_z$.

The unperturbed eigenstates are $\ket{\uparrow}$ and $\ket{\downarrow}$ with energies $E_\uparrow^{(0)} = -\frac{1}{2}\gamma B_z$ and $E_\downarrow^{(0)} = +\frac{1}{2}\gamma B_z$.

For the first-order energy correction to the spin-up state:
\begin{equation}
    E_\uparrow^{(1)} = \bra{\uparrow}(-\gamma B_x S_x)\ket{\uparrow} = -\gamma B_x \bra{\uparrow}S_x\ket{\uparrow} = 0
\end{equation}
since $S_x$ has no diagonal matrix elements in the $S_z$ basis. Similarly, $E_\downarrow^{(1)} = 0$.

The first-order state correction is:
\begin{equation}
    \ket{\uparrow^{(1)}} = \frac{\bra{\downarrow}(-\gamma B_x S_x)\ket{\uparrow}}{E_\uparrow^{(0)} - E_\downarrow^{(0)}}\ket{\downarrow} = \frac{-\gamma B_x (1/2)}{-\gamma B_z}\ket{\downarrow} = \frac{B_x}{2B_z}\ket{\downarrow}
\end{equation}

The perturbed spin-up state tilts toward the $x$-direction:
\begin{equation}
    \ket{\uparrow'} \approx \ket{\uparrow} + \frac{B_x}{2B_z}\ket{\downarrow}
\end{equation}
This makes physical sense: the spin aligns with the total field direction.
\end{example}

\begin{example}{Zeeman Effect on Hydrogen Ground State}
Consider a hydrogen atom in its ground state ($1s$, with $\ell = 0$) placed in a uniform magnetic field $\vec{B} = B\hat{z}$. The perturbation is:
\begin{equation}
    V = -\vec{\mu} \cdot \vec{B} = \frac{e}{2m_e}(\vec{L} + 2\vec{S}) \cdot \vec{B} = \frac{eB}{2m_e}(L_z + 2S_z)
\end{equation}

For the $1s$ state, $\ell = 0$ so $L_z\ket{1s} = 0$. The unperturbed ground state is two-fold degenerate due to spin: $\ket{1s,\uparrow}$ and $\ket{1s,\downarrow}$.

However, these states are eigenstates of $S_z$, so the perturbation is already diagonal:
\begin{align}
    E_{1s,\uparrow}^{(1)} &= \frac{eB}{2m_e}\bra{1s,\uparrow}2S_z\ket{1s,\uparrow} = \frac{eB}{2m_e} = \mu_B B \\
    E_{1s,\downarrow}^{(1)} &= \frac{eB}{2m_e}\bra{1s,\downarrow}2S_z\ket{1s,\downarrow} = -\mu_B B
\end{align}
where $\mu_B = \frac{e}{2m_e} = \frac{1}{2}$ in atomic units is the Bohr magneton.

The magnetic field splits the ground state by $\Delta E = 2\mu_B B$. For a 1 Tesla field, this is about $1.16 \times 10^{-4}$ eV.

\textit{Context:} Magnetic field control of atomic states is foundational to atomic clocks, MRI, and quantum sensing. Trapped ion qubits often use Zeeman sublevels as computational basis states, where precise control of this splitting enables gate operations.
\end{example}

\subsection{Second-Order Corrections}

The second-order energy correction involves virtual transitions:
\begin{equation}
    E_n^{(2)} = \sum_{m \neq n} \frac{|\bra{m^{(0)}}V\ket{n^{(0)}}|^2}{E_n^{(0)} - E_m^{(0)}}
\end{equation}

\begin{physicalinsight}
Second-order corrections arise from ``virtual transitions'' to intermediate states. The ground state always shifts down in energy (since all denominators are negative), reflecting the system's ability to lower its energy by mixing with excited states.
\end{physicalinsight}

\begin{example}{Anharmonic Oscillator: Quartic Perturbation}
Consider a harmonic oscillator with a quartic perturbation:
\begin{equation}
    H = \frac{p^2}{2m} + \frac{1}{2}m\omega^2 x^2 + \lambda x^4
\end{equation}

The unperturbed energies are $E_n^{(0)} = \omega(n + \frac{1}{2})$. For the ground state ($n = 0$), we need $\bra{0}x^4\ket{0}$.

Using the ladder operator representation $x = \sqrt{\frac{1}{2\omega}}(a + a^\dagger)$ (with $m=1$ in atomic units):
\begin{equation}
    x^4 = \left(\frac{1}{2\omega}\right)^2 (a + a^\dagger)^4
\end{equation}

The first-order correction requires terms with equal numbers of $a$ and $a^\dagger$:
\begin{equation}
    E_0^{(1)} = \lambda\bra{0}x^4\ket{0} = \lambda\left(\frac{1}{2\omega}\right)^2 \bra{0}(a + a^\dagger)^4\ket{0}
\end{equation}

Expanding and keeping terms that return to $\ket{0}$:
\begin{equation}
    \bra{0}(a + a^\dagger)^4\ket{0} = \bra{0}(a a^\dagger a a^\dagger + a a^\dagger a^\dagger a + a^\dagger a a a^\dagger + \text{2 more})\ket{0} = 3
\end{equation}

Therefore:
\begin{equation}
    E_0^{(1)} = \frac{3\lambda}{4\omega^2}
\end{equation}

For the second-order correction, $x^4$ connects $\ket{0}$ to $\ket{2}$ and $\ket{4}$:
\begin{align}
    \bra{2}x^4\ket{0} &= \left(\frac{1}{2\omega}\right)^2 \sqrt{2} \cdot 3 = \frac{3\sqrt{2}}{4\omega^2} \\
    \bra{4}x^4\ket{0} &= \left(\frac{1}{2\omega}\right)^2 \sqrt{24} = \frac{\sqrt{24}}{4\omega^2}
\end{align}

The second-order correction is:
\begin{equation}
    E_0^{(2)} = \frac{|\bra{2}x^4\ket{0}|^2}{E_0^{(0)} - E_2^{(0)}} + \frac{|\bra{4}x^4\ket{0}|^2}{E_0^{(0)} - E_4^{(0)}} = -\frac{21\lambda^2}{8\omega^5}
\end{equation}

The negative second-order shift reflects the general principle that the ground state energy is pushed down by coupling to excited states.
\end{example}

\begin{example}{Quadratic Stark Effect in Hydrogen}
Working in atomic units, for a hydrogen atom in a weak electric field $\mathcal{E}$ along the $z$-axis, the perturbation is:
\begin{equation}
    V = \mathcal{E}z = \mathcal{E}r\cos\theta
\end{equation}

For the ground state $\ket{1s}$, which has definite parity (even under $\vec{r} \to -\vec{r}$), the first-order shift vanishes:
\begin{equation}
    E_1^{(1)} = \mathcal{E}\bra{1s}z\ket{1s} = 0
\end{equation}
since $z$ is odd under parity. Physically, the spherically symmetric $1s$ state has no permanent electric dipole moment.

The second-order correction requires summing over all intermediate states:
\begin{equation}
    E_1^{(2)} = \sum_{n\ell m \neq 1s} \frac{|\mathcal{E}\bra{n\ell m}z\ket{1s}|^2}{E_1 - E_n}
\end{equation}

The selection rule $\Delta\ell = \pm 1$ from the angular integration means only $p$-states contribute. After detailed calculation:
\begin{equation}
    E_1^{(2)} = -\frac{9}{4}\mathcal{E}^2 = -\frac{1}{2}\alpha_H \mathcal{E}^2
\end{equation}
where $\alpha_H = 9/2$ is the ground state polarizability of hydrogen in atomic units.

The quadratic dependence on field strength is characteristic of induced (rather than permanent) dipole moments. The polarizability $\alpha_H = 4.5$ a.u.\ is quite small, but for Rydberg states with $n \gg 1$, the polarizability scales as $n^7$, making electric field control extremely effective.

\textit{Context:} Electric field control of atoms underlies optical tweezers, atom traps, and Rydberg atom quantum simulators. The enormous polarizability of Rydberg atoms makes Stark shifts a primary tool for control and readout in Rydberg-based quantum computing. In precision atomic clocks, understanding and compensating for Stark shifts from blackbody radiation is essential for reaching $10^{-18}$ fractional accuracy.
\end{example}

\section{Degenerate Perturbation Theory}

When unperturbed states are degenerate, standard perturbation theory fails because the denominators $E_n^{(0)} - E_m^{(0)}$ vanish for states within the degenerate subspace. We must use degenerate perturbation theory.

\subsection{The Degenerate Subspace}

For a $g$-fold degenerate level with energy $E^{(0)}$, we form the perturbation matrix:
\begin{equation}
    W_{ij} = \bra{\psi_i^{(0)}}V\ket{\psi_j^{(0)}}
\end{equation}
within the degenerate subspace. The first-order energy corrections are the eigenvalues of this $g \times g$ matrix.

\subsection{Good Quantum Numbers}

The eigenvectors of $W$ give the correct zeroth-order states:
\begin{equation}
    \ket{n,\alpha} = \sum_{i=1}^g c_i^{(\alpha)}\ket{\psi_i^{(0)}}
\end{equation}
where $\alpha$ labels the eigenstates within the formerly degenerate subspace. These eigenvectors correspond to ``good quantum numbers'' that remain well-defined in the presence of the perturbation.

\begin{keyidea}{Lifting Degeneracy}
Perturbations often break symmetries, lifting degeneracies. The perturbation selects a preferred basis within the degenerate subspace, namely the basis that diagonalizes the perturbation.
\end{keyidea}

\begin{example}{Two Coupled Spins with Symmetry-Breaking Perturbation}
Consider two spin-1/2 particles with exchange coupling and a perturbation that treats them asymmetrically:
\begin{equation}
    H = J\vec{S}_1 \cdot \vec{S}_2 + \epsilon(S_{1z} - S_{2z})
\end{equation}

The unperturbed Hamiltonian $H_0 = J\vec{S}_1 \cdot \vec{S}_2$ has eigenstates in the coupled basis: the singlet $\ket{0,0}$ with $E = -\frac{3}{4}J\hbar^2$ and the triplet $\ket{1,m}$ with $E = \frac{1}{4}J\hbar^2$.

The triplet is three-fold degenerate. The perturbation $V = \epsilon(S_{1z} - S_{2z})$ breaks the exchange symmetry. In the triplet subspace:
\begin{align}
    \ket{1,1} &= \ket{\uparrow\uparrow} \\
    \ket{1,0} &= \frac{1}{\sqrt{2}}(\ket{\uparrow\downarrow} + \ket{\downarrow\uparrow}) \\
    \ket{1,-1} &= \ket{\downarrow\downarrow}
\end{align}

Computing the perturbation matrix elements:
\begin{align}
    \bra{1,1}V\ket{1,1} &= \epsilon\bra{\uparrow\uparrow}(S_{1z} - S_{2z})\ket{\uparrow\uparrow} = \epsilon(\tfrac{\hbar}{2} - \tfrac{\hbar}{2}) = 0 \\
    \bra{1,0}V\ket{1,0} &= \frac{\epsilon}{2}[\bra{\uparrow\downarrow} + \bra{\downarrow\uparrow}](S_{1z} - S_{2z})[\ket{\uparrow\downarrow} + \ket{\downarrow\uparrow}] = 0 \\
    \bra{1,-1}V\ket{1,-1} &= 0
\end{align}

All diagonal elements vanish! The off-diagonal elements:
\begin{equation}
    \bra{1,1}V\ket{1,0} = \bra{1,0}V\ket{1,-1} = 0
\end{equation}
but
\begin{equation}
    \bra{\uparrow\downarrow}(S_{1z} - S_{2z})\ket{\uparrow\downarrow} = \hbar, \quad \bra{\downarrow\uparrow}(S_{1z} - S_{2z})\ket{\downarrow\uparrow} = -\hbar
\end{equation}

The perturbation matrix in the $\{\ket{1,1}, \ket{1,0}, \ket{1,-1}\}$ basis is zero, but this is because these are already the good states. The perturbation does not mix states with different $m$, and within each $m$ subspace the matrix is diagonal.

However, if we had worked in the uncoupled basis $\{\ket{\uparrow\downarrow}, \ket{\downarrow\uparrow}\}$ for $m=0$:
\begin{equation}
    W = \epsilon\hbar\begin{pmatrix} 1 & 0 \\ 0 & -1 \end{pmatrix}
\end{equation}

The eigenstates are $\ket{\uparrow\downarrow}$ and $\ket{\downarrow\uparrow}$ with energies $\pm\epsilon\hbar$. The perturbation breaks the triplet-singlet structure and makes the uncoupled basis the ``good'' basis.
\end{example}

\begin{example}{Linear Stark Effect in Hydrogen ($n=2$)}
Working in atomic units, the $n=2$ level of hydrogen is four-fold degenerate (ignoring spin): one $2s$ state and three $2p$ states. An electric field $\mathcal{E}$ along $z$ introduces the perturbation $V = \mathcal{E}z$.

The states are $\ket{2,0,0}$ (2s), $\ket{2,1,0}$ (2p$_z$), $\ket{2,1,1}$ (2p$_+$), and $\ket{2,1,-1}$ (2p$_-$).

Since $[L_z, z] = 0$, the perturbation conserves $m$. The $m = \pm 1$ states cannot mix with $m = 0$ states, so the $4 \times 4$ problem separates:
\begin{itemize}
    \item $m = +1$: $\bra{2,1,1}V\ket{2,1,1} = 0$ (no shift)
    \item $m = -1$: $\bra{2,1,-1}V\ket{2,1,-1} = 0$ (no shift)
    \item $m = 0$: $2 \times 2$ subspace of $\ket{2s}$ and $\ket{2p_z}$
\end{itemize}

For the $m = 0$ subspace, diagonal elements vanish by parity:
\begin{equation}
    \bra{2s}z\ket{2s} = \bra{2p_z}z\ket{2p_z} = 0
\end{equation}

The crucial off-diagonal element:
\begin{equation}
    \bra{2s}z\ket{2p_z} = \int_0^\infty r^2 dr \int d\Omega\, R_{20}(r)Y_0^0 \cdot r\cos\theta \cdot R_{21}(r)Y_1^0
\end{equation}

Using explicit wave functions (in atomic units where $a_0 = 1$):
\begin{align}
    R_{20} &= \frac{1}{4\sqrt{2}}\left(2 - r\right)e^{-r/2} \\
    R_{21} &= \frac{1}{4\sqrt{6}}r\,e^{-r/2}
\end{align}

After integration: $\bra{2s}z\ket{2p_z} = -3$.

The perturbation matrix in the $\{\ket{2s}, \ket{2p_z}\}$ basis:
\begin{equation}
    W = \mathcal{E}\begin{pmatrix} 0 & -3 \\ -3 & 0 \end{pmatrix}
\end{equation}

Eigenvalues: $E^{(1)} = \pm 3\mathcal{E}$. Eigenstates:
\begin{equation}
    \ket{\pm} = \frac{1}{\sqrt{2}}(\ket{2s} \mp \ket{2p_z})
\end{equation}

The linear Stark effect gives energy shifts proportional to $\mathcal{E}$, not $\mathcal{E}^2$! This is because the degenerate $2s$ and $2p$ states can hybridize to form states with permanent electric dipole moments.

\textit{Context:} The linear Stark effect demonstrates how degeneracy enables qualitatively different physics. Rydberg atoms, with their near-degenerate high-$n$ manifolds, exhibit enormous electric dipole moments ($\sim n^2$ a.u.), making them exquisitely sensitive to electric fields. This sensitivity is exploited in Rydberg-atom-based electric field sensors and in creating strong dipole-dipole interactions for quantum simulation.
\end{example}

\subsection{Choice of Basis: Coupled vs.\ Uncoupled vs.\ Full Diagonalization}

A recurring theme in perturbation theory is the choice of basis. When multiple quantum numbers could label states, choosing the right basis can dramatically simplify calculations.

\begin{keyidea}{Choosing the Right Basis}
The ``right'' basis for perturbation theory is one in which the perturbation is diagonal or block-diagonal. This basis corresponds to the quantum numbers that remain good (conserved) in the presence of the perturbation.
\end{keyidea}

Consider a system with two angular momenta $\vec{L}$ and $\vec{S}$. We can work in:
\begin{itemize}
    \item \textbf{Uncoupled basis:} $\ket{\ell, m_\ell, s, m_s}$, eigenstates of $L^2, L_z, S^2, S_z$
    \item \textbf{Coupled basis:} $\ket{\ell, s, j, m_j}$, eigenstates of $L^2, S^2, J^2, J_z$
\end{itemize}

The choice depends on which operators commute with the perturbation:
\begin{itemize}
    \item If $[V, J_z] = 0$ but $[V, L_z] \neq 0$: use coupled basis
    \item If $[V, L_z] = 0$ and $[V, S_z] = 0$: use uncoupled basis
    \item If neither is diagonal: must diagonalize explicitly
\end{itemize}

\begin{example}{Hydrogen Fine Structure: Comparing Basis Choices}
The fine structure Hamiltonian for hydrogen includes the spin-orbit term:
\begin{equation}
    H_{SO} = \frac{1}{2m_e^2c^2}\frac{1}{r}\frac{dV}{dr}\vec{L}\cdot\vec{S} = \xi(r)\vec{L}\cdot\vec{S}
\end{equation}

\textbf{Attempt 1: Uncoupled basis} $\ket{n,\ell,m_\ell,m_s}$

In this basis, $\vec{L}\cdot\vec{S} = L_xS_x + L_yS_y + L_zS_z$ has off-diagonal elements. For example, $L_+S_-$ connects $\ket{m_\ell, m_s}$ to $\ket{m_\ell+1, m_s-1}$. The perturbation matrix is not diagonal, and we would need to diagonalize a $(2\ell+1) \times 2$ dimensional matrix for each $\ell$.

\textbf{Attempt 2: Coupled basis} $\ket{n,\ell,j,m_j}$

Using $\vec{J} = \vec{L} + \vec{S}$:
\begin{equation}
    \vec{L}\cdot\vec{S} = \frac{1}{2}(J^2 - L^2 - S^2)
\end{equation}

In the coupled basis, this is diagonal:
\begin{equation}
    \bra{n,\ell,j,m_j}\vec{L}\cdot\vec{S}\ket{n,\ell,j,m_j} = \frac{\hbar^2}{2}[j(j+1) - \ell(\ell+1) - s(s+1)]
\end{equation}

No diagonalization needed! The first-order energy shift is:
\begin{equation}
    E_{SO}^{(1)} = \frac{\hbar^2}{2}[j(j+1) - \ell(\ell+1) - \tfrac{3}{4}]\langle\xi(r)\rangle_{n\ell}
\end{equation}

For $\ell > 0$, $j = \ell \pm \frac{1}{2}$:
\begin{align}
    j = \ell + \tfrac{1}{2}: &\quad E_{SO}^{(1)} = \frac{\ell\hbar^2}{2}\langle\xi\rangle \\
    j = \ell - \tfrac{1}{2}: &\quad E_{SO}^{(1)} = -\frac{(\ell+1)\hbar^2}{2}\langle\xi\rangle
\end{align}

The coupled basis is the natural choice because $J^2$ and $J_z$ commute with $\vec{L}\cdot\vec{S}$, making $j$ and $m_j$ good quantum numbers.

\textit{Context:} Basis choice is not just mathematical convenience; it reflects physical insight about which quantities are conserved. The same principle governs LS-coupling versus jj-coupling in multi-electron atoms, molecular orbital versus valence bond descriptions in chemistry, and product versus entangled bases in quantum information. In quantum computing, gate design requires choosing bases adapted to the available control Hamiltonians.
\end{example}

\section{Time-Dependent Perturbation Theory}

For time-dependent Hamiltonians $H = H_0 + V(t)$, we need different techniques. In Chapter~\ref{ch:time}, we developed the general framework for time-dependent evolution, including the Dyson series (Section~4.8) and the interaction picture (Section~4.9). Here we apply these tools in the perturbative regime, where the time-dependent part $V(t)$ is small compared to $H_0$.

\subsection{The Interaction Picture}

Recall from Section~4.9 that the interaction picture provides a hybrid between the Schr\"odinger and Heisenberg pictures, transferring the ``trivial'' evolution under $H_0$ to the operators while states evolve only due to the perturbation. In the interaction picture, states evolve according to:
\begin{equation}
    i\frac{\partial}{\partial t}\ket{\psi_I(t)} = V_I(t)\ket{\psi_I(t)}
\end{equation}
with $V_I(t) = e^{iH_0t}V(t)e^{-iH_0t}$.

As we showed in Chapter~\ref{ch:time}, the formal solution involves the time-ordered exponential (Dyson series):
\begin{equation}
    \ket{\psi_I(t)} = \mathcal{T}\exp\left(-i\int_0^t V_I(t')\,dt'\right)\ket{\psi_I(0)}
\end{equation}
Time ordering is essential when $[V_I(t_1), V_I(t_2)] \neq 0$ for different times, reflecting the causal structure of quantum evolution. In perturbation theory, we truncate this series at low order.

\subsection{Transition Probabilities}

To first order, the amplitude for transition from state $\ket{i}$ to state $\ket{f}$ is:
\begin{equation}
    c_{fi}(t) = -i\int_0^t dt' \bra{f}V_I(t')\ket{i} = -i\int_0^t dt'\, e^{i\omega_{fi}t'}\bra{f}V(t')\ket{i}
\end{equation}
where $\omega_{fi} = E_f - E_i$.

\begin{example}{Rabi Oscillations in a Driven Two-Level System}
\label{ex:rabi_perturbative}
Consider a spin-1/2 in a static field $B_0\hat{z}$ driven by an oscillating transverse field:
\begin{equation}
    H = -\gamma B_0 S_z - \gamma B_1 \cos(\omega t) S_x
\end{equation}

This is precisely the NMR system we analyzed exactly in Section~4.9 of Chapter~\ref{ch:time}. There, we transformed to the rotating frame and applied the rotating wave approximation to obtain the effective time-independent Hamiltonian. Here we take a complementary approach: we treat the oscillating field perturbatively, valid when $\Omega_1 \ll |\omega_0 - \omega|$ (far from resonance) or for short times.

Let $\omega_0 = \gamma B_0$ (Larmor frequency) and $\Omega_1 = \gamma B_1$ (Rabi frequency). The unperturbed states are $\ket{\uparrow}$ and $\ket{\downarrow}$ with energy splitting $\omega_0$.

Starting in $\ket{\downarrow}$, the transition amplitude to $\ket{\uparrow}$ is:
\begin{equation}
    c_{\uparrow}(t) = -i\int_0^t dt'\, e^{i\omega_0 t'}(-\gamma B_1\cos\omega t')\bra{\uparrow}S_x\ket{\downarrow}
\end{equation}

Using $\bra{\uparrow}S_x\ket{\downarrow} = 1/2$ and $\cos\omega t' = \frac{1}{2}(e^{i\omega t'} + e^{-i\omega t'})$:
\begin{equation}
    c_{\uparrow}(t) = \frac{i\Omega_1}{4}\int_0^t dt'\left(e^{i(\omega_0+\omega)t'} + e^{i(\omega_0-\omega)t'}\right)
\end{equation}

Near resonance ($\omega \approx \omega_0$), the first term oscillates rapidly and averages to zero (rotating wave approximation). The second term gives:
\begin{equation}
    c_{\uparrow}(t) \approx \frac{i\Omega_1}{4}\frac{e^{i(\omega_0-\omega)t} - 1}{i(\omega_0-\omega)} = \frac{\Omega_1}{2(\omega_0-\omega)}\sin\left(\frac{(\omega_0-\omega)t}{2}\right)e^{i(\omega_0-\omega)t/2}
\end{equation}

The transition probability:
\begin{equation}
    P_{\uparrow}(t) = |c_{\uparrow}(t)|^2 = \frac{\Omega_1^2}{(\omega_0-\omega)^2}\sin^2\left(\frac{(\omega_0-\omega)t}{2}\right)
\end{equation}

At exact resonance ($\omega = \omega_0$), this formula breaks down, but the exact solution gives:
\begin{equation}
    P_{\uparrow}(t) = \sin^2\left(\frac{\Omega_1 t}{2}\right)
\end{equation}

The system oscillates coherently between $\ket{\downarrow}$ and $\ket{\uparrow}$ at the Rabi frequency $\Omega_1$. Perturbation theory is valid when $\Omega_1 \ll |\omega_0 - \omega|$ or for short times $t \ll 2\pi/\Omega_1$.

Compare this perturbative result with the exact rotating-frame solution from Section~4.9. The exact on-resonance result $P_{\uparrow}(t) = \sin^2(\Omega_1 t/2)$ agrees with what we would obtain by summing the perturbation series to all orders. The perturbative approach is particularly valuable for understanding the off-resonant response and for systems where the driving is genuinely weak.
\end{example}

\begin{example}{Electric Dipole Transitions in Hydrogen}
Consider a hydrogen atom interacting with an oscillating electric field $\vec{\mathcal{E}}(t) = \mathcal{E}_0\hat{z}\cos(\omega t)$. The perturbation is:
\begin{equation}
    V(t) = e\mathcal{E}_0 z \cos(\omega t)
\end{equation}

The transition amplitude from state $\ket{i} = \ket{n_i,\ell_i,m_i}$ to $\ket{f} = \ket{n_f,\ell_f,m_f}$ is:
\begin{equation}
    c_{fi}(t) \propto \bra{f}z\ket{i}\int_0^t dt'\, e^{i\omega_{fi}t'}\cos(\omega t')
\end{equation}

The matrix element $\bra{f}z\ket{i}$ imposes \textbf{selection rules}:
\begin{equation}
    z = r\cos\theta = r\sqrt{\frac{4\pi}{3}}Y_1^0(\theta,\phi)
\end{equation}

The angular integration:
\begin{equation}
    \int d\Omega\, Y_{\ell_f}^{m_f*}Y_1^0 Y_{\ell_i}^{m_i}
\end{equation}
vanishes unless:
\begin{align}
    \Delta m &= m_f - m_i = 0 \\
    \Delta\ell &= \ell_f - \ell_i = \pm 1
\end{align}

These are the \textbf{electric dipole selection rules}. Additionally, parity must change since $z$ is odd.

For the Lyman-$\alpha$ transition $2p \to 1s$:
\begin{equation}
    \bra{1s}z\ket{2p_z} = \bra{1,0,0}r\cos\theta\ket{2,1,0} = \frac{2^7}{3^5}a_0 \approx 0.74\, a_0
\end{equation}

\textit{Context:} Selection rules determine which transitions can be driven by lasers, the foundation of laser cooling, optical pumping, and quantum state preparation. ``Forbidden'' transitions (electric quadrupole, magnetic dipole) have much smaller matrix elements but are used for ultra-narrow clock transitions because their slower decay rates give longer coherence times. The matrix elements directly determine oscillator strengths used throughout astrophysical spectroscopy.
\end{example}

\section{Fermi's Golden Rule}

\subsection{Transition Rates to a Continuum}

When transitions occur to a continuum of final states, we compute the transition rate rather than the transition probability. Starting from time-dependent perturbation theory with a periodic perturbation $V(t) = Ve^{-i\omega t} + V^\dagger e^{i\omega t}$, the transition rate to a continuum of final states is:

\begin{equation}
    \Gamma_{i \to f} = 2\pi|\bra{f}V\ket{i}|^2\rho(E_f)
\end{equation}
where $\rho(E_f)$ is the density of final states at the energy determined by energy conservation: $E_f = E_i + \omega$ (absorption) or $E_f = E_i - \omega$ (emission).

\begin{mathematicalaside}{Derivation of Fermi's Golden Rule}
The transition amplitude at time $t$ is $c_f(t) = -i\bra{f}V\ket{i}\frac{e^{i(\omega_{fi}-\omega)t}-1}{i(\omega_{fi}-\omega)}$. The probability $|c_f(t)|^2 \propto \frac{\sin^2[(\omega_{fi}-\omega)t/2]}{(\omega_{fi}-\omega)^2}$ becomes sharply peaked at $\omega_{fi} = \omega$ as $t$ increases. In the limit $t \to \infty$, this approaches $\pi t\,\delta(\omega_{fi}-\omega)$, giving a constant transition rate.
\end{mathematicalaside}

\begin{example}{Ionization from a Bound State}
Consider a particle initially bound in a one-dimensional well, perturbed by an oscillating electric field that can ionize it into continuum states.

For a particle in the ground state of an infinite square well of width $L$:
\begin{equation}
    \psi_1(x) = \sqrt{\frac{2}{L}}\sin\left(\frac{\pi x}{L}\right), \quad E_1 = \frac{\pi^2}{2L^2}
\end{equation}

An oscillating field $V(t) = E_0 x\cos(\omega t)$ can eject the particle into continuum states. For final states with energy $E_f = k_f^2/2$, energy conservation requires:
\begin{equation}
    \omega = E_f - E_1 = \frac{k_f^2}{2} - \frac{\pi^2}{2L^2}
\end{equation}

The matrix element $\bra{k_f}x\ket{1}$ couples the bound state to the continuum. The density of states for a free particle is $\rho(E) = \frac{L}{2\pi}\sqrt{\frac{2}{E}}$, and Fermi's golden rule gives:
\begin{equation}
    \Gamma = 2\pi|E_0|^2|\bra{k_f}x\ket{1}|^2\rho(E_f)
\end{equation}

This ionization rate increases with field strength as $E_0^2$ and depends on the overlap between bound and continuum wave functions.
\end{example}

\begin{example}{Spontaneous Emission: Hydrogen Lyman-$\alpha$}
For spontaneous emission, an excited atom decays by emitting a photon into the electromagnetic vacuum. The transition rate for electric dipole radiation is:
\begin{equation}
    A_{fi} = \frac{\omega_{fi}^3}{3\pi\epsilon_0\hbar c^3}|\bra{f}\vec{d}\ket{i}|^2
\end{equation}
where $\vec{d} = e\vec{r}$ is the electric dipole operator.

For the Lyman-$\alpha$ transition ($2p \to 1s$) in hydrogen:
\begin{itemize}
    \item Transition energy: $\hbar\omega = 10.2$ eV (wavelength 121.6 nm)
    \item Dipole matrix element: $|\bra{1s}er\ket{2p}|^2 = e^2\left(\frac{2^7}{3^5}a_0\right)^2$
\end{itemize}

The spontaneous emission rate:
\begin{equation}
    A_{2p\to 1s} = \frac{\omega^3}{3\pi\epsilon_0\hbar c^3}\cdot \frac{e^2 2^{15} a_0^2}{3^{10}} = 6.27 \times 10^8\text{ s}^{-1}
\end{equation}

The lifetime of the $2p$ state is $\tau = 1/A = 1.6$ ns.

For comparison, Rydberg states with principal quantum number $n$ have:
\begin{equation}
    A_n \propto \frac{1}{n^3}
\end{equation}
A Rydberg state with $n = 50$ has a lifetime of order milliseconds, a factor of $10^6$ longer than low-lying states. This suppressed spontaneous emission enables long coherence times in Rydberg-atom quantum computing.

\textit{Context:} Spontaneous emission rates determine atomic lifetimes, laser threshold conditions, and quantum efficiency of light sources. The Lyman-$\alpha$ line is the most important spectral line in astrophysics, used for mapping cosmic hydrogen distribution and studying the epoch of reionization. In cavity QED, the Purcell effect modifies spontaneous emission rates, enabling single-photon sources for quantum communication.
\end{example}

\section{The Sudden Approximation}

When a perturbation is applied suddenly (faster than characteristic system timescales), the wave function has no time to adjust. We use the sudden approximation.

\subsection{Overlap Probabilities}

If the Hamiltonian suddenly changes from $H_0$ to $H'$ at $t = 0$, the system's state $\ket{\psi}$ is unchanged, but it is no longer an eigenstate of the new Hamiltonian. The probability of finding the system in eigenstate $\ket{\phi_n'}$ of $H'$ is:
\begin{equation}
    P_n = |\braket{\phi_n'}{\psi}|^2
\end{equation}

\begin{example}{Sudden Change of Harmonic Oscillator Frequency}
Consider a particle in the ground state of a harmonic oscillator with frequency $\omega_1$. At $t = 0$, the frequency suddenly changes to $\omega_2$.

The initial state is:
\begin{equation}
    \psi_0^{(1)}(x) = \left(\frac{m\omega_1}{\pi\hbar}\right)^{1/4}\exp\left(-\frac{m\omega_1 x^2}{2\hbar}\right)
\end{equation}

The new ground state is:
\begin{equation}
    \psi_0^{(2)}(x) = \left(\frac{m\omega_2}{\pi\hbar}\right)^{1/4}\exp\left(-\frac{m\omega_2 x^2}{2\hbar}\right)
\end{equation}

The probability of remaining in the ground state is:
\begin{equation}
    P_0 = |\braket{\psi_0^{(2)}}{\psi_0^{(1)}}|^2 = \left|\int_{-\infty}^{\infty}dx\,\psi_0^{(2)*}(x)\psi_0^{(1)}(x)\right|^2
\end{equation}

The integral of two Gaussians:
\begin{equation}
    \braket{\psi_0^{(2)}}{\psi_0^{(1)}} = \left(\frac{m\omega_1}{\pi\hbar}\right)^{1/4}\left(\frac{m\omega_2}{\pi\hbar}\right)^{1/4}\sqrt{\frac{2\pi\hbar}{m(\omega_1+\omega_2)}} = \frac{2\sqrt{\omega_1\omega_2}}{\omega_1+\omega_2}
\end{equation}

Therefore:
\begin{equation}
    P_0 = \frac{4\omega_1\omega_2}{(\omega_1+\omega_2)^2}
\end{equation}

For a small change $\omega_2 = \omega_1(1+\epsilon)$:
\begin{equation}
    P_0 \approx 1 - \frac{\epsilon^2}{4} + O(\epsilon^3)
\end{equation}

For a doubling of frequency ($\omega_2 = 2\omega_1$): $P_0 = 8/9 \approx 0.89$.

The particle is most likely to remain in the ground state, but there is a finite probability of excitation to higher states (only even states, by parity).
\end{example}

\begin{example}{Sudden Nuclear Charge Change: Beta Decay}
In nuclear $\beta^-$ decay, a neutron converts to a proton, suddenly increasing the nuclear charge from $Z$ to $Z+1$. Consider an electron initially in the ground state of a hydrogen-like atom.

The initial wave function is:
\begin{equation}
    \psi_{1s}^{(Z)}(r) = \frac{1}{\sqrt{\pi}}\left(\frac{Z}{a_0}\right)^{3/2}e^{-Zr/a_0}
\end{equation}

After the decay, the electron finds itself in the Coulomb potential of charge $Z+1$. The probability of remaining in the $1s$ state of the new atom:
\begin{equation}
    P_{1s} = |\braket{\psi_{1s}^{(Z+1)}}{\psi_{1s}^{(Z)}}|^2
\end{equation}

The overlap integral:
\begin{equation}
    \braket{\psi_{1s}^{(Z+1)}}{\psi_{1s}^{(Z)}} = \frac{1}{\pi}\left(\frac{Z(Z+1)}{a_0^2}\right)^{3/2}\int_0^\infty 4\pi r^2 dr\, e^{-(2Z+1)r/a_0}
\end{equation}

\begin{equation}
    = 4\left(\frac{Z(Z+1)}{a_0^2}\right)^{3/2}\frac{2a_0^3}{(2Z+1)^3} = \frac{8[Z(Z+1)]^{3/2}}{(2Z+1)^3}
\end{equation}

For tritium decay ($Z = 1 \to Z = 2$):
\begin{equation}
    P_{1s} = \left(\frac{8(1)(2)^{3/2}}{27}\right)^2 = \left(\frac{16\sqrt{2}}{27}\right)^2 \approx 0.70
\end{equation}

There is a 30\% probability that the electron is shaken into an excited state or ionized by the sudden change in nuclear charge.

\textit{Context:} The sudden approximation applies to fast quenches in ultracold atomic gases, where trap parameters can be changed faster than atomic motion timescales. In condensed matter, sudden quenches are used to study non-equilibrium dynamics and thermalization. The beta decay calculation was an early confirmation of quantum mechanical overlap predictions.
\end{example}

\section{The Adiabatic Approximation}

When changes occur slowly compared to system dynamics, the system remains in an instantaneous eigenstate. We developed the adiabatic theorem in detail in Section~4.10 of Chapter~\ref{ch:time}, including the derivation of the Berry phase for two-level systems. Here we summarize the key results and place them in the context of perturbation theory.

\subsection{The Adiabatic Theorem}

Recall from Section~4.10 that if a system starts in the $n$th eigenstate of $H(0)$ and $H(t)$ changes slowly enough that transitions to other states are negligible, the system remains in the $n$th instantaneous eigenstate of $H(t)$:
\begin{equation}
    \ket{\psi(t)} = e^{i\gamma_n(t)}e^{-\frac{i}{\hbar}\int_0^t E_n(t')dt'}\ket{n(t)}
\end{equation}

The adiabatic condition requires that the rate of change be slow compared to level spacings:
\begin{equation}
    \hbar\left|\frac{\bra{m}\dot{H}\ket{n}}{(E_n - E_m)^2}\right| \ll 1 \quad \text{for all } m \neq n
\end{equation}

This condition can be understood through time-dependent perturbation theory: the transition amplitude to state $\ket{m}$ involves the matrix element $\bra{m}\dot{H}\ket{n}$ divided by the energy denominator $(E_n - E_m)$. When the adiabatic condition is satisfied, these transition amplitudes remain small for all times.

\subsection{Berry Phase}

As derived in Section~4.10.2, the geometric phase $\gamma_n(t)$ acquired during adiabatic evolution depends only on the path taken through parameter space, not on the rate of traversal. The key result for two-level systems is that the Berry phase equals half the solid angle subtended by the path on the Bloch sphere:
\begin{equation}
    \gamma_{\text{geom}} = -\frac{1}{2}\Omega
\end{equation}
where $\Omega$ is the solid angle enclosed by the closed path traced by the Hamiltonian's parameters.

For a general cyclic evolution in parameter space:
\begin{equation}
    \gamma_n = i\oint \bra{n(R)}\nabla_R\ket{n(R)} \cdot dR
\end{equation}
This Berry connection $\vec{A}_n(R) = i\bra{n(R)}\nabla_R\ket{n(R)}$ plays a role analogous to the vector potential in electromagnetism, with the Berry curvature acting like a magnetic field in parameter space.

\begin{physicalinsight}
The Berry phase is topological in nature: it depends on the geometry of the parameter space and the structure of eigenstate degeneracies. As we showed in Chapter~\ref{ch:time} through the Bloch cube visualization, the factor of $1/2$ arises from the spinorial nature of spin-1/2 wavefunctions. The Berry phase underlies phenomena from the Aharonov-Bohm effect to topological insulators.
\end{physicalinsight}

\section{Variational Methods}

The variational principle provides an alternative approach to approximation that does not require a small parameter.

\subsection{The Variational Principle}

For any normalized trial wave function $\ket{\psi_{\text{trial}}}$:
\begin{equation}
    E_0 \leq \frac{\bra{\psi_{\text{trial}}}H\ket{\psi_{\text{trial}}}}{\braket{\psi_{\text{trial}}}{\psi_{\text{trial}}}}
\end{equation}
where $E_0$ is the true ground state energy. The equality holds only when $\ket{\psi_{\text{trial}}}$ is the exact ground state.

\subsection{The Rayleigh-Ritz Method}

We parameterize trial functions and minimize the energy expectation value with respect to the parameters.

\begin{example}{Gaussian Trial Function for Harmonic Oscillator}
Consider the harmonic oscillator $H = \frac{p^2}{2m} + \frac{1}{2}m\omega^2x^2$ with a Gaussian trial function:
\begin{equation}
    \psi_\alpha(x) = \left(\frac{2\alpha}{\pi}\right)^{1/4}e^{-\alpha x^2}
\end{equation}
where $\alpha > 0$ is the variational parameter.

The kinetic energy:
\begin{equation}
    \langle T \rangle = -\frac{\hbar^2}{2m}\int_{-\infty}^{\infty}dx\,\psi_\alpha^* \frac{d^2\psi_\alpha}{dx^2}
\end{equation}

Using $\frac{d^2\psi_\alpha}{dx^2} = (4\alpha^2x^2 - 2\alpha)\psi_\alpha$:
\begin{equation}
    \langle T \rangle = \frac{\hbar^2}{2m}(2\alpha - 4\alpha^2\langle x^2\rangle) = \frac{\hbar^2}{2m}\left(2\alpha - 4\alpha^2 \cdot \frac{1}{4\alpha}\right) = \frac{\hbar^2\alpha}{2m}
\end{equation}

The potential energy:
\begin{equation}
    \langle V \rangle = \frac{1}{2}m\omega^2\langle x^2\rangle = \frac{1}{2}m\omega^2 \cdot \frac{1}{4\alpha} = \frac{m\omega^2}{8\alpha}
\end{equation}

Total energy:
\begin{equation}
    E(\alpha) = \frac{\hbar^2\alpha}{2m} + \frac{m\omega^2}{8\alpha}
\end{equation}

Minimizing: $\frac{dE}{d\alpha} = \frac{\hbar^2}{2m} - \frac{m\omega^2}{8\alpha^2} = 0$, giving:
\begin{equation}
    \alpha_{\min} = \frac{m\omega}{2\hbar}
\end{equation}

The minimum energy:
\begin{equation}
    E_{\min} = \frac{\hbar^2}{2m}\cdot\frac{m\omega}{2\hbar} + \frac{m\omega^2}{8}\cdot\frac{2\hbar}{m\omega} = \frac{\hbar\omega}{4} + \frac{\hbar\omega}{4} = \frac{\hbar\omega}{2}
\end{equation}

This is exactly the ground state energy! The Gaussian trial function with optimal $\alpha$ is the exact ground state wave function. This is a special case where the variational method gives the exact answer.
\end{example}

\begin{example}{Helium Ground State}
The helium atom has two electrons in the Coulomb field of a $Z = 2$ nucleus:
\begin{equation}
    H = -\frac{\hbar^2}{2m}(\nabla_1^2 + \nabla_2^2) - \frac{2e^2}{4\pi\epsilon_0}\left(\frac{1}{r_1} + \frac{1}{r_2}\right) + \frac{e^2}{4\pi\epsilon_0}\frac{1}{r_{12}}
\end{equation}

The electron-electron repulsion makes this problem unsolvable analytically. We use a trial function where each electron occupies a hydrogen-like $1s$ orbital with effective charge $Z_{\text{eff}}$:
\begin{equation}
    \psi(r_1, r_2) = \frac{Z_{\text{eff}}^3}{\pi a_0^3}e^{-Z_{\text{eff}}(r_1 + r_2)/a_0}
\end{equation}

The effective charge accounts for screening: each electron partially shields the other from the nucleus.

The energy expectation value can be computed analytically:
\begin{equation}
    E(Z_{\text{eff}}) = \left(Z_{\text{eff}}^2 - 2Z \cdot Z_{\text{eff}} + \frac{5}{8}Z_{\text{eff}}\right)\frac{e^2}{4\pi\epsilon_0 a_0}
\end{equation}

For helium ($Z = 2$):
\begin{equation}
    E(Z_{\text{eff}}) = \left(Z_{\text{eff}}^2 - 4Z_{\text{eff}} + \frac{5}{8}Z_{\text{eff}}\right)\text{Ry} = \left(Z_{\text{eff}}^2 - \frac{27}{8}Z_{\text{eff}}\right)\text{Ry}
\end{equation}

Minimizing: $\frac{dE}{dZ_{\text{eff}}} = 2Z_{\text{eff}} - \frac{27}{8} = 0$, giving $Z_{\text{eff}} = \frac{27}{16} = 1.6875$.

The variational energy:
\begin{equation}
    E_{\min} = -\left(\frac{27}{16}\right)^2 \text{Ry} = -2.848\text{ Ry} = -77.5 \text{ eV}
\end{equation}

The experimental value is $-78.975$ eV, so the simple one-parameter variational calculation gives 98\% of the correct binding energy.

\textit{Context:} Variational methods are the foundation of computational quantum chemistry. Hartree-Fock, density functional theory (DFT), and coupled-cluster methods all build on variational principles. Variational quantum eigensolvers (VQE) are leading algorithms for near-term quantum computers, using parameterized quantum circuits as trial wave functions. Tensor network methods in condensed matter physics also rely on variational optimization.
\end{example}

\section{WKB Approximation}

The WKB (Wentzel-Kramers-Brillouin) approximation provides semiclassical wave functions valid when the potential varies slowly on the scale of the de Broglie wavelength.

\subsection{The WKB Wave Function}

In classically allowed regions where $E > V(x)$, the local momentum is $p(x) = \sqrt{2m(E - V(x))}$. The WKB wave function is:
\begin{equation}
    \psi(x) \approx \frac{C}{\sqrt{p(x)}}\exp\left(\pm\frac{i}{\hbar}\int^x p(x')dx'\right)
\end{equation}

In classically forbidden regions where $E < V(x)$, $p(x)$ becomes imaginary:
\begin{equation}
    \psi(x) \approx \frac{C}{\sqrt{|p(x)|}}\exp\left(\pm\frac{1}{\hbar}\int^x |p(x')|dx'\right)
\end{equation}

\subsection{Bohr-Sommerfeld Quantization}

For bound states between classical turning points $a$ and $b$, the quantization condition is:
\begin{equation}
    \oint p\,dx = 2\int_a^b p(x)\,dx = 2\pi\hbar\left(n + \frac{1}{2}\right)
\end{equation}

The $\frac{1}{2}$ arises from connection formulas at the turning points.

\begin{example}{WKB for the Harmonic Oscillator}
For $V(x) = \frac{1}{2}m\omega^2x^2$, the classical turning points at energy $E$ are $x = \pm\sqrt{2E/m\omega^2}$.

The local momentum: $p(x) = \sqrt{2m(E - \frac{1}{2}m\omega^2x^2)} = \sqrt{2mE}\sqrt{1 - \frac{m\omega^2x^2}{2E}}$.

The quantization integral:
\begin{equation}
    \int_{-a}^{a} p(x)\,dx = \sqrt{2mE}\int_{-a}^{a}\sqrt{1 - x^2/a^2}\,dx = \sqrt{2mE} \cdot \frac{\pi a}{2}
\end{equation}
where $a = \sqrt{2E/m\omega^2}$.

Substituting:
\begin{equation}
    \sqrt{2mE} \cdot \frac{\pi}{2}\sqrt{\frac{2E}{m\omega^2}} = \frac{\pi E}{\omega}
\end{equation}

The quantization condition $\frac{2\pi E}{\omega} = 2\pi\hbar(n + \frac{1}{2})$ gives:
\begin{equation}
    E_n = \hbar\omega\left(n + \frac{1}{2}\right)
\end{equation}

WKB gives the exact harmonic oscillator spectrum! This is another special case where the semiclassical approximation is exact.
\end{example}

\begin{example}{WKB for Hydrogen: Radial Quantization}
Working in atomic units, the radial Schr\"odinger equation for hydrogen can be written as a one-dimensional problem:
\begin{equation}
    -\frac{1}{2}\frac{d^2u}{dr^2} + V_{\text{eff}}(r)u = Eu
\end{equation}
where $u = rR(r)$ and
\begin{equation}
    V_{\text{eff}}(r) = -\frac{1}{r} + \frac{\ell(\ell+1)}{2r^2}
\end{equation}

The classical turning points $r_1$ and $r_2$ satisfy $E = V_{\text{eff}}(r)$. The WKB quantization:
\begin{equation}
    \int_{r_1}^{r_2}\sqrt{2(E - V_{\text{eff}}(r))}\,dr = \pi\left(n_r + \frac{1}{2}\right)
\end{equation}
where $n_r = 0, 1, 2, \ldots$ is the radial quantum number.

This integral can be evaluated using contour integration. The result is:
\begin{equation}
    \frac{1}{\sqrt{-2E}} - \left(\ell + \frac{1}{2}\right) = \left(n_r + \frac{1}{2}\right)
\end{equation}

Defining $n = n_r + \ell + 1$:
\begin{equation}
    E_n = -\frac{1}{2n^2}
\end{equation}
which is the exact hydrogen spectrum in atomic units (where the ground state energy is $-1/2$ Hartree $= -1$ Rydberg).

WKB reproduces the exact hydrogen spectrum.

\textit{Context:} Semiclassical methods bridge quantum and classical descriptions, essential for understanding the quantum-classical correspondence. WKB tunneling rates govern alpha decay, scanning tunneling microscopy, and Josephson junction dynamics. The method extends to classically chaotic systems through trace formulas connecting periodic orbits to quantum spectra.
\end{example}

\section{The Complete Hydrogen Atom}

Having developed perturbation theory techniques, we now apply them systematically to understand the real hydrogen atom beyond the idealized Coulomb solution. This synthesis demonstrates how perturbation theory connects textbook quantum mechanics to precision atomic physics.

\subsection{The Hierarchy of Energy Scales}

The hydrogen atom exhibits a beautiful hierarchy of energy scales, each probing different physics:

\begin{center}
\begin{tabular}{lcc}
\hline
\textbf{Effect} & \textbf{Energy Scale} & \textbf{Origin} \\
\hline
Gross structure & $\text{Ry} \approx 13.6$ eV & Coulomb interaction \\
Fine structure & $\alpha^2 \text{Ry} \sim 10^{-4}$ eV & Relativity + spin-orbit \\
Lamb shift & $\alpha^3 \text{Ry} \sim 10^{-6}$ eV & QED vacuum fluctuations \\
Hyperfine structure & $(m_e/m_p)\alpha^2 \text{Ry} \sim 10^{-7}$ eV & Nuclear magnetic moment \\
\hline
\end{tabular}
\end{center}

Here $\alpha = e^2/(4\pi\epsilon_0\hbar c) \approx 1/137$ is the fine structure constant.

\begin{physicalinsight}
The hierarchy arises because different interactions involve different combinations of fundamental constants. The fine structure constant $\alpha$ controls relativistic corrections, the electron-to-proton mass ratio governs nuclear effects, and higher powers of $\alpha$ indicate quantum electrodynamic corrections. This pattern of hierarchical scales, where each level of structure involves smaller energy scales, recurs throughout physics and underlies the effective field theory approach to modern particle physics.
\end{physicalinsight}

\subsection{Fine Structure in Detail}

The fine structure Hamiltonian consists of three terms:
\begin{equation}
    H_{\text{fine}} = H_{\text{rel}} + H_{SO} + H_{\text{Darwin}}
\end{equation}

\textbf{1. Relativistic kinetic energy correction:}

The relativistic kinetic energy $T = \sqrt{p^2c^2 + m^2c^4} - mc^2$ expands as:
\begin{equation}
    T = \frac{p^2}{2m} - \frac{p^4}{8m^3c^2} + \cdots
\end{equation}

The first-order perturbation:
\begin{equation}
    E_{\text{rel}}^{(1)} = -\frac{1}{2mc^2}\langle\psi_{n\ell m}|\frac{p^4}{4m^2}|\psi_{n\ell m}\rangle
\end{equation}

Using $p^2/2m = E_n - V(r)$ and the known expectation values $\langle r^{-1}\rangle$ and $\langle r^{-2}\rangle$:
\begin{equation}
    E_{\text{rel}}^{(1)} = -\frac{E_n^2}{2mc^2}\left(\frac{4n}{\ell + 1/2} - 3\right)
\end{equation}

\textbf{2. Spin-orbit coupling:}

From the electron's rest frame, the nucleus appears to orbit, creating a magnetic field that couples to the electron spin:
\begin{equation}
    H_{SO} = \frac{1}{2m_e^2c^2}\frac{1}{r}\frac{dV}{dr}\vec{L}\cdot\vec{S} = \frac{e^2}{8\pi\epsilon_0 m_e^2c^2 r^3}\vec{L}\cdot\vec{S}
\end{equation}

Using the coupled basis $\ket{n,\ell,j,m_j}$ where $\vec{L}\cdot\vec{S} = \frac{1}{2}(J^2 - L^2 - S^2)$:
\begin{equation}
    E_{SO}^{(1)} = \frac{e^2}{16\pi\epsilon_0 m_e^2c^2}\langle r^{-3}\rangle_{n\ell}[j(j+1) - \ell(\ell+1) - \tfrac{3}{4}]\hbar^2
\end{equation}

For $\ell > 0$, this gives different energies for $j = \ell + \frac{1}{2}$ and $j = \ell - \frac{1}{2}$.

\textbf{3. Darwin term:}

The Darwin term arises from the ``zitterbewegung'' of the electron (rapid oscillation at the Compton wavelength scale) and affects only $s$-states:
\begin{equation}
    H_{\text{Darwin}} = \frac{\pi\hbar^2}{2m_e^2c^2}\frac{e^2}{4\pi\epsilon_0}\delta^3(\vec{r})
\end{equation}

For $s$-states: $E_{\text{Darwin}}^{(1)} = \frac{\pi\hbar^2 e^2}{8\pi\epsilon_0 m_e^2 c^2}|\psi_{n00}(0)|^2$.

\textbf{Combined fine structure:}

Remarkably, all three contributions combine to give a simple formula that depends only on $n$ and $j$:
\begin{equation}
    E_{\text{fine}} = -\frac{\alpha^2\,\text{Ry}}{n^3}\left(\frac{1}{j + 1/2} - \frac{3}{4n}\right)
\end{equation}

where $\text{Ry} \approx 13.6$~eV is the Rydberg energy. This is the Sommerfeld fine structure formula. Using $E_n = -\text{Ry}/n^2$, it can equivalently be written $E_{\text{fine}} = (\alpha^2 E_n / n)\left[1/(j+1/2) - 3/(4n)\right]$, which is negative because $E_n < 0$.

For $n = 2$:
\begin{itemize}
    \item $2s_{1/2}$ ($\ell = 0$, $j = 1/2$): $E_{\text{fine}} = -\frac{5\alpha^2}{64}\,\text{Ry}$
    \item $2p_{1/2}$ ($\ell = 1$, $j = 1/2$): $E_{\text{fine}} = -\frac{5\alpha^2}{64}\,\text{Ry}$
    \item $2p_{3/2}$ ($\ell = 1$, $j = 3/2$): $E_{\text{fine}} = -\frac{\alpha^2}{64}\,\text{Ry}$
\end{itemize}

The $2s_{1/2}$ and $2p_{1/2}$ states are predicted to be degenerate, a degeneracy broken only by the Lamb shift, which requires quantum electrodynamics.

\textit{Context:} The fine structure constant $\alpha$ governs the strength of electromagnetic interactions and appears throughout atomic physics. Spin-orbit coupling is the mechanism underlying spintronic devices and topological insulators. In artificial atoms (quantum dots), the ``fine structure'' splitting is tunable, enabling engineered spin-photon interfaces for quantum networks.

\subsection{The Zeeman Effect: Weak, Strong, and Intermediate Fields}

When hydrogen is placed in a magnetic field, the choice of perturbative approach depends on the field strength relative to fine structure splitting.

\textbf{Weak-field Zeeman effect (Anomalous Zeeman effect):}

When $\mu_B B \ll E_{\text{fine}}$, the fine structure dominates. We treat $H_{\text{Zeeman}} = \frac{\mu_B}{\hbar}(L_z + 2S_z)B$ as a perturbation on states labeled by $\ket{n,\ell,j,m_j}$.

Since $J_z = L_z + S_z$ but not $L_z$ and $S_z$ individually are good quantum numbers, we use:
\begin{equation}
    \langle L_z + 2S_z\rangle = \langle J_z\rangle + \langle S_z\rangle = \hbar m_j + \langle S_z\rangle
\end{equation}

The projection theorem gives $\langle S_z\rangle = m_j\hbar \cdot \frac{j(j+1) + s(s+1) - \ell(\ell+1)}{2j(j+1)}$.

The energy shift:
\begin{equation}
    E_{\text{Zeeman}}^{(1)} = \mu_B B m_j g_j
\end{equation}
where the Land'e $g$-factor is:
\begin{equation}
    g_j = 1 + \frac{j(j+1) + s(s+1) - \ell(\ell+1)}{2j(j+1)}
\end{equation}

For $j = \ell + \frac{1}{2}$: $g_j = 1 + \frac{1}{2\ell+1}$.
For $j = \ell - \frac{1}{2}$: $g_j = 1 - \frac{1}{2\ell+1}$.

\textbf{Strong-field Zeeman effect (Paschen-Back effect):}

When $\mu_B B \gg E_{\text{fine}}$, the magnetic field dominates. The good quantum numbers are $m_\ell$ and $m_s$ (uncoupled basis), and fine structure becomes a perturbation:
\begin{equation}
    E = E_n + \mu_B B(m_\ell + 2m_s) + E_{\text{fine}}^{(1)}(m_\ell, m_s)
\end{equation}

\textbf{Intermediate fields:}

When $\mu_B B \sim E_{\text{fine}}$, neither basis diagonalizes the total perturbation. We must diagonalize the combined $H_{\text{fine}} + H_{\text{Zeeman}}$ within the degenerate subspace.

\begin{example}{$n = 2$ Hydrogen in an Intermediate Magnetic Field}
Consider the $n = 2$ states (ignoring the $2p_{3/2}$ for simplicity). The $2s_{1/2}$ and $2p_{1/2}$ states each have $m_j = \pm 1/2$.

In the basis $\{|2s_{1/2}, +\tfrac{1}{2}\rangle, |2s_{1/2}, -\tfrac{1}{2}\rangle, |2p_{1/2}, +\tfrac{1}{2}\rangle, |2p_{1/2}, -\tfrac{1}{2}\rangle\}$, the total perturbation matrix is:

\begin{equation}
    H' = \begin{pmatrix}
    E_s + \mu_B B g_s \cdot \frac{1}{2} & 0 & 0 & 0 \\
    0 & E_s - \mu_B B g_s \cdot \frac{1}{2} & 0 & 0 \\
    0 & 0 & E_p + \mu_B B g_p \cdot \frac{1}{2} & 0 \\
    0 & 0 & 0 & E_p - \mu_B B g_p \cdot \frac{1}{2}
    \end{pmatrix}
\end{equation}

where $g_s = 2$ (pure spin) and $g_p = 2/3$ for $2p_{1/2}$.

Without the Lamb shift, $E_s = E_p$, and states with the same $m_j$ would cross. The Lamb shift ($E_s > E_p$ by about 1058 MHz) creates an avoided crossing, and at intermediate fields, states are superpositions of $s$ and $p$ character.

\textit{Context:} Magneto-optical traps (MOTs) exploit Zeeman shifts combined with laser detuning to create position-dependent forces that cool and trap atoms. This technique launched the ultracold atom revolution. Zeeman slowers decelerate atomic beams for loading MOTs. In astrophysics, the Paschen-Back regime is relevant for white dwarf and neutron star atmospheres, where magnetic fields reach $10^8$--$10^{11}$ T.
\end{example}

\subsection{Hyperfine Structure}

The proton has a magnetic moment $\vec{\mu}_p = g_p\frac{e}{2m_p}\vec{I}$ where $\vec{I}$ is the nuclear spin ($I = 1/2$ for hydrogen) and $g_p \approx 5.59$. This creates a magnetic field that couples to the electron spin.

The hyperfine Hamiltonian is:
\begin{equation}
    H_{\text{hf}} = A\vec{I}\cdot\vec{S}
\end{equation}
where $A$ depends on the electron wave function at the nucleus.

For the ground state:
\begin{equation}
    A_{1s} = \frac{2}{3}\mu_0 g_p\frac{e^2}{4\pi m_e m_p}|\psi_{1s}(0)|^2 = \frac{8}{3}g_p\frac{m_e}{m_p}\alpha^2 E_1
\end{equation}

Coupling $\vec{I}$ and $\vec{S}$ gives total spin $\vec{F} = \vec{I} + \vec{S}$:
\begin{equation}
    \vec{I}\cdot\vec{S} = \frac{1}{2}(F^2 - I^2 - S^2) = \frac{\hbar^2}{2}[F(F+1) - I(I+1) - S(S+1)]
\end{equation}

For hydrogen ($I = S = 1/2$), $F = 0$ or $F = 1$:
\begin{align}
    F = 1 &: \quad E_{\text{hf}} = +\frac{A\hbar^2}{4} \\
    F = 0 &: \quad E_{\text{hf}} = -\frac{3A\hbar^2}{4}
\end{align}

The hyperfine splitting of the ground state:
\begin{equation}
    \Delta E_{\text{hf}} = A\hbar^2 = 5.87 \times 10^{-6}\text{ eV} \quad \Leftrightarrow \quad \nu = 1420.4\text{ MHz} \quad \Leftrightarrow \quad \lambda = 21\text{ cm}
\end{equation}

\textit{Context:} The 21 cm hydrogen line is radio astronomy's most important spectral line. It maps the distribution of neutral hydrogen throughout the universe, probes the cosmic ``dark ages'' before the first stars, and tests cosmological models. Closer to home, the hyperfine transition in cesium-133 defines the SI second, and hyperfine states in trapped ions (such as $^{171}$Yb$^+$ and $^{133}$Ba$^+$) serve as qubit states in leading quantum computing platforms. The hydrogen maser, based on stimulated emission of the 21 cm transition, remains one of the most stable frequency references available.

\subsection{Summary: When Different Methods Apply}

The hydrogen atom illustrates how to choose among perturbative approaches:

\begin{center}
\begin{tabular}{ll}
\hline
\textbf{Situation} & \textbf{Method} \\
\hline
Non-degenerate level, small perturbation & Standard perturbation theory \\
Degenerate level, perturbation breaks degeneracy & Degenerate perturbation theory \\
Competing perturbations of similar size & Full matrix diagonalization \\
Perturbation commutes with some $H_0$ eigenbasis & Choose that basis \\
Oscillating perturbation, transitions & Time-dependent perturbation theory \\
Transition to continuum & Fermi's golden rule \\
Fast change & Sudden approximation \\
Slow change & Adiabatic theorem \\
No small parameter & Variational method \\
\hline
\end{tabular}
\end{center}

The art of perturbation theory lies in recognizing which regime applies and choosing the appropriate basis to simplify calculations.

\section{Chapter Summary}

Perturbation theory provides powerful tools for approximate solutions throughout quantum mechanics and beyond. Time-independent perturbation theory handles energy shifts and state mixing in static systems, with first-order shifts determined by expectation values and second-order shifts by virtual transitions to intermediate states. Degenerate perturbation theory addresses situations when symmetries are broken, requiring diagonalization within the degenerate subspace to find the appropriate zeroth-order states and good quantum numbers. Time-dependent perturbation theory describes transitions between states driven by external fields. Fermi's golden rule gives transition rates to continua of final states, underlying phenomena from spontaneous emission to scattering. The sudden and adiabatic approximations handle the limiting cases of fast and slow parameter changes. Variational methods provide upper bounds on ground state energies without requiring a small expansion parameter. The WKB approximation treats semiclassical systems with slowly varying potentials, connecting quantum mechanics to classical trajectories.

The hydrogen atom, with its hierarchy of corrections from gross structure through fine structure to hyperfine structure, exemplifies how these methods work together. Choice of basis (coupled versus uncoupled, weak-field versus strong-field) reflects physical insight about which quantum numbers remain good under the perturbation at hand. These same principles extend to multi-electron atoms, molecules, solids, and nuclei, making perturbation theory one of the most widely used tools in theoretical physics.

\input{book_problems/ch09_problems.tex}

\section*{References and Further Reading}
\addcontentsline{toc}{section}{References and Further Reading}

\begin{description}
\item[Sakurai, J.~J., and Napolitano, J.] \emph{Modern Quantum Mechanics}, 3rd ed. Cambridge University Press, 2017. Chapter 5 covers time-independent and time-dependent perturbation theory at the level of this chapter; the standard graduate reference for both the formalism and the hydrogen fine-structure application.

\item[Cohen-Tannoudji, C., Diu, B., and Lalo\"e, F.] \emph{Quantum Mechanics}, Vol.~2. Wiley-VCH, 1977. Chapters XI--XIII work through stationary, degenerate, and time-dependent perturbation theory with worked examples; the most thorough textbook treatment, particularly clear on degenerate perturbation theory and good quantum numbers.

\item[Born, M., and Fock, V.] ``Beweis des Adiabatensatzes.'' \emph{Zeitschrift f\"ur Physik} \textbf{51}, 165--180 (1928). \href{https://doi.org/10.1007/BF01343193}{doi:10.1007/BF01343193}. The original proof of the adiabatic theorem; the pre-Berry foundation for slow parameter changes.

\item[Berry, M.~V.] ``Quantal phase factors accompanying adiabatic changes.'' \emph{Proceedings of the Royal Society A} \textbf{392}, 45--57 (1984). \href{https://doi.org/10.1098/rspa.1984.0023}{doi:10.1098/rspa.1984.0023}. The geometric phase that the standard adiabatic theorem misses; essential reading after this chapter for anyone interested in topological phases.

\item[Dirac, P.~A.~M.] ``The quantum theory of the emission and absorption of radiation.'' \emph{Proceedings of the Royal Society A} \textbf{114}, 243--265 (1927). \href{https://doi.org/10.1098/rspa.1927.0039}{doi:10.1098/rspa.1927.0039}. The paper in which time-dependent perturbation theory and what came to be called Fermi's golden rule were first applied to spontaneous emission.

\item[Landau, L.~D., and Lifshitz, E.~M.] \emph{Quantum Mechanics: Non-Relativistic Theory}, 3rd ed. Pergamon Press, 1977. Chapter VI on perturbation theory and Chapter VII on the WKB approximation are characteristically terse and physically motivated; recommended for the variational and semiclassical methods touched on at the end of this chapter.
\end{description}

%% file: book_problems/ch09_problems.tex
\section{Problems}
\setcounter{hwproblem}{0}

\noindent\textbf{Note:} Throughout these problems, we use atomic units ($\hbar = m_e = e = 1$) unless otherwise specified.

\problem{Non-Degenerate Perturbation Theory: Infinite Square Well}
A particle in the ground state of an infinite square well of width $L$ (walls at $x = 0$ and $x = L$) experiences a perturbation $V(x) = V_0 \sin(\pi x/L)$.
\begin{enumerate}[label=(\alph*)]
    \item Calculate the first-order energy correction $E_1^{(1)}$ for the ground state.
    \item Calculate the first-order energy correction $E_2^{(1)}$ for the first excited state. Why is this qualitatively different from part (a)?
    \item Calculate the second-order energy correction to the ground state energy. Which excited state contributes most?
    \item For what value of $V_0$ relative to $E_1^{(0)}$ does perturbation theory break down? Give a quantitative criterion.
\end{enumerate}

\problem{Non-Degenerate Perturbation Theory: Anharmonic Oscillator}
A harmonic oscillator has a cubic perturbation: $H = p^2/(2m) + m\omega^2 x^2/2 + \lambda x^3$. Use $x = (x_0/\sqrt{2})(a + a^\dagger)$ with $x_0 = \sqrt{\hbar/(m\omega)}$.
\begin{enumerate}[label=(\alph*)]
    \item Show that $E_n^{(1)} = \lambda\bra{n}x^3\ket{n}$ vanishes for every $n$.
    \item Calculate $(a + a^\dagger)^3\ket{0}$ iteratively.
    \item Compute the matrix elements $\bra{1}x^3\ket{0}$ and $\bra{3}x^3\ket{0}$.
    \item Calculate the second-order ground state energy correction $E_0^{(2)}$ and express it in terms of $\lambda$, $x_0$, and $\hbar\omega$.
    \item Although the potential is unbounded below for large $|x|$, the perturbation series gives useful results for small $\lambda$. Explain why physically.
\end{enumerate}

\problem{Quadratic Stark Effect in Hydrogen}
The hydrogen ground state $\ket{1s}$ (energy $E_1 = -1/2$ in atomic units) is placed in a uniform electric field $\mathcal{E}$ along $z$, giving perturbation $V = \mathcal{E}z$.
\begin{enumerate}[label=(\alph*)]
    \item Show that $E_1^{(1)} = \mathcal{E}\bra{1s}z\ket{1s} = 0$. What symmetry is responsible?
    \item The second-order correction is $E_1^{(2)} = \sum_{n\ell m} |\mathcal{E}\bra{n,\ell,m}z\ket{1s}|^2/(E_1 - E_n)$. Explain why only $p$-states ($\ell=1$, $m=0$) contribute. Cite two selection rules.
    \item The exact result is $E_1^{(2)} = -\frac{9}{4}\mathcal{E}^2$, giving polarizability $\alpha_H = 9/2$ in atomic units. Express $\alpha_H$ in SI units and calculate its numerical value.
    \item Show that replacing all $E_n$ in the denominator with $E_2$ gives an upper bound on $|E_1^{(2)}|$. What polarizability does this yield, and how does it compare to the exact value?
\end{enumerate}

\problem{Degenerate Perturbation Theory: Four-State System}
A system with four states and unperturbed energies $E_1^{(0)} = E_3^{(0)} = 0$ and $E_2^{(0)} = E_4^{(0)} = 10$ has perturbation matrix:
\begin{equation*}
    V \;\approx \; \begin{pmatrix} 2 & 0.3 & 1 & 0 \\ 0.3 & -1 & 0 & 2 \\ 1 & 0 & -2 & 0.5 \\ 0 & 2 & 0.5 & 3 \end{pmatrix}
\end{equation*}
\begin{enumerate}[label=(\alph*)]
    \item Identify the two degenerate subspaces and extract the $2 \times 2$ perturbation submatrices.
    \item Diagonalize the $E^{(0)} = 0$ subspace to find $E_\pm^{(1)}$ and the correct zeroth-order eigenstates.
    \item Diagonalize the $E^{(0)} = 10$ subspace to find its first-order corrections.
    \item Sketch the energy level diagram showing how the two degenerate pairs split. Are any levels still degenerate?
    \item Estimate whether the second-order corrections from off-diagonal elements between subspaces are negligible.
\end{enumerate}

\problem{Two Coupled Spins with Perturbation}
Two spin-1/2 particles interact via $H = J\vec{S}_1 \cdot \vec{S}_2 + \epsilon S_{1z}$ where $\epsilon \ll J$ is a perturbation.
\begin{enumerate}[label=(\alph*)]
    \item Show that the unperturbed energies are $E_{\text{singlet}} = -3J/4$ and $E_{\text{triplet}} = J/4$ (three-fold degenerate).
    \item Construct the $3 \times 3$ perturbation matrix for the triplet subspace using the Clebsch-Gordan decomposition of triplet states.
    \item Diagonalize this matrix to find the first-order splittings of the three triplet states.
    \item Does the perturbation mix singlet and triplet? Calculate $\bra{0,0}S_{1z}\ket{1,0}$ to check. At what order in perturbation theory does this mixing enter?
\end{enumerate}

\problem{Linear Stark Effect in Hydrogen $n = 2$}
The $n=2$ level of hydrogen (ignoring spin) is 4-fold degenerate: $\ket{2s}$, $\ket{2p_z}$, $\ket{2p_+}$, $\ket{2p_-}$. An electric field $\mathcal{E}$ along $z$ gives perturbation $V = \mathcal{E}z$.
\begin{enumerate}[label=(\alph*)]
    \item Using selection rules $\Delta m = 0$ and $\Delta\ell = \pm 1$, show that the only nonzero matrix element within $n=2$ is $\bra{2s}z\ket{2p_z}$.
    \item Verify that $\bra{2s}z\ket{2p_z} = -3$ (atomic units) using the given radial wave functions $R_{20}(r) = \frac{1}{2\sqrt{2}}(2-r)e^{-r/2}$ and $R_{21}(r) = \frac{1}{2\sqrt{6}}re^{-r/2}$.
    \item Write out the full $4 \times 4$ perturbation matrix. Show that it is block-diagonal in $m$.
    \item Find the four first-order energy corrections and the correct zeroth-order eigenstates.
    \item The states $\ket{\pm} = (\ket{2s} \mp \ket{2p_z})/\sqrt{2}$ are $sp$-hybrid orbitals with permanent dipole moments. Calculate $\bra{\pm}z\ket{\pm}$. How is this possible for parity eigenstates?
\end{enumerate}

\problem{Deriving Fermi's Golden Rule}
For a constant perturbation $V$ suddenly switched on at $t=0$, the first-order transition probability is:
\begin{equation*}
    P_{i \to f}(t) = \frac{|\bra{f}V\ket{i}|^2}{\hbar^2}\frac{\sin^2(\omega_{fi}t/2)}{(\omega_{fi}/2)^2}
\end{equation*}
where $\omega_{fi} = (E_f - E_i)/\hbar$.
\begin{enumerate}[label=(\alph*)]
    \item Sketch $P_{i \to f}$ as a function of $\omega_{fi}$ for fixed $t$. What is the peak height and width? How do these scale with $t$?
    \item For a continuum of final states with density $\rho(E_f)$, show that the total probability is $P(t) = \frac{2\pi}{\hbar}|\bra{f}V\ket{i}|^2\rho(E_i)t$ by using $\int_{-\infty}^{\infty} \frac{\sin^2(xt/2)}{(x/2)^2}dx = 2\pi t$.
    \item Derive the transition rate $\Gamma = dP/dt$ (Fermi's golden rule): $\Gamma = \frac{2\pi}{\hbar}|\bra{f}V\ket{i}|^2\rho(E_i)$. Explain why a constant rate requires a continuum.
    \item Show that Fermi's golden rule requires both $t$ large (sinc-squared acts like delta function) and $t$ small ($P \ll 1$), but these conditions are compatible when $\Gamma \ll \Delta E / \hbar$.
\end{enumerate}

\problem{Sudden and Adiabatic: Harmonic Oscillator Frequency Change}
A harmonic oscillator in its ground state with frequency $\omega_1$ suddenly changes to frequency $\omega_2 = 4\omega_1$.
\begin{enumerate}[label=(\alph*)]
    \item \textbf{Sudden limit:} Calculate the probability of remaining in the ground state and the probability of reaching the first excited state. Use parity to argue that only even-$n$ final states are accessible.
    \item \textbf{Adiabatic limit:} What is the final-state probability and energy? What is the adiabatic phase?
    \item The energy expectation immediately after the change is $\langle H' \rangle = \bra{\psi_0^{(1)}}H'\ket{\psi_0^{(1)}}$ where $H' = p^2/(2m) + m\omega_2^2 x^2/2$. Calculate $\langle H' \rangle$ in both limits and interpret.
    \item Prove that in the sudden case, the particle can only be found in even-$n$ states of the new oscillator (parity conservation).
\end{enumerate}

\problem{Variational Method: Helium Ground State}
Two electrons in helium interact via $H = -\nabla_1^2/2 - \nabla_2^2/2 - 2/r_1 - 2/r_2 + 1/r_{12}$ (atomic units).
\begin{enumerate}[label=(\alph*)]
    \item Use a trial wave function $\psi(r_1, r_2) = (Z^3/\pi)^{1/2}e^{-Zr_1} (Z^3/\pi)^{1/2}e^{-Zr_2}$ where $Z$ is the variational parameter. Calculate $\langle H \rangle$ as a function of $Z$.
    \item Show that the energy is $E(Z) = -2Z^2 + (5/8)Z$ (the electron-electron repulsion integral yields $5Z/8$).
    \item Minimize to find the optimal $Z_{\text{eff}}$ and the resulting ground state energy. Compare to the experimental value of $-5.81$ eV.
    \item Explain physically why $Z_{\text{eff}} < 2$. What does the second term in $E(Z)$ represent?
\end{enumerate}

\problem{WKB Approximation: Tunneling and Quantization}
\begin{enumerate}[label=(\alph*)]
    \item For a potential $V(x) = |x|$ (linear potential), find the WKB quantization condition $\oint p(x)dx = 2\pi n\hbar$ for bound states. What are the allowed energies?
    \item For potential $V(x) = |x|^\alpha$ with $\alpha > 0$, calculate the leading-order WKB spectrum and identify which potentials give equally spaced levels.
    \item A particle with energy $E$ incident on a step potential $V(x) = V_0$ (for $x > 0$) has WKB transmission coefficient $T \approx e^{-2\gamma}$ where $\gamma = \int_0^d \sqrt{2m(V_0-E)}/\hbar\,dx$ is the tunneling exponent. For a rectangular barrier of height $V_0 = 5$ eV and width $d = 1$ \AA, calculate the transmission for $E = 4$ eV.
\end{enumerate}

\problem{Born Approximation and Scattering Amplitude}
The scattering amplitude in the Born approximation is $f(\vec{q}) = -\frac{m}{4\pi}(2\pi)^3 \tilde{V}(\vec{q})$ where $\tilde{V}$ is the Fourier transform of the potential and $\vec{q} = \vec{k}_i - \vec{k}_f$ is the momentum transfer.
\begin{enumerate}[label=(\alph*)]
    \item For a Yukawa potential $V(r) = -\frac{g^2}{4\pi}e^{-\mu r}/r$, calculate the Fourier transform $\tilde{V}(\vec{q})$.
    \item Find the scattering amplitude $f(\theta)$ and the differential cross section $d\sigma/d\Omega = |f(\theta)|^2$ in terms of the scattering angle $\theta$.
    \item For low-energy scattering ($\mu \ll k$), show that the $s$-wave phase shift is $\delta_0 \approx -k a_s$ where $a_s$ is the scattering length. Relate $a_s$ to the potential.
\end{enumerate}

\problem{Rotating Wave Approximation and Rabi Oscillations}
A two-level atom with transition frequency $\omega_0$ is driven by an oscillating field $\vec{E}(t) = E_0 \cos(\omega t) \hat{z}$, producing perturbation $V(t) = -d E_0 \cos(\omega t)$ where $d$ is the dipole moment.
\begin{enumerate}[label=(\alph*)]
    \item Decompose $V(t)$ into rotating and counter-rotating terms: $V = V^{\text{rot}} e^{-i\omega t} + \text{c.c.}$ plus counter-rotating terms oscillating at $2\omega$.
    \item In the rotating frame with $U = e^{-i\omega_0 t \sigma_z/2}$, show that the counter-rotating terms oscillate at $2\omega$ (far off-resonance) and can be dropped if $|\omega - \omega_0| \ll \omega$. This is the rotating wave approximation.
    \item In the RWA, the Hamiltonian becomes $H_{\text{RWA}} = \Delta\sigma_z/2 + \Omega\sigma_x$ where $\Delta = \omega - \omega_0$ is the detuning and $\Omega = dE_0$ is the Rabi frequency.
    \item For resonant driving ($\Delta = 0$), show that the population oscillates between states: $P_e(t) = \sin^2(\Omega t/2)$. At what time does the system achieve maximum transition probability (a $\pi$ pulse)?
\end{enumerate}

%% file: chapters/ch10_entanglement_bell.tex
\chapter{Entanglement and Bell's Theorem}
\label{ch:entanglement_bell}

\section{Introduction}

When a qubit is in the state $\ket{+x}$ and we measure $\sigma_z$, we get $+1$ or $-1$ with equal probability. Why? One natural explanation is that the qubit actually \emph{has} a definite $z$-spin value; we simply don't know what it is. The state $\ket{+x}$ would then be an incomplete description, reflecting our ignorance rather than any genuine indeterminacy in nature.

This is not a naive idea. It is a precise physical hypothesis called \emph{realism}: that measurement reveals pre-existing values rather than creating them. Einstein held this view, and it captures a deep intuition about how the physical world works.

In this chapter, we will show that realism works perfectly well for a single qubit. We will construct an explicit model in which hidden Bloch cubes encode definite answers to multiple spin measurements simultaneously. The Bloch cube provides a natural taxonomy of measurement directions---face, edge, and corner---and hidden variable models can be built for each. The quantum mechanical probabilities emerge from our ignorance of which cubes the particle carries.

The model fails only when we consider \emph{entangled} systems and impose \emph{locality}, the requirement that a measurement on one particle cannot instantaneously affect a distant particle. Bell's theorem proves that no model can be simultaneously realistic and local while reproducing the predictions of quantum mechanics. The measurement directions that maximally violate Bell's inequality turn out to be faces and edges of the Bloch cube---a connection that makes the geometry of the violation visible. This is not a philosophical statement; it is a mathematical theorem with experimental consequences, and the experiments have spoken decisively.

\section{Hidden Variables for a Single Qubit}

\subsection{What Realism Requires}

Consider a spin-1/2 particle in state $\ket{+x}$. Quantum mechanics tells us nothing definite about the outcome of a $\sigma_z$ measurement, assigning equal probabilities to $\pm 1$. A realist interpretation says that this uncertainty reflects our ignorance: the particle possesses a definite value of $\sigma_z$, determined by some variable $\lambda$ that quantum mechanics fails to track.

More ambitiously, a realist would like every spin component to have a definite value simultaneously. The Bloch cube from Chapter~1 provides a natural language for making this idea precise.

\subsection{Hidden Bloch Cubes}
\label{sec:hidden_bloch_cubes}

Recall that the Bloch cube has six faces corresponding to the states $\ket{\pm x}$, $\ket{\pm y}$, and $\ket{\pm z}$. A quantum state like $\ket{+z}$ specifies a definite outcome for $\sigma_z$ measurements but leaves $\sigma_x$ and $\sigma_y$ undetermined. One Bloch cube, oriented with $\ket{+z}$ facing up, tells us the answer to the $Z$ question but says nothing definite about $X$ or $Y$.

Now consider your viewing angle. Looking straight down at the top face, you see $\ket{+z}$ and nothing else. The four side faces ($\ket{\pm x}$ and $\ket{\pm y}$) are hidden by perspective. But tilt your line of sight toward one of the four upper corners, and two side faces come into view. The corner where the $+x$, $+y$, and $+z$ faces meet reveals both $\ket{+x}$ and $\ket{+y}$. Rotate the cube $90^\circ$ about the vertical axis and you see the corner where $-x$, $+y$, and $+z$ meet, revealing $\ket{-x}$ and $\ket{+y}$. Another $90^\circ$ shows $\ket{-x}$ and $\ket{-y}$; another shows $\ket{+x}$ and $\ket{-y}$. The four corners of the top face correspond to exactly four combinations of hidden answers:
\begin{equation}
    (\ket{+x}, \ket{+y}), \quad (\ket{-x}, \ket{+y}), \quad (\ket{-x}, \ket{-y}), \quad (\ket{+x}, \ket{-y})
\end{equation}

A hidden variable theory says that the particle's complete state is specified not just by the visible top face but by the particular corner you would see if you tilted your view. Each corner encodes definite answers to the $X$ and $Y$ questions that quantum mechanics refuses to provide. The complete description of a single particle is therefore a triple:
\begin{equation}
    \bigl(\underbrace{\ket{+z}}_{\text{visible}}, \;\underbrace{\ket{\pm x}}_{\text{hidden}}, \;\underbrace{\ket{\pm y}}_{\text{hidden}}\bigr)
\end{equation}
When we measure $\sigma_z$, we read the visible face and get $+1$ with certainty. When we measure $\sigma_x$ or $\sigma_y$, we read the corresponding hidden answer encoded in the corner. There is no randomness in any individual measurement. Each outcome is predetermined. The apparent randomness of quantum mechanics arises because we cannot see which corner the cube is oriented toward.

\subsection{Reproducing Quantum Predictions}

For this picture to work, the hidden corners must be distributed in a way that reproduces the quantum probabilities. Consider an ensemble of particles all prepared in $\ket{+z}$. Every particle has the same visible face, but the corner orientation varies from particle to particle:

\begin{center}
\begin{tabular}{cc|c}
\textbf{Hidden $X$ answer} & \textbf{Hidden $Y$ answer} & \textbf{Probability} \\ \hline
$\ket{+x}$ & $\ket{+y}$ & $p_1$ \\
$\ket{+x}$ & $\ket{-y}$ & $p_2$ \\
$\ket{-x}$ & $\ket{+y}$ & $p_3$ \\
$\ket{-x}$ & $\ket{-y}$ & $p_4$
\end{tabular}
\end{center}

Quantum mechanics predicts that measuring $\sigma_x$ on $\ket{+z}$ gives $\pm 1$ with equal probability, and likewise for $\sigma_y$. This requires:
\begin{align}
    P(+x) &= p_1 + p_2 = \tfrac{1}{2} \\
    P(+y) &= p_1 + p_3 = \tfrac{1}{2}
\end{align}
Together with the normalization $p_1 + p_2 + p_3 + p_4 = 1$ and the requirement that all $p_i \geq 0$, these three equations in four unknowns appear to leave a one-parameter family of solutions. But a symmetry argument pins down the answer completely.

Notice what the marginal constraints already imply. From $p_1 + p_2 = \tfrac{1}{2}$ and $p_1 + p_3 = \tfrac{1}{2}$ we get $p_2 = p_3$. Substituting into the normalization gives $p_4 = p_1$. So opposite corners of the top face automatically carry equal probability: the diagonal pair $(\ket{+x}, \ket{+y})$ and $(\ket{-x}, \ket{-y})$ each have weight $p_1$, while $(\ket{+x}, \ket{-y})$ and $(\ket{-x}, \ket{+y})$ each have weight $\tfrac{1}{2} - p_1$. The two diagonal pairs sum to $2p_1$ and $1 - 2p_1$ respectively, and the sole remaining freedom is the value of $p_1$.

The symmetry of $\ket{+z}$ fixes it. The state $\ket{+z}$ is invariant under rotations about the $z$-axis. A $90^\circ$ rotation cycles the four top-face corners and interchanges the two diagonal pairs. Since the quantum state does not change under this rotation, the probability distribution over hidden corners must not change either: the two diagonal-pair sums must be equal. Setting $2p_1 = 1 - 2p_1$ gives $p_1 = \tfrac{1}{4}$, and therefore:
\begin{equation}
    p_1 = p_2 = p_3 = p_4 = \tfrac{1}{4}
    \label{eq:uniform_hv}
\end{equation}
The uniform distribution is not merely the simplest solution; it is the unique one consistent with the rotational symmetry of the quantum state. The two hidden answers are independent and uniformly random.

This model is satisfying, but limited: it only handles measurements along the three Bloch cube face directions $X$, $Y$, and $Z$. Spin can be measured along \emph{any} direction. Can the hidden-variable picture be extended to other axes?

\subsection{Edge States of the Bloch Cube}
\label{sec:edge_states}

The Bloch cube defines a natural hierarchy of measurement directions. The six \textbf{face centers} $\pm\hat{x}$, $\pm\hat{y}$, $\pm\hat{z}$ are the directions we have already used. The eight \textbf{body-diagonal} (corner) directions $(\pm\hat{x} \pm \hat{y} \pm \hat{z})/\sqrt{3}$ admit an analogous hidden-variable construction not pursued here. Between these lie the twelve \textbf{edge midpoints}: each edge of the cube connects two adjacent face centers, and its midpoint defines a spin measurement axis at $45^\circ$ from both faces.

In the $xz$-plane, the two edge directions are:
\begin{equation}
    \hat{e}_+ = \frac{\hat{z} + \hat{x}}{\sqrt{2}} \quad (45^\circ \text{ from } \hat{z}), \qquad
    \hat{e}_- = \frac{\hat{x} - \hat{z}}{\sqrt{2}} \quad (135^\circ \text{ from } \hat{z})
    \label{eq:edge_directions}
\end{equation}
The direction $\hat{e}_+$ points from the center of the cube toward the midpoint of the edge shared by the $+z$ and $+x$ faces. It lies exactly halfway between two face axes. The direction $\hat{e}_-$ points toward the midpoint of the edge shared by the $+x$ and $-z$ faces, diametrically opposite to $(\hat{z} - \hat{x})/\sqrt{2}$.

For a particle in the state $\ket{+z}$, quantum mechanics predicts (recall that the probability of obtaining $+1$ along $\hat{n}$ is $\cos^2(\theta/2)$, where $\theta$ is the angle between $\hat{n}$ and the prepared state):
\begin{align}
    P(+1 \,|\, \hat{e}_+) &= \cos^2(22.5^\circ) \approx 0.854 \label{eq:edge_prob_close} \\
    P(+1 \,|\, \hat{e}_-) &= \cos^2(67.5^\circ) \approx 0.146 \label{eq:edge_prob_far}
\end{align}
The edge direction $\hat{e}_+$ is close to $\hat{z}$ (only $45^\circ$ away), so the particle is very likely to be found spin-up along it. The edge direction $\hat{e}_-$ is far from $\hat{z}$ ($135^\circ$), so spin-up is unlikely. These probabilities---$\cos^2(22.5^\circ) \approx 85\%$ and its complement $\sin^2(22.5^\circ) \approx 15\%$---will reappear when we test Bell's inequality.

Now try to extend the face model to handle edge measurements.  The face model tells us the outcome of a $\sigma_x$ measurement is simply $\epsilon_x$, the hidden face value carried by the particle.  But notice that this is the same as computing $\text{sign}(\hat{x} \cdot \vec{\lambda})$, where $\vec{\lambda} = (\epsilon_x, \epsilon_y, \epsilon_z)$ is the hidden triple viewed as a vector.  Indeed, $\hat{x} \cdot \vec{\lambda} = \epsilon_x = \pm 1$, whose sign is just $\epsilon_x$ itself.  The same holds for the other two face directions: $\text{sign}(\hat{y} \cdot \vec{\lambda}) = \epsilon_y$ and $\text{sign}(\hat{z} \cdot \vec{\lambda}) = \epsilon_z$.  For the three face directions, ``read the hidden face value'' and ``compute the sign of the dot product with the hidden triple'' are the same instruction.

This reformulation suggests a natural way to extend the model beyond face directions.  For an arbitrary measurement direction $\hat{n}$, define the outcome function:
\begin{equation}
    A(\hat{n}) = \text{sign}(\hat{n} \cdot \vec{\lambda})
    \label{eq:sign_rule}
\end{equation}
This rule agrees with the face model whenever $\hat{n}$ is a face direction, and it provides an answer---the sign of $\hat{n} \cdot \vec{\lambda}$---for any other direction as well.  It is the simplest generalization of the face-reading rule to the full unit sphere: project the hidden triple onto the measurement axis, and report the sign.

Applying Eq.~\eqref{eq:sign_rule} to the edge direction $\hat{e}_+$ gives:
\begin{equation}
    A(\hat{e}_+) = \text{sign}\!\left(\frac{\epsilon_x + \epsilon_z}{\sqrt{2}}\right) = \text{sign}(\epsilon_x + \epsilon_z)
\end{equation}
When $\epsilon_x = \epsilon_z$ (both $+1$ or both $-1$), the sign is definite: $+1$ or $-1$ respectively. But when $\epsilon_x = -\epsilon_z$, the argument is zero and the sign is undefined. The edge direction $\hat{e}_+$ lies exactly on the boundary between the hemispheres defined by these hidden triples. For half of the four relevant face configurations, the model has no answer.

This is not a problem that can be patched by adopting a convention for $\text{sign}(0)$. The deeper issue is geometric: the face model partitions the unit sphere using planes perpendicular to $\hat{x}$, $\hat{y}$, and $\hat{z}$, and the edge directions sit exactly on these partition boundaries. The discrete model is too coarse to resolve directions halfway between its grid points.

The model that resolves this limitation---Bell's continuous construction---handles all directions, including edges, faces, corners, and everything in between. But the fact that the face model breaks down \emph{precisely at the edge directions} will prove deeply significant: these are exactly the directions where Bell's inequality is maximally violated.

\subsection{Bell's Continuous Model}
\label{sec:bells_continuous_model}

\subsubsection*{From faces to edges to all directions}

The face model assigned a definite $\pm 1$ answer to each of three face directions. We have just seen that it cannot cleanly handle the twelve edge directions. A more ambitious realist would assign a definite $\pm 1$ answer to \emph{every} direction on the unit sphere simultaneously---faces, edges, corners, and the continuum of directions between them. Mathematically, such an assignment is a function:
\begin{equation}
    f: S^2 \to \{+1, -1\}
\end{equation}
that maps each measurement direction $\hat{n}$ to a definite outcome $f(\hat{n}) = \pm 1$. The face model is just $f$ sampled at three points: $f(\hat{x}), f(\hat{y}), f(\hat{z})$. In the continuous limit, $f$ is defined everywhere.

The complete hidden-variable description of a particle in state $\ket{+z}$ is the pair $(\ket{+z}, f)$, where the quantum state tells us $f(\hat{z}) = +1$ with certainty and the function $f$ fills in the rest. A particle carrying a specific $f$ has no randomness at all: every spin question has a predetermined answer, just as a particle carrying a specific set of hidden face values had definite answers to all three face questions. The apparent randomness of quantum mechanics arises from not knowing which $f$ the particle carries.

The ensemble is described by a probability distribution $\rho[f]$ over such functions, constrained so that the average of $f(\hat{n})$ over the ensemble reproduces the quantum prediction for every direction:
\begin{equation}
    P(+1 | \hat{n}) = \sum_{f:\, f(\hat{n}) = +1} \rho[f] = \frac{1 + \hat{n} \cdot \hat{s}}{2}
    \label{eq:general_hv_constraint}
\end{equation}
Here $\hat{s}$ is the Bloch vector of the prepared quantum state (e.g., $\hat{s} = \hat{z}$ for $\ket{+z}$), $\hat{n}$ is the measurement direction---the question we choose to ask---and $f(\hat{n}) = \pm 1$ is the predetermined answer that the hidden function $f$ assigns to that question. The sum runs over all functions $f$ in the ensemble that happen to answer $+1$ for the direction $\hat{n}$, weighted by their probabilities $\rho[f]$. This is a weaker requirement than it might seem. Individual functions can have answer patterns that look nothing like the quantum probability distribution. A particular $f$ might assign $+1$ to two nearly parallel directions and $-1$ to a direction between them. That is perfectly acceptable, because the model only needs to get the statistics right on average, not direction by direction for each particle. Just as in the face model, where one particular hidden triple gave $\sigma_x = +1$ and $\sigma_y = +1$ even though the quantum state $\ket{+z}$ has no preference between these, each $f$ carries its own idiosyncratic assignment; the quantum probabilities emerge only from averaging over the ensemble.

There is a subtlety here that strengthens the realist position considerably. Many different distributions $\rho[f]$ over many different kinds of functions can satisfy Eq.~\eqref{eq:general_hv_constraint}. The situation is exactly analogous to the completely mixed state $\rho = I/2$ of a single qubit. That density matrix can be decomposed as an equal mixture of $\ket{+z}$ and $\ket{-z}$, or equally well as a mixture of $\ket{+x}$ and $\ket{-x}$, or as a uniform distribution over all pure states on the Bloch sphere. No experiment can distinguish these decompositions; the density matrix captures everything observable, and the choice of decomposition is physically meaningless. In the same way, no single-qubit experiment can distinguish between different hidden variable models that all reproduce the same quantum statistics. The realist posits that each particle carries a definite $f$, but the specific $f$ is forever inaccessible---only the ensemble average $\rho[f]$ has observable consequences. This means that single-qubit realism is not merely compatible with quantum mechanics; it is \emph{unfalsifiable} by single-qubit experiments. We can neither confirm nor deny the existence of the hidden variables. The realist can always claim they are there, and no measurement on a single particle can prove otherwise.

\subsubsection*{Bell's sign construction}

The space of all functions $f: S^2 \to \{+1, -1\}$ is enormous. Bell's insight was to find a specific, geometrically natural family of such functions that suffices to reproduce all quantum predictions.

The face model motivates the construction. In Section~\ref{sec:edge_states}, we showed that the face model's measurement rule can be rewritten as $A(\hat{n}) = \text{sign}(\hat{n} \cdot \vec{\lambda})$, where $\vec{\lambda} = (\epsilon_x, \epsilon_y, \epsilon_z)$ is the hidden triple viewed as a vector. That vector takes only eight discrete values, one for each combination of face signs. Bell's insight was to promote this to a continuous construction: let the hidden variable be a unit vector $\hat{\lambda}$ that can point anywhere on the sphere, and define:
\begin{equation}
    f_{\hat{\lambda}}(\hat{n}) = \text{sign}(\hat{n} \cdot \hat{\lambda})
\end{equation}
This single vector $\hat{\lambda}$ compresses the entire function $f$ into one object. It divides the unit sphere into two hemispheres: every direction $\hat{n}$ within $90^\circ$ of $\hat{\lambda}$ gets $f = +1$, and every direction on the opposite side gets $f = -1$. One vector, one clean geometric rule, and the answer to every conceivable measurement is determined. Crucially, the hemisphere boundary is now a great circle that can be oriented in any direction, so there is no preferred grid of directions along which the model might fail. The edge-direction ambiguity of the face model---where $\text{sign}(0)$ was undefined---is a set of measure zero in the continuous model: for almost every $\hat{\lambda}$, the dot product $\hat{e}_+ \cdot \hat{\lambda}$ is nonzero.

The quantum probabilities emerge from averaging over the ensemble, exactly as in the face model. There, each particle carried a definite triple $(\epsilon_x, \epsilon_y, \epsilon_z)$, and the probability $P(+x) = 1/2$ arose from the uniform distribution over the four possible triples. In the continuous model, each particle carries a definite $\hat{\lambda}$, and the ensemble is described by a probability density $\rho(\hat{\lambda})$ over the unit sphere. The probability of obtaining $+1$ along direction $\hat{n}$ is the integral of $\rho(\hat{\lambda})$ over all $\hat{\lambda}$ whose hemispheres include $\hat{n}$:
\begin{equation}
    P(+1 | \hat{n}) = \int_{\hat{n} \cdot \hat{\lambda} > 0} \rho(\hat{\lambda})\, d\Omega
\end{equation}
In 1966, John Bell showed that for any quantum state with Bloch vector $\hat{s}$, there exists a density $\rho(\hat{\lambda})$ such that this integral reproduces
\begin{equation}
    P(+1 | \hat{n}) = \frac{1 + \hat{n} \cdot \hat{s}}{2} = \cos^2\frac{\theta}{2}
\end{equation}
exactly, for every measurement direction $\hat{n}$---faces, edges, corners, and all directions in between. The Bloch cube face model is a discrete special case of this construction, restricted to three measurement axes.

\subsection{The Lesson}

The hidden variable program for a single qubit is in an extraordinarily strong position. Not only can we construct explicit models that reproduce all quantum predictions---from the discrete face model to Bell's continuous construction---but the realist hypothesis is \emph{unfalsifiable} by any single-particle experiment. Many different hidden variable models, with different $f$'s and different $\rho[f]$'s, all produce the same quantum statistics. No measurement can reveal which model is correct, or even whether any such model is ``really'' at work behind the scenes. The realist can always maintain that particles carry hidden variables, and no single-qubit experiment can prove otherwise.

This seems like a decisive victory for realism. If one had only single-particle quantum mechanics, one might reasonably conclude that quantum mechanics is simply an incomplete statistical theory---useful for making predictions, but missing a deeper deterministic layer in which every question has a definite answer. This is precisely what Einstein, Podolsky, and Rosen argued in 1935.

The story does not end here. When we turn to entangled systems and demand locality, the hidden variable program collapses---and it collapses completely. Bell's theorem does not merely rule out one particular model, or one particular choice of $f$, or one particular $\rho[f]$. It rules out \emph{every} local hidden variable model simultaneously, regardless of the structure of $\lambda$, regardless of how the hidden variables are constructed, regardless of the distribution over the ensemble. That is what makes Bell's result so remarkable.

\section{Bipartite Entanglement}

\subsection{Product States vs.\ Entangled States}

For a two-qubit system in $\hilbert_A \otimes \hilbert_B$, a product state has the form:
\begin{equation}
    \ket{\psi} = \ket{\phi}_A \otimes \ket{\chi}_B
\end{equation}
In a product state, each subsystem has a definite quantum state of its own. Measuring subsystem $A$ tells us nothing about subsystem $B$ that we didn't already know.

An entangled state cannot be written in product form. The simplest example is:
\begin{equation}
    \ket{\Phi^+} = \frac{1}{\sqrt{2}}(\ket{00} + \ket{11})
\end{equation}
Here, neither qubit has a definite state individually. The state of the whole is more than the sum of its parts.

\subsection{The Four Bell States}

The maximally entangled Bell basis consists of four orthonormal states:
\begin{align}
    \ket{\Phi^+} &= \frac{1}{\sqrt{2}}(\ket{00} + \ket{11}) \\
    \ket{\Phi^-} &= \frac{1}{\sqrt{2}}(\ket{00} - \ket{11}) \\
    \ket{\Psi^+} &= \frac{1}{\sqrt{2}}(\ket{01} + \ket{10}) \\
    \ket{\Psi^-} &= \frac{1}{\sqrt{2}}(\ket{01} - \ket{10})
\end{align}
These form a complete orthonormal basis for the two-qubit Hilbert space. We encountered these states in Chapter 2; they will now play a central role in understanding the foundations of quantum mechanics.

\begin{physicalinsight}
Consider the state $\ket{\Phi^+}$. If Alice measures her qubit in the computational basis and obtains $\ket{0}$, the state collapses to $\ket{00}$, so Bob's qubit is certainly in state $\ket{0}$. If Alice obtains $\ket{1}$, Bob's qubit is certainly in state $\ket{1}$. The outcomes are perfectly correlated, even if Alice and Bob are light-years apart. 

This correlation is not, by itself, mysterious. One could imagine the qubits carried pre-determined values, each qubit with its own hidden Bloch cube packed at the source, like a pair of gloves shipped to different cities. Bell's theorem shows that no such local explanation can account for all quantum predictions.
\end{physicalinsight}

\section{The EPR Paradox and Local Realism}

\subsection{The EPR Argument}

Having seen that realism works perfectly for a single qubit, Einstein, Podolsky, and Rosen (EPR) argued in 1935 that it should work for entangled systems too. Their reasoning was as follows.

Consider two particles prepared in the singlet state:
\begin{equation}
    \ket{\Psi^-} = \frac{1}{\sqrt{2}}(\ket{01} - \ket{10})
\end{equation}
where $\ket{0}$ and $\ket{1}$ represent spin-up and spin-down along the $z$-axis. Alice and Bob each take one particle and travel to distant locations.

If Alice measures spin along the $z$-axis, she can predict with certainty what Bob will obtain if he also measures along $z$: the outcomes are perfectly anti-correlated. But Alice could equally well measure along the $x$-axis, and then she could predict Bob's $x$-spin with certainty.

EPR introduced a criterion of physical reality: ``If, without in any way disturbing a system, we can predict with certainty the value of a physical quantity, then there exists an element of physical reality corresponding to this physical quantity.''

Since Alice's measurement cannot instantaneously affect Bob's distant particle (by the principle of locality), and since Alice can predict either Bob's $z$-spin or $x$-spin with certainty depending on her choice, EPR concluded that both spin components must have definite values prior to measurement. But quantum mechanics does not assign definite values to non-commuting observables like $S_x$ and $S_z$. Therefore, EPR argued, quantum mechanics must be an incomplete description of physical reality.

The argument is compelling. We have just seen that a single qubit \emph{can} carry definite values for all spin components, encoded in a hidden Bloch cube, with quantum mechanics emerging as a statistical description of which hidden values appear in the ensemble. EPR were simply extending this to the entangled case: each particle would carry its own set of hidden cubes, with the additional requirement of locality ensuring that Alice's measurement choice does not disturb Bob's.

The face-cube model makes this extension concrete. Imagine the source prepares a pair of Bloch cubes, one for Alice and one for Bob, with every face anti-correlated:
\begin{center}
\begin{tabular}{l|cc}
& \textbf{Alice's cube} & \textbf{Bob's cube} \\ \hline
$Z$ (visible) & $\ket{-z}$ & $\ket{+z}$ \\
$X$ (hidden) & $\ket{+x}$ & $\ket{-x}$ \\
$Y$ (hidden) & $\ket{-y}$ & $\ket{+y}$
\end{tabular}
\end{center}
Both cubes sit with their $Z$ faces on top. Looking straight down, you see only the $Z$ answers: Alice has spin-down, Bob has spin-up. The $X$ and $Y$ answers are on the side faces---present but hidden by perspective. If Alice measures $\sigma_z$, she reads her top face and gets $-1$; if she measures $\sigma_x$ instead, she reads the hidden side face and gets $+1$. Either way, the answer was there before she looked. Bob's cube, being anti-correlated face by face, guarantees that whenever they measure along the same axis, they get opposite results. The source packs each pair with one of the four possible anti-correlated face-value assignments, each with probability $1/4$, and the singlet-state statistics emerge from not knowing which pair was packed.

\subsection{Local Realism}

The EPR argument rests on two assumptions that together constitute \textit{local realism}:

\textbf{Realism:} Physical properties have definite values that exist independent of observation. When we measure an observable, we are revealing a pre-existing value, not creating one. As we showed in Section~\ref{sec:hidden_bloch_cubes}, the hidden-cube model makes this assumption concrete and reproduces all single-qubit quantum predictions.

\textbf{Locality:} An action performed at one location cannot instantaneously influence physical reality at a distant location. Influences propagate at most at the speed of light.

These assumptions seem eminently reasonable. Realism captures the intuition that the moon exists whether or not anyone is looking at it, and we have demonstrated that it is entirely compatible with single-particle quantum mechanics. Locality is built into the structure of special relativity. For nearly thirty years after EPR, most physicists assumed that quantum mechanics could eventually be supplemented by hidden variables that would restore a local realistic description.

\subsection{Local Hidden Variable Theories}

A local hidden variable theory for two particles extends the single-qubit model in a natural way. The source prepares \emph{two} Bloch cubes, one for Alice's particle and one for Bob's, each with a definite set of hidden face values $\vec{\lambda} = (\epsilon_x, \epsilon_y, \epsilon_z)$, and sends them off in opposite directions. When Alice measures spin along $\vec{a}$, she evaluates $A(\vec{a}) = \text{sign}(\vec{a} \cdot \vec{\lambda}_A)$. Bob does the same with his hidden triple: $B(\vec{b}) = \text{sign}(\vec{b} \cdot \vec{\lambda}_B)$.

The crucial constraint is \emph{locality}: Alice's cube determines her outcome for \emph{any} measurement she might choose, independent of what Bob does with his. Alice's setting $\vec{a}$ does not appear in Bob's outcome function, and vice versa. Both cubes can be correlated (the source can pack them according to any joint distribution $\rho(\vec{\lambda}_A, \vec{\lambda}_B)$), but once the particles separate, each outcome depends only on the local measurement setting and the local hidden triple.

This is a straightforward generalization of the model that worked so well for a single qubit. The question is whether such a model can reproduce the quantum mechanical predictions for entangled states.

\section{Bell's Theorem}

In 1964, John Bell proved a remarkable theorem: no local hidden variable theory can reproduce all the predictions of quantum mechanics. The hidden variable approach that worked flawlessly for a single qubit fails irrecoverably when applied to entangled pairs with a locality constraint.

\subsection{The Scenario}

Consider the following experimental setup. A source prepares pairs of spin-1/2 particles in the singlet state $\ket{\Psi^-}$ and sends one to Alice and one to Bob. Alice chooses to measure spin along one of two directions, $\vec{a}$ or $\vec{a}'$. Bob independently chooses to measure along $\vec{b}$ or $\vec{b}'$. Each measurement yields $+1$ or $-1$.

We define the correlation function:
\begin{equation}
    E(\vec{a}, \vec{b}) = \langle A(\vec{a}) \cdot B(\vec{b}) \rangle
\end{equation}
This is the average value of the product of outcomes when Alice measures along $\vec{a}$ and Bob measures along $\vec{b}$.

\subsection{The Continuous Hidden Variable Model}
\label{sec:continuous_two_particle}

Recall that Bell's single-qubit model (Section~\ref{sec:bells_continuous_model}) used a continuous hidden variable $\hat{\lambda}$, a unit vector distributed over the sphere, with outcome rule $A(\hat{n}) = \text{sign}(\hat{n} \cdot \hat{\lambda})$. Bell showed that a suitable probability density $\rho(\hat{\lambda})$ reproduces all single-qubit quantum predictions exactly.

For two entangled particles, we try the natural extension: give each pair a shared hidden variable $\hat{\lambda}$, and impose anti-correlation by taking $A(\hat{a}) = \text{sign}(\hat{a} \cdot \hat{\lambda})$ and $B(\hat{b}) = -\text{sign}(\hat{b} \cdot \hat{\lambda})$. For the singlet state, the simplest choice is a uniform distribution $\rho(\hat{\lambda}) = 1/4\pi$.

The correlation function is:
\begin{equation}
    E(\hat{a}, \hat{b}) = -\int \frac{d\Omega}{4\pi}\, \text{sign}(\hat{a} \cdot \hat{\lambda})\, \text{sign}(\hat{b} \cdot \hat{\lambda})
    \label{eq:continuous_correlation}
\end{equation}
This integral has an elegant geometric interpretation. The planes perpendicular to $\hat{a}$ and $\hat{b}$ divide the unit sphere into four regions. In two of them, the signs of $\hat{a} \cdot \hat{\lambda}$ and $\hat{b} \cdot \hat{\lambda}$ agree; in the other two, they disagree. The fraction of the sphere where they \emph{disagree} is $\theta/\pi$, where $\theta$ is the angle between $\hat{a}$ and $\hat{b}$. The fraction where they agree is $(\pi - \theta)/\pi$. Therefore:
\begin{equation}
    E(\theta) = -\left[\frac{\pi - \theta}{\pi} - \frac{\theta}{\pi}\right] = -1 + \frac{2\theta}{\pi}
    \label{eq:triangle_correlation}
\end{equation}
This is a straight line from $E = -1$ at $\theta = 0$ to $E = +1$ at $\theta = \pi$.

Compare the two results:
\begin{center}
\begin{tabular}{l|ccc}
& $\theta = 0$ & $\theta = \pi/2$ & $\theta = \pi$ \\ \hline
Local hidden variable model (straight line) & $-1$ & $0$ & $+1$ \\
Quantum mechanics ($-\cos\theta$) & $-1$ & $0$ & $+1$
\end{tabular}
\end{center}

\begin{figure}[ht]
\centering
\begin{tikzpicture}
\begin{axis}[
    width=12cm, height=8cm,
    xlabel={Angle $\theta$ between measurement directions},
    ylabel={Correlation $E(\theta)$},
    xmin=0, xmax=180,
    ymin=-1.15, ymax=1.15,
    xtick={0,22.5,45,67.5,90,112.5,135,157.5,180},
    xticklabels={$0$,$22.5^\circ$,$45^\circ$,$67.5^\circ$,$90^\circ$,$112.5^\circ$,$135^\circ$,$157.5^\circ$,$180^\circ$},
    x tick label style={rotate=45, anchor=east, font=\small},
    ytick={-1,-0.707,0,0.707,1},
    yticklabels={$-1$,$-1/\sqrt{2}$,$0$,$1/\sqrt{2}$,$1$},
    grid=major,
    grid style={gray!30},
    legend pos=south east,
    legend style={font=\small, fill=white, fill opacity=0.9, draw=gray!50},
    every axis plot/.append style={thick},
    clip=false,
]
\addplot[blue!80!black, very thick, domain=0:180, samples=200] {-cos(x)};
\addlegendentry{Quantum: $-\cos\theta$}

\addplot[red!70!black, thick, dashed, domain=0:180] {-1 + 2*x/180};
\addlegendentry{Local HV: $-1+2\theta/\pi$}

\draw[blue!40, thick, decorate, decoration={brace, amplitude=5pt, mirror}]
    (axis cs:0,-1.08) -- (axis cs:30,-1.08)
    node[midway, below=6pt, font=\footnotesize, blue!60!black] {stickier};
\draw[blue!40, thick, decorate, decoration={brace, amplitude=5pt}]
    (axis cs:150,1.08) -- (axis cs:180,1.08)
    node[midway, above=6pt, font=\footnotesize, blue!60!black] {stickier};

\end{axis}
\end{tikzpicture}
\caption{Comparison of the correlation function $E(\theta)$ for the local hidden variable model and quantum mechanics. The quantum prediction $-\cos\theta$ (solid blue) is \emph{stickier} near perfect anti-correlation ($\theta=0$) and perfect correlation ($\theta=180^\circ$) than the local hidden variable model (dashed red), which departs linearly. Both agree at $\theta = 0^\circ$, $90^\circ$, and $180^\circ$.}
\label{fig:correlation_comparison}
\end{figure}
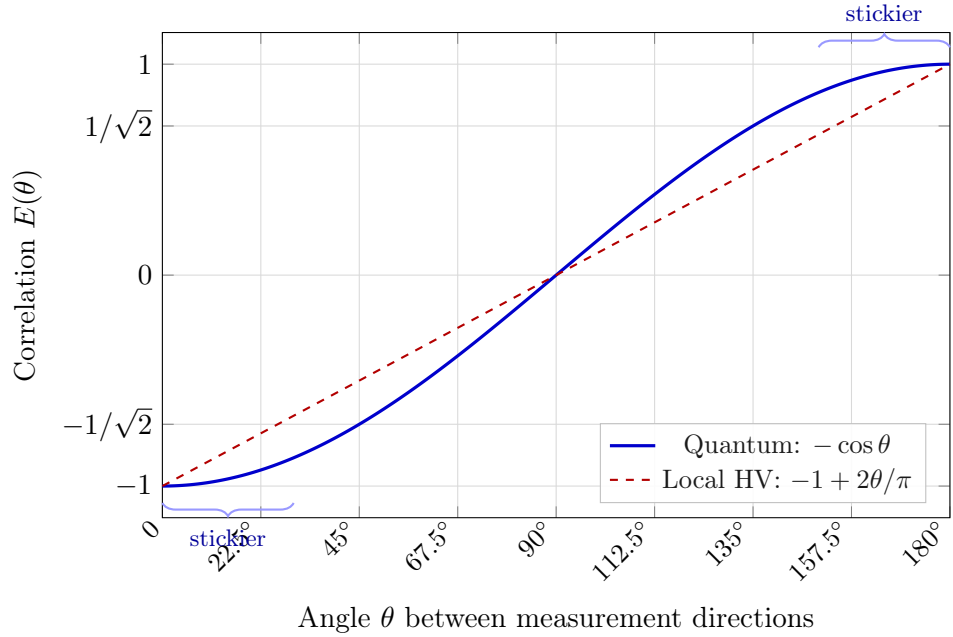

Both agree at $\theta = 0$, $\pi/2$, and $\pi$, as shown in Fig.~\ref{fig:correlation_comparison}. But the discrepancy between them is physically significant.

The crucial difference appears near $\theta = 0$. The quantum correlation $-\cos\theta \approx -1 + \theta^2/2$ is flat at perfect anti-correlation; the curve barely moves away from $-1$ for small $\theta$. The local model $-1 + 2\theta/\pi$ departs linearly, with a constant slope $2/\pi \approx 0.64$. Quantum correlations are \emph{stickier} near perfect correlation than any local model can produce. This is Bell's original observation from his 1964 paper: the quantum cosine clings to $-1$ at small angles more tightly than the linear function, and clings to $+1$ at large angles more tightly than the linear function. The straight line is the best a local model with a uniform distribution can do, and it is not good enough.

Recall that for a single qubit, Bell's continuous model reproduced all quantum predictions exactly. For two entangled qubits, the same construction fails. The problem is the locality constraint, which forces the correlation function to be a weighted average of sign products---and no such average can produce the curvature of $-\cos\theta$.

\subsection{The CHSH Inequality}

The continuous model fails to reproduce the quantum correlation for entangled pairs. But could some cleverer local model succeed, perhaps with a non-uniform distribution or a more complicated hidden variable space? The CHSH inequality shows that the answer is no: \emph{every} local hidden variable theory, regardless of its internal structure, is bounded by the same constraint.

The generality of the argument rests on exactly two properties that any local hidden variable theory must have:
\begin{enumerate}
\item \textbf{Bounded outcomes.} Each measurement yields $A = \pm 1$ or $B = \pm 1$. This is not an assumption about hidden variables; it is a fact about spin-1/2: the eigenvalues of any spin component are $\pm\hbar/2$, and we label these $\pm 1$.
\item \textbf{Locality.} Alice's outcome $A(\vec{a}, \lambda)$ depends on her measurement setting $\vec{a}$ and the shared hidden variable $\lambda$, but \emph{not} on Bob's setting $\vec{b}$. Likewise, $B(\vec{b}, \lambda)$ does not depend on $\vec{a}$.
\end{enumerate}
Everything else is unrestricted. The hidden variable $\lambda$ can be anything: a discrete triple, a point on the unit sphere, the complete works of Shakespeare, or an arbitrarily complicated instruction set. The functions $A(\vec{a}, \lambda)$ and $B(\vec{b}, \lambda)$ can depend on their arguments in any way they like, as long as each output is $\pm 1$. The distribution $\rho(\lambda)$ can be anything normalizable. The CHSH argument treats $A$ and $B$ as black boxes and never looks inside.

For each particle pair carrying a particular $\lambda$, Alice could measure along $\vec{a}$ or $\vec{a}'$, and Bob could measure along $\vec{b}$ or $\vec{b}'$. Consider the combination:
\begin{equation}
    S(\lambda) = A(\vec{a}, \lambda) \bigl[B(\vec{b}, \lambda) - B(\vec{b}', \lambda)\bigr] + A(\vec{a}', \lambda) \bigl[B(\vec{b}, \lambda) + B(\vec{b}', \lambda)\bigr]
\end{equation}
This is where boundedness does its work. Since $B(\vec{b}, \lambda)$ and $B(\vec{b}', \lambda)$ are each exactly $\pm 1$, they either agree or disagree. If they agree, $B(\vec{b}) - B(\vec{b}') = 0$ and $B(\vec{b}) + B(\vec{b}') = \pm 2$. If they disagree, $B(\vec{b}) - B(\vec{b}') = \pm 2$ and $B(\vec{b}) + B(\vec{b}') = 0$. In either case, one bracket vanishes and the other has magnitude~2. The surviving bracket is multiplied by an $A$ value of magnitude~1, so $|S(\lambda)| = 2$ for every single $\lambda$.

Table~\ref{tab:chsh_combinations} displays all sixteen combinations explicitly. Every row gives $|S| = 2$, confirming the per-$\lambda$ bound by exhaustion.

\begin{table}[ht]
\centering
\caption{All sixteen combinations of the four $\pm 1$ outcomes and the resulting CHSH value $S(\lambda)$. In each row, exactly one of the two brackets vanishes while the other has magnitude~2, giving $|S| = 2$ regardless of the combination. The ``vanishes/survives'' columns show which bracket is zero and which is $\pm 2$.}
\label{tab:chsh_combinations}
\renewcommand{\arraystretch}{1.15}
\begin{tabular}{cccc|rc|rc|c}
\toprule
$A$ & $A'$ & $B$ & $B'$ & $B{-}B'$ & & $B{+}B'$ & & $S$ \\
\midrule
$+1$ & $+1$ & $+1$ & $+1$ & $0$ & & $+2$ & $\leftarrow$ & $+2$ \\
$+1$ & $+1$ & $+1$ & $-1$ & $+2$ & $\leftarrow$ & $0$ & & $+2$ \\
$+1$ & $+1$ & $-1$ & $+1$ & $-2$ & $\leftarrow$ & $0$ & & $-2$ \\
$+1$ & $+1$ & $-1$ & $-1$ & $0$ & & $-2$ & $\leftarrow$ & $-2$ \\
\addlinespace
$+1$ & $-1$ & $+1$ & $+1$ & $0$ & & $+2$ & $\leftarrow$ & $-2$ \\
$+1$ & $-1$ & $+1$ & $-1$ & $+2$ & $\leftarrow$ & $0$ & & $+2$ \\
$+1$ & $-1$ & $-1$ & $+1$ & $-2$ & $\leftarrow$ & $0$ & & $-2$ \\
$+1$ & $-1$ & $-1$ & $-1$ & $0$ & & $-2$ & $\leftarrow$ & $+2$ \\
\addlinespace
$-1$ & $+1$ & $+1$ & $+1$ & $0$ & & $+2$ & $\leftarrow$ & $+2$ \\
$-1$ & $+1$ & $+1$ & $-1$ & $+2$ & $\leftarrow$ & $0$ & & $-2$ \\
$-1$ & $+1$ & $-1$ & $+1$ & $-2$ & $\leftarrow$ & $0$ & & $+2$ \\
$-1$ & $+1$ & $-1$ & $-1$ & $0$ & & $-2$ & $\leftarrow$ & $-2$ \\
\addlinespace
$-1$ & $-1$ & $+1$ & $+1$ & $0$ & & $+2$ & $\leftarrow$ & $-2$ \\
$-1$ & $-1$ & $+1$ & $-1$ & $+2$ & $\leftarrow$ & $0$ & & $-2$ \\
$-1$ & $-1$ & $-1$ & $+1$ & $-2$ & $\leftarrow$ & $0$ & & $+2$ \\
$-1$ & $-1$ & $-1$ & $-1$ & $0$ & & $-2$ & $\leftarrow$ & $+2$ \\
\bottomrule
\end{tabular}
\end{table}

This is the critical step. For any particular $\lambda$, the quantity $S$ is exactly $\pm 2$---not approximately, not on average, but for every individual instruction set the source could produce. No cleverness in choosing $\lambda$ or designing the functions $A$ and $B$ can make $|S(\lambda)|$ exceed~2, because the bound follows from nothing more than $\pm 1$ arithmetic.

Since $|S(\lambda)| \leq 2$ holds for every $\lambda$ individually, averaging over any distribution $\rho(\lambda)$ preserves the bound:
\begin{equation}
    |\langle S \rangle| = \biggl| \int \rho(\lambda)\, S(\lambda)\, d\lambda \biggr| \leq \int \rho(\lambda)\, |S(\lambda)|\, d\lambda \leq 2
\end{equation}
Writing this in terms of correlation functions:
\begin{equation}
    \bigl| E(\vec{a}, \vec{b}) - E(\vec{a}, \vec{b}') + E(\vec{a}', \vec{b}) + E(\vec{a}', \vec{b}') \bigr| \leq 2
\end{equation}
This is the \textbf{CHSH inequality}, named after Clauser, Horne, Shimony, and Holt. Any local hidden variable theory must satisfy it, regardless of the structure of $\lambda$, regardless of the form of $A$ and $B$, and regardless of the distribution $\rho(\lambda)$. The entire argument used only two facts: that outcomes are $\pm 1$, and that Alice's outcome does not depend on Bob's setting. The sign construction, the hemisphere picture, the uniform distribution---none of these played any role. They were useful for building intuition about specific models, but the inequality transcends every particular model.

\subsection{Quantum Violation: Faces and Edges}
\label{sec:chsh_violation}

For the singlet state $\ket{\Psi^-}$, quantum mechanics predicts:
\begin{equation}
    E(\vec{a}, \vec{b}) = -\vec{a} \cdot \vec{b} = -\cos\theta_{ab}
\end{equation}
where $\theta_{ab}$ is the angle between directions $\vec{a}$ and $\vec{b}$.

\begin{example}{Derivation of the Quantum Correlation}
Let Alice measure spin along $\vec{a} = \hat{z}$ and Bob measure along a direction $\vec{b}$ that makes angle $\theta$ with $\hat{z}$. 

In the singlet state, if Alice obtains $+1$ (spin up along $\hat{z}$), then Bob's particle is in state $\ket{1}$ (spin down along $\hat{z}$). Bob's measurement along $\vec{b}$ then yields $+1$ with probability $\sin^2(\theta/2)$ and $-1$ with probability $\cos^2(\theta/2)$, giving expected value $\sin^2(\theta/2) - \cos^2(\theta/2) = -\cos\theta$.

Similarly, if Alice obtains $-1$, Bob's expected value is $+\cos\theta$.

Since Alice gets $\pm 1$ with equal probability:
\begin{equation}
    E(\vec{a}, \vec{b}) = \frac{1}{2}(+1)(-\cos\theta) + \frac{1}{2}(-1)(+\cos\theta) = -\cos\theta
\end{equation}
\end{example}

The measurement directions that maximally violate the CHSH inequality have a natural interpretation in the Bloch cube taxonomy introduced in Section~\ref{sec:edge_states}. Alice measures along two \textbf{face} directions:
\begin{equation}
    \hat{a} = \hat{z} \quad (0^\circ), \qquad \hat{a}' = \hat{x} \quad (90^\circ)
\end{equation}
Bob measures along two \textbf{edge} directions:
\begin{equation}
    \hat{b} = \frac{\hat{z} + \hat{x}}{\sqrt{2}} \quad (45^\circ), \qquad \hat{b}' = \frac{\hat{x} - \hat{z}}{\sqrt{2}} \quad (135^\circ)
\end{equation}
In the $xz$-plane, these four directions alternate face--edge--face--edge at $0^\circ$, $45^\circ$, $90^\circ$, and $135^\circ$ from the $z$-axis. Bob's edge directions sit at the midpoints of the edges connecting Alice's two face directions on the Bloch cube.

\begin{figure}[ht]
\centering
\begin{tikzpicture}[scale=2.8]
\draw[gray!40, thin] (0,0) circle (1);

\draw[gray!50, ->] (-1.25,0) -- (1.35,0) node[right, font=\small] {$\hat{x}$};
\draw[gray!50, ->] (0,-0.3) -- (0,1.35) node[above, font=\small] {$\hat{z}$};

\draw[-{Stealth[length=3mm]}, blue!80!black, very thick] (0,0) -- (0,1)
    node[above right, font=\small\bfseries, blue!80!black] {$\hat{a} = \hat{z}\;(0^\circ)$};
\draw[-{Stealth[length=3mm]}, blue!80!black, very thick] (0,0) -- (1,0)
    node[right, font=\small\bfseries, blue!80!black, yshift=2pt] {$\hat{a}' = \hat{x}\;(90^\circ)$};

\draw[-{Stealth[length=3mm]}, red!70!black, very thick, dashed] (0,0) -- ({cos(45)},{sin(45)})
    node[above right, font=\small\bfseries, red!70!black] {$\hat{b}\;(45^\circ)$};
\draw[-{Stealth[length=3mm]}, red!70!black, very thick, dashed] (0,0) -- ({cos(135)},{sin(135)})
    node[above left, font=\small\bfseries, red!70!black] {$\hat{b}'\;(135^\circ)$};

\draw[green!50!black, thick, <->] ({0.55*cos(0)},{0.55*sin(0)}) arc (0:45:0.55);
\node[font=\scriptsize, green!50!black] at ({0.67*cos(22.5)},{0.67*sin(22.5)}) {$45^\circ$};

\draw[green!50!black, thick, <->] ({0.45*cos(45)},{0.45*sin(45)}) arc (45:90:0.45);
\node[font=\scriptsize, green!50!black] at ({0.57*cos(67.5)},{0.57*sin(67.5)}) {$45^\circ$};

\draw[green!50!black, thick, <->] ({0.55*cos(90)},{0.55*sin(90)}) arc (90:135:0.55);
\node[font=\scriptsize, green!50!black] at ({0.67*cos(112.5)},{0.67*sin(112.5)}) {$45^\circ$};

\draw[orange!70!black, thick, <->, dashed] ({0.35*cos(0)},{0.35*sin(0)}) arc (0:135:0.35);
\node[font=\scriptsize, orange!70!black, anchor=east] at ({0.22*cos(67.5)},{0.22*sin(67.5)}) {$135^\circ$};

\node[font=\small, blue!80!black, anchor=west] at (-1.2,-0.55) {\textbf{---} Alice (faces)};
\node[font=\small, red!70!black, anchor=west] at (-1.2,-0.72) {\textbf{- -} Bob (edges)};

\end{tikzpicture}
\caption{The four CHSH measurement directions in the $xz$-plane. Alice measures along two face directions of the Bloch cube ($\hat{a} = \hat{z}$ and $\hat{a}' = \hat{x}$, solid blue). Bob measures along two edge directions ($\hat{b}$ at $45^\circ$ and $\hat{b}'$ at $135^\circ$, dashed red). Adjacent directions are separated by $45^\circ$, producing the three-close-one-far pattern: the pairs $(\hat{a}, \hat{b})$, $(\hat{a}', \hat{b})$, and $(\hat{a}', \hat{b}')$ are each $45^\circ$ apart, while $(\hat{a}, \hat{b}')$ is $135^\circ$ apart.}
\label{fig:chsh_directions}
\end{figure}
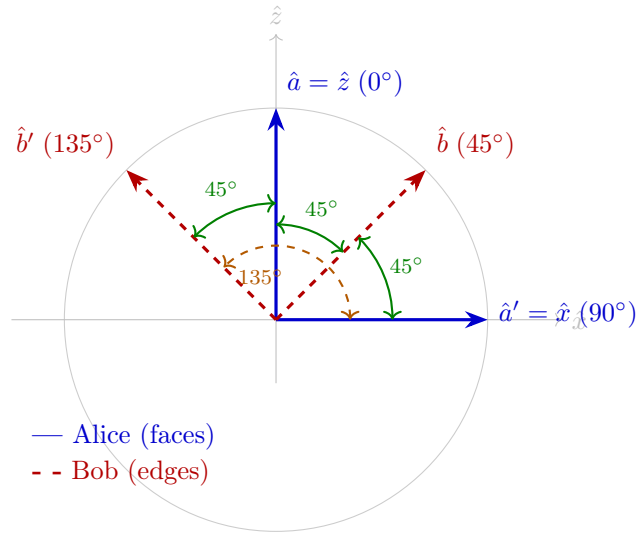

The geometry produces a crucial asymmetry (Fig.~\ref{fig:chsh_directions}). Three of the four pairs are neighboring face--edge combinations, each separated by $45^\circ$:
\begin{equation}
    \theta_{ab} = \theta_{a'b} = \theta_{a'b'} = 45^\circ
\end{equation}
The fourth pair---Alice's face direction $\hat{z}$ and Bob's far edge direction $(\hat{x} - \hat{z})/\sqrt{2}$---are $135^\circ$ apart:
\begin{equation}
    \theta_{ab'} = 135^\circ
\end{equation}
This three-close-one-far pattern is not a choice we impose; it is forced by the face--edge geometry of the Bloch cube.

The quantum correlations follow:
\begin{equation}
    E(\vec{a}, \vec{b}) = E(\vec{a}', \vec{b}) = E(\vec{a}', \vec{b}') = -\cos 45^\circ = -\frac{1}{\sqrt{2}}, \quad E(\vec{a}, \vec{b}') = -\cos 135^\circ = +\frac{1}{\sqrt{2}}
\end{equation}

The CHSH combination gives:
\begin{equation}
    S = E(\vec{a}, \vec{b}) - E(\vec{a}, \vec{b}') + E(\vec{a}', \vec{b}) + E(\vec{a}', \vec{b}') = -\frac{1}{\sqrt{2}} - \frac{1}{\sqrt{2}} - \frac{1}{\sqrt{2}} - \frac{1}{\sqrt{2}} = -2\sqrt{2}
\end{equation}

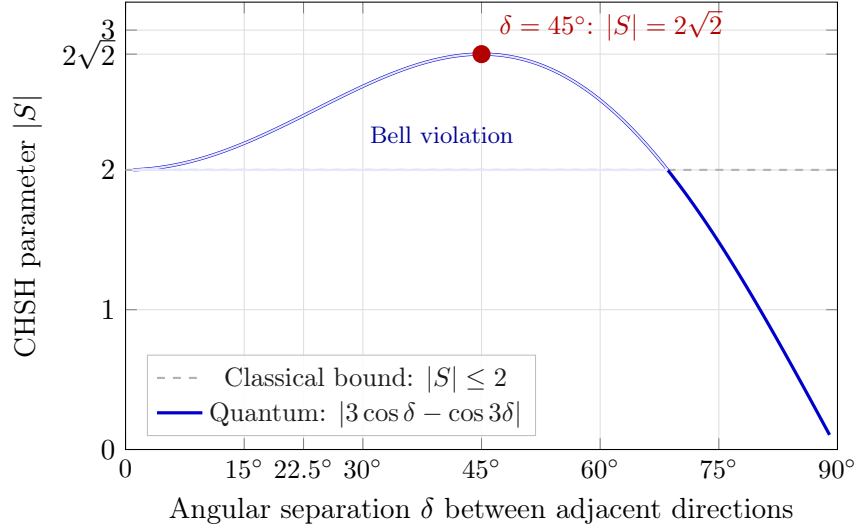
\begin{figure}[ht]
\centering
\begin{tikzpicture}
\begin{axis}[
    width=11cm, height=7.5cm,
    xlabel={Angular separation $\delta$ between adjacent directions},
    ylabel={CHSH parameter $|S|$},
    xmin=0, xmax=90,
    ymin=0, ymax=3.2,
    xtick={0,15,22.5,30,45,60,75,90},
    xticklabels={$0$,$15^\circ$,$22.5^\circ$,$30^\circ$,$45^\circ$,$60^\circ$,$75^\circ$,$90^\circ$},
    x tick label style={font=\small},
    ytick={0,1,2,2.828,3},
    yticklabels={$0$,$1$,$2$,$2\sqrt{2}$,$3$},
    grid=major,
    grid style={gray!25},
    legend pos=south west,
    legend style={font=\small, fill=white, fill opacity=0.9, draw=gray!50},
    every axis plot/.append style={thick},
]
\addplot[gray!60, thick, dashed, domain=0:90] {2};
\addlegendentry{Classical bound: $|S| \leq 2$}

\addplot[blue!80!black, very thick, domain=1:89, samples=200] {abs(3*cos(x) - cos(3*x))};
\addlegendentry{Quantum: $|3\cos\delta - \cos 3\delta|$}

\addplot[only marks, mark=*, mark size=3pt, red!70!black] coordinates {(45, 2.828)};
\node[font=\small, red!70!black, anchor=south west] at (axis cs:46,2.85) {$\delta = 45^\circ$: $|S| = 2\sqrt{2}$};

\addplot[blue!10, forget plot, domain=1:68.5, samples=100]
    {abs(3*cos(x) - cos(3*x))} |- (axis cs:68.5,2) -- (axis cs:1,2) -- cycle;

\node[font=\footnotesize, blue!60!black] at (axis cs:40,2.25) {Bell violation};

\end{axis}
\end{tikzpicture}
\caption{The CHSH parameter $|S|$ as a function of the angular separation $\delta$ between adjacent measurement directions, assuming four equally spaced directions at angles $0$, $\delta$, $2\delta$, $3\delta$ in the $xz$-plane. The quantum prediction (solid blue) exceeds the classical bound of~2 (dashed gray) for a range of angles centered on $\delta = 45^\circ$, where the violation is maximal at $|S| = 2\sqrt{2} \approx 2.83$. The shaded region marks the Bell violation. The optimal $45^\circ$ spacing corresponds to the face--edge separation on the Bloch cube.}
\label{fig:chsh_violation}
\end{figure}

Therefore $|S| = 2\sqrt{2} \approx 2.83$, which exceeds the local hidden variable bound of~2 (Fig.~\ref{fig:chsh_violation}).

\begin{keyidea}{Bell's Theorem}
No local hidden variable theory can reproduce all predictions of quantum mechanics. The CHSH inequality, which any local realistic theory must satisfy, is violated by quantum mechanical predictions for entangled states. Experiments confirm the quantum predictions.

The Bloch cube makes the geometry of the violation visible. The optimal CHSH measurement directions are two faces and two edges of the Bloch cube, and the violation lives in the face--edge asymmetry: three pairs at $45^\circ$ and one at $135^\circ$. The $45^\circ$ angle that defines an edge direction is the same angle that drives the maximal quantum violation.
\end{keyidea}

\subsection{Where the Single-Qubit Model Breaks Down}

It is instructive to see precisely why the hidden variable strategy that worked for one qubit fails for two.

For a single qubit, Bell's model used a hidden variable $\hat{\lambda}$ to determine all measurement outcomes through $A(\hat{n}) = \text{sign}(\hat{n} \cdot \hat{\lambda})$. One might try to extend this by giving each particle its own hidden variable: Alice's particle carries $\hat{\lambda}_A$ and Bob's carries $\hat{\lambda}_B$, with outcomes $A(\vec{a}) = \text{sign}(\vec{a} \cdot \hat{\lambda}_A)$ and $B(\vec{b}) = \text{sign}(\vec{b} \cdot \hat{\lambda}_B)$, both prepared at the source.

This model is explicitly local: Alice's outcome depends only on $\vec{a}$ and $\vec{\lambda}_A$, and Bob's only on $\vec{b}$ and $\vec{\lambda}_B$. The correlations arise solely from the joint distribution $\rho(\vec{\lambda}_A, \vec{\lambda}_B)$ established at the source.

The problem is that no distribution $\rho(\vec{\lambda}_A, \vec{\lambda}_B)$ can produce the quantum correlation $E(\vec{a}, \vec{b}) = -\cos\theta_{ab}$ for all measurement directions simultaneously. The proof is Bell's theorem: any such local model satisfies $|S| \leq 2$, but quantum mechanics demands $|S| = 2\sqrt{2}$. The quantum correlations are too strong to arise from shared classical information, no matter how cleverly the hidden variables at the source are prepared.

\section{GHZ States: A Starker Contradiction}
\label{sec:ghz}

Bell's theorem involves statistical predictions: local hidden variables and quantum mechanics give different probabilities. Greenberger, Horne, and Zeilinger (GHZ) discovered an even more dramatic conflict involving definite predictions, not just probabilities.

\subsection{The GHZ State}

Consider three qubits in the state:
\begin{equation}
    \ket{\text{GHZ}} = \frac{1}{\sqrt{2}}(\ket{000} - \ket{111})
\end{equation}
This is a maximally entangled state of three particles. A source prepares this state and sends one qubit to Alice, one to Bob, and one to Charlie.

\subsection{Quantum Predictions for GHZ}

Each experimenter can measure their qubit in either the $X$ basis (eigenstates of $\sigma_x$) or the $Y$ basis (eigenstates of $\sigma_y$), obtaining outcomes $\pm 1$.

The quantum predictions for specific measurement combinations can be computed using the operator eigenvalue approach. Consider the operator $X_1 X_2 X_3 = \sigma_x \otimes \sigma_x \otimes \sigma_x$. Acting on the GHZ state:
\begin{equation}
    X_1 X_2 X_3 \ket{\text{GHZ}} = \frac{1}{\sqrt{2}}\bigl(\sigma_x\ket{0} \otimes \sigma_x\ket{0} \otimes \sigma_x\ket{0} - \sigma_x\ket{1} \otimes \sigma_x\ket{1} \otimes \sigma_x\ket{1}\bigr) = \frac{1}{\sqrt{2}}(\ket{111} - \ket{000}) = -\ket{\text{GHZ}}
\end{equation}
So $\ket{\text{GHZ}}$ is an eigenstate of $X_1 X_2 X_3$ with eigenvalue $-1$, meaning the product of the three $X$ outcomes is $-1$ on every run, not just on average. By the same method (using $\sigma_y\ket{0} = -i\ket{1}$ and $\sigma_y\ket{1} = i\ket{0}$), one finds that $\ket{\text{GHZ}}$ is also an eigenstate of the operators $X_1 Y_2 Y_3$, $Y_1 X_2 Y_3$, and $Y_1 Y_2 X_3$, each with eigenvalue $+1$. The four predictions are:
\begin{align}
    \langle X_1 Y_2 Y_3 \rangle &= +1 \\
    \langle Y_1 X_2 Y_3 \rangle &= +1 \\
    \langle Y_1 Y_2 X_3 \rangle &= +1 \\
    \langle X_1 X_2 X_3 \rangle &= -1
\end{align}

The first three equations say that when we measure $XYY$, $YXY$, or $YYX$, the product of the three outcomes is always $+1$. The fourth says that when we measure $XXX$, the product is always $-1$. Crucially, all four are eigenvalue equations: the outcomes are definite, not merely statistical.

\subsection{The Contradiction with Local Realism}

Suppose local hidden variables govern the outcomes. Each particle carries predetermined values $x_i = \pm 1$ (its hidden answer to an $X$ measurement) and $y_i = \pm 1$ (its hidden answer to a $Y$ measurement). All six values are determined at the source and carried locally to the three detectors.

The first three quantum predictions require:
\begin{align}
    x_1 y_2 y_3 &= +1 \\
    y_1 x_2 y_3 &= +1 \\
    y_1 y_2 x_3 &= +1
\end{align}

Multiply these three equations:
\begin{equation}
    (x_1 y_2 y_3)(y_1 x_2 y_3)(y_1 y_2 x_3) = (+1)(+1)(+1) = +1
\end{equation}

The left side simplifies to:
\begin{equation}
    x_1 x_2 x_3 \cdot y_1^2 y_2^2 y_3^2 = x_1 x_2 x_3 \cdot (+1) = x_1 x_2 x_3
\end{equation}
since each $y_i = \pm 1$.

Therefore the hidden variables predict $x_1 x_2 x_3 = +1$: the predetermined $X$ answers from the three particles must multiply to $+1$.

But quantum mechanics predicts $\langle X_1 X_2 X_3 \rangle = -1$: the product is always $-1$.

This is not a statistical disagreement. It is an outright contradiction. No assignment of predetermined values to the three particles can simultaneously satisfy all four quantum predictions. The hidden variable program, which worked flawlessly for one qubit and failed only statistically for two, now fails on every single run of a three-particle experiment.

\begin{keyidea}{The GHZ Argument}
The GHZ state provides an ``all-or-nothing'' refutation of local hidden variables. Unlike Bell's original argument, which requires statistical analysis of many runs, the GHZ argument shows a logical contradiction from a single set of measurements. No assignment of predetermined values to the three particles can reproduce all four quantum predictions simultaneously.
\end{keyidea}

\section{Experimental Tests}
\label{sec:experimental_tests}

Bell's theorem and the GHZ argument are not philosophical speculations---they make quantitative predictions that differ from any local hidden variable theory. Over more than four decades, a series of increasingly sophisticated experiments has confirmed the quantum predictions and ruled out local realism.

\subsection{Bell Inequality Tests with Photon Pairs}

The first experimental test of a Bell inequality was performed by Freedman and Clauser in 1972 using polarization-entangled photon pairs produced by an atomic cascade source. The measured correlations violated the CHSH inequality, favoring quantum mechanics over local hidden variables.

The most influential early experiments were those of Aspect, Dalibard, and Roger (1982), who introduced rapid switching of polarizer settings during the flight of the photons. This addressed the \emph{locality loophole}: the concern that if the measurement settings were chosen in advance, some local signal could communicate Alice's setting to Bob's detector in time to influence the result. By switching settings after the photons had separated, the experiment ensured that no light-speed signal could carry the information. The measured correlations violated the CHSH inequality in agreement with quantum predictions.

However, each early experiment left open at least one loophole:

\textbf{The locality loophole:} Were the measurement settings truly chosen with spacelike separation, so that no signal could connect Alice's choice to Bob's outcome? Aspect's switching experiment was a major step. Weihs et~al.\ (1998) strengthened this by using quantum random number generators to select settings and separating the detectors by 400~meters, ensuring strict spacelike separation.

\textbf{The detection loophole:} Photon detectors are imperfect---typical efficiencies were 10--20\% in early experiments. If only a small fraction of emitted pairs are detected, one might worry that the detected subset is not representative of the full ensemble. A sufficiently biased subset could violate the CHSH inequality even if the full ensemble did not. Rowe et~al.\ (2001) closed this loophole using trapped ions, which can be detected with near-unit efficiency, though the ions were not spacelike-separated.

The definitive experiments came in 2015, when three independent groups closed \emph{both} loopholes simultaneously. Hensen et~al.\ used entangled electron spins in nitrogen-vacancy centers in diamond, separated by 1.3~km. Giustina et~al.\ and Shalm et~al.\ used high-efficiency superconducting photon detectors with spacelike-separated measurement stations. All three experiments confirmed the quantum predictions and violated the CHSH inequality with high statistical significance.

The experimental verdict is unambiguous: local hidden variable theories are ruled out.

\subsection{The GHZ Experiment}

The GHZ argument is theoretically sharper than CHSH---an all-or-nothing contradiction on every run, rather than a statistical violation accumulated over many runs. But it is experimentally more demanding, requiring entanglement of at least three particles.

The first observation of three-photon GHZ entanglement was achieved by Bouwmeester et~al.\ (1999) in Zeilinger's group in Vienna. The technique used two pairs of polarization-entangled photons produced by pulsed parametric down-conversion. A polarizing beam splitter combined one photon from each pair, and detection of a ``trigger'' photon in a way that erased which-pair information projected the remaining three photons into a GHZ-entangled state.

The full nonlocality test was reported by Pan et~al.\ (2000) in \textit{Nature}. The experiment followed the logic of the GHZ argument directly. Three experiments measured the polarization correlations corresponding to $XYY$, $YXY$, and $YYX$ on the three-photon GHZ state. In each case, the product of the three outcomes was found to be $+1$, consistent with the quantum eigenvalue predictions. Under the assumption of local realism, these three results predict that the $XXX$ measurement must also yield product $+1$. The fourth experiment measured $XXX$ and found the product to be $-1$---in agreement with quantum mechanics and in direct contradiction with local realism.

The measured error rate was approximately 15\% per measurement configuration. The experiment required fourfold coincidence detection (three GHZ photons plus the trigger), and entanglement was confirmed only when all detectors fired, leaving the detection loophole open. Subsequent experiments have extended GHZ tests to four, six, and even ten entangled photons, and have been performed on platforms such as superconducting qubits and trapped ions where detection efficiencies are much higher.

Zeilinger shared the 2022 Nobel Prize in Physics, in part for this experimental program demonstrating multi-photon entanglement and its foundational implications.

\begin{physicalinsight}
The progression of experimental difficulty mirrors the theoretical progression. Testing CHSH requires entangling two particles; testing GHZ requires three. Closing all loopholes simultaneously---spacelike separation \emph{and} high detection efficiency---took decades of technological development in source brightness, detector efficiency, and random number generation. In every case, the quantum predictions have been confirmed.
\end{physicalinsight}

\section{What Bell's Theorem Means}

The experiments described in the previous section have spoken decisively: the quantum predictions are correct, and local hidden variable theories are ruled out. What does this imply for our understanding of reality?

\subsection{The Assumptions}

Bell's theorem derives a contradiction from three assumptions. \textbf{Realism} holds that measurement outcomes are determined by pre-existing properties. \textbf{Locality} requires that the outcome at one location cannot depend on the measurement setting at a distant location for spacelike-separated measurements. \textbf{Freedom of choice} asserts that the measurement settings can be chosen independently of the hidden variables.

Since experiments violate the Bell inequality, at least one assumption must be false.

\subsection{Giving Up Locality}

One response is to accept that quantum mechanics involves genuine nonlocality: what happens at one location can instantaneously influence what happens at a distant location.

This does not permit faster-than-light signaling. Although Alice's measurement instantaneously affects the quantum state at Bob's location, Bob cannot detect this without knowing Alice's result, which must arrive by classical (slower-than-light) communication. Quantum mechanics maintains ``peaceful coexistence'' with special relativity in this sense.

Nevertheless, the nonlocality is real: the joint probability distribution for Alice's and Bob's outcomes cannot be explained by any local mechanism. The universe exhibits correlations that cannot be accounted for by information traveling from a common source.

\subsection{Giving Up Realism}

An alternative is to abandon realism: measurement outcomes are not revealing pre-existing values but are created in the act of measurement. On this view, it is simply meaningless to ask what Bob's spin ``really is'' before he measures it.

This is the orthodox Copenhagen interpretation. It maintains locality in the sense that no physical influence travels between the particles, because there is no underlying reality to be influenced. The correlations arise from the entangled quantum state, which is not itself a local object.

Note the irony: the single-qubit hidden variable models of Section~2 gave every qubit definite answers to all spin questions, and Bell's continuous model extended this to every direction on the sphere. The Copenhagen interpretation says we should not have taken that success too seriously. What seemed like evidence for an underlying deterministic reality was merely the fact that single-particle statistics happen to be reproducible by hidden variables. The failure only becomes apparent in the richer correlational structure of entangled systems, where shared hidden variables cannot produce correlations as strong as those quantum mechanics predicts.

\subsection{Giving Up Freedom of Choice}

A more radical option is ``superdeterminism'': perhaps the measurement settings are not truly free but are correlated with the hidden variables. If the universe conspired to always choose settings that produce the observed correlations, no violation of locality or realism would be required.

Most physicists reject this option as undermining the foundations of experimental science. If we cannot choose measurement settings independently, we cannot trust any experiment.

\subsection{Interpretations of Quantum Mechanics}

Different interpretations of quantum mechanics respond to Bell's theorem in different ways:

\textbf{Copenhagen interpretation:} Accepts that quantum mechanics is complete and that asking about pre-measurement values is meaningless. The ``collapse'' of the wavefunction is not a physical process but an update of our knowledge.

\textbf{Many-worlds interpretation:} All outcomes occur in different branches of a universal wavefunction. There is no collapse and no nonlocality; the appearance of definite outcomes arises from decoherence and the observer's own branching.

\textbf{Bohmian mechanics:} A deterministic, nonlocal hidden variable theory. Particles have definite positions at all times, guided by a ``pilot wave'' that depends on the configuration of all particles. Explicitly nonlocal, it reproduces quantum predictions exactly.

\textbf{Collapse theories (GRW):} Physical collapse mechanisms cause spontaneous localization, modifying the Schr\"{o}dinger equation. The nonlocality is explicit in the collapse dynamics.

Bell's theorem does not adjudicate between these interpretations. They all reproduce the experimental predictions. What it does rule out is any interpretation that is both local and realistic in the senses defined above.

\subsection{The Significance for Physics}

Bell's theorem transforms what had been a philosophical debate into an empirical question. The EPR argument seemed to show that quantum mechanics was incomplete and must be supplemented by hidden variables. Bell showed that such a supplementation, if local, leads to testable predictions that differ from quantum mechanics. Experiments have decided the matter: the quantum predictions are correct.

\textbf{On the nature of reality:} The classical picture of a world with definite properties existing independently of observation, with influences propagating locally, cannot be the whole story. Quantum mechanics reveals a world that is, at a fundamental level, stranger than our everyday intuitions suggest.

\textbf{On the completeness of quantum mechanics:} In a precise sense, quantum mechanics cannot be ``completed'' by adding local hidden variables. Whatever deeper theory might underlie quantum mechanics, it cannot restore a classical worldview.

\textbf{On the unity of physics:} The tension between quantum nonlocality and relativistic locality remains an open problem. Quantum field theory manages to combine quantum mechanics with special relativity, but the interpretation of this combination, especially regarding entanglement and measurement, is not fully understood.

\section{Chapter Summary}

The Bloch cube defines a natural hierarchy of measurement directions---faces, edges, and corners---that organizes the entire hidden variable story. For a single qubit, hidden variable models succeed completely at every level of this hierarchy: the face model reproduces predictions for the three face directions, an analogous corner model handles the four body diagonals, and Bell's 1966 continuous model extends to all directions on the sphere. Single-particle quantum randomness is fully compatible with realism, and the realist hypothesis is unfalsifiable by any single-particle experiment.

Entangled states cannot be written as products of individual subsystem states. The Bell states form a complete basis for two-qubit systems and exhibit perfect correlations between distant measurements.

The EPR argument, emboldened by the success of single-qubit realism, attempted to extend hidden variables to entangled systems while maintaining locality. Bell's theorem proves this is impossible. The CHSH inequality provides a quantitative test: local theories satisfy $|S| \leq 2$, while quantum mechanics predicts $|S| = 2\sqrt{2}$ for optimal measurements on singlet states. The optimal measurement directions are two faces and two edges of the Bloch cube, arranged in the $xz$-plane at $0^\circ$, $45^\circ$, $90^\circ$, and $135^\circ$. The face--edge geometry produces the asymmetry---three pairs at $45^\circ$, one at $135^\circ$---that drives the maximal violation.

The $45^\circ$ angle that defines an edge direction is the same angle that appears in the CHSH game's winning probability $\cos^2(22.5^\circ) \approx 85\%$, exceeding the classical limit of $75\%$.

GHZ states provide an even starker contradiction: three sets of hidden face values make a definite prediction ($x_1 x_2 x_3 = +1$) that quantum mechanics contradicts ($-1$) in every single run, not just statistically.

Experimental tests confirm quantum mechanical predictions. At least one of the assumptions underlying local realism must be abandoned. How we interpret this result depends on our preferred interpretation of quantum mechanics, but no interpretation can restore a fully local and realistic worldview.

\input{book_problems/ch10_problems.tex}

\section*{References and Further Reading}
\addcontentsline{toc}{section}{References and Further Reading}

\begin{description}
\item[Bell, J.~S.] ``On the Einstein Podolsky Rosen paradox.'' \emph{Physics Physique Fizika} \textbf{1}, 195--200 (1964). \href{https://doi.org/10.1103/PhysicsPhysiqueFizika.1.195}{doi:10.1103/PhysicsPhysiqueFizika.1.195}. The original theorem. Three pages, no advanced machinery, and the source of everything in this chapter from Section~\ref{sec:bells_continuous_model} onward.

\item[Bell, J.~S.] ``On the problem of hidden variables in quantum mechanics.'' \emph{Reviews of Modern Physics} \textbf{38}, 447--452 (1966). \href{https://doi.org/10.1103/RevModPhys.38.447}{doi:10.1103/RevModPhys.38.447}. Written before the 1964 paper but published later; constructs the single-qubit hidden variable model used in Section~\ref{sec:hidden_bloch_cubes} and dismantles von Neumann's earlier no-go ``proof.''

\item[Einstein, A., Podolsky, B., and Rosen, N.] ``Can quantum-mechanical description of physical reality be considered complete?'' \emph{Physical Review} \textbf{47}, 777--780 (1935). \href{https://doi.org/10.1103/PhysRev.47.777}{doi:10.1103/PhysRev.47.777}. The argument Bell's theorem refutes; the cleanest statement of the realism-plus-locality position the experiments later ruled out.

\item[Clauser, J.~F., Horne, M.~A., Shimony, A., and Holt, R.~A.] ``Proposed experiment to test local hidden-variable theories.'' \emph{Physical Review Letters} \textbf{23}, 880--884 (1969). \href{https://doi.org/10.1103/PhysRevLett.23.880}{doi:10.1103/PhysRevLett.23.880}. The CHSH inequality in its experimentally testable form; the inequality whose violation was measured by every experiment cited in Section~\ref{sec:experimental_tests}.

\item[Mermin, N.~D.] ``Hidden variables and the two theorems of John Bell.'' \emph{Reviews of Modern Physics} \textbf{65}, 803--815 (1993). \href{https://doi.org/10.1103/RevModPhys.65.803}{doi:10.1103/RevModPhys.65.803}. The masterful pedagogical exposition; covers both Bell's inequality and the Bell--Kochen--Specker theorem with characteristic clarity. The companion piece ``Quantum mysteries revisited,'' \emph{Am.\ J.\ Phys.}\ \textbf{58}, 731--734 (1990), \href{https://doi.org/10.1119/1.16503}{doi:10.1119/1.16503}, gives the simplest accessible version of the GHZ argument used in Section~\ref{sec:ghz}.

\item[Hensen, B., \emph{et al.}] ``Loophole-free Bell inequality violation using electron spins separated by 1.3 kilometres.'' \emph{Nature} \textbf{526}, 682--686 (2015). \href{https://doi.org/10.1038/nature15759}{doi:10.1038/nature15759}. The first experiment to close both the locality and detection loopholes simultaneously; the definitive experimental refutation of local realism. Companion 2015 results by Giustina \emph{et~al.}\ (\href{https://doi.org/10.1103/PhysRevLett.115.250401}{PRL \textbf{115}, 250401}) and Shalm \emph{et~al.}\ (\href{https://doi.org/10.1103/PhysRevLett.115.250402}{PRL \textbf{115}, 250402}) close the loopholes using entangled photons.
\end{description}

%% file: book_problems/ch10_problems.tex
\section{Problems}
\setcounter{hwproblem}{0}

\problem{Hidden Bloch Cubes: The Face Model}
A particle is prepared in the state $\ket{+z}$. In the face model, the complete hidden-variable description is a triple $(\ket{+z}, \ket{\epsilon_x}, \ket{\epsilon_y})$ where $\epsilon_x, \epsilon_y = \pm 1$ are hidden answers to the $X$ and $Y$ questions. The four corners of the top face are equally probable: $p_1 = p_2 = p_3 = p_4 = 1/4$.
\begin{enumerate}[label=(\alph*)]
    \item Verify that this uniform distribution reproduces $P(+x) = 1/2$ and $P(+y) = 1/2$, as quantum mechanics requires.
    \item The face model assigns a measurement rule $A(\hat{n}) = \text{sign}(\hat{n} \cdot \vec{\lambda})$, where $\vec{\lambda} = (\epsilon_x, \epsilon_y, \epsilon_z)$. Show that this correctly returns $\epsilon_x$ when measuring along $\hat{x}$ and $\epsilon_z$ when measuring along $\hat{z}$.
    \item Now consider the edge direction $\hat{e}_+ = (\hat{z} + \hat{x})/\sqrt{2}$. Show that the face model predicts $P(+1 \,|\, \hat{e}_+) = 3/4$ for a particle in $\ket{+z}$.
    \item Quantum mechanics predicts $P(+1 \,|\, \hat{e}_+) = \cos^2(22.5^\circ) \approx 0.854$. Compare this with your answer from (c). Does the face model succeed or fail at edge directions?
    \item The face model works with four equally-weighted hidden states. Explain physically why a finite discrete model cannot reproduce a continuous function like $\cos^2(\theta/2)$ for arbitrary angles $\theta$.
\end{enumerate}

\problem{Bell's Continuous Hidden Variable Model}
Bell's continuous model assigns each particle a hidden unit vector $\hat{\lambda}$ on the sphere, with the measurement rule $f_{\hat{\lambda}}(\hat{n}) = \text{sign}(\hat{n} \cdot \hat{\lambda})$. This divides the sphere into two hemispheres separated by the great circle perpendicular to $\hat{n}$.
\begin{enumerate}[label=(\alph*)]
    \item For a particle prepared in $\ket{+z}$ (Bloch vector $\hat{s} = \hat{z}$), the probability of $+1$ along direction $\hat{n}$ at polar angle $\theta$ from $\hat{z}$ must satisfy $P(+1 \,|\, \hat{n}) = (1 + \cos\theta)/2$. Show that this equals $\cos^2(\theta/2)$.
    \item Consider $\hat{n} = \hat{z}$. The measurement hemisphere is $\hat{n} \cdot \hat{\lambda} > 0$, i.e., the upper hemisphere. Show geometrically that $P(+1 \,|\, \hat{z}) = 1$ requires all the probability weight $\rho(\hat{\lambda})$ to lie in the upper hemisphere.
    \item Now consider $\hat{n}$ at angle $\theta$ from $\hat{z}$. The great circle perpendicular to $\hat{n}$ carves out a fraction of the upper hemisphere that gives $f = -1$. Sketch the geometry for $\theta = 45^\circ$ and explain qualitatively why $P(+1) < 1$ but $P(+1) > 1/2$.
    \item Bell's model reproduces all single-qubit quantum predictions for any direction $\hat{n}$. Does this mean hidden variables are the correct description of a single qubit? Explain what ''unfalsifiable'' means in this context.
\end{enumerate}

\problem{Bell States and Entanglement}
The four Bell states are:
\begin{align*}
    \ket{\Phi^+} &= \frac{1}{\sqrt{2}}(\ket{00} + \ket{11}) &
    \ket{\Phi^-} &= \frac{1}{\sqrt{2}}(\ket{00} - \ket{11}) \\
    \ket{\Psi^+} &= \frac{1}{\sqrt{2}}(\ket{01} + \ket{10}) &
    \ket{\Psi^-} &= \frac{1}{\sqrt{2}}(\ket{01} - \ket{10})
\end{align*}
\begin{enumerate}[label=(\alph*)]
    \item Show that the four Bell states are orthonormal.
    \item Compute the reduced density matrix $\rho_A = \text{Tr}_B(\ket{\Phi^+}\bra{\Phi^+})$. Repeat for $\ket{\Psi^-}$. What do you notice?
    \item For $\ket{\Phi^+}$, suppose Alice measures $\sigma_z$ and obtains $+1$. What is Bob's post-measurement state? Repeat for the case where Alice obtains $-1$.
    \item For $\ket{\Psi^-}$ (the singlet), suppose Alice measures along an arbitrary direction $\hat{n}$ and obtains $+1$. Show that Bob's post-measurement state is the spin-down state along $\hat{n}$.
    \item The singlet state has perfect anti-correlation in every basis simultaneously. Explain why this does not contradict locality.
\end{enumerate}

\problem{The EPR Argument and Locality}
Alice and Bob share the singlet state $\ket{\Psi^-} = \frac{1}{\sqrt{2}}(\ket{01} - \ket{10})$ and are separated by a large distance.
\begin{enumerate}[label=(\alph*)]
    \item Alice measures $\sigma_z$ on her qubit. With what probability does she obtain each outcome? For each outcome, what can she predict about Bob's $\sigma_z$ result with certainty?
    \item Alice instead measures $\sigma_x$. For each outcome, what can she predict about Bob's $\sigma_x$ result?
    \item EPR's locality assumption: Alice's choice of measurement cannot instantaneously affect Bob's particle. Using this assumption, argue that Bob's particle must have had definite values of both $\sigma_z$ and $\sigma_x$ before either measurement.
    \item Quantum mechanics says $\sigma_z$ and $\sigma_x$ cannot simultaneously have definite values. What logical conclusion did EPR draw?
    \item Identify the assumption in the EPR argument that Bell's theorem ultimately shows must be abandoned.
\end{enumerate}

\problem{The CHSH Inequality}
In a local hidden variable theory, Alice's outcome $A(\vec{a}, \lambda) = \pm 1$ depends on her setting $\vec{a}$ and a shared hidden variable $\lambda$, but not on Bob's setting $\vec{b}$. Similarly, $B(\vec{b}, \lambda) = \pm 1$ is independent of $\vec{a}$.
\begin{enumerate}[label=(\alph*)]
    \item For a single value of $\lambda$, define $S(\lambda) = A(\vec{a})[B(\vec{b}) - B(\vec{b}')] + A(\vec{a}')[B(\vec{b}) + B(\vec{b}')]$. Show that if $B(\vec{b}) = B(\vec{b}')$, then $|S(\lambda)| = 2$.
    \item Show that if $B(\vec{b}) = -B(\vec{b}')$, then $|S(\lambda)| = 2$ as well.
    \item Since $B(\vec{b})$ and $B(\vec{b}')$ are each $\pm 1$, one of the two cases above must hold. Conclude that $|S(\lambda)| = 2$ for every $\lambda$.
    \item Averaging over $\rho(\lambda)$, derive the CHSH inequality: $|E(\vec{a}, \vec{b}) - E(\vec{a}, \vec{b}') + E(\vec{a}', \vec{b}) + E(\vec{a}', \vec{b}')| \leq 2$.
    \item Could a cleverly chosen distribution $\rho(\lambda)$ violate this bound? Explain why or why not.
\end{enumerate}

\problem{Quantum Violation with Faces and Edges}
For the singlet state $\ket{\Psi^-}$, the quantum correlation function is $E(\vec{a}, \vec{b}) = -\cos\theta_{ab}$, where $\theta_{ab}$ is the angle between the measurement directions.
\begin{enumerate}[label=(\alph*)]
    \item Alice measures along two face directions $\hat{a} = \hat{z}$ and $\hat{a}' = \hat{x}$. Bob measures along two edge directions $\hat{b} = (\hat{z} + \hat{x})/\sqrt{2}$ and $\hat{b}' = (\hat{x} - \hat{z})/\sqrt{2}$. Compute the four angles $\theta_{ab}$, $\theta_{ab'}$, $\theta_{a'b}$, $\theta_{a'b'}$.
    \item Calculate the CHSH combination $S = E(\vec{a}, \vec{b}) - E(\vec{a}, \vec{b}') + E(\vec{a}', \vec{b}) + E(\vec{a}', \vec{b}')$ using $E(\theta) = -\cos\theta$.
    \item Verify that $|S| = 2\sqrt{2}$, violating the CHSH bound. Explain the role of the face-edge geometry in producing maximal violation.
    \item For a local hidden variable model with $\hat{\lambda}$ uniformly distributed on the sphere, show that $E_{\text{local}}(\theta) = -1 + 2\theta/\pi$.
    \item Compare the behaviors of the local and quantum correlations near $\theta = 0$. Which departs more slowly from perfect anti-correlation, and what does this tell you about quantum ''stickiness''?
\end{enumerate}

\problem{Concurrence and Entanglement Measures}
The concurrence $C(\rho)$ is an entanglement measure for two-qubit states defined by $C = \max(0, \lambda_1 - \lambda_2 - \lambda_3 - \lambda_4)$, where $\lambda_i$ are the eigenvalues of $R = \sqrt{\sqrt{\rho}\tilde{\rho}\sqrt{\rho}}$ in decreasing order, and $\tilde{\rho} = (\sigma_y \otimes \sigma_y)\rho^*(\sigma_y \otimes \sigma_y)$.
\begin{enumerate}[label=(\alph*)]
    \item For a pure state $\ket{\psi}_{AB}$, show that the concurrence equals $C = 2|\alpha\gamma - \beta\delta|$ where $\ket{\psi} = \alpha\ket{00} + \beta\ket{01} + \gamma\ket{10} + \delta\ket{11}$.
    \item Compute the concurrence for the Bell state $\ket{\Phi^+}$.
    \item Compute the concurrence for a separable state $\ket{\psi}_A \otimes \ket{\phi}_B$. What does this tell you about separable states?
    \item For the mixed state $\rho = p\ket{\Phi^+}\bra{\Phi^+} + (1-p)\frac{\identity}{4}$, the concurrence is $C = \max(0, 2p/2 - 1/2)$. For what value of $p$ does entanglement disappear?
    \item Relate the concurrence to the entanglement of formation $E_f(\rho) = h\!\left(\frac{1 + \sqrt{1 - C^2}}{2}\right)$, where $h(x) = -x\log_2 x - (1-x)\log_2(1-x)$.
\end{enumerate}

\problem{Teleportation and Superdense Coding}
\begin{enumerate}[label=(\alph*)]
    \item In quantum teleportation, Alice holds half of a Bell pair and the unknown state $\ket{\psi}$ on her qubit. She performs a Bell measurement on these two qubits, projecting onto one of four Bell states with equal probability $1/4$. If the measurement outcome is $\ket{\Psi^+}$, what unitary operation must Bob apply to his qubit to recover $\ket{\psi}$?
    \item Verify that after Bob applies the corrective unitary, his state is indeed $\ket{\psi}$ (up to global phase).
    \item In superdense coding, Alice and Bob share a Bell pair. Alice applies one of four unitary operations to her qubit: $I$, $\sigma_x$, $\sigma_z$, or $i\sigma_y$. For each operation, determine the resulting two-qubit state. Are they orthogonal?
    \item Why does superdense coding allow Alice to transmit two classical bits to Bob by sending only one qubit?
    \item List the two key differences between teleportation and superdense coding in terms of resource use (pre-shared entanglement, classical bits, and qubits sent).
\end{enumerate}

\problem{Entanglement Entropy of Bipartite Pure States}
\begin{enumerate}[label=(\alph*)]
    \item For a pure bipartite state $\ket{\psi}_{AB}$ with Schmidt decomposition $\ket{\psi} = \sum_i \sqrt{\lambda_i}\,\ket{u_i}_A\ket{v_i}_B$, define the entanglement entropy as $S_E = -\sum_i \lambda_i\log_2\lambda_i$.
    \item Show that $S_E = S_A = S_B$, where $S_A = -\text{Tr}(\rho_A\log_2\rho_A)$ is the von Neumann entropy of the reduced density matrix on subsystem $A$.
    \item Compute $S_E$ for the Bell state $\ket{\Phi^+}$. Is it maximally entangled?
    \item Compute $S_E$ for a separable state $\ket{\psi}_A\ket{\phi}_B$.
    \item For the state $\ket{\psi} = \frac{1}{\sqrt{2}}(\ket{00} + \ket{11})$, what is the entanglement entropy? Compare it with $\ket{\psi}' = \frac{1}{\sqrt{2}}\ket{0}(\ket{0} + \ket{1})$. Which is more entangled?
\end{enumerate}

\problem{The GHZ State and Three-Body Correlations}
The GHZ state for three qubits is:
\begin{equation*}
    \ket{\text{GHZ}} = \frac{1}{\sqrt{2}}(\ket{000} - \ket{111})
\end{equation*}
\begin{enumerate}[label=(\alph*)]
    \item Verify that $X_1 Y_2 Y_3 \ket{\text{GHZ}} = +\ket{\text{GHZ}}$, $Y_1 X_2 Y_3 \ket{\text{GHZ}} = +\ket{\text{GHZ}}$, and $Y_1 Y_2 X_3 \ket{\text{GHZ}} = +\ket{\text{GHZ}}$.
    \item Verify that $X_1 X_2 X_3 \ket{\text{GHZ}} = -\ket{\text{GHZ}}$.
    \item Assume local hidden variables with predetermined values $x_i = \pm 1$ and $y_i = \pm 1$ for each qubit. Show that the eigenvalue equations $x_1 y_2 y_3 = +1$, $y_1 x_2 y_3 = +1$, $y_1 y_2 x_3 = +1$, and $x_1 x_2 x_3 = -1$ are mutually incompatible.
    \item Explain why the GHZ contradiction is stronger than the CHSH inequality: one is deterministic while the other is statistical.
    \item Can you prepare a state of three qubits such that all three of $X_1 Y_2 Y_3$, $Y_1 X_2 Y_3$, and $Y_1 Y_2 X_3$ measure to $+1$ while $X_1 X_2 X_3$ measures to $+1$ (not $-1$)? Why or why not?
\end{enumerate}

\problem{Werner States and the Entanglement Threshold}
A Werner state is a mixture of a Bell pair and white noise:
\begin{equation*}
    \rho_W(p) = p\ket{\Psi^-}\bra{\Psi^-} + (1-p)\frac{\identity_4}{4}
\end{equation*}
where $\ket{\Psi^-} = \frac{1}{\sqrt{2}}(\ket{01} - \ket{10})$ and $p \in [0,1]$.
\begin{enumerate}[label=(\alph*)]
    \item Show that $\rho_W(p)$ is a valid density matrix for all $p \in [0,1]$.
    \item Compute the concurrence $C(p)$ of the Werner state. For what value $p_c$ does entanglement disappear?
    \item Compute the CHSH expectation value $\langle S \rangle$ for measurements along the optimal face-edge geometry used in Problem 6. Show that $\langle S \rangle = 2\sqrt{2}p$.
    \item Determine the critical value $p_{\text{CHSH}}$ below which the Werner state cannot violate the CHSH inequality, even with optimal measurements.
    \item Compare $p_c$ (entanglement threshold) with $p_{\text{CHSH}}$ (CHSH violation threshold). What does the difference tell you about the relationship between entanglement and nonlocality?
\end{enumerate}

\problem{No-Cloning Theorem}
\begin{enumerate}[label=(\alph*)]
    \item State the no-cloning theorem: it is impossible to construct a unitary operation that copies an arbitrary unknown quantum state.
    \item Prove the no-cloning theorem by assuming a hypothetical ''cloning'' unitary $U$ satisfying $U\ket{\psi}\ket{0} = \ket{\psi}\ket{\psi}$ for any state $\ket{\psi}$. Apply $U$ to an arbitrary superposition $\ket{\phi} = \frac{1}{\sqrt{2}}(\ket{0} + \ket{1})$ and derive a contradiction with unitarity.
    \item Explain why classical information can be copied (e.g., one can always copy a bit) but quantum information cannot.
    \item Show that the no-cloning theorem is consistent with the linearity of quantum mechanics. (Hint: consider whether the cloning operation is linear in the state.)
    \item If one is allowed to make multiple copies of a state provided one knows in advance what that state is (i.e., the state is classical knowledge, not quantum data), can one clone? Reconcile your answer with the no-cloning theorem.
\end{enumerate}

\problem{Mermin Inequality for Three Qubits}
\begin{enumerate}[label=(\alph*)]
    \item The Mermin inequality for three qubits states that any local hidden variable model predicts $|M| \leq 4$, where $M = \langle A_1 B_1 C_1 \rangle + \langle A_1 B_2 C_2 \rangle + \langle A_2 B_1 C_2 \rangle - \langle A_2 B_2 C_1 \rangle$ for dichotomic observables $A_i, B_i, C_i \in \{+1, -1\}$.
    \item Derive the Mermin bound by writing $M = A_1(B_1 C_1 + B_2 C_2) + A_2(B_1 C_2 - B_2 C_1)$ and using the fact that $(C_1 \pm C_2) \in \{0, \pm 2\}$ for dichotomic observables.
    \item For the GHZ state $\ket{\text{GHZ}}$ and measurements $A_1 = \sigma_z$, $A_2 = \sigma_x$, $B_1 = \sigma_z$, $B_2 = \sigma_x$, $C_1 = \sigma_z$, $C_2 = \sigma_x$, compute $\langle A_i B_j C_k \rangle$.
    \item Calculate $M$ for the GHZ state with the measurements in part (c). Does it violate the Mermin inequality?
    \item Compare the structure of the Mermin inequality to the CHSH inequality: how many measurement choices are involved in each, and how many parties?
\end{enumerate}

%% file: chapters/ch11_dirac_equation.tex
\chapter{The Dirac Equation}
\label{ch:dirac_equation}

\section{Introduction}

\textbf{Conventions.} Throughout this chapter we work in natural units $\hbar = c = 1$ and use the metric signature $g_{\mu\nu} = \mathrm{diag}(+1,-1,-1,-1)$ (mostly minus), except where indicated otherwise. Greek indices $\mu, \nu, \dots$ run over $0,1,2,3$ and Latin indices $i, j, \dots$ over the spatial components $1,2,3$.

The Schr\"odinger equation $i\partial_t\psi = H\psi$ is first-order in time but, through $H = -\nabla^2/2m$, second-order in space. This asymmetry is fundamentally non-relativistic: special relativity treats time and space on equal footing, so a relativistic wave equation should treat their derivatives symmetrically.

\subsection{The Klein-Gordon Equation}

The most obvious fix is to promote $E^2 = p^2 + m^2$ directly to an operator equation. Using $E \to i\partial_t$ and $\vec{p} \to -i\nabla$ gives
\begin{equation}
    (\partial_t^2 - \nabla^2 + m^2)\phi = 0
\end{equation}
This is the Klein-Gordon equation. Its plane wave solutions $\phi = e^{i(\vec{k}\cdot\vec{x} - \omega t)}$ satisfy $\omega^2 = k^2 + m^2$, giving the correct relativistic dispersion. But the equation is second-order in time, which ruins the probability interpretation. To see why, we derive the conserved density by multiplying by $\phi^*$ and subtracting the complex conjugate:
\begin{equation}
    \phi^*\partial_t^2\phi - \phi\partial_t^2\phi^* = \phi^*\nabla^2\phi - \phi\nabla^2\phi^*
\end{equation}
The left side equals $\partial_t(\phi^*\partial_t\phi - \phi\partial_t\phi^*)$ and the right side is $\nabla\cdot(\phi^*\nabla\phi - \phi\nabla\phi^*)$. This gives the continuity equation $\partial_t\rho + \nabla\cdot\vec{j} = 0$ with
\begin{equation}
    \rho = \frac{i}{2m}(\phi^*\partial_t\phi - \phi\partial_t\phi^*)
\end{equation}
where the factor of $1/2m$ is chosen so that $\rho \to |\phi|^2$ in the non-relativistic limit. For a positive-frequency plane wave with $\omega = +\sqrt{k^2 + m^2}$, this gives $\rho = \omega/m > 0$. But for a negative-frequency solution with $\omega \to -\omega$, we get $\rho = -\omega/m < 0$. Since a general solution superposes both, $\rho$ can take either sign and cannot be interpreted as a probability density.

The root cause is that a second-order equation requires two pieces of initial data ($\phi$ and $\partial_t\phi$), so the conserved density necessarily involves $\partial_t\phi$ and is not positive-definite. (In quantum field theory the Klein-Gordon equation is rehabilitated: $\rho$ becomes a charge density, which can naturally be negative. But as a single-particle equation, it fails.)

\subsection{Dirac's Insight}

Dirac's response (1928) was to go the other direction: instead of promoting the Schr\"odinger equation to second-order in both derivatives, demote the spatial part to first-order. This means writing $i\partial_t\psi = H\psi$ with
\begin{equation}
    H = -i\alpha\partial_x + \beta m
\end{equation}
where $\alpha$ and $\beta$ are constants to be determined. Applying $H$ twice must recover the relativistic dispersion, $H^2 = p^2 + m^2$, and expanding $H^2 = (-i\alpha\partial_x + \beta m)^2 = \alpha^2 p^2 + \{\alpha,\beta\}mp + \beta^2 m^2$ shows this requires $\alpha^2 = 1$, $\beta^2 = 1$, and $\{\alpha, \beta\} = 0$. No ordinary numbers satisfy all three conditions. The crucial realization is that $\alpha$ and $\beta$ can be \emph{matrices}---the simplest choice being $2 \times 2$ matrices like the Pauli matrices. This forces $\psi$ to be a multi-component object (a spinor rather than a scalar), and spin-1/2, antiparticles, and the electron's magnetic moment all emerge as consequences.

Following our earlier approach, we will build up to the full Dirac equation by first considering discrete versions, providing both mathematical insight and connections to condensed matter physics.

\begin{keyidea}{From Discrete to Continuous}
By starting with discrete time and space, we can see how the Dirac equation emerges naturally from requiring a first-order difference equation that respects relativistic symmetries. This approach also reveals deep connections to topological insulators and graphene.
\end{keyidea}

A word on interpretation as we proceed. In the discrete and condensed-matter models---the 1D lattice, the SSH chain, graphene---the two-component spinor encodes \emph{sublattice pseudospin}: the two components label which atom in the unit cell the electron occupies, and the Pauli matrices $\sigma_x, \sigma_y, \sigma_z$ act on that sublattice degree of freedom. Real electron spin is a separate, spectator degree of freedom suppressed in those models. In the full relativistic Dirac equation, by contrast, the Pauli matrices act on \emph{genuine spin-1/2}, and the four-component spinor additionally accommodates particle and antiparticle solutions. The mathematical structure---anticommuting matrices whose squares and mutual anticommutators are fixed by the metric---is identical in both settings, which is why the condensed-matter analogy illuminates the algebra so cleanly. But the physical meaning of the internal degree of freedom changes at the transition from lattice to relativistic theory, and that distinction should be kept in mind throughout the chapter.

\section{Discrete Time and Space Dirac Equation}

\subsection{The Discrete Setting}

Consider a particle on a 1D lattice with spacing $a = 1$ and discrete time steps $\Delta t = 1$. We also set $\hbar = c = 1$. Position: $x = n$ and time: $t = m$.

The discrete Schr\"odinger equation is:
\begin{equation}
    i(\psi_n(m+1) - \psi_n(m)) = (H\psi)_n(m)
\end{equation}

\subsection{Relativistic Constraints}

As discussed in the introduction, we want a wave equation that is first-order in both time and space, so that (1) the probability density $\rho = \psi^\dagger\psi$ is guaranteed positive, and (2) the equation respects the relativistic symmetry between time and space. Squaring the equation must recover the relativistic dispersion $E^2 = p^2 + m^2$, which forces the coefficients of the spatial and mass terms to be anticommuting matrices rather than ordinary numbers. On our discrete lattice, this means the Hamiltonian must involve Pauli matrices acting on a two-component spinor.

\subsection{The Discrete Dirac Hamiltonian}

To write the Hamiltonian we work in the position representation, where the state at time $m$ is specified by a two-component spinor $\psi_n(m)$ at each site $n$ (we will usually suppress the time index and write $\psi_n$, since the Hamiltonian acts at a single time step). The action of the Hamiltonian on this wavefunction is
\begin{equation}
    (H\psi)_n = \frac{1}{2i}\sigma_x(\psi_{n+1} - \psi_{n-1}) + m\sigma_z\,\psi_n
\end{equation}
The first term is a nearest-neighbor hopping: the value of $\psi$ at site $n$ is coupled to its neighbors at $n \pm 1$ through $\sigma_x$. The second term is an on-site mass that acts through $\sigma_z$.

\begin{physicalinsight}
The appearance of two components is not accidental; relativity requires doubling the degrees of freedom to accommodate both positive and negative energy solutions. On the lattice, this manifests as the two sublattices.
\end{physicalinsight}

\begin{example}{Explicit Spinor Evolution}
Consider the discrete Dirac equation with $m = 0$ (massless case). Writing the spinor as $\psi_n = \begin{pmatrix} \psi_n^A \\ \psi_n^B \end{pmatrix}$, the Hamiltonian becomes
\begin{equation}
    H\psi_n = \frac{1}{2i}\begin{pmatrix} 0 & 1 \\ 1 & 0 \end{pmatrix}\begin{pmatrix} \psi_{n+1}^A - \psi_{n-1}^A \\ \psi_{n+1}^B - \psi_{n-1}^B \end{pmatrix} = \frac{1}{2i}\begin{pmatrix} \psi_{n+1}^B - \psi_{n-1}^B \\ \psi_{n+1}^A - \psi_{n-1}^A \end{pmatrix}
\end{equation}
The two components couple to each other through the hopping terms. Component $A$ at site $n$ evolves based on the values of component $B$ at neighboring sites, and vice versa. This mixing between components is the essence of how the Dirac equation generates spin dynamics.
\end{example}

\subsection{Dispersion Relation}

To find the energy spectrum, we look for stationary states of the form $\psi_n = u\, e^{ikn}$, where $u = \begin{pmatrix} u_1 \\ u_2 \end{pmatrix}$ is a constant spinor and $k$ is the lattice momentum. Substituting into the Hamiltonian,
\begin{equation}
    H(u\,e^{ikn}) = \frac{1}{2i}\sigma_x\bigl(u\,e^{ik(n+1)} - u\,e^{ik(n-1)}\bigr) + m\sigma_z\,u\,e^{ikn}
    = e^{ikn}\left[\frac{e^{ik} - e^{-ik}}{2i}\sigma_x + m\sigma_z\right]u
\end{equation}
Recognizing $(e^{ik} - e^{-ik})/2i = \sin k$, the eigenvalue equation $Hu = Eu$ becomes
\begin{equation}
    \bigl[\sin(k)\,\sigma_x + m\,\sigma_z\bigr]u = Eu
\end{equation}
or in matrix form,
\begin{equation}
    \begin{pmatrix} m & \sin k \\ \sin k & -m \end{pmatrix}\begin{pmatrix} u_1 \\ u_2 \end{pmatrix} = E\begin{pmatrix} u_1 \\ u_2 \end{pmatrix}
\end{equation}
Setting $\det(H - EI) = 0$ gives $(m - E)(-m - E) - \sin^2 k = 0$, so
\begin{equation}
    E(k) = \pm E_k, \qquad E_k \equiv \sqrt{\sin^2(k) + m^2}
\end{equation}

To find the eigenstates, take the first row of the matrix equation: $(m - E)u_1 + \sin(k)\,u_2 = 0$. For the positive-energy branch $E = +E_k$, this gives $u_2 = (E_k - m)u_1/\sin k$, so the unnormalized eigenvector is proportional to $\begin{pmatrix} \sin k \\ E_k - m \end{pmatrix}$. Equivalently, using the second row to eliminate in the other direction, we can write it as $\begin{pmatrix} E_k + m \\ \sin k \end{pmatrix}$, which avoids a singularity at $k = 0$. Normalizing,
\begin{equation}
    u_+(k) = \frac{1}{\sqrt{2E_k(E_k + m)}}\begin{pmatrix} E_k + m \\ \sin k \end{pmatrix}
\end{equation}
where we used $|u_+|^2 = (E_k + m)^2 + \sin^2 k = 2E_k(E_k + m)$. The negative-energy eigenstate is orthogonal to $u_+$:
\begin{equation}
    u_-(k) = \frac{1}{\sqrt{2E_k(E_k + m)}}\begin{pmatrix} -\sin k \\ E_k + m \end{pmatrix}
\end{equation}

In the non-relativistic limit $k \ll 1$ and $m \gg k$, we have $E_k \approx m + k^2/2m$ and $\sin k \approx k$, so $u_+$ approaches $\begin{pmatrix} 1 \\ k/2m \end{pmatrix}$---predominantly upper component, with a small lower component of order $v/c$. In the massless limit $m = 0$, we have $E_k = |\sin k|$ and the eigenstates become $u_\pm = \frac{1}{\sqrt{2}}\begin{pmatrix} 1 \\ \pm\text{sgn}(\sin k) \end{pmatrix}$, which are eigenstates of $\sigma_x$---each component propagates independently in one direction. We will see in section~\ref{sec:chiral_symmetry} that this property is called chirality.

For $k \ll 1$: $E \approx \pm\sqrt{k^2 + m^2}$, recovering the relativistic energy-momentum relation.

Here is an \href{https://claude.ai/public/artifacts/ef6aa097-a196-4178-9d0b-a2f4708766cc}{interactive simulation} that diagonalizes this Hamiltonian on a finite lattice and visualizes the eigenstates as Bloch vectors.

\begin{figure}
    \centering
    \includegraphics[width=1\linewidth]{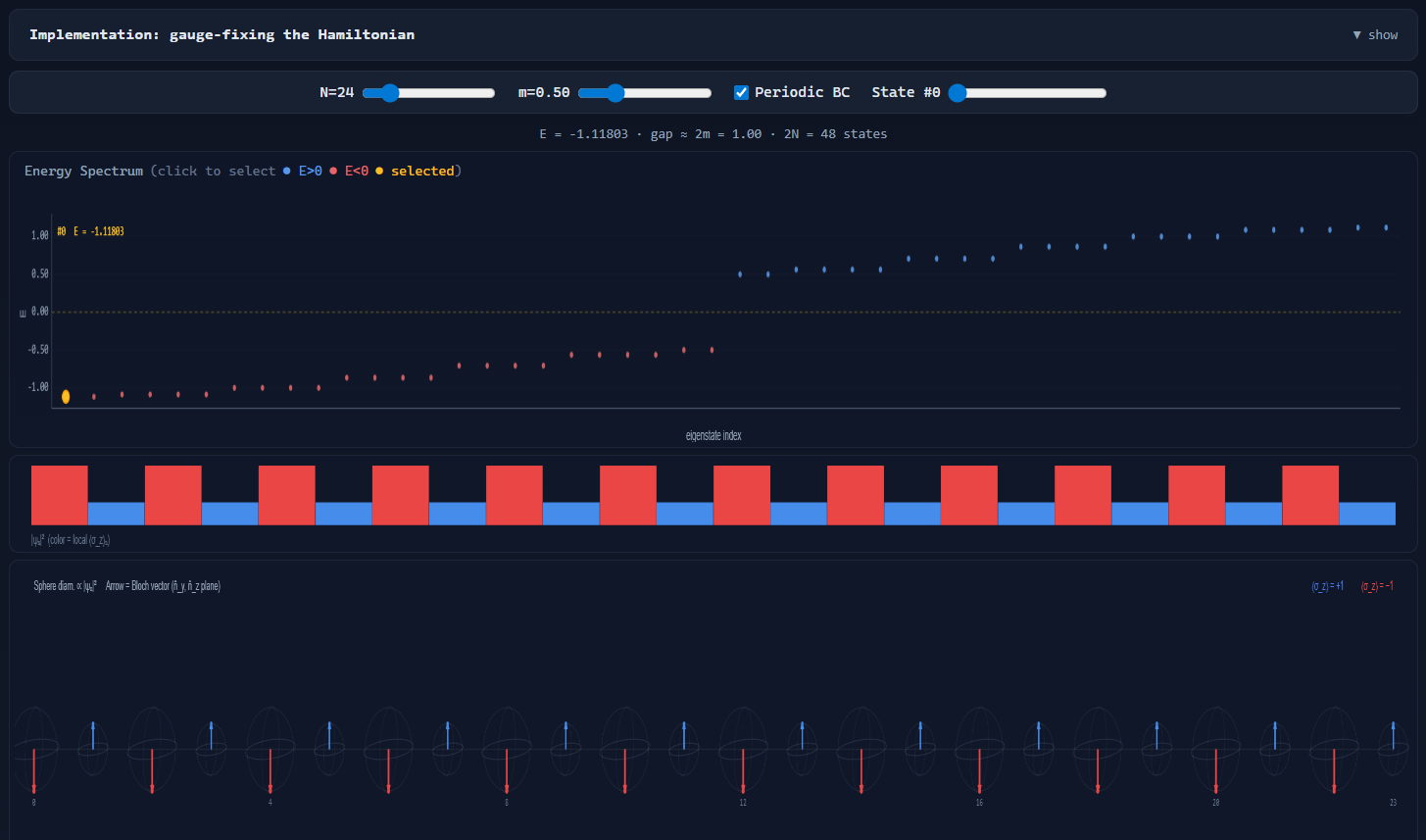}
    \caption{Energy eigenstates of discrete Dirac Hamiltonian.}
    \label{fig:placeholder}
\end{figure}

\section{Discrete Space, Continuous Time}

\subsection{The Lattice Dirac Equation}

With continuous time but discrete space:
\begin{equation}
    i\frac{\partial\psi_n}{\partial t} = \frac{1}{2i}\sigma_x(\psi_{n+1} - \psi_{n-1}) + m\sigma_z\psi_n
\end{equation}

\subsection{Connection to Condensed Matter}

This lattice Dirac equation is not merely a mathematical stepping stone; it directly describes real physical systems. In graphene, the honeycomb lattice structure leads to linear band crossings at two inequivalent points in the Brillouin zone. Near these Dirac points, electrons behave as massless Dirac fermions with $m = 0$. The effective speed of light is replaced by the Fermi velocity $v_F \approx 1/300$.

At the edges of topological insulators, protected surface states arise that obey effective Dirac equations. The bulk of a topological insulator is gapped, but the boundary hosts gapless states that cannot be removed without closing the bulk gap. These edge states are robust against disorder because they are protected by time-reversal symmetry.

The Su-Schrieffer-Heeger (SSH) model provides another realization. This model describes electrons hopping on a dimerized chain with alternating bond strengths. In the topological phase, zero-energy states appear at the ends of the chain, directly analogous to the edge states in the lattice Dirac equation.

\begin{example}{Graphene as a Dirac Material}
In graphene's honeycomb lattice, low-energy excitations near the Dirac points obey:
\begin{equation}
    H_{\text{graphene}} = v_F(\sigma_x k_x + \sigma_y k_y)
\end{equation}
where $v_F \approx 1/300$ is the Fermi velocity. This makes graphene a ``relativistic'' material despite non-relativistic electron speeds.
\end{example}

\subsection{Chiral Symmetry}
\label{sec:chiral_symmetry}

Working in momentum space, we linearize near $k = 0$ (setting $\sin k \approx k$), giving the Hamiltonian as the $2\times 2$ matrix
\begin{equation}
    H(k) = k\,\sigma_x + m\,\sigma_z = \begin{pmatrix} m & k \\ k & -m \end{pmatrix}
\end{equation}
The structure is transparent in the $\sigma_z$ eigenbasis: the kinetic term $k\,\sigma_x$ is purely off-diagonal, coupling sublattice $A$ to $B$ and $B$ to $A$. The mass term $m\sigma_z$ is diagonal, assigning opposite on-site energies to the two sublattices.

Chiral symmetry is the statement that $H(k)$ is purely off-diagonal---equivalently, that $\sigma_z$ anticommutes with $H(k)$. Computing directly,
\begin{equation}
    \{H(k),\,\sigma_z\} = k\{\sigma_x,\sigma_z\} + m\{\sigma_z,\sigma_z\} = 0 + 2m\,I
\end{equation}
The first term vanishes because $\sigma_x$ and $\sigma_z$ anticommute; the second is zero only when $m = 0$. So chiral symmetry holds if and only if $m = 0$. When $m \neq 0$ the mass term is diagonal and commutes with $\sigma_z$, breaking the symmetry. Physically, a nonzero mass makes the two sublattices inequivalent.

Chiral symmetry constrains the spectrum. If $|\chi\rangle$ is an eigenstate of $H(k)$ with energy $E$, then $\sigma_z|\chi\rangle$ is an eigenstate with energy $-E$:
\begin{equation}
    H(k)\,\sigma_z|\chi\rangle = -\sigma_z H(k)|\chi\rangle = -E\bigl(\sigma_z|\chi\rangle\bigr)
\end{equation}
Every nonzero energy eigenvalue is paired with its negative. This is already visible in the dispersion $E(k) = \pm\sqrt{k^2 + m^2}$: the two branches are exactly $\pm E_k$. A zero-energy state satisfies $E = -E = 0$ and needs no partner.

This pairing is what protects zero-energy edge states on a finite chain. Such a state sits at exactly $E = 0$ by the symmetry of the spectrum. Any perturbation that preserves chiral symmetry must preserve the $\pm E$ pairing, so it cannot shift a zero mode to nonzero energy without simultaneously producing a partner for it. On a finite chain with boundaries, the number of unpaired zero modes is a topological invariant tied to the bulk winding number, so no chirally symmetric perturbation can remove them. They can only disappear if two zero modes from opposite ends hybridize, or if the chiral symmetry is broken by a mass term.

\section{The Continuous Dirac Equation in 1+1D}

\subsection{Taking the Continuum Limit}

As $a \to 0$:
\begin{equation}
    \frac{\psi_{n+1} - \psi_{n-1}}{2} \to \frac{\partial\psi}{\partial x}
\end{equation}

The continuum 1+1D Dirac equation becomes:
\begin{equation}
    i\frac{\partial\psi}{\partial t} = \left(-i\sigma_x\frac{\partial}{\partial x} + m\sigma_z\right)\psi
\end{equation}

\begin{example}{Plane Wave Solutions in 1+1D}
Consider a plane wave solution $\psi(x,t) = u e^{i(kx - \omega t)}$ where $u = \begin{pmatrix} u_1 \\ u_2 \end{pmatrix}$ is a constant spinor. Substituting into the 1+1D Dirac equation gives
\begin{equation}
    \omega \begin{pmatrix} u_1 \\ u_2 \end{pmatrix} = \begin{pmatrix} m & k \\ k & -m \end{pmatrix}\begin{pmatrix} u_1 \\ u_2 \end{pmatrix}
\end{equation}
The eigenvalue equation $\det(H - E) = 0$ gives $(m - E)(-m - E) - k^2 = 0$, which yields $E^2 = k^2 + m^2$. This is exactly the relativistic energy-momentum relation.
\end{example}

\subsection{Covariant Form}

The 1+1D Dirac equation as written above treats time and space asymmetrically: the left-hand side has a time derivative while the right-hand side has a spatial derivative, with different matrices ($I$ and $\sigma_x$) multiplying each. This obscures the fact that special relativity demands time and space be treated on equal footing. To make this symmetry manifest, we want to write the equation in a form where $\partial_t$ and $\partial_x$ appear in a single object $\partial_\mu = (\partial_t, \partial_x)$, contracted with a set of matrices $\gamma^\mu$ that encode the algebraic structure.

To do this, multiply both sides of the 1+1D Dirac equation by $\sigma_z$ from the left:
\begin{equation}
    i\sigma_z\partial_t\psi = -i\sigma_z\sigma_x\,\partial_x\psi + m\sigma_z^2\psi = -i(i\sigma_y)\partial_x\psi + m\psi = \sigma_y\partial_x\psi + m\psi
\end{equation}
where we used $\sigma_z\sigma_x = i\sigma_y$. Rearranging into the form $i\gamma^\mu\partial_\mu\psi - m\psi = 0$, we read off $\gamma^0 = \sigma_z$ and $\gamma^1 = i\sigma_y$:
\begin{equation}
    (i\gamma^\mu\partial_\mu - m)\psi = 0
\end{equation}
This is the covariant form. By ``covariant'' we mean that the equation takes the same form in every inertial frame: under a Lorentz transformation the derivative $\partial_\mu$ and the spinor $\psi$ each transform, but because $\gamma^\mu\partial_\mu$ contracts a matrix-valued four-vector with a derivative four-vector, the structure $i\gamma^\mu\partial_\mu\psi - m\psi = 0$ is preserved. The Hamiltonian form, by contrast, treats $\partial_t$ and $\partial_x$ differently, so Lorentz invariance is present but hidden.

But what constraints must the $\gamma^\mu$ matrices satisfy? The answer comes from demanding the correct dispersion relation. Every solution of the Dirac equation should also satisfy the relativistic energy-momentum relation $E^2 = p^2 + m^2$, which in operator form is the Klein-Gordon equation $(\partial_\mu\partial^\mu + m^2)\psi = 0$. To check this, act on the Dirac equation with the operator $(i\gamma^\nu\partial_\nu + m)$:
\begin{equation}
    (i\gamma^\nu\partial_\nu + m)(i\gamma^\mu\partial_\mu - m)\psi = (-\gamma^\nu\gamma^\mu\partial_\nu\partial_\mu - m^2)\psi = 0
\end{equation}
\begin{mathematicalaside}{Covariant and Contravariant Indices}
In special relativity a four-vector carries either an upper (contravariant) or lower (covariant) index. The spacetime position is $x^\mu = (t, \vec{x})$ with an upper index; the associated lower-index object is $x_\mu = (t, -\vec{x})$. The metric tensor $g_{\mu\nu} = \mathrm{diag}(+1,-1,-1,-1)$ converts between them: $v_\mu = g_{\mu\nu}v^\nu$ (lowering), while its inverse $g^{\mu\nu}$ raises, $v^\mu = g^{\mu\nu}v_\nu$. A repeated upper-lower pair is summed by the Einstein convention, $v^\mu w_\mu \equiv \sum_\mu v^\mu w_\mu$, and the result is a Lorentz scalar. The structure is directly analogous to bras and kets: $v^\mu$ is like a ket $|v\rangle$, the metric maps it to the covariant dual $v_\mu$ which is like the bra $\langle v|$, and the contraction $v^\mu w_\mu$ is the inner product.
\end{mathematicalaside}
For this to reduce to $(\partial_\mu\partial^\mu + m^2)\psi = 0$, we need $\gamma^\nu\gamma^\mu\partial_\nu\partial_\mu = g^{\mu\nu}\partial_\mu\partial_\nu$. Since partial derivatives commute ($\partial_\nu\partial_\mu = \partial_\mu\partial_\nu$), only the symmetric part of $\gamma^\nu\gamma^\mu$ contributes, so the condition is
\begin{equation}
    \tfrac{1}{2}(\gamma^\mu\gamma^\nu + \gamma^\nu\gamma^\mu) = g^{\mu\nu} \qquad \text{i.e.} \qquad \{\gamma^\mu, \gamma^\nu\} = 2g^{\mu\nu}
\end{equation}
This is the \emph{Clifford algebra}\label{eq:clifford}, and it is not an assumption we impose by hand---it is forced on us by the requirement that the first-order Dirac equation be consistent with the second-order Klein-Gordon equation. Any set of matrices satisfying this relation will produce the correct relativistic dispersion. The covariant form also makes Lorentz invariance manifest, since $\gamma^\mu\partial_\mu$ transforms as a scalar when $\psi$ transforms appropriately (as we will see in the section on Lorentz covariance). This would be difficult to verify starting from the Hamiltonian form, where time and space play different roles.

\begin{mathematicalaside}{Properties of the Clifford Algebra}
The relation $\{\gamma^\mu, \gamma^\nu\} = 2g^{\mu\nu}$ implies several useful properties.

\textbf{Squares and anticommutativity.} Taking $\mu = \nu$ gives $(\gamma^\mu)^2 = g^{\mu\mu}I$ (no sum): $(\gamma^0)^2 = +I$ and $(\gamma^i)^2 = -I$. For $\mu \neq \nu$, the anticommutator vanishes, so $\gamma^\mu\gamma^\nu = -\gamma^\nu\gamma^\mu$: distinct gamma matrices anticommute.

\textbf{Independent elements.} In $d$ spacetime dimensions, products of gamma matrices generate a $2^d$-dimensional algebra. In 3+1D ($d=4$) the 16 independent elements are $I$, $\gamma^\mu$ (4), $\gamma^\mu\gamma^\nu$ with $\mu < \nu$ (6), $\gamma^\mu\gamma^\nu\gamma^\rho$ (4), and $\gamma^5 \equiv i\gamma^0\gamma^1\gamma^2\gamma^3$ (1). These correspond precisely to the scalar, vector, antisymmetric tensor, pseudovector, and pseudoscalar bilinears of the Dirac theory.

\textbf{Chirality.} The matrix $\gamma^5$ anticommutes with all $\gamma^\mu$: $\{\gamma^5, \gamma^\mu\} = 0$. It therefore commutes with $H$ only when $m = 0$, which is the relativistic generalization of the chiral symmetry we found on the lattice. Its eigenvalues $\pm 1$ define left- and right-handed Weyl spinors.

\textbf{Trace identities.} Since $(\gamma^\mu)^2 = \pm I$ and $\gamma^\mu = -(\gamma^\nu\gamma^\mu)\gamma^\nu/g^{\nu\nu}$ for any fixed $\nu \neq \mu$, one can show $\mathrm{tr}(\gamma^\mu) = 0$, and more generally that the trace of any odd number of gamma matrices vanishes. The key even-product traces are $\mathrm{tr}(\gamma^\mu\gamma^\nu) = 4g^{\mu\nu}$ and $\mathrm{tr}(\gamma^\mu\gamma^\nu\gamma^\rho\gamma^\sigma) = 4(g^{\mu\nu}g^{\rho\sigma} - g^{\mu\rho}g^{\nu\sigma} + g^{\mu\sigma}g^{\nu\rho})$. These identities are the workhorses of QED Feynman diagram calculations.

\textbf{Representation dimension.} In $d$ dimensions the minimal matrix representation has dimension $2^{\lfloor d/2 \rfloor}$. In 1+1D that gives $2\times 2$ matrices (the Pauli matrices suffice); in 3+1D it gives $4\times 4$ matrices, forcing the Dirac spinor to have four components. Different explicit choices of $4\times 4$ matrices satisfying the algebra---the Dirac, Weyl, and Majorana representations---are all related by similarity transformations and describe the same physics.
\end{mathematicalaside}

\begin{example}{Verifying the Gamma Matrix Algebra in 1+1D}
The gamma matrices must satisfy $\{\gamma^\mu, \gamma^\nu\} = 2g^{\mu\nu}$ where $g = \text{diag}(1, -1)$. Let us check: $\gamma^0 = \sigma_z$ and $\gamma^1 = i\sigma_y$. We need $(\gamma^0)^2 = 1$, $(\gamma^1)^2 = -1$, and $\{\gamma^0, \gamma^1\} = 0$. Indeed,
\begin{align}
    (\gamma^0)^2 &= \sigma_z^2 = I \\
    (\gamma^1)^2 &= (i\sigma_y)^2 = -\sigma_y^2 = -I \\
    \gamma^0\gamma^1 + \gamma^1\gamma^0 &= \sigma_z(i\sigma_y) + (i\sigma_y)\sigma_z = i(\sigma_z\sigma_y + \sigma_y\sigma_z) = 0
\end{align}
using the anticommutation of Pauli matrices.
\end{example}

\section{The Dirac Equation in 3+1D}

\subsection{The Need for Four Components}

In 3+1D, we need three anticommuting spatial matrices plus one for mass. The Clifford algebra $\{\gamma^\mu, \gamma^\nu\} = 2g^{\mu\nu}$ requires matrices satisfying $(\gamma^0)^2 = I$, $(\gamma^i)^2 = -I$ for $i = 1,2,3$, and $\gamma^\mu\gamma^\nu = -\gamma^\nu\gamma^\mu$ for $\mu \neq \nu$. No set of 2\ensuremath{\times}2 matrices can satisfy all these conditions because the Pauli matrices plus identity span the space of 2\ensuremath{\times}2 Hermitian matrices, but we need four mutually anticommuting matrices. The minimal representation therefore requires 4\ensuremath{\times}4 matrices, giving the Dirac spinor four components.

\subsection{The Standard Representation}

The gamma matrices can be written compactly as tensor products of Pauli matrices:
\begin{equation}
    \gamma^0 = \sigma_z \otimes I, \qquad \gamma^i = i\sigma_y \otimes \sigma^i
\end{equation}
The first (left) tensor factor acts on particle/antiparticle space; the second (right) factor acts on spin space. Expanding the tensor products gives the standard block form:
\begin{align}
    \gamma^0 &= \begin{pmatrix} I & 0 \\ 0 & -I \end{pmatrix}, \qquad
    \gamma^i = \begin{pmatrix} 0 & \sigma^i \\ -\sigma^i & 0 \end{pmatrix}
\end{align}
This structure also clarifies the connection to the 1+1D case: the 1+1D gamma matrices $\gamma^0 = \sigma_z$ and $\gamma^1 = i\sigma_y$ are precisely the left tensor factors, with the spin space being trivial.

The Clifford algebra (equation~\ref{eq:clifford}), $\{\gamma^\mu, \gamma^\nu\} = 2g^{\mu\nu}I$, is straightforward to verify in the tensor product form: $(\gamma^0)^2 = \sigma_z^2 \otimes I = I$, $(\gamma^i)^2 = (i\sigma_y)^2 \otimes (\sigma^i)^2 = (-I)\otimes I = -I$, and anticommutativity between distinct $\gamma^\mu$ follows from the Pauli anticommutation relations in each factor.

\begin{example}{Explicit Verification of the Clifford Algebra}
Let us verify that $\{\gamma^1, \gamma^2\} = 0$. We compute
\begin{equation}
    \gamma^1\gamma^2 = \begin{pmatrix} 0 & \sigma^1 \\ -\sigma^1 & 0 \end{pmatrix}\begin{pmatrix} 0 & \sigma^2 \\ -\sigma^2 & 0 \end{pmatrix} = \begin{pmatrix} -\sigma^1\sigma^2 & 0 \\ 0 & -\sigma^1\sigma^2 \end{pmatrix}
\end{equation}
and similarly
\begin{equation}
    \gamma^2\gamma^1 = \begin{pmatrix} -\sigma^2\sigma^1 & 0 \\ 0 & -\sigma^2\sigma^1 \end{pmatrix}
\end{equation}
Since $\sigma^1\sigma^2 = i\sigma^3 = -\sigma^2\sigma^1$, we have $\gamma^1\gamma^2 = -\gamma^2\gamma^1$, confirming $\{\gamma^1, \gamma^2\} = 0$.
\end{example}

\subsection{The Full Dirac Equation}

The Dirac equation in covariant form:
\begin{equation}
    (i\gamma^\mu\partial_\mu - m)\psi = 0
\end{equation}

Or explicitly:
\begin{equation}
    i\frac{\partial\psi}{\partial t} = \left(-i\vec{\alpha}\cdot\nabla + \beta m\right)\psi
\end{equation}
where $\alpha^i = \gamma^0\gamma^i$ and $\beta = \gamma^0$.

\begin{example}{Computing the Alpha Matrices}
From the gamma matrices, we find
\begin{equation}
    \alpha^i = \gamma^0\gamma^i = \begin{pmatrix} I & 0 \\ 0 & -I \end{pmatrix}\begin{pmatrix} 0 & \sigma^i \\ -\sigma^i & 0 \end{pmatrix} = \begin{pmatrix} 0 & \sigma^i \\ \sigma^i & 0 \end{pmatrix}
\end{equation}
These satisfy $(\alpha^i)^2 = I$ and $\alpha^i\alpha^j + \alpha^j\alpha^i = 2\delta^{ij}I$ for $i \neq j$. Combined with $\beta = \gamma^0$, we have $\alpha^i\beta + \beta\alpha^i = 0$, which ensures that the Hamiltonian $H = \vec{\alpha}\cdot\vec{p} + \beta m$ correctly reproduces the relativistic dispersion when squared.
\end{example}

\section{Solutions of the Dirac Equation}

\subsection{Plane Wave Solutions}

For a free particle with four-momentum $p^\mu = (E, \vec{p})$, we try the plane wave ansatz
\begin{equation}
    \psi(x) = u(p)e^{-ip\cdot x}
\end{equation}
where $p\cdot x = Et - \vec{p}\cdot\vec{x}$ and $u(p)$ is a constant four-component spinor. Substituting into $(i\gamma^\mu\partial_\mu - m)\psi = 0$, the derivative acts only on the exponential: $\partial_\mu e^{-ip\cdot x} = -ip_\mu e^{-ip\cdot x}$. The exponential cancels, leaving the algebraic condition
\begin{equation}
    (\gamma^\mu p_\mu - m)u(p) = 0
\end{equation}
This $4\times 4$ matrix equation has nontrivial solutions only when $\det(\gamma^\mu p_\mu - m) = 0$, which enforces the relativistic dispersion $E^2 = \vec{p}^2 + m^2$.

\subsection{Positive and Negative Energy Solutions}

For each momentum $\vec{p}$, the Dirac equation admits four independent solutions. Two of these correspond to positive energy $E = +\sqrt{\vec{p}^2 + m^2}$ and describe a particle with spin up or spin down. The remaining two correspond to negative energy $E = -\sqrt{\vec{p}^2 + m^2}$ and similarly come in spin up and spin down varieties. The existence of negative energy solutions was initially troubling, as it seemed to imply that particles could radiate away unlimited amounts of energy by transitioning to ever more negative energy states.

\begin{keyidea}{The Dirac Sea}
Dirac interpreted negative energy states as a filled ``sea.'' Holes in this sea behave as antiparticles with positive energy and opposite charge, predicting the positron before its discovery.
\end{keyidea}

\begin{example}{Energy of a Relativistic Electron}
Consider a particle with momentum $p = m$ (a moderately relativistic case). The energy is
\begin{equation}
    E = \pm\sqrt{p^2 + m^2} = \pm\sqrt{2}\,m
\end{equation}
The positive energy solution describes a free particle with kinetic energy $(\sqrt{2} - 1)m$. The negative energy solution, in modern interpretation, describes an antiparticle state.
\end{example}

\subsection{Spinor Solutions}

The standard normalization convention is $u^{(r)\dagger}u^{(s)} = 2E\,\delta^{rs}$, chosen because $2E$ transforms as the time component of a four-vector, making the normalization Lorentz covariant. Equivalently, $\bar{u}^{(r)}u^{(s)} = 2m\,\delta^{rs}$ where $\bar{u} = u^\dagger\gamma^0$. In the rest frame ($\vec{p} = 0$, $E = m$), this requires $u^\dagger u = 2m$, giving
\begin{equation}
    u^{(1)} = \sqrt{2m}\begin{pmatrix} 1 \\ 0 \\ 0 \\ 0 \end{pmatrix}, \quad
    u^{(2)} = \sqrt{2m}\begin{pmatrix} 0 \\ 1 \\ 0 \\ 0 \end{pmatrix}
\end{equation}

\begin{example}{Boosted Spinors}
For a particle moving with momentum $\vec{p} = p\hat{z}$, we solve $(\gamma^\mu p_\mu - m)u = 0$ directly. Using the standard representation with $p^\mu = (E, 0, 0, p)$ and $p_\mu = (E, 0, 0, -p)$,
\begin{equation}
    \gamma^\mu p_\mu = \gamma^0 E - \gamma^3 p = \begin{pmatrix} E & -p\sigma^3 \\ p\sigma^3 & -E \end{pmatrix}
\end{equation}
where we used the block forms $\gamma^0 = \mathrm{diag}(I,-I)$ and $\gamma^3 = \bigl(\begin{smallmatrix}0&\sigma^3\\-\sigma^3&0\end{smallmatrix}\bigr)$. Writing $u = \bigl(\begin{smallmatrix}\phi\\\chi\end{smallmatrix}\bigr)$, the equation $(\gamma^\mu p_\mu - m)u = 0$ splits into two coupled equations:
\begin{align*}
    (E - m)\phi &= p\sigma^3\chi \\
    p\sigma^3\phi &= (E + m)\chi
\end{align*}
The second equation gives $\chi = \frac{p\sigma^3}{E+m}\phi$. For spin up, $\phi = \bigl(\begin{smallmatrix}1\\0\end{smallmatrix}\bigr)$, and since $\sigma^3\bigl(\begin{smallmatrix}1\\0\end{smallmatrix}\bigr) = \bigl(\begin{smallmatrix}1\\0\end{smallmatrix}\bigr)$, we get $\chi = \frac{p}{E+m}\bigl(\begin{smallmatrix}1\\0\end{smallmatrix}\bigr)$. Assembling and normalizing gives
\begin{equation}
    u^{(1)}(p) = \sqrt{E + m}\begin{pmatrix} 1 \\ 0 \\ \frac{p}{E + m} \\ 0 \end{pmatrix}
\end{equation}
where $E = \sqrt{p^2 + m^2}$. In the non-relativistic limit $p \ll m$, $E \approx m$ and the lower component $p/(E+m) \approx p/2m \ll 1$, so the spinor is predominantly upper. In the ultra-relativistic limit $p \gg m$, $E \approx p$ and $p/(E+m) \approx 1$, so upper and lower components are comparable.
\end{example}

\section{Properties of the Dirac Equation}

\subsection{Current Conservation}

The Dirac equation implies conservation of probability current:
\begin{equation}
    \partial_\mu j^\mu = 0
\end{equation}
where $j^\mu = \bar{\psi}\gamma^\mu\psi$ and $\bar{\psi} = \psi^\dagger\gamma^0$.

\begin{example}{The Probability Current Components:}

The time component of the current is $j^0 = \bar{\psi}\gamma^0\psi = \psi^\dagger\gamma^0\gamma^0\psi = \psi^\dagger\psi$, which is manifestly positive. This resolves the problem that plagued the Klein-Gordon equation. The spatial components are $j^i = \bar{\psi}\gamma^i\psi = \psi^\dagger\gamma^0\gamma^i\psi = \psi^\dagger\alpha^i\psi$. For a plane wave solution, the current $\vec{j}$ points in the direction of the particle's momentum, as expected for probability flow.
\end{example}

\subsection{Spin and Magnetic Moment}

\subsubsection{Spin-1/2 from the Matrix Structure}

Ordinary orbital angular momentum $\vec{L} = \vec{r}\times\vec{p}$ is not conserved for the Dirac Hamiltonian $H = \vec{\alpha}\cdot\vec{p} + \beta m$. This is not a defect of the theory; it is a clue. In non-relativistic quantum mechanics, $\vec{L}$ fails to be conserved precisely when there is an additional angular momentum degree of freedom---spin---that must be included. The question is whether the Dirac equation supplies a natural candidate for that extra piece.

The answer is yes. Each $4\times 4$ matrix in the Dirac theory is built from $2\times 2$ blocks. Define the spin operator
\begin{equation}
    \vec{S} = \frac{1}{2}\vec{\Sigma}, \qquad \Sigma^i = \begin{pmatrix}\sigma^i & 0 \\ 0 & \sigma^i\end{pmatrix}
\end{equation}
This is the unique rotationally natural $4\times 4$ operator that (a) acts within each $2\times 2$ block without mixing upper and lower components, and (b) satisfies the angular momentum algebra $[S^i, S^j] = i\epsilon^{ijk}S^k$. Its eigenvalues are $\pm\frac{1}{2}$, inherited from $\sigma^i$. The total angular momentum is $\vec{J} = \vec{L} + \vec{S}$.

\begin{example}{$[H, J^z] = 0$: Spin-1/2 is Not a Postulate}
We show that $[H, L^z] \neq 0$ but $[H, J^z] = [H, L^z + S^z] = 0$, establishing that $\vec{J}$ is the conserved total angular momentum and that the Dirac equation necessarily describes spin-1/2 particles.

\textbf{Step 1: $[H, L^z]$ is nonzero.} With $L^z = xp_y - yp_x$ and using $[p_i, x_j] = -i\delta_{ij}$, the mass term $\beta m$ commutes with $L^z$. For the kinetic terms, using $[p_x, x] = -i$ and $[p_y, y] = -i$ (all other pairs commute):
\begin{align}
    [p_x,\, xp_y - yp_x] &= [p_x, x]p_y = -ip_y \\
    [p_y,\, xp_y - yp_x] &= -[p_y, y]p_x = +ip_x
\end{align}
Therefore $[H, L^z] = \alpha^x(-ip_y) + \alpha^y(ip_x) = i(\alpha^y p_x - \alpha^x p_y) \neq 0$.

\textbf{Step 2: $[H, S^z]$ exactly cancels.} We need $[\alpha^i, \Sigma^z]$. Using $\alpha^i = \begin{pmatrix}0 & \sigma^i \\ \sigma^i & 0\end{pmatrix}$:
\begin{align}
    [\alpha^i, \Sigma^z] &= \begin{pmatrix}0 & [\sigma^i, \sigma^z] \\ {}[\sigma^i, \sigma^z] & 0\end{pmatrix}
\end{align}
From the Pauli commutation relation $[\sigma^i, \sigma^j] = 2i\epsilon^{ijk}\sigma^k$:
\begin{align}
    [\sigma^x, \sigma^z] = -2i\sigma^y \quad&\Rightarrow\quad [\alpha^x, \Sigma^z] = -2i\alpha^y \\
    [\sigma^y, \sigma^z] = +2i\sigma^x \quad&\Rightarrow\quad [\alpha^y, \Sigma^z] = +2i\alpha^x
\end{align}
Assembling:
\begin{align}
    [H, S^z] &= \tfrac{1}{2}p_x[\alpha^x,\Sigma^z] + \tfrac{1}{2}p_y[\alpha^y,\Sigma^z] \notag\\
    &= \tfrac{1}{2}p_x(-2i\alpha^y) + \tfrac{1}{2}p_y(+2i\alpha^x) = -i(\alpha^y p_x - \alpha^x p_y)
\end{align}
This is exactly $-[H, L^z]$. Therefore $[H, J^z] = [H, L^z] + [H, S^z] = 0$. The same argument applies to $J^x$ and $J^y$. Spin-1/2 emerges from the four-component structure of the Dirac equation---it is not put in by hand.
\end{example}

\subsubsection{The Magnetic Moment and $g = 2$}

To see how a magnetic moment emerges from the Dirac equation, include an electromagnetic field via minimal coupling $\partial_\mu \to \partial_\mu + ieA_\mu$, or equivalently $\vec{p} \to \vec{\Pi} \equiv \vec{p} - e\vec{A}$. Writing $\psi = \bigl(\begin{smallmatrix}\phi\\\chi\end{smallmatrix}\bigr)$ and using the standard-representation Hamiltonian form, the Dirac equation becomes
\begin{align*}
    (E - m - e\Phi)\phi &= \vec{\sigma}\cdot\vec{\Pi}\,\chi \\
    (E + m - e\Phi)\chi &= \vec{\sigma}\cdot\vec{\Pi}\,\phi
\end{align*}
In the non-relativistic limit, write $E = m + \varepsilon$ with $\varepsilon \ll m$, and neglect $e\Phi$ relative to $2m$ in the second equation: $\chi \approx \frac{\vec{\sigma}\cdot\vec{\Pi}}{2m}\phi$. Substituting into the first equation gives
\begin{equation}
    \varepsilon\phi = \left[\frac{(\vec{\sigma}\cdot\vec{\Pi})^2}{2m} + e\Phi\right]\phi
\end{equation}
The key step is expanding $(\vec{\sigma}\cdot\vec{\Pi})^2$ using the identity $(\vec{\sigma}\cdot\vec{A})(\vec{\sigma}\cdot\vec{B}) = \vec{A}\cdot\vec{B} + i\vec{\sigma}\cdot(\vec{A}\times\vec{B})$:
\begin{equation}
    (\vec{\sigma}\cdot\vec{\Pi})^2 = \vec{\Pi}^2 + i\vec{\sigma}\cdot(\vec{\Pi}\times\vec{\Pi})
\end{equation}
Since $\vec{\Pi} = \vec{p} - e\vec{A}$ involves an operator $\vec{p} = -i\nabla$, the cross product $\vec{\Pi}\times\vec{\Pi}$ does not vanish: the commutator $[\Pi_i, \Pi_j] = -ie(\partial_i A_j - \partial_j A_i) = -ie\epsilon_{ijk}B_k$ gives $i\vec{\sigma}\cdot(\vec{\Pi}\times\vec{\Pi}) = -e\vec{\sigma}\cdot\vec{B}$. Therefore
\begin{equation}
    \varepsilon\phi = \left[\frac{(\vec{p} - e\vec{A})^2}{2m} - \frac{e}{2m}\vec{\sigma}\cdot\vec{B} + e\Phi\right]\phi
\end{equation}
This is the Pauli equation. The magnetic interaction term is $-\frac{e}{2m}\vec{\sigma}\cdot\vec{B} = -2\cdot\frac{e}{2m}\vec{S}\cdot\vec{B}$ (since $\vec{S} = \frac{1}{2}\vec{\sigma}$), giving the magnetic moment
\begin{equation}
    \vec{\mu} = \frac{e}{2m}\vec{\Sigma}, \qquad \vec{\Sigma} = \begin{pmatrix} \vec{\sigma} & 0 \\ 0 & \vec{\sigma} \end{pmatrix}
\end{equation}
with $g = 2$: the magnetic moment is twice the classical expectation $e\vec{S}/m$.

\begin{example}{The Electron g-Factor from the $\vec{\sigma}$ Identity}
The general form of the magnetic moment interaction is $-g\frac{e}{2m}\vec{S}\cdot\vec{B}$, where $g$ is the g-factor and $\vec{S} = \frac{1}{2}\vec{\sigma}$. The $g=2$ prediction follows directly from the Dirac algebra. The term $i\vec{\sigma}\cdot(\vec{\Pi}\times\vec{\Pi})$ arises from the non-commutativity of $\vec{\Pi}$: using $[\Pi_i,\Pi_j] = -ie\epsilon_{ijk}B_k$,
\begin{equation}
    i\sigma_k(\vec{\Pi}\times\vec{\Pi})_k = i\sigma_k\,\epsilon_{kij}\,\tfrac{1}{2}[\Pi_i,\Pi_j] = i\sigma_k\,\epsilon_{kij}\,\tfrac{1}{2}(-ie\epsilon_{ijk}B_k) = -e\sigma_k B_k
\end{equation}
So $(\vec{\sigma}\cdot\vec{\Pi})^2 = \vec{\Pi}^2 - e\vec{\sigma}\cdot\vec{B}$, and the Hamiltonian contains $-\frac{e}{2m}\vec{\sigma}\cdot\vec{B} = -\frac{e}{m}\vec{S}\cdot\vec{B} = -g\frac{e}{2m}\vec{S}\cdot\vec{B}$ with $g=2$. This factor of 2 is a direct consequence of the first-order structure of the Dirac equation: it would not appear in a na\"ive second-order theory. Experimentally, $g \approx 2.002319$; the small deviation is explained by QED loop corrections.
\end{example}

\section{Lorentz Covariance}

\subsection{Spinor Transformations}

Under a Lorentz transformation $x^\mu \to \Lambda^\mu_\nu x^\nu$:
\begin{equation}
    \psi(x) \to S(\Lambda)\psi(\Lambda^{-1}x)
\end{equation}
where $S(\Lambda)$ is the spinor representation.

\begin{example}{Rotation of a Dirac Spinor}
Under a rotation by angle $\theta$ about the $z$-axis, the spinor transformation matrix is
\begin{equation}
    S(R_z(\theta)) = \exp\left(-\frac{i\theta}{2}\Sigma^3\right) = \begin{pmatrix} e^{-i\theta/2} & 0 & 0 & 0 \\ 0 & e^{i\theta/2} & 0 & 0 \\ 0 & 0 & e^{-i\theta/2} & 0 \\ 0 & 0 & 0 & e^{i\theta/2} \end{pmatrix}
\end{equation}
A rotation by $2\pi$ gives $S(R_z(2\pi)) = -I$, so the spinor changes sign. This is the hallmark of spin-1/2: a $360^\circ$ rotation produces a phase of $-1$, and only a $720^\circ$ rotation returns the spinor to its original value.
\end{example}

\subsection{Bilinear Covariants}

From a Dirac spinor $\psi$, we can construct quantities that transform in well-defined ways under Lorentz transformations. The scalar $\bar{\psi}\psi$ is invariant under Lorentz transformations and appears in the mass term of the Lagrangian. The vector $\bar{\psi}\gamma^\mu\psi$ transforms as a four-vector and gives the probability current. The antisymmetric tensor $\bar{\psi}\sigma^{\mu\nu}\psi$, where $\sigma^{\mu\nu} = \frac{i}{2}[\gamma^\mu, \gamma^\nu]$, has six independent components and relates to the electromagnetic field coupling. The pseudovector (or axial vector) $\bar{\psi}\gamma^\mu\gamma^5\psi$ transforms as a four-vector but picks up an extra sign under parity. The pseudoscalar $\bar{\psi}\gamma^5\psi$, where $\gamma^5 = i\gamma^0\gamma^1\gamma^2\gamma^3$, is invariant under proper Lorentz transformations but changes sign under parity.

\begin{example}{The Pseudoscalar and Parity}
Under parity, spatial coordinates flip sign: $(t, \vec{x}) \to (t, -\vec{x})$. The spinor transforms as $\psi \to \gamma^0\psi$. Therefore
\begin{equation}
    \bar{\psi}\gamma^5\psi \to (\psi^\dagger\gamma^0\gamma^0)\gamma^5(\gamma^0\psi) = \psi^\dagger\gamma^5\gamma^0\psi = -\psi^\dagger\gamma^0\gamma^5\psi = -\bar{\psi}\gamma^5\psi
\end{equation}
using $\{\gamma^5, \gamma^0\} = 0$. The pseudoscalar changes sign under parity, which is why it is called ``pseudo.''
\end{example}

\section{The Dirac Equation in Electromagnetic Fields}

\subsection{Minimal Coupling}

Including electromagnetic fields via minimal coupling:
\begin{equation}
    (i\gamma^\mu D_\mu - m)\psi = 0
\end{equation}
where $D_\mu = \partial_\mu + ieA_\mu$.

\begin{example}{The Dirac Equation in a Uniform Magnetic Field}
Consider a magnetic field $\vec{B} = B\hat{z}$ with vector potential $\vec{A} = \frac{1}{2}\vec{B} \times \vec{r} = \frac{B}{2}(-y, x, 0)$. The Dirac Hamiltonian becomes
\begin{equation}
    H = \vec{\alpha}\cdot(\vec{p} - e\vec{A}) + \beta m
\end{equation}
In the non-relativistic limit, this reduces to the Pauli Hamiltonian with the interaction term $-\vec{\mu}\cdot\vec{B} = -g\frac{e}{2m}\vec{S}\cdot\vec{B}$. With $g = 2$, the spin-up and spin-down states split by $\Delta E = \frac{eB}{m}$, giving the electron spin resonance frequency.
\end{example}

\subsection{The Hydrogen Atom Revisited}

The Dirac equation applied to the hydrogen atom yields the fine structure splitting automatically, without needing to add it as a perturbation. The spin-orbit coupling, which in the non-relativistic treatment must be introduced by hand, emerges naturally from the relativistic formulation. The energy levels depend on both $n$ and $j$ (total angular momentum), leading to the fine structure formula
\begin{equation}
    E_{n,j} = m\left[1 + \left(\frac{\alpha}{n - (j + 1/2) + \sqrt{(j+1/2)^2 - \alpha^2}}\right)^2\right]^{-1/2}
\end{equation}
where $\alpha \approx 1/137$ is the fine structure constant. This formula agrees with experiment to high precision. A small remaining discrepancy, the Lamb shift, is explained by quantum electrodynamic effects involving virtual photon exchange.

\begin{example}{Fine Structure Splitting in Hydrogen}
For the $n = 2$ level, the Dirac equation predicts that states with $j = 1/2$ (the $2S_{1/2}$ and $2P_{1/2}$ states) are degenerate, while $j = 3/2$ (the $2P_{3/2}$ state) has higher energy. The splitting between $j = 1/2$ and $j = 3/2$ is approximately $\frac{\alpha^2}{16}\times 13.6$ eV $\approx 4.5 \times 10^{-5}$ eV, corresponding to a wavelength splitting of about 0.016 nm in the Balmer-$\alpha$ line at 656 nm. This prediction matched experimental observations and confirmed the validity of the relativistic treatment.
\end{example}

\section{Modern Applications}

\subsection{Topological Insulators}

The surface states of 3D topological insulators obey an effective 2D Dirac equation:
\begin{equation}
    H_{\text{surface}} = v_F(\vec{\sigma} \times \vec{k})_z
\end{equation}
These surface states have several remarkable properties. The spin direction is locked perpendicular to the momentum, creating a ``spin texture'' in momentum space. Backscattering is forbidden because reversing momentum would require flipping spin, which is protected by time-reversal symmetry. This leads to robust metallic conduction on the surface even when the bulk is insulating.

\begin{example}{Spin-Momentum Locking}
For a surface state with momentum $\vec{k} = k(\cos\phi, \sin\phi)$, the Hamiltonian gives
\begin{equation}
    H = v_F(k_x\sigma_y - k_y\sigma_x) = v_F k(-\sin\phi\,\sigma_x + \cos\phi\,\sigma_y)
\end{equation}
The eigenstates have spin pointing perpendicular to $\vec{k}$ in the surface plane. An electron moving in the $+x$ direction has spin pointing in the $+y$ direction, while an electron moving in the $-x$ direction has spin pointing in the $-y$ direction. To backscatter (reverse $k_x$), the spin would need to flip, but non-magnetic impurities cannot flip spin, so backscattering is suppressed.
\end{example}

\subsection{Weyl Semimetals}

Materials with separated Weyl points host chiral fermions obeying:
\begin{equation}
    H_{\text{Weyl}} = \pm v_F\vec{\sigma}\cdot\vec{k}
\end{equation}
The $\pm$ sign defines the chirality of the Weyl node. Unlike Dirac points, which can gap out, Weyl points are topologically stable because they act as monopoles of Berry curvature in momentum space. They can only be removed by annihilating with a Weyl point of opposite chirality.

\begin{example}{Chiral Anomaly in Weyl Semimetals}
In parallel electric and magnetic fields, charge is pumped between Weyl nodes of opposite chirality. When $\vec{E} \parallel \vec{B}$, the zeroth Landau level at each Weyl point carries current in a direction determined by the chirality. This leads to a chiral current
\begin{equation}
    \frac{\partial n_R}{\partial t} - \frac{\partial n_L}{\partial t} = \frac{e^2}{2\pi^2}\vec{E}\cdot\vec{B}
\end{equation}
where $n_R$ and $n_L$ are the densities at right- and left-handed nodes. This chiral anomaly, a quantum field theory phenomenon, manifests as negative magnetoresistance in Weyl semimetals: the resistance decreases when magnetic field is applied parallel to the current.
\end{example}

\section{Chapter Summary}

The Dirac equation unifies quantum mechanics and special relativity in a remarkably elegant framework. We have seen that discrete versions of the equation connect directly to condensed matter physics, describing electrons in graphene and edge states in topological materials. The requirement of relativistic covariance forces the wave function to be a spinor with two components in 1+1 dimensions and four components in 3+1 dimensions.

The equation's predictive power is striking. It naturally incorporates spin-1/2 without any additional assumptions and predicts the electron's magnetic moment with the correct g-factor of 2. The existence of negative energy solutions led Dirac to propose the filled sea interpretation, which correctly predicted the existence of antiparticles before the positron was discovered experimentally.

When applied to the hydrogen atom, the Dirac equation automatically produces the fine structure splitting that must be added perturbatively in non-relativistic treatments. The relativistic framework provides the foundation for quantum electrodynamics and modern quantum field theory.

Contemporary applications extend beyond particle physics to condensed matter systems. Surface states of topological insulators obey effective Dirac equations with spin-momentum locking and topological protection. Weyl semimetals host chiral fermions that exhibit the quantum chiral anomaly. These connections demonstrate that the Dirac equation, nearly a century after its discovery, remains central to our understanding of quantum phenomena across multiple fields of physics.

This relativistic framework prepares us for exploring symmetries and conservation laws in the next chapter.

\input{book_problems/ch11_problems.tex}

\section*{References and Further Reading}
\addcontentsline{toc}{section}{References and Further Reading}

\begin{description}
\item[Dirac, P.~A.~M.] ``The quantum theory of the electron.'' \emph{Proceedings of the Royal Society A} \textbf{117}, 610--624 (1928). \href{https://doi.org/10.1098/rspa.1928.0023}{doi:10.1098/rspa.1928.0023}. The original paper. Reads remarkably well today and is the cleanest introduction to the linearization argument that produces the gamma matrices.

\item[Bjorken, J.~D., and Drell, S.~D.] \emph{Relativistic Quantum Mechanics}. McGraw-Hill, 1964. The classic graduate textbook on the Dirac equation as a single-particle wave equation, before second quantization. The natural sequel to this chapter.

\item[Greiner, W.] \emph{Relativistic Quantum Mechanics: Wave Equations}, 3rd ed. Springer, 2000. Worked-example-heavy presentation; particularly useful for spinor algebra, Lorentz covariance, and the hydrogen atom solution.

\item[Peskin, M.~E., and Schroeder, D.~V.] \emph{An Introduction to Quantum Field Theory}. Westview Press, 1995. Chapter 3 develops the Dirac equation as the prelude to QFT and explains how the negative-energy solutions are reinterpreted as antiparticles via second quantization.

\item[Hasan, M.~Z., and Kane, C.~L.] ``Colloquium: Topological insulators.'' \emph{Reviews of Modern Physics} \textbf{82}, 3045--3067 (2010). \href{https://doi.org/10.1103/RevModPhys.82.3045}{doi:10.1103/RevModPhys.82.3045}. The standard accessible review; Section IV connects the surface Dirac Hamiltonian discussed in this chapter to its experimental signatures.

\item[Armitage, N.~P., Mele, E.~J., and Vishwanath, A.] ``Weyl and Dirac semimetals in three-dimensional solids.'' \emph{Reviews of Modern Physics} \textbf{90}, 015001 (2018). \href{https://doi.org/10.1103/RevModPhys.90.015001}{doi:10.1103/RevModPhys.90.015001}. Comprehensive review of Weyl and Dirac semimetals; the natural follow-up to the Weyl-semimetal section above and the chiral anomaly example.
\end{description}

%% file: book_problems/ch11_problems.tex
\section{Problems}
\setcounter{hwproblem}{0}

\problem{Klein-Gordon Probability Current}
The Klein-Gordon equation is $(\partial_t^2 - \nabla^2 + m^2)\phi = 0$.
\begin{enumerate}[label=(\alph*)]
    \item Multiply the Klein-Gordon equation by $\phi^*$ and subtract the complex conjugate to derive the continuity equation $\partial_t \rho + \nabla\cdot\vec{j} = 0$ with $\rho = \frac{i}{2m}(\phi^*\partial_t\phi - \phi\,\partial_t\phi^*)$ and identify $\vec{j}$.
    \item For the positive-frequency plane wave $\phi = e^{i(\vec{k}\cdot\vec{x} - \omega t)}$ with $\omega = +\sqrt{k^2 + m^2}$, show that $\rho = \omega/m > 0$.
    \item For the negative-frequency solution $\phi = e^{i(\vec{k}\cdot\vec{x} + \omega t)}$, show that $\rho < 0$. Why does this rule out $\rho$ as a probability density?
    \item In the non-relativistic limit $k \ll m$, show that $\omega \approx m + k^2/2m$ and that $\rho \approx |\phi|^2$.
    \item Why is the Klein-Gordon equation second-order in time, and how does this lead to the problem with $\rho$? How does the Dirac equation avoid this issue?
\end{enumerate}

\problem{Discrete Dirac Dispersion}
The discrete 1D Dirac Hamiltonian acts on two-component spinors via $(H\psi)_n = \frac{1}{2i}\sigma_x(\psi_{n+1} - \psi_{n-1}) + m\sigma_z\,\psi_n$.
\begin{enumerate}[label=(\alph*)]
    \item Substitute the Bloch ansatz $\psi_n = u\,e^{ikn}$ and show that the eigenvalue equation reduces to $[\sin(k)\,\sigma_x + m\,\sigma_z]\,u = E\,u$.
    \item Write this as a $2\times 2$ matrix equation and use $\det(H - EI) = 0$ to derive $E(k) = \pm\sqrt{\sin^2(k) + m^2}$.
    \item Show that the normalized positive-energy eigenstate is $u_+(k) = \frac{1}{\sqrt{2E_k(E_k + m)}}\begin{pmatrix} E_k + m \\ \sin k \end{pmatrix}$ where $E_k = \sqrt{\sin^2 k + m^2}$.
    \item In the limit $k \ll 1$ and $m \gg k$, expand $E_k$ and show that $E_k \approx m + k^2/2m$.
    \item For $m = 0$, find the two eigenstates and show each is an eigenstate of $\sigma_x$. What does this tell you about the two spinor components?
\end{enumerate}

\problem{Chiral Symmetry and Zero Modes}
Working in momentum space with the linearized Hamiltonian $H(k) = k\,\sigma_x + m\,\sigma_z$:
\begin{enumerate}[label=(\alph*)]
    \item Compute $\{\sigma_z, H(k)\}$ and show that it equals $2m\,I$. Conclude that chiral symmetry holds if and only if $m = 0$.
    \item Assuming $m = 0$, show that if $|\chi\rangle$ is an eigenstate of $H(k)$ with energy $E$, then $\sigma_z|\chi\rangle$ is an eigenstate with energy $-E$.
    \item A zero-energy state satisfies $E = 0$. What is its partner under the pairing argument? Why can chiral symmetry accommodate unpaired zero modes?
    \item On a finite lattice, what boundary condition is required to support a zero mode at one end of the chain?
    \item Explain why a small mass perturbation breaks chiral symmetry and generically lifts zero modes away from $E = 0$.
\end{enumerate}

\problem{Clifford Algebra in 3+1D}
In the standard representation, the gamma matrices are $\gamma^0 = \begin{pmatrix} I & 0 \\ 0 & -I \end{pmatrix}$ and $\gamma^i = \begin{pmatrix} 0 & \sigma^i \\ -\sigma^i & 0 \end{pmatrix}$ for $i = 1, 2, 3$.
\begin{enumerate}[label=(\alph*)]
    \item Verify directly that $(\gamma^0)^2 = I$, $(\gamma^i)^2 = -I$, and $\{\gamma^0, \gamma^i\} = 0$.
    \item Verify that $\{\gamma^1, \gamma^2\} = 0$ using the block matrix form and the Pauli anticommutation relation.
    \item Show that the Clifford algebra $\{\gamma^\mu, \gamma^\nu\} = 2g^{\mu\nu}I$ is forced by requiring that every solution of the Dirac equation also satisfies the Klein-Gordon equation.
    \item Define $\gamma^5 = i\gamma^0\gamma^1\gamma^2\gamma^3$. Show that $\{\gamma^5, \gamma^\mu\} = 0$ for any $\mu$.
    \item The Weyl representation diagonalizes $\gamma^5$. Given that $(\gamma^5)^2 = I$, what are its eigenvalues and multiplicities? What do these define?
\end{enumerate}

\problem{Spin-1/2 from the Dirac Equation}
The Dirac Hamiltonian is $H = \vec{\alpha}\cdot\vec{p} + \beta m$ where $\alpha^i = \begin{pmatrix} 0 & \sigma^i \\ \sigma^i & 0 \end{pmatrix}$ and $\beta = \begin{pmatrix} I & 0 \\ 0 & -I \end{pmatrix}$. The spin operator is $\vec{S} = \frac{1}{2}\vec{\Sigma}$ with $\Sigma^i = \begin{pmatrix} \sigma^i & 0 \\ 0 & \sigma^i \end{pmatrix}$.
\begin{enumerate}[label=(\alph*)]
    \item Show that $[L^z, H] \neq 0$ by computing the relevant commutators. What does this say about orbital angular momentum conservation?
    \item Using Pauli commutation relations, compute $[\alpha^x, \Sigma^z]$ and $[\alpha^y, \Sigma^z]$.
    \item Show that $[H, S^z] = -[H, L^z]$. Conclude that $[H, J^z] = 0$ where $\vec{J} = \vec{L} + \vec{S}$ is the total angular momentum.
    \item From $(\Sigma^i)^2 = I$ and $\{\Sigma^i, \Sigma^j\} = 2\delta^{ij}I$, show that $\vec{S}^2 = \frac{3}{4}I$ and extract the spin quantum number.
    \item Explain in two sentences why spin-1/2 is a prediction of the Dirac equation rather than an assumption.
\end{enumerate}

\problem{Lorentz Covariance of the Dirac Equation}
\begin{enumerate}[label=(\alph*)]
    \item Under a Lorentz transformation $x^\mu \to \Lambda^\mu_\nu x^\nu$, the spinor transforms as $\psi(x) \to S(\Lambda)\psi(\Lambda^{-1}x)$. Show that the Dirac equation $(i\gamma^\mu\partial_\mu - m)\psi = 0$ is covariant, i.e., has the same form in all inertial frames.
    \item Verify that a boost in the $z$-direction by rapidity $\eta$ gives $S(\text{boost}) = \exp(-i\eta\gamma^0\gamma^3/2)$. What is the relationship between $\eta$ and the velocity $v$?
    \item Under a rotation by angle $\theta$ about the $z$-axis, the spinor transformation is $S(R_z(\theta)) = \exp(-i\theta\Sigma^3/2)$. Show that a $2\pi$ rotation gives $S(R_z(2\pi)) = -I$ (the spinor acquires a sign).
    \item Explain why this sign change is not a problem for physical observables (which are bilinear in $\psi$).
    \item Define the bilinear covariants: the scalar $\bar{\psi}\psi$, the vector $\bar{\psi}\gamma^\mu\psi$, and the pseudoscalar $\bar{\psi}\gamma^5\psi$. Under parity ($\vec{x} \to -\vec{x}$), how does each transform?
\end{enumerate}

\problem{Positive-Energy Projector and Antiparticles}
\begin{enumerate}[label=(\alph*)]
    \item For a free particle with momentum $\vec{p}$, the positive-energy projector is $\Lambda_+(\vec{p}) = \frac{\gamma^\mu p_\mu + m}{2m}$ where $p^\mu = (E, \vec{p})$ with $E = \sqrt{\vec{p}^2 + m^2}$. Show that $\Lambda_+^2 = \Lambda_+$ (idempotent) and that $\text{Tr}(\Lambda_+) = 2$ (two-dimensional).
    \item Verify that $(\gamma^\mu p_\mu - m)\Lambda_+(\vec{p}) = 0$, confirming that the positive-energy spinors lie in the range of $\Lambda_+$.
    \item The negative-energy projector is $\Lambda_-(\vec{p}) = I - \Lambda_+(\vec{p})$. Verify that it projects onto negative-energy states.
    \item In the rest frame ($\vec{p} = 0$, $E = m$), show that $\Lambda_+ = \frac{\gamma^0 + 1}{2}$ and that the two states in its range are the spin-up and spin-down spinors in the standard representation.
    \item Define the charge-conjugation operator $C = -i\gamma^2\gamma^0$ such that $C^{-1}\gamma^\mu C = -\gamma^{\mu T}$. Show that $C^{-1}(\gamma^\mu p_\mu - m)C = -(\gamma^\mu p_\mu + m)$. What does this mean physically?
\end{enumerate}

\problem{Gamma Matrices in Different Representations}
\begin{enumerate}[label=(\alph*)]
    \item The standard (Dirac) representation uses $\gamma^0 = \begin{pmatrix} I & 0 \\ 0 & -I \end{pmatrix}$ and $\gamma^i = \begin{pmatrix} 0 & \sigma^i \\ -\sigma^i & 0 \end{pmatrix}$. The Weyl representation diagonalizes $\gamma^5$ as $\gamma^5 = \begin{pmatrix} -I & 0 \\ 0 & I \end{pmatrix}$. Show that a change of representation is a similarity transformation: $\gamma^{\mu}_{\text{Weyl}} = U^{-1}\gamma^{\mu}_{\text{Dirac}}U$ for some unitary $U$.
    \item Compute the $U$ that transforms from the Dirac to the Weyl representation.
    \item In the Weyl representation, the Dirac spinor decomposes into right- and left-handed Weyl spinors. Write the Weyl representation Dirac equation in block form and interpret the upper and lower blocks.
    \item Why is the Weyl representation useful for massless particles? What happens when you add a mass term?
    \item Verify that the Clifford algebra $\{\gamma^\mu, \gamma^\nu\} = 2g^{\mu\nu}I$ holds in the Weyl representation.
\end{enumerate}

\problem{Helicity Eigenstates of Massless Dirac Fermions}
For a massless Dirac particle, the Hamiltonian is $H = \vec{\alpha}\cdot\vec{p}$.
\begin{enumerate}[label=(\alph*)]
    \item The helicity operator is $h = \vec{S}\cdot\hat{p} = \frac{1}{2}\vec{\Sigma}\cdot\hat{p}$ where $\hat{p} = \vec{p}/|\vec{p}|$. Show that $[H, h] = 0$ for massless particles.
    \item For a particle moving along $\hat{z}$ with momentum $p$, show that the helicity eigenvalue equation reduces to $\sigma^z u = \pm u$ where $u$ is a two-component spinor.
    \item Solve for the helicity eigenstates: the right-handed state $u_R$ (helicity $+1/2$) and left-handed state $u_L$ (helicity $-1/2$).
    \item Verify that $u_R$ and $u_L$ are orthogonal and that they are eigenstates of $\gamma^5 = i\gamma^0\gamma^1\gamma^2\gamma^3$ in the massless limit.
    \item Show that the massless Dirac equation $i\gamma^\mu\partial_\mu\psi = 0$ decouples into separate equations for $u_R$ and $u_L$. What does this tell you about Weyl spinors?
\end{enumerate}

\problem{Foldy-Wouthuysen Non-Relativistic Limit}
\begin{enumerate}[label=(\alph*)]
    \item In the non-relativistic limit, write $\psi = e^{-imt}\tilde{\psi}$ where $\tilde{\psi}$ is slowly varying. Decompose the spinor as $\tilde{\psi} = \begin{pmatrix}\phi\\\chi\end{pmatrix}$ where $\chi = O(v/c)\phi$.
    \item From the Dirac equation in the presence of an electromagnetic field, show that in the non-relativistic limit $\chi \approx \frac{\vec{\alpha}\cdot\vec{\Pi}}{2m}\phi$ where $\vec{\Pi} = \vec{p} - e\vec{A}$.
    \item Substitute this back into the upper component equation to derive the Pauli equation: $i\partial_t\phi = \left[\frac{(\vec{p} - e\vec{A})^2}{2m} + e\Phi - \frac{e}{2m}\vec{\sigma}\cdot\vec{B}\right]\phi$.
    \item Extract the magnetic moment interaction term and show that the g-factor is $g = 2$.
    \item Explain why the Foldy-Wouthuysen transformation is useful for understanding the connection between relativistic Dirac theory and non-relativistic quantum mechanics.
\end{enumerate}

\problem{Zitterbewegung: The Electron's Trembling Motion}
\begin{enumerate}[label=(\alph*)]
    \item For a free Dirac electron, the velocity is $\vec{v} = d\vec{x}/dt = \vec{\alpha}$. Compute $d\vec{\alpha}/dt$ using the Heisenberg equation $\dot{O} = i[H, O]$ where $H = \vec{\alpha}\cdot\vec{p} + \beta m$.
    \item Show that the center-of-mass position $\vec{x}_{\text{cm}} = \vec{x} - \vec{v}t$ satisfies $d\vec{x}_{\text{cm}}/dt = \vec{v} - \vec{\alpha} = -i[\vec{v}, H]/E$ where $E = \sqrt{\vec{p}^2 + m^2}$.
    \item The fluctuation of the velocity about its mean is $\delta\vec{v} = \vec{v}(t) - \langle\vec{v}\rangle$. Show that this oscillates with frequency $\omega_z = 2m$ (in natural units) and amplitude $\sim v/c$.
    \item Explain physically why this oscillation (Zitterbewegung) arises and why it is suppressed in the non-relativistic limit.
    \item Estimate the Compton wavelength $\lambda_C = h/(mc)$ and the zitterbewegung amplitude. Why is this effect unobservable in non-relativistic systems?
\end{enumerate}

\problem{Charge Conjugation and the Dirac Sea}
\begin{enumerate}[label=(\alph*)]
    \item The charge conjugation operator $C$ acting on the Dirac equation satisfies $C^{-1}\gamma^\mu C = -\gamma^{\mu T}$ (charge-conjugated gamma matrices). Define $C = -i\gamma^2\gamma^0$ in the standard representation and verify this property.
    \item Under charge conjugation, $\psi \to C\bar{\psi}^T = -i\gamma^2\gamma^0(\psi^\dagger\gamma^0)^T$. Show that the Dirac equation for $\psi$ maps to the Dirac equation for the conjugate spinor with opposite charge $-e$.
    \item In Dirac's interpretation, a hole in the filled negative-energy sea is interpreted as a positron (antiparticle with positive energy $+E$ and charge $+e$). Explain how the hole's energy and charge relate to the missing negative-energy electron.
    \item Consider a positive-energy electron state $u_+(\vec{p})$ from problem 5 (boosted spinors). What is the corresponding conjugate spinor $v_-(-\vec{p})$ for a positron with momentum $-\vec{p}$? Show that $v_-$ is the charge-conjugated version of $u_+$.
    \item While the Dirac sea interpretation is historically important, modern quantum field theory replaces it with second quantization. Briefly explain why second quantization is more satisfactory.
\end{enumerate}

%% file: chapters/ch12_renormalization_group.tex
\chapter{The Renormalization Group}
\label{ch:renormalization_group}

\section{Introduction: The Problem of Scale}

At the beginning of this course, we started with the simplest quantum system: a single qubit living in a two-dimensional Hilbert space. We then built up systematically: tensor products gave us multiple qubits, lattice systems connected discrete sites, and finally we took continuum limits to recover the familiar Schr\"odinger equation in continuous space. Throughout this progression, a fundamental question lurked beneath the surface: when we describe physics at different scales, are we looking at the same theory, or do the rules themselves change?

The renormalization group (RG) provides a profound answer: physical theories transform under changes of scale. Rather than asking ``what is the true theory,'' we should ask ``how do theories flow as we coarse-grain our description?'' This perspective resolves the discrete-versus-continuous tension that motivated our entire pedagogical arc. The continuum limit is not simply the limit as the lattice spacing goes to zero; it represents a particular fixed point in the space of all possible theories. The RG is not a group in the mathematical sense, but rather a semigroup of scale transformations that maps a theory to an effective theory at longer length scales. By systematically integrating out short-distance degrees of freedom, it reveals which properties of a system are robust under coarse-graining and which are sensitive to microscopic details.

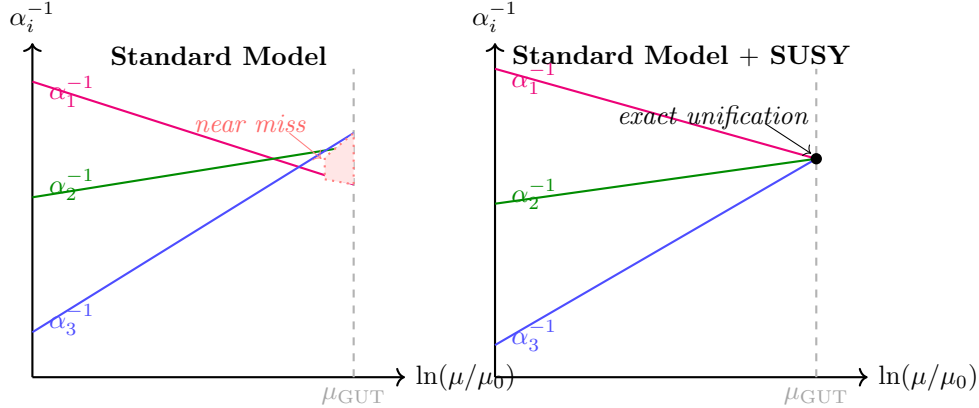
\begin{figure}[htbp]
\centering
\begin{tikzpicture}[scale=0.85,
    ax/.style={->, thick},
    gl/.style={dashed, gray!60, thick}]
%
\begin{scope}[xshift=0cm]
  \draw[ax] (0,0) -- (5.8,0) node[right, font=\small] {$\ln(\mu/\mu_0)$};
  \draw[ax] (0,0) -- (0,5.2) node[above, font=\small] {$\alpha_i^{-1}$};
  \draw[gl] (5.0,0) -- (5.0,4.8);
  \node[below, font=\small, gray!70] at (5.0,0) {$\mu_{\mathrm{GUT}}$};
  \draw[thick, magenta!80!red]  (0,4.6) -- (5.0,3.0);
  \draw[thick, green!55!black]  (0,2.8) -- (5.0,3.6);
  \draw[thick, blue!70]         (0,0.7) -- (5.0,3.8);
  \fill[red!10] (4.55,3.08) -- (5.0,3.0) -- (5.0,3.8) -- (4.55,3.42) -- cycle;
  \draw[red!50, thick, dotted] (4.55,3.08) -- (5.0,3.0) -- (5.0,3.8) -- (4.55,3.42) -- cycle;
  \node[right, font=\small, magenta!80!red]  at (0.1,4.45) {$\alpha_1^{-1}$};
  \node[right, font=\small, green!55!black]  at (0.1,3.00) {$\alpha_2^{-1}$};
  \node[right, font=\small, blue!70]         at (0.1,0.90) {$\alpha_3^{-1}$};
  \node[font={\small\bfseries}] at (2.9,5.0) {Standard Model};
  \node[font=\small, red!60] at (3.4,3.95) {\textit{near miss}};
  \draw[->, red!50] (4.0,3.80) -- (4.50,3.40);
\end{scope}
%
\begin{scope}[xshift=7.2cm]
  \draw[ax] (0,0) -- (5.8,0) node[right, font=\small] {$\ln(\mu/\mu_0)$};
  \draw[ax] (0,0) -- (0,5.2) node[above, font=\small] {$\alpha_i^{-1}$};
  \draw[gl] (5.0,0) -- (5.0,4.8);
  \node[below, font=\small, gray!70] at (5.0,0) {$\mu_{\mathrm{GUT}}$};
  \draw[thick, magenta!80!red]  (0,4.8) -- (5.0,3.4);
  \draw[thick, green!55!black]  (0,2.7) -- (5.0,3.4);
  \draw[thick, blue!70]         (0,0.5) -- (5.0,3.4);
  \fill (5.0,3.4) circle (2.5pt);
  \node[right, font=\small, magenta!80!red]  at (0.1,4.65) {$\alpha_1^{-1}$};
  \node[right, font=\small, green!55!black]  at (0.1,2.85) {$\alpha_2^{-1}$};
  \node[right, font=\small, blue!70]         at (0.1,0.65) {$\alpha_3^{-1}$};
  \node[font={\small\bfseries}] at (2.9,5.0) {Standard Model $+$ SUSY};
  \node[font=\small] at (3.4,4.05) {\textit{exact unification}};
  \draw[->] (4.2,3.95) -- (4.92,3.48);
\end{scope}
\end{tikzpicture}
\caption{Schematic of running coupling constants. Each of the three fundamental forces has a dimensionless coupling constant that changes with the energy scale $\mu$ at which it is probed. Plotting the inverse couplings $\alpha_i^{-1}$ against $\ln\mu$, each runs as a straight line whose slope is set by the structure of that force. Left: in the Standard Model the three lines converge toward $\mu \sim 10^{16}\,\text{GeV}$ but fall slightly short of a common point (shaded near-miss region). Right: a supersymmetric extension of the theory modifies the slopes just enough that all three lines meet exactly, suggesting a single unified interaction at high energy.}
\label{fig:running_couplings}
\end{figure}

A single example from high-energy physics makes the payoff vivid. The fine-structure constant $\alpha \approx 1/137$ appears throughout atomic physics and feels like a fixed property of nature. But the RG reveals that $\alpha$ is a function of the energy scale $\mu$ at which you probe the interaction: approximately $1/137$ at atomic energies, but measurably larger ($\approx 1/128$) at the energy of the $Z$ boson ($\mu \approx 91\,\text{GeV}$). The coupling constant \emph{runs}. All three fundamental forces exhibit running couplings, and as Figure~\ref{fig:running_couplings} shows, evolving them upward in energy using the RG flow equations reveals a remarkable convergence near $\mu \sim 10^{16}\,\text{GeV}$, suggesting that the electromagnetic, weak, and strong forces may be three low-energy faces of a single unified interaction. This convergence is entirely invisible at any fixed energy scale. It is a statement about flow, about how the effective description of nature changes as you change the resolution at which you examine it. We will not derive this result, as doing so requires quantum field theory, but it sets the ambition: the RG reveals structure in physics that would otherwise be completely hidden.

Before turning to the RG itself, it is worth recalling the two broad classes of phase transition. In a \emph{first-order} transition (melting ice, boiling water), the order parameter jumps discontinuously and the two phases coexist at the transition point; the correlation length remains finite. In a \emph{second-order} (or \emph{continuous}) transition, the order parameter vanishes continuously and the correlation length $\xi$ diverges. It is the second-order transitions that exhibit the scale invariance the RG is designed to exploit.

The correlation length is defined through the connected spin-spin correlation function
\begin{equation}
    G(r) = \langle s_i\, s_{i+r} \rangle - \langle s_i \rangle\langle s_{i+r} \rangle,
\end{equation}
which measures how much the spin at site $i+r$ ``knows about'' the spin at site $i$ beyond any uniform magnetization. Far from the critical point, $G(r)$ decays exponentially,
\begin{equation}
    G(r) \sim e^{-r/\xi},
    \label{eq:correlation_decay}
\end{equation}
defining the correlation length $\xi$: the characteristic distance over which spins are statistically correlated. As the temperature approaches the critical temperature $T_c$ from either side, $\xi$ diverges as $\xi \sim |T - T_c|^{-\nu}$, where $\nu$ is the correlation-length critical exponent. At the critical point itself, $\xi$ diverges, removing the only length scale from the problem. The correlation function must then decay as a power law rather than an exponential, and the question is: what power? The answer begins with the simplest continuum model of a fluctuating order parameter, a free (Gaussian) scalar field $\phi(\mathbf{x})$ that plays the role of a coarse-grained, continuous version of the discrete spin $s_i$. The correlation function for this field is $G(\mathbf{r}) = \langle \phi(\mathbf{x})\,\phi(\mathbf{x}+\mathbf{r}) \rangle - \langle \phi \rangle^2$, the direct continuum analog of the spin-spin correlator defined above. At the critical point it satisfies
\begin{equation}
    -\nabla^2 G(\mathbf{r}) = \delta^d(\mathbf{r}),
\end{equation}
i.e., $G(\mathbf{r})$ is the Green's function of the Laplacian in $d$ dimensions. Fourier transforming both sides gives $k^2 \tilde{G}(\mathbf{k}) = 1$, so $\tilde{G}(\mathbf{k}) = 1/k^2$. The real-space correlation function is then
\begin{equation}
    G(\mathbf{r}) = \int \frac{d^d k}{(2\pi)^d}\,\frac{e^{i\mathbf{k}\cdot\mathbf{r}}}{k^2}.
\end{equation}
The $r$-dependence can be extracted without evaluating the integral explicitly. Substitute $\mathbf{k} = \mathbf{q}/r$, so that $d^d k = r^{-d}\,d^d q$, $k^2 = q^2/r^2$, and $\mathbf{k}\cdot\mathbf{r} = \mathbf{q}\cdot\hat{r}$:
\begin{equation}
    G(r) = r^{-(d-2)} \int \frac{d^d q}{(2\pi)^d}\,\frac{e^{i\mathbf{q}\cdot\hat{r}}}{q^2}.
\end{equation}
The remaining integral is a pure number independent of $r$ (it depends only on the direction $\hat{r}$, and after angular averaging becomes a numerical constant), so
\begin{equation}
    G(r) \sim r^{-(d-2)} \qquad \text{(free field)}.
\end{equation}
This is the prediction of mean-field theory, in which fluctuations are treated as non-interacting. In an interacting system the exponent receives a correction:
\begin{equation}
    G(r) \sim r^{-(d-2+\eta)},
\end{equation}
where $\eta$ is the \emph{anomalous dimension}. It measures how much the true critical correlations deviate from the free-field prediction; $\eta = 0$ in mean-field theory, but takes the value $\eta = 1/4$ in the 2D Ising model.

The RG is equally central to classical and quantum statistical mechanics. Near a continuous phase transition, a magnet approaching its Curie temperature, helium approaching its superfluid transition, a binary mixture near its critical point, $\xi$ diverges and fluctuations exist on all length scales at once. This is the hallmark of an RG fixed point. The striking consequence is \emph{universality}: the critical exponents that govern the power-law singularities of thermodynamic quantities depend only on symmetry and spatial dimensionality, not on microscopic details. A magnet and a liquid-gas system near their respective critical points can obey identical power laws despite having nothing microscopically in common. Kenneth Wilson was awarded the Nobel Prize in Physics in 1982 for making this connection precise.

\begin{keyidea}{Scale Transformations and Effective Theories}
The renormalization group is a semigroup of scale transformations that maps theories to effective theories at longer length scales. Whether discrete or continuous descriptions are ``more fundamental'' depends on which RG fixed point you are near, resolving a question posed in Chapter~\ref{ch:foundations}.
\end{keyidea}

This chapter develops the RG through the Ising model, a lattice of binary spins interacting with their neighbors. The Ising Hamiltonian $H = -J\sum s_i^z s_{i+1}^z$ is a perfectly valid quantum-mechanical operator, but because it is diagonal in the $z$-basis, the entire analysis can be carried out using classical statistical mechanics: no superpositions or operator ordering enter. This is deliberate: working in a setting where classical methods suffice isolates the core logic of coarse-graining in the simplest possible way. The procedure, called decimation, integrates out alternating spins exactly and yields a recursion relation for the effective coupling. Fixed points of this recursion correspond to phases of matter, and their stability determines whether a phase transition occurs. The goal is a concrete, fully worked example that builds the intuition underlying both the quantum field theory of running couplings and the statistical mechanics of critical phenomena.

\section{Block Spin Renormalization: The Ising Model}

\subsection{The One-Dimensional Ising Model}

We begin with the simplest example that captures the essential physics: spins on a one-dimensional lattice. The Hamiltonian is:
\begin{equation}
    H = -J\sum_{i=1}^{N-1} s_i s_{i+1}
\end{equation}
where $s_i = \pm 1$ are classical spins and $J > 0$ favors alignment. These values are the eigenvalues of $\hat{\sigma}^z$, familiar throughout this course, but here each spin simply takes a definite classical value: there is no superposition.

What is new is temperature. The system does not sit in a single configuration; instead it samples all $2^N$ configurations $\{s_1, \ldots, s_N\}$ with probabilities set by the Boltzmann weight $e^{-\beta H}$, where $\beta \equiv 1/(k_B T)$ is the inverse temperature. Lower-energy configurations are exponentially more probable; at high temperature $\beta \to 0$ and all configurations become equally likely. The operator $e^{-\beta H}$ should look familiar: it has exactly the same structure as the time-evolution operator $e^{-i\hat{H}t/\hbar}$ from Chapter~\ref{ch:time}, with the replacement $it/\hbar \to \beta$. This is not a coincidence. The normalization factor that makes this a valid probability distribution is the partition function
\begin{equation}
    Z = \mathrm{Tr}\left(e^{-\beta \hat{H}}\right),
\end{equation}
and the thermal density operator
\begin{equation}
    \hat{\rho}_{\text{th}} = \frac{e^{-\beta \hat{H}}}{Z}
\end{equation}
assigns probability $p_n = e^{-\beta E_n}/Z$ to each energy eigenstate $\ket{n}$ and gives expectation values via $\langle A \rangle = \mathrm{Tr}(\hat{\rho}_{\text{th}}\hat{A})$. You have seen the density operator formalism for qubits; here it extends to systems with many degrees of freedom.
For the classical Ising model the Hamiltonian is diagonal in the spin-configuration basis, so there are no off-diagonal coherences and the trace is just a sum over configurations:
\begin{align}
    Z &= \sum_{\{s_i\}} e^{-\beta H} \nonumber\\
      &= \sum_{\{s_i\}} \exp\!\left(\beta J \sum_{i=1}^{N-1} s_i s_{i+1}\right) \nonumber\\
      &= \sum_{\{s_i\}} \prod_{i=1}^{N-1} e^{\beta J s_i s_{i+1}}.
\end{align}
The last step uses the fact that the exponential of a sum equals the product of exponentials. Because $H$ is a sum of independent bond terms, $e^{-\beta H}$ factorizes into one factor per bond. This factorization is what makes decimation possible: each spin $s_i$ appears in only a finite number of bond factors, so summing over $s_i$ affects only those factors and leaves the rest of the product unchanged.
All thermodynamic quantities follow from $Z$: the average energy is $-\partial \ln Z/\partial \beta$, the free energy is $F = -k_B T \ln Z$, and so on. The decimation procedure below amounts to performing part of this sum analytically.

\subsection{Decimation: Integrating Out Every Other Spin}

The central idea is simple: we have a chain of $N$ spins, and we want to trade it for a shorter chain that captures the same long-distance physics. The way to do this is to perform part of the sum in $Z$ exactly, specifically to sum over every other spin while leaving the remaining spins untouched. This is called \emph{decimation}, because we are removing (``decimating'') a fraction of the degrees of freedom.

Label the spins by site: $s_1, s_2, s_3, s_4, \ldots$ We will sum over the even-site spins $s_2, s_4, \ldots$ and keep the odd-site spins $s_1, s_3, s_5, \ldots$ The question is whether the result of that partial sum can itself be written as an Ising model, just with a different coupling constant and a different lattice spacing. If yes, we have a map from one Ising model to another: that map is the RG transformation.

To see how this works, focus on a single even-site spin $s_2$ sitting between its two odd-site neighbors $s_1$ and $s_3$. In the full partition function, $s_2$ appears only in the two bond factors $e^{\beta J s_1 s_2}$ and $e^{\beta J s_2 s_3}$, so summing over $s_2 = \pm 1$ affects only those two factors:
\begin{align}
    \sum_{s_2=\pm 1} e^{\beta J s_1 s_2}\, e^{\beta J s_2 s_3}
    &= e^{\beta J(s_1+s_3)} + e^{-\beta J(s_1+s_3)} \nonumber\\
    &= 2\cosh\bigl[\beta J(s_1 + s_3)\bigr].
\end{align}
Define the dimensionless coupling $K \equiv \beta J$; then $s_1$ and $s_3$ each equal $\pm 1$, and the argument of $\cosh$ takes only two values:
\begin{align}
    s_1 = s_3:\quad &2\cosh(2K),\label{eq:dec_parallel}\\
    s_1 = -s_3:\quad &2\cosh(0) = 2.\label{eq:dec_antiparallel}
\end{align}
Now ask: can this be written in the form $A\,e^{K' s_1 s_3}$ for some constants $A$ and $K'$? The temperature is fixed throughout, $\beta = 1/k_B T$ does not change when we decimate. What changes is the effective coupling $J'$ between surviving spins, and it is the dimensionless combination $K' \equiv \beta J'$ (same $\beta$, new $J'$) that the RG maps. Since $s_1 s_3 = +1$ when the spins are parallel and $-1$ when they are antiparallel, the ansatz gives two conditions:
\begin{align}
    A\,e^{+K'} &= 2\cosh(2K), \label{eq:match1}\\
    A\,e^{-K'} &= 2. \label{eq:match2}
\end{align}
Multiplying equations~\eqref{eq:match1} and~\eqref{eq:match2} gives $A^2 = 4\cosh(2K)$, so
\begin{equation}
    A = 2[\cosh(2K)]^{1/2}.
\end{equation}
Dividing equation~\eqref{eq:match1} by equation~\eqref{eq:match2} gives $e^{2K'} = \cosh(2K)$, so the RG recursion relation is
\begin{equation}
    K' = \frac{1}{2}\ln[\cosh(2K)].
    \label{eq:rg_recursion}
\end{equation}
The same factor $A$ appears for every decimated spin, so after integrating out all even sites the partition function becomes
\begin{equation}
    Z = A^{N/2}\, Z',
\end{equation}
where $Z'$ is the partition function of an Ising model on the remaining $N/2$ spins with effective coupling $J' = K'/\beta = k_B T \cdot K'$. The factor $A^{N/2}$ contributes an additive constant to the free energy per site but does not affect correlations or critical behavior. The physics we care about is entirely encoded in the recursion relation~\eqref{eq:rg_recursion}, which tells us how the coupling flows when we double the lattice spacing by removing every other spin.

\begin{physicalinsight}
Decimation does not change the temperature. The heat bath is external, and $\beta = 1/k_B T$ is the same before and after summing over the even-site spins. What changes is the effective coupling $J' = k_B T \cdot K'$ between the surviving spins. Since $K' < K$ for any finite $K$, coarse-graining always weakens the effective interaction. The fixed point $K^* = 0$ means the effective coupling has been driven to zero: the long-distance physics looks completely disordered, as a paramagnet would at any temperature. This is not because the temperature rose, but because repeated coarse-graining washes out short-range correlations, leaving no tendency for neighboring block spins to align. The fixed point $K^* = \infty$ is the only alternative, approached only as $T \to 0$ where thermal fluctuations vanish entirely.

Because the partition function of the 1D Ising model depends only on the single dimensionless combination $K = J/k_B T$, the flow $K \to K'$ can be reinterpreted as holding $J$ fixed and letting the effective temperature rise, $T \to T' = J/(k_B K') > T$. In this reading, coarse-graining ``heats'' the system, and the disordered fixed point $K^* = 0$ is literally the $T \to \infty$ limit. The two interpretations, renormalizing $J$ at fixed $T$ or renormalizing $T$ at fixed $J$, are equivalent because only the ratio $K$ is physical.

This equivalence is special to the 1D model and does not generalize. In any system with more than one independent coupling, the RG flow changes each coupling at a different rate and can generate entirely new interactions absent from the original Hamiltonian. Temperature is a single overall scale factor on the Hamiltonian; it cannot account for the differential running of multiple couplings simultaneously. Moreover, in the 2D Ising model the critical fixed point sits at a finite value $K^* \neq 0$, so the flow near criticality does not head toward infinite temperature at all. And in quantum systems at $T = 0$ the RG is perfectly well-defined even though there are no thermal fluctuations whatsoever. In all these cases, RG is a statement about coarse-graining, not about temperature.
\end{physicalinsight}

\subsection{Flow Diagram and Fixed Points}

Define the dimensionless coupling $K = \beta J$. The RG equation becomes:
\begin{equation}
    K' = \frac{1}{2}\ln[\cosh(2K)]
\end{equation}

For small $K$ (high temperature):
\begin{equation}
    K' \approx \frac{1}{2}\ln\left[1 + \frac{(2K)^2}{2}\right] \approx K^2
\end{equation}

This exhibits two fixed points: $K^* = 0$ (vanishing effective coupling, fully disordered) where $K' = 0$ exactly, and $K^* = \infty$ (zero temperature, perfect order) where $K' = K$ asymptotically.

The flow diagram shows that all finite temperatures flow toward $K^* = 0$ under repeated coarse-graining. The one-dimensional Ising model has no phase transition at finite temperature.

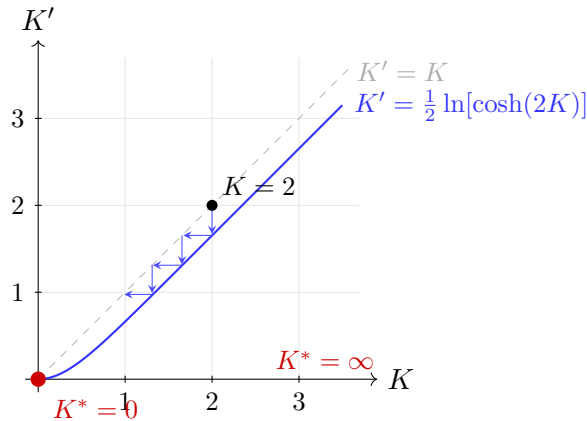
\begin{figure}[hbt!]
\centering
\begin{tikzpicture}[scale=1.15,
    arr/.style={->, >=stealth, blue!70, thin}]

    \foreach \k in {1,2,3} {
        \draw[gray!20, thin] (\k, 0) -- (\k, 3.7);
        \draw[gray!20, thin] (0, \k) -- (3.7, \k);
    }

    \draw[->] (-0.15,0) -- (3.9,0) node[right] {$K$};
    \draw[->] (0,-0.15) -- (0,3.9) node[above] {$K'$};

    \foreach \k in {1,2,3} {
        \draw (\k, 0.04) -- (\k,-0.04) node[below, font=\small] {$\k$};
        \draw (0.04,\k) -- (-0.04,\k) node[left,  font=\small] {$\k$};
    }

    \draw[dashed, gray!55] (0,0) -- (3.6,3.6);
    \node[gray!65, right, font=\small] at (3.55,3.55) {$K'=K$};

    \draw[thick, blue!80] plot[domain=0:3.5, samples=200] (\x, {0.5*ln(cosh(2*\x))});
    \node[blue!80, right, font=\small] at (3.52,3.15)
        {$K' = \tfrac{1}{2}\ln[\cosh(2K)]$};

    \draw[arr] (2,2)     -- (2,1.67);
    \draw[arr] (2,1.654) -- (1.67,1.654);
    \draw[arr] (1.654,1.654) -- (1.654,1.32);
    \draw[arr] (1.654,1.311) -- (1.32,1.311);
    \draw[arr] (1.311,1.311) -- (1.311,0.99);
    \draw[arr] (1.311,0.975) -- (0.99,0.975);

    \fill[black] (2,2) circle (1.8pt);
    \node[above right, font=\small] at (2,2) {$K=2$};

    \fill[red!80!black] (0,0) circle (2.5pt);
    \node[red!80!black, below right, font=\small] at (0.05,-0.12) {$K^*=0$};

    \node[red!80!black, font=\small] at (3.3,0.22) {$K^*=\infty$};

\end{tikzpicture}
\caption{RG flow diagram for the 1D Ising model. The blue curve shows $K' = \frac{1}{2}\ln[\cosh(2K)]$, which lies strictly below the diagonal $K'=K$ for all finite $K>0$. The cobweb construction illustrates successive RG steps starting from $K=2$: drop vertically to the curve to read off $K'$, then move horizontally back to the diagonal to set the new starting point. The sequence converges to the disordered fixed point $K^*=0$; the only other fixed point, $K^*=\infty$, is approached only at $T=0$.}
\label{fig:ising_1d_flow}
\end{figure}

\begin{example}{Why 1D Has No Phase Transition:}
The absence of a nontrivial fixed point has a physical interpretation. Consider a 1D chain at temperature $T$. A single broken bond (a domain wall) costs energy $2J$ but gains entropy $\sim k_B \ln N$ from being anywhere in the chain. For large $N$, entropy always wins at any finite $T$, destroying long-range order.

The RG makes this manifest: at any finite temperature, coarse-graining eventually averages over domain walls, washing out correlations. Only at $T = 0$, where thermal fluctuations vanish, can order persist. This connects to our earlier discussion of finite vs.\ infinite systems: thermodynamic phases require the limit $N \to \infty$, but 1D systems are ``too small'' in the transverse direction for this limit to produce phase transitions.
\end{example}

\section{Two-Dimensional Ising Model and Critical Behavior}

\subsection{The Square Lattice}

For the 2D Ising model on a square lattice:
\begin{equation}
    H = -J\sum_{\langle ij \rangle} s_i s_j
\end{equation}
where the sum runs over nearest-neighbor pairs. Unlike the 1D case, this model exhibits a genuine second-order phase transition at a critical temperature $T_c$ that can be determined by an elegant argument due to Kramers and Wannier (1941). The Kramers-Wannier argument constructs two exact representations of the same partition function, one organized around the low-temperature ordered phase and one around the high-temperature disordered phase, then shows that the two are related by an exact transformation that pins down $T_c$.

\paragraph{Heuristic picture.}
Before working through the details, here is the idea in one sentence: both the low-temperature and high-temperature representations of $Z$ turn out to be sums over closed loops drawn on the lattice, weighted by a per-bond factor that depends on $K = \beta J$. The critical temperature is then determined by asking when the two loop gases describe the same system.

At low temperature, the spins are nearly all aligned and excitations above the ground state are islands of flipped spins. What costs energy is the \emph{boundary} of each island: every bond crossing the boundary is a broken bond and pays an energetic price in the Boltzmann factor. These boundaries are necessarily closed curves, because an island is always fully enclosed. The partition function becomes a sum over all possible collections of such closed curves, with each bond on a curve contributing a small weight $e^{-2K}$. When $T$ is low, $K$ is large, $e^{-2K}$ is tiny, and long curves are strongly suppressed, so $Z$ is dominated by the empty configuration plus a few short loops.

At high temperature, the spins are nearly independent and correlations are rare. An exact rewriting of each bond factor (carried out below) expresses $e^{Ks_is_j}$ as an uncorrelated piece plus a small correlated piece proportional to $\tanh K$. Selecting the correlated piece on some subset of bonds and then averaging over spin configurations, only subsets forming \emph{closed loops} survive the average: an open path has dangling endpoints whose spins average to zero, while a closed loop visits each spin an even number of times and so cannot vanish. The partition function is again a sum over closed loops, but now the per-bond weight is $\tanh K$, which is small when $K$ is small.

So both expansions give $Z$ as a loop gas on the square lattice, differing only in their per-bond weight: $e^{-2K}$ at low $T$, $\tanh K$ at high $T$. As $T$ rises, $e^{-2K}$ grows from $0$ toward $1$ while $\tanh K$ shrinks from $1$ toward $0$, and the two curves must cross somewhere. Kramers and Wannier observed that at the crossing the two loop gases are mathematically identical: the system maps onto itself. If the Ising model has a single sharp transition between ordered and disordered phases, that transition can only sit at the self-dual point where the two weights coincide,
\begin{equation}
    \tanh K_c = e^{-2K_c},
    \label{eq:self_dual_heuristic}
\end{equation}
and solving this equation (shown near the end of this section) gives $\sinh(2K_c) = 1$, so $k_BT_c/J \approx 2.269$. The critical temperature drops out of a one-line self-consistency condition, without ever computing a thermodynamic quantity directly. The derivations that follow establish the two loop-gas representations carefully and justify why the self-dual point must be the transition; readers content with the heuristic can skim them and rejoin the argument at equation~\eqref{eq:duality}.

\paragraph{Low-temperature expansion.}
At low temperature, the dominant configurations are nearly ordered. A square lattice with $N$ sites has $2N$ bonds under periodic boundary conditions. Writing $K = \beta J$, a configuration with $b$ broken bonds (bonds where $s_i s_j = -1$) has Boltzmann weight
\begin{equation}
    e^{K\sum_{\langle ij\rangle}s_is_j} = e^{K(2N-b)}\,e^{-Kb} = e^{2KN}\,e^{-2Kb}.
\end{equation}
Every spin configuration maps to a collection of broken bonds, and the broken bonds form the boundaries of flipped domains. On the square lattice, these boundaries are closed loops on the dual lattice (a square lattice whose sites sit at the centers of the original faces, with bonds crossing the original bonds). Crucially, global spin flip leaves the loop collection unchanged, so the map from spin configurations to loop collections is exactly 2-to-1. The partition function is therefore
\begin{equation}
    Z = 2e^{2KN}\sum_{\mathcal{P}}\left(e^{-2K}\right)^{|\mathcal{P}|},
    \label{eq:ising_lowT}
\end{equation}
where the sum runs over all collections $\mathcal{P}$ of closed loops on the dual lattice and $|\mathcal{P}|$ is their total bond length. The empty collection (ground state) contributes~1.

\begin{figure}[htb]
\centering
\begin{tikzpicture}[scale=1.05]

\begin{scope}

  \fill[blue!10] (0.75,0.75) rectangle (2.25,2.25);

  \foreach \x in {0,1,2,3} {
    \foreach \y in {0,1,2,3,4} { \draw[gray!50] (\x,\y)--(\x+1,\y); }
  }
  \foreach \x in {0,1,2,3,4} {
    \foreach \y in {0,1,2,3} { \draw[gray!50] (\x,\y)--(\x,\y+1); }
  }

  \draw[red!80!black,very thick] (0,1)--(1,1);
  \draw[red!80!black,very thick] (0,2)--(1,2);
  \draw[red!80!black,very thick] (2,1)--(3,1);
  \draw[red!80!black,very thick] (2,2)--(3,2);
  \draw[red!80!black,very thick] (1,0)--(1,1);
  \draw[red!80!black,very thick] (2,0)--(2,1);
  \draw[red!80!black,very thick] (1,2)--(1,3);
  \draw[red!80!black,very thick] (2,2)--(2,3);

  \foreach \y in {0,1,2,3,4} { \node[fill=white,inner sep=1.5pt] at (0,\y) {$+$}; }
  \foreach \y in {0,3,4}      { \node[fill=white,inner sep=1.5pt] at (1,\y) {$+$}; }
  \foreach \y in {0,3,4}      { \node[fill=white,inner sep=1.5pt] at (2,\y) {$+$}; }
  \foreach \y in {0,1,2,3,4} { \node[fill=white,inner sep=1.5pt] at (3,\y) {$+$}; }
  \foreach \y in {0,1,2,3,4} { \node[fill=white,inner sep=1.5pt] at (4,\y) {$+$}; }
  \node[blue!80!black,fill=white,inner sep=1.5pt] at (1,1) {$-$};
  \node[blue!80!black,fill=white,inner sep=1.5pt] at (2,1) {$-$};
  \node[blue!80!black,fill=white,inner sep=1.5pt] at (1,2) {$-$};
  \node[blue!80!black,fill=white,inner sep=1.5pt] at (2,2) {$-$};

  \node[font=\small,below] at (2.0,-0.35) {(a)};
\end{scope}

\begin{scope}[xshift=6.5cm]

  \fill[blue!10] (0.75,0.75) rectangle (2.25,2.25);

  \foreach \x in {0,1,2,3} {
    \foreach \y in {0,1,2,3,4} { \draw[gray!30] (\x,\y)--(\x+1,\y); }
  }
  \foreach \x in {0,1,2,3,4} {
    \foreach \y in {0,1,2,3} { \draw[gray!30] (\x,\y)--(\x,\y+1); }
  }

  \draw[red!55,very thick] (0,1)--(1,1);
  \draw[red!55,very thick] (0,2)--(1,2);
  \draw[red!55,very thick] (2,1)--(3,1);
  \draw[red!55,very thick] (2,2)--(3,2);
  \draw[red!55,very thick] (1,0)--(1,1);
  \draw[red!55,very thick] (2,0)--(2,1);
  \draw[red!55,very thick] (1,2)--(1,3);
  \draw[red!55,very thick] (2,2)--(2,3);

  \foreach \x in {0,1,2} {
    \foreach \y in {0,1,2,3} { \draw[blue!30,dashed,thin] (\x+0.5,\y+0.5)--(\x+1.5,\y+0.5); }
  }
  \foreach \x in {0,1,2,3} {
    \foreach \y in {0,1,2} { \draw[blue!30,dashed,thin] (\x+0.5,\y+0.5)--(\x+0.5,\y+1.5); }
  }
  \foreach \x in {0,1,2,3} {
    \foreach \y in {0,1,2,3} { \fill[blue!60] (\x+0.5,\y+0.5) circle (2pt); }
  }

  \draw[blue!80!black,very thick] (0.5,0.5)--(2.5,0.5)--(2.5,2.5)--(0.5,2.5)--cycle;
  \node[blue!80!black,font=\footnotesize] at (1.5,0.22) {$|\mathcal{P}|=8$};

  \foreach \y in {0,1,2,3,4} { \node[gray!50,fill=white,inner sep=1.5pt] at (0,\y) {$+$}; }
  \foreach \y in {0,3,4}      { \node[gray!50,fill=white,inner sep=1.5pt] at (1,\y) {$+$}; }
  \foreach \y in {0,3,4}      { \node[gray!50,fill=white,inner sep=1.5pt] at (2,\y) {$+$}; }
  \foreach \y in {0,1,2,3,4} { \node[gray!50,fill=white,inner sep=1.5pt] at (3,\y) {$+$}; }
  \foreach \y in {0,1,2,3,4} { \node[gray!50,fill=white,inner sep=1.5pt] at (4,\y) {$+$}; }
  \node[blue!40,fill=white,inner sep=1.5pt] at (1,1) {$-$};
  \node[blue!40,fill=white,inner sep=1.5pt] at (2,1) {$-$};
  \node[blue!40,fill=white,inner sep=1.5pt] at (1,2) {$-$};
  \node[blue!40,fill=white,inner sep=1.5pt] at (2,2) {$-$};

  \node[font=\small,below] at (2.0,-0.35) {(b)};
\end{scope}
\end{tikzpicture}
\caption{Low-temperature expansion: domain walls as closed loops on the dual lattice. (a)~A $2\times 2$ block of flipped spins ($-$, blue, shaded) surrounded by $+$ spins. The eight broken bonds (thick red) form the boundary of the flipped domain. (b)~The dual lattice (dashed lines, sites at half-integer positions shown as filled circles) overlaid on the same configuration. Each broken bond on the original lattice is crossed by exactly one dual bond. The eight broken bonds map to a single closed loop of perimeter $|\mathcal{P}|=8$ on the dual lattice (bold blue), contributing weight $(e^{-2K})^8$ in equation~\eqref{eq:ising_lowT}. Global spin flip ($+\leftrightarrow -$ everywhere) produces the same loop collection, accounting for the factor of~$2$ in equation~\eqref{eq:ising_lowT}.}
\label{fig:kramers_wannier_domain}
\end{figure}
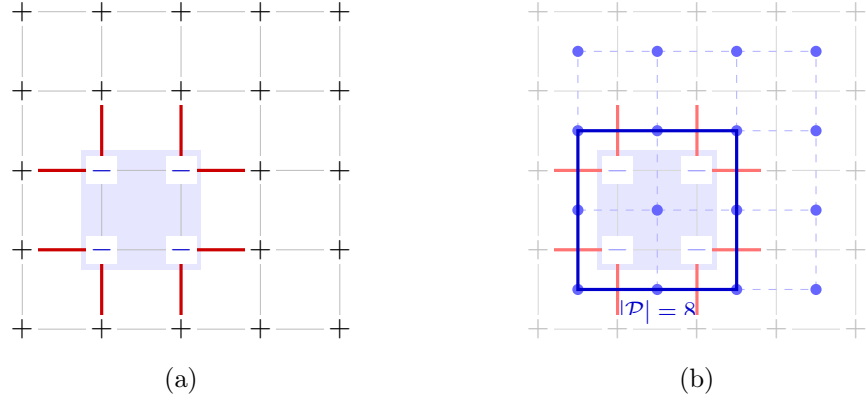

\paragraph{High-temperature expansion.}
At high temperature, $K = \beta J$ is small, and we seek an expansion in powers of $K$ (equivalently, in powers of $\tanh K$, which is also small). The starting point is an exact rewriting of each bond factor $e^{Ks_is_j}$ that takes advantage of the fact that $s_is_j$ takes only the two values $\pm 1$. For either sign,
\begin{equation}
    e^{\pm K} = \cosh K \pm \sinh K = \cosh K\,(1 \pm \tanh K),
\end{equation}
and the two cases combine into a single expression by noting that the second term carries the same sign as $s_is_j$:
\begin{equation}
    e^{Ks_is_j} = \cosh K\,(1 + \tanh K\cdot s_is_j).
    \label{eq:bond_factor_identity}
\end{equation}
This identity is exact, not a truncated Taylor series; it holds because $(s_is_j)^2 = 1$ collapses all higher powers. The partition function is a sum over spin configurations of a product of one such factor per bond:
\begin{align}
    Z &= \sum_{\{s_i\}}\prod_{\langle ij\rangle}e^{Ks_is_j} \nonumber\\
      &= (\cosh K)^{2N}\sum_{\{s_i\}}\prod_{\langle ij\rangle}\left(1 + \tanh K\cdot s_is_j\right),
\end{align}
where $(\cosh K)^{2N}$ pulls out one $\cosh K$ from each of the $2N$ bonds on the square lattice with periodic boundary conditions.

Now ``expanding the product'' means multiplying out the $2N$ factors in parentheses. Each factor offers two choices: either the $1$ term or the $\tanh K\cdot s_is_j$ term. A term in the fully expanded product is specified by going through all $2N$ bonds and recording, for each one, which of the two choices was made. Let $\mathcal{B}$ denote the subset of bonds for which the $\tanh K\cdot s_is_j$ term was chosen; the remaining bonds contribute factors of $1$. Multiplying everything together, the fully expanded product is
\begin{equation}
    \prod_{\langle ij\rangle}\left(1 + \tanh K\cdot s_is_j\right) = \sum_{\mathcal{B}}(\tanh K)^{|\mathcal{B}|}\prod_{\langle ij\rangle\in\mathcal{B}}s_is_j,
    \label{eq:bond_subset_expansion}
\end{equation}
where $|\mathcal{B}|$ is the number of bonds in $\mathcal{B}$ and the sum runs over all $2^{2N}$ subsets of bonds. The empty subset ($\mathcal{B} = \emptyset$) contributes $1$; the full subset contributes $(\tanh K)^{2N}\prod_{\langle ij\rangle}s_is_j$. Every intermediate possibility appears exactly once. This is the same bookkeeping one uses to expand $(1+a)(1+b)(1+c) = 1 + a + b + c + ab + ac + bc + abc$: each term is labeled by the subset of factors from which the non-unit piece was taken.

Substituting equation~\eqref{eq:bond_subset_expansion} back into $Z$ and exchanging the order of the sum over spin configurations with the sum over bond subsets:
\begin{equation}
    Z = (\cosh K)^{2N}\sum_{\mathcal{B}}(\tanh K)^{|\mathcal{B}|}\sum_{\{s_i\}}\prod_{\langle ij\rangle\in\mathcal{B}}s_is_j.
\end{equation}
Each bond in $\mathcal{B}$ contributes one factor of $s_i$ and one factor of $s_j$ to the product. Collecting these factors by site, the spin product becomes
\begin{equation}
    \prod_{\langle ij\rangle\in\mathcal{B}}s_is_j = \prod_i s_i^{n_i},
\end{equation}
where $n_i$ is the number of bonds in $\mathcal{B}$ incident on site $i$ (the degree of site $i$ within $\mathcal{B}$). The spin sum then factorizes over sites because different $s_i$ are independent:
\begin{equation}
    \sum_{\{s_i\}}\prod_i s_i^{n_i} = \prod_i\left(\sum_{s_i=\pm 1}s_i^{n_i}\right).
\end{equation}
Each site sum is elementary:
\begin{equation}
    \sum_{s=\pm 1}s^n = \begin{cases} 2 & n \text{ even},\\ 0 & n \text{ odd}. \end{cases}
\end{equation}
A single odd $n_i$ anywhere on the lattice kills the entire term. Only subsets $\mathcal{B}$ in which \emph{every} site has even degree survive, and each surviving term contributes $2^N$ (one factor of $2$ per site). The subsets with all even degrees are exactly the unions of closed loops on the original lattice: a closed loop enters and leaves each site it visits the same number of times, giving even degree at every vertex, while any open path has odd degree at its endpoints and is killed by the spin sum. The partition function becomes
\begin{equation}
    Z = 2^N(\cosh K)^{2N}\sum_{\mathcal{P}'}(\tanh K)^{|\mathcal{P}'|},
    \label{eq:ising_highT}
\end{equation}
where $\mathcal{P}'$ runs over all collections of closed loops on the original lattice and $|\mathcal{P}'|$ is their total bond length. At small $K$, $\tanh K$ is small, so the sum is dominated by the shortest loops (on the square lattice, the elementary plaquettes of perimeter~$4$), with longer loops contributing higher-order corrections in $\tanh K$.

\begin{figure}[htb]
\centering
\begin{tikzpicture}[scale=1.2]

\begin{scope}

  \foreach \x in {0,1,2} {
    \foreach \y in {0,1,2,3} { \draw[gray!45] (\x,\y)--(\x+1,\y); }
  }
  \foreach \x in {0,1,2,3} {
    \foreach \y in {0,1,2} { \draw[gray!45] (\x,\y)--(\x,\y+1); }
  }

  \draw[green!55!black,very thick] (0,0)--(1,0)--(1,1)--(0,1)--cycle;

  \foreach \x in {0,1,2,3} {
    \foreach \y in {0,1,2,3} { \fill[black] (\x,\y) circle (2pt); }
  }

  \node[green!55!black,font=\footnotesize,below left]  at (0,0) {$n{=}2$};
  \node[green!55!black,font=\footnotesize,below right] at (1,0) {$n{=}2$};
  \node[green!55!black,font=\footnotesize,above right] at (1,1) {$n{=}2$};
  \node[green!55!black,font=\footnotesize,above left]  at (0,1) {$n{=}2$};

  \node[green!55!black,font=\small] at (1.5,-0.65) {closed loop};
  \node[font=\small] at (1.5,-1.05) {all $n_i$ even $\Rightarrow$ survives};

\end{scope}

\begin{scope}[xshift=5.5cm]

  \foreach \x in {0,1,2} {
    \foreach \y in {0,1,2,3} { \draw[gray!45] (\x,\y)--(\x+1,\y); }
  }
  \foreach \x in {0,1,2,3} {
    \foreach \y in {0,1,2} { \draw[gray!45] (\x,\y)--(\x,\y+1); }
  }

  \draw[red!70!black,very thick] (0,0)--(2,0);
  \draw[red!70!black,very thick] (1,0)--(1,1);

  \foreach \x in {0,1,2,3} {
    \foreach \y in {0,1,2,3} { \fill[black] (\x,\y) circle (2pt); }
  }

  \node[red!70!black,font=\footnotesize,below left]  at (0,0) {$n{=}1$};
  \node[red!70!black,font=\footnotesize,below]        at (1,0) {$n{=}3$};
  \node[red!70!black,font=\footnotesize,below right]  at (2,0) {$n{=}1$};
  \node[red!70!black,font=\footnotesize,above right]  at (1,1) {$n{=}1$};

  \node[red!70!black,font=\small] at (1.5,-0.65) {open path};
  \node[font=\small] at (1.5,-1.05) {odd $n_i$ present $\Rightarrow$ vanishes};

\end{scope}
\end{tikzpicture}
\caption{High-temperature expansion: the closed-loop condition. A bond subset $\mathcal{B}$ contributes to $Z$ only if every site $i$ has even degree $n_i$ (the number of bonds in $\mathcal{B}$ incident on $i$), because $\sum_{s=\pm 1}s^n = 0$ for $n$ odd. Left: a unit square (four bonds, green) has $n_i=2$ at all four corners; the spin sums all yield~$2$, contributing $(\tanh K)^4 \cdot 2^N$ to $Z$. Right: an L-shaped path of three bonds (red) has $n_i=1$ at its three endpoints and $n_i=3$ at the junction; the spin sum at each of these sites vanishes, so this subset contributes zero. Only bond subsets forming unions of closed loops on the original lattice survive, giving equation~\eqref{eq:ising_highT}.}
\label{fig:highT_loop}
\end{figure}
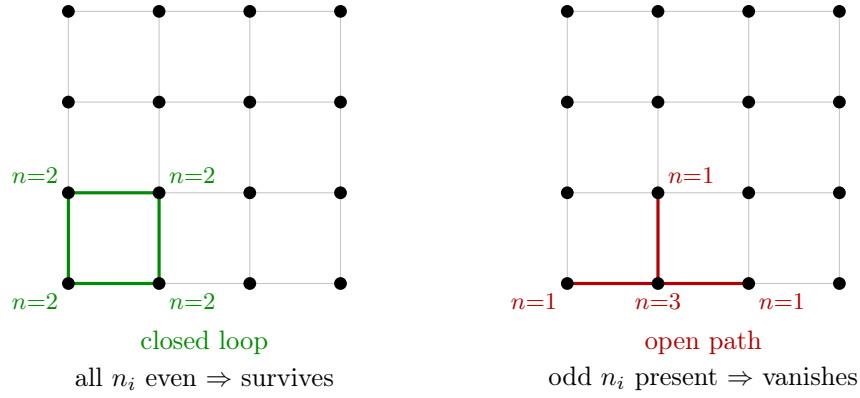

\paragraph{Duality and the critical point.}
Comparing equations~\eqref{eq:ising_lowT} and~\eqref{eq:ising_highT}: both have the form of a prefactor times a weighted sum over closed loops on a square lattice, with per-bond weights $e^{-2K}$ and $\tanh K$ respectively. On the square lattice the dual lattice is again a square lattice, so the two loop sums are structurally identical. Define $K^*$ by
\begin{equation}
    \tanh K^* = e^{-2K}.
    \label{eq:duality}
\end{equation}
The loop sum in~\eqref{eq:ising_lowT} at coupling $K$ then equals the loop sum in~\eqref{eq:ising_highT} at coupling $K^*$, giving the exact relation
\begin{equation}
    Z(K) = \frac{2e^{2KN}}{2^N(\cosh K^*)^{2N}}\,Z(K^*).
    \label{eq:duality_relation}
\end{equation}
The prefactor is a smooth function of $K$, so any non-analyticity in $Z(K)$ at $K_c$ implies a non-analyticity in $Z(K^*)$ at the dual value $K_c^*$. The map~\eqref{eq:duality} sends small $K$ (high temperature, disordered) to large $K^*$ (low temperature, ordered) and vice versa, so it relates the two sides of the transition. If the model has a unique phase transition, the transition point must be its own image under the duality, i.e., $K_c = K_c^*$. Setting $K = K^*$ in~\eqref{eq:duality} and using $\tanh K = (e^{2K}-1)/(e^{2K}+1)$:
\begin{align}
    \frac{e^{2K_c}-1}{e^{2K_c}+1} &= e^{-2K_c} \nonumber\\
    e^{2K_c} - 1 &= 1 + e^{-2K_c} \nonumber\\
    e^{2K_c} - e^{-2K_c} &= 2 \nonumber\\
    \sinh(2K_c) &= 1,
\end{align}
so $2K_c = \ln(1+\sqrt{2})$, giving
\begin{equation}
    \frac{k_B T_c}{J} = \frac{1}{K_c} = \frac{2}{\ln(1+\sqrt{2})} \approx 2.269.
    \label{eq:ising_2d_tc}
\end{equation}
Kramers and Wannier obtained this result in 1941, three years before Onsager's complete exact solution.

\subsection{Kadanoff Block Spin and the Scaling Hypothesis}

The 1D decimation worked because the calculation is exact: summing over every other spin produces a perfect Ising model with a new coupling, and nothing else. The 2D case is more demanding. Integrating out a fraction of spins on a square lattice generates not just a renormalized nearest-neighbor coupling but also next-nearest-neighbor couplings, four-spin couplings, and infinitely many higher-order terms. Tracking all of these exactly is intractable.

Kadanoff (1966) cut through this difficulty with a physical argument rather than a calculation. Near a critical point the system is scale-invariant: fluctuations occur on every length scale simultaneously and there is no characteristic microscopic scale. If we group $L \times L$ spins into blocks and assign each block a single block spin
\begin{equation}
    S_I = \mathrm{sign}\!\left(\sum_{i \in I} s_i\right),
\end{equation}
then near criticality the system of block spins should look statistically identical to the original system, rescaled in temperature and field. All the complicated higher-order couplings generated by the block-spin procedure shrink to zero under further coarse-graining near the critical point; they shift $T_c$ slightly but do not change the universality class.

This is Kadanoff's \emph{scaling hypothesis}: the only effect of a block-spin step with rescaling factor $b = L$ on the two relevant couplings is
\begin{align}
    t' &= b^{y_t}\, t, \label{eq:scaling_t}\\
    h' &= b^{y_h}\, h, \label{eq:scaling_h}
\end{align}
where $t = (T - T_c)/T_c$ is the reduced temperature, $h$ is a dimensionless external field, and $y_t > 0$, $y_h > 0$ are called the \emph{thermal} and \emph{magnetic scaling exponents}. Both are positive, meaning both $t$ and $h$ grow under coarse-graining: they are the relevant operators at the 2D Ising fixed point. The fixed point itself sits at $t = h = 0$ (exactly at $T_c$, zero field), and the scaling hypothesis asserts that near this point the RG acts as a pure rescaling of the two relevant parameters. Kadanoff could not derive this from the microscopic model, but it is consistent with the exact solution and captures the essential physics: a scale-invariant system has no intrinsic length, so scale transformations can only multiply relevant couplings by powers of $b$.

More generally, near any RG fixed point the linearized flow assigns an eigenvalue $b^{y}$ to each coupling in the Hamiltonian. The sign of $y$ classifies operators into three categories.

\begin{tcolorbox}[enhanced, title={Classification of Operators}, colback=blue!5, colframe=blue!50!black]
\textbf{Relevant operators} have $y > 0$ and grow under coarse-graining. They drive the system away from the fixed point and determine which phase the system is in. An example is the temperature deviation $t = (T-T_c)/T_c$.

\textbf{Irrelevant operators} have $y < 0$ and shrink under coarse-graining. They flow toward zero at the fixed point and do not affect universal critical behavior. An example is microscopic lattice details.

\textbf{Marginal operators} have $y = 0$ and remain unchanged by RG to leading order. They require higher-order analysis to determine whether they are marginally relevant or marginally irrelevant. Marginal operators are rare and special; they arise, for example, in certain four-dimensional quantum field theories studied in particle physics.
\end{tcolorbox}

With this classification in hand, Kadanoff's scaling hypothesis asserts that $t$ and $h$ are the only relevant operators at the 2D Ising fixed point, and that all the complicated couplings generated by block-spinning are irrelevant.

\subsection{Linearized RG, Scaling Relations, and Critical Exponents}

The scaling hypothesis can be written as a linearized RG matrix. Near the fixed point $(t, h) = (0, 0)$, the transformation acts as
\begin{equation}
    \begin{pmatrix} t' \\ h' \end{pmatrix}
    = \begin{pmatrix} b^{y_t} & 0 \\ 0 & b^{y_h} \end{pmatrix}
    \begin{pmatrix} t \\ h \end{pmatrix}.
\end{equation}
The matrix is diagonal because the Ising Hamiltonian is symmetric under $h \to -h$, so temperature deviations and field deviations cannot mix under RG. The eigenvalues $b^{y_t}$ and $b^{y_h}$ are both greater than~1, confirming that $t$ and $h$ are relevant.

\textbf{Correlation length exponent $\nu$.} The correlation length $\xi$ is a physical length. When we coarse-grain by a factor $b$, the lattice spacing changes from $a$ to $ba$, but the physical correlation length $\xi a$ does not change (the RG reorganizes degrees of freedom without altering the physics). Expressed in units of the new lattice spacing, the correlation length is therefore $\xi' = \xi/b$. Since $\xi$ depends on $t$, and one RG step maps $t \to b^{y_t} t$, we can write $\xi(t)/b = \xi(b^{y_t} t)$, or equivalently
\begin{equation}
    \xi(t) = b\, \xi(b^{y_t} t).
\end{equation}
This is a functional equation for $\xi(t)$ that we can solve by a judicious choice of $b$. Set $b = |t|^{-1/y_t}$ so that the argument of $\xi$ on the right becomes $\pm 1$. Then $\xi(t) = |t|^{-1/y_t}\,\xi(\pm 1) \sim |t|^{-1/y_t}$, where $\xi(\pm 1)$ is a nonuniversal constant. Comparing with the definition $\xi \sim |t|^{-\nu}$ gives
\begin{equation}
    \boxed{\nu = \frac{1}{y_t}}.
\end{equation}

\textbf{Free energy scaling.} The partition function $Z$ is invariant under a single RG step: decimation or block-spinning reorganizes the sum over spin configurations, grouping terms together, but it does not change the total. Every original configuration still contributes, just counted through the intermediate (coarse-grained) variables. The total free energy $F = -k_B T \ln Z$ is therefore also invariant.

Now write $F$ in two different ways. Before the RG step, the system has $N$ sites and is described by the reduced couplings $(t, h)$. Writing the free energy as $N$ times a free energy density $f(t,h)$:
\begin{equation}
    F = N\,f(t,h).
\end{equation}
After a block-spin step with rescaling factor $b$ in $d$ spatial dimensions, each block contains $b^d$ original sites and is replaced by a single new effective site. The number of effective sites is therefore
\begin{equation}
    N' = \frac{N}{b^d},
\end{equation}
and the new couplings are $t' = b^{y_t}t$ and $h' = b^{y_h}h$ from the scaling hypothesis. The same total free energy, now written in terms of the new description, is
\begin{equation}
    F = N'\,f(t', h') = \frac{N}{b^d}\,f(b^{y_t}t,\,b^{y_h}h).
\end{equation}
Setting these two expressions for $F$ equal and canceling the common factor $N$:
\begin{equation}
    f(t,h) = b^{-d}\,f(b^{y_t}t,\,b^{y_h}h).
    \label{eq:free_energy_scaling}
\end{equation}
This is the central scaling form for the free energy density. It says that $f$ at couplings $(t,h)$ equals $f$ at the renormalized couplings $(b^{y_t}t, b^{y_h}h)$, up to the geometric factor $b^{-d}$ that accounts for each effective site representing $b^d$ original sites.

\textbf{Standard critical exponents.} Before extracting consequences from the scaling form, we introduce the four standard exponents that label the power-law singularities of thermodynamic quantities near $T_c$. Each is defined by tuning either the reduced temperature $t = (T - T_c)/T_c$ at zero field, or the field $h$ at $t = 0$:
\begin{align}
    m(t, 0) &\sim |t|^{\beta} \qquad (t \to 0^{-}),\\
    \chi(t, 0) &\sim |t|^{-\gamma},\\
    m(0, h) &\sim h^{1/\delta},\\
    C(t, 0) &\sim |t|^{-\alpha}.
\end{align}
Here $m = -\partial f/\partial h$ is the magnetization (the order parameter), $\chi = \partial m/\partial h$ is the zero-field susceptibility (the linear response of $m$ to a small applied field), and $C \sim -\partial^2 f/\partial t^2$ is the specific heat at zero field. The exponent $\beta$ governs how the spontaneous magnetization vanishes as $T \to T_c^-$; $\gamma$ governs the divergence of the susceptibility, which becomes infinite at $T_c$ because fluctuations are correlated across all length scales; $\delta$ fixes the shape of the critical isotherm at $t = 0$; and $\alpha$ governs the singularity of the specific heat (with $\alpha = 0$ corresponding to a logarithmic divergence rather than a power law). Together with the correlation-length exponent $\nu$ already obtained, these are the five standard exponents characterizing a continuous transition.

Equation~\eqref{eq:free_energy_scaling}, combined with the definitions of the standard critical exponents above, determines all four of $\beta$, $\gamma$, $\delta$, and $\alpha$ in terms of $y_t$, $y_h$, and $d$. The strategy is the same in every case: differentiate both sides the appropriate number of times, apply the chain rule to pick up factors of $b^{y_t}$ or $b^{y_h}$, then make a clever choice of $b$ that eliminates one of the two arguments on the right-hand side.

\textbf{Magnetization exponent $\beta$.} The magnetization is defined by
\begin{equation}
    m(t,h) = -\frac{\partial f(t,h)}{\partial h}.
\end{equation}
Differentiate both sides of equation~\eqref{eq:free_energy_scaling} with respect to $h$. The left-hand side gives $-m(t,h)$. On the right-hand side, $h$ enters only through the rescaled argument $h' = b^{y_h}h$, so the chain rule gives
\begin{align}
    \frac{\partial}{\partial h}\Bigl[b^{-d}\,f(b^{y_t}t,\,b^{y_h}h)\Bigr]
    &= b^{-d}\,\frac{\partial f}{\partial h'}\bigg|_{(b^{y_t}t,\,b^{y_h}h)}\cdot\frac{\partial h'}{\partial h} \nonumber\\
    &= b^{-d}\cdot b^{y_h}\,\frac{\partial f}{\partial h'}\bigg|_{(b^{y_t}t,\,b^{y_h}h)} \nonumber\\
    &= -\,b^{y_h - d}\,m(b^{y_t}t,\,b^{y_h}h).
\end{align}
Equating the two sides (the minus signs cancel):
\begin{equation}
    m(t,h) = b^{y_h - d}\,m(b^{y_t}t,\,b^{y_h}h).
    \label{eq:m_scaling}
\end{equation}
Now set $h = 0$ on both sides and choose $b$ to eliminate $t$ from the argument on the right. The choice $b = |t|^{-1/y_t}$ does the job, because then $b^{y_t}t = |t|^{-1}\,t = \mathrm{sgn}(t) = \pm 1$. Substituting:
\begin{equation}
    m(t,0) = \bigl(|t|^{-1/y_t}\bigr)^{y_h - d}\,m(\pm 1, 0) = |t|^{(d - y_h)/y_t}\,m(\pm 1, 0).
\end{equation}
The quantity $m(\pm 1, 0)$ is a nonuniversal number (the magnetization at the reference point), not a function of $t$. The $t$-dependence is therefore entirely in the prefactor $|t|^{(d - y_h)/y_t}$. Comparing with the conventional definition $m \sim |t|^{\beta}$ (below $T_c$):
\begin{equation}
    \boxed{\beta = \frac{d - y_h}{y_t} = \nu(d - y_h),}
\end{equation}
where in the last step we used $\nu = 1/y_t$.

\textbf{Susceptibility exponent $\gamma$.} The zero-field susceptibility is
\begin{equation}
    \chi(t,h) = \frac{\partial m}{\partial h} = -\frac{\partial^2 f}{\partial h^2},
\end{equation}
so it requires a second derivative with respect to $h$. Differentiate equation~\eqref{eq:m_scaling} once more with respect to $h$, and apply the chain rule to the $b^{y_h}h$ argument again. A second factor of $b^{y_h}$ comes down:
\begin{equation}
    \chi(t,h) = b^{y_h - d}\cdot b^{y_h}\,\chi(b^{y_t}t,\,b^{y_h}h) = b^{2y_h - d}\,\chi(b^{y_t}t,\,b^{y_h}h).
\end{equation}
Set $h = 0$ and again choose $b = |t|^{-1/y_t}$:
\begin{equation}
    \chi(t, 0) = \bigl(|t|^{-1/y_t}\bigr)^{2y_h - d}\,\chi(\pm 1, 0) = |t|^{-(2y_h - d)/y_t}\,\chi(\pm 1, 0).
\end{equation}
Comparing with the definition $\chi \sim |t|^{-\gamma}$:
\begin{equation}
    \boxed{\gamma = \frac{2y_h - d}{y_t} = \nu(2y_h - d).}
\end{equation}

\textbf{Critical isotherm exponent $\delta$.} Exactly at the critical temperature, $t = 0$, and the magnetization depends only on $h$. The conventional definition is $m \sim h^{1/\delta}$, so we extract $\delta$ by following the $h$-dependence of $m$ at $t = 0$.

Return to the magnetization scaling relation~\eqref{eq:m_scaling} and set $t = 0$:
\begin{equation}
    m(0,h) = b^{y_h - d}\,m(0,\,b^{y_h}h).
\end{equation}
This time, rather than eliminating $t$, we eliminate $h$ from the right-hand side. Choose $b = |h|^{-1/y_h}$ so that $b^{y_h}h = |h|^{-1}h = \pm 1$:
\begin{equation}
    m(0,h) = \bigl(|h|^{-1/y_h}\bigr)^{y_h - d}\,m(0, \pm 1) = |h|^{(d - y_h)/y_h}\,m(0, \pm 1).
\end{equation}
Comparing with $m \sim h^{1/\delta}$ reads off $1/\delta = (d - y_h)/y_h$, so
\begin{equation}
    \boxed{\delta = \frac{y_h}{d - y_h}.}
\end{equation}

\textbf{Specific heat exponent $\alpha$.} The specific heat (at fixed field $h = 0$) comes from a second derivative of the free energy density with respect to temperature:
\begin{equation}
    C(t) \sim -\frac{\partial^2 f(t,0)}{\partial t^2}.
\end{equation}
Differentiate equation~\eqref{eq:free_energy_scaling} twice with respect to $t$. Each derivative acts only through the rescaled argument $t' = b^{y_t}t$, and each application of the chain rule pulls down a factor of $b^{y_t}$:
\begin{align}
    \frac{\partial f(t,h)}{\partial t} &= b^{-d}\cdot b^{y_t}\,\frac{\partial f}{\partial t'}\bigg|_{(b^{y_t}t,\,b^{y_h}h)} = b^{y_t - d}\,\frac{\partial f}{\partial t'}\bigg|_{(b^{y_t}t,\,b^{y_h}h)},\\
    \frac{\partial^2 f(t,h)}{\partial t^2} &= b^{y_t - d}\cdot b^{y_t}\,\frac{\partial^2 f}{\partial t'^2}\bigg|_{(b^{y_t}t,\,b^{y_h}h)} = b^{2y_t - d}\,\frac{\partial^2 f}{\partial t'^2}\bigg|_{(b^{y_t}t,\,b^{y_h}h)}.
\end{align}
Therefore
\begin{equation}
    C(t,h) \sim b^{2y_t - d}\,C(b^{y_t}t,\,b^{y_h}h).
\end{equation}
Set $h = 0$ and choose $b = |t|^{-1/y_t}$:
\begin{equation}
    C(t, 0) \sim \bigl(|t|^{-1/y_t}\bigr)^{2y_t - d}\,C(\pm 1, 0) = |t|^{-(2y_t - d)/y_t}\,C(\pm 1, 0) = |t|^{d/y_t - 2}\,C(\pm 1, 0).
\end{equation}
Comparing with $C \sim |t|^{-\alpha}$ gives $-\alpha = d/y_t - 2$, or
\begin{equation}
    \boxed{\alpha = 2 - \frac{d}{y_t} = 2 - d\nu.}
\end{equation}
This last relation, $2 - \alpha = d\nu$, is the \emph{hyperscaling relation}. It connects the specific heat exponent to the correlation length exponent and the spatial dimension, and it is a nontrivial prediction of the scaling hypothesis (it fails above the upper critical dimension, where mean-field exponents take over and fluctuations no longer determine the critical behavior).

These are not independent results. All four exponents ($\beta$, $\gamma$, $\delta$, $\alpha$) follow from just two numbers ($y_t$ and $y_h$) through the free energy scaling~\eqref{eq:free_energy_scaling}. This is the power of the RG: instead of calculating each exponent separately, the scaling hypothesis reduces everything to two eigenvalues of the linearized RG transformation.

\begin{example}{2D Ising Critical Exponents:}
Onsager's exact solution (1944) gives $\nu = 1$ and $\beta = 1/8$ for $d = 2$. From these, the scaling exponents are $y_t = 1/\nu = 1$ and $y_h = d - \beta/\nu = 2 - 1/8 = 15/8$. All remaining exponents follow:
\begin{align}
    \gamma &= \nu(2y_h - d) = 1\!\left(\tfrac{15}{4} - 2\right) = \frac{7}{4},\\
    \delta &= \frac{y_h}{d - y_h} = \frac{15/8}{1/8} = 15,\\
    \alpha &= 2 - d\nu = 0 \quad (\text{logarithmic divergence}).
\end{align}
These are exact, not approximate. Onsager's solution proceeds via the transfer matrix method: the 2D partition function is rewritten as a trace over products of $2^L \times 2^L$ matrices (one per row of the lattice), reducing the statistical mechanics problem to finding the largest eigenvalue of this matrix. The calculation is a tour de force of algebra, and we do not reproduce it here, but the key point is that the critical exponents can be extracted analytically. The values happen to be simple fractions, a consequence of the unusually high degree of symmetry at the 2D Ising fixed point, which constrains the exponents far more tightly than in a generic critical system. Kadanoff's framework guarantees that exponents $y_t$ and $y_h$ must exist and that all other critical exponents follow from them through scaling relations; the numerical values require either Onsager's exact solution or a more refined calculation such as the $\varepsilon$-expansion.
\end{example}

\section{Universality and RG Flows}

The classification of operators into relevant, irrelevant, and marginal has a powerful consequence: at criticality, irrelevant operators vanish, leaving only relevant and marginal operators. Systems with the same relevant operators exhibit identical critical behavior regardless of microscopic details.

\begin{physicalinsight}
Universality explains why liquid-gas critical points, ferromagnetic transitions, and binary alloy phase separations all have the same critical exponents. They share the same RG fixed point structure despite vastly different microscopic physics.
\end{physicalinsight}

What does it mean for two different theories to flow to the same fixed point? Consider the space of all possible Hamiltonians as a high-dimensional ``theory space,'' where each point represents a different set of coupling constants. The RG defines flow lines in this space: trajectories along which theories evolve as we coarse-grain.

A fixed point is a theory that remains unchanged under RG transformations. Near each fixed point, we can linearize the RG flow and identify directions (eigenvectors) that grow (relevant), shrink (irrelevant), or remain unchanged (marginal). The \textit{basin of attraction} of a fixed point consists of all theories that flow toward it under repeated coarse-graining.

\begin{example}{Universality in the 2D Ising Model:}
Consider three different microscopic models:
\begin{enumerate}
    \item Square lattice Ising model: $H_1 = -J\sum_{\langle ij\rangle_{\text{square}}} s_i s_j$
    \item Triangular lattice Ising model: $H_2 = -J\sum_{\langle ij\rangle_{\text{tri}}} s_i s_j$
    \item Square lattice with next-nearest neighbors: $H_3 = -J\sum_{\langle ij\rangle} s_i s_j - J_2\sum_{\langle\langle ij\rangle\rangle} s_i s_j$
\end{enumerate}

All three have different microscopic physics: different coordination numbers, different numbers of neighbors, different short-range structures. Yet at their critical temperatures, they exhibit \textit{identical} critical exponents: $\beta = 1/8$, $\gamma = 7/4$, $\nu = 1$, $\delta = 15$.

Why? The lattice geometry and next-nearest neighbor couplings correspond to irrelevant operators. Under RG flow toward the critical point, these differences wash out. All three theories lie in the basin of attraction of the same fixed point: the 2D Ising fixed point. At long distances, they become indistinguishable.
\end{example}

This has profound implications. Once we know the fixed point structure, we can predict critical behavior for entire classes of systems without solving each microscopic model, giving the RG tremendous predictive power. Measurements on completely different physical systems (magnets, fluids, alloys) can test the same theoretical predictions. The theory space perspective shifts our thinking: rather than asking about ``the theory'' of a system, we think of theories as points in theory space that flow under scale transformations. The long-distance physics depends only on which fixed point you flow toward, not where you started.

Two theories flow to the same fixed point if and only if they have the same set of relevant operators. Irrelevant operators specify \textit{which} microscopic realization you are studying, but they do not affect universal critical behavior. This is why condensed matter physicists can study critical phenomena using simple lattice models even though real materials have enormously complex microscopic structure. The complexity lives in irrelevant operators that vanish at long distances under RG flow.

\begin{figure}[htbp]
\centering
\begin{tikzpicture}[scale=1.4]
    \draw[->] (-0.5,0) -- (6.5,0) node[right] {Relevant operator};
    \draw[->] (0,-0.5) -- (0,5.5) node[above] {Irrelevant operator};
    
    \draw[thick, purple] plot[smooth] coordinates {
        (1.500,4.000)
        (1.444,3.895)
        (1.387,3.797)
        (1.307,3.677)
        (1.244,3.595)
        (1.157,3.493)
        (1.088,3.424)
        (1.018,3.359)
        (0.919,3.280)
        (0.842,3.225)
        (0.735,3.158)
        (0.651,3.112)
        (0.534,3.055)
        (0.443,3.016)
        (0.348,2.980)
        (0.217,2.936)
        (0.114,2.905)
        (-0.030,2.868)
        (-0.142,2.842)
        (-0.298,2.810)
    };
    \draw[->, thick, purple] (0.816,3.207) -- (0.707,3.142);
    \fill[purple] (1.5,4.0) circle (2pt);
    
    \draw[thick, green!60!black] plot[smooth] coordinates {
        (4.500,4.000)
        (4.556,3.895)
        (4.613,3.797)
        (4.693,3.677)
        (4.756,3.595)
        (4.843,3.493)
        (4.912,3.424)
        (4.982,3.359)
        (5.081,3.280)
        (5.158,3.225)
        (5.265,3.158)
        (5.349,3.112)
        (5.466,3.055)
        (5.557,3.016)
        (5.652,2.980)
        (5.783,2.936)
        (5.886,2.905)
        (6.030,2.868)
        (6.142,2.842)
        (6.298,2.810)
    };
    \draw[->, thick, green!60!black] (5.184,3.207) -- (5.293,3.142);
    \fill[green!60!black] (4.5,4.0) circle (2pt);
    
    \draw[thick, orange] plot[smooth] coordinates {
        (2.000,0.800)
        (1.938,0.994)
        (1.871,1.166)
        (1.801,1.318)
        (1.726,1.453)
        (1.646,1.573)
        (1.561,1.679)
        (1.453,1.790)
        (1.356,1.871)
        (1.254,1.943)
        (1.144,2.006)
        (1.029,2.063)
        (0.905,2.113)
        (0.747,2.165)
        (0.607,2.203)
        (0.457,2.237)
        (0.298,2.267)
        (0.129,2.294)
        (-0.050,2.317)
        (-0.280,2.342)
    };
    \draw[->, thick, orange] (1.232,1.956) -- (1.144,2.006);
    \fill[orange] (2.0,0.8) circle (2pt);
    
    \draw[thick, cyan!70!blue] plot[smooth] coordinates {
        (3.100,2.800)
        (3.106,2.766)
        (3.113,2.735)
        (3.120,2.709)
        (3.127,2.685)
        (3.137,2.660)
        (3.146,2.641)
        (3.155,2.625)
        (3.164,2.611)
        (3.175,2.598)
        (3.188,2.585)
        (3.200,2.575)
        (3.212,2.567)
        (3.225,2.559)
        (3.239,2.552)
        (3.257,2.545)
        (3.273,2.540)
        (3.291,2.536)
        (3.309,2.531)
        (3.332,2.527)
    };
    \draw[->, thick, cyan!70!blue] (3.179,2.594) -- (3.188,2.585);
    \fill[cyan!70!blue] (3.1,2.8) circle (2pt);
    
    \draw[thick, magenta!70] plot[smooth] coordinates {
        (1.800,3.000)
        (1.740,2.954)
        (1.678,2.912)
        (1.595,2.865)
        (1.525,2.831)
        (1.452,2.801)
        (1.355,2.766)
        (1.274,2.742)
        (1.188,2.719)
        (1.075,2.694)
        (0.979,2.676)
        (0.853,2.656)
        (0.746,2.642)
        (0.634,2.629)
        (0.486,2.614)
        (0.362,2.603)
        (0.230,2.594)
        (0.057,2.583)
        (-0.089,2.575)
        (-0.282,2.567)
    };
    \draw[->, thick, magenta!70] (1.051,2.690) -- (0.954,2.672);
    \fill[magenta!70] (1.8,3.0) circle (2pt);
    
    \draw[thick, brown!70] plot[smooth] coordinates {
        (4.200,2.000)
        (4.260,2.046)
        (4.322,2.088)
        (4.405,2.135)
        (4.475,2.169)
        (4.548,2.199)
        (4.645,2.234)
        (4.726,2.258)
        (4.812,2.281)
        (4.925,2.306)
        (5.021,2.324)
        (5.147,2.344)
        (5.254,2.358)
        (5.366,2.371)
        (5.514,2.386)
        (5.638,2.397)
        (5.770,2.406)
        (5.943,2.417)
        (6.089,2.425)
        (6.282,2.433)
    };
    \draw[->, thick, brown!70] (4.949,2.310) -- (5.046,2.328);
    \fill[brown!70] (4.2,2.0) circle (2pt);
    
    \fill[red] (3.0,2.5) circle (3pt);
    
    \fill[red!10, opacity=0.3] (0.3,0.3) rectangle (6.2,5.2);
    
    \draw[->, very thick, blue!70, line width=1.5pt] (3, 4.2) -- (3, 3.3);
    \draw[->, very thick, blue!70, line width=1.5pt] (3, 0.8) -- (3, 1.7);
    \draw[->, very thick, red!70, line width=1.5pt] (2.0, 2.5) -- (1.2, 2.5);
    \draw[->, very thick, red!70, line width=1.5pt] (4.0, 2.5) -- (4.8, 2.5);
    
    \node[anchor=west, font=\footnotesize, align=left] at (6.8, 2.8) {
        \textbf{Flow trajectories:}\\[2pt]
        \textcolor{purple}{\rule{0.4cm}{0.08cm}} Square lattice\\
        \textcolor{green!60!black}{\rule{0.4cm}{0.08cm}} Triangular\\
        \textcolor{orange}{\rule{0.4cm}{0.08cm}} With NNN\\
        \textcolor{cyan!70!blue}{\rule{0.4cm}{0.08cm}} Near critical\\
        \textcolor{magenta!70}{\rule{0.4cm}{0.08cm}} Left sector\\
        \textcolor{brown!70}{\rule{0.4cm}{0.08cm}} Right sector\\[4pt]
        \textbf{Fixed point:}\\[2pt]
        \textcolor{red}{$\bullet$} Critical FP\\[4pt]
        \textbf{Eigenlines:}\\[2pt]
        \textcolor{blue!70}{\rule{0.4cm}{0.08cm}} Irrelevant (attractive)\\
        \textcolor{red!70}{\rule{0.4cm}{0.08cm}} Relevant (repulsive)
    };
    
\end{tikzpicture}
\caption{Theory space and RG flows near the critical fixed point. Different microscopic models start at different points (colored dots) and flow along curved trajectories computed from the linearized RG equations. The fixed point acts as a saddle: flows are attracted along the irrelevant direction (vertical, blue arrows) but repelled along the relevant direction (horizontal, red arrows). Trajectories veer toward the fixed point as irrelevant operators decay, then diverge along the relevant direction to different phases. All theories in the shaded basin eventually pass through the same critical behavior.}
\label{fig:theory_space_flow}
\end{figure}
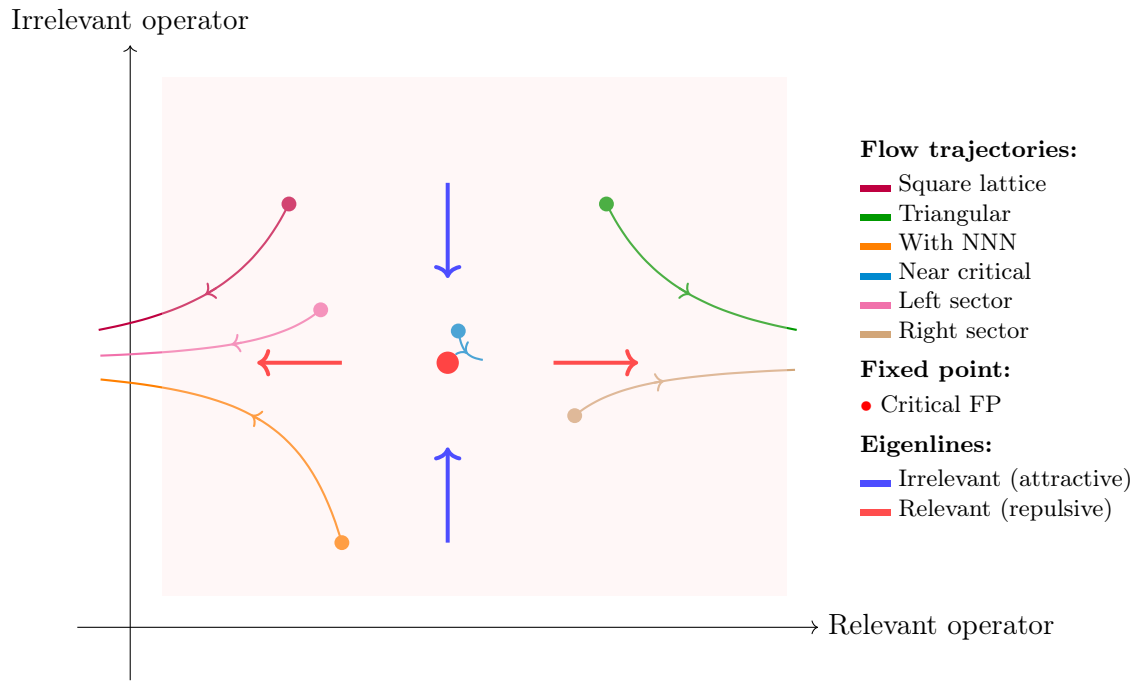

\begin{example}{Identifying Relevant and Irrelevant Operators:}
Consider the 2D Ising model near criticality. We can write down many possible terms in the Hamiltonian:
\begin{equation}
    H = -J\sum_{\langle ij\rangle} s_i s_j - J_2 \sum_{\langle\langle ij\rangle\rangle} s_i s_j - h\sum_i s_i - h_4 \sum_{\langle ijkl\rangle} s_i s_j s_k s_l + \ldots
\end{equation}

How important is each term? The answer is a property of the \emph{fixed point}, not something that can be read off from the Hamiltonian by inspection. One must either solve the model exactly or perform an explicit RG calculation to determine the scaling exponents $y$ for each coupling.

For the 2D Ising model, Onsager's exact solution provides the answer. The fixed point has exactly two relevant directions: the temperature deviation $t \sim (T - T_c)/T_c$ with $y_t = 1 > 0$, and the magnetic field $h$ with $y_h = 15/8 > 0$. Since the fixed point has only two relevant directions, every other coupling must have $y < 0$ and is therefore irrelevant. The next-nearest-neighbor coupling $J_2$ and the four-spin interaction $h_4$ both fall into this category: they flow to zero under coarse-graining and do not affect critical behavior.

Physically this is plausible. Under a block-spin step with block size $b$, the block spin averages over $b^d$ microscopic spins. Details that distinguish nearest-neighbor from next-nearest-neighbor geometry, or that correlate four spins around a plaquette, are short-range structures that get washed out by this averaging. But plausibility is not proof: the classification of operators requires a calculation, not an intuitive argument. This is why we can ignore lattice details when computing critical exponents. The microscopic differences between square, triangular, and hexagonal lattices all correspond to irrelevant operators that vanish under RG flow to the critical point.
\end{example}

This is the proper definition of a universality class: it is the set of all theories that flow to the same fixed point under RG. Two theories are equivalent, in the sense of sharing the same long-distance physics, if and only if they belong to the same universality class. The equivalence is one of RG fate, not of microscopic structure. Away from criticality this equivalence is coarser, both theories flow to the same trivial fixed point and agree on which phase they inhabit, but the universality class still leaves its mark in the shared structure of corrections to scaling and in how each theory approaches the transition. The fixed point is the organizing principle; the basin of attraction and the flow geometry are what partition theory space into equivalence classes.

\section{Quantum Phase Transitions}

\subsection{Zero-Temperature Criticality}

Classical phase transitions are driven by thermal fluctuations. At high temperature, entropy dominates and disordered phases prevail; at low temperature, energy dominates and ordered phases emerge. The critical point separating them occurs at a specific temperature $T_c$, determined by the competition between energy and entropy.

Quantum mechanics introduces a fundamentally different mechanism. At absolute zero, thermal fluctuations vanish entirely, yet a system can still undergo a phase transition if a non-thermal parameter, a coupling constant, an applied field, a chemical doping, is tuned through a critical value. These are \emph{quantum phase transitions} (QPTs), driven not by thermal energy but by quantum fluctuations arising from the uncertainty principle. The ground state of a quantum system is not simply the lowest-energy classical configuration; it is a superposition spread over many configurations by zero-point motion. As a control parameter is varied, this ground-state wave function can change qualitatively, undergoing a sharp, nonanalytic transition at $T = 0$.

\subsection{The Transverse-Field Ising Model}

The paradigmatic model for a QPT is the transverse-field Ising chain,
\begin{equation}
    H = -J\sum_i \sigma_i^z \sigma_{i+1}^z - h\sum_i \sigma_i^x,
    \label{eq:tfim}
\end{equation}
which students of this course will recognize immediately. The first term favors ferromagnetic alignment along $z$; the transverse field along $x$ introduces quantum fluctuations. Because $\sigma_i^z$ and $\sigma_i^x$ do not commute, the two terms are in direct competition and cannot be simultaneously minimized.

\begin{figure}[htb]
\centering
\includegraphics[width=0.75\textwidth]{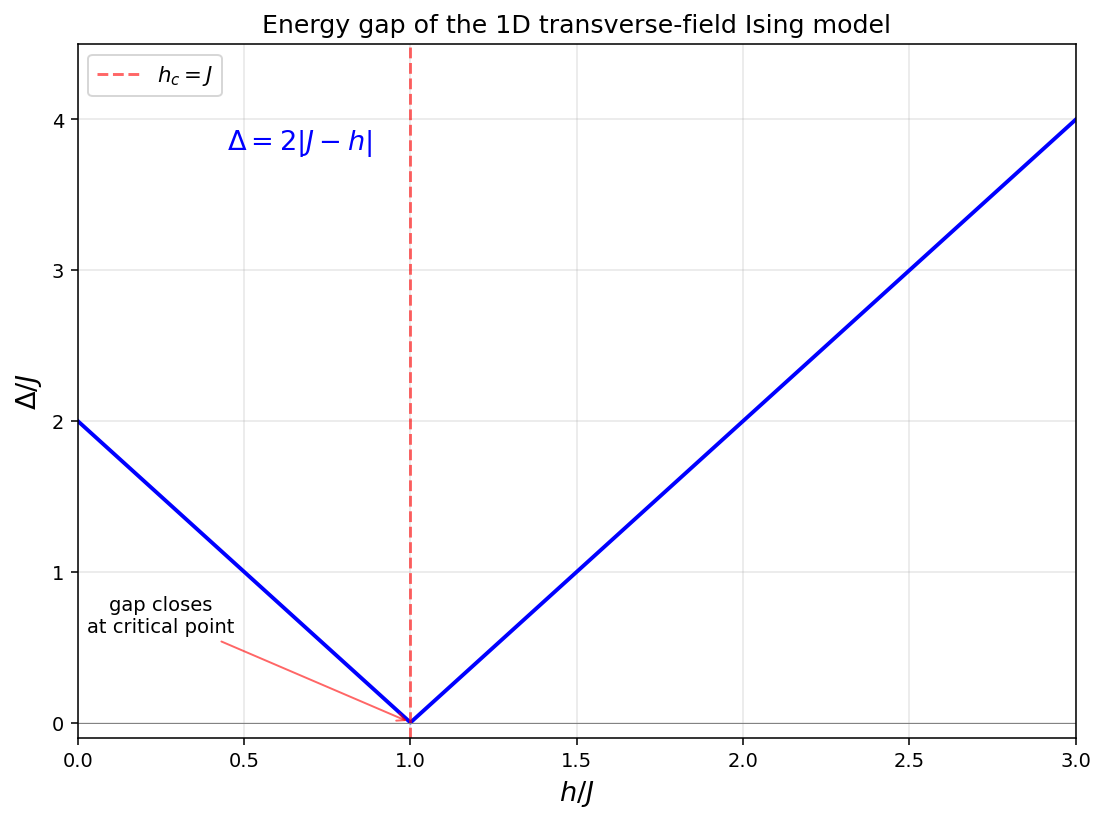}
\caption{Energy gap $\Delta$ of the 1D transverse-field Ising model as a function of the transverse field $h$, in units of the Ising coupling $J$. The exact result $\Delta = 2|J - h|$ takes the value $2J$ at $h = 0$ (the cost to flip a spin against its ferromagnetically aligned neighbors), decreases linearly with slope $-2J$ until it closes at the quantum critical point $h_c = J$, and then grows linearly with slope $+2J$ as $\Delta = 2(h - J)$ for $h > J$, asymptoting to $2h$ in the deep paramagnetic regime (the Zeeman cost to flip a spin against the transverse field). The linear closing reflects the critical exponent product $z\nu = 1$.}
\label{fig:tfim_gap}
\end{figure}

The limiting cases are transparent. When $h = 0$ the ground state is twofold degenerate: all spins aligned along $+z$ or all along $-z$. This is the ferromagnetic (ordered) phase, with $\langle \sigma_i^z \rangle \neq 0$. When $h \gg J$ the transverse field dominates and each spin settles into $|{+}x\rangle = (|{\uparrow}\rangle + |{\downarrow}\rangle)/\sqrt{2}$, giving $\langle \sigma_i^z \rangle = 0$. This is the paramagnetic (disordered) phase. The terminology refers to the order parameter $\langle\sigma_i^z\rangle$: the state is ``disordered'' in the sense that it has no spontaneous magnetization along $z$ and respects the $\mathbb{Z}_2$ symmetry $\sigma_i^z \to -\sigma_i^z$ generated by $\prod_i \sigma_i^x$. The ground state itself is of course a perfectly definite product state $|{+}x\rangle^{\otimes N}$; ``disorder'' here is a statement about correlations of the order parameter, not about the wavefunction. Between these extremes, at a critical value $h_c = J$ (the exact value follows from mapping to the 2D classical Ising model via the transfer matrix, discussed below), there is a quantum critical point. The spatial correlation length diverges as $\xi \sim |h - h_c|^{-\nu}$, exactly as at a classical critical point. But a quantum critical point has a feature absent from classical transitions: the energy gap $\Delta$ between the ground state and first excited state also closes,
\begin{equation}
    \Delta \sim \xi^{-z} \sim |h - h_c|^{z\nu}.
\end{equation}
The relation $\Delta \sim \xi^{-z}$ defines the \emph{dynamical critical exponent} $z$; we state it here without proof, as its derivation requires the transfer matrix formalism. Figure~\ref{fig:tfim_gap} shows the exact gap $\Delta = 2|J - h|$ obtained from the Jordan-Wigner solution of the 1D transverse-field Ising model, displaying the characteristic V-shape: the gap decreases linearly from $2J$ at $h = 0$, closes at the quantum critical point $h_c = J$, and then grows linearly as $2h$ for $h > J$. The linear closing is the signature $\Delta \sim |h - h_c|^{z\nu}$ with $z\nu = 1$, and for this model the exact values are $z = 1$ and $\nu = 1$ separately.

A closing gap means a diverging timescale: the inverse gap $\hbar/\Delta$ sets the characteristic time for quantum fluctuations between the two competing ground states. The dynamical exponent describes how this timescale grows relative to the spatial correlation length, with correlation time $\xi_\tau \sim \xi^z$. At a classical critical point, only spatial correlations diverge. At a quantum critical point, spatial and temporal correlations diverge together, and $z$ controls their relative rate. For the 1D transverse-field Ising model, the exact solution gives $z = 1$: space and imaginary time scale identically at the critical point.

\subsection{The Quantum-Classical Correspondence}

The deepest structural insight into QPTs is a mapping between quantum systems in $d$ spatial dimensions and classical systems in $d+1$ dimensions, where the extra dimension is imaginary time.

The connection runs through a substitution already noted in this chapter. Chapter~\ref{ch:time} introduced the time-evolution operator $e^{-i\hat{H}t/\hbar}$, and the Ising model section above observed that the Boltzmann factor $e^{-\beta\hat{H}}$ has the same mathematical form with $t$ replaced by $-i\beta\hbar$. Make this substitution explicit: define imaginary time $\tau$ by $t = -i\tau$, where $\tau$ is real and has units of time. The evolution operator becomes
\begin{equation}
    e^{-i\hat{H}t/\hbar}\big|_{t=-i\tau} = e^{-\hat{H}\tau/\hbar}.
\end{equation}
This is identical to the Boltzmann factor when $\tau = \beta\hbar = \hbar/(k_B T)$. The partition function is therefore
\begin{equation}
    Z = \mathrm{Tr}(e^{-\beta\hat{H}}) = \mathrm{Tr}\!\left(e^{-\hat{H}\tau/\hbar}\Big|_{\tau=\beta\hbar}\right),
\end{equation}
which is quantum evolution in imaginary time over a total ``length'' $\beta\hbar$, with periodic boundary conditions (since the trace closes the path). At finite temperature, this imaginary-time direction is compact with circumference $\hbar/(k_B T)$. As $T \to 0$, $\beta \to \infty$ and the imaginary-time direction becomes infinitely long, adding a genuine new spatial-like dimension to the problem.

\paragraph{Why this is legitimate.} The substitution $t \to -i\tau$ may look like an arbitrary trick, but no physics is being analytically continued. Both $e^{-i\hat{H}t/\hbar}$ and $e^{-\beta\hat{H}}$ are well-defined operators built from the same Hermitian Hamiltonian $\hat{H}$, with the same spectrum and the same eigenstates. They are simply different functions of $\hat{H}$. The partition function $Z = \mathrm{Tr}(e^{-\beta\hat{H}})$ is a number we want to compute; the observation is that this number happens to equal $\mathrm{Tr}(e^{-i\hat{H}t/\hbar})$ evaluated at the formal value $t = -i\beta\hbar$. We are identifying two expressions, not deforming time itself.

\paragraph{Where the extra dimension comes from.} To compute the trace, insert resolutions of the identity. Split $\beta = N\Delta\tau$ into $N$ small slices and use $e^{-\beta\hat{H}} = (e^{-\Delta\tau\hat{H}})^N$. Between each pair of factors, insert $\sum_n \ket{n}\bra{n} = \mathbb{I}$ in some convenient basis. The trace becomes
\begin{equation}
    Z = \sum_{n_1,\ldots,n_N} \bra{n_1}e^{-\Delta\tau\hat{H}}\ket{n_N}\bra{n_N}e^{-\Delta\tau\hat{H}}\ket{n_{N-1}}\cdots\bra{n_2}e^{-\Delta\tau\hat{H}}\ket{n_1}.
\end{equation}
Each intermediate index $n_k$ labels the state at imaginary time $\tau_k = k\Delta\tau$; the original Hilbert space has been replaced by an array of $N$ copies of itself, indexed by a discrete time slice. The trace forces $n_{N+1} = n_1$, giving periodic boundary conditions in $\tau$. The ``extra dimension'' is just the bookkeeping label for these intermediate sums.

\paragraph{A worked example: one qubit in a transverse field.} Take $\hat{H} = -h\sigma^x$ on a single qubit. Then $Z = \mathrm{Tr}(e^{\beta h \sigma^x}) = 2\cosh(\beta h)$, which we can also obtain directly from the eigenvalues $\pm h$. Now compute it the imaginary-time way. Slice $\beta = N\Delta\tau$ and insert $\sigma^z$-eigenstates $\ket{s}$ with $s = \pm 1$ between slices:
\begin{equation}
    Z = \sum_{s_1,\ldots,s_N} \prod_{k=1}^{N} \bra{s_{k+1}}e^{\Delta\tau\, h\sigma^x}\ket{s_k}, \qquad s_{N+1} = s_1.
\end{equation}
The single-slice matrix element is elementary. Expand $e^{\Delta\tau\, h\sigma^x} = \cosh(\Delta\tau\, h)\mathbb{I} + \sinh(\Delta\tau\, h)\sigma^x$ and use $\bra{s'}\mathbb{I}\ket{s} = \delta_{s's}$, $\bra{s'}\sigma^x\ket{s} = \delta_{s',-s}$:
\begin{equation}
    \bra{s'}e^{\Delta\tau\, h\sigma^x}\ket{s} = \begin{cases}\cosh(\Delta\tau\, h) & s' = s,\\ \sinh(\Delta\tau\, h) & s' = -s.\end{cases}
\end{equation}
Both cases can be written in the unified form $A\,e^{K s' s}$, with $A^2 = \cosh(\Delta\tau\, h)\sinh(\Delta\tau\, h) = \tfrac{1}{2}\sinh(2\Delta\tau\, h)$ and $e^{2K} = \coth(\Delta\tau\, h)$. Substituting:
\begin{equation}
    Z = A^N \sum_{\{s_k\}} \exp\!\left(K\sum_{k=1}^N s_{k+1}s_k\right).
\end{equation}
This is exactly the partition function of a \emph{classical} 1D Ising chain of $N$ spins with periodic boundary conditions and coupling $K$. A single qubit in 0 spatial dimensions has become a classical Ising model in 1 dimension, with the new dimension being imaginary time. Summing this classical chain by the transfer-matrix method (a standard exercise) reproduces $Z = 2\cosh(\beta h)$ exactly in the $N \to \infty$ limit; the equivalence is not approximate.

The same construction applied to the 1D transverse-field Ising chain (now with both $\sigma^z\sigma^z$ and $\sigma^x$ terms) generates a 2D classical Ising model on a square lattice, where one direction is the original spatial chain and the other is imaginary time.

For the 1D transverse-field Ising model, this Euclidean path-integral representation maps onto the 2D classical Ising model, solved exactly by Onsager in 1944. The QPT at $h_c$ in the quantum model corresponds precisely to the thermal transition at $T_c$ in the classical model, and the two systems share the same universality class and the same RG fixed point. The full derivation parallels the single-qubit calculation above but with two basis labels per site (spatial position and time slice) and an additional $\sigma^z\sigma^z$ factor in each slice; the essential mechanism is the same $t \to -i\tau$ substitution that converts quantum dynamics into classical statistical mechanics in one higher dimension. More generally, a $d$-dimensional quantum system at a QPT belongs to the universality class of a $(d+z)$-dimensional classical system, with $z$ controlling the anisotropy between spatial and imaginary-time directions.

This correspondence means that the entire RG machinery developed for classical critical phenomena, relevant and irrelevant operators, universality, scaling relations among critical exponents, applies directly to quantum phase transitions. The two subjects are facets of the same mathematical structure.

\section{From Spins to Fields}
\label{sec:spins_to_fields}

The RG framework developed so far rests on Kadanoff's scaling hypothesis, which assumes that scaling exponents $y_t$ and $y_h$ exist at the critical fixed point and lets all other critical exponents follow from them. But Kadanoff's hypothesis takes those two numbers as inputs, supplied by Onsager's exact solution in 2D, or by experiment and numerical simulation in 3D. The remainder of this chapter closes that gap. A continuum description of the order parameter, combined with a momentum-space implementation of the RG, produces a beta function whose fixed points and eigenvalues are the scaling exponents Kadanoff's hypothesis required. The key insight, due to Wilson and Fisher (1972), is that working in $d = 4 - \varepsilon$ dimensions converts a degenerate marginal situation into a calculable interacting fixed point. Setting $\varepsilon = 1$ gives quantitative predictions for three-dimensional critical exponents.

The earlier sections worked with discrete spins $s_i = \pm 1$ on a lattice. After many rounds of block-spin coarse-graining, the lattice spacing becomes irrelevant and the order parameter varies smoothly over large distances. At that point it is natural to replace the discrete spin at each site with a continuous real number $\phi(\mathbf{x})$: the coarse-grained magnetization, averaged over a small region around $\mathbf{x}$. In the ordered phase $\langle\phi\rangle \neq 0$; in the disordered phase $\langle\phi\rangle = 0$. This is the same field that appeared in the free-Gaussian discussion of the correlation function in the chapter introduction.

\subsection{The Partition Function as a Functional Integral}

For a lattice system the partition function sums over all spin configurations weighted by $e^{-\beta H}$. For a continuous field the analogous object is
\begin{equation}
    Z = \int \mathcal{D}\phi\; e^{-S_E[\phi]},
    \label{eq:Z_field}
\end{equation}
where $\int\mathcal{D}\phi$ means ``sum over every possible function $\phi(\mathbf{x})$
''  and $S_E[\phi]$ is the \emph{Euclidean action}, which plays the role of energy: configurations with small $S_E$ are exponentially favored, exactly as low-energy spin configurations are favored by $e^{-\beta H}$. The subscript $E$ refers to the quantum-classical correspondence developed earlier in this chapter: the thermal density matrix $e^{-\beta H}$ is mathematically equivalent to the quantum time-evolution operator evaluated at imaginary time. Expressing the partition function as a functional integral produces weights $e^{-S_E}$ that are real and positive like Boltzmann factors.

\subsection{\texorpdfstring{The $\phi^4$ Action}{The phi-fourth Action}}

The action $S_E[\phi]$ must respect the same symmetries as the underlying lattice model: spin-flip symmetry $\phi\to -\phi$ (so only even powers appear), and translation and rotation invariance. The simplest action with these properties is
\begin{equation}
    S_E[\phi] = \int d^dx \left[\frac{1}{2}(\nabla\phi)^2 + \frac{1}{2}m^2\phi^2 + \frac{\lambda}{4!}\phi^4 + \cdots\right].
    \label{eq:phi4_action}
\end{equation}
Each term has a clear physical meaning. The gradient term $(\nabla\phi)^2$ penalizes spatial variation: a uniform field costs nothing, rapid fluctuations are suppressed. This is the continuum analog of the Ising nearest-neighbor coupling. The mass term $\frac{1}{2}m^2\phi^2$ controls ordering: when $m^2 > 0$ the only minimum is $\phi = 0$ (disordered); when $m^2 < 0$ the minimum shifts to $\phi \neq 0$ (ordered). Near criticality, $m^2 \propto T - T_c$, so tuning $m^2$ through zero is tuning through the phase transition. The quartic term $\frac{\lambda}{4!}\phi^4$ with $\lambda > 0$ stabilizes the action when $m^2 < 0$, creating the double-well potential that describes the two ordered states. This theory is called $\phi^4$ theory.

Equation~\eqref{eq:phi4_action} is the Landau-Ginzburg free energy familiar from statistical mechanics. The field theory is not a new object: it is the continuum limit of the block-spin description, now equipped with tools to perform the RG explicitly.

\section{Dimensional Analysis and the Upper Critical Dimension}
\label{sec:dimensional_analysis}

Before doing any calculation, dimensional analysis already reveals the most important qualitative fact about $\phi^4$ theory. The calculation is short but deserves to be worked out explicitly, because its conclusion drives everything that follows.

\paragraph{How many dimensionful quantities are there?}
Only one. The action~\eqref{eq:phi4_action} contains no $\hbar$, no $c$, no particle mass; the only ingredients are the field $\phi$ and the couplings $m^2$ and $\lambda$, integrated over space. The sole dimensionful scale in the problem is the short-distance cutoff, the lattice spacing $a$ (or equivalently the cutoff momentum $\Lambda = 1/a$). Every dimensionful quantity in the theory can therefore be expressed as a pure number times a power of $a$. We write $[X] = a^p$ to mean that $X$ has dimensions of length to the $p$; equivalently, $[X] = \mathrm{mass}^{-p}$ since inverse length plays the role of mass here.

This is a weaker convention than the atomic units used throughout the rest of this course. In atomic units one sets $\hbar = m_e = e = 1$ (with $4\pi\epsilon_0 = 1$), collapsing mechanical problems to a single scale, the Bohr radius $a_0$; note that $c = 1/\alpha \approx 137$ is \emph{not} unity. In high-energy physics one instead sets $\hbar = c = 1$, making length and inverse mass interchangeable. The $\phi^4$ theory has neither $\hbar$ nor $c$ nor $m_e$ to begin with; the unit of length is whatever cutoff scale you chose when regulating the theory. In all three conventions the structural point is the same: fixing enough fundamental constants to unity leaves exactly one independent dimension, and every coupling and field is labeled by its power of that one dimension.

\paragraph{The dimension of the field.}
Because $S_E$ appears in $e^{-S_E}$, it must be a pure number, $[S_E] = a^0$. The integration measure contributes $[d^dx] = a^d$, so every term in the integrand must carry dimension $a^{-d}$ to cancel.

Start with the gradient term, which has the simplest structure. A derivative $\nabla$ has dimension $a^{-1}$ (it is $\partial/\partial x$), so
\begin{equation}
    [(\nabla\phi)^2] = a^{-2}\,[\phi]^2.
\end{equation}
Setting this equal to $a^{-d}$ gives $[\phi]^2 = a^{-(d-2)}$, so
\begin{equation}
    [\phi] = a^{-(d-2)/2}.
    \label{eq:phi_dim}
\end{equation}
In $d=4$, for instance, $[\phi] = a^{-1}$, i.e., $\phi$ has the dimension of inverse length (or of mass). This is the \emph{canonical} dimension of the field, the dimension demanded by requiring the kinetic term alone to be dimensionless.

\paragraph{The dimension of $m^2$.}
The mass term has dimension $[m^2]\cdot[\phi^2] = [m^2]\cdot a^{-(d-2)}$. Setting this to $a^{-d}$ gives
\begin{equation}
    [m^2] = a^{-2}, \qquad [m] = a^{-1},
\end{equation}
so $m$ is an inverse length (equivalently, a mass). This matches the intuition from the correlation-function discussion at the start of this chapter: $m$ is the inverse correlation length away from criticality, and the phase transition is approached by tuning $m^2 \to 0$.

\paragraph{The dimension of $\lambda$.}
The quartic term has dimension $[\lambda]\cdot[\phi^4] = [\lambda]\cdot a^{-2(d-2)}$. Setting this to $a^{-d}$:
\begin{equation}
    [\lambda] = a^{-d}\cdot a^{2(d-2)} = a^{d-4}.
\end{equation}
Converting to mass units via $[X] = \mathrm{mass}^{-p}$ when $[X] = a^p$,
\begin{equation}
    [\lambda] = \mathrm{mass}^{4-d} \equiv \mathrm{mass}^\varepsilon, \qquad \varepsilon \equiv 4-d.
    \label{eq:lambda_dim}
\end{equation}
This is the crucial result. From the operator classification developed earlier, a coupling with positive mass dimension grows under coarse-graining (relevant); one with negative mass dimension shrinks (irrelevant). The $\phi^4$ coupling $\lambda$ therefore changes character depending on dimension:

\begin{center}
\begin{tabular}{lcc}
\hline
Dimension & $[\lambda]$ & Classification \\
\hline
$d > 4$ & negative & irrelevant \\
$d = 4$ & dimensionless & marginal \\
$d < 4$ & positive & relevant \\
\hline
\end{tabular}
\end{center}

Above $d = 4$ interactions are irrelevant: the long-distance physics is governed by the free-field (Gaussian) fixed point, and mean-field exponents ($\nu = 1/2$, $\eta = 0$) are exact. Below $d = 4$ interactions are relevant: they grow under coarse-graining and drive the system to a new interacting fixed point with different exponents. At $d = 4$ exactly, $\lambda$ is dimensionless and the situation is marginal, with quantum fluctuations determining the outcome. This makes $d = 4$ the \emph{upper critical dimension} of $\phi^4$ theory, the boundary below which mean-field theory fails.

\begin{keyidea}{Upper Critical Dimension}
The failure of mean-field theory in three dimensions is not an accident. The interaction coupling $\lambda$ is classically relevant for $d < 4$, meaning fluctuations grow as one coarse-grains, pushing the system away from the mean-field (Gaussian) fixed point toward a new fixed point with different critical exponents. The dimension $d = 4$ is the boundary: above it, mean-field theory is exact; below it, fluctuations matter.
\end{keyidea}

\section{Aside: Cutoffs in Single-Particle Quantum Mechanics}
\label{sec:cutoffs_qm_comparison}

The preceding dimensional analysis may feel unfamiliar, but the underlying logic appeared much earlier in this course. Chapter~\ref{ch:lattice_qft} constructed the continuum Schr\"odinger equation as the limit of a tight-binding chain with lattice spacing $a$, and the continuum limit was recovered by taking $a \to 0$ at fixed $\hbar^2/(2ma^2)$. That lattice spacing played exactly the role the cutoff $a = 1/\Lambda$ plays here in $\phi^4$ theory: it is the shortest distance the theory can resolve, and physics at scales $\gg a$ is described by the continuum effective theory. Dimensional analysis in the continuum is, in both cases, bookkeeping for how quantities scale relative to the one microscopic length the regulator supplies.

The shared logic runs deeper than just ``there is a lattice somewhere.'' Several features transfer directly.

\paragraph{Continuum theories need cutoffs to be well-defined.} In single-particle QM, this is usually invisible, because the standard problems (harmonic oscillator, hydrogen atom, particle in a box) are finite without any regulator. But the moment one considers singular potentials, the infinities of continuum QM become manifest. A delta-function potential $V(x) = g\,\delta(x)$ in one dimension is perfectly integrable and produces a single bound state, but the analogous contact interaction in two or three dimensions requires a short-distance cutoff to define at all; the bare coupling has to be tuned with the cutoff to give a finite physical binding energy. Likewise the attractive inverse-square potential $V(r) = -\hbar^2 g/(2mr^2)$ exhibits the ``fall to the center'' pathology for $g$ above a critical value and requires a self-adjoint extension, a regulator in disguise. Field theory generalizes this: every loop integral $\int d^dk/(k^2 + m^2)^n$ diverges at large $k$ unless cut off at $|k| < \Lambda = 1/a$. The lattice is implicit in one-body QM and explicit in $\phi^4$, but the same principle applies: the continuum is a limit, and the regulator is what makes the limit sensible.

\paragraph{Dimensional analysis classifies perturbations identically in both settings.} The operator classification of Section~\ref{sec:dimensional_analysis} (relevant, marginal, irrelevant, according to the sign of the mass dimension) is not specific to statistical field theory. It applies, implicitly, to perturbations of single-particle QM as well. Consider a 1D Hamiltonian $\hat{H}_0 = -\nabla^2/(2m)$ perturbed by a short-range potential $V(x)$ of characteristic range $a$ and strength $g_n$ with $[g_n] = a^{n-1}$. The dimensionless coupling at a resolution $\ell \gg a$ is $g_n \ell^{1-n}$, which grows when $n > 1$, stays fixed when $n = 1$, and shrinks when $n < 1$. Translated back to familiar examples:

\begin{itemize}
\item A delta function $V = g\,\delta(x)$ has $[g] = a^0$ in $d = 1$ (the integral $\int V\,dx$ must give an energy, and $\int\delta(x)\,dx = 1$ is dimensionless in the sense of having no $a$ dependence once $\delta(x)$ is written in units of the cutoff). A little more carefully, the dimensionless ratio controlling the physics is $g\,(m/\hbar^2)\,a$, which \emph{shrinks} as $a \to 0$. The 1D delta function is \emph{irrelevant} in the RG sense.
\item An inverse-square potential $V = -\hbar^2 g/(2mr^2)$ has dimensionless $g$. No power of the cutoff appears in the dimensionless coupling; it is \emph{marginal}. This is precisely the potential whose subtle behavior (fall-to-the-center, required self-adjoint extension) reflects marginality: logarithmic running of the effective coupling replaces simple power-law scaling, and the physics depends sensitively on the regulator.
\item A $1/r^4$ potential has $[g] = a^2$; its dimensionless coupling grows as $a \to 0$, so it is \emph{relevant}. The short-distance physics dominates, and the continuum theory is singular without a cutoff.
\end{itemize}

The trichotomy is exactly the one we found for $\phi^4$. Relevant perturbations drive the low-energy physics away from the free Hamiltonian; irrelevant perturbations become progressively less important at long distances; marginal perturbations sit on the boundary and require more careful analysis.

\paragraph{Where the analogy genuinely breaks.} Two important differences keep the parallel from being complete.

First, in most single-particle problems the continuum limit is painless. A particle in a smooth confining potential is already well-defined in the continuum; the lattice is a numerical convenience, not a conceptual necessity. Finite-difference solvers for the Schr\"odinger equation use a grid spacing because computers demand it, not because the physics demands it. In $\phi^4$ theory the situation is qualitatively different: even perturbation theory at one loop diverges without a cutoff. The cutoff is mandatory, and the central question of renormalization is whether one can remove it in a controlled way.

Second, the RG flow in a non-interacting single-particle problem is trivial. Coarse-graining a particle in a smooth potential just produces a coarser grid with renormalized effective-mass and effective-potential parameters; the flow has no fixed points other than the free theory. The entire apparatus of the Wilson-Fisher fixed point, the running coupling, and nontrivial critical exponents exists because $\phi^4$ is an \emph{interacting} theory whose coupling $\lambda$ generates a nontrivial flow competing with classical scaling. Single-particle QM is the base case of a theory where everything is classical and the flow is geometric; interacting field theory is where the flow first becomes dynamical.

\begin{physicalinsight}
The lattice spacing introduced in Chapter~\ref{ch:lattice_qft} to discretize the Schr\"odinger equation, and the cutoff $\Lambda = 1/a$ introduced here to regulate $\phi^4$ theory, are the same conceptual object. Both set the shortest length scale the theory resolves; both make dimensional analysis well-defined by supplying the one unit of length relative to which couplings are measured; both classify perturbations into relevant, marginal, and irrelevant by how their dimensionless versions behave under coarse-graining. The content of the renormalization group is the extension of this classification from the kinematic setting of single-particle QM, where the classification reduces to dimensional analysis of a potential, to the dynamical setting of interacting field theory, where quantum fluctuations generate nontrivial flow between couplings.
\end{physicalinsight}

\section{The Renormalization Group Flow}
\label{sec:rg_flow_continuum}

The momentum-space RG is the continuum analog of decimation. Every field configuration can be decomposed into Fourier modes at different wavevectors. The RG proceeds in two steps: first integrate out the short-wavelength (high-momentum) modes between $\Lambda/b$ and $\Lambda$, then rescale momenta by $k \to bk$ to restore the original cutoff. Repeating this generates a flow in the space of coupling constants, exactly as for the lattice models considered earlier.

\paragraph{The RG ``time'' $\ell$.}
It is convenient to parametrize the sequence of RG steps continuously. Define
\begin{equation}
    \ell \equiv \ln b,
\end{equation}
so that one RG step with rescaling factor $b$ advances $\ell$ by $\ln b$, and an infinitesimal step advances $\ell$ by $d\ell$. The cutoff after $\ell$ steps is $\Lambda(\ell) = \Lambda_0 e^{-\ell}$: $\ell$ increases as we coarse-grain to longer distances. Every coupling in the theory becomes a function of $\ell$, and the RG flow is specified by a set of differential equations, one per coupling.

\paragraph{Dimensionless couplings.}
The couplings $m^2$ and $\lambda$ carry powers of the cutoff, as established in Section~\ref{sec:dimensional_analysis}: $[m^2] = \Lambda^2$ and $[\lambda] = \Lambda^{\varepsilon}$ where $\varepsilon = 4 - d$. Because the cutoff itself flows under RG, the bare couplings flow trivially just from this dimensional scaling. To isolate the non-trivial, fluctuation-driven part of the flow we work with dimensionless versions:
\begin{align}
    t &\equiv m^2/\Lambda^2, \label{eq:t_dimless}\\
    g &\equiv \lambda\,\Lambda^{-\varepsilon}\,K_d, \label{eq:g_dimless}
\end{align}
where $K_d = S_{d-1}/(2\pi)^d$ is a conventional numerical normalization (the surface area of the unit sphere in $d$ dimensions divided by $(2\pi)^d$) that absorbs the angular integrals appearing in the one-loop calculation below, making the coefficients of the beta function come out as simple integers. With these definitions $t$ and $g$ are pure numbers.

\paragraph{Classical part of the flow.}
Part of the flow follows immediately from dimensional analysis and requires no calculation. Holding the bare coupling $\lambda$ fixed while lowering the cutoff $\Lambda \to \Lambda\,e^{-d\ell}$, the dimensionless coupling defined in~\eqref{eq:g_dimless} changes as
\begin{equation}
    \frac{dg}{d\ell}\bigg|_{\text{classical}} = \varepsilon\,g,
\end{equation}
simply because $g \propto \Lambda^{-\varepsilon}$ and $d\ln\Lambda/d\ell = -1$. This is the relevant scaling that Kadanoff's hypothesis assigned to $\lambda$ with exponent $\varepsilon$, now derived from the action rather than assumed.

\paragraph{Fluctuation corrections.}
Integrating out the momentum shell $\Lambda e^{-d\ell} < |k| < \Lambda$ generates additional contributions. The mechanism is the following: in perturbation theory the $\phi^4$ vertex couples four fields, and contracting two pairs of fields from two different vertices across the momentum shell produces an effective correction to the remaining four-field (coupling) term. The correction is an integral over the shell, and the resulting flow equation at one-loop order is
\begin{equation}
    \frac{dg}{d\ell} = \varepsilon\,g - 3\,g^2 + \cdots,
    \label{eq:dg_dl}
\end{equation}
which we state without proof: the calculation that produces the coefficient $3$ is a standard one-loop exercise in quantum field theory, requiring the Feynman rules for $\phi^4$ theory and evaluation of an integral over the momentum shell. The factor of $3$ is a combinatorial count of the ways the four external legs of two $\phi^4$ vertices can be paired so that two pairs are contracted across the shell and two pairs remain external. Two features of equation~\eqref{eq:dg_dl} deserve comment. The correction is \emph{quadratic} in $g$ because the diagram involves two vertices (one factor of $\lambda$ each). It is \emph{negative} because integrating out high-momentum modes reduces the strength of the interaction at long wavelengths: short-wavelength fluctuations screen the bare coupling. The sign is what makes a non-trivial fixed point possible.

\paragraph{The beta function.}
The right-hand side of equation~\eqref{eq:dg_dl} is, by definition, the \emph{beta function}:
\begin{equation}
    \beta(g) \equiv \frac{dg}{d\ell} = \varepsilon g - 3g^2.
    \label{eq:beta_function}
\end{equation}
The name comes from quantum field theory, where $\beta(g)$ describes how a running coupling evolves with scale. A zero of $\beta$ is a fixed point of the RG flow, and the sign of $\beta'(g)$ at a fixed point determines whether the flow is attractive ($\beta' < 0$) or repulsive ($\beta' > 0$). The competition between the classical term $\varepsilon g$ and the fluctuation correction $-3g^2$ is what determines the fixed-point structure analyzed next.

\section{The Wilson-Fisher Fixed Point}
\label{sec:wf_fixed_point}

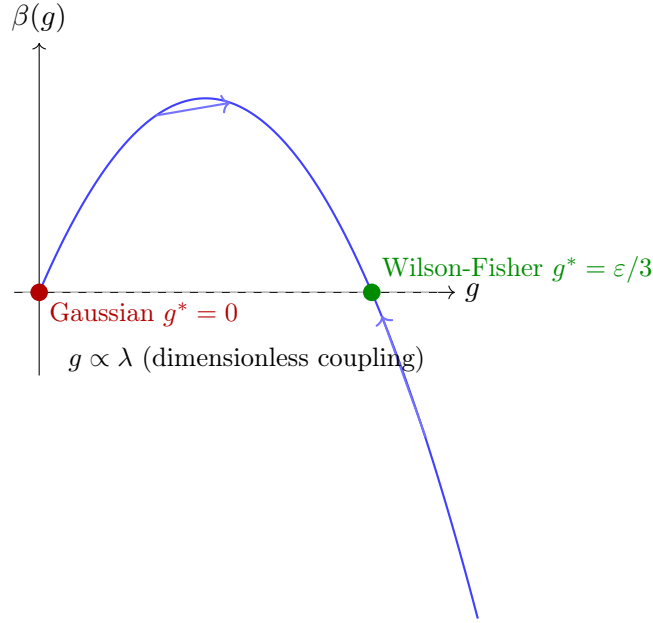
\begin{figure}[htbp]
\centering
\begin{tikzpicture}[scale=1.1]
    \draw[->] (-0.3,0) -- (5.0,0) node[right] {$g$};
    \draw[->] (0,-1.0) -- (0,3.0) node[above] {$\beta(g)$};
    \draw[dashed,gray!40] (-0.2,0) -- (4.8,0);
    \draw[thick, blue!75] plot[domain=0:0.44,samples=100] ({12*\x},{28*(\x - 3*\x*\x)});
    \fill[red!70!black] (0,0) circle (3pt)
        node[below right, font=\small] {Gaussian $g^*=0$};
    \fill[green!55!black] (4,0) circle (3pt)
        node[above right, font=\small, green!55!black] {Wilson-Fisher $g^*=\varepsilon/3$};
    \draw[->,blue!55,thick] (1.4,{28*(0.117-3*0.117*0.117)}) -- (2.3,{28*(0.192-3*0.192*0.192)});
    \draw[->,blue!55,thick] (4.65,{28*(0.387-3*0.387*0.387)}) -- (4.12,{28*(0.343-3*0.343*0.343)});
    \node[font=\small] at (2.5,-0.8) {$g \propto \lambda$ (dimensionless coupling)};
\end{tikzpicture}
\caption{The beta function $\beta(g) = \varepsilon g - 3g^2$ for $\phi^4$ theory in $d = 4-\varepsilon$ (shown for $\varepsilon = 1$, i.e., $d = 3$). The classical term $\varepsilon g$ drives $g$ upward; the fluctuation term $-3g^2$ drives it downward. The Gaussian fixed point is unstable for $d < 4$; the Wilson-Fisher fixed point at $g^* = \varepsilon/3$ is stable.}
\label{fig:beta_function}
\end{figure}

Setting $dg/d\ell = 0$ in~\eqref{eq:beta_function} gives two fixed points: $g^* = 0$ and $g^* = \varepsilon/3$. Figure~\ref{fig:beta_function} shows their structure.

The \textbf{Gaussian fixed point} at $g^* = 0$ is the free, non-interacting theory. Above four dimensions ($\varepsilon < 0$), it is stable: any small interaction is driven back to zero by the $-3g^2$ term. Below four dimensions ($\varepsilon > 0$), it is unstable: the $\varepsilon g$ term wins at small $g$ and interactions grow.

The \textbf{Wilson-Fisher fixed point} at $g^* = \varepsilon/3$ exists only for $\varepsilon > 0$. It is stable: the beta function has negative slope there, so any nearby value of $g$ flows back toward $g^*$. This is the fixed point governing critical behavior in $d < 4$.

\subsection{Critical Exponents From the Fixed Point}

Once the Wilson-Fisher fixed point is located, the critical exponents follow by linearizing the RG flow around it, exactly the procedure applied to Kadanoff's scaling hypothesis earlier. The linearized flow has two eigenvalues. The first,
\begin{equation}
    y_t = 2 - \frac{\varepsilon}{3} + O(\varepsilon^2),
    \label{eq:yt_wf}
\end{equation}
is the thermal scaling exponent: it is positive (temperature deviation is relevant), and it is the number that Kadanoff's hypothesis required as input. The correlation length exponent follows directly:
\begin{equation}
    \nu = \frac{1}{y_t} \approx \frac{1}{2} + \frac{\varepsilon}{12}.
    \label{eq:nu_wf}
\end{equation}
At $\varepsilon = 0$ (four dimensions), $\nu = 1/2$: mean-field theory. At $\varepsilon = 1$ (three dimensions), $\nu \approx 0.583$, compared to the exact 3D Ising value $\nu = 0.630$. The second eigenvalue $y_g = -\varepsilon$ is negative, confirming the fixed point is stable in the coupling direction.

The magnetic scaling exponent $y_h \approx 3 - \varepsilon/2$ follows from how an external magnetic field rescales under the RG. All other critical exponents ($\beta$, $\gamma$, $\delta$, $\alpha$) then follow from $y_t$ and $y_h$ through the scaling relations derived earlier in this chapter. The $\varepsilon$ expansion thus converts Kadanoff's hypothesis into a calculation.

\begin{physicalinsight}
The Wilson-Fisher fixed point exists because of a competition between two tendencies. Classical scaling drives the coupling $g$ upward (interactions are relevant below four dimensions). Fluctuations drive $g$ downward (short-wavelength modes, when integrated out, weaken the effective long-wavelength coupling). These two effects balance at $g^* = \varepsilon/3$. The resulting fixed point is not mean-field ($g^* = 0$) but it is close to mean-field when $\varepsilon$ is small, which is why the $\varepsilon$ expansion works.
\end{physicalinsight}

\subsection{Universality from Symmetry and Dimension}

The Wilson-Fisher fixed point at $g^* = \varepsilon/3$ has no free parameters. It depends only on the spatial dimension $d$ and the $\mathbb{Z}_2$ symmetry of the $\phi^4$ action. This is the field-theory explanation of the universality observed earlier: any microscopic model with $\mathbb{Z}_2$ symmetry in $d < 4$ flows to the same fixed point under coarse-graining, regardless of lattice type, interaction range, or coupling strength. The fixed point and its eigenvalues are determined by symmetry and dimension alone.

The irrelevant operators discussed earlier in this chapter now have explicit identities. The $\phi^6$ term has mass dimension $6 - 2d$, which is negative for $d > 3$: it is irrelevant at the Wilson-Fisher fixed point in three dimensions. Lattice anisotropy, next-neighbor interactions, the precise shape of the lattice all map onto irrelevant operators that flow to zero under repeated coarse-graining. They shift $T_c$ but leave the critical exponents unchanged.

When the symmetry group changes, so does the universality class. An $N$-component order parameter with $O(N)$ symmetry gives a beta function
\begin{equation}
    \frac{dg}{d\ell} = \varepsilon g - \frac{N+8}{3}\,g^2,
    \label{eq:ON_beta}
\end{equation}
with fixed point $g^* = 3\varepsilon/(N+8)$. The Ising model ($N=1$), the XY model for superfluids ($N=2$), and the Heisenberg model for isotropic ferromagnets ($N=3$) all fit this framework with different values of $N$, producing distinct universality classes with different exponents, consistent with what is observed experimentally.

\section{Chapter Summary}

The RG resolves the tension between discrete and continuous descriptions that we encountered in Chapter~\ref{ch:foundations}. When we started with qubits and built up to the Schr\"odinger equation, we were implicitly performing an RG flow. The lattice quantum field theory in Chapter~\ref{ch:lattice_qft} provided a regularization, a well-defined theory at all scales. Taking the continuum limit meant flowing to a particular fixed point where the lattice spacing became irrelevant. This perspective transforms how we think about fundamental physics. Rather than asking ``is space fundamentally discrete or continuous?'', we should ask ``which RG fixed point describes nature at the scales we observe?'' Quantum field theory works not because spacetime is truly continuous, but because the continuum fixed point accurately describes physics at length scales much larger than any underlying cutoff.

The shift from fixed theories to flows in theory space represents one of the deepest insights in modern theoretical physics. Rather than asking ``what is the theory at all scales?'' the RG teaches us to ask ``how do theories transform under changes of scale?'' Fixed points organize the space of all theories, with a small number of relevant operators determining which fixed point a system flows toward, while the vast majority of microscopic details are washed out as irrelevant.

Three pillars of the RG framework have emerged. First, the scaling hypothesis: near a critical fixed point, the linearized RG acts as a pure rescaling of a small number of relevant couplings, and all critical exponents follow from the corresponding eigenvalues. Second, universality: systems flowing to the same fixed point share identical long-distance physics regardless of microscopic detail, with the universality class determined by symmetry and dimension alone. Third, the quantum-classical correspondence: a $d$-dimensional quantum system at a quantum critical point belongs to the universality class of a $(d+z)$-dimensional classical system, so the entire RG machinery applies equally to thermal and quantum transitions.

These pillars are not independent assumptions but consequences of a single calculation. The Landau-Ginzburg free energy is the Euclidean action for a scalar field $\phi(\mathbf{x})$, and writing the partition function as a functional integral $Z = \int\mathcal{D}\phi\,e^{-S_E[\phi]}$ places it on the same footing as a quantum field theory. Momentum-space RG produces the beta function $dg/d\ell = \varepsilon g - 3g^2$, whose Wilson-Fisher fixed point at $g^* = \varepsilon/3$ governs critical behavior in $d < 4$. The scaling exponents $y_t \approx 2 - \varepsilon/3$ and $y_h \approx 3 - \varepsilon/2$ are not inputs from experiment but eigenvalues of the linearized flow, and setting $\varepsilon = 1$ gives quantitative predictions for three-dimensional critical behavior. This unifies statistical mechanics, quantum field theory, and condensed matter physics, and provides a satisfying resolution to questions about the nature of the continuum that motivated our journey from qubits to quantum fields.

\input{book_problems/ch12_problems.tex}

\section*{References and Further Reading}
\addcontentsline{toc}{section}{References and Further Reading}

\begin{description}
\item[Cardy, J.] \emph{Scaling and Renormalization in Statistical Physics}. Cambridge Lecture Notes in Physics, Cambridge University Press, 1996. The standard graduate introduction to the RG and critical phenomena, emphasizing scaling and field-theoretic methods. Chapters~3, 5, and 7 cover Kadanoff scaling, the $\varepsilon$-expansion, and quantum critical phenomena, respectively.

\item[Goldenfeld, N.] \emph{Lectures on Phase Transitions and the Renormalization Group}. Frontiers in Physics, Addison-Wesley, 1992. A pedagogical companion to Cardy with extensive worked examples; particularly clear on the conceptual foundations of universality and the meaning of fixed points.

\item[Sachdev, S.] \emph{Quantum Phase Transitions}, 2nd ed. Cambridge University Press, 2011. The definitive reference on quantum critical phenomena. Chapter~5 covers the transverse-field Ising model in detail, including the exact Jordan-Wigner solution used to obtain the gap shown in Figure~\ref{fig:tfim_gap}.

\item[Wilson, K.~G., and Kogut, J.] ``The renormalization group and the $\varepsilon$-expansion.'' \emph{Physics Reports} \textbf{12}, 75--199 (1974). \href{https://doi.org/10.1016/0370-1573(74)90023-4}{doi:10.1016/0370-1573(74)90023-4}. The classic review by the architect of the modern RG, written shortly after Wilson's Nobel-winning work. Demanding but rewarding.

\item[Kadanoff, L.~P.] ``Scaling laws for Ising models near $T_c$.'' \emph{Physics Physique Fizika} \textbf{2}, 263--272 (1966). \href{https://doi.org/10.1103/PhysicsPhysiqueFizika.2.263}{doi:10.1103/PhysicsPhysiqueFizika.2.263}. The original block-spin paper that introduced the scaling hypothesis used in this chapter.

\item[Kramers, H.~A., and Wannier, G.~H.; and Onsager, L.] ``Statistics of the two-dimensional ferromagnet,'' \emph{Physical Review} \textbf{60}, 252--262 (1941), \href{https://doi.org/10.1103/PhysRev.60.252}{doi:10.1103/PhysRev.60.252}; and ``Crystal statistics. I. A two-dimensional model with an order-disorder transition,'' \emph{Physical Review} \textbf{65}, 117--149 (1944), \href{https://doi.org/10.1103/PhysRev.65.117}{doi:10.1103/PhysRev.65.117}. The duality argument that locates $T_c$ exactly, followed three years later by Onsager's full solution; together they are the source of the 2D Ising critical exponents quoted in this chapter.
\end{description}

%% file: book_problems/ch12_problems.tex
\section{Problems}
\setcounter{hwproblem}{0}

\problem{1D Ising Decimation and RG Flow}
Consider the one-dimensional Ising model with $H = -J\sum_{i=1}^{N-1} s_i s_{i+1}$ where $s_i = \pm 1$ and $J > 0$. The partition function is $Z = \sum_{\{s_i\}} e^{-\beta H}$.
\begin{enumerate}[label=(\alph*)]
    \item Show that $Z = \sum_{\{s_i\}} \prod_{i=1}^{N-1} e^{K s_i s_{i+1}}$ where $K = \beta J$.
    \item Focus on the even-site spin $s_2$ between $s_1$ and $s_3$. Perform the sum $\sum_{s_2} e^{K s_1 s_2} e^{K s_2 s_3}$ for the cases $s_1 = s_3$ and $s_1 = -s_3$ separately.
    \item Express the result as $A\,e^{K' s_1 s_3}$ and determine the RG recursion: $K' = \frac{1}{2}\ln[\cosh(2K)]$.
    \item Show that $K^* = 0$ is a stable fixed point by expanding for small $K$. Verify that $K' < K$ for $0 < K < 1$.
    \item Argue that $K^* = \infty$ is also a fixed point. Is it stable or unstable?
    \item Starting from $K = 1.0$, iterate the recursion numerically to compute $K'$, $K''$, $K'''$. In which direction does the flow proceed?
\end{enumerate}

\problem{Critical Correlations and Anomalous Dimension}
At a critical point, the free-field correlation function in $d$ spatial dimensions is $G(\mathbf{r}) = \int \frac{d^d k}{(2\pi)^d}\frac{e^{i\mathbf{k}\cdot\mathbf{r}}}{k^2}$.
\begin{enumerate}[label=(\alph*)]
    \item Use the substitution $\mathbf{k} = \mathbf{q}/r$ to show that $G(r) \sim r^{-(d-2)}$.
    \item Evaluate the exponent for $d = 1, 2, 3, 4$. For which dimension does the correlator grow with distance?
    \item The interacting critical correlator is $G(r) \sim r^{-(d-2+\eta)}$ where $\eta$ is the anomalous dimension. For the 2D Ising model, $\eta = 1/4$. What is the power-law exponent?
    \item Away from criticality, correlations decay as $G(r) \sim e^{-r/\xi}$ with $\xi \sim |t|^{-\nu}$. Explain why power-law decay emerges precisely when $\xi$ diverges.
\end{enumerate}

\problem{Scaling Relations and Critical Exponents}
The scaling hypothesis asserts that under a block-spin transformation with rescaling factor $b$, the free energy density satisfies $f(t, h) = b^{-d}f(b^{y_t}t, b^{y_h}h)$ where $t = (T - T_c)/T_c$ is the reduced temperature, $h$ is the field, and $y_t, y_h$ are RG exponents.
\begin{enumerate}[label=(\alph*)]
    \item Set $b = |t|^{-1/y_t}$ and use $\xi(t) = b\,\xi(b^{y_t}t)$ to derive $\nu = 1/y_t$.
    \item Differentiate $f$ once with respect to $h$ to obtain the magnetization $m(t,h) = b^{y_h-d}m(b^{y_t}t, b^{y_h}h)$. Derive $\beta = (d - y_h)/y_t$.
    \item Differentiate twice to obtain the susceptibility and derive $\gamma = (2y_h - d)/y_t$.
    \item Differentiate twice with respect to $t$ to derive $\alpha = 2 - d/y_t$. Verify the hyperscaling relation $2 - \alpha = d\nu$.
    \item For the 2D Ising model with $\nu = 1$ and $\beta = 1/8$, extract $y_t$ and $y_h$. Compute $\gamma$, $\delta = y_h/(d - y_h)$, and $\alpha$.
    \item Verify the Rushbrooke inequality $\alpha + 2\beta + \gamma \geq 2$.
\end{enumerate}

\problem{Kramers-Wannier Duality and the Critical Temperature}
The 2D Ising partition function can be expanded in low temperature (variable $u = e^{-2K}$) or high temperature (variable $v = \tanh K$) forms, related by duality: $e^{-2K^*} = \tanh K$.
\begin{enumerate}[label=(\alph*)]
    \item Explain physically what $u = e^{-2K}$ counts. Why is it small at low temperature?
    \item Show that the duality maps high temperature (small $K$) to low temperature (large $K^*$) and vice versa.
    \item Argue that a unique phase transition must occur at the self-dual point $K_c = K_c^*$.
    \item From the self-duality condition $e^{-2K_c} = \tanh K_c$, derive $\sinh(2K_c) = 1$ and show that $\frac{k_B T_c}{J} = \frac{2}{\ln(1 + \sqrt{2})} \approx 2.269$.
    \item The 1D Ising model is self-dual at $K = 0$ (not finite temperature). Explain why this is consistent with no phase transition at finite $T$.
\end{enumerate}

\problem{Quantum Phase Transition in the Transverse-Field Ising Model}
The transverse-field Ising model has Hamiltonian $H = -J\sum_i \sigma_i^z\sigma_{i+1}^z - h\sum_i \sigma_i^x$.
\begin{enumerate}[label=(\alph*)]
    \item For $h = 0$, describe the ground state(s) and compute $\langle\sigma_i^z\rangle$.
    \item For $h \gg J$, find the ground state and show that $\langle\sigma_i^z\rangle = 0$.
    \item Why does the non-commutativity of the two terms in $H$ lead to a quantum phase transition at $T = 0$?
    \item The quantum-classical correspondence maps a $d$-dimensional quantum system at a QPT onto a $(d+z)$-dimensional classical system, where $z$ is the dynamical critical exponent. For the 1D transverse-field Ising model with $z = 1$, what classical model results? What universality class?
    \item The energy gap closes as $\Delta \sim |h - h_c|^{z\nu}$. Using $z = 1$ and $\nu = 1$, compute the gap exponent and explain what this means physically.
\end{enumerate}

\problem{Ising Magnetization at $T = 0$ with Transverse Field}
\begin{enumerate}[label=(\alph*)]
    \item For the transverse-field Ising model with $J = 1$ and varying field $h$, the ground state magnetization is $m(h) = \langle\sigma^z\rangle$. At $h = 0$, $m(0) = 1$. At the quantum critical point $h_c = 1$, $m(h_c) = 0$. Show that $m(h) \sim (1 - h)^\beta$ near $h_c$ with $\beta = 1/8$ (matching 2D Ising).
    \item Compute $m(h)$ numerically (or analytically for the 1D chain) at $h = 0.5, 0.8, 0.95, 0.99$ and verify the power-law scaling near $h_c = 1$.
    \item Near the critical point, the magnetization susceptibility $\chi = \partial m/\partial h$ diverges. Show that $\chi \sim |h - h_c|^{-\gamma}$ with $\gamma = 7/4$ (2D Ising).
    \item Compare the magnetization curves for different values of $J$ (the coupling strength). Does the critical field $h_c$ depend on $J$? How does the critical exponent $\beta$ depend on $J$?
    \item Explain in physical terms why increasing the transverse field $h$ reduces the ferromagnetic magnetization.
\end{enumerate}

\problem{RG Fixed-Point Linearization for Transverse-Field Ising}
\begin{enumerate}[label=(\alph*)]
    \item Near the quantum critical point of the transverse-field Ising model, the RG flow in the $(h, J)$ plane can be linearized around the critical point. Write the RG transformation as $h' = b^{y_h}h$, $J' = b^{y_J}J$ (or equivalently, $K_h' = b^{y_h}K_h$, $K_J' = b^{y_J}K_J$).
    \item At the critical point $(h_c, J_c) = (1, 1)$ (up to rescaling), the RG scaling exponents for the transverse-field Ising model are related to those of the 2D classical Ising model. Using the quantum-classical correspondence with $z = 1$, relate $y_h$ and $y_J$ to the classical exponents $y_t$ and $y_h$.
    \item The correlation length along the $h$ direction diverges as $\xi_h \sim (h - h_c)^{-\nu_h}$. Show that $\nu_h = 1/y_h$ and relate it to the classical exponents.
    \item Near the critical point, sketch the RG flow trajectories in the $(h, J)$ plane. Identify the directions of relevant, irrelevant, and marginal perturbations.
    \item A perturbation away from criticality (e.g., a small magnetic field $B\sigma^z$) is relevant or irrelevant depending on its sign (which direction it points in coupling space). Explain what this means for the long-distance physics.
\end{enumerate}

\problem{Correlation Length Critical Exponent from RG}
\begin{enumerate}[label=(\alph*)]
    \item The correlation length $\xi$ is the scale at which correlations decay. In terms of the lattice spacing $a$, after a block-spin transformation with rescaling factor $b$, the correlation length becomes $\xi' = \xi/b$ (measured in units of the new lattice spacing $ba$).
    \item If the reduced temperature $t = (T - T_c)/T_c$ transforms as $t' = b^{y_t}t$, show that $\xi(t) = b\,\xi(b^{y_t}t)$. By choosing $b = |t|^{-1/y_t}$, derive $\xi \sim |t|^{-\nu}$ with $\nu = 1/y_t$.
    \item For the 2D Ising model, $y_t = 1$ (the scaling exponent of temperature), so $\nu = 1$. Compute the correlation length at $t = 0.01$ (i.e., 1
    \item As $T \to T_c$, the correlation length diverges. Explain physically why arbitrarily large spatial structures appear at the critical point and why the system becomes scale-invariant.
    \item Compute the correlation length for a different universality class (e.g., the Ising model in a different dimension) and compare with the 2D result.
\end{enumerate}

\problem{Finite-Size Scaling Collapse}
At a critical point, finite-size systems of linear size $L$ exhibit scaling: $m(T, L) = L^{-\beta/\nu}f_m(|T - T_c|L^{1/\nu})$ where $f_m$ is a universal scaling function.
\begin{enumerate}[label=(\alph*)]
    \item Simulate the 1D or 2D Ising model at various temperatures near $T_c$ and system sizes $L$. Compute the magnetization $m(T, L)$ for each combination.
    \item Plot $m(T, L) \cdot L^{\beta/\nu}$ versus $|T - T_c|L^{1/\nu}$ on a log-log scale (using literature values $\beta = 1/8$, $\nu = 1$ for the 2D Ising model). Verify that the data from different system sizes collapse onto a single universal curve.
    \item The quality of the collapse is a test of the critical exponents. If the exponents are incorrect, the data will not collapse. Vary $\beta$ and $\nu$ slightly and show how the collapse deteriorates.
    \item Extract the universal scaling function $f_m(x)$ from the collapsed data and compare with theoretical predictions or other simulations.
    \item Use the finite-size scaling method to estimate $T_c$ from your simulation data (the temperature at which the collapse is best).
\end{enumerate}

\problem{Kosterlitz-Thouless Transition: Qualitative Picture}
\begin{enumerate}[label=(\alph*)]
    \item The 2D XY model (classical Heisenberg model with spins in the plane) has the Hamiltonian $H = -J\sum_{\langle i,j \rangle}\cos(\theta_i - \theta_j)$ where $\theta_i \in [0, 2\pi)$ is the angle of the spin at site $i$.
    \item At low temperature, spins are quasi-aligned, supporting long-range order and power-law correlations. At high temperature, spins are disordered. However, unlike the Ising model, the 2D XY model has a different transition mechanism: the Kosterlitz-Thouless (KT) transition.
    \item The key role is played by topological defects called vortices, which are point defects where the phase wind by $2\pi$ around a plaquette. Describe qualitatively what happens as temperature increases: (1) at $T < T_{KT}$, vortices are bound in pairs (vortex-antivortex pairs) with opposite winding; (2) at $T > T_{KT}$, vortices unbind and proliferate.
    \item The KT transition is characterized by a different universality class than the Ising model. The correlation length diverges as $\xi \sim \exp(C|T - T_{KT}|^{-1/2})$ (essential singularity) rather than power-law, and there is no conventional order parameter with a critical exponent.
    \item Sketch the phase diagram and the vortex density as a function of temperature. Explain why the transition is called ''topological.''
    \item How does the 2D XY model differ from the 2D Ising model in terms of symmetry and the nature of the phase transition? What role does the continuous symmetry play?
\end{enumerate}

\problem{Quantum Hall Plateau: RG Argument}
\begin{enumerate}[label=(\alph*)]
    \item In the quantum Hall regime, electrons in a strong perpendicular magnetic field $B$ form discrete Landau levels. The quantum Hall conductance is quantized in units of $e^2/h$. A plateau at conductance $\sigma = \nu e^2/h$ (for filling fraction $\nu$) arises from the insulating incompressible state.
    \item Disorder in the system broadens the Landau levels into bands and can cause transitions between plateaus. A key insight from RG is that the random network of extended states at energy $E_c$ (separating the localized states of different plateaus) is a critical two-dimensional electron system.
    \item At the transition (critical point), the localization length diverges: $\xi \sim |E - E_c|^{-\nu_c}$ with a critical exponent $\nu_c \approx 2.3$ (for the integer quantum Hall effect).
    \item The quantum Hall plateau is broad in energy (or magnetic field) because only a small window around $E_c$ supports extended states. States far from $E_c$ are localized, contributing nothing to the conductance.
    \item Sketch the density of states $\rho(E)$ and the localization length $\xi(E)$ across a Landau level. Indicate the plateau region (localized, $\sigma = \nu e^2/h$) and the critical point (extended, metallic).
    \item Explain qualitatively why the quantum Hall conductance is precisely quantized ($\nu e^2/h$) and why it is insensitive to disorder at the plateau centers. What role does the topological robustness play?
\end{enumerate}

\problem{Conformal Invariance and Operator Scaling Dimensions}
\begin{enumerate}[label=(\alph*)]
    \item At a critical point, the theory is scale-invariant and, in many cases (especially in 2D), has a larger symmetry: conformal invariance. This means the theory is invariant under angle-preserving transformations (conformal maps).
    \item In conformal field theory (CFT), every operator has a scaling dimension $\Delta$ such that under a scale transformation $x \to \lambda x$, the operator transforms as $\mathcal{O}(x) \to \lambda^{-\Delta}\mathcal{O}(\lambda x)$.
    \item For the 2D Ising model at criticality, the primary operators include the identity $I$ (dimension $\Delta = 0$), the energy density $\epsilon$ (dimension $\Delta = 1$), the magnetization $\sigma$ (dimension $\Delta = 1/8$), and the disorder operator $\mu$ (dimension $\Delta = 1/8$).
    \item The correlation function of two identical primary operators has the form $\langle\mathcal{O}(x)\mathcal{O}(0)\rangle \sim |x|^{-2\Delta}$. Explain how this relates to the power-law exponent in the problem on critical correlations.
    \item For the 2D Ising model, compute $\Delta_\sigma = 1/8$ from the known anomalous dimension $\eta = 1/4$ (using $\Delta = (d - 2 + \eta)/2$ for the Ising magnetization in $d$ dimensions). Verify that the resulting correlation exponent matches the literature.
    \item Explain why conformal invariance constrains the form of correlators and why it is a powerful tool for exact solutions in 2D.
\end{enumerate}